\documentclass[11pt,oneside]{book}

\usepackage[utf8]{inputenc}
\usepackage[T1]{fontenc}
\usepackage{lmodern}
\usepackage[a4paper,margin=1in]{geometry}
\usepackage{microtype}
\usepackage{amsmath,amssymb,amsfonts}
\usepackage{mathtools}
\usepackage{bm}
\usepackage{booktabs}
\usepackage{array}
\usepackage{tabularx}
\usepackage{multirow}
\usepackage{longtable}
\usepackage{xcolor}
\usepackage{graphicx}
\usepackage{tikz}
\usepackage{pgfplots}
\usepackage{tcolorbox}
\usepackage{listings}
\usepackage{enumitem}
\usepackage{needspace}
\usepackage{morewrites}
\usepackage[authoryear,round]{natbib}
\definecolor{boxblue}{RGB}{235,244,255}
\definecolor{deepblue}{RGB}{30,85,160}
\usepackage[colorlinks=true,linkcolor=blue!60!black,citecolor=blue!60!black,urlcolor=blue!60!black]{hyperref}

\usepackage{pdfpages}

\usetikzlibrary{arrows.meta,positioning,calc,fit,shapes.geometric,shapes.misc,decorations.pathreplacing,backgrounds}
\pgfplotsset{compat=1.18}
\tcbuselibrary{skins,breakable}

\newcolumntype{Y}{>{\centering\arraybackslash}X}
\newcolumntype{L}{>{\raggedright\arraybackslash}X}
\newcolumntype{R}{>{\raggedleft\arraybackslash}X}

\colorlet{deepblue}{blue!60!black}
\colorlet{deepred}{red!60!black}
\colorlet{deepgreen}{green!50!black}
\colorlet{boxblue}{blue!60!black}
\colorlet{boxgreen}{green!50!black}
\colorlet{boxred}{red!60!black}

\definecolor{codebg}{RGB}{248,248,248}
\definecolor{codeframe}{RGB}{210,210,210}
\definecolor{commentgreen}{RGB}{70,130,70}
\definecolor{keywordblue}{RGB}{0,70,150}
\definecolor{stringred}{RGB}{160,60,60}

\lstdefinestyle{pythonstyle}{
	language=Python,basicstyle=\ttfamily\small,
	keywordstyle=\color{keywordblue}\bfseries,
	commentstyle=\color{commentgreen},
	stringstyle=\color{stringred},
	showstringspaces=false,breaklines=true,
	frame=single,rulecolor=\color{codeframe},
	backgroundcolor=\color{codebg},tabsize=4,
	numbers=left,numberstyle=\tiny\color{gray},
	xleftmargin=1.2em,framexleftmargin=1.0em,captionpos=b}

\newtcolorbox{keybox}[1][Key idea]{colback=blue!4,colframe=blue!60!black,title=\textbf{#1},fonttitle=\bfseries,arc=2mm,boxrule=0.8pt}
\newtcolorbox{keyidea}[1][Key Idea]{colback=blue!4,colframe=blue!55!black,title=\textbf{#1},fonttitle=\bfseries,arc=2mm,boxrule=0.8pt}
\newtcolorbox{warningbox}[1][Warning]{colback=red!4,colframe=red!65!black,title=\textbf{#1},fonttitle=\bfseries,arc=2mm,boxrule=0.8pt}
\newtcolorbox{researchbox}[1][Research note]{colback=green!4,colframe=green!50!black,title=\textbf{#1},fonttitle=\bfseries,arc=2mm,boxrule=0.8pt}
\newtcolorbox{pitfallbox}[1][Pitfall]{colback=red!4,colframe=red!60!black,title=\textbf{#1},fonttitle=\bfseries,arc=2mm,boxrule=0.8pt}
\newtcolorbox{intuitionbox}[1][Intuition]{colback=green!3,colframe=green!40!black,title=\textbf{#1},fonttitle=\bfseries,arc=2mm,boxrule=0.8pt}
\newtcolorbox{examplebox}[1][Example]{colback=orange!4,colframe=orange!60!black,title=\textbf{#1},fonttitle=\bfseries,arc=2mm,boxrule=0.8pt}
\newtcolorbox{definitionbox}[1][Definition]{colback=blue!3,colframe=blue!50!black,title=\textbf{#1},fonttitle=\bfseries,arc=2mm,boxrule=0.8pt}
\newtcolorbox{definition}[1][Definition]{colback=blue!3,colframe=blue!50!black,title=\textbf{#1},fonttitle=\bfseries,arc=2mm,boxrule=0.8pt}
\newtcolorbox{algobox}[1][Algorithm]{colback=gray!3,colframe=black!65,title=\textbf{#1},fonttitle=\bfseries,arc=2mm,boxrule=0.8pt}
\newtcolorbox{mathbox}[1][]{colback=gray!3,colframe=gray!50!black,title=\textbf{#1},fonttitle=\bfseries,arc=2mm,boxrule=0.8pt}
\newtcolorbox{frontierbox}{colback=purple!4,colframe=purple!70!black,arc=2mm,boxrule=0.8pt}
\newtcolorbox{practicebox}{colback=orange!5,colframe=orange!70!black,title=\textbf{Practical note},fonttitle=\bfseries,arc=2mm,boxrule=0.8pt}
\newtcolorbox{implementationbox}[1][Implementation]{colback=cyan!3,colframe=cyan!50!black,title=\textbf{#1},fonttitle=\bfseries,arc=2mm,boxrule=0.8pt}
\newtcolorbox{notebox}[1][Note]{colback=yellow!6,colframe=yellow!60!black,title=\textbf{#1},fonttitle=\bfseries,arc=2mm,boxrule=0.8pt}

\providecommand{\E}{\mathbb{E}}
\providecommand{\R}{\mathbb{R}}
\providecommand{\Prob}{\mathbb{P}}

\providecommand{\one}{\mathbf{1}}
\providecommand{\Var}{\operatorname{Var}}
\providecommand{\CVaR}{\mathrm{CVaR}}
\providecommand{\KL}{\operatorname{KL}}
\providecommand{\IQM}{\mathrm{IQM}}
\providecommand{\given}{\mid}
\providecommand{\clip}{\operatorname{clip}}
\providecommand{\grad}{\nabla}

\providecommand{\sg}{\operatorname{stopgrad}}
\providecommand{\se}{\mathrm{SE}}

\providecommand{\Normal}{\mathcal{N}}
\providecommand{\Sset}{\mathcal{S}}
\providecommand{\Sspace}{\mathcal{S}}
\providecommand{\Aset}{\mathcal{A}}

\providecommand{\A}{\mathcal{A}}
\providecommand{\D}{\mathcal{D}}
\providecommand{\B}{\mathcal{B}}

\DeclareMathOperator*{\argmax}{arg\,max}
\DeclareMathOperator*{\argmin}{arg\,min}

\title{\textbf{Deep Reinforcement Learning}\\
	\large From First Principles to Reasoning Models}
\author{Ghoshana Bista, PhD\\
	\small Université Côte d'Azur, LEAT Laboratory}
\date{2026}

\begin{document}
\frontmatter

\newgeometry{margin=0pt}
\begin{titlepage}
    \thispagestyle{empty}
    \noindent
    \includegraphics[width=\paperwidth,height=\paperheight]{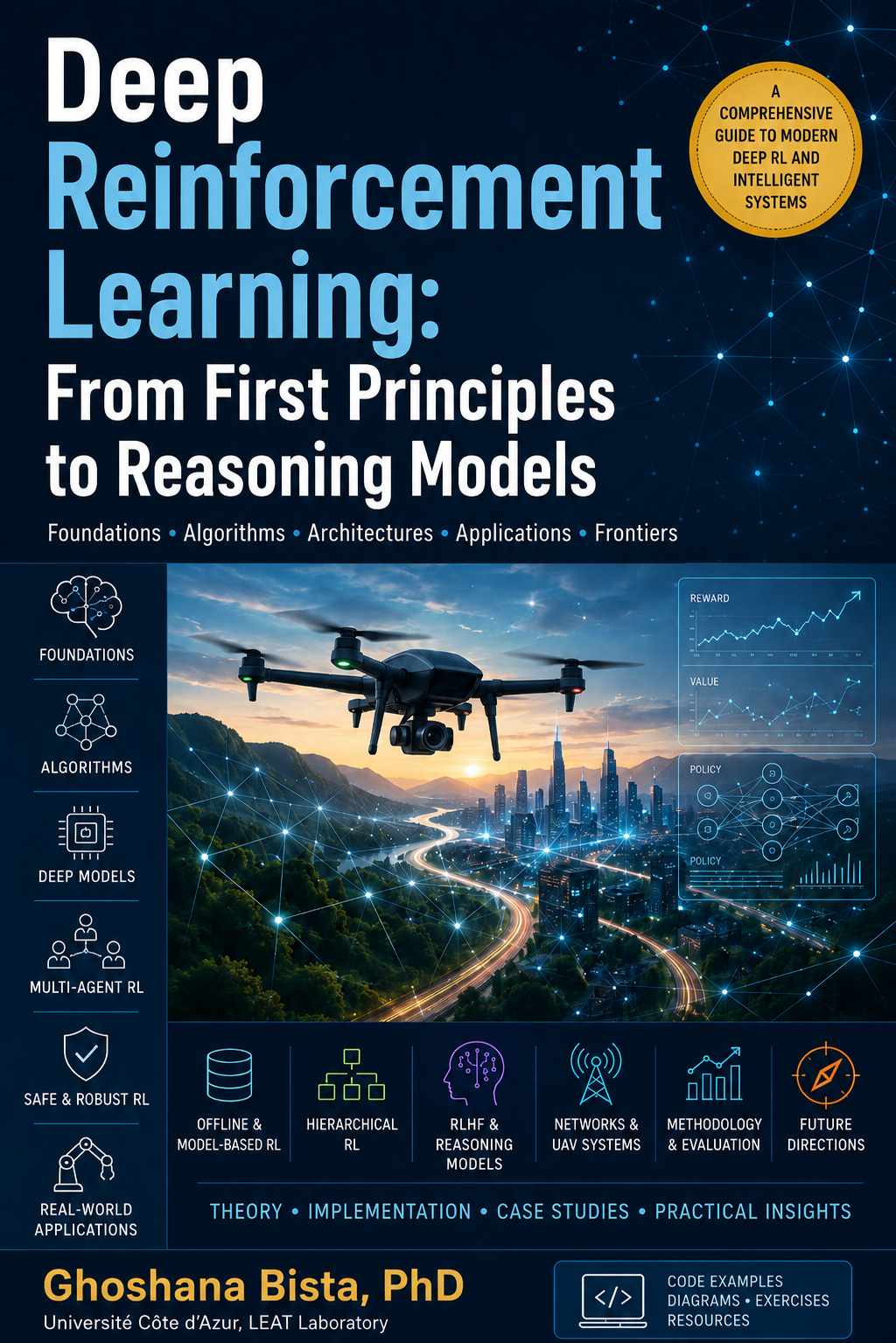}
\end{titlepage}
\restoregeometry
\clearpage
\cleardoublepage
\thispagestyle{empty}
\null
\cleardoublepage
%\tableofcontents
\chapter*{Copyright and License}
\addcontentsline{toc}{chapter}{Copyright and License}

\noindent
Copyright \textcopyright\ 2026 Ghoshana Bista.

\medskip

\noindent
This book is licensed under the \textbf{Creative Commons Attribution 4.0 International License (CC BY 4.0)}.

\medskip

\noindent
You are free to share and adapt this work for any purpose, provided that appropriate credit is given, a link to the license is provided, and any changes made are indicated.

\medskip

\noindent
To view a copy of this license, visit:
\texttt{https://creativecommons.org/licenses/by/4.0/}
\mainmatter
\cleardoublepage
\thispagestyle{empty}

\vspace*{0.22\textheight}

\begin{center}
{\Huge\bfseries Dedication}
\end{center}

\vspace{1.5cm}

\begin{flushright}
\itshape
To my wife,\\
for her patience, encouragement, and belief.\\[0.5cm]
To every teacher, colleague, and student\\
who helped shape this journey.\\[0.5cm]
And to all readers who continue exploring,\\
questioning, and building intelligent systems.
\end{flushright}

\cleardoublepage
\chapter*{Preface}
\addcontentsline{toc}{chapter}{Preface}

Deep reinforcement learning has grown from a specialized research topic into one of the central frameworks of modern artificial intelligence. What began with dynamic programming, temporal-difference learning, and tabular control has expanded into a rich ecosystem of value-based learning, actor-critic methods, model-based reinforcement learning, offline reinforcement learning, multi-agent systems, safe reinforcement learning, reinforcement learning from human feedback, and reinforcement learning for reasoning-oriented AI systems.

This book was written to provide a structured path through that landscape.

Its goal is not only to explain algorithms, but also to explain why they exist, what problems they solve, where they fail, and how they connect to real systems. Throughout the book, reinforcement learning is treated not merely as a collection of update rules, but as a framework for sequential decision-making under uncertainty, delayed consequences, constraints, and limited information.

The book is intentionally hybrid in style. It is part textbook, part research-oriented guide, and part systems perspective. The early chapters build the classical foundations: reinforcement learning, Markov decision processes, dynamic programming, Monte Carlo learning, temporal-difference learning, and the transition from tabular methods to deep reinforcement learning. The middle chapters develop the major deep reinforcement learning families, including DQN, improved value-based methods, policy gradients, actor-critic methods, PPO, SAC, model-based RL, MuZero, offline RL, and sequence-modeling approaches. The later chapters broaden the perspective toward multi-agent learning, hierarchical control, safe reinforcement learning, RLHF, reasoning models, communication networks, UAV systems, implementation pipelines, experimental methodology, failure analysis, and future research directions.

A recurring theme of the book is that reinforcement learning becomes most meaningful when placed in real decision systems. For this reason, UAV-assisted communication networks, SD-WAN traffic engineering, safe control, and reasoning-based AI systems appear throughout as running examples. These examples help connect mathematical ideas to engineering realities such as partial observability, conflicting objectives, safety constraints, deployment drift, and evaluation under uncertainty.

This book is written for advanced students, researchers, and engineers. It assumes some comfort with probability, linear algebra, calculus, and programming, but it does not assume prior mastery of reinforcement learning. Readers who are new to the area may proceed from Chapter 1 onward in order. Readers already familiar with the basics may move directly to later chapters according to their interests.

The most important aim of this book is clarity. Deep reinforcement learning is often presented either too abstractly, with equations disconnected from intuition, or too operationally, with code disconnected from concepts. This book tries to bridge those two views. Each chapter therefore aims to connect intuition, mathematics, implementation, and application.

No single book can settle a field as active as this one. Reinforcement learning continues to evolve rapidly, especially in the areas of world models, reasoning systems, safe control, multi-agent learning, and human-aligned AI. Still, the central questions remain remarkably consistent: how should an intelligent system act, how should it learn from consequences, how should it balance short-term and long-term objectives, and how should it remain reliable under uncertainty and constraints?

If this book helps readers move from isolated algorithm names to a coherent understanding of the field, then it will have achieved its purpose.

\vspace{1cm}

\noindent
Ghoshana Bista\\
Université Côte d'Azur, LEAT Laboratory\\
2026

\chapter*{Foreword}
\addcontentsline{toc}{chapter}{Foreword}

Deep reinforcement learning has evolved from a specialized research topic into one of the central frameworks of modern artificial intelligence. Its influence now extends far beyond classical control and game-playing benchmarks into robotics, communication systems, safe decision-making, large-scale multi-agent coordination, and reasoning-oriented AI. As the field has expanded, so too has the need for resources that do more than present isolated algorithms. Readers increasingly need a structured and coherent view of how the foundations connect to the modern landscape.

This book responds to that need in a thoughtful and ambitious way. It takes the reader from the first principles of reinforcement learning and Bellman-based reasoning to the major algorithmic families of deep reinforcement learning, including value-based methods, actor-critic methods, model-based learning, offline reinforcement learning, multi-agent learning, hierarchical reinforcement learning, and safe reinforcement learning. Just as importantly, it extends the discussion toward modern topics such as reinforcement learning from human feedback, reasoning models, and real-world cyber-physical systems.

One of the strengths of this book is that it does not treat reinforcement learning as a purely abstract mathematical subject. Instead, it repeatedly connects theory to practical systems, especially in the context of UAV-assisted communication networks, SD-WAN control, and safety-critical intelligent systems. This makes the text especially valuable for readers who want not only conceptual clarity, but also a realistic understanding of where reinforcement learning succeeds, where it struggles, and how it is evaluated in practice.

Ghoshana Bista brings to this work a perspective shaped by both theoretical depth and systems-oriented research. The result is a book that is broad in scope, careful in structure, and highly relevant to the current evolution of the field. It will be useful to graduate students, researchers, and practitioners who seek a guided path through one of the most dynamic areas of AI.

I am pleased to recommend this book to readers interested in the foundations, development, and future of deep reinforcement learning.

\vspace{1cm}

\noindent
\textbf{Sudip Lama}\\
Principal Software Engineering Manager\\
Microsoft

\chapter*{How to Use This Book}
\addcontentsline{toc}{chapter}{How to Use This Book}

This book is designed as a hybrid text: it is both a structured introduction to deep reinforcement learning and a research-oriented guide to modern reinforcement-learning systems. It can therefore be read in more than one way.

Some readers will want a progressive path from the basics of reinforcement learning to modern deep reinforcement learning algorithms. Others may already know the foundations and want to focus on offline RL, multi-agent RL, safe RL, RLHF, reasoning models, or communication-network applications. This section explains how to navigate the book effectively.

\section*{Who this book is for}
This book is written primarily for:
\begin{itemize}[leftmargin=*]
    \item graduate students and PhD students studying reinforcement learning, machine learning, control, robotics, networking, or AI systems;
    \item researchers who want a structured overview from classical RL foundations to modern 2025--2026 research directions;
    \item engineers and practitioners building decision-making systems in robotics, UAVs, communication networks, safe control, and AI-assisted operations;
    \item advanced self-learners who want both conceptual understanding and practical implementation guidance.
\end{itemize}

The book assumes some familiarity with probability, linear algebra, calculus, and programming. However, it does not assume prior mastery of reinforcement learning.

\section*{How the book is organized}
The book is divided into seven parts.

\begin{enumerate}[leftmargin=*]
    \item \textbf{Before Deep Reinforcement Learning} introduces reinforcement learning, Markov decision processes, and the classical foundations of value learning.
    \item \textbf{The Birth of Deep Reinforcement Learning} explains why tabular and classical methods are not enough, then develops DQN and improved value-based methods.
    \item \textbf{Policy Gradients and Actor-Critic Methods} covers direct policy optimization, REINFORCE, actor-critic learning, PPO, and SAC.
    \item \textbf{Planning, Models, and Offline Learning} develops model-based RL, MuZero, offline RL, and sequence-modeling approaches.
    \item \textbf{Multi-Agent, Hierarchical, and Safe DRL} covers MARL, hierarchical RL, and safe reinforcement learning.
    \item \textbf{DRL in Modern AI Systems} extends the discussion to RLHF, reasoning models, and communication/networked systems.
    \item \textbf{How to Build and Evaluate DRL Systems} focuses on implementation, evaluation methodology, common failure modes, and future research directions.
\end{enumerate}

\section*{Suggested reading paths}

\subsection*{Path A: First full introduction}
If you are new to reinforcement learning, read the book in order:
\begin{center}
Chapters 1--11
\end{center}
This path gives the cleanest progression from intuition to core deep RL algorithms.

\subsection*{Path B: Foundations plus modern methods}
If you want both the basics and the most important modern algorithmic families, read:
\begin{center}
Chapters 1--18
\end{center}
This path covers the core of classical RL, DQN, actor-critic methods, model-based RL, offline RL, MARL, HRL, and safe RL.

\subsection*{Path C: RL for language models and modern AI systems}
If your main interest is RLHF, reasoning models, and modern AI pipelines, start with:
\begin{center}
Chapters 1--11, then Chapters 19--20, and finally Chapters 22--25
\end{center}
This path gives the algorithmic foundations first, then moves into language-model alignment, reasoning-oriented RL, implementation, evaluation, and failure analysis.

\subsection*{Path D: Networks, UAVs, and cyber-physical systems}
If you are most interested in communication systems, UAVs, SD-WAN, and safe control, read:
\begin{center}
Chapters 1--11, then Chapters 16--18, 21--25
\end{center}
This path emphasizes multi-agent learning, hierarchical control, safety, networking applications, system implementation, and evaluation methodology.

\subsection*{Path E: Research-oriented fast track}
If you already know the basics of RL and want the research-heavy parts, begin with:
\begin{center}
Chapters 12--25
\end{center}
Use Chapters 1--11 as reference chapters when needed.

\section*{How each chapter is designed}
Most chapters follow a common structure:
\begin{itemize}[leftmargin=*]
    \item a short overview explaining why the topic matters;
    \item the main concepts and mathematical foundations;
    \item practical interpretation and common pitfalls;
    \item implementation-oriented code blocks;
    \item domain examples, often based on UAVs, networks, or safe control;
    \item exercises and a short bridge to the next chapter.
\end{itemize}

The intention is to connect four levels of understanding:
\begin{center}
\textit{intuition \(\rightarrow\) mathematics \(\rightarrow\) implementation \(\rightarrow\) application}
\end{center}

\section*{How to read the code}
The code blocks in this book are designed to be educational and modular. Some are fully usable small examples, while others are compact research-oriented sketches meant to illustrate an idea clearly rather than serve as drop-in software packages.

Readers should therefore treat the listings as:
\begin{itemize}[leftmargin=*]
    \item minimal conceptual implementations;
    \item templates for experimentation;
    \item bridges between mathematical updates and real code.
\end{itemize}

\section*{A note on notation}
The notation is kept as consistent as possible across chapters, but some later chapters introduce additional symbols specific to offline RL, sequence modeling, MARL, safe RL, RLHF, or reasoning systems. In general:
\begin{itemize}[leftmargin=*]
    \item \(s\) denotes a state or observation;
    \item \(a\) denotes an action;
    \item \(r\) denotes a reward;
    \item \(\pi\) denotes a policy;
    \item \(V\) and \(Q\) denote value functions;
    \item \(\gamma\) denotes the discount factor.
\end{itemize}

Later chapters extend this language to costs, constraints, options, multiple agents, latent skills, preference rewards, and verifier-based objectives.

\section*{What makes this book different}
This book is not only about benchmark algorithms. It is also about how reinforcement learning behaves in real decision systems. Throughout the text, the discussion repeatedly returns to:
\begin{itemize}[leftmargin=*]
    \item delayed consequences and long-horizon reasoning;
    \item partial observability and uncertainty;
    \item reward design and reward failure;
    \item safety constraints and action filtering;
    \item deployment realism, evaluation, and reproducibility;
    \item communication networks, UAV systems, and modern AI agents.
\end{itemize}

The aim is to help the reader move beyond isolated algorithm names and toward a coherent systems-level understanding of deep reinforcement learning.

\section*{Final advice to the reader}
Do not try to memorize every algorithm immediately. Focus first on the recurring ideas:
\begin{itemize}[leftmargin=*]
    \item Bellman consistency;
    \item bootstrapping;
    \item policy improvement;
    \item exploration versus exploitation;
    \item approximation and instability;
    \item constraints, safety, and evaluation.
\end{itemize}

If those ideas become clear, the later chapters will feel like connected extensions rather than disconnected techniques.

This book is best read actively: derive equations, run small code examples, compare algorithms, question assumptions, and relate each method to a real decision-making problem. Reinforcement learning is ultimately learned best in the same way it operates: through interaction, feedback, and refinement.
\cleardoublepage
\thispagestyle{empty}
\null
\cleardoublepage
	\tableofcontents
      
	\part{Before Deep Reinforcement Learning}
\chapter{What Is Reinforcement Learning?}
\label{ch:what_is_rl}

\section*{Chapter Overview}
\addcontentsline{toc}{section}{Chapter Overview}

This chapter introduces reinforcement learning as the study of learning to act through interaction. The goal is to build intuition before the mathematical formalism of Markov decision processes. We start from the agent--environment loop, then explain states, actions, rewards, returns, policies, value functions, exploration, and the transition from classical reinforcement learning to deep reinforcement learning.

\begin{tcolorbox}[
	title={Main idea of this chapter},
	colback=blue!4,
	colframe=blue!60!black,
	fonttitle=\bfseries,
	arc=2mm,
	boxrule=0.8pt
	]
	Reinforcement learning is not only prediction. It is sequential decision-making under uncertainty. An agent acts, receives feedback, and improves its future behavior by learning from consequences.
\end{tcolorbox}

\section*{Learning Objectives}
\addcontentsline{toc}{section}{Learning Objectives}

After reading this chapter, the reader should be able to:
\begin{enumerate}[leftmargin=*]
	\item explain the basic reinforcement learning interaction loop;
	\item distinguish agent, environment, state, action, reward, return, policy, and value function;
	\item explain why reinforcement learning differs from supervised and unsupervised learning;
	\item describe why delayed consequences make RL different from one-step prediction;
	\item understand the exploration--exploitation dilemma;
	\item explain why deep neural networks allow RL to scale beyond small tabular problems;
	\item connect the running UAV network-control example to states, actions, rewards, and constraints;
	\item prepare for the MDP formalism introduced in Chapter~2.
\end{enumerate}

\section{Why Reinforcement Learning Matters}

Reinforcement learning is one of the most important ideas in artificial intelligence because it studies a problem that ordinary prediction cannot fully solve: how an intelligent system should act.

A supervised learning model predicts an output from an input. A classifier predicts whether an image contains a cat or a dog. A regression model predicts tomorrow's electricity price. A language model predicts the next token. These tasks are extremely powerful, but they are not yet the full problem of intelligence. An intelligent system must often choose actions, observe the consequences, and adapt its future behavior.

Reinforcement learning, usually abbreviated as RL, studies this kind of learning. An agent interacts with an environment. At each time step, the agent observes the current situation, chooses an action, receives a reward signal, and moves to a new situation. The goal is not merely to choose actions that produce immediate reward. The goal is to learn behavior that maximizes long-term cumulative reward \citep{sutton2018reinforcement}.

This idea is simple, but it creates a very different learning problem. The agent does not passively receive a fixed dataset. Its actions influence the data it will see later. A bad decision can change future states. A good decision may require short-term sacrifice. The same action may have different effects in different contexts. Therefore, RL is not only about prediction. It is about sequential decision-making.

This is why RL became a central framework for problems such as robotics, game playing, autonomous driving, communication networks, resource allocation, recommendation systems, and, more recently, alignment and reasoning in large language models. The field is historically rooted in dynamic programming, optimal control, psychology, neuroscience, and machine learning. Sutton and Barto's reinforcement learning framework became one of the foundations of modern AI, and their contributions were recognized by the 2024 ACM A.M. Turing Award \citep{acm2024turing}.

In this book, we are especially interested in deep reinforcement learning, or DRL. DRL combines reinforcement learning with deep neural networks. Classical RL can solve small problems using tables. DRL allows agents to learn from images, high-dimensional sensor data, complex network measurements, and large-scale trajectories. This shift transformed RL from a mainly theoretical and small-scale framework into a practical tool for complex AI systems.

\section{The Basic Idea: Learning by Interaction}

The simplest way to understand reinforcement learning is to imagine a child learning to ride a bicycle. Nobody gives the child a complete mathematical model of balance, friction, steering, and body dynamics. Instead, the child tries actions, observes consequences, falls, adjusts, and gradually improves. The feedback is not a clean label like ``the correct handlebar angle is 7 degrees.'' The feedback is experiential: staying balanced is good; falling is bad; turning too sharply has consequences; moving too slowly makes balance difficult.

An RL agent learns in a similar way. It does not need a labeled dataset of correct actions. It needs an environment where it can act and receive feedback.

A basic RL problem contains five essential elements:
\begin{enumerate}
	\item \textbf{Agent} --- the learner or decision-maker.
	\item \textbf{Environment} --- the external system with which the agent interacts.
	\item \textbf{State or observation} --- information describing the current situation.
	\item \textbf{Action} --- a decision made by the agent.
	\item \textbf{Reward} --- a numerical signal that evaluates the immediate consequence of the action.
\end{enumerate}

At time step $t$, the agent observes a state $S_t$, chooses an action $A_t$, receives a reward $R_{t+1}$, and reaches a new state $S_{t+1}$. This interaction repeats over time.

The agent's objective is to learn a strategy that produces high long-term return. The word ``long-term'' is essential. Reinforcement learning is not the same as greedily maximizing the next reward. Many intelligent behaviors require delayed gratification. A robot may need to move away from a target to avoid an obstacle. A UAV may need to leave a high-demand area to recharge. A network controller may need to reroute traffic before congestion becomes visible. A language model trained for reasoning may need to produce intermediate reasoning steps before reaching a final answer.

Thus, RL is the study of learning under delayed consequences.

\section{The Agent--Environment Loop}

The agent--environment loop is the central picture of reinforcement learning.

At each discrete time step:
\begin{enumerate}
	\item The environment provides the agent with state $S_t$.
	\item The agent selects action $A_t$ according to its policy.
	\item The environment responds with reward $R_{t+1}$.
	\item The environment transitions to a new state $S_{t+1}$.
	\item The cycle repeats.
\end{enumerate}

In compact form,
\begin{equation}
	S_t \rightarrow A_t \rightarrow (R_{t+1},S_{t+1}).
\end{equation}
The complete interaction sequence is called a trajectory or episode:
\begin{equation}
	S_0,A_0,R_1,S_1,A_1,R_2,S_2,A_2,R_3,\ldots
\end{equation}

\begin{figure}[t]
	\centering
	\begin{tikzpicture}[
		node distance=3.2cm,
		box/.style={draw, rounded corners, thick, minimum width=3.0cm, minimum height=1.1cm, align=center},
		arrow/.style={-{Latex[length=3mm]}, thick}
		]
		\node[box] (agent) {Agent\\Policy $\pi$};
		\node[box, right=of agent] (env) {Environment\\Dynamics $p$};
		
		\draw[arrow] (agent.north east) .. controls +(0.8,1.0) and +(-0.8,1.0) ..
		node[above, align=center] {action\\$A_{t}$} (env.north west);
		
		\draw[arrow] (env.south west) .. controls +(-0.8,-1.0) and +(0.8,-1.0) ..
		node[below, align=center] {state/reward\\$S_{t+1},R_{t+1}$} (agent.south east);
		
		\node[below=2.6cm of $(agent)!0.5!(env)$, align=center]
		{The agent acts, the environment responds,\\and learning improves future decisions.};
	\end{tikzpicture}
	% FIX 2: changed \cite to \citep to match natbib author-year style used everywhere else
	\caption{Basic agent--environment interaction loop in reinforcement learning. At time $t$, the agent selects action $A_{t}$; the environment returns reward $R_{t+1}$ and next state $S_{t+1}$ \citep{sutton2018reinforcement,puterman1994markov}.}
	\label{fig:agent_environment_loop}
\end{figure}

In many problems, an episode has a natural end. A game ends when one player wins or loses. A robot episode ends when the robot reaches the goal or falls. A UAV mission episode may end after a fixed simulation time, battery depletion, or mission completion. In continuing tasks, there may be no natural terminal state; the agent acts indefinitely, as in traffic control, data-center cooling, or continuous network management.

A key assumption in many RL formulations is that the current state contains all necessary information for decision-making. This is called the Markov property. If the Markov property holds, then the future depends on the present state and action, not on the full history:
\begin{equation}
	\Prob(S_{t+1},R_{t+1}\mid S_0,A_0,\ldots,S_t,A_t)=\Prob(S_{t+1},R_{t+1}\mid S_t,A_t).
\end{equation}
When this assumption holds, the problem can be formalized as a Markov Decision Process, or MDP \citep{puterman1994markov}. MDPs are the mathematical foundation of classical reinforcement learning. Later chapters will study MDPs deeply, but % FIX 3: removed awkward self-reference "Chapter~\ref{ch:what_is_rl}"
this chapter only needs the main intuition: the state should summarize everything the agent needs to choose good actions.

In real systems, the agent often does not observe the full true state. A drone may not know the exact future user mobility. A network controller may not observe all hidden traffic patterns. A robot may only see part of the room. These are partially observable problems. In such cases, the agent may need memory, recurrent networks, belief estimation, or history-based policies.

\section{Reinforcement Learning Versus Supervised and Unsupervised Learning}

To understand reinforcement learning clearly, it helps to compare it with supervised and unsupervised learning.

\subsection{Supervised Learning}
In supervised learning, we have input-output examples,
\begin{equation}
	(x_1,y_1),(x_2,y_2),\ldots,(x_n,y_n).
\end{equation}
The model learns a mapping from $x$ to $y$. For example, given medical features, predict disease risk. Given an image, predict its class. Given a sentence, predict its translation. The important point is that supervised learning receives the correct or target answer during training.

\subsection{Unsupervised Learning}
In unsupervised learning, the model receives data without labels. It may learn clusters, latent representations, density models, or compressed descriptions. The goal is not direct action but structure discovery.

\subsection{Reinforcement Learning}
In reinforcement learning, the agent does not receive the correct action. It receives reward after acting. The reward may be delayed, sparse, noisy, or incomplete.

This creates several unique difficulties:
\begin{itemize}
	\item The agent must explore because it does not know which actions are good.
	\item The agent's actions affect future data.
	\item Feedback may arrive long after the important decision.
	\item The same action may be good in one state and bad in another.
	\item Training can be unstable because the data distribution changes as the policy changes.
\end{itemize}

\begin{figure}[t]
	\centering
	\resizebox{\textwidth}{!}{%
		\begin{tikzpicture}[
			box/.style={draw, rounded corners, thick, minimum width=4.0cm, minimum height=1.2cm, align=center},
			smallbox/.style={draw, rounded corners, minimum width=3.8cm, minimum height=0.75cm, align=center},
			arrow/.style={-{Latex[length=2.5mm]}, thick},
			node distance=0.75cm
			]
			\node[box] (sup) {Supervised Learning};
			\node[smallbox, below=of sup] (supdata) {Input $x$ + target $y$};
			\node[smallbox, below=of supdata] (supgoal) {Learn prediction $f(x)\approx y$};
			\node[box, right=1.8cm of sup] (unsup) {Unsupervised Learning};
			\node[smallbox, below=of unsup] (unsupdata) {Unlabeled data $x$};
			\node[smallbox, below=of unsupdata] (unsupgoal) {Discover structure or representation};
			\node[box, right=2.1cm of unsup] (rl) {Reinforcement Learning};
			\node[smallbox, below=of rl] (rldata) {Interaction: $S_t,A_t,R_{t+1}$};
			\node[smallbox, below=of rldata] (rlgoal) {Learn actions maximizing return};
			\draw[arrow] (sup) -- (supdata);
			\draw[arrow] (supdata) -- (supgoal);
			\draw[arrow] (unsup) -- (unsupdata);
			\draw[arrow] (unsupdata) -- (unsupgoal);
			\draw[arrow] (rl) -- (rldata);
			\draw[arrow] (rldata) -- (rlgoal);
		\end{tikzpicture}%
	}
	\caption{Supervised learning learns from labeled examples, unsupervised learning discovers structure in data, and reinforcement learning learns through action and feedback. Unlike supervised learning, RL does not usually receive the correct action as a training label \citep{sutton2018reinforcement}.}
	\label{fig:learning_paradigms}
\end{figure}
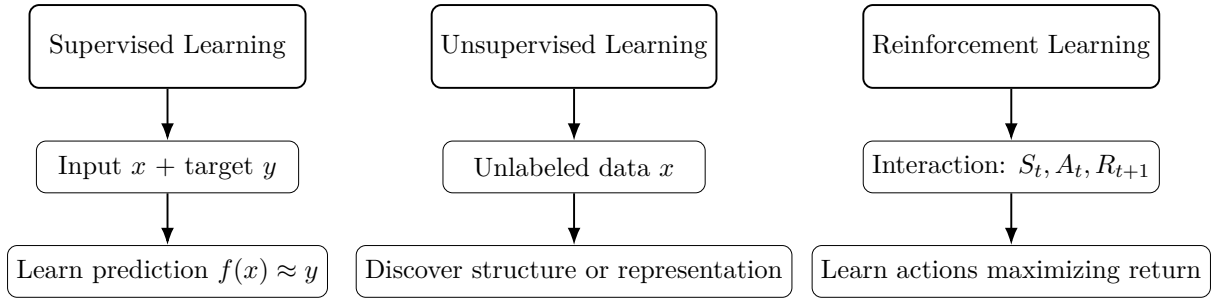

This is why RL is often harder than supervised learning. It is not only a learning problem; it is a control problem. Table~\ref{tab:learning_paradigms} summarizes the difference.

\begin{table}[t]
	\centering
	\caption{Comparison of supervised learning, unsupervised learning, and reinforcement learning.}
	\label{tab:learning_paradigms}
	\begin{tabular}{p{3.2cm}p{4.0cm}p{4.5cm}p{2.7cm}}
		\toprule
		Learning paradigm & Training signal & Main question & Example \\
		\midrule
		Supervised learning & Correct label or target & What should I predict? & Classify an image \\
		Unsupervised learning & No explicit label & What structure exists in the data? & Cluster users \\
		Reinforcement learning & Reward from interaction & What action should I take over time? & Control a drone \\
		\bottomrule
	\end{tabular}
\end{table}

\section{Rewards, Returns, and Delayed Consequences}

The reward is the immediate feedback signal. It tells the agent whether something good or bad happened after an action. However, the objective of RL is usually not to maximize immediate reward only. The objective is to maximize return.

The return is the cumulative discounted reward from time $t$ onward:
\begin{equation}
	G_t=R_{t+1}+\gamma R_{t+2}+\gamma^2R_{t+3}+\cdots.
\end{equation}
Equivalently,
\begin{equation}
	G_t=\sum_{k=0}^{\infty}\gamma^k R_{t+k+1}.
\end{equation}
Here $\gamma$ is the discount factor, where $0\le\gamma\le1$.

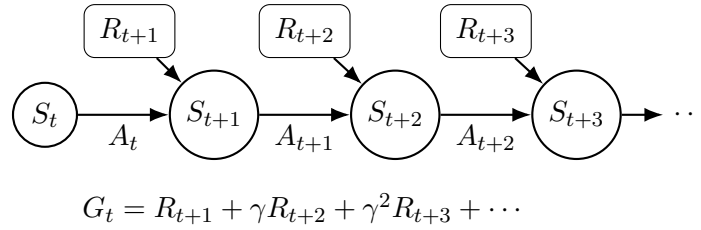
\begin{figure}[t]
	\centering
	\begin{tikzpicture}[
		time/.style={circle, draw, thick, minimum size=0.8cm},
		reward/.style={rectangle, draw, rounded corners, minimum width=1.2cm, minimum height=0.7cm, align=center},
		arrow/.style={-{Latex[length=2.5mm]}, thick},
		node distance=1.2cm
		]
		\node[time] (s0) {$S_t$};
		\node[time, right=of s0] (s1) {$S_{t+1}$};
		\node[time, right=of s1] (s2) {$S_{t+2}$};
		\node[time, right=of s2] (s3) {$S_{t+3}$};
		\node[right=0.55cm of s3] (dots) {$\cdots$};
		\node[reward, above=0.75cm of $(s0)!0.5!(s1)$] (r1) {$R_{t+1}$};
		\node[reward, above=0.75cm of $(s1)!0.5!(s2)$] (r2) {$R_{t+2}$};
		\node[reward, above=0.75cm of $(s2)!0.5!(s3)$] (r3) {$R_{t+3}$};
		\draw[arrow] (s0) -- node[below] {$A_t$} (s1);
		\draw[arrow] (s1) -- node[below] {$A_{t+1}$} (s2);
		\draw[arrow] (s2) -- node[below] {$A_{t+2}$} (s3);
		\draw[arrow] (s3) -- (dots);
		\draw[arrow] (r1) -- (s1);
		\draw[arrow] (r2) -- (s2);
		\draw[arrow] (r3) -- (s3);
		\node[below=0.9cm of $(s1)!0.5!(s2)$, align=center] {$G_t=R_{t+1}+\gamma R_{t+2}+\gamma^2R_{t+3}+\cdots$};
	\end{tikzpicture}
	\caption{Reward is immediate feedback, while return is the discounted accumulation of future rewards. This distinction is central to sequential decision-making and delayed consequences \citep{sutton2018reinforcement}.}
	\label{fig:reward_return}
\end{figure}

The discount factor controls how much the agent cares about future rewards. If $\gamma=0$, the agent only cares about immediate reward. If $\gamma$ is close to 1, the agent cares strongly about future reward.

\subsection{Example: Short-Term Reward Can Be Misleading}
Consider an autonomous UAV serving high-priority users. At one moment, the UAV can stay above a dense user cluster and receive high immediate reward. However, its battery is low. If it stays, it may soon lose power and fail the mission. If it moves to a charging station, the immediate reward decreases, but the long-term return may be higher. A greedy controller may choose the first option. A reinforcement learning agent should learn the second option if the future reward justifies the short-term cost.

\subsection{Reward Design}
Reward design is one of the most important and dangerous parts of RL. The reward tells the agent what to optimize. If the reward is poorly designed, the agent may learn behavior that is mathematically optimal but practically wrong. This problem is often called reward hacking or specification gaming % FIX 5: added citation for reward hacking / specification gaming
\citep{krakovna2020specification,amodei2016concrete}.

Therefore, a good RL system often needs multiple evaluation metrics, constraints, safety checks, and human inspection. In real systems, reward is never just a number. It is a design choice with scientific and ethical consequences.

\section{Policies: The Behavior of an Agent}

A policy defines how the agent chooses actions. A deterministic policy maps each state to one action,
\begin{equation}
	A_t=\pi(S_t),
\end{equation}
while a stochastic policy gives a probability distribution over actions,
\begin{equation}
	\pi(a\mid s)=\Prob(A_t=a\mid S_t=s).
\end{equation}

Stochastic policies are important for exploration. If the agent always chooses the same action, it may never discover better alternatives. Stochastic policies are also important in partially observable environments, games, and multi-agent systems, where predictable behavior may be exploited.

In deep reinforcement learning, the policy is often represented by a neural network. The input is the state or observation, and the output is either an action or a distribution over actions.

\section{Value Functions: Predicting Long-Term Usefulness}

A value function estimates how good a state or action is in terms of expected long-term return. The state-value function under policy $\pi$ is
\begin{equation}
	V^\pi(s)=\E_\pi[G_t\mid S_t=s].
\end{equation}
The action-value function under policy $\pi$ is
\begin{equation}
	Q^\pi(s,a)=\E_\pi[G_t\mid S_t=s,A_t=a].
\end{equation}

These two functions are central because they convert delayed consequences into learnable predictions. If the agent can estimate $Q(s,a)$, then it can choose actions with high expected return.

\begin{figure}[t]
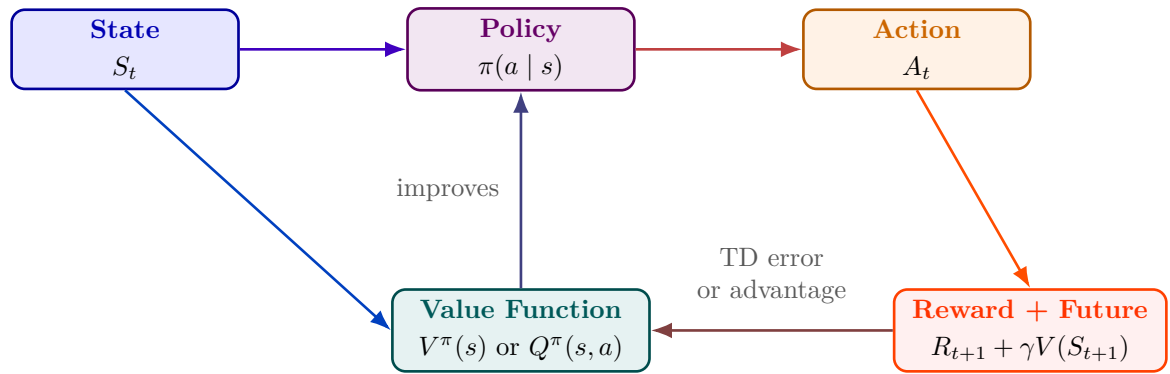

	\centering

	\caption{A policy selects actions, while a value function estimates
		long-term usefulness. Many RL algorithms learn by comparing predicted
		value with reward plus discounted future value, producing a learning
		signal such as a temporal-difference error or an advantage estimate
		\citep{sutton1988learning,williams1992simple}.}
	\label{fig:policy_value_learning}
\end{figure}

Value functions are connected by Bellman equations. The Bellman idea is that the value of the present can be written as immediate reward plus discounted value of the future:
\begin{align}
	V^\pi(s)&=\E_\pi\left[R_{t+1}+\gamma V^\pi(S_{t+1})\mid S_t=s\right],\\
	V^*(s)&=\max_a\E\left[R_{t+1}+\gamma V^*(S_{t+1})\mid S_t=s,A_t=a\right].
\end{align}
This recursive structure is one of the deepest ideas in reinforcement learning. % FIX 4: added bridge sentence
A fuller treatment of both equations, including the policy-evaluation and policy-improvement operators that connect them, appears in Chapter~2.

\section{Exploration Versus Exploitation}

Every RL agent faces the exploration--exploitation dilemma. Exploitation means choosing the action that currently seems best. Exploration means trying actions whose value is uncertain.

\begin{figure}[t]
	\centering
	\begin{tikzpicture}[
		box/.style={draw, rounded corners, thick, minimum width=3.0cm, minimum height=1.0cm, align=center},
		arrow/.style={-{Latex[length=2.5mm]}, thick}
		]
		\node[box] (explore) at (0,0) {Exploration\\Try uncertain actions};
		\node[box] (exploit) at (6,0) {Exploitation\\Use best-known action};
		\draw[thick] (explore.east) -- (exploit.west);
		\draw[arrow] (3,0.45) -- +(0.01,0) node[above] {balance};
		\node[align=center] (risk) at (0,-1.8) {Risk:\\unsafe or costly trials};
		\node[align=center] (trap) at (6,-1.8) {Risk:\\suboptimal behavior};
		\draw[arrow] (risk.north) -- (explore.south);
		\draw[arrow] (trap.north) -- (exploit.south);
	\end{tikzpicture}
	\caption{The exploration--exploitation dilemma. An RL agent must explore to discover better actions, but excessive exploration can be inefficient or unsafe. Exploitation uses current knowledge but can trap the agent in suboptimal behavior \citep{sutton2018reinforcement}.}
	\label{fig:exploration_exploitation}
\end{figure}
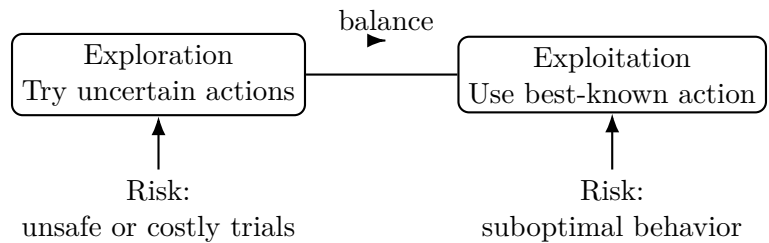

If the agent only exploits, it may get stuck with a suboptimal behavior. If it explores too much, it may waste time or cause unsafe behavior. A simple exploration method is $\epsilon$-greedy action selection. With probability $1-\epsilon$, the agent chooses the best-known action. With probability $\epsilon$, it chooses a random action.

More advanced exploration methods include entropy regularization, parameter noise, curiosity-driven exploration, intrinsic motivation, optimistic initialization, uncertainty-aware exploration, count-based or pseudo-count exploration, and information-gain-based exploration.

In safety-critical systems, exploration is especially difficult. A drone cannot freely crash thousands of times in the real world. A medical decision system cannot test dangerous treatments randomly. A network controller cannot intentionally overload a production network. This is one major reason for safe RL, offline RL, simulation-based training, and formal safety filters.

\section{From Classical RL to Deep Reinforcement Learning}

Classical reinforcement learning was developed long before modern deep learning became dominant. Early RL methods used tables, linear function approximators, dynamic programming, Monte Carlo updates, and temporal-difference learning.

A tabular Q-learning agent stores one value for every state-action pair, $Q(s,a)$. This works if the number of states and actions is small. But many real problems are far too large.

Deep reinforcement learning solves this problem by replacing tables with neural networks:
\begin{equation}
	Q(s,a;\theta)\approx Q^*(s,a),
\end{equation}
where $\theta$ represents the neural network parameters.

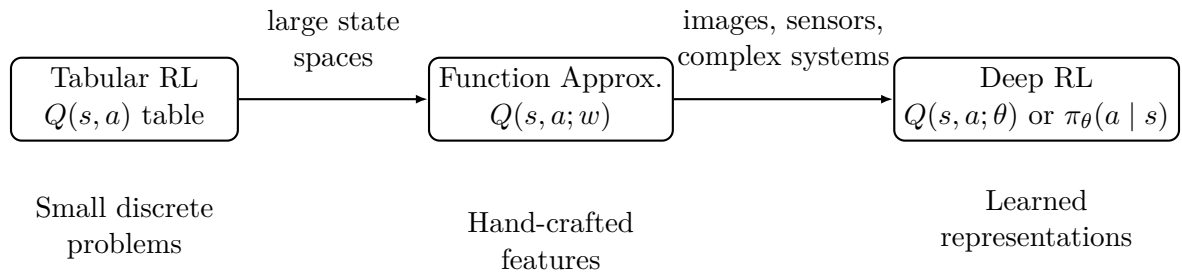
\begin{figure}[t]
	\centering
	\begin{tikzpicture}[
		box/.style={draw, rounded corners, thick, minimum width=3.0cm, minimum height=1.0cm, align=center},
		arrow/.style={-{Latex[length=1.6mm]}, thick},
		]
		\node[box] (tabular) {Tabular RL\\$Q(s,a)$ table};
		\node[box, right=2.5cm of tabular] (linear) {Function Approx.\\$Q(s,a;w)$};
		\node[box, right=2.9cm of linear] (deep) {Deep RL\\$Q(s,a;\theta)$ or $\pi_\theta(a\mid s)$};
		
		\draw[arrow] (tabular.east) -- node[midway, yshift=22pt, align=center] {large state\\spaces} (linear.west);
		\draw[arrow] (linear.east) -- node[midway, yshift=22pt, align=center] {images, sensors,\\complex systems} (deep.west);
		
		\node[below=0.6cm of tabular, align=center] {Small discrete\\problems};
		\node[below=0.8cm of linear, align=center] {Hand-crafted\\features};
		\node[below=0.5cm of deep, align=center] {Learned\\representations};
	\end{tikzpicture}
	\caption{The transition from tabular RL to deep reinforcement learning. Deep neural networks replace tables or simple feature-based approximators, allowing RL to scale to high-dimensional observations such as images, sensor streams, and network states \citep{mnih2015human}.}
	\label{fig:classical_to_deep_rl}
\end{figure}

This idea made it possible to learn policies from high-dimensional inputs such as images and sensor streams. The major public breakthrough came when DeepMind showed that a deep Q-network could learn to play many Atari games from raw pixels \citep{mnih2015human}. After DQN, the field expanded rapidly. AlphaGo combined deep policy networks, deep value networks, reinforcement learning, self-play, and tree search to defeat elite human Go players \citep{silver2016mastering}.

DRL appeared as a sequence of solutions to specific difficulties: high-dimensional observations, continuous actions, unstable training, sparse rewards, poor sample efficiency, unsafe exploration, multi-agent coordination, transfer to real systems, and alignment with human goals.

\section{Why DRL Became Important for Modern AI}

Deep reinforcement learning became important because it connects perception, decision-making, and long-term optimization. Deep learning is strong at representation learning. Reinforcement learning is strong at sequential decision-making. DRL combines them.

\begin{figure}[t]
	\centering
	\resizebox{\textwidth}{!}{%
		\begin{tikzpicture}[
			event/.style={draw, rounded corners, thick, minimum width=2.55cm, minimum height=0.95cm, align=center},
			arrow/.style={-{Latex[length=2.5mm]}, thick},
			node distance=0.65cm
			]
			\node[event] (bellman) {1957\\Bellman\\Dynamic Programming};
			\node[event, right=of bellman] (td) {1988\\TD Learning};
			\node[event, right=of td] (qlearn) {1992\\Q-learning\\REINFORCE};
			\node[event, right=of qlearn] (dqn) {2015\\DQN\\Atari};
			\node[event, right=of dqn] (alphago) {2016\\AlphaGo};
			\node[event, right=of alphago] (muzero) {2019--2021\\MuZero, Offline RL,\\Decision Transformer};
			\node[event, right=of muzero] (rlhf) {2022--2024\\RLHF and\\LLM Alignment};
			\node[event, right=of rlhf] (reasoning) {2025--2026\\RL for Reasoning\\World Models};
			\node[event, below=1.1cm of dqn] (ppo) {2017--2018\\PPO, SAC,\\Rainbow};
			\draw[arrow] (bellman) -- (td);
			\draw[arrow] (td) -- (qlearn);
			\draw[arrow] (qlearn) -- (dqn);
			\draw[arrow] (dqn) -- (alphago);
			\draw[arrow] (alphago) -- (muzero);
			\draw[arrow] (muzero) -- (rlhf);
			\draw[arrow] (rlhf) -- (reasoning);
			\draw[arrow] (dqn.south) -- (ppo.north);
			\draw[arrow] (ppo.east) -| (muzero.south);
		\end{tikzpicture}%
	}
	\caption{A simplified historical timeline of reinforcement learning and deep reinforcement learning, from dynamic programming and temporal-difference learning to DQN, AlphaGo, offline RL, RLHF, and reinforcement learning for reasoning-oriented AI systems \citep{bellman1957dynamic,sutton1988learning,watkins1992q,mnih2015human,silver2016mastering,ouyang2022training}.}
	\label{fig:rl_historical_timeline}
\end{figure}
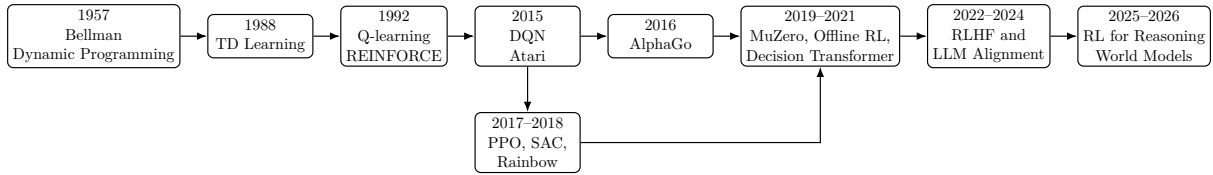

This combination made it possible to train agents that process raw sensory input, learn long-term strategies, adapt through interaction, optimize complex objectives, coordinate with other agents, and operate in uncertain environments.

\begin{figure}[t]
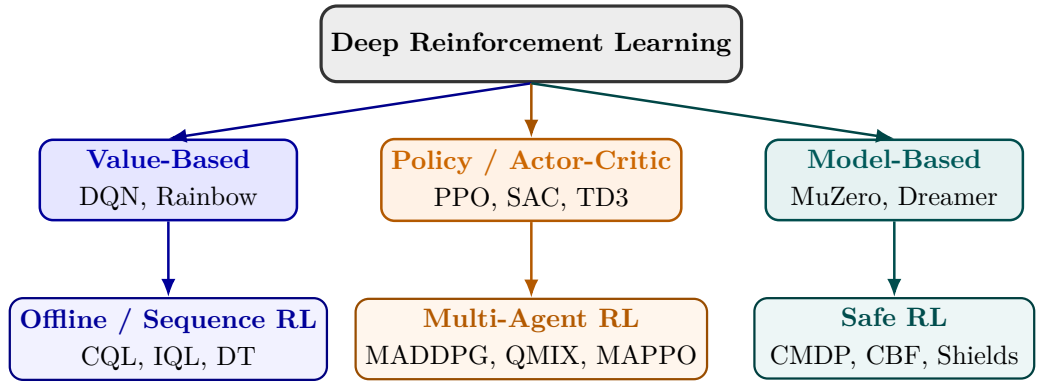

	\centering

	\caption{Major families of deep reinforcement learning algorithms. The
		boundaries are not strict: modern systems often combine value learning,
		policy optimisation, world models, offline datasets, multi-agent
		coordination, and safety mechanisms.}
	\label{fig:drl_algorithm_families}
\end{figure}

Games became an early benchmark because they offer clear rules, measurable success, and large-scale simulation. Robotics requires continuous control, perception, contact dynamics, safety, and sample efficiency. Networks are dynamic systems with changing traffic, users, channels, failures, and constraints. From 2022 onward, reinforcement learning became central in language-model alignment through reinforcement learning from human feedback \citep{ouyang2022training}. Later, RL became increasingly important for reasoning models, where rewards may come from verifiable outcomes such as correct mathematical answers, executable code tests, or successful task completion \citep{jaech2024openai,deepseek2025r1}.

\section{A Running Example: Autonomous UAV Network Control}

Throughout this book, we will use several examples, but one running example will be especially useful: autonomous UAV-assisted wireless network control.

Imagine a set of UAVs acting as flying base stations. They serve users with different service requirements: high-priority users need very low latency, video users need high throughput, and IoT users need reliable but low-rate communication. Each UAV must decide where to move, which users to serve, how much bandwidth to allocate, when to recharge, how to avoid collisions, how to coordinate with other UAVs, and how to respond to SDN controller guidance.

\begin{figure}[t]
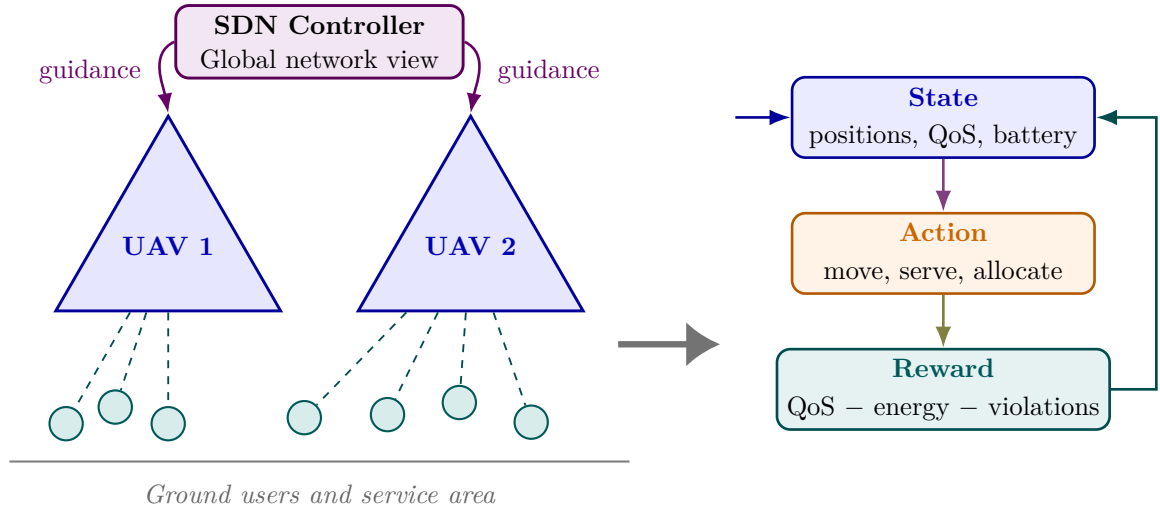

	\centering

	\caption{A UAV-assisted wireless network can be formulated as a reinforcement
		learning problem. The agent observes network state, chooses movement and
		resource-allocation actions, and receives rewards based on QoS, energy
		consumption, and safety violations. This running example will be used
		throughout the book.}
	\label{fig:uav_rl_problem}
\end{figure}

This is a natural reinforcement learning problem because the system is sequential, dynamic, uncertain, and constrained.

\subsection{State}
The state may include UAV positions, UAV battery levels, user positions, user priority classes, channel quality, SINR, throughput, latency, traffic load, charging station locations, and neighboring UAV information.

\subsection{Action}
The action may include movement, hovering, bandwidth allocation, user association, transmission-power control, and following or rejecting SDN guidance.

\subsection{Reward}
The reward may combine QoS satisfaction, latency reduction, throughput improvement, energy efficiency, collision avoidance, fairness, and safety constraint satisfaction.

A simple reward could be
\begin{equation}
	r_t=w_{\mathrm{qos}}R_{\mathrm{qos}}-w_{\mathrm{energy}}C_{\mathrm{energy}}-w_{\mathrm{collision}}C_{\mathrm{collision}}-w_{\mathrm{violation}}C_{\mathrm{violation}}.
\end{equation}

\begin{figure}[t]
	\centering
	\begin{tikzpicture}[
		term/.style={
			draw=#1!60!black, rounded corners=5pt,
			line width=0.9pt,
			minimum width=3.4cm, minimum height=1.05cm,
			align=center, fill=#1!12, font=\small
		},
		hub/.style={
			draw=black!75, rounded corners=6pt,
			line width=1.5pt,
			minimum width=3.6cm, minimum height=1.15cm,
			align=center, fill=black!6,
			font=\small\bfseries
		},
		carrow/.style={
			-{Latex[length=2.8mm, width=2.2mm]},
			line width=1.1pt, color=#1!70!black
		},
		eqbox/.style={
			draw=black!30, rounded corners=4pt,
			fill=black!4, line width=0.7pt,
			inner xsep=10pt, inner ysep=6pt,
			align=center, font=\small
		},
		]
		\node[hub] (reward) {Total Reward $r_t$};
		
		\node[term=teal,   above left=1.6cm  and 3.6cm of reward] (qos)
		{\textbf{QoS Gain}\\[2pt]$+\,w_q\,R_{\mathrm{QoS}}$};
		\node[term=orange, above right=1.6cm and 3.6cm of reward] (energy)
		{\textbf{Energy Cost}\\[2pt]$-\,w_e\,C_{\mathrm{energy}}$};
		\node[term=red,    below left=1.6cm  and 3.6cm of reward] (safe)
		{\textbf{Safety Violation}\\[2pt]$-\,w_s\,C_{\mathrm{violation}}$};
		\node[term=violet, below right=1.6cm and 3.6cm of reward] (fair)
		{\textbf{Fairness / Priority}\\[2pt]$+\,w_f\,R_{\mathrm{fair}}$};
		
		\draw[carrow=teal]   (qos)    -- (reward);
		\draw[carrow=orange] (energy) -- (reward);
		\draw[carrow=red]    (safe)   -- (reward);
		\draw[carrow=violet] (fair)   -- (reward);
		
		\node[eqbox, below=1.6cm of reward] (eq)
		{$r_t \;=\; w_q\,R_{\mathrm{QoS}}
			\;-\; w_e\,C_{\mathrm{energy}}
			\;-\; w_s\,C_{\mathrm{violation}}
			\;+\; w_f\,R_{\mathrm{fair}}$};
		
		\draw[dotted, thick, black!40] (reward.south) -- (eq.north);
	\end{tikzpicture}
	\caption{Reward design often combines several competing objectives. In UAV
		network control, maximising QoS alone may waste energy or violate safety
		constraints, so the reward must reflect the real mission objective rather
		than a single metric.}
	\label{fig:uav_reward_design}
\end{figure}
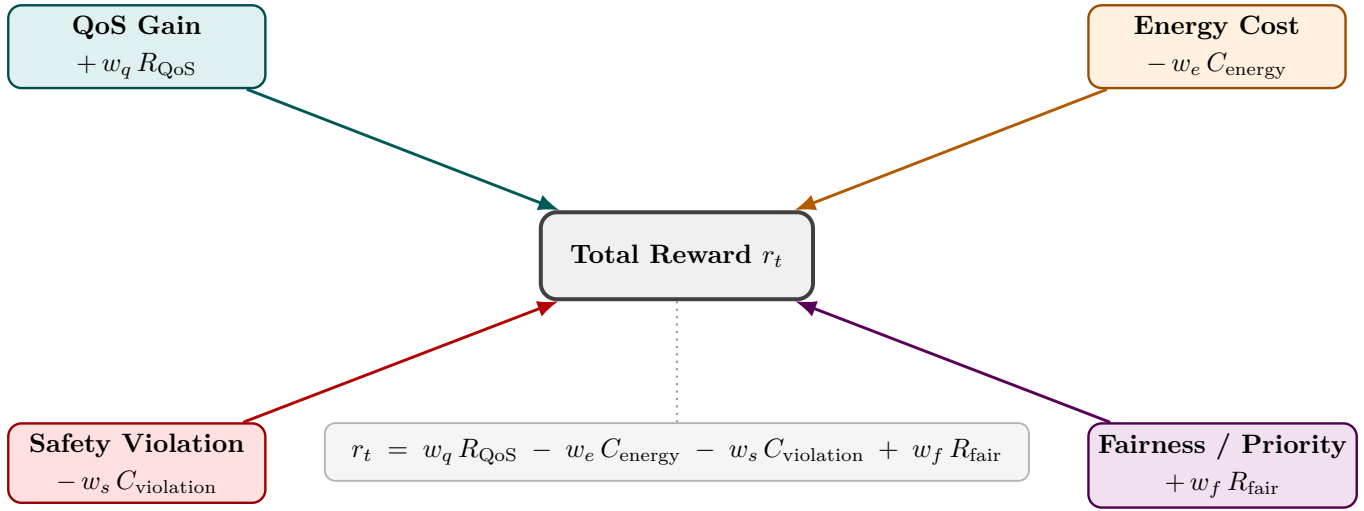

However, this reward must be designed carefully. If $w_{\mathrm{qos}}$ is too high, UAVs may serve users aggressively and waste energy. If $w_{\mathrm{energy}}$ is too high, UAVs may save battery but provide poor service. If safety penalties are too weak, the learned policy may violate constraints.

This example shows why DRL is powerful but difficult. The problem is not only choosing an algorithm. The full system requires modeling, reward design, safety design, evaluation metrics, and careful experimental methodology.

\begin{tcolorbox}[
	title={Historical Box 1.1: From Bellman to Reasoning Models},
	colback=gray!5,
	colframe=black!70,
	fonttitle=\bfseries,
	arc=2mm,
	boxrule=0.8pt
	]
	The history of reinforcement learning can be read as a gradual expansion of the same basic idea: learning good decisions from consequences. Bellman's dynamic programming formalized sequential optimal decision-making. Temporal-difference learning showed how to learn predictions from incomplete experience. Q-learning showed how to learn action values off-policy. Deep Q-Networks connected these ideas to high-dimensional perception. AlphaGo showed the power of combining deep learning, reinforcement learning, self-play, and search. RLHF then brought reinforcement learning into language-model alignment, and recent reasoning-oriented systems use reinforcement learning signals to improve multi-step problem solving. The central question remains the same: how should an intelligent system improve its future behavior using feedback?
\end{tcolorbox}

\section{Key Takeaways}

\begin{itemize}
	\item Reinforcement learning is about learning to act through interaction.
	\item The agent observes a state, chooses an action, receives reward, and transitions to a new state.
	\item The objective is not immediate reward but long-term return.
	\item A policy defines the agent's behavior.
	\item A value function estimates expected long-term return.
	\item The Bellman idea connects present value to immediate reward and future value.
	\item Exploration is necessary because the agent must discover good actions, but exploration can be dangerous in real systems.
	\item Deep reinforcement learning replaces tables with neural networks, allowing RL to scale to high-dimensional observations and complex decision problems.
	\item Modern DRL connects classical RL, deep learning, planning, world models, multi-agent systems, safe control, and language-model alignment.
\end{itemize}

\section{Exercises}

\subsection*{Conceptual Exercises}
\begin{enumerate}
	\item Explain the difference between supervised learning and reinforcement learning using your own example.
	\item Why is immediate reward not enough for intelligent behavior?
	\item Give one example where a short-term penalty can lead to long-term benefit.
	\item What is the exploration--exploitation dilemma?
	\item Why can poor reward design lead to unsafe or unwanted behavior?
\end{enumerate}

\subsection*{Mathematical Exercises}
\begin{enumerate}
	\item Suppose an agent receives rewards $1,1,1,1,\ldots$ forever and $\gamma=0.9$. Compute the return $G_0$. (Hint: use the geometric series $\sum_{k=0}^\infty r^k = 1/(1-r)$ for $|r|<1$.)
	% FIX 6: added hint for geometric series so first-time readers are not blocked
	\item Suppose an agent receives rewards $2,0,5$ over a three-step episode and $\gamma=0.5$. Compute $G_0$.
	\item Write the difference between $V^\pi(s)$ and $Q^\pi(s,a)$ in words and equations.
	\item If $\gamma=0$, what kind of behavior does the agent optimize? If $\gamma$ is close to 1, what changes?
	% FIX 6: added fifth math exercise to match 5/5/5 symmetry with other exercise blocks
	\item Suppose $\gamma=0.95$ and an agent receives rewards $10, 0, 0, 5$ over a four-step episode. Compute $G_0$.
\end{enumerate}

\subsection*{Research Thinking Exercises}
\begin{enumerate}
	\item In a UAV network, what could go wrong if reward only measures throughput?
	\item In a medical decision system, why is online exploration dangerous?
	\item In language-model training, why might human preference reward be imperfect?
	\item Design a reward function for a drone delivery problem. Include at least three terms and explain why each term matters.
	\item Choose one real-world system you know. Describe its state, action, reward, and safety constraints.
\end{enumerate}

\section*{Looking Ahead to Chapter 2: Food for Thought}
% FIX 1: heading now matches \addcontentsline below (both title case)
\addcontentsline{toc}{section}{Looking Ahead to Chapter 2: Food for Thought}

Before moving to the mathematical foundation of reinforcement learning, it is useful to pause and ask what exactly must be formalized. Chapter~1 introduced reinforcement learning as learning through interaction. However, if we want to design algorithms, prove properties, or implement agents, intuition alone is not enough. We need a mathematical language for decision-making.

Chapter~2 introduces this language through the Markov decision process. The goal is to answer a simple but fundamental question:

\begin{quote}
	How can we describe an agent that acts, receives rewards, and faces uncertain future consequences?
\end{quote}

To prepare for that question, consider the following points.

\subsection*{Questions to Ponder}

\begin{enumerate}
	\item What information does an agent really need in order to make a good decision?
	\item Is the current observation enough, or does the agent need memory of the past?
	\item When an agent takes an action, is the next state deterministic, or only probabilistic?
	\item Can two different actions give the same immediate reward but very different long-term consequences?
	\item How should an agent compare a small reward now with a larger reward later?
	\item What does it mean for one policy to be better than another policy?
	\item If the environment is unknown, how can the agent learn from experience?
	\item If the environment is known, can the agent compute the best behavior directly?
	\item Why do value functions appear naturally when we care about future rewards?
	\item What is the minimum mathematical model needed to connect states, actions, rewards, and future outcomes?
\end{enumerate}

\subsection*{A Simple Thought Experiment}

Imagine a robot standing in a room. It can move left, right, forward, or backward. Somewhere in the room there is a charging station. The robot receives a positive reward when it reaches the charger and a negative reward when it hits a wall.

At first, this sounds simple. But several questions appear immediately:

\begin{itemize}
	\item What is the robot's state: its position, its battery level, its distance to the charger, or all of them?
	\item What are the possible actions?
	\item Is movement always successful, or can the robot slip?
	\item Should the robot care only about the next reward, or about the total reward until the end of the task?
	\item How can the robot decide whether a state is good or bad before it reaches the charger?
\end{itemize}

These questions lead directly to the mathematical objects of Chapter~2: states, actions, transition probabilities, rewards, returns, policies, and value functions.

\subsection*{Main Idea for the Next Chapter}

The key idea of Chapter~2 is that reinforcement learning problems can be described as sequential decision processes. A decision made now changes the future, and the future affects the value of the present decision.

This is why reinforcement learning needs more than ordinary prediction. It needs a mathematical framework for time, uncertainty, reward, and action. The Markov decision process provides that framework.

\begin{quote}
	Chapter~1 explained what reinforcement learning is.
	Chapter~2 explains how to write it mathematically.
\end{quote}

% ============================================================
% TWO NEW BIBLIOGRAPHY ENTRIES NEEDED IN SHARED BIBLIOGRAPHY:
%
% \bibitem[Krakovna et~al.(2020)]{krakovna2020specification}
% Krakovna, V., Uesato, J., Mikulik, V., Martic, M., Everitt, T.,
% Kumar, R., Ziegler, Z., Leike, J., and Legg, S. 2020.
% \newblock Specification gaming: the flip side of AI ingenuity.
% \newblock DeepMind Blog, \url{https://deepmind.com/blog/article/Specification-gaming-the-flip-side-of-AI-ingenuity}.
%
% \bibitem[Amodei et~al.(2016)]{amodei2016concrete}
% Amodei, D., Olah, C., Steinhardt, J., Christiano, P., Schulman, J.,
% and Man\'e, D. 2016.
% \newblock Concrete problems in AI safety.
% \newblock arXiv:1606.06565.
% ============================================================
	% ============================================================
% Chapter 2 -- Markov Decision Processes
% Included by the main book file with \include{chapter2}
% ============================================================
\chapter{Markov Decision Processes}
\label{ch:mdp}

\section*{Chapter Overview}
\addcontentsline{toc}{section}{Chapter Overview}

Chapter~1 introduced reinforcement learning as learning through interaction. The agent observes a situation, takes an action, receives reward, and then faces a new situation. That description is intuitive, but it is not yet enough to design algorithms. To build reinforcement learning methods, we need a precise mathematical language for states, actions, rewards, uncertainty, and time.

The Markov decision process, or MDP, provides this language. It is one of the central mathematical models behind dynamic programming, optimal control, and reinforcement learning \citep{bellman1957dynamic,howard1960dynamic,puterman1994markov,sutton2018reinforcement}. An MDP formalizes the idea that an agent makes sequential decisions in an uncertain environment, and that the quality of a decision depends not only on its immediate reward but also on its future consequences.

\begin{keybox}{Main idea of this chapter}
	An MDP is a mathematical model of an agent that repeatedly asks:
	\begin{quote}
		Given the current state, what action should I choose so that the long-term consequence is as good as possible?
	\end{quote}
	The key point is that the present state must contain enough information for predicting the future. This is the Markov property.
\end{keybox}

\section*{Learning Objectives}
\addcontentsline{toc}{section}{Learning Objectives}

After reading this chapter, the reader should be able to:
\begin{enumerate}[leftmargin=*]
	\item define an MDP formally using states, actions, transition probabilities, rewards, and a discount factor;
	\item explain the Markov property and why it is an assumption about state representation;
	\item distinguish rewards, returns, value functions, and policies;
	\item derive the Bellman expectation equations for a fixed policy;
	\item derive the Bellman optimality equations for optimal control;
	\item understand why Bellman equations are recursive consistency equations;
	\item map a real problem, such as robot navigation or UAV network control, into the MDP framework;
	\item recognize common modeling failures when the MDP assumption is not appropriate.
\end{enumerate}

\section{From Intuition to Mathematical Modeling}

Reinforcement learning begins with interaction, but algorithms require formal objects. Consider a robot in a room. It must move to a charging station while avoiding walls. We can describe the task in ordinary language, but a learning algorithm needs precise answers to the following questions:

\begin{itemize}[leftmargin=*]
	\item What exactly is the robot's \emph{state}?
	\item What actions are available?
	\item What happens after each action?
	\item Is the next state deterministic or random?
	\item What reward is obtained?
	\item Should the robot optimize immediate reward or long-term reward?
\end{itemize}

An MDP answers these questions by converting a decision problem into a tuple of mathematical objects. This does not mean that every real problem is perfectly Markovian. Rather, the MDP is a modeling framework. Its usefulness depends on how well the chosen state captures the information needed for decision-making.

\begin{figure}[t]
	\centering
	\begin{tikzpicture}[
		>=Latex,
		node distance=2.0cm and 2.0cm,
		every node/.style={font=\small},
		box/.style={
			draw, rounded corners=4pt, thick,
			minimum width=2.50cm, minimum height=1.1cm,
			align=center, font=\small\sffamily
		},
		problembox/.style={box, fill=blue!10,    draw=blue!60!black},
		modelbox/.style  ={box, fill=orange!15,  draw=orange!70!black},
		algobox/.style   ={box, fill=green!12,   draw=green!50!black},
		policybox/.style ={box, fill=purple!12,  draw=purple!60!black},
		arrow/.style={->, very thick, draw=black!70},
		annotation/.style={
			draw=gray!50, dashed, rounded corners=3pt,
			fill=gray!5, align=center,
			font=\footnotesize, inner sep=4pt
		}
		]
		
		\node[problembox] (problem) {Real decision\\problem};
		\node[modelbox,   right=of problem] (model)     {MDP model};
		\node[algobox,    right=of model]   (algorithm) {RL / DP\\algorithm};
		\node[policybox,  right=of algorithm] (policy)  {Policy\,$\pi$};
		
		\draw[arrow] (problem)   -- node[above, font=\footnotesize\itshape] {formalize}     (model);
		\draw[arrow] (model)     -- node[above, font=\footnotesize\itshape] {solve / learn} (algorithm);
		\draw[arrow] (algorithm) -- node[above, font=\footnotesize\itshape] {outputs}       (policy);
		
		\node[annotation, below=1.0cm of model] (annot) {%
			states, actions, transitions,\\
			rewards, discount factor
		};
		\draw[draw=gray!60, dotted, thick] (model.south) -- (annot.north);
	\end{tikzpicture}
	\caption{The role of an MDP in reinforcement learning. A real decision problem is first formalized as a sequential decision model. Algorithms then use this formal model, or data generated from it, to learn a policy $\pi$.}
	\label{fig:mdp_role}
\end{figure}
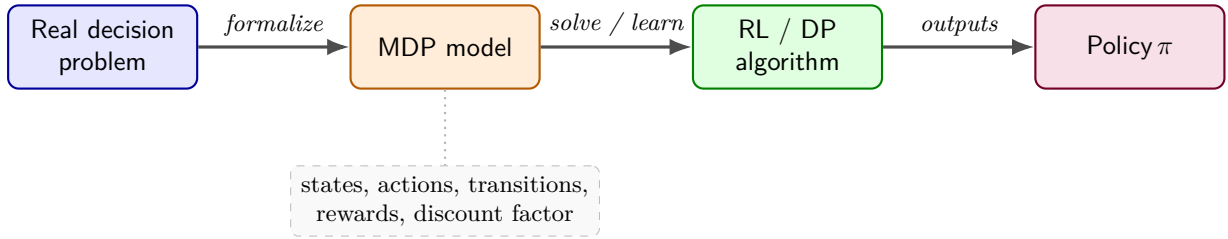

\section{The Markov Property}

The word ``Markov'' means that the future is conditionally independent of the past given the present. In reinforcement learning, this means that once the current state and action are known, the entire previous history should not provide additional information for predicting the next state and reward.

Let the history up to time $t$ be
\begin{equation}
	H_t = (S_0,A_0,R_1,S_1,A_1,R_2,\ldots,S_t).
\end{equation}
A process is Markov if
\begin{equation}
	\Prob(S_{t+1}=s', R_{t+1}=r \mid H_t, A_t=a)
	=
	\Prob(S_{t+1}=s', R_{t+1}=r \mid S_t=s, A_t=a).
	\label{eq:markov_property}
\end{equation}

Equation~\eqref{eq:markov_property} is not just a mathematical detail. It is a statement about what information the state contains. If the state is poorly designed, the Markov property may fail.

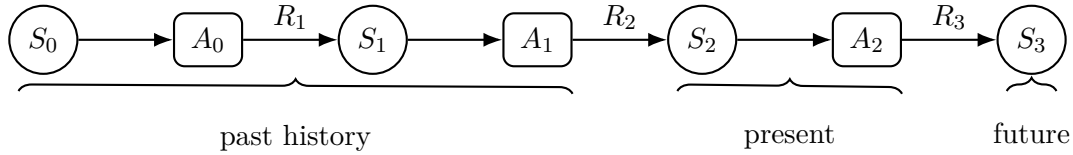
\begin{figure}[t]
	\centering
	\begin{tikzpicture}[
		state/.style={circle, draw, thick, minimum size=0.9cm},
		action/.style={rectangle, draw, rounded corners, thick, minimum width=0.9cm, minimum height=0.7cm, align=center},
		arrow/.style={-{Latex[length=2.5mm]}, thick},
		node distance=1.25cm
		]
		\node[state] (s0) {$S_0$};
		\node[action, right=of s0] (a0) {$A_0$};
		\node[state, right=of a0] (s1) {$S_1$};
		\node[action, right=of s1] (a1) {$A_1$};
		\node[state, right=of a1] (s2) {$S_2$};
		\node[action, right=of s2] (a2) {$A_2$};
		\node[state, right=of a2] (s3) {$S_3$};
		
		\draw[arrow] (s0) -- (a0);
		\draw[arrow] (a0) -- node[above] {$R_1$} (s1);
		\draw[arrow] (s1) -- (a1);
		\draw[arrow] (a1) -- node[above] {$R_2$} (s2);
		\draw[arrow] (s2) -- (a2);
		\draw[arrow] (a2) -- node[above] {$R_3$} (s3);
		
		\draw[decorate, decoration={brace, amplitude=5pt}, thick] ($(s0.south west)+(0,-0.35)$) -- node[below=8pt] {past history} ($(a1.south east)+(0,-0.35)$);
		\draw[decorate, decoration={brace, amplitude=5pt}, thick] ($(s2.south west)+(0,-0.35)$) -- node[below=8pt] {present} ($(a2.south east)+(0,-0.35)$);
		\draw[decorate, decoration={brace, amplitude=5pt}, thick] ($(s3.south west)+(0,-0.35)$) -- node[below=8pt] {future} ($(s3.south east)+(0,-0.35)$);
	\end{tikzpicture}
	\caption{The Markov property says that, for prediction of the next state and reward, the current state and action summarize all relevant information from the past. If this is not true, the chosen state representation is incomplete.}
	\label{fig:markov_property}
\end{figure}

\begin{practicebox}{When the Markov property fails}
	Suppose a robot observes only its current camera image but not its velocity. Two identical images may correspond to different hidden velocities, leading to different next states after the same action. The observation alone is not Markov. A better state might include recent images, velocity estimates, or a recurrent memory. This motivates partially observable MDPs and memory-based policies \citep{kaelbling1998planning}.
\end{practicebox}

\section{Formal Definition of a Markov Decision Process}

A finite discounted MDP is often defined as a tuple
\begin{equation}
	\mathcal{M} = (\Sset, \Aset, p, r, \gamma),
	\label{eq:mdp_tuple_simple}
\end{equation}
where:
\begin{itemize}[leftmargin=*]
	\item $\Sset$ is the set of states;
	\item $\Aset$ is the set of actions;
	\item $p(s',r\mid s,a)$ is the transition-reward probability distribution;
	\item $r(s,a)$ or $r(s,a,s')$ is the expected immediate reward;
	\item $\gamma \in [0,1]$ is the discount factor.
\end{itemize}

Some texts include an initial-state distribution $\rho_0(s)$ and a horizon $T$ as part of the task definition. This is often useful in episodic problems:
\begin{equation}
	\mathcal{M} = (\Sset, \Aset, p, r, \rho_0, \gamma, T).
\end{equation}
The exact tuple varies slightly across authors, but the conceptual content is the same \citep{puterman1994markov,sutton2018reinforcement,szepesvari2010algorithms}.

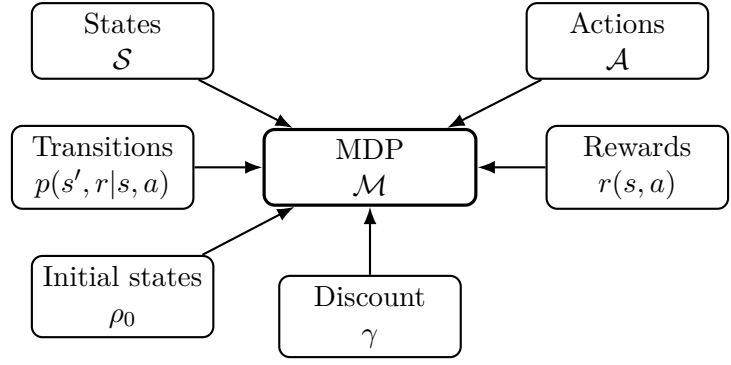
\begin{figure}[t]
	\centering
	\begin{tikzpicture}[
		component/.style={draw, rounded corners, thick, minimum width=2.4cm, minimum height=0.85cm, align=center},
		center/.style={draw, rounded corners, very thick, minimum width=2.8cm, minimum height=1.0cm, align=center},
		arrow/.style={-{Latex[length=2.3mm]}, thick},
		node distance=0.9cm
		]
		\node[center] (mdp) {MDP\\$\mathcal{M}$};
		\node[component, above left=of mdp] (s) {States\\$\Sset$};
		\node[component, above right=of mdp] (a) {Actions\\$\Aset$};
		\node[component, left=of mdp] (p) {Transitions\\$p(s',r|s,a)$};
		\node[component, right=of mdp] (r) {Rewards\\$r(s,a)$};
		\node[component, below=of mdp] (g) {Discount\\$\gamma$};
		\node[component, below left=of mdp] (rho) {Initial states\\$\rho_0$};
		
		\draw[arrow] (s) -- (mdp);
		\draw[arrow] (a) -- (mdp);
		\draw[arrow] (p) -- (mdp);
		\draw[arrow] (r) -- (mdp);
		\draw[arrow] (g) -- (mdp);
		\draw[arrow] (rho) -- (mdp);
	\end{tikzpicture}
	\caption{Core components of an MDP. Most reinforcement learning algorithms can be understood as different ways of using or estimating these components.}
	\label{fig:mdp_tuple}
\end{figure}

\subsection{States}

A state is a representation of the current situation. In a board game, the state may be the current board configuration. In robot navigation, it may include position, velocity, sensor readings, and battery level. In a wireless network, it may include traffic load, channel quality, queue lengths, user locations, and available resources.

The most important question is not whether the state is large or small, but whether it is informative enough. A state should ideally be sufficient for predicting future outcomes under actions.

\subsection{Actions}

An action is a decision available to the agent. Actions can be discrete or continuous. In a grid world, actions may be $\{\text{up},\text{down},\text{left},\text{right}\}$. In robotics, actions may be continuous torques. In UAV network control, actions may include movement directions, transmission power, user association, and bandwidth allocation.

\subsection{Transition Probabilities}

The transition kernel describes the environment dynamics:
\begin{equation}
	p(s'\mid s,a) = \Prob(S_{t+1}=s' \mid S_t=s, A_t=a).
\end{equation}
If reward is included jointly, the kernel is
\begin{equation}
	p(s',r\mid s,a) = \Prob(S_{t+1}=s', R_{t+1}=r \mid S_t=s, A_t=a).
\end{equation}
This kernel can be deterministic or stochastic. A deterministic transition is a special case where one next state has probability one.

\begin{figure}[t]
	\centering
	\begin{tikzpicture}[
		state/.style={circle, draw=blue!70, thick, fill=blue!10, minimum size=0.9cm, font=\bfseries},
		action/.style={rectangle, draw=orange!80, rounded corners, thick, fill=orange!15, minimum width=1.2cm, minimum height=0.7cm, align=center, font=\bfseries},
		arrow/.style={-{Latex[length=2.3mm]}, thick},
		prob/.style={font=\small\bfseries}
		]
		\node[state] (s) at (0,0) {$s$};
		\node[action] (a) at (2,0) {$a$};
		\node[state, draw=green!60!black, fill=green!10] (s1) at (5,1.4) {$s'_1$};
		\node[state, draw=green!60!black, fill=green!10] (s2) at (5,0) {$s'_2$};
		\node[state, draw=green!60!black, fill=green!10] (s3) at (5,-1.4) {$s'_3$};
		
		\draw[arrow, color=blue!60] (s) -- (a);
		\draw[arrow, color=green!60!black] (a) -- node[above, prob, color=red!70!black] {$0.60$} (s1);
		\draw[arrow, color=green!60!black] (a) -- node[above, prob, color=red!70!black] {$0.30$} (s2);
		\draw[arrow, color=green!60!black] (a) -- node[below, prob, color=red!70!black] {$0.10$} (s3);
		
		\node[below=0.3cm of s3, anchor=north, align=center, font=\small, color=gray!70!black] {$p(s'\mid s,a)$ is a probability distribution over next states.};
	\end{tikzpicture}
	\caption{A transition kernel maps a state-action pair to a probability distribution over next states. This is how an MDP represents uncertainty in the environment.}
	\label{fig:transition_kernel}
\end{figure}

\subsection{Rewards}

The reward function gives immediate feedback. There are several equivalent forms:
\begin{align}
	r(s,a) &= \E[R_{t+1}\mid S_t=s,A_t=a], \\
	r(s,a,s') &= \E[R_{t+1}\mid S_t=s,A_t=a,S_{t+1}=s'].
\end{align}

The reward is part of the model, not a fact of nature. In many applications, reward design is a modeling choice. A good reward should reflect the real objective of the task. A bad reward can produce behavior that is optimal for the formula but undesirable in the real system.

\subsection{Discount Factor}

The discount factor $\gamma$ controls how future rewards are weighted:
\begin{equation}
	G_t = \sum_{k=0}^{\infty}\gamma^k R_{t+k+1}.
\end{equation}
When $\gamma$ is close to zero, the agent is short-sighted. When $\gamma$ is close to one, the agent values future rewards more strongly.

Discounting also has mathematical advantages. In infinite-horizon problems with bounded rewards and $\gamma<1$, the return is finite and the Bellman operator becomes a contraction, which gives strong convergence guarantees for dynamic programming methods \citep{puterman1994markov,bertsekas2012dynamic}.

\section{Trajectories, Episodes, and Horizons}

A trajectory is a sequence generated by interaction between a policy and an environment:
\begin{equation}
	\tau = (S_0,A_0,R_1,S_1,A_1,R_2,\ldots).
\end{equation}
A finite episode terminates after some time $T$:
\begin{equation}
	\tau = (S_0,A_0,R_1,\ldots,S_T).
\end{equation}

MDPs may be categorized by their horizon:
\begin{itemize}[leftmargin=*]
	\item \textbf{Finite-horizon MDP}: the task lasts for a fixed number of steps $T$.
	\item \textbf{Episodic MDP}: the task ends when a terminal state is reached.
	\item \textbf{Infinite-horizon discounted MDP}: the task continues indefinitely but future rewards are discounted.
	\item \textbf{Average-reward MDP}: the objective is long-run average reward instead of discounted return.
\end{itemize}

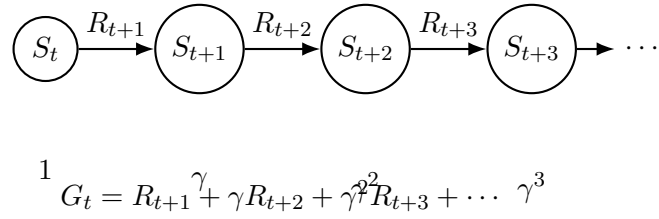
\begin{figure}[t]
	\centering
	\begin{tikzpicture}[
		tick/.style={circle, draw, thick, minimum size=0.7cm},
		arrow/.style={-{Latex[length=2.4mm]}, thick},
		node distance=1.0cm
		]
		\node[tick] (s0) {$S_t$};
		\node[tick, right=of s0] (s1) {$S_{t+1}$};
		\node[tick, right=of s1] (s2) {$S_{t+2}$};
		\node[tick, right=of s2] (s3) {$S_{t+3}$};
		\node[right=0.5cm of s3] (dots) {$\cdots$};
		
		\draw[arrow] (s0) -- node[above] {$R_{t+1}$} (s1);
		\draw[arrow] (s1) -- node[above] {$R_{t+2}$} (s2);
		\draw[arrow] (s2) -- node[above] {$R_{t+3}$} (s3);
		\draw[arrow] (s3) -- (dots);
		
		\node[below=0.9cm of s0] {$1$};
		\node[below=0.9cm of s1] {$\gamma$};
		\node[below=0.9cm of s2] {$\gamma^2$};
		\node[below=0.9cm of s3] {$\gamma^3$};
		
		\node[below=1.6cm of $(s1)!0.5!(s2)$, align=center]
		{$G_t = R_{t+1}+\gamma R_{t+2}+\gamma^2 R_{t+3}+\cdots$};
	\end{tikzpicture}
	\caption{Discounted return assigns geometrically decreasing weights to future rewards. The discount factor balances immediate and delayed consequences.}
	\label{fig:discounted_return}
\end{figure}

\section{Policies}

A policy defines how the agent chooses actions. A stochastic policy is written as
\begin{equation}
	\pi(a\mid s) = \Prob(A_t=a\mid S_t=s).
\end{equation}
A deterministic policy is a special case:
\begin{equation}
	a = \pi(s).
\end{equation}

Policies may also be classified as stationary or non-stationary. A stationary policy uses the same decision rule at every time step:
\begin{equation}
	\pi_t(a\mid s)=\pi(a\mid s), \quad \forall t.
\end{equation}
A non-stationary policy may change with time:
\begin{equation}
	\pi_t(a\mid s).
\end{equation}
For infinite-horizon discounted MDPs under standard assumptions, there exists an optimal deterministic stationary policy \citep{puterman1994markov}. This is one reason why stationary policies are central in reinforcement learning.

\section{Value Functions}

A value function measures the expected long-term return. The state-value function of a policy $\pi$ is
\begin{equation}
	V^\pi(s) = \E_\pi[G_t \mid S_t=s].
	\label{eq:v_def}
\end{equation}
The action-value function is
\begin{equation}
	Q^\pi(s,a) = \E_\pi[G_t \mid S_t=s,A_t=a].
	\label{eq:q_def}
\end{equation}

These definitions are simple, but they are the foundation of many RL algorithms. If an agent knows $Q^\pi(s,a)$, it can compare actions. If it knows $V^\pi(s)$, it can evaluate how promising a state is under the current policy.

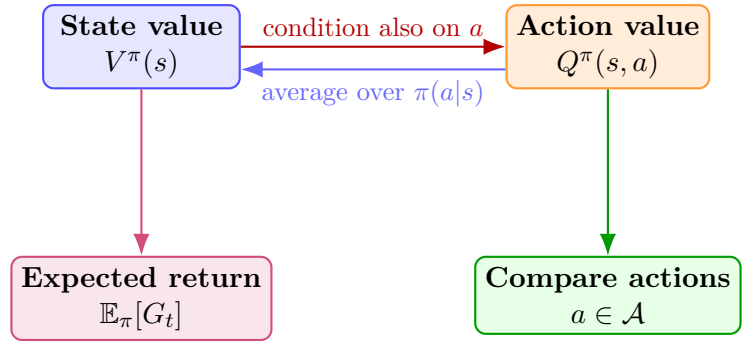
\begin{figure}[t]
	\centering
	\begin{tikzpicture}[
		box/.style={draw, rounded corners, thick, minimum width=2.6cm, minimum height=0.9cm, align=center},
		arrow/.style={-{Latex[length=2.9mm]}, thick},
		node distance=2.2cm and 3.5cm
		]
		\node[box, draw=blue!70, fill=blue!10, font=\bfseries] (s) {State value\\$V^\pi(s)$};
		\node[box, draw=orange!80, fill=orange!15, font=\bfseries, right=of s] (qa) {Action value\\$Q^\pi(s,a)$};
		\node[box, draw=green!60!black, fill=green!10, font=\bfseries, below=of qa] (choice) {Compare actions\\$a\in\mathcal{A}$};
		\node[box, draw=purple!70, fill=purple!10, font=\bfseries, below=of s] (future) {Expected return\\$\mathbb{E}_\pi[G_t]$};
		
		\draw[arrow, color=red!70!black] (s) -- node[above, font=\small, color=red!70!black] {condition also on $a$} (qa);
		\draw[arrow, color=green!60!black] (qa) -- (choice);
		\draw[arrow, color=purple!70] (s) -- (future);
		\draw[arrow, color=blue!60] ([yshift=-3mm]qa.west) -- node[below, font=\small, color=blue!60] {average over $\pi(a|s)$} ([yshift=-3mm]s.east);
	\end{tikzpicture}
	\caption{State values evaluate states, while action values evaluate state-action pairs. The two functions are related by averaging $Q^\pi(s,a)$ over the policy's action probabilities.}
	\label{fig:value_q_relation}
\end{figure}

The relation between $V^\pi$ and $Q^\pi$ is
\begin{equation}
	V^\pi(s) = \sum_{a\in\Aset}\pi(a\mid s)Q^\pi(s,a),
	\label{eq:v_from_q}
\end{equation}
for discrete actions. The action-value function can be written as
\begin{equation}
	Q^\pi(s,a) = \sum_{s',r}p(s',r\mid s,a)\left[r + \gamma V^\pi(s')\right].
	\label{eq:q_from_v}
\end{equation}

\section{The Bellman Expectation Equations}

The Bellman idea is that the value of the present equals immediate reward plus discounted value of the future. This recursive decomposition was central to dynamic programming \citep{bellman1957dynamic,howard1960dynamic} and remains central in RL.

Starting from the definition of return,
\begin{align}
	G_t &= R_{t+1} + \gamma R_{t+2} + \gamma^2R_{t+3}+\cdots \\
	&= R_{t+1}+\gamma G_{t+1}.
\end{align}
Taking expectation under policy $\pi$ gives the Bellman expectation equation:
\begin{equation}
	V^\pi(s)
	=
	\sum_{a}\pi(a\mid s)\sum_{s',r}p(s',r\mid s,a)
	\left[r+\gamma V^\pi(s')\right].
	\label{eq:bellman_expectation_v}
\end{equation}
For action values,
\begin{equation}
	Q^\pi(s,a)
	=
	\sum_{s',r}p(s',r\mid s,a)
	\left[r+\gamma\sum_{a'}\pi(a'\mid s')Q^\pi(s',a')\right].
	\label{eq:bellman_expectation_q}
\end{equation}

\begin{figure}[t]
	\centering
	\begin{tikzpicture}[
		state/.style={circle, draw=blue!70, thick, fill=blue!10, minimum size=0.9cm, font=\bfseries},
		action/.style={rectangle, draw=orange!80, rounded corners, thick, fill=orange!15, minimum width=0.9cm, minimum height=0.7cm, align=center, font=\bfseries},
		nextstate/.style={circle, draw=green!60!black, thick, fill=green!10, minimum size=0.9cm, font=\bfseries},
		arrow/.style={-{Latex[length=2.2mm]}, thick},
		labelnode/.style={font=\small\bfseries}
		]
		\node[state] (s) at (0,0) {$s$};
		\node[action] (a1) at (2,1.2) {$a_1$};
		\node[action] (a2) at (2,-1.2) {$a_2$};
		\node[nextstate] (sp1) at (5,2.0) {$s'_1$};
		\node[nextstate] (sp2) at (5,0.7) {$s'_2$};
		\node[nextstate] (sp3) at (5,-0.7) {$s'_3$};
		\node[nextstate] (sp4) at (5,-2.0) {$s'_4$};
		
		\draw[arrow, color=blue!60] (s) -- node[above left, labelnode, color=red!70!black] {$\pi(a_1|s)$} (a1);
		\draw[arrow, color=blue!60] (s) -- node[below left, labelnode, color=red!70!black] {$\pi(a_2|s)$} (a2);
		\draw[arrow, color=green!60!black] (a1) -- node[above, labelnode, color=purple!70] {$p,r$} (sp1);
		\draw[arrow, color=green!60!black] (a1) -- node[below, labelnode, color=purple!70] {$p,r$} (sp2);
		\draw[arrow, color=green!60!black] (a2) -- node[above, labelnode, color=purple!70] {$p,r$} (sp3);
		\draw[arrow, color=green!60!black] (a2) -- node[below, labelnode, color=purple!70] {$p,r$} (sp4);
		
		\node[below=0.3cm of sp4, anchor=north, align=center, font=\small, color=gray!70!black] {Bellman backup: combine immediate reward with discounted value of possible next states.};
	\end{tikzpicture}
	\caption{A Bellman expectation backup. The value of a state under policy $\pi$ is computed by averaging over actions selected by the policy and next states generated by the transition model.}
	\label{fig:bellman_backup_expectation}
\end{figure}

\begin{keybox}{Why Bellman equations matter}
	Bellman equations turn a long-term problem into a one-step consistency condition. Instead of directly reasoning about an entire future trajectory, the agent asks whether the current value estimate agrees with immediate reward plus discounted next-state value.
\end{keybox}

\section{Optimal Value Functions and the Bellman Optimality Equations}

The goal of control is to find a policy that maximizes expected return. A policy $\pi$ is better than or equal to a policy $\pi'$ if
\begin{equation}
	V^\pi(s) \geq V^{\pi'}(s), \quad \forall s\in\Sset.
\end{equation}
An optimal policy $\pi^*$ satisfies
\begin{equation}
	V^{\pi^*}(s) = V^*(s) = \max_\pi V^\pi(s).
\end{equation}
The optimal action-value function is
\begin{equation}
	Q^*(s,a) = \max_\pi Q^\pi(s,a).
\end{equation}

The Bellman optimality equation for $V^*$ is
\begin{equation}
	V^*(s)
	=
	\max_{a\in\Aset}\sum_{s',r}p(s',r\mid s,a)
	\left[r+\gamma V^*(s')\right].
	\label{eq:bellman_optimality_v}
\end{equation}
The Bellman optimality equation for $Q^*$ is
\begin{equation}
	Q^*(s,a)
	=
	\sum_{s',r}p(s',r\mid s,a)
	\left[r+\gamma\max_{a'}Q^*(s',a')\right].
	\label{eq:bellman_optimality_q}
\end{equation}

Once $Q^*$ is known, an optimal greedy policy is
\begin{equation}
	\pi^*(s) \in \argmax_{a\in\Aset} Q^*(s,a).
	\label{eq:greedy_optimal_policy}
\end{equation}

\begin{figure}[t]
	\centering
	\begin{tikzpicture}[
		box/.style={draw, rounded corners, thick, minimum width=3.0cm, minimum height=0.9cm, align=center},
		arrow/.style={-{Latex[length=2.4mm]}, thick},
		node distance=0.9cm
		]
		\node[box] (policy_eval) {Policy evaluation\\compute $V^\pi$};
		\node[box, right=of policy_eval] (policy_improve) {Policy improvement\\make $\pi$ greedy};
		\node[box, below=of policy_eval] (prediction) {Prediction\\How good is $\pi$?};
		\node[box, below=of policy_improve] (control) {Control\\What is optimal?};
		
		\draw[arrow] (policy_eval) -- (policy_improve);
		\draw[arrow] (policy_improve) .. controls +(0,1.0) and +(0,1.0) .. (policy_eval);
		\draw[arrow] (policy_eval) -- (prediction);
		\draw[arrow] (policy_improve) -- (control);
		
		\node[below=1.2cm of $(prediction)!0.5!(control)$, align=center]
		{Bellman expectation equations support prediction; Bellman optimality equations support control.};
	\end{tikzpicture}
	\caption{Prediction and control in MDPs. Policy evaluation estimates the value of a fixed policy. Policy improvement changes the policy toward actions with higher value. Repeating these ideas leads to dynamic programming and many RL algorithms.}
	\label{fig:prediction_control}
\end{figure}

\section{A Small Grid-World MDP}

A grid world is a useful example because the state and action spaces are easy to see. Suppose a robot moves in a small grid. The goal is to reach a charging station. Hitting a wall gives negative reward. Reaching the charger gives positive reward. Every ordinary step gives a small cost, encouraging shorter paths.

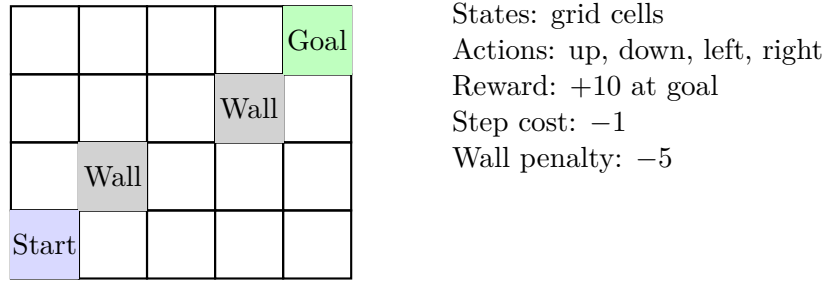
\begin{figure}[t]
	\centering
	\begin{tikzpicture}[scale=0.9]
		\foreach \x in {0,1,2,3,4} {
			\foreach \y in {0,1,2,3} {
				\draw[thick] (\x,\y) rectangle +(1,1);
			}
		}
		\fill[gray!35] (1,1) rectangle +(1,1);
		\fill[gray!35] (3,2) rectangle +(1,1);
		\fill[green!25] (4,3) rectangle +(1,1);
		\fill[blue!15] (0,0) rectangle +(1,1);
		
		\node at (0.5,0.5) {Start};
		\node at (4.5,3.5) {Goal};
		\node at (1.5,1.5) {Wall};
		\node at (3.5,2.5) {Wall};
		
		\node[align=left, right=1.2cm] at (5,2.8) {States: grid cells\\Actions: up, down, left, right\\Reward: $+10$ at goal\\Step cost: $-1$\\Wall penalty: $-5$};
	\end{tikzpicture}
	\caption{A small grid-world MDP. Each cell can be treated as a state, and movement commands are actions. If movement is noisy, the transition model assigns probabilities to possible next cells.}
	\label{fig:gridworld_mdp}
\end{figure}

This example illustrates how an MDP is specified:
\begin{itemize}[leftmargin=*]
	\item $\Sset$ is the set of grid cells;
	\item $\Aset=\{\text{up},\text{down},\text{left},\text{right}\}$;
	\item $p(s'\mid s,a)$ describes the probability of moving to a neighboring cell;
	\item $r(s,a,s')$ gives step cost, wall penalty, or goal reward;
	\item $\gamma$ controls whether the robot prefers fast goal-reaching or longer-term reward.
\end{itemize}

If movement is deterministic, taking action ``right'' always moves the robot one cell right unless blocked. If movement is stochastic, ``right'' might move right with probability $0.8$, slip upward with probability $0.1$, and slip downward with probability $0.1$. This small change makes the task more realistic and shows why transition probabilities matter.

\section{Model-Based and Model-Free Views}

An MDP contains a transition model and reward model. If these are known, the agent can use dynamic programming methods such as policy iteration or value iteration \citep{bellman1957dynamic,howard1960dynamic,puterman1994markov}. If these are unknown, the agent must learn from interaction. This distinction leads to model-based and model-free reinforcement learning.

\begin{table}[t]
	\centering
	\caption{Known-model and unknown-model perspectives on MDPs.}
	\label{tab:model_based_model_free}
	\begin{tabular}{p{0.28\textwidth}p{0.30\textwidth}p{0.30\textwidth}}
		\toprule
		Setting & What is available? & Typical methods \\
		\midrule
		Known MDP model & $p(s',r\mid s,a)$ is known or can be queried exactly & Dynamic programming, planning, value iteration, policy iteration \\
		Sample-based model & Samples from the environment are available; model may be estimated & Model-based RL, Dyna, learned dynamics, planning with learned model \\
		Model-free RL & No explicit transition model is learned or used for planning & Monte Carlo, TD learning, SARSA, Q-learning, policy gradients \\
		\bottomrule
	\end{tabular}
\end{table}

The same MDP formalism supports both views. The difference is whether the agent knows, estimates, or ignores the transition model.

\section{MDP Assumptions and Practical Modeling Issues}

The MDP framework is powerful, but it relies on modeling assumptions. In practice, three issues are especially important.

\subsection{State Representation}

The Markov property depends on the state. If important information is missing, learning may become unstable or suboptimal. For example, a UAV controller that observes only current user demand but not battery level cannot make good charging decisions. Battery level must be part of the state.

\subsection{Stationarity}

Standard MDPs assume that the transition and reward functions do not change over time. In real systems, user behavior, network traffic, weather, or adversarial conditions may change. If the environment changes, the agent may need continual learning, adaptive policies, robust RL, or non-stationary models.

\subsection{Partial Observability}

When the agent cannot observe the full state, the problem is better modeled as a partially observable Markov decision process, or POMDP \citep{kaelbling1998planning}. A POMDP introduces observations and belief states. In deep RL, recurrent networks and memory mechanisms are often used to handle partial observability.

\begin{warningbox}{Important modeling warning}
	An MDP is not automatically correct just because we write down states and actions. The state representation must contain the information needed for future prediction. If it does not, the agent is solving a different problem from the one we intended.
\end{warningbox}

\section{Running Example: UAV-Assisted Network Control as an MDP}

The UAV network example from Chapter~1 can now be written more formally. Suppose multiple UAVs serve ground users with different QoS requirements. At each time step, the system state might include
\begin{equation}
	S_t = \big( X_t^{\mathrm{uav}}, B_t, X_t^{\mathrm{user}}, C_t, L_t, Q_t, I_t \big),
\end{equation}
where $X_t^{\mathrm{uav}}$ denotes UAV positions, $B_t$ battery levels, $X_t^{\mathrm{user}}$ user positions, $C_t$ channel conditions, $L_t$ latency, $Q_t$ queue or QoS state, and $I_t$ interference information.

An action may include movement, service association, and resource allocation:
\begin{equation}
	A_t = \big( A_t^{\mathrm{move}}, A_t^{\mathrm{assoc}}, A_t^{\mathrm{bw}}, A_t^{\mathrm{power}} \big).
\end{equation}

A reward might be
\begin{equation}
	R_{t+1}
	=
	w_q R_{t+1}^{\mathrm{QoS}}
	- w_e C_{t+1}^{\mathrm{energy}}
	- w_s C_{t+1}^{\mathrm{safety}}
	- w_f C_{t+1}^{\mathrm{fairness}}.
	\label{eq:uav_mdp_reward}
\end{equation}

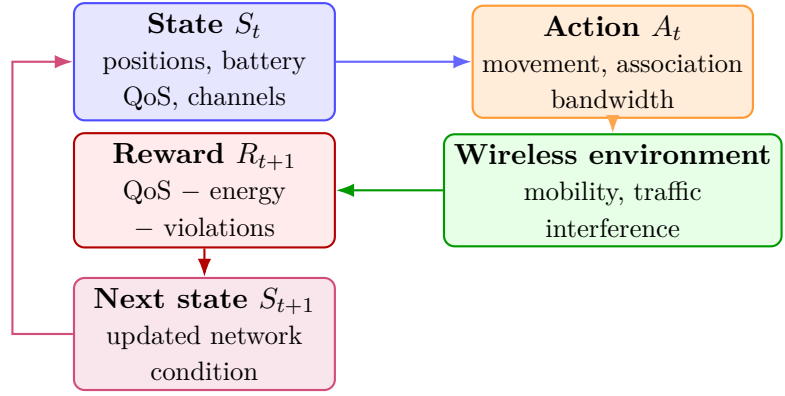
\begin{figure}[t]
	\centering
	\begin{tikzpicture}[
		block/.style={draw, rounded corners, thick, minimum width=3.45cm, minimum height=1.08cm, align=center},
		arrow/.style={-{Latex[length=2.4mm]}, thick}
		]
		\node[block, draw=blue!70, fill=blue!10, font=\bfseries] (state) at (0,2.4) {State $S_t$\\{\normalfont\small positions, battery}\\{\normalfont\small QoS, channels}};
		\node[block, draw=orange!80, fill=orange!15, font=\bfseries] (action) at (5.4,2.4) {Action $A_t$\\{\normalfont\small movement, association}\\{\normalfont\small bandwidth}};
		\node[block, draw=green!60!black, fill=green!10, font=\bfseries] (env) at (5.4,0.7) {Wireless environment\\{\normalfont\small mobility, traffic}\\{\normalfont\small interference}};
		\node[block, draw=red!70!black, fill=red!8, font=\bfseries] (reward) at (0,0.7) {Reward $R_{t+1}$\\{\normalfont\small QoS $-$ energy}\\{\normalfont\small $-$ violations}};
		\node[block, draw=purple!70, fill=purple!10, font=\bfseries] (next) at (0,-1.2) {Next state $S_{t+1}$\\{\normalfont\small updated network}\\{\normalfont\small condition}};
		
		\draw[arrow, color=blue!60] (state) -- (action);
		\draw[arrow, color=orange!70] (action) -- (env);
		\draw[arrow, color=green!60!black] (env) -- (reward);
		\draw[arrow, color=red!70!black] (reward) -- (next);
		\draw[arrow, color=purple!70] (next.west) -- +(-0.8,0) |- (state.west);
	\end{tikzpicture}
	\caption{A UAV-assisted wireless network as an MDP. The state summarizes the network condition, the action controls mobility and resource allocation, the environment evolves stochastically, and the reward evaluates QoS, energy, and safety.}
	\label{fig:uav_mdp_loop}
\end{figure}

This example also shows why the MDP formulation can become difficult in real systems. The state may be high-dimensional, the action may mix discrete and continuous choices, the transition model may be unknown, and rewards may involve conflicting objectives. These difficulties are exactly why deep RL, multi-agent RL, and safe RL become necessary in later chapters.

\section{From MDPs to Algorithms}

The MDP formalism prepares the ground for algorithms. If the model is known, dynamic programming can compute value functions. If the model is unknown, reinforcement learning estimates values or policies from sampled experience.

The next chapter studies three classical routes:
\begin{enumerate}[leftmargin=*]
	\item \textbf{Dynamic programming}: assumes known dynamics and uses Bellman equations directly;
	\item \textbf{Monte Carlo learning}: estimates value from complete sampled returns;
	\item \textbf{Temporal-difference learning}: learns from one-step bootstrapped targets.
\end{enumerate}

All three are built on the same MDP foundation.

\section{Key Takeaways}

\begin{itemize}[leftmargin=*]
	\item An MDP formalizes sequential decision-making under uncertainty.
	\item The Markov property says that the current state and action contain all information needed to predict the next state and reward.
	\item A discounted MDP is commonly written as $(\Sset,\Aset,p,r,\gamma)$.
	\item A policy maps states to actions or action probabilities.
	\item The return is the discounted sum of future rewards.
	\item The value function $V^\pi(s)$ evaluates states under a policy.
	\item The action-value function $Q^\pi(s,a)$ evaluates actions in states under a policy.
	\item Bellman expectation equations describe the value of a fixed policy.
	\item Bellman optimality equations describe the value of optimal behavior.
	\item MDP modeling choices matter: poor state design, non-stationarity, and partial observability can break the assumptions.
\end{itemize}

\section*{Looking Ahead to Chapter 3:  Food for Thought}
\addcontentsline{toc}{section}{Looking Ahead to Chapter 3: Food for Thought}

Chapter~2 gave the mathematical model. Chapter~3 asks how to compute or learn useful values and policies.

Before moving on, consider the following questions:
\begin{enumerate}[leftmargin=*]
	\item If the transition model $p(s',r\mid s,a)$ is known, can we compute the optimal policy without trial and error?
	\item If the transition model is unknown, how can an agent estimate values from experience?
	\item Why might waiting until the end of an episode be inefficient for learning?
	\item What does it mean to update a value estimate using another value estimate?
	\item Why is the Bellman equation both a definition of consistency and a practical learning target?
	\item How does Q-learning use the Bellman optimality equation without knowing the transition probabilities?
\end{enumerate}

\begin{quote}
	Chapter~2 defines the world mathematically. Chapter~3 explains how an agent begins to solve it.
\end{quote}

\section{Exercises}

\subsection*{Conceptual Exercises}
\begin{enumerate}[leftmargin=*]
	\item Explain the Markov property in your own words. Give one example where it holds and one where it fails.
	\item Why is the state representation a modeling choice rather than a fixed truth?
	\item What is the difference between a transition probability and a reward function?
	\item Explain the difference between prediction and control in MDPs.
	\item Why is an MDP useful even when the true environment is more complex than the model?
\end{enumerate}

\subsection*{Mathematical Exercises}
\begin{enumerate}[leftmargin=*]
	\item Let an agent receive rewards $2,2,2,\ldots$ forever with $\gamma=0.8$. Compute $G_0$.
	\item For a policy $\pi$, show that $V^\pi(s)=\sum_a\pi(a\mid s)Q^\pi(s,a)$.
	\item Starting from $G_t=R_{t+1}+\gamma G_{t+1}$, derive the Bellman expectation equation for $V^\pi(s)$.
	\item Write the Bellman optimality equation for $Q^*(s,a)$ and explain each term.
	\item Suppose $\gamma=0$. What does the Bellman optimality equation become?
\end{enumerate}

\subsection*{Modeling Exercises}
\begin{enumerate}[leftmargin=*]
	\item Model a robot vacuum cleaner as an MDP. Define states, actions, rewards, and terminal conditions.
	\item Model a simple SDN routing problem as an MDP. What should be included in the state?
	\item For a UAV network, explain why battery level must be part of the state.
	\item Give an example of a reward function that could cause reward hacking.
	\item Choose a real system and explain whether it is better modeled as an MDP or a POMDP.
\end{enumerate}

%\section{Chapter References}
%
%Bellman introduced dynamic programming as a general framework for sequential optimization \citep{bellman1957dynamic}. Howard developed policy iteration for dynamic programming and Markov processes \citep{howard1960dynamic}. Puterman's book remains a standard reference on Markov decision processes \citep{puterman1994markov}. Sutton and Barto provide the standard reinforcement learning treatment of MDPs, value functions, and Bellman equations \citep{sutton2018reinforcement}. Bertsekas and Tsitsiklis connect dynamic programming with approximate and neuro-dynamic programming \citep{bertsekas1996neurodynamic,bertsekas2012dynamic}. Szepesvari gives a concise algorithmic treatment of reinforcement learning and MDPs \citep{szepesvari2010algorithms}. Kaelbling, Littman, and Cassandra provide a classic survey of planning and acting under partial observability \citep{kaelbling1998planning}.
	\chapter{Dynamic Programming, Monte Carlo, and Temporal-Difference Learning}
\label{ch:dp_mc_td}
\chaptermark{DP, MC, and TD Learning}

\section*{Chapter Overview}
\addcontentsline{toc}{section}{Chapter Overview}

Chapter~2 introduced the Markov decision process (MDP) as the mathematical model behind reinforcement learning. We defined states, actions, transition probabilities, rewards, policies, returns, and value functions. This chapter answers the next question:

\begin{quote}
	Once a reinforcement learning problem is written as an MDP, how can an agent actually learn good values or good behavior?
\end{quote}

The answer begins with three classical families of methods:

\begin{enumerate}[leftmargin=*]
	\item \textbf{Dynamic programming (DP)}: compute value functions and optimal policies when the model of the environment is known.
	\item \textbf{Monte Carlo (MC) learning}: estimate value functions from complete sampled episodes when the model is unknown.
	\item \textbf{Temporal-difference (TD) learning}: learn from sampled experience by combining Monte Carlo sampling with dynamic-programming-style bootstrapping.
\end{enumerate}

These methods are not old material that can be skipped. They are the conceptual engine behind modern deep reinforcement learning. Deep Q-Networks replace tabular Q-values with neural networks, but the learning target is still a temporal-difference target. Actor-critic methods use TD critics. PPO uses advantage estimates often built from TD residuals. Multi-agent value decomposition, offline RL, and safe RL all inherit ideas from DP, MC, and TD learning.

\begin{intuitionbox}{Main message of the chapter}
	Dynamic programming teaches us how to solve an MDP if we know the model. Monte Carlo teaches us how to learn from complete experience without a model. Temporal-difference learning teaches us how to learn online from incomplete experience. Modern DRL is largely built by scaling these three ideas with function approximation, replay, optimization, and deep neural networks.
\end{intuitionbox}

\section*{Learning Objectives}
\addcontentsline{toc}{section}{Learning Objectives}

After reading this chapter, the reader should be able to:
\begin{enumerate}[leftmargin=*]
	\item distinguish prediction from control in reinforcement learning;
	\item explain generalized policy iteration as the interaction between policy evaluation and policy improvement;
	\item derive dynamic-programming updates for policy evaluation, policy iteration, and value iteration;
	\item explain why dynamic programming requires a known transition and reward model;
	\item define Monte Carlo prediction and control using sampled returns;
	\item derive the TD(0) update and interpret the temporal-difference error;
	\item compare SARSA, Q-learning, and Expected SARSA using their bootstrap targets;
	\item explain how eligibility traces and TD$(\lambda)$ connect one-step TD and Monte Carlo learning;
	\item implement basic tabular DP, MC, and TD algorithms in a small grid-world;
	\item connect classical value-learning methods to UAV-assisted network control and modern deep reinforcement learning.
\end{enumerate}

\section{Why this chapter matters}

Dynamic programming, Monte Carlo learning, and temporal-difference learning are the three basic ways to estimate value functions and improve policies. They are often introduced as separate algorithms, but they should be understood as different answers to one central question:

\begin{quote}
	How can an agent estimate the long-term value of a state or action?
\end{quote}

The difficulty is that value is not directly observed. The agent observes immediate rewards. But the value of a state is the expected return from that state, and return depends on the future. Therefore, value learning is a problem of assigning long-term consequences to earlier decisions.

In Chapter~2, the state-value function was defined as
\begin{equation}
	V^{\pi}(s) = \mathbb{E}_{\pi}\left[G_t \mid S_t=s\right],
\end{equation}
where
\begin{equation}
	G_t = R_{t+1} + \gamma R_{t+2} + \gamma^2 R_{t+3} + \cdots .
\end{equation}
The action-value function was defined as
\begin{equation}
	Q^{\pi}(s,a) = \mathbb{E}_{\pi}\left[G_t \mid S_t=s, A_t=a\right].
\end{equation}

The challenge is to compute or estimate these quantities. If the model is known, dynamic programming can use Bellman equations directly. If the model is unknown but full episodes are available, Monte Carlo methods can average sampled returns. If the model is unknown and learning must happen step by step, temporal-difference learning can update values using reward plus an estimate of future value.

Historically, dynamic programming comes from Bellman's work on sequential decision-making and optimality principles \citep{bellman1957dynamic}. Policy iteration was developed in the dynamic-programming tradition and is commonly associated with Howard's work on Markov processes \citep{howard1960dynamic}. Temporal-difference learning was formalized by Sutton \citep{sutton1988learning}. Q-learning was introduced by Watkins and analyzed by Watkins and Dayan \citep{watkins1989learning,watkins1992q}. SARSA emerged from on-line action-value learning work by Rummery and Niranjan and was later named according to the transition tuple $(S_t,A_t,R_{t+1},S_{t+1},A_{t+1})$ \citep{rummery1994online,sutton2018reinforcement}.

\begin{figure}[t]
	\centering
	\begin{tikzpicture}[
		box/.style={draw, rounded corners, thick, minimum width=3.4cm, minimum height=0.9cm, align=center, font=\small},
		arrow/.style={-{Latex[length=2.5mm]}, thick}
		]
		\node[box, draw=blue!70, fill=blue!10] (dp) at (0,0) {Dynamic Programming\\{\footnotesize Known model $p(s',r|s,a)$}};
		\node[box, draw=orange!80, fill=orange!15] (mc) at (6,0) {Monte Carlo\\{\footnotesize Complete sampled episodes}};
		\node[box, draw=green!60!black, fill=green!10] (td) at (6,-3) {Temporal Difference\\{\footnotesize One-step sampled transitions}};
		
		\draw[arrow, color=red!70!black] (dp) -- node[above=5pt, font=\scriptsize, color=red!70!black, align=center] {less model\\knowledge} (mc);
		\draw[arrow, color=purple!70] (mc) -- node[right=5pt, font=\scriptsize, color=purple!70, align=center, pos=0.6] {less waiting\\for full returns} (td);
		
		\node[above=12pt of dp, font=\scriptsize, color=blue!70] {Exact Bellman expectations};
		\node[above=12pt of mc, font=\scriptsize, color=orange!80] {Sampled returns $G_t$};
		\node[below=12pt of td, font=\scriptsize, color=green!60!black] {Bootstrapped $R_{t+1}+\gamma V(S_{t+1})$};
	\end{tikzpicture}
	\caption{Dynamic programming, Monte Carlo learning, and temporal-difference learning form a spectrum. DP assumes a known model, MC samples complete returns, and TD bootstraps from one-step transitions.}
	\label{fig:dp_mc_td_spectrum}
\end{figure}
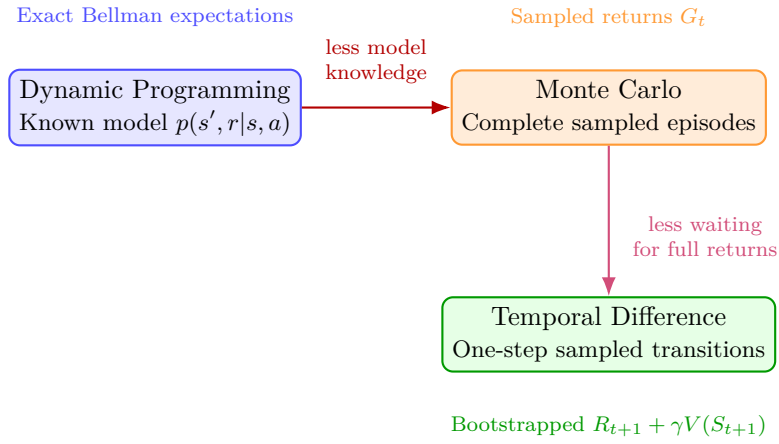

\section{Prediction versus control}

Before studying algorithms, we must distinguish two tasks: \textbf{prediction} and \textbf{control}.

\subsection{Prediction}

Prediction means evaluating a fixed policy. The policy is already given. The question is:

\begin{quote}
	If the agent follows policy $\pi$, how good are the states or state-action pairs?
\end{quote}

For state values, prediction estimates $V^{\pi}(s)$. For action values, prediction estimates $Q^{\pi}(s,a)$.

Examples:
\begin{itemize}[leftmargin=*]
	\item Given a robot navigation policy, estimate the expected time or reward from each room.
	\item Given a network routing policy, estimate the expected long-term QoS from each traffic state.
	\item Given a UAV placement policy, estimate how valuable each UAV configuration is.
\end{itemize}

\subsection{Control}

Control means finding a better policy. The question is:

\begin{quote}
	How should the agent change its behavior to obtain higher return?
\end{quote}

Control usually alternates between two operations:
\begin{enumerate}[leftmargin=*]
	\item evaluate the current policy;
	\item improve the policy using the estimated values.
\end{enumerate}

This alternating process is the basis of generalized policy iteration.

\begin{table}[t]
	\centering
	\begin{tabular}{p{0.22\textwidth}p{0.34\textwidth}p{0.34\textwidth}}
		\toprule
		Task & Main question & Typical output \\
		\midrule
		Prediction & How good is this policy? & $V^{\pi}$ or $Q^{\pi}$ \\
		Control & What policy should the agent use? & improved policy $\pi'$ or optimal policy $\pi^*$ \\
		\bottomrule
	\end{tabular}
	\caption{Prediction evaluates a given policy, while control improves or learns a policy. Almost all classical RL algorithms can be understood through this distinction.}
	\label{tab:prediction_control}
\end{table}

\section{Generalized policy iteration}

Generalized policy iteration (GPI) is one of the most important conceptual patterns in reinforcement learning \citep{sutton2018reinforcement}. It describes the interaction between policy evaluation and policy improvement.

Policy evaluation makes the value function consistent with the current policy. Policy improvement makes the policy greedy or more nearly greedy with respect to the current value function. These two processes push each other. The value function tracks the policy; the policy moves toward actions suggested by the value function.

\begin{figure}[t]
	\centering
	\begin{tikzpicture}[
		box/.style={draw, rounded corners, thick, minimum width=3.7cm, minimum height=1.1cm, align=center},
		arrow/.style={-{Latex[length=2.7mm]}, thick},
		node distance=2.5cm
		]
		\node[box, draw=blue!70, fill=blue!10, font=\bfseries] (eval) {Policy Evaluation\\{\normalfont\small Estimate $V^{\pi}$ or $Q^{\pi}$}};
		\node[box, draw=orange!80, fill=orange!15, font=\bfseries, right=of eval] (improve) {Policy Improvement\\{\normalfont\small Make $\pi$ greedy or near-greedy}};
		
		\draw[arrow, color=green!60!black] (eval.north east) .. controls +(0.8,1.1) and +(-0.8,1.1) .. node[above, font=\small, color=green!60!black] {values guide behavior} (improve.north west);
		\draw[arrow, color=red!70!black] (improve.south west) .. controls +(-0.8,-1.1) and +(0.8,-1.1) .. node[below, font=\small, color=red!70!black] {new behavior changes values} (eval.south east);
		
		\node[below=2.0cm of $(eval)!0.5!(improve)$, align=center, font=\small, color=gray!70!black] {When both processes stabilize, the policy is greedy with respect to its own value function.\\In finite discounted MDPs, this corresponds to optimality under standard assumptions.};
	\end{tikzpicture}
	\caption{Generalized policy iteration. Evaluation estimates the value of a policy; improvement changes the policy using the value estimate. Dynamic programming, Monte Carlo control, SARSA, Q-learning, and many actor-critic methods can be interpreted as instances of this pattern.}
	\label{fig:gpi_loop}
\end{figure}

The important point is that GPI does not require full exact evaluation before improvement. Policy iteration performs exact or near-exact evaluation, then improvement. Value iteration truncates evaluation to one Bellman optimality backup. Monte Carlo and TD methods perform evaluation from sampled experience while the policy changes gradually. Modern actor-critic algorithms also follow a form of GPI: the critic evaluates, and the actor improves.

\section{Dynamic programming with a known model}

Dynamic programming solves MDPs by using the model of the environment. The model is the transition and reward distribution
\begin{equation}
	p(s',r \mid s,a) = \Pr(S_{t+1}=s', R_{t+1}=r \mid S_t=s, A_t=a).
\end{equation}
If this model is known, then Bellman equations can be turned into iterative algorithms.

DP is powerful because it provides the cleanest form of the Bellman idea. It is also limited because most real environments do not provide exact transition probabilities. Still, DP is essential because it gives the target structure that later sample-based algorithms approximate.

\subsection{Bellman expectation backup}

For a fixed policy $\pi$, the Bellman expectation equation is
\begin{equation}
	V^{\pi}(s)
	= \sum_a \pi(a|s) \sum_{s',r} p(s',r|s,a)\left[r + \gamma V^{\pi}(s')\right].
	\label{eq:bellman_expectation_v_ch3}
\end{equation}
This equation says that the value of state $s$ is the expected immediate reward plus the discounted value of the next state, averaged over the policy and the environment dynamics.

The corresponding action-value form is
\begin{equation}
	Q^{\pi}(s,a)
	= \sum_{s',r} p(s',r|s,a)\left[r + \gamma \sum_{a'} \pi(a'|s')Q^{\pi}(s',a')\right].
	\label{eq:bellman_expectation_q_ch3}
\end{equation}

\begin{figure}[t]
	\centering
	\begin{tikzpicture}[
		state/.style={circle, draw=blue!70, thick, fill=blue!10, minimum size=0.9cm, align=center, font=\bfseries},
		leaf/.style={circle, draw=green!60!black, thick, fill=green!10, minimum size=0.75cm, align=center, font=\bfseries},
		arrow/.style={-{Latex[length=2.2mm]}, thick}
		]
		\node[state] (s) at (0,0) {$s$};
		\node[leaf] (s1) at (5,1.5) {$s'_1$};
		\node[leaf] (s2) at (5,0) {$s'_2$};
		\node[leaf] (s3) at (5,-1.5) {$s'_3$};
		
		\draw[arrow, color=red!70!black] (s) -- node[above, sloped, font=\scriptsize, color=red!70!black] {$p(s'_1,r_1|s,a)$} (s1);
		\draw[arrow, color=red!70!black] (s) -- node[above, font=\scriptsize, color=red!70!black] {$p(s'_2,r_2|s,a)$} (s2);
		\draw[arrow, color=red!70!black] (s) -- node[below, sloped, font=\scriptsize, color=red!70!black] {$p(s'_3,r_3|s,a)$} (s3);
		
		\node[right=0.4cm of s1, font=\small, color=purple!70] {$r_1 + \gamma V(s'_1)$};
		\node[right=0.4cm of s2, font=\small, color=purple!70] {$r_2 + \gamma V(s'_2)$};
		\node[right=0.4cm of s3, font=\small, color=purple!70] {$r_3 + \gamma V(s'_3)$};
		
		\node at (2.5,-2.5) [align=center, font=\small, color=gray!70!black] {A Bellman backup averages immediate reward plus discounted successor value.};
	\end{tikzpicture}
	\caption{Bellman backup diagram. The value of state $s$ under action $a$ is computed by averaging reward-plus-discounted-value over all possible next states, weighted by transition probabilities.}
	\label{fig:bellman_backup}
\end{figure}

\subsection{Policy evaluation}

Policy evaluation estimates $V^{\pi}$ for a fixed policy $\pi$. Starting from an arbitrary value function $V_0$, iterative policy evaluation applies
\begin{equation}
	V_{k+1}(s)
	\leftarrow
	\sum_a \pi(a|s)\sum_{s',r}p(s',r|s,a)
	\left[r + \gamma V_k(s')\right].
	\label{eq:iterative_policy_evaluation}
\end{equation}
For finite discounted MDPs, repeated application of this update converges to $V^{\pi}$ under standard assumptions \citep{puterman1994markov,sutton2018reinforcement}.

\begin{algobox}{Algorithm 3.1: Iterative policy evaluation}
	\begin{enumerate}[leftmargin=*]
		\item Initialize $V(s)$ arbitrarily for all states $s$.
		\item Repeat until the maximum change is small:
		\begin{enumerate}
			\item For each state $s$, compute
			\[
			V_{new}(s) = \sum_a \pi(a|s)\sum_{s',r}p(s',r|s,a)
			\left[r+\gamma V(s')\right].
			\]
			\item Replace $V(s)$ by $V_{new}(s)$.
		\end{enumerate}
		\item Return $V$.
	\end{enumerate}
\end{algobox}

\subsection{Policy improvement}

Once we have $V^{\pi}$, we can improve the policy by choosing actions that maximize one-step lookahead:
\begin{equation}
	\pi'(s) \in \arg\max_a \sum_{s',r}p(s',r|s,a)\left[r + \gamma V^{\pi}(s')\right].
	\label{eq:policy_improvement}
\end{equation}
The policy improvement theorem states that if $\pi'$ is greedy with respect to $V^{\pi}$, then $\pi'$ is at least as good as $\pi$ under standard conditions \citep{puterman1994markov,sutton2018reinforcement}.

Intuitively, if the value function tells us what the future is worth under the old policy, then choosing the action with the best immediate reward plus future value cannot make the first step worse. Repeating this process leads toward an optimal policy.

\subsection{Policy iteration}

Policy iteration alternates policy evaluation and policy improvement:
\begin{equation}
	\pi_0 \xrightarrow{\text{evaluate}} V^{\pi_0}
	\xrightarrow{\text{improve}} \pi_1
	\xrightarrow{\text{evaluate}} V^{\pi_1}
	\xrightarrow{\text{improve}} \cdots
\end{equation}
For finite discounted MDPs, policy iteration converges to an optimal policy in a finite number of policy improvements because there are finitely many deterministic policies \citep{howard1960dynamic,puterman1994markov}.

\begin{algobox}{Algorithm 3.2: Policy iteration}
	\begin{enumerate}[leftmargin=*]
		\item Initialize a policy $\pi$ arbitrarily.
		\item \textbf{Policy evaluation}: compute or approximate $V^{\pi}$.
		\item \textbf{Policy improvement}: for each state $s$, set
		\[
		\pi(s) \leftarrow \arg\max_a \sum_{s',r}p(s',r|s,a)[r+\gamma V^{\pi}(s')].
		\]
		\item If the policy is unchanged, stop. Otherwise return to evaluation.
	\end{enumerate}
\end{algobox}

\subsection{Value iteration}

Policy iteration may be expensive because policy evaluation can require many sweeps. Value iteration combines evaluation and improvement by repeatedly applying the Bellman optimality backup:
\begin{equation}
	V_{k+1}(s)
	\leftarrow
	\max_a \sum_{s',r}p(s',r|s,a)\left[r + \gamma V_k(s')\right].
	\label{eq:value_iteration}
\end{equation}
After convergence, an optimal policy is extracted greedily:
\begin{equation}
	\pi^*(s) \in \arg\max_a \sum_{s',r}p(s',r|s,a)\left[r + \gamma V^*(s')\right].
\end{equation}

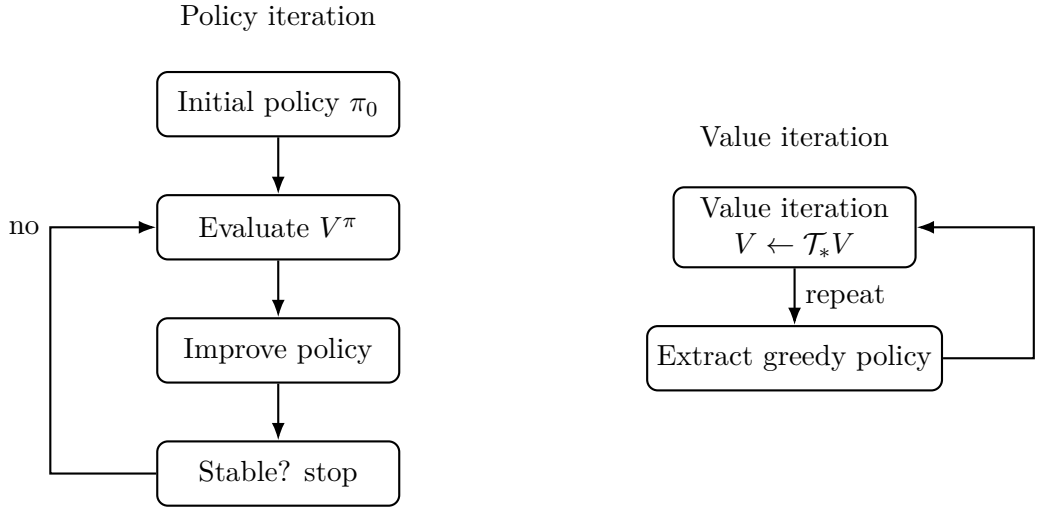
\begin{figure}[t]
	\centering
	\begin{tikzpicture}[
		box/.style={draw, rounded corners, thick, minimum width=3.2cm, minimum height=0.85cm, align=center},
		arrow/.style={-{Latex[length=2.3mm]}, thick},
		node distance=0.75cm
		]
		\node[box] (pi0) {Initial policy $\pi_0$};
		\node[box, below=of pi0] (eval) {Evaluate $V^{\pi}$};
		\node[box, below=of eval] (improve) {Improve policy};
		\node[box, below=of improve] (stop) {Stable? stop};
		\draw[arrow] (pi0) -- (eval);
		\draw[arrow] (eval) -- (improve);
		\draw[arrow] (improve) -- (stop);
		\draw[arrow] (stop.west) -- +(-1.4,0) |- node[left] {no} (eval.west);
		\node[box, right=3.6cm of eval] (vi) {Value iteration\\$V \leftarrow \mathcal{T}_* V$};
		\node[box, below=of vi] (greedy) {Extract greedy policy};
		\draw[arrow] (vi) -- node[right] {repeat} (greedy);
		\draw[arrow] (greedy.east) -- +(1.2,0) |- (vi.east);
		\node[above=0.4cm of pi0] {Policy iteration};
		\node[above=0.4cm of vi] {Value iteration};
	\end{tikzpicture}
	\caption{Policy iteration alternates policy evaluation and policy improvement. Value iteration performs Bellman optimality backups directly and extracts a greedy policy after convergence. Both are dynamic-programming methods that require a known model.}
	\label{fig:policy_vs_value_iteration}
\end{figure}

\subsection{The Bellman operator view}

It is useful to define the Bellman expectation operator $\mathcal{T}^{\pi}$:
\begin{equation}
	(\mathcal{T}^{\pi}V)(s)
	= \sum_a \pi(a|s)\sum_{s',r}p(s',r|s,a)[r+\gamma V(s')].
\end{equation}
The Bellman optimality operator is
\begin{equation}
	(\mathcal{T}_*V)(s)
	= \max_a \sum_{s',r}p(s',r|s,a)[r+\gamma V(s')].
\end{equation}
For discounted finite MDPs with $0\leq \gamma <1$, these operators are contractions in the max norm. This means that repeated application moves value estimates closer to their fixed point. This contraction property is one reason Bellman methods are so mathematically powerful \citep{bertsekas1996neurodynamic,puterman1994markov,sutton2018reinforcement}.

\section{Monte Carlo learning}

Dynamic programming requires the model. Monte Carlo learning removes this requirement. Instead of averaging over all possible next states using $p(s',r|s,a)$, Monte Carlo methods estimate value functions by averaging sampled returns from experience.

The basic Monte Carlo estimator for $V^{\pi}(s)$ is
\begin{equation}
	V(s) \leftarrow \text{average of observed returns following visits to } s.
\end{equation}
If a state $s$ is visited many times and the returns are sampled under policy $\pi$, then the sample average converges to the expected return under standard statistical assumptions.

\begin{figure}[t]
	\centering
	\begin{tikzpicture}[
		state/.style={circle, draw, thick, minimum size=0.75cm},
		reward/.style={rectangle, draw, rounded corners, minimum width=0.8cm, minimum height=0.55cm, align=center},
		arrow/.style={-{Latex[length=2.1mm]}, thick},
		node distance=0.85cm
		]
		\node[state] (s0) {$s$};
		\node[state, right=of s0] (s1) {$s_1$};
		\node[state, right=of s1] (s2) {$s_2$};
		\node[state, right=of s2] (s3) {$s_T$};
		\node[reward, above=0.65cm of $(s0)!0.5!(s1)$] (r1) {$r_1$};
		\node[reward, above=0.65cm of $(s1)!0.5!(s2)$] (r2) {$r_2$};
		\node[reward, above=0.65cm of $(s2)!0.5!(s3)$] (r3) {$r_T$};
		\draw[arrow] (s0) -- (s1);
		\draw[arrow] (s1) -- (s2);
		\draw[arrow] (s2) -- node[above] {$\cdots$} (s3);
		\node[below=1.1cm of $(s1)!0.5!(s2)$, align=center] {$G_t = r_1 + \gamma r_2 + \cdots + \gamma^{T-t-1}r_T$\\MC waits for the complete return.};
	\end{tikzpicture}
	\caption{Monte Carlo learning estimates value by observing complete returns from sampled episodes. It does not require a model, but it must wait until enough future rewards are known.}
	\label{fig:mc_episode_return}
\end{figure}

\subsection{First-visit and every-visit Monte Carlo prediction}

There are two common Monte Carlo prediction variants.

\textbf{First-visit MC} averages returns only after the first time a state is visited in an episode. \textbf{Every-visit MC} averages returns after every visit to the state. In many episodic tasks, both converge to the correct value under appropriate assumptions, though their finite-sample behavior may differ \citep{sutton2018reinforcement}.

Let $N(s)$ be the number of observed returns from state $s$. An incremental mean update is
\begin{equation}
	V(s) \leftarrow V(s) + \frac{1}{N(s)}\left(G_t - V(s)\right).
	\label{eq:mc_incremental_mean}
\end{equation}
More generally, we can use a constant step size $\alpha$:
\begin{equation}
	V(s) \leftarrow V(s) + \alpha\left(G_t - V(s)\right).
	\label{eq:mc_constant_alpha}
\end{equation}
The constant step-size version is useful in non-stationary environments because it gives more weight to recent data.

\begin{algobox}{Algorithm 3.3: First-visit Monte Carlo prediction}
	\begin{enumerate}[leftmargin=*]
		\item Input: policy $\pi$ to evaluate.
		\item Initialize $V(s)$ arbitrarily and returns list or counters for each state.
		\item Generate an episode following $\pi$: $S_0,A_0,R_1,\ldots,S_T$.
		\item For each state $S_t$ appearing for the first time in the episode:
		\begin{enumerate}
			\item Compute $G_t = R_{t+1}+\gamma R_{t+2}+\cdots+\gamma^{T-t-1}R_T$.
			\item Update $V(S_t)$ toward $G_t$.
		\end{enumerate}
		\item Repeat for many episodes.
	\end{enumerate}
\end{algobox}

\subsection{Monte Carlo control}

Monte Carlo control uses returns to learn action values $Q(s,a)$ and improve the policy. The action-value function is important because without a model, the agent cannot easily perform one-step lookahead using $V(s')$. If the agent knows $Q(s,a)$, it can choose actions directly.

A simple MC control update is
\begin{equation}
	Q(S_t,A_t) \leftarrow Q(S_t,A_t) + \alpha\left(G_t - Q(S_t,A_t)\right).
\end{equation}
Then the policy is made greedy or $\epsilon$-greedy with respect to $Q$.

A deterministic greedy policy is
\begin{equation}
	\pi(s) \in \arg\max_a Q(s,a).
\end{equation}
An $\epsilon$-greedy policy chooses the greedy action with high probability and random exploratory actions with small probability:
\begin{equation}
	\pi(a|s)=
	\begin{cases}
		1-\epsilon + \frac{\epsilon}{|\mathcal{A}(s)|}, & a \in \arg\max_{a'} Q(s,a'),\\
		\frac{\epsilon}{|\mathcal{A}(s)|}, & \text{otherwise.}
	\end{cases}
\end{equation}

\begin{warningbox}{Why Monte Carlo control needs exploration}
	If the policy becomes greedy too early, some actions may never be tried. Then their values cannot be estimated. Monte Carlo control needs some mechanism ensuring that useful state-action pairs are sampled. Exploring starts is one theoretical solution; $\epsilon$-greedy exploration is more practical.
\end{warningbox}

\section{Temporal-difference learning}

Temporal-difference learning combines ideas from Monte Carlo learning and dynamic programming. Like Monte Carlo, TD learns from sampled experience and does not require a model. Like dynamic programming, TD bootstraps: it updates estimates using other estimates.

For policy evaluation, the TD(0) update is
\begin{equation}
	V(S_t) \leftarrow V(S_t) + \alpha\left[R_{t+1}+\gamma V(S_{t+1}) - V(S_t)\right].
	\label{eq:td0_update}
\end{equation}
The quantity
\begin{equation}
	\delta_t = R_{t+1}+\gamma V(S_{t+1}) - V(S_t)
	\label{eq:td_error}
\end{equation}
is the \textbf{temporal-difference error}. It measures the difference between the old estimate and a one-step improved estimate.

\begin{figure}[t]
	\centering
	\begin{tikzpicture}[
		box/.style={draw, rounded corners, thick, minimum width=3.6cm, minimum height=0.9cm, align=center},
		arrow/.style={-{Latex[length=2.3mm]}, thick},
		node distance=1.2cm
		]
		\node[box] (mc) {Monte Carlo target\\$G_t$};
		\node[box, below=of mc] (td) {TD target\\$R_{t+1}+\gamma V(S_{t+1})$};
		\node[box, right=2.7cm of mc] (wait) {Waits until episode end};
		\node[box, right=2.7cm of td] (online) {Updates after one step};
		\draw[arrow] (mc) -- (wait);
		\draw[arrow] (td) -- (online);
		\node[below=1.0cm of $(td)!0.5!(online)$, align=center] {MC uses a sampled complete return. TD uses a sampled reward plus a bootstrapped value estimate.};
	\end{tikzpicture}
	\caption{Monte Carlo and TD targets. MC uses the full return $G_t$, while TD(0) uses a one-step bootstrapped target. TD can learn online before the episode ends.}
	\label{fig:mc_vs_td_target}
\end{figure}

\subsection{Why TD learning is important}

TD learning is one of the central ideas of reinforcement learning \citep{sutton1988learning}. It is important for several reasons:

\begin{itemize}[leftmargin=*]
	\item It learns online from incomplete episodes.
	\item It does not require a model.
	\item It can be more data-efficient than pure Monte Carlo learning.
	\item It provides the learning target used in Q-learning, SARSA, DQN, and many actor-critic algorithms.
\end{itemize}

TD learning also introduces a subtle bias-variance tradeoff. The TD target has lower variance because it uses only one sampled reward and a value estimate, but it is biased when the current value estimate is inaccurate. Monte Carlo targets are unbiased samples of the return under the policy, but they may have high variance.

\section{SARSA, Q-learning, and Expected SARSA}

For control without a model, action values are often more useful than state values. The agent can choose actions directly from $Q(s,a)$.

\subsection{SARSA: on-policy TD control}

SARSA is an on-policy TD control algorithm. It updates $Q(S_t,A_t)$ using the action actually selected at the next state:
\begin{equation}
	Q(S_t,A_t) \leftarrow Q(S_t,A_t)
	+ \alpha\left[R_{t+1}+\gamma Q(S_{t+1},A_{t+1}) - Q(S_t,A_t)\right].
	\label{eq:sarsa_update}
\end{equation}
The name SARSA comes from the tuple
\begin{equation}
	(S_t,A_t,R_{t+1},S_{t+1},A_{t+1}).
\end{equation}

SARSA is on-policy because it evaluates and improves the same behavior policy. If the behavior policy is $\epsilon$-greedy, then SARSA learns the value of that exploratory policy. This often makes SARSA more conservative in risky environments because it accounts for the possibility of exploratory actions.

\subsection{Q-learning: off-policy TD control}

Q-learning is an off-policy TD control algorithm \citep{watkins1992q}. It updates toward the greedy action at the next state, regardless of which action the behavior policy actually takes:
\begin{equation}
	Q(S_t,A_t) \leftarrow Q(S_t,A_t)
	+ \alpha\left[R_{t+1}+\gamma \max_{a'}Q(S_{t+1},a') - Q(S_t,A_t)\right].
	\label{eq:q_learning_update}
\end{equation}

Q-learning is off-policy because it can behave exploratorily while learning about the greedy target policy. Under standard tabular assumptions, including sufficient exploration and appropriate step-size conditions, Q-learning converges to optimal action values in finite discounted MDPs \citep{watkins1992q,jaakkola1994convergence}.

\subsection{Expected SARSA}

Expected SARSA replaces the sampled next action value with an expectation under the current policy:
\begin{equation}
	Q(S_t,A_t) \leftarrow Q(S_t,A_t)
	+ \alpha\left[R_{t+1}+\gamma \sum_{a'}\pi(a'|S_{t+1})Q(S_{t+1},a') - Q(S_t,A_t)\right].
	\label{eq:expected_sarsa_update}
\end{equation}
Expected SARSA can reduce variance relative to SARSA because it averages over next actions instead of sampling one next action. It can behave on-policy or off-policy depending on the policy used inside the expectation \citep{sutton2018reinforcement,vanseijen2009theoretical}.

\begin{figure}[t]
	\centering
	\begin{tikzpicture}[
		box/.style={draw, rounded corners, thick, minimum width=4.0cm, minimum height=1.0cm, align=center},
		note/.style={align=center, font=\small},
		node distance=0.85cm
		]
		\node[box, draw=blue!70, fill=blue!10, font=\bfseries] (sarsa) {SARSA target\\{\normalfont\small $R_{t+1}+\gamma Q(S_{t+1},A_{t+1})$}};
		\node[box, draw=orange!80, fill=orange!15, font=\bfseries, below=of sarsa] (qlearn) {Q-learning target\\{\normalfont\small $R_{t+1}+\gamma \max_{a'}Q(S_{t+1},a')$}};
		\node[box, draw=green!60!black, fill=green!10, font=\bfseries, below=of qlearn] (expected) {Expected SARSA target\\{\normalfont\small $R_{t+1}+\gamma \sum_{a'}\pi(a'|S_{t+1})Q(S_{t+1},a')$}};
		
		\node[note, right=1.2cm of sarsa, color=blue!70] {uses actual\\next action};
		\node[note, right=1.2cm of qlearn, color=orange!80] {uses greedy\\next action};
		\node[note, right=1.2cm of expected, color=green!60!black] {uses expected\\next action value};
	\end{tikzpicture}
	\caption{SARSA, Q-learning, and Expected SARSA differ mainly in the bootstrap target. SARSA uses the action actually taken, Q-learning uses the greedy next action, and Expected SARSA averages over next actions under a policy.}
	\label{fig:sarsa_q_expected_targets}
\end{figure}
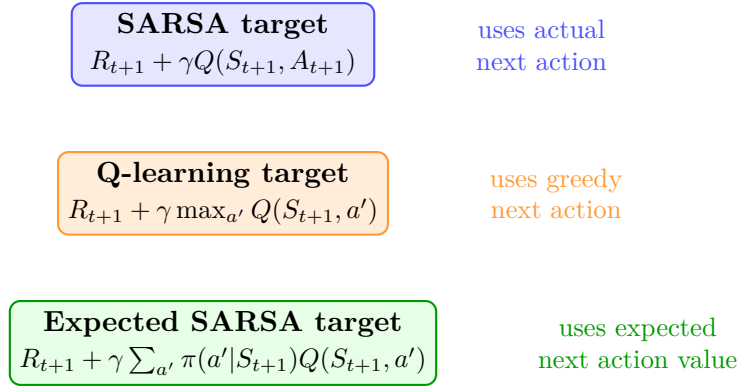

\section[Eligibility traces and TD(lambda)]{Eligibility traces and TD$(\lambda)$}

TD(0) uses a one-step bootstrap target. Monte Carlo uses the full return. Eligibility traces provide a continuum between these extremes.

An $n$-step return is
\begin{equation}
	G_t^{(n)} = R_{t+1}+\gamma R_{t+2}+\cdots+\gamma^{n-1}R_{t+n}+\gamma^n V(S_{t+n}).
\end{equation}
For $n=1$, this is the TD(0) target. As $n$ approaches the episode length, it approaches the Monte Carlo return.

The $\lambda$-return combines all $n$-step returns with geometrically decaying weights:
\begin{equation}
	G_t^{\lambda} = (1-\lambda)\sum_{n=1}^{\infty}\lambda^{n-1}G_t^{(n)},
	\quad 0\leq\lambda\leq 1.
\end{equation}
When $\lambda=0$, the method behaves like TD(0). When $\lambda$ is close to 1, it behaves more like Monte Carlo learning.

Eligibility traces provide an efficient backward-view implementation. For state values, a simple accumulating trace is
\begin{equation}
	e_t(s) = \gamma\lambda e_{t-1}(s) + \mathbf{1}\{S_t=s\}.
\end{equation}
Then the update is
\begin{equation}
	V(s) \leftarrow V(s) + \alpha \delta_t e_t(s).
\end{equation}
Eligibility traces are historically important because they connect one-step TD, multi-step returns, and Monte Carlo ideas in one framework \citep{sutton1988learning,singh1996reinforcement,sutton2018reinforcement}.

\begin{figure}[t]
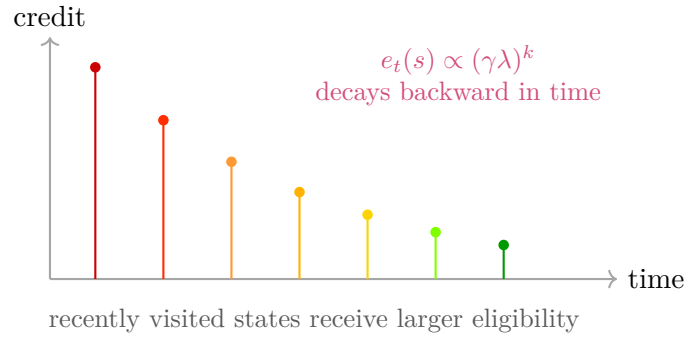

	\centering

	\caption{Eligibility traces assign credit backward through recently visited states. The trace decays roughly according to $\gamma\lambda$, allowing TD errors to update multiple earlier states rather than only the immediately preceding one.}
	\label{fig:eligibility_traces}
\end{figure}

\section{Comparing DP, MC, and TD}

The three families differ in what they require and when they update.

\begin{table}[t]
	\centering
	\begin{tabular}{p{0.18\textwidth}p{0.22\textwidth}p{0.23\textwidth}p{0.23\textwidth}}
		\toprule
		Method & Requires model? & Update target & Main strength \\
		\midrule
		Dynamic programming & Yes & Expected Bellman backup & Exact planning when model is known \\
		Monte Carlo & No & Complete sampled return $G_t$ & Simple, unbiased return samples for episodic tasks \\
		Temporal difference & No & Bootstrapped target $R+\gamma V(S')$ & Online, incremental, central to modern RL \\
		\bottomrule
	\end{tabular}
	\caption{Comparison of dynamic programming, Monte Carlo learning, and temporal-difference learning.}
	\label{tab:dp_mc_td_comparison}
\end{table}

\begin{figure}[t]
	\centering
	%
	\caption{Conceptual map of classical value-learning methods. DP, MC, TD, multi-step TD, and TD$(\lambda)$ are not isolated tricks; they form a connected design space.}
	\label{fig:classical_rl_map}
\end{figure}

\section{Python implementation from zero}

This section provides compact but complete Python code for a small grid-world. The goal is not to build a library. The goal is to make the algorithms concrete.

\subsection{A small grid-world environment}

We use a rectangular grid. The agent starts somewhere, moves up, down, left, or right, receives a step penalty, and receives a terminal reward at the goal. Walls block movement. This simple environment is enough to test DP, MC, TD(0), SARSA, and Q-learning.

\begin{lstlisting}[style=pythonstyle,caption={A minimal tabular grid-world environment.},label={lst:gridworld}]
import random
from collections import defaultdict

class GridWorld:
    def __init__(self, width=5, height=5, start=(0, 0), goal=(4, 4), walls=None):
        self.width = width
        self.height = height
        self.start = start
        self.goal = goal
        self.walls = set(walls or [])
        self.actions = [0, 1, 2, 3]  # up, down, left, right
        self.action_delta = {
            0: (0, -1),
            1: (0, 1),
            2: (-1, 0),
            3: (1, 0),
        }
        self.state = start

    def states(self):
        out = []
        for y in range(self.height):
            for x in range(self.width):
                s = (x, y)
                if s not in self.walls:
                    out.append(s)
        return out

    def is_terminal(self, s):
        return s == self.goal

    def reset(self):
        self.state = self.start
        return self.state

    def step(self, action):
        if self.is_terminal(self.state):
            return self.state, 0.0, True

        dx, dy = self.action_delta[action]
        x, y = self.state
        ns = (x + dx, y + dy)

        if (
            ns[0] < 0 or ns[0] >= self.width or
            ns[1] < 0 or ns[1] >= self.height or
            ns in self.walls
        ):
            ns = self.state

        reward = 10.0 if ns == self.goal else -1.0
        done = ns == self.goal
        self.state = ns
        return ns, reward, done
\end{lstlisting}

\subsection{Policy evaluation with a known model}

For DP, we need a model. In this deterministic grid-world, the model can be computed by temporarily applying each action.

\Needspace{14\baselineskip}
\begin{lstlisting}[style=pythonstyle,caption={Iterative policy evaluation for a known deterministic grid-world model.},label={lst:policy_eval}]
def one_step_model(env, state, action):
    old_state = env.state
    env.state = state
    next_state, reward, done = env.step(action)
    env.state = old_state
    return next_state, reward, done


def iterative_policy_evaluation(env, policy, gamma=0.99, theta=1e-6):
    V = {s: 0.0 for s in env.states()}

    while True:
        delta = 0.0
        new_V = V.copy()

        for s in env.states():
            if env.is_terminal(s):
                new_V[s] = 0.0
                continue

            v = 0.0
            for a, prob in policy[s].items():
                ns, r, done = one_step_model(env, s, a)
                v += prob * (r + gamma * V[ns] * (not done))

            delta = max(delta, abs(v - V[s]))
            new_V[s] = v

        V = new_V
        if delta < theta:
            break

    return V
\end{lstlisting}

\subsection{Value iteration}

Value iteration uses the Bellman optimality backup.

\begin{lstlisting}[style=pythonstyle,caption={Value iteration for the grid-world.},label={lst:value_iteration}]
def value_iteration(env, gamma=0.99, theta=1e-6):
    V = {s: 0.0 for s in env.states()}

    while True:
        delta = 0.0
        new_V = V.copy()

        for s in env.states():
            if env.is_terminal(s):
                new_V[s] = 0.0
                continue

            action_values = []
            for a in env.actions:
                ns, r, done = one_step_model(env, s, a)
                action_values.append(r + gamma * V[ns] * (not done))

            best_v = max(action_values)
            delta = max(delta, abs(best_v - V[s]))
            new_V[s] = best_v

        V = new_V
        if delta < theta:
            break

    policy = {}
    for s in env.states():
        if env.is_terminal(s):
            policy[s] = None
            continue

        values = []
        for a in env.actions:
            ns, r, done = one_step_model(env, s, a)
            values.append(r + gamma * V[ns] * (not done))

        policy[s] = max(range(len(values)), key=lambda i: values[i])

    return V, policy
\end{lstlisting}

\subsection{Monte Carlo prediction}

Now suppose the model is not known. We estimate values from complete episodes.

\begin{lstlisting}[style=pythonstyle,caption={First-visit Monte Carlo prediction.},label={lst:mc_prediction}]
def generate_episode(env, policy, max_steps=200):
    episode = []
    s = env.reset()

    for _ in range(max_steps):
        action_probs = policy[s]
        actions, probs = zip(*action_probs.items())
        a = random.choices(actions, weights=probs, k=1)[0]
        ns, r, done = env.step(a)
        episode.append((s, a, r))
        s = ns
        if done:
            break

    return episode


def first_visit_mc_prediction(env, policy, episodes=5000, gamma=0.99):
    V = defaultdict(float)
    N = defaultdict(int)

    for _ in range(episodes):
        episode = generate_episode(env, policy)
        G = 0.0
        visited = set()

        for t in reversed(range(len(episode))):
            s, a, r = episode[t]
            G = r + gamma * G

            if s not in visited:
                visited.add(s)
                N[s] += 1
                V[s] += (G - V[s]) / N[s]

    return dict(V)
\end{lstlisting}

\subsection{TD(0) prediction}

TD(0) updates after each step.

\begin{lstlisting}[style=pythonstyle,caption={TD(0) prediction.},label={lst:td0_prediction}]
def td0_prediction(env, policy, episodes=5000, alpha=0.1, gamma=0.99, max_steps=200):
    V = defaultdict(float)

    for _ in range(episodes):
        s = env.reset()

        for _ in range(max_steps):
            actions, probs = zip(*policy[s].items())
            a = random.choices(actions, weights=probs, k=1)[0]
            ns, r, done = env.step(a)

            target = r if done else r + gamma * V[ns]
            V[s] += alpha * (target - V[s])

            s = ns
            if done:
                break

    return dict(V)
\end{lstlisting}

\subsection{SARSA and Q-learning}

The following helper implements $\epsilon$-greedy action selection.

\begin{lstlisting}[style=pythonstyle,caption={Epsilon-greedy action selection.},label={lst:epsilon_greedy}]
def epsilon_greedy_action(Q, state, actions, epsilon):
    if random.random() < epsilon:
        return random.choice(actions)
    return max(actions, key=lambda a: Q[(state, a)])
\end{lstlisting}

SARSA uses the next action actually selected.

\begin{lstlisting}[style=pythonstyle,caption={SARSA control.},label={lst:sarsa}]
def sarsa(env, episodes=10000, alpha=0.1, gamma=0.99, epsilon=0.1, max_steps=200):
    Q = defaultdict(float)

    for _ in range(episodes):
        s = env.reset()
        a = epsilon_greedy_action(Q, s, env.actions, epsilon)

        for _ in range(max_steps):
            ns, r, done = env.step(a)
            na = epsilon_greedy_action(Q, ns, env.actions, epsilon)

            target = r if done else r + gamma * Q[(ns, na)]
            Q[(s, a)] += alpha * (target - Q[(s, a)])

            s, a = ns, na
            if done:
                break

    return Q
\end{lstlisting}

Q-learning uses the greedy next action in the target.

\begin{lstlisting}[style=pythonstyle,caption={Q-learning control.},label={lst:q_learning}]
def q_learning(env, episodes=10000, alpha=0.1, gamma=0.99, epsilon=0.1, max_steps=200):
    Q = defaultdict(float)

    for _ in range(episodes):
        s = env.reset()

        for _ in range(max_steps):
            a = epsilon_greedy_action(Q, s, env.actions, epsilon)
            ns, r, done = env.step(a)

            best_next = max(Q[(ns, na)] for na in env.actions)
            target = r if done else r + gamma * best_next
            Q[(s, a)] += alpha * (target - Q[(s, a)])

            s = ns
            if done:
                break

    return Q
\end{lstlisting}

\subsection{Extracting and printing a greedy policy}

\begin{lstlisting}[style=pythonstyle,caption={Printing a greedy policy as arrows.},label={lst:print_policy}]
def greedy_policy_from_Q(env, Q):
    arrows = {0: '^', 1: 'v', 2: '<', 3: '>'}
    policy = {}
    for s in env.states():
        if env.is_terminal(s):
            policy[s] = 'G'
        else:
            best_a = max(env.actions, key=lambda a: Q[(s, a)])
            policy[s] = arrows[best_a]
    return policy


def print_grid_policy(env, policy):
    for y in range(env.height):
        row = []
        for x in range(env.width):
            s = (x, y)
            if s in env.walls:
                row.append('#')
            else:
                row.append(policy.get(s, '.'))
        print(' '.join(row))
\end{lstlisting}

\subsection{Minimal experiment script}

\begin{lstlisting}[style=pythonstyle,caption={Running SARSA and Q-learning on the grid-world.},label={lst:run_experiment}]
if __name__ == '__main__':
    walls = {(1, 1), (1, 2), (3, 1)}
    env = GridWorld(width=5, height=5, start=(0, 0), goal=(4, 4), walls=walls)

    Q_sarsa = sarsa(env, episodes=5000, alpha=0.1, gamma=0.95, epsilon=0.1)
    pi_sarsa = greedy_policy_from_Q(env, Q_sarsa)
    print('SARSA greedy policy:')
    print_grid_policy(env, pi_sarsa)

    Q_q = q_learning(env, episodes=5000, alpha=0.1, gamma=0.95, epsilon=0.1)
    pi_q = greedy_policy_from_Q(env, Q_q)
    print('\nQ-learning greedy policy:')
    print_grid_policy(env, pi_q)
\end{lstlisting}

\begin{figure}[t]
	\centering
	\begin{tikzpicture}[
		cell/.style={draw, minimum size=0.72cm, align=center},
		wall/.style={draw, fill=gray!35, minimum size=0.72cm},
		goal/.style={draw, fill=green!20, minimum size=0.72cm, align=center},
		start/.style={draw, fill=blue!12, minimum size=0.72cm, align=center}
		]
		\foreach \x in {0,...,4}{
			\foreach \y in {0,...,4}{
				\node[cell] at (\x,-\y) {};
			}
		}
		\node[start] at (0,0) {S};
		\node[wall] at (1,-1) {\#};
		\node[wall] at (1,-2) {\#};
		\node[wall] at (3,-1) {\#};
		\node[goal] at (4,-4) {G};
		\draw[-{Latex[length=2.2mm]}, thick] (0,0) -- (0,-1);
		\draw[-{Latex[length=2.2mm]}, thick] (0,-1) -- (0,-2);
		\draw[-{Latex[length=2.2mm]}, thick] (0,-2) -- (0,-3);
		\draw[-{Latex[length=2.2mm]}, thick] (0,-3) -- (1,-3);
		\draw[-{Latex[length=2.2mm]}, thick] (1,-3) -- (2,-3);
		\draw[-{Latex[length=2.2mm]}, thick] (2,-3) -- (3,-3);
		\draw[-{Latex[length=2.2mm]}, thick] (3,-3) -- (4,-3);
		\draw[-{Latex[length=2.2mm]}, thick] (4,-3) -- (4,-4);
		\node[below=0.5cm] at (2,-4.8) {A small grid-world for testing DP, MC, TD, SARSA, and Q-learning.};
	\end{tikzpicture}
	\caption{Example grid-world. Walls block movement, the start state is marked S, and the goal is marked G. This environment is deliberately simple so that algorithmic differences can be inspected clearly.}
	\label{fig:gridworld_example}
\end{figure}

\section{UAV and networking interpretation}

The classical algorithms in this chapter are not only for grid-worlds. They provide the foundation for real network-control problems.

In a UAV-assisted wireless network, a state may include UAV positions, battery levels, user distribution, SINR, throughput, latency, and traffic load. Actions may include motion, user association, resource allocation, routing, or SDN-guided decisions. Rewards may combine QoS satisfaction, energy cost, fairness, and safety penalties.

Dynamic programming would require a complete model of user mobility, wireless channel transitions, traffic arrivals, battery dynamics, and reward distribution. This is usually unrealistic. Monte Carlo learning could evaluate policies from full simulated episodes, but long episodes may make learning slow and high variance. TD learning can update online from each transition, which is one reason TD-style algorithms are attractive for control problems.

\begin{figure}[t]
	\centering
	\begin{tikzpicture}[
		box/.style={draw, rounded corners, thick, minimum width=3.2cm, minimum height=0.85cm, align=center},
		arrow/.style={-{Latex[length=2.2mm]}, thick},
		node distance=0.8cm
		]
		\node[box] (state) {Network state\\UAVs, users, QoS, battery};
		\node[box, right=of state] (action) {Control action\\move, associate, allocate};
		\node[box, right=of action] (reward) {Reward\\QoS $-$ energy $-$ violations};
		\node[box, below=of action] (td) {TD update\\learn from transition};
		\draw[arrow] (state) -- (action);
		\draw[arrow] (action) -- (reward);
		\draw[arrow] (reward) |- (td);
		\draw[arrow] (td) -| (state);
		\node[below=1.0cm of td, align=center] {The same TD error idea can evaluate and improve policies in simulated network-control systems.};
	\end{tikzpicture}
	\caption{Interpretation of TD learning for UAV-assisted network control. A transition produces a reward and a next state; the TD error updates value estimates without requiring a full analytic environment model.}
	\label{fig:uav_td_interpretation}
\end{figure}

\section{Practical pitfalls and research lessons}

\subsection{Pitfall 1: confusing reward with value}

Reward is immediate. Value is long-term. A state with low immediate reward can still have high value if it leads to good future outcomes. This distinction is crucial in robotics, networking, and resource allocation.

\subsection{Pitfall 2: using DP when the model is not actually known}

Dynamic programming requires transition probabilities and rewards. In a simulator, we may be able to call \texttt{step()}, but that is not the same as knowing $p(s',r|s,a)$. A step function gives samples. DP needs expectations.

\subsection{Pitfall 3: trusting Monte Carlo estimates too early}

Monte Carlo returns can have high variance. In long-horizon tasks, early estimates may be extremely noisy. This is why averaging, many episodes, and variance-reduction methods matter.

\subsection{Pitfall 4: forgetting that TD bootstraps}

TD learning can be efficient, but it updates using its own current predictions. If the value function is badly initialized or function approximation is unstable, bootstrapping can propagate errors.

\subsection{Pitfall 5: ignoring exploration}

Control algorithms need exploration. Without enough exploration, Q-values for untried actions remain wrong. With too much exploration, the system may perform poorly or unsafely. This tradeoff becomes even more important in deep RL and safe RL.

\subsection{Pitfall 6: assuming tabular convergence transfers to neural networks}

Tabular Q-learning has convergence guarantees under classical assumptions. Deep Q-learning does not automatically inherit these guarantees. Function approximation, bootstrapping, and off-policy learning can create instability, a phenomenon often called the deadly triad \citep{sutton2018reinforcement}.

\begin{warningbox}{The deadly triad}
	Instability can arise when three ingredients are combined: function approximation, bootstrapping, and off-policy learning. Modern DQN stabilizes this combination using experience replay and target networks, but the underlying issue begins with classical TD learning.
\end{warningbox}

\section{Key Takeaways}

\begin{itemize}[leftmargin=*]
	\item Dynamic programming solves MDPs using a known model.
	\item Policy evaluation estimates the value of a fixed policy.
	\item Policy improvement makes the policy greedy with respect to values.
	\item Policy iteration alternates evaluation and improvement.
	\item Value iteration applies Bellman optimality backups directly.
	\item Monte Carlo learning estimates values from complete sampled returns.
	\item TD learning updates from one-step sampled transitions using bootstrapping.
	\item SARSA is on-policy TD control.
	\item Q-learning is off-policy TD control.
	\item Expected SARSA averages over next-action values.
	\item Eligibility traces connect TD(0), multi-step learning, and Monte Carlo learning.
	\item Modern deep RL scales these ideas using neural networks, replay, optimization, and function approximation.
\end{itemize}

\section*{Looking Ahead to Chapter 4:  Food for Thought}
\addcontentsline{toc}{section}{Looking Ahead to Chapter 4:  Food for Thought}

Chapter~3 introduced the first major families of reinforcement learning algorithms: dynamic programming, Monte Carlo learning, and temporal-difference learning. These methods are foundational. They explain how an agent can evaluate a policy, improve a policy, learn from complete episodes, or update its estimates step by step from partial experience.

However, these methods also reveal an important limitation. Most of the algorithms in this chapter were presented in tabular form. They assume that the agent can store a value for every state, or for every state--action pair. This assumption is useful for understanding the mathematics, but it quickly becomes unrealistic in large problems.

Before moving to deep reinforcement learning, it is useful to pause and ask why classical reinforcement learning was not enough.

Before moving on, consider the following questions:

\begin{enumerate}[leftmargin=*]
	\item What happens when the number of states is too large to store in a table?
	\item What happens when the state is not a simple symbol, but an image, a sensor vector, a network measurement, or a continuous physical position?
	\item Can Q-learning still work if the agent has never seen exactly the same state before?
	\item If two states are different but very similar, should the agent learn completely separate values for them?
	\item How can an agent generalize from past experience to new but related situations?
	\item Why is a table a poor representation for problems involving pixels, robot joints, UAV coordinates, traffic matrices, or wireless channel measurements?
	\item What kind of function could replace a table?
	\item What makes neural networks attractive as value-function or policy approximators?
	\item Why can combining reinforcement learning with neural networks make training unstable?
	\item What new problems appear when the value function itself is learned by a deep model?
\end{enumerate}

\subsection*{A Simple Thought Experiment}

Imagine a small grid-world with only sixteen cells. In this case, tabular Q-learning is easy: the agent can store one value for each state--action pair. If there are sixteen states and four actions, the Q-table contains only sixty-four values.

Now imagine a UAV moving in a three-dimensional environment. Its state may include its $x$, $y$, and $z$ coordinates, velocity, battery level, neighboring UAV positions, user locations, channel quality, traffic demand, latency, SINR, and throughput. Even if each variable is discretized coarsely, the number of possible states grows explosively.

Now imagine that the UAV observes a camera image or a wireless heatmap. The state is no longer a small index. It is a high-dimensional input. A table is no longer meaningful.

This thought experiment exposes the central difficulty:

\begin{quote}
	Classical RL tells us how to learn values and policies, but it does not by itself solve the problem of representation.
\end{quote}

\subsection*{From Tables to Functions}

The natural next step is to replace a table with a function approximator. Instead of storing
\[
Q(s,a)
\]
for every possible state--action pair, we approximate it using parameters:
\[
Q(s,a;\theta) \approx Q^*(s,a),
\]
where $\theta$ represents learnable parameters.

In early reinforcement learning, this function approximator might be linear. In deep reinforcement learning, it is usually a neural network. This change is simple to write, but profound in consequence. The agent is no longer just memorizing values. It is learning a representation that may generalize across states.

For example, a neural network may learn that two different images correspond to similar game situations, or that two different UAV positions produce similar QoS outcomes. This ability to generalize is one of the main reasons deep learning became important for RL.

\subsection*{The Central Tension}

The transition from classical RL to deep RL creates a powerful but dangerous combination.

On one hand, neural networks allow RL to scale to large and complex problems. They can process images, continuous states, high-dimensional measurements, and learned representations.

On the other hand, reinforcement learning already involves bootstrapping, delayed reward, non-stationary data, and exploration. Adding nonlinear function approximation makes the learning process more powerful, but also more unstable.

This tension is one of the main themes of deep reinforcement learning:

\begin{quote}
	Deep networks give RL the ability to generalize, but they also make stability much harder.
\end{quote}

\subsection*{Main Idea for the Next Chapter}

Chapter~4 begins the transition from classical reinforcement learning to deep reinforcement learning. The key question is:

\begin{quote}
	How can reinforcement learning scale beyond small tabular problems?
\end{quote}

To answer this, we will study the curse of dimensionality, function approximation, representation learning, and the historical path that led to Deep Q-Networks. The goal is not only to understand that neural networks can replace tables, but also to understand why this replacement was necessary, why it was difficult, and why it changed the field.

\begin{quote}
	Chapter~3 showed how agents learn from experience. \\
	Chapter~4 asks how they can do so when the world is too large for a table.
\end{quote}

\section{Exercises}

\subsection*{Conceptual exercises}

\begin{enumerate}[leftmargin=*]
	\item Explain the difference between dynamic programming, Monte Carlo learning, and temporal-difference learning in your own words.
	\item Why does dynamic programming require a model?
	\item Why can Monte Carlo learning have high variance?
	\item Why can TD learning update before an episode ends?
	\item Explain generalized policy iteration using one paragraph.
	\item Why is SARSA called an on-policy method?
	\item Why is Q-learning called an off-policy method?
	\item In what kind of environment might SARSA be safer than Q-learning?
\end{enumerate}

\subsection*{Mathematical exercises}

\begin{enumerate}[leftmargin=*]
	\item Given rewards $2, -1, 4$ and $\gamma=0.5$, compute the Monte Carlo return $G_0$.
	\item Suppose $V(S_t)=3$, $R_{t+1}=1$, $V(S_{t+1})=5$, and $\gamma=0.9$. Compute the TD error.
	\item Write the Bellman expectation update for $V^{\pi}(s)$.
	\item Write the value-iteration update for $V(s)$.
	\item Compare the SARSA target and Q-learning target mathematically.
	\item For $\lambda=0$, what does TD$(\lambda)$ become? For $\lambda\approx 1$, what does it resemble?
\end{enumerate}

\subsection*{Implementation exercises}

\begin{enumerate}[leftmargin=*]
	\item Modify the grid-world so that every step gives reward $-0.01$ instead of $-1$. How does the learned policy change?
	\item Add a cliff region with reward $-100$. Compare SARSA and Q-learning.
	\item Implement Expected SARSA using Listing~\ref{lst:q_learning} as a starting point.
	\item Plot the episode return over training for SARSA and Q-learning.
	\item Add stochastic movement: with probability $0.1$, the agent moves in a random direction. Which algorithm is more stable?
\end{enumerate}

\subsection*{Research thinking exercises}

\begin{enumerate}[leftmargin=*]
	\item In UAV network control, what would be the difference between a DP solution and a TD solution?
	\item Why is a simulator step function not the same as a full transition model?
	\item In an SDN routing problem, what might be a state, action, reward, and TD error?
	\item Explain why off-policy learning is useful but potentially unstable with function approximation.
	\item How does this chapter prepare us for Deep Q-Networks?
\end{enumerate}

% \section{Chapter References}
%
% Bellman's dynamic programming work introduced the optimality principle underlying Bellman backups \citep{bellman1957dynamic}. Howard's work on Markov processes shaped policy iteration \citep{howard1960dynamic}. Puterman provides a standard treatment of Markov decision processes and dynamic programming theory \citep{puterman1994markov}. Sutton's temporal-difference paper formalized TD learning \citep{sutton1988learning}, and Sutton and Barto give the standard reinforcement learning treatment of Monte Carlo methods, TD learning, SARSA, eligibility traces, and generalized policy iteration \citep{sutton2018reinforcement}. Watkins introduced Q-learning and Watkins and Dayan analyzed its convergence in the tabular setting \citep{watkins1989learning,watkins1992q}. Bertsekas and Tsitsiklis connect dynamic programming with approximation and neuro-dynamic programming \citep{bertsekas1996neuro}. Expected SARSA is discussed theoretically and empirically by van Seijen and coauthors \citep{vanseijen2009theoretical}.
	\part{The Birth of Deep Reinforcement Learning}
%\addcontentsline{toc}{part}{Part II -- The Birth of Deep Reinforcement Learning}

\chapter{Why Classical RL Was Not Enough}
\label{ch:classical_to_deep}
\chaptermark{Why Classical RL Was Not Enough}

\begin{keybox}{Chapter goal}
Classical dynamic programming, Monte Carlo learning, SARSA, and Q-learning explain how agents learn values and improve policies. However, their simplest forms assume that values can be stored in tables. Chapter~4 explains why that assumption breaks in realistic problems, why function approximation becomes necessary, why neural networks became attractive, and why this transition created the instability problems that later motivated DQN.
\end{keybox}

\section*{Chapter Overview}
\addcontentsline{toc}{section}{Chapter Overview}

This chapter explains why classical reinforcement learning, despite its mathematical elegance, was not enough for realistic large-scale decision problems. The main limitation is not the Bellman framework itself. The limitation is the representation. Tabular methods assume that values can be stored explicitly for each state or state-action pair. That assumption breaks in continuous, high-dimensional, perceptual, and large multi-agent domains. Function approximation becomes necessary, and this in turn motivates the transition from classical RL to deep RL.

\section{Why this chapter matters}

The first three chapters built the foundation of reinforcement learning. We introduced Markov decision processes, value functions, Bellman equations, dynamic programming, Monte Carlo learning, temporal-difference learning, SARSA, and Q-learning. These ideas are not obsolete. They remain the conceptual backbone of deep reinforcement learning. DQN is a deep version of Q-learning. Actor-critic methods combine policy learning with learned value functions. Offline RL often turns Bellman updates into supervised regression. Model-based RL still uses Bellman-style planning and value backups.

However, classical RL in its cleanest textbook form often uses tables. A value table stores one value per state, or one value per state-action pair:
\begin{equation}
    V(s), \qquad Q(s,a).
\end{equation}
This is perfect for explaining the mathematics. It is not enough for large-scale intelligence.

A table is a memory device. It does not know that two states are similar unless they are literally the same table index. If an agent has never visited a state before, a tabular method has no built-in way to infer its value from similar states. This is acceptable in a small grid-world. It is disastrous for images, robot sensors, UAV trajectories, network measurements, medical records, and language-model interactions.

The birth of deep reinforcement learning was driven by this mismatch: RL had powerful learning rules, but classical tabular representations could not scale to high-dimensional perception and control. Deep learning supplied a possible solution: learn representations and approximate value functions or policies directly from complex inputs. DQN later showed that a convolutional neural network could learn action values from raw Atari frames, using Q-learning-style targets with stabilization mechanisms such as experience replay and a target network \citep{mnih2013atari,mnih2015human}.

This chapter is the bridge between classical RL and deep RL. The main question is:
\begin{quote}
    How can reinforcement learning survive when the state space is too large, too continuous, or too perceptual for a table?
\end{quote}

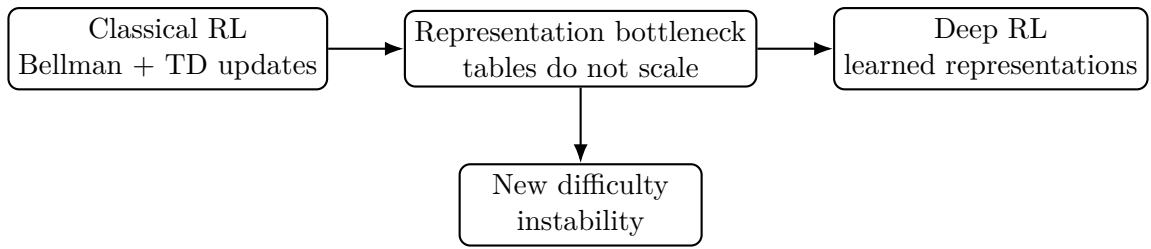
\begin{figure}[htbp]
    \centering
    \begin{tikzpicture}[
        box/.style={draw, rounded corners, thick, minimum width=3.2cm, minimum height=0.9cm, align=center},
        arrow/.style={-{Latex[length=2.5mm]}, thick},
        node distance=1.0cm
    ]
        \node[box] (classical) {Classical RL\\Bellman + TD updates};
        \node[box, right=of classical] (bottleneck) {Representation bottleneck\\tables do not scale};
        \node[box, right=of bottleneck] (deep) {Deep RL\\learned representations};
        \node[box, below=of bottleneck] (problem) {New difficulty\\instability};
        \draw[arrow] (classical) -- (bottleneck);
        \draw[arrow] (bottleneck) -- (deep);
        \draw[arrow] (bottleneck) -- (problem);
    \end{tikzpicture}
    \caption{Chapter~4 is the bridge from classical RL to deep RL. The mathematical update rules of classical RL remain important, but tabular representations cannot scale. Deep neural networks address representation, but they introduce new stability issues.}
    \label{fig:ch4_bridge}
\end{figure}

\section{The hidden assumption behind tabular RL}

A tabular method assumes that each state, or each state-action pair, can be indexed explicitly. For a finite MDP with $|\Sset|$ states and $|\A|$ actions, a Q-table contains
\begin{equation}
    |\Sset|\,|\A|
\end{equation}
entries. If $|\Sset|$ and $|\A|$ are small, this is efficient and clear. If $|\Sset|$ is enormous, the table becomes impossible.

The problem is not only memory. It is also learning. A Q-table treats every state-action pair as independent. The update
\begin{equation}
    Q(s,a) \leftarrow Q(s,a) + \alpha\left[r + \gamma \max_{a'}Q(s',a') - Q(s,a)\right]
\end{equation}
changes only one entry. It does not directly update nearby or similar states. This is useful for theoretical clarity but inefficient in large domains.

Consider the difference between two representations:
\begin{itemize}[leftmargin=*]
    \item In a table, state $s_1$ and state $s_2$ are unrelated unless they are exactly the same index.
    \item In a function approximator, state $s_1$ and state $s_2$ may share features, so learning about one can affect predictions for the other.
\end{itemize}

This ability to share information is called \emph{generalization}. Without generalization, an agent must visit almost everything it needs to know. With generalization, an agent can learn from partial experience and extrapolate to related situations.

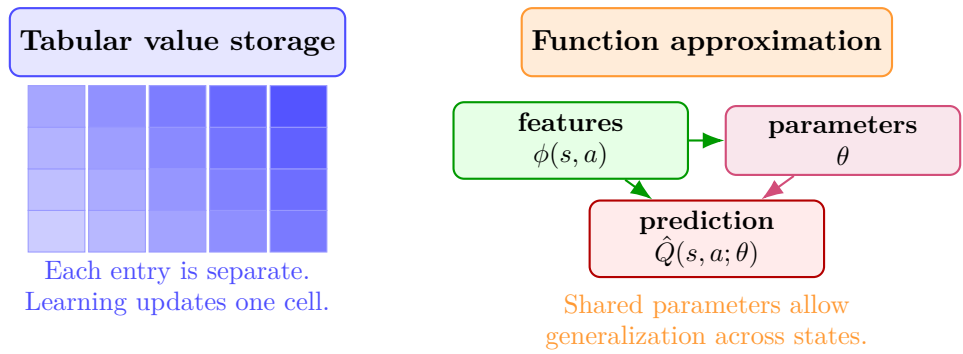
\begin{figure}[htbp]
    \centering
    \begin{tikzpicture}[
        cell/.style={draw, minimum width=0.75cm, minimum height=0.55cm, align=center},
        box/.style={draw, rounded corners, thick, minimum width=3.1cm, minimum height=0.9cm, align=center},
        arrow/.style={-{Latex[length=2.9mm]}, thick}
    ]
        \node[box, draw=blue!70, fill=blue!10, font=\bfseries] (tabtitle) at (0,2.2) {Tabular value storage};
        \foreach \i in {0,...,4} {
            \foreach \j in {0,...,3} {
                \pgfmathsetmacro{\shade}{20 + \i*8 + \j*5}
                \node[cell, fill=blue!\shade!white, draw=blue!40] at (0.8*\i-1.6,0.55*\j-0.3) {};
            }
        }
        \node[align=center, font=\small, color=blue!70] at (0,-1.05) {Each entry is separate.\\Learning updates one cell.};

        \node[box, draw=orange!80, fill=orange!15, font=\bfseries] (functitle) at (7,2.2) {Function approximation};
        \node[box, draw=green!60!black, fill=green!10, font=\small\bfseries] (features) at (5.2,0.9) {features\\{\normalfont\small $\phi(s,a)$}};
        \node[box, draw=purple!70, fill=purple!10, font=\small\bfseries] (params) at (8.8,0.9) {parameters\\{\normalfont\small $\theta$}};
        \node[box, draw=red!70!black, fill=red!8, font=\small\bfseries] (pred) at (7,-0.4) {prediction\\{\normalfont\small $\hat Q(s,a;\theta)$}};

        \draw[arrow, color=green!60!black] (features) -- (params);
        \draw[arrow, color=purple!70] (params) -- (pred);
        \draw[arrow, color=green!60!black] (features) -- (pred);

        \node[align=center, font=\small, color=orange!80] at (7,-1.5) {Shared parameters allow\\generalization across states.};
    \end{tikzpicture}
    \caption{A table stores separate values for separate state-action pairs. A function approximator maps features to values using shared parameters. The central advantage is generalization; the central risk is approximation error and instability.}
    \label{fig:tabular_vs_function}
\end{figure}

\section{The curse of dimensionality}

The curse of dimensionality describes the explosive growth of the state space as the number of variables increases. Suppose a state is described by $d$ variables, and each variable is discretized into $m$ bins. The number of discrete states is
\begin{equation}
    |\Sset| = m^d.
\end{equation}
If each state has $|\A|$ actions, the Q-table has
\begin{equation}
    m^d |\A|
\end{equation}
entries. This exponential growth is the most basic reason tabular RL cannot scale.

\begin{practicebox}{Numerical example}
Suppose a robot state has position $(x,y)$, velocity $(v_x,v_y)$, battery level, and distance-to-obstacle. This is already $d=6$ variables. If each variable is discretized into only $20$ bins and the agent has $5$ actions, then the Q-table has
\[
20^6 \times 5 = 320,000,000
\]
entries. This is before adding camera images, lidar readings, communication load, uncertainty estimates, or other agents.
\end{practicebox}

\begin{figure}[htbp]
    \centering
    \begin{tikzpicture}
        \begin{axis}[
            width=0.85\textwidth,
            height=0.42\textwidth,
            ymode=log,
            xlabel={Number of state variables $d$},
            ylabel={Number of discrete states $m^d$},
            legend pos=north west,
            grid=both,
            ymin=1,
            ymax=1e18
        ]
            \addplot+[mark=*] coordinates {(1,10) (2,100) (3,1000) (4,10000) (5,100000) (6,1000000) (8,100000000) (10,10000000000) (12,1000000000000) (15,1000000000000000)};
            \addlegendentry{$m=10$ bins per variable}
            \addplot+[mark=square*] coordinates {(1,20) (2,400) (3,8000) (4,160000) (5,3200000) (6,64000000) (8,25600000000) (10,10240000000000) (12,4096000000000000)};
            \addlegendentry{$m=20$ bins per variable}
        \end{axis}
    \end{tikzpicture}
    \caption{The curse of dimensionality. Even coarse discretization becomes impossible as the number of state variables grows. The vertical axis is logarithmic.}
    \label{fig:curse_dimensionality}
\end{figure}
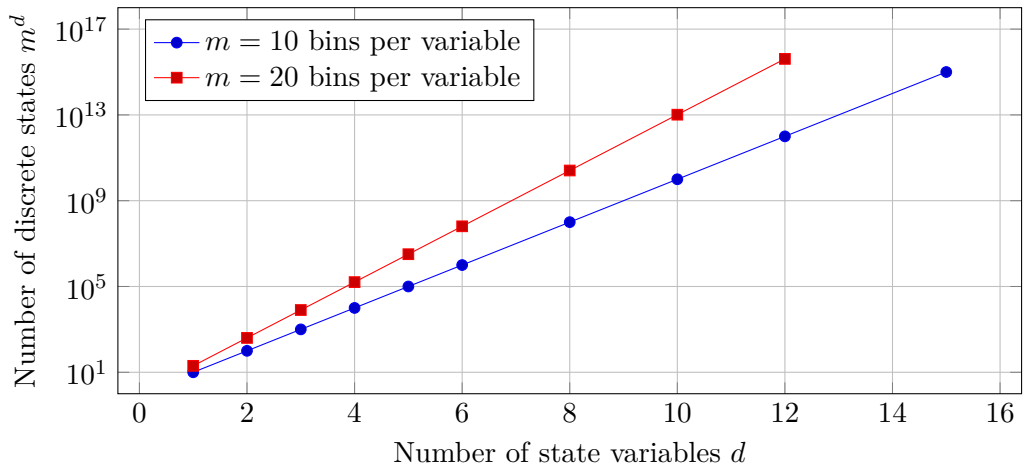

This is not a minor engineering inconvenience. It changes the nature of the problem. Tabular RL assumes repeated visits to the same state-action pairs. In high-dimensional spaces, exact revisits may be rare or impossible. Therefore, the agent must learn patterns that transfer across related states.

\Needspace{8\baselineskip}
\begin{lstlisting}[style=pythonstyle,caption={A small calculation showing how Q-table size explodes with state dimension.},label={lst:state_explosion}]
def q_table_entries(num_bins_per_variable, num_variables, num_actions):
    return (num_bins_per_variable ** num_variables) * num_actions

for d in [2, 4, 6, 8, 10, 12, 15, 20]:
    entries = q_table_entries(
        num_bins_per_variable=20,
        num_variables=d,
        num_actions=5,
    )
    print(f"d={d:2d}: {entries:.3e} Q-values")

# Even with coarse discretization, the table becomes impossible.
# d=10 already gives about 5.12e13 state-action values.
\end{lstlisting}

\section{Continuous states, continuous actions, and high-dimensional observations}

The state space may fail to be tabular in three different ways.

\subsection{Continuous states}

Many physical systems have continuous state variables. A UAV position might be $(x,y,z) \in \R^3$. A robot arm state might include joint angles and velocities. A wireless controller may observe continuous signal-to-noise ratios, queue lengths, traffic demands, and channel measurements.

A continuous state space cannot be stored exactly in a finite table. We can discretize, but discretization creates a trade-off:
\begin{itemize}[leftmargin=*]
    \item coarse discretization loses important detail;
    \item fine discretization creates too many states.
\end{itemize}

\subsection{Continuous actions}

Some environments also have continuous action spaces. A drone may choose a velocity vector, a transmit power, or a bandwidth allocation. A robot may choose torque values. A network controller may choose traffic-splitting ratios.

Value-based methods can handle continuous actions only with difficulty because the greedy action requires solving
\begin{equation}
    \arg\max_{a \in \A} Q(s,a).
\end{equation}
If $\A$ is discrete and small, this is just a maximum over a list. If $\A$ is continuous, it becomes an optimization problem inside every Bellman backup. This is one reason later chapters move toward policy-gradient and actor-critic methods.

\subsection{High-dimensional observations}

The strongest motivation for deep RL came from perceptual inputs. A raw image is not a small state symbol. Even an $84\times84$ grayscale image has $7056$ pixels. If multiple frames are stacked, the observation dimension is much larger. DQN showed that a convolutional network could process raw Atari frames and output action values \citep{mnih2013atari,mnih2015human}. The Arcade Learning Environment provided a standardized platform for such experiments \citep{bellemare2013ale,machado2018revisiting}.

\begin{figure}[htbp]
    \centering
    \begin{tikzpicture}[
        box/.style={draw, rounded corners, thick, minimum width=2.4cm, minimum height=0.8cm, align=center},
        arrow/.style={-{Latex[length=2.5mm]}, thick},
        node distance=0.7cm
    ]
        \node[box] (pixels) {Raw observation\\pixels or sensors};
        \node[box, right=of pixels] (features) {Representation\\features};
        \node[box, right=of features] (value) {Value or policy\\$Q_\theta$ or $\pi_\theta$};
        \node[box, right=of value] (action) {Action\\control};
        \draw[arrow] (pixels) -- (features);
        \draw[arrow] (features) -- (value);
        \draw[arrow] (value) -- (action);
        \node[below=0.8cm of features, align=center] {Deep learning enters RL mainly by learning\\representations from high-dimensional observations.};
    \end{tikzpicture}
    \caption{High-dimensional observations require representation learning. Deep RL uses neural networks to convert raw inputs into features useful for value estimation or policy learning.}
    \label{fig:perception_to_action}
\end{figure}

\section{Generalization: the missing ingredient}

Generalization means using experience from one part of the state-action space to improve predictions elsewhere. It is the reason supervised learning works at scale, and it is the reason function approximation becomes essential in RL.

Suppose a UAV has learned that serving a dense cluster of high-priority users from location $(x,y,z)$ gives high reward. If a similar cluster appears nearby, a good agent should not start from zero. It should transfer knowledge. A tabular method cannot do this unless the two situations share the same index or engineered aggregation. A learned function can.

A function approximator replaces the table by a parameterized mapping:
\begin{equation}
    \hat V(s;\theta) \approx V^\pi(s), \qquad
    \hat Q(s,a;\theta) \approx Q^\pi(s,a), \qquad
    \pi_\theta(a|s) \approx \pi(a|s).
\end{equation}
The parameters $\theta$ are shared across many states and actions. Updating $\theta$ using one transition can change predictions for many inputs. This is both powerful and risky.

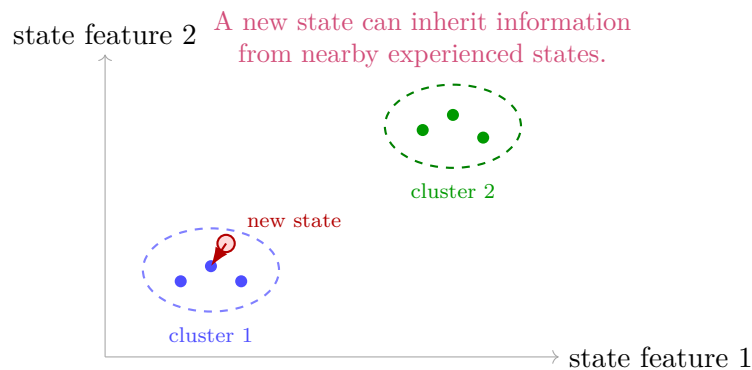
\begin{figure}[htbp]
    \centering
    \begin{tikzpicture}[
        point/.style={circle, fill, inner sep=1.6pt},
        newpoint/.style={circle, draw=red!70!black, thick, fill=red!15, inner sep=2.3pt},
        arrow/.style={-{Latex[length=2.5mm]}, thick}
    ]
        \draw[->, color=gray!70] (0,0) -- (6,0) node[right, color=black] {state feature 1};
        \draw[->, color=gray!70] (0,0) -- (0,4) node[above, color=black] {state feature 2};

        \foreach \x/\y in {1/1,1.4/1.2,1.8/1.0} {
            \node[point, fill=blue!70] at (\x,\y) {};
        }
        \foreach \x/\y in {4.2/3.0,4.6/3.2,5.0/2.9} {
            \node[point, fill=green!60!black] at (\x,\y) {};
        }

        \node[newpoint] (new) at (1.6,1.5) {};
        \draw[arrow, red!70!black] (1.6,1.5) -- (1.4,1.2);

        \draw[dashed, thick, blue!50] (1.4,1.15) ellipse (0.9 and 0.55);
        \draw[dashed, thick, green!50!black] (4.6,3.05) ellipse (0.9 and 0.55);

        \node[font=\scriptsize, color=blue!70] at (1.4,0.3) {cluster 1};
        \node[font=\scriptsize, color=green!60!black] at (4.6,2.2) {cluster 2};
        \node[font=\scriptsize, color=red!70!black] at (2.5,1.8) {new state};

        \node[align=center, font=\small, color=purple!70, fill=white, inner sep=2pt] at (4.2,4.2) {A new state can inherit information\\from nearby experienced states.};
    \end{tikzpicture}
    \caption{Generalization allows learning from similar states. In a table, unseen states have no value unless visited. With features or neural representations, knowledge can transfer across nearby or structurally similar situations.}
    \label{fig:generalization}
\end{figure}

\section{From tables to function approximation}

The tabular value function is a special case of function approximation. A table can be viewed as a parameter vector with one parameter per state or state-action pair. The tabular representation is exact but does not generalize.

A general approximator has the form
\begin{equation}
    \hat v(s;\theta) \approx v_\pi(s)
\end{equation}
or
\begin{equation}
    \hat q(s,a;\theta) \approx q_\pi(s,a).
\end{equation}
The learning problem becomes parameter estimation. A common value-prediction objective is mean-squared value error:
\begin{equation}
    J(\theta) = \frac{1}{2}\E_{s\sim d}\left[\left(v_\pi(s)-\hat v(s;\theta)\right)^2\right],
\end{equation}
where $d$ is a state distribution. In practice, $v_\pi(s)$ is unknown, so TD methods use bootstrapped targets such as
\begin{equation}
    y_t = R_{t+1} + \gamma \hat v(S_{t+1};\theta).
\end{equation}
This gives a semi-gradient update:
\begin{equation}
    \theta \leftarrow \theta + \alpha \delta_t \nabla_\theta \hat v(S_t;\theta),
\end{equation}
where
\begin{equation}
    \delta_t = R_{t+1} + \gamma \hat v(S_{t+1};\theta) - \hat v(S_t;\theta).
\end{equation}
Sutton and Barto treat function approximation as a core topic because it is the practical route from small finite MDPs to large problems \citep{sutton2018reinforcement}. Convergence properties become more delicate with approximation. For linear TD prediction under suitable conditions, important convergence results were established by Tsitsiklis and Van Roy \citep{tsitsiklis1997analysis}. For off-policy control with approximation, instability is a major concern, as shown by classic counterexamples \citep{baird1995residual}.

\section{Linear approximation, basis functions, and feature engineering}

Before deep learning became dominant, many RL systems used hand-designed features. A linear value approximator has the form
\begin{equation}
    \hat v(s;\theta) = \theta^\top \phi(s),
\end{equation}
where $\phi(s)$ is a feature vector. Similarly,
\begin{equation}
    \hat q(s,a;\theta) = \theta^\top \phi(s,a).
\end{equation}
Features may include distances, velocities, tile-coded indicators, radial basis functions, Fourier features, or domain-specific measurements.

Linear approximation is less expressive than deep neural networks, but it is easier to analyze. It also makes the role of representation very clear. If the features are good, linear RL can work well. If the features are poor, the algorithm cannot recover information that is not represented.

\begin{figure}[htbp]
    \centering
    \begin{tikzpicture}[
        box/.style={draw, rounded corners, thick, minimum width=2.7cm, minimum height=0.75cm, align=center},
        arrow/.style={-{Latex[length=2.4mm]}, thick},
        node distance=0.65cm
    ]
        \node[box] (state) {state $s$};
        \node[box, right=of state] (features) {hand-designed features\\$\phi(s)$};
        \node[box, right=of features] (linear) {linear predictor\\$\theta^\top\phi(s)$};
        \node[box, right=of linear] (value) {value estimate\\$\hat V(s)$};
        \draw[arrow] (state) -- (features);
        \draw[arrow] (features) -- (linear);
        \draw[arrow] (linear) -- (value);
        \node[below=0.9cm of features, align=center] {Before deep RL, success often depended on\\manual representation design.};
    \end{tikzpicture}
    \caption{Feature-based value approximation. The designer chooses $\phi(s)$, and learning adjusts $\theta$. Deep RL reduces manual feature engineering by learning representations, but it does not remove the need for careful modeling and evaluation.}
    \label{fig:linear_approximation}
\end{figure}

\Needspace{12\baselineskip}
\begin{lstlisting}[style=pythonstyle,caption={Linear value approximation with semi-gradient TD(0). This code is intentionally small to show the idea clearly.},label={lst:linear_td}]
import numpy as np

class LinearValueFunction:
    def __init__(self, num_features, alpha=0.05):
        self.theta = np.zeros(num_features, dtype=np.float64)
        self.alpha = alpha

    def value(self, features):
        return float(np.dot(self.theta, features))

    def update_td0(self, phi_s, reward, phi_next, gamma, done):
        v_s = self.value(phi_s)
        v_next = 0.0 if done else self.value(phi_next)
        td_target = reward + gamma * v_next
        td_error = td_target - v_s
        self.theta += self.alpha * td_error * phi_s
        return td_error

# In practice, phi_s must be designed or learned.
# For a robot, features might contain position, velocity, distance to goal,
# battery level, obstacle distance, or tile-coded indicators.
\end{lstlisting}

\section{Fitted value iteration and the supervised-learning view of RL}

One of the most useful ways to understand the transition from tabular RL to deep RL is to view value learning as repeated supervised learning. The Bellman equation defines targets. A regression model fits those targets.

For Q-learning-style control, the target is
\begin{equation}
    y_i = r_i + \gamma \max_{a'} \hat Q(s_i',a';\theta^-),
\end{equation}
where $\theta^-$ may be the previous parameters or a slowly updated target. The approximator is trained to minimize
\begin{equation}
    \mathcal{L}(\theta) = \frac{1}{N}\sum_{i=1}^N \left(y_i - \hat Q(s_i,a_i;\theta)\right)^2.
\end{equation}
This is supervised regression, but the targets are generated from the agent's own current value estimates. That makes RL different from ordinary supervised learning: the labels are moving targets.

Fitted Q-Iteration (FQI) formalizes this batch view. It repeatedly constructs regression targets from a dataset of transitions and fits a function approximator to those targets \citep{ernst2005fqi}. Neural Fitted Q-Iteration (NFQ) used multilayer perceptrons for data-efficient Q-function learning before the DQN breakthrough \citep{riedmiller2005nfq}.

\begin{figure}[htbp]
    \centering
    \begin{tikzpicture}[
        box/.style={draw, rounded corners, thick, minimum width=3.0cm, minimum height=0.8cm, align=center},
        arrow/.style={-{Latex[length=2.5mm]}, thick},
        node distance=0.8cm
    ]
        \node[box] (data) {transition dataset\\$(s,a,r,s')$};
        \node[box, right=of data] (target) {Bellman target\\$y=r+\gamma\max_{a'}Q(s',a')$};
        \node[box, right=of target] (regress) {supervised regression\\fit $Q_\theta(s,a)$};
        \node[box, below=of target] (repeat) {repeat with updated\\value estimates};

        \draw[arrow] (data) -- (target);
        \draw[arrow] (target) -- (regress);
        \draw[arrow] (regress.south) |- (repeat.east);
        \draw[arrow] (repeat.north) -- (target.south);
    \end{tikzpicture}
    \caption{Fitted value iteration turns RL into repeated supervised regression. The difficulty is that the regression targets depend on current or previous value estimates, so the target distribution changes during learning.}
    \label{fig:fqi_loop}
\end{figure}
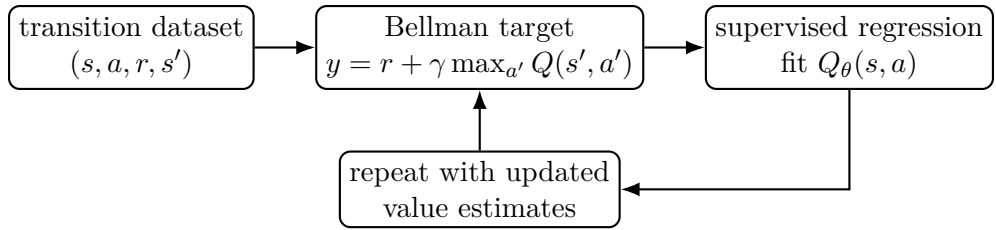

\Needspace{18\baselineskip}
\begin{lstlisting}[style=pythonstyle,caption={A minimal Fitted Q-Iteration skeleton using a generic regressor. The regressor could be a tree, random forest, linear model, or neural network.},label={lst:fqi_skeleton}]
import numpy as np

class FittedQIteration:
    def __init__(self, regressor_factory, actions, gamma=0.99):
        self.regressor = regressor_factory()
        self.actions = list(actions)
        self.gamma = gamma
        self.is_fitted = False

    def _sa_features(self, states, actions):
        actions = np.asarray(actions).reshape(-1, 1)
        return np.concatenate([states, actions], axis=1)

    def predict_q(self, states, actions):
        if not self.is_fitted:
            return np.zeros(len(states), dtype=np.float64)
        x = self._sa_features(states, actions)
        return self.regressor.predict(x)

    def fit(self, transitions, num_iterations=50):
        states, actions, rewards, next_states, dones = transitions

        for _ in range(num_iterations):
            next_q_values = []
            for a in self.actions:
                a_batch = np.full(len(next_states), a)
                q_next_a = self.predict_q(next_states, a_batch)
                next_q_values.append(q_next_a)

            max_next_q = np.max(np.stack(next_q_values, axis=1), axis=1)
            targets = rewards + self.gamma * (1.0 - dones) * max_next_q

            x_train = self._sa_features(states, actions)
            self.regressor.fit(x_train, targets)
            self.is_fitted = True
\end{lstlisting}

\section{Neural networks before DQN}

It is tempting to tell the history as if deep RL appeared suddenly in 2015. That is not accurate. Neural networks and reinforcement learning had been connected much earlier. Sutton and Barto note that function approximation in RL has roots in early neural-network learning systems \citep{sutton2018reinforcement}. Experience replay also predates DQN; Lin proposed using stored experiences for reinforcement learning in the early 1990s \citep{lin1992self}. Neural Fitted Q-Iteration used neural networks for Q-function approximation in 2005 \citep{riedmiller2005nfq}.

What changed around the early 2010s was not just one idea. Several conditions aligned:
\begin{itemize}[leftmargin=*]
    \item deep convolutional networks became much stronger after major computer-vision successes, including the ImageNet breakthrough of Krizhevsky, Sutskever, and Hinton \citep{krizhevsky2012imagenet};
    \item GPU computation became more available;
    \item the Arcade Learning Environment provided a diverse benchmark for general agents \citep{bellemare2013ale};
    \item DQN combined Q-learning, convolutional networks, experience replay, and target networks in a scalable way \citep{mnih2013atari,mnih2015human}.
\end{itemize}

Thus, the birth of modern deep RL was not simply “replace the table with a neural network.” Earlier attempts had shown that this could be unstable. The breakthrough required representation learning and stabilization.

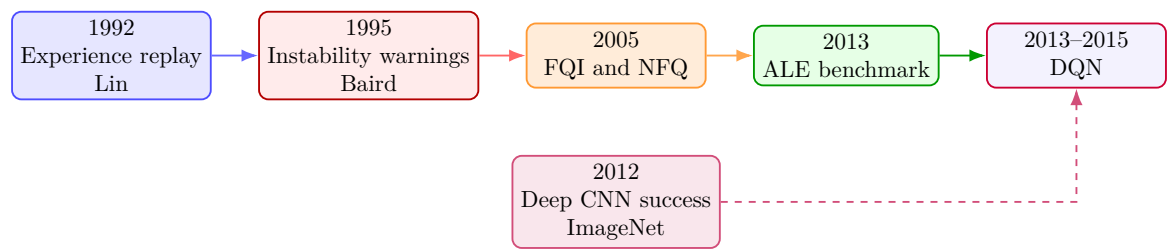
\begin{figure}[htbp]
    \centering
    \resizebox{0.96\textwidth}{!}{%
        \begin{tikzpicture}[
            event/.style={draw, rounded corners, thick, minimum width=2.7cm, minimum height=0.8cm, align=center, font=\small},
            arrow/.style={-{Latex[length=2.5mm]}, thick},
            node distance=0.7cm
        ]
            \node[event, draw=blue!70, fill=blue!10] (lin) {1992\\Experience replay\\Lin};
            \node[event, draw=red!70!black, fill=red!8, right=of lin] (baird) {1995\\Instability warnings\\Baird};
            \node[event, draw=orange!80, fill=orange!15, right=of baird] (fqi) {2005\\FQI and NFQ};
            \node[event, draw=purple!70, fill=purple!10, below=1.0cm of fqi] (imagenet) {2012\\Deep CNN success\\ImageNet};
            \node[event, draw=green!60!black, fill=green!10, right=of fqi] (ale) {2013\\ALE benchmark};
            \node[event, draw=red!80!blue, fill=red!5!blue!5, right=of ale] (dqn) {2013--2015\\DQN};

            \draw[arrow, color=blue!60] (lin) -- (baird);
            \draw[arrow, color=red!60] (baird) -- (fqi);
            \draw[arrow, color=orange!70] (fqi) -- (ale);
            \draw[arrow, color=green!60!black] (ale) -- (dqn);
            \draw[arrow, color=purple!70, dashed] (imagenet) -| (dqn.south);
        \end{tikzpicture}%
    }
    \caption{DQN was built on a long chain of ideas: replay, function approximation, warnings about instability, fitted value methods, convolutional networks, and benchmark environments.}
    \label{fig:pre_dqn_history}
\end{figure}

\section{Why combining RL and neural networks is hard}

Deep learning usually assumes a fixed dataset with relatively stable targets. Reinforcement learning violates this in several ways.

\subsection{The data distribution changes}

In supervised learning, the training data are often treated as samples from a fixed distribution. In RL, the data distribution depends on the policy. As the policy changes, the states visited by the agent change. Therefore, the training distribution is non-stationary.

\subsection{The target changes}

In supervised learning, a label such as “cat” or “dog” does not change when the model changes. In value-based RL, the target often contains the model's own future prediction:
\begin{equation}
    y_t = r_t + \gamma \max_{a'} Q(s_{t+1},a';\theta^-).
\end{equation}
If the value estimator changes, the target changes. This creates a moving-target problem.

\subsection{Errors can bootstrap into future errors}

Bootstrapping is powerful because it allows learning from incomplete experience. But it also means that a value estimate is trained using another value estimate. If the future estimate is wrong, the error can be propagated backward.

\subsection{Exploration creates correlated data}

Consecutive transitions are highly correlated. If the agent moves through a corridor, many samples look similar. Neural networks trained on highly correlated online samples can overfit recent experience and forget earlier cases. Experience replay reduces this correlation by sampling older transitions from a buffer \citep{lin1992self,mnih2015human}.

\begin{figure}[htbp]
    \centering
    \begin{tikzpicture}[
        box/.style={draw, rounded corners, thick, minimum width=3.2cm, minimum height=0.85cm, align=center, font=\small},
        arrow/.style={-{Latex[length=2.5mm]}, thick},
        node distance=0.8cm
    ]
        \node[box, draw=blue!70, fill=blue!10, font=\small\bfseries] (policy) {policy changes};
        \node[box, draw=orange!80, fill=orange!15, font=\small\bfseries, right=of policy] (data) {visited states change};
        \node[box, draw=green!60!black, fill=green!10, font=\small\bfseries, right=of data] (targets) {Bellman targets change};
        \node[box, draw=red!70!black, fill=red!8, font=\small\bfseries, below=1.2cm of targets] (network) {network parameters change};

        \draw[arrow, color=blue!60] (policy) -- (data);
        \draw[arrow, color=orange!70] (data) -- (targets);
        \draw[arrow, color=green!60!black] (targets) -- (network);
        \draw[arrow, color=red!70!black] (network.west) -| (policy.south);
    \end{tikzpicture}
    \caption{The moving-target problem in deep RL. The policy determines data, the data determine Bellman targets, the targets update the network, and the network changes the policy. This feedback loop is why stability is harder than in ordinary supervised learning.}
    \label{fig:moving_targets}
\end{figure}
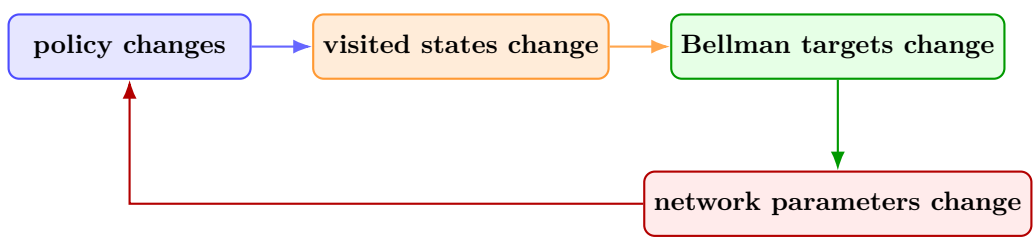

\section{The deadly triad}

Sutton and Barto identify a dangerous combination in reinforcement learning: function approximation, bootstrapping, and off-policy learning \citep{sutton2018reinforcement}. This is often called the \emph{deadly triad}. When all three are present, value learning can diverge.

\begin{itemize}[leftmargin=*]
    \item \textbf{Function approximation}: values are represented by a shared approximator rather than a table.
    \item \textbf{Bootstrapping}: updates use other learned value estimates as targets.
    \item \textbf{Off-policy learning}: the agent learns about one policy using data generated by another policy.
\end{itemize}

Q-learning is off-policy and bootstrapped. DQN adds nonlinear function approximation. This means deep Q-learning lives near the deadly triad. That does not mean it always fails, but it explains why stabilizing mechanisms are needed. Baird's counterexample showed that even linear function approximation can diverge in off-policy TD-style learning \citep{baird1995residual}. Later analyses studied TD learning with function approximation more rigorously \citep{tsitsiklis1997analysis}. Van Hasselt and colleagues investigated the deadly triad in deep Q-learning contexts \citep{vanhasselt2018deadly}.

\begin{figure}[htbp]
    \centering
    \begin{tikzpicture}[
        tri/.style={circle, draw, thick, minimum size=2.5cm, align=center},
        center/.style={draw, rounded corners, thick, minimum width=2.3cm, minimum height=0.8cm, align=center}
    ]
        \node[tri] (fa) at (0,2.4) {Function\\approximation};
        \node[tri] (boot) at (-2.7,-1.6) {Bootstrapping};
        \node[tri] (off) at (2.7,-1.6) {Off-policy\\learning};
        \node[center, fill=red!6] at (0,-0.15) {Risk of\\divergence};
        \draw[thick] (fa) -- (boot);
        \draw[thick] (boot) -- (off);
        \draw[thick] (off) -- (fa);
    \end{tikzpicture}
    \caption{The deadly triad: function approximation, bootstrapping, and off-policy learning. Each component is useful, but their combination can produce instability in value learning.}
    \label{fig:deadly_triad}
\end{figure}
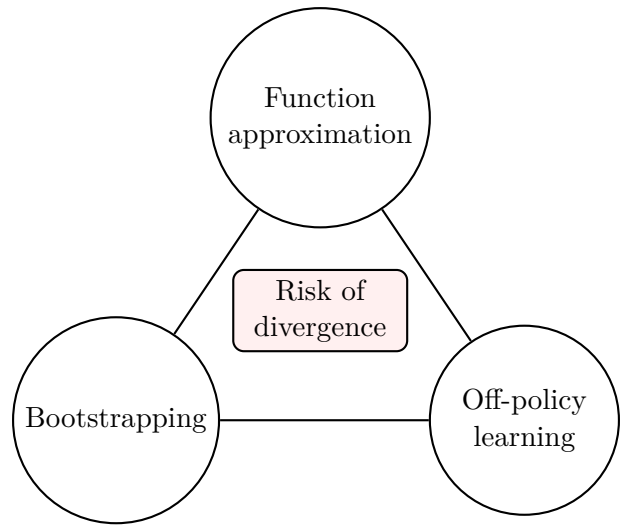

\begin{warningbox}{Important distinction}
The deadly triad is not an argument against deep RL. It is an explanation of why naive deep RL can fail. Modern algorithms do not avoid all three ingredients completely; instead, they control the damage using replay, target networks, conservative updates, trust regions, entropy regularization, distributional learning, double estimators, normalization, and careful evaluation.
\end{warningbox}

\section{Experience replay and target networks as stabilization ideas}

DQN's success did not come from neural approximation alone. Two stabilization mechanisms were especially important \citep{mnih2013atari,mnih2015human}.

\subsection{Experience replay}

Experience replay stores transitions in a buffer:
\begin{equation}
    \D = \{(s_i,a_i,r_i,s_i',d_i)\}_{i=1}^N.
\end{equation}
Training samples are drawn from this buffer, rather than directly from the most recent transition. Replay helps because it
\begin{itemize}[leftmargin=*]
    \item reuses data;
    \item breaks short-term temporal correlations;
    \item mixes old and new experience;
    \item makes the update look more like minibatch supervised learning.
\end{itemize}
Prioritized replay later sampled more informative transitions more often \citep{schaul2016prioritized}.

\subsection{Target networks}

A target network keeps a delayed copy of the value network. The online network has parameters $\theta$. The target network has parameters $\theta^-$. The DQN target is
\begin{equation}
    y_t = r_t + \gamma \max_{a'} Q(s_{t+1},a';\theta^-).
\end{equation}
The online network is updated to match this target, while $\theta^-$ changes only periodically or slowly. This reduces target drift.

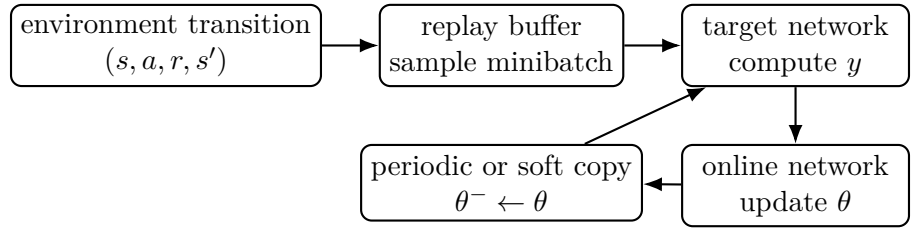
\begin{figure}[htbp]
    \centering
    \begin{tikzpicture}[
        box/.style={draw, rounded corners, thick, minimum width=3.0cm, minimum height=0.8cm, align=center},
        arrow/.style={-{Latex[length=2.5mm]}, thick},
        node distance=0.75cm
    ]
        \node[box] (env) {environment transition\\$(s,a,r,s')$};
        \node[box, right=of env] (buffer) {replay buffer\\sample minibatch};
        \node[box, right=of buffer] (target) {target network\\compute $y$};
        \node[box, below=of target] (online) {online network\\update $\theta$};
        \node[box, below=of buffer] (copy) {periodic or soft copy\\$\theta^- \leftarrow \theta$};
        \draw[arrow] (env) -- (buffer);
        \draw[arrow] (buffer) -- (target);
        \draw[arrow] (target) -- (online);
        \draw[arrow] (online) -- (copy);
        \draw[arrow] (copy) -- (target);
    \end{tikzpicture}
    \caption{Replay and target networks stabilize deep value learning. Replay improves data reuse and reduces correlation; target networks reduce moving-target instability.}
    \label{fig:replay_target}
\end{figure}

\Needspace{18\baselineskip}
\begin{lstlisting}[style=pythonstyle,caption={A simple replay buffer. This is the memory mechanism that later becomes central in DQN.},label={lst:replay_buffer}]
from collections import deque
import random
import numpy as np

class ReplayBuffer:
    def __init__(self, capacity):
        self.buffer = deque(maxlen=capacity)

    def add(self, state, action, reward, next_state, done):
        self.buffer.append((state, action, reward, next_state, done))

    def sample(self, batch_size):
        batch = random.sample(self.buffer, batch_size)
        states, actions, rewards, next_states, dones = zip(*batch)
        return (
            np.asarray(states),
            np.asarray(actions),
            np.asarray(rewards, dtype=np.float32),
            np.asarray(next_states),
            np.asarray(dones, dtype=np.float32),
        )

    def __len__(self):
        return len(self.buffer)
\end{lstlisting}

\section{Running example: UAV network control beyond tabular RL}

Consider again a UAV-assisted wireless network. A simplified tabular state might include:
\begin{equation}
    s = (x, y, z, b, \ell, q),
\end{equation}
where $(x,y,z)$ is UAV position, $b$ is battery level, $\ell$ is traffic load, and $q$ is a QoS indicator. If each variable is discretized into 20 bins, the state space already has $20^6$ states. If the action space has 10 actions, the Q-table has $640$ million entries.

A more realistic state may include:
\begin{itemize}[leftmargin=*]
    \item positions and velocities of multiple UAVs;
    \item battery states;
    \item user locations and priority classes;
    \item SINR, throughput, delay, jitter, and packet loss;
    \item traffic demand history;
    \item charging station availability;
    \item SDN controller recommendations;
    \item local map or heatmap observations.
\end{itemize}
This state is not a table index. It is a structured, high-dimensional observation. Deep RL becomes attractive because a neural network can process such observations and learn useful internal features.

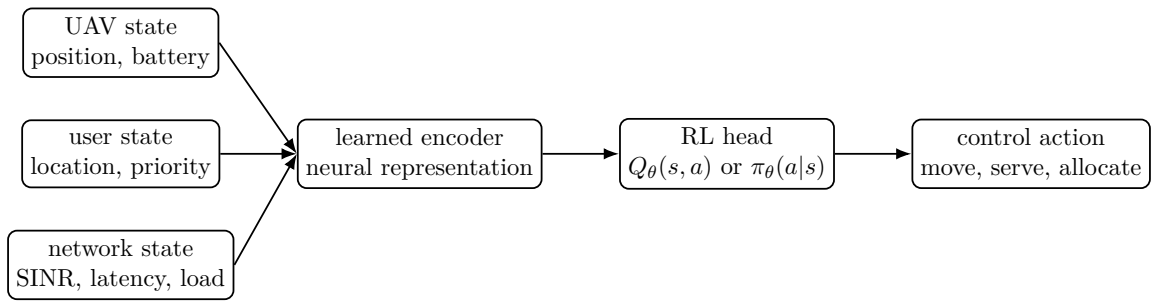
\begin{figure}[htbp]
    \centering
    \resizebox{0.95\textwidth}{!}{%
        \begin{tikzpicture}[
            block/.style={draw, rounded corners, thick, minimum width=2.7cm, minimum height=0.75cm, align=center},
            arrow/.style={-{Latex[length=2.4mm]}, thick},
            node distance=0.65cm
        ]
            \node[block] (pos) {UAV state\\position, battery};
            \node[block, below=of pos] (users) {user state\\location, priority};
            \node[block, below=of users] (qos) {network state\\SINR, latency, load};
            \node[block, right=1.2cm of users] (encoder) {learned encoder\\neural representation};
            \node[block, right=1.2cm of encoder] (head) {RL head\\$Q_\theta(s,a)$ or $\pi_\theta(a|s)$};
            \node[block, right=1.2cm of head] (action) {control action\\move, serve, allocate};
            \draw[arrow] (pos.east) -- (encoder.west);
            \draw[arrow] (users.east) -- (encoder.west);
            \draw[arrow] (qos.east) -- (encoder.west);
            \draw[arrow] (encoder) -- (head);
            \draw[arrow] (head) -- (action);
        \end{tikzpicture}%
    }
    \caption{In UAV network control, the state is naturally structured and high-dimensional. A learned encoder can combine mobility, user, and network features before producing a value estimate or policy.}
    \label{fig:uav_deep_state}
\end{figure}

\section{Python implementation: from table to neural approximator}

This section gives minimal code to show the conceptual transition. The goal is not to implement full DQN yet. That is Chapter~5. Here the purpose is to see how a value table becomes a parameterized neural function.

\subsection{Tabular Q-function}

\Needspace{14\baselineskip}
\begin{lstlisting}[style=pythonstyle,caption={A tabular Q-function. This is simple but requires a finite state index.},label={lst:tabular_q}]
import numpy as np

class TabularQ:
    def __init__(self, num_states, num_actions):
        self.q = np.zeros((num_states, num_actions), dtype=np.float32)

    def value(self, state_index, action):
        return self.q[state_index, action]

    def greedy_action(self, state_index):
        return int(np.argmax(self.q[state_index]))

    def update(self, s, a, r, s_next, done, alpha=0.1, gamma=0.99):
        next_value = 0.0 if done else np.max(self.q[s_next])
        target = r + gamma * next_value
        error = target - self.q[s, a]
        self.q[s, a] += alpha * error
        return error
\end{lstlisting}

\subsection{Neural Q-function}

The neural version accepts a feature vector rather than a state index. For discrete actions, a common architecture outputs one Q-value per action:
\begin{equation}
    Q_\theta(s,\cdot) = \left[Q_\theta(s,a_1),\ldots,Q_\theta(s,a_{|\A|})\right].
\end{equation}

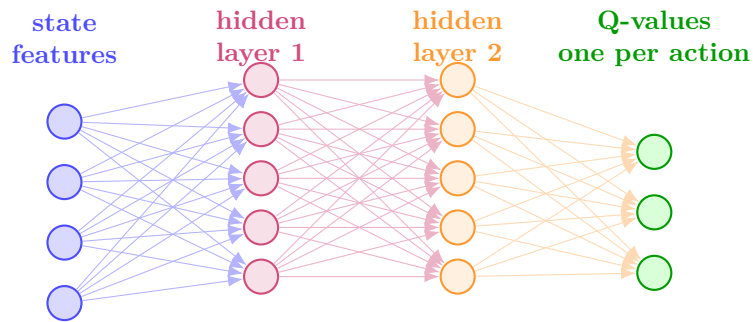
\begin{figure}[htbp]
    \centering
    \begin{tikzpicture}[
        neuron/.style={circle, draw, thick, minimum size=0.45cm},
        arrow/.style={-{Latex[length=2mm]}},
        layerlabel/.style={align=center, font=\small\bfseries}
    ]
        \foreach \i in {1,...,4} {\node[neuron, draw=blue!70, fill=blue!15] (i\i) at (0,1.2-0.8*\i) {};}
        \node[layerlabel, color=blue!70] at (0,1.5) {state\\features};

        \foreach \i in {1,...,5} {\node[neuron, draw=purple!70, fill=purple!12] (h\i) at (2.6,1.6-0.65*\i) {};}
        \node[layerlabel, color=purple!70] at (2.6,1.5) {hidden\\layer 1};

        \foreach \i in {1,...,5} {\node[neuron, draw=orange!80, fill=orange!12] (g\i) at (5.2,1.6-0.65*\i) {};}
        \node[layerlabel, color=orange!80] at (5.2,1.5) {hidden\\layer 2};

        \foreach \i in {1,...,3} {\node[neuron, draw=green!60!black, fill=green!15] (o\i) at (7.8,0.8-0.8*\i) {};}
        \node[layerlabel, color=green!60!black] at (7.8,1.5) {Q-values\\one per action};

        \foreach \i in {1,...,4} {\foreach \j in {1,...,5} {\draw[arrow, color=blue!30] (i\i) -- (h\j);}}
        \foreach \i in {1,...,5} {\foreach \j in {1,...,5} {\draw[arrow, color=purple!30] (h\i) -- (g\j);}}
        \foreach \i in {1,...,5} {\foreach \j in {1,...,3} {\draw[arrow, color=orange!30] (g\i) -- (o\j);}}
    \end{tikzpicture}
    \caption{A neural Q-function for discrete actions. The input is a feature vector or encoded observation. The output contains one value estimate for each action.}
    \label{fig:neural_q_network}
\end{figure}

\Needspace{22\baselineskip}
\begin{lstlisting}[style=pythonstyle,caption={A minimal PyTorch Q-network. Chapter 5 will add replay, target networks, and full DQN training.},label={lst:pytorch_q_network}]
import torch
import torch.nn as nn
import torch.nn.functional as F

class QNetwork(nn.Module):
    def __init__(self, obs_dim, num_actions, hidden_dim=128):
        super().__init__()
        self.net = nn.Sequential(
            nn.Linear(obs_dim, hidden_dim),
            nn.ReLU(),
            nn.Linear(hidden_dim, hidden_dim),
            nn.ReLU(),
            nn.Linear(hidden_dim, num_actions),
        )

    def forward(self, obs):
        return self.net(obs)

# Example usage:
obs_dim = 12          # e.g., UAV/network features
num_actions = 6       # e.g., move N/S/E/W/up/down
q_net = QNetwork(obs_dim, num_actions)

obs_batch = torch.randn(32, obs_dim)
q_values = q_net(obs_batch)       # shape: [32, num_actions]
actions = torch.argmax(q_values, dim=1)
\end{lstlisting}

\subsection{One neural Q-learning update}

A single minibatch Q-learning update minimizes
\begin{equation}
    \left(y_i - Q_\theta(s_i,a_i)\right)^2,
\end{equation}
where
\begin{equation}
    y_i = r_i + \gamma(1-d_i)\max_{a'} Q_{\theta^-}(s_i',a').
\end{equation}
The target network $Q_{\theta^-}$ is optional in this small code fragment but essential in DQN-style training.

\Needspace{22\baselineskip}
\begin{lstlisting}[style=pythonstyle,caption={One neural Q-learning minibatch update using an online network and a target network.},label={lst:neural_q_update}]
def q_learning_update(q_net, target_net, optimizer, batch, gamma=0.99):
    states, actions, rewards, next_states, dones = batch

    states = torch.as_tensor(states, dtype=torch.float32)
    actions = torch.as_tensor(actions, dtype=torch.long)
    rewards = torch.as_tensor(rewards, dtype=torch.float32)
    next_states = torch.as_tensor(next_states, dtype=torch.float32)
    dones = torch.as_tensor(dones, dtype=torch.float32)

    q_all = q_net(states)
    q_sa = q_all.gather(1, actions.view(-1, 1)).squeeze(1)

    with torch.no_grad():
        next_q = target_net(next_states).max(dim=1).values
        target = rewards + gamma * (1.0 - dones) * next_q

    loss = F.mse_loss(q_sa, target)

    optimizer.zero_grad()
    loss.backward()
    torch.nn.utils.clip_grad_norm_(q_net.parameters(), max_norm=10.0)
    optimizer.step()

    return float(loss.item())
\end{lstlisting}

\section{Why this chapter is not yet DQN}

At this point, it may seem that we already have DQN. Not quite. We have the ingredients for neural Q-learning, but DQN is a specific, historically important system:
\begin{itemize}[leftmargin=*]
    \item high-dimensional visual input;
    \item convolutional neural network;
    \item replay buffer;
    \item target network;
    \item Q-learning target;
    \item epsilon-greedy exploration;
    \item standardized Atari evaluation.
\end{itemize}
Chapter~5 studies DQN in detail. Chapter~4's role is to explain why DQN was needed, what problems it solved, and why it was not obvious that it would work.

\begin{figure}[htbp]
    \centering
    \begin{tikzpicture}[
        box/.style={draw, rounded corners, thick, minimum width=2.7cm, minimum height=0.8cm, align=center, font=\small},
        arrow/.style={-{Latex[length=3.5mm]}, thick},
        node distance=1.6cm
    ]
        \node[box, draw=blue!70, fill=blue!10, font=\small\bfseries] (tabular) {Tabular\\Q-learning};
        \node[box, draw=orange!80, fill=orange!15, font=\small\bfseries, right=of tabular] (approx) {Function\\approximation};
        \node[box, draw=purple!70, fill=purple!10, font=\small\bfseries, right=of approx] (neural) {Neural\\Q-learning};
        \node[box, draw=green!60!black, fill=green!10, font=\small\bfseries, right=of neural] (dqn) {DQN};

        \draw[arrow, color=red!70!black] (tabular) -- node[above, align=center, font=\scriptsize, color=red!70!black] {too many\\states} (approx);
        \draw[arrow, color=red!70!black] (approx) -- node[above, font=\scriptsize, color=red!70!black, align=center] {learn\\features} (neural);
        \draw[arrow, color=red!70!black] (neural) -- node[above, align=center, font=\scriptsize, color=red!70!black] {stabilize\\training} (dqn);
    \end{tikzpicture}
    \caption{Conceptual route to DQN. DQN is not merely neural Q-learning; it is neural Q-learning stabilized and applied to high-dimensional perceptual control.}
    \label{fig:route_to_dqn}
\end{figure}
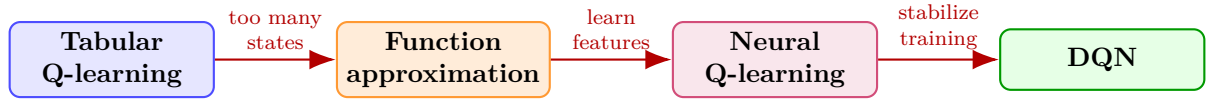

\section{Practical checklist: when classical RL is not enough}

A tabular method is usually not enough when one or more of the following conditions hold:
\begin{enumerate}[leftmargin=*]
    \item The state space is continuous.
    \item The observation is high-dimensional, such as images or sensor arrays.
    \item Exact repeated visits to the same state are rare.
    \item The agent must generalize to unseen situations.
    \item Manual discretization destroys important structure.
    \item The action space is continuous or very large.
    \item The environment is partially observable and requires memory.
    \item Multiple agents make the effective environment non-stationary.
    \item The task requires transfer across scenarios.
    \item The reward is sparse, delayed, or expensive to obtain.
\end{enumerate}

A function approximator is useful when:
\begin{enumerate}[leftmargin=*]
    \item similar states should have similar values;
    \item the representation can be learned or engineered;
    \item approximation error is acceptable compared with tabular impossibility;
    \item the training process can be stabilized.
\end{enumerate}

A deep neural network becomes attractive when:
\begin{enumerate}[leftmargin=*]
    \item observations are high-dimensional;
    \item useful features are hard to design manually;
    \item a large amount of experience or simulation is available;
    \item generalization matters more than exact tabular convergence.
\end{enumerate}

\section{Key takeaways}

\begin{itemize}[leftmargin=*]
    \item Classical RL provides the mathematical foundation of modern deep RL, but tabular representations do not scale to realistic state spaces.
    \item The curse of dimensionality makes naive discretization impossible for continuous or high-dimensional problems.
    \item Function approximation replaces value tables with parameterized models such as linear predictors, trees, or neural networks.
    \item Generalization is the core reason to use function approximation: learning about one state can improve predictions about similar states.
    \item Fitted value iteration shows that RL can be viewed as repeated supervised regression with Bellman-generated targets.
    \item Neural networks were explored in RL before DQN; DQN succeeded because it combined neural representation learning with stabilization mechanisms.
    \item Deep RL is difficult because the data distribution changes, the targets move, and errors can bootstrap into future errors.
    \item The deadly triad explains why function approximation, bootstrapping, and off-policy learning can be unstable when combined.
    \item Experience replay and target networks are not cosmetic details; they are stabilization tools that made deep Q-learning practical.
\end{itemize}

\section{Exercises}

\subsection*{Conceptual exercises}

\begin{enumerate}[leftmargin=*]
    \item Explain why a tabular Q-function cannot generalize to unseen states.
    \item What is the curse of dimensionality? Give a numerical example using at least five state variables.
    \item Why is discretization not a complete solution for continuous control problems?
    \item Explain the difference between a table and a function approximator.
    \item Why can neural networks help with high-dimensional observations?
    \item Why is deep RL usually less stable than supervised learning?
    \item What are the three components of the deadly triad?
    \item Why are experience replay and target networks useful in deep Q-learning?
\end{enumerate}

\subsection*{Mathematical exercises}

\begin{enumerate}[leftmargin=*]
    \item Suppose a state has $d=8$ variables, each discretized into $m=15$ bins, and there are $6$ actions. How many Q-values are required?
    \item Derive the semi-gradient TD(0) update for a linear value function $\hat v(s;\theta)=\theta^\top\phi(s)$.
    \item For a Q-network outputting one value per action, write the loss for a minibatch of $N$ transitions.
    \item Explain why the Bellman target is a moving target when $Q$ is approximated by a neural network.
    \item Compare the supervised regression loss in FQI with the tabular Q-learning update.
\end{enumerate}

\subsection*{Implementation exercises}

\begin{enumerate}[leftmargin=*]
    \item Modify Listing~\ref{lst:state_explosion} to compute memory usage in gigabytes if each Q-value is stored as a 32-bit float.
    \item Implement a simple feature map $\phi(s)$ for a two-dimensional continuous state using radial basis functions.
    \item Modify Listing~\ref{lst:pytorch_q_network} to use three hidden layers instead of two.
    \item Add a soft target-network update to Listing~\ref{lst:neural_q_update}:
    \[
    \theta^- \leftarrow \tau\theta + (1-\tau)\theta^-.
    \]
    \item Build a small continuous-state grid-world and compare tabular discretization with a neural Q approximator.
\end{enumerate}

\subsection*{Research thinking exercises}

\begin{enumerate}[leftmargin=*]
    \item In UAV network control, which variables should be treated as raw observations and which should be engineered features?
    \item What would be dangerous about using a neural approximator for safety-critical control without a safety filter?
    \item How could approximation error create unfair service among users in a wireless network?
    \item In what cases might a simple linear approximator be preferable to a deep network?
    \item Design an evaluation protocol to test whether a learned value function generalizes to unseen traffic loads.
\end{enumerate}

\section*{Looking Ahead to Chapter 5: Food for Thought}
\addcontentsline{toc}{section}{Looking Ahead to Chapter 5: Food for Thought}

Chapter~4 explained why classical RL was not enough. A table cannot represent values for high-dimensional images, continuous robot states, wireless network measurements, or complex multi-agent systems. Function approximation is necessary, and neural networks are powerful approximators.

But necessity is not the same as success. Simply replacing a Q-table with a neural network can fail. The agent's data are correlated, its targets move, and bootstrapped errors can reinforce themselves. The key historical question is therefore:

\begin{quote}
    What made deep Q-learning work when naive neural Q-learning was unstable?
\end{quote}

\subsection*{Questions to ponder}

\begin{enumerate}[leftmargin=*]
    \item Why is learning from raw pixels harder than learning from a compact state vector?
    \item Why does a convolutional neural network make sense for Atari frames?
    \item Why does DQN output one Q-value per discrete action?
    \item Why does experience replay make online RL more like supervised learning?
    \item Why does a target network reduce instability?
    \item Why is epsilon-greedy exploration still used even with a neural network?
    \item What problems remained after DQN?
\end{enumerate}

\subsection*{Main idea for the next chapter}

Chapter~5 studies Deep Q-Networks in detail. It explains the Atari breakthrough, the DQN architecture, the loss function, replay buffers, target networks, preprocessing, training loops, and limitations. The central idea is:

\begin{quote}
    DQN made deep reinforcement learning practical by combining Q-learning with representation learning and stabilization mechanisms.
\end{quote}
	\chapter[Deep Q-Networks]{Deep Q-Networks and the Atari Breakthrough}
\label{ch:dqn}
\chaptermark{Deep Q-Networks}

\begin{keybox}{Chapter goal}
Chapter~4 explained why classical tabular reinforcement learning could not scale to large, continuous, or perceptual state spaces. Chapter~5 studies the first widely recognized breakthrough that made deep reinforcement learning visible: the Deep Q-Network, or DQN. The chapter explains how DQN transformed Q-learning into an end-to-end visual control algorithm by combining convolutional neural networks, experience replay, target networks, reward clipping, frame stacking, and epsilon-greedy exploration. It also explains why DQN worked, why it was unstable without its stabilizers, how to implement it in PyTorch, and what limitations motivated the next generation of value-based deep RL algorithms.
\end{keybox}

\section*{Chapter Overview}
\addcontentsline{toc}{section}{Chapter Overview}

\begin{enumerate}[leftmargin=*]
    \item Why DQN matters
    \item The Atari benchmark and the Arcade Learning Environment
    \item From Q-learning to neural Q-learning
    \item The central difficulty: bootstrapping with a moving neural approximator
    \item DQN as an architecture and as a training system
    \item Input preprocessing: pixels, grayscale, resizing, frame stacking, and action repeat
    \item The convolutional Q-network
    \item Experience replay: learning from remembered transitions
    \item Target networks: slowing down the moving target
    \item The DQN loss and semi-gradient update
    \item Epsilon-greedy exploration and evaluation protocols
    \item Algorithm box: Deep Q-learning with experience replay
    \item Python implementation: replay buffer, network, update, and training loop
    \item Practical engineering details and debugging checklist
    \item What DQN achieved and what it did not solve
    \item Explicit limitations: what DQN cannot do well
    \item Running example: DQN ideas for UAV and network control
    \item Worked mini-example: DQN for a UAV service-control task
    \item Key takeaways
    \item Exercises
    \item Looking ahead to Chapter 6
\end{enumerate}

\section{Why DQN matters}

Deep Q-Networks are important not because they were the first attempt to combine neural networks and reinforcement learning. Earlier work had already studied neural fitted Q-iteration, experience replay, and function approximation in reinforcement learning \citep{lin1992self,riedmiller2005nfq,gordon1995stable,bertsekas1996neurodynamic}. DQN is important because it demonstrated, in a convincing and reusable way, that a single deep RL system could learn useful control policies directly from high-dimensional visual input across multiple Atari 2600 games \citep{mnih2013atari,mnih2015human,bellemare2013ale}.

The core achievement was conceptual and engineering-based at the same time. Conceptually, DQN showed that the Bellman optimality target from Q-learning could be combined with a convolutional neural network that receives raw pixels. Engineering-wise, DQN showed that this combination becomes much more stable when two mechanisms are added:
\begin{itemize}[leftmargin=*]
    \item \textbf{experience replay}, which stores past transitions and trains on randomly sampled minibatches;
    \item \textbf{a target network}, which computes the bootstrap target using a delayed copy of the online Q-network.
\end{itemize}
These two ideas did not eliminate all instability, but they made deep Q-learning practical enough to create a major breakthrough.

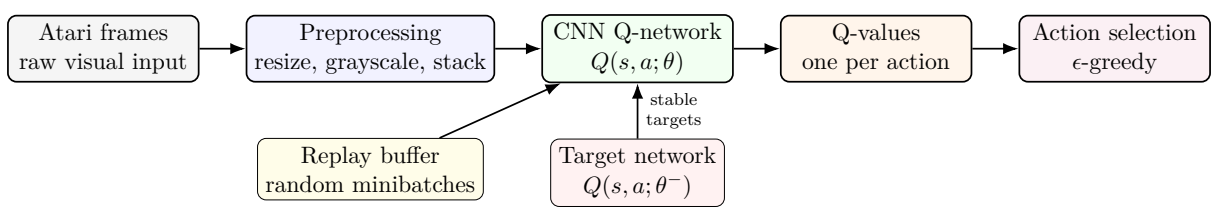
\begin{figure}[htbp]
    \centering
    \resizebox{\textwidth}{!}{%
        \begin{tikzpicture}[
            box/.style={draw, rounded corners, thick, minimum width=3.1cm, minimum height=0.9cm, align=center},
            smallbox/.style={draw, rounded corners, minimum width=2.6cm, minimum height=0.75cm, align=center},
            arrow/.style={-{Latex[length=2.5mm]}, thick},
            node distance=0.75cm
        ]
            \node[box, fill=gray!8] (pixels) {Atari frames\\raw visual input};
            \node[box, right=of pixels, fill=blue!5] (pre) {Preprocessing\\resize, grayscale, stack};
            \node[box, right=of pre, fill=green!5] (cnn) {CNN Q-network\\$Q(s,a;\theta)$};
            \node[box, right=of cnn, fill=orange!8] (qvals) {Q-values\\one per action};
            \node[box, right=of qvals, fill=purple!6] (act) {Action selection\\$\epsilon$-greedy};
            \node[smallbox, below=0.9cm of cnn, fill=red!5] (target) {Target network\\$Q(s,a;\theta^-)$};
            \node[smallbox, below=0.9cm of pre, fill=yellow!10] (replay) {Replay buffer\\random minibatches};
            \draw[arrow] (pixels) -- (pre);
            \draw[arrow] (pre) -- (cnn);
            \draw[arrow] (cnn) -- (qvals);
            \draw[arrow] (qvals) -- (act);
            \draw[arrow] (replay) -- (cnn);
            \draw[arrow] (target) -- node[right, font=\scriptsize, align=center] {stable\\targets} (cnn);
        \end{tikzpicture}%
    }
    \caption{DQN is not only a neural network. It is a training system that combines pixel preprocessing, a convolutional Q-network, experience replay, a delayed target network, and epsilon-greedy exploration. The key achievement was to make neural Q-learning stable enough for high-dimensional visual control \citep{mnih2013atari,mnih2015human}.}
    \label{fig:ch5_dqn_system}
\end{figure}

Before DQN, deep learning had already transformed supervised learning, especially after the success of convolutional neural networks on large-scale image classification \citep{lecun1998gradient,krizhevsky2012imagenet}. Reinforcement learning, however, was harder. The data distribution changes as the agent learns. Targets depend on the current network. Feedback is delayed. Consecutive samples are highly correlated. Exploration changes what data the agent collects. DQN mattered because it showed one workable recipe for this difficult setting.

\begin{researchbox}{Historical interpretation}
DQN should be read as a bridge between two traditions: Q-learning from classical RL \citep{watkins1992q,sutton2018reinforcement} and convolutional representation learning from deep learning \citep{lecun1998gradient,krizhevsky2012imagenet}. Its success did not come from replacing RL theory. It came from embedding Bellman-style temporal-difference learning inside a deep visual architecture and adding stabilizing mechanisms that made training feasible.
\end{researchbox}

\section{The Atari benchmark and the Arcade Learning Environment}

The Arcade Learning Environment, or ALE, provided a standard benchmark for training and evaluating agents on Atari 2600 games \citep{bellemare2013ale,machado2018revisiting}. Atari games were attractive for several reasons:
\begin{itemize}[leftmargin=*]
    \item they require sequential decision-making;
    \item observations are high-dimensional images;
    \item action spaces are discrete;
    \item rewards are game scores;
    \item the emulator allows large-scale repeated training;
    \item many games can be tested using the same interface.
\end{itemize}

Atari was not chosen because it perfectly represents the physical world. It was chosen because it was complex enough to test perception, control, and delayed reward, while still being cheap enough to simulate millions of steps. This made it an ideal bridge benchmark between toy grid-worlds and real robots.

\begin{figure}[htbp]
    \centering
    \begin{tikzpicture}[
        game/.style={draw, rounded corners, thick, minimum width=2.7cm, minimum height=1.0cm, align=center},
        center/.style={draw, rounded corners, very thick, minimum width=3.8cm, minimum height=1.0cm, align=center},
        arrow/.style={-{Latex[length=2.5mm]}, thick},
        node distance=0.85cm
    ]
        \node[center, fill=blue!5] (ale) {Arcade Learning Environment\\standard RL interface};
        \node[game, above left=of ale, fill=gray!7] (breakout) {Breakout\\paddle + ball};
        \node[game, above right=of ale, fill=gray!7] (pong) {Pong\\two-player reflexes};
        \node[game, below left=of ale, fill=gray!7] (seaquest) {Seaquest\\resource + survival};
        \node[game, below right=of ale, fill=gray!7] (space) {Space Invaders\\shooting + avoidance};
        \draw[arrow] (breakout) -- (ale);
        \draw[arrow] (pong) -- (ale);
        \draw[arrow] (seaquest) -- (ale);
        \draw[arrow] (space) -- (ale);
        \node[below=2.2cm of ale, align=center] {Same learning algorithm, same interface, many different games.};
    \end{tikzpicture}
    \caption{The Arcade Learning Environment turned Atari 2600 games into a common RL benchmark. DQN used the same architecture and core hyperparameters across many games, which made its success more convincing than solving one hand-designed task \citep{bellemare2013ale,mnih2015human,machado2018revisiting}.}
    \label{fig:ch5_ale}
\end{figure}

The original DQN line of work had two landmark publications. The 2013 paper demonstrated deep Q-learning on seven Atari games and introduced the essential idea of learning from pixels using a convolutional network and experience replay \citep{mnih2013atari}. The 2015 \emph{Nature} paper expanded the system to a larger set of games and added a more developed architecture and target-network stabilization \citep{mnih2015human}.

Modern Atari evaluation protocols are more careful than early ones. Later work emphasized issues such as sticky actions, human starts, random no-op starts, evaluation length, and reporting practices \citep{machado2018revisiting,hessel2018rainbow}. The lesson is important for this book: benchmark success is not only about the algorithm. It also depends on evaluation protocol.

\section{From Q-learning to neural Q-learning}

Classical Q-learning learns an action-value function using the update
\begin{equation}
    Q(s_t,a_t) \leftarrow Q(s_t,a_t) + \alpha\left[y_t - Q(s_t,a_t)\right],
\end{equation}
where the one-step target is
\begin{equation}
    y_t = r_{t+1} + \gamma \max_{a'} Q(s_{t+1},a').
\end{equation}
This update is tabular: it modifies the entry corresponding to $(s_t,a_t)$.

DQN replaces the table with a neural network:
\begin{equation}
    Q(s,a;\theta) \approx Q^*(s,a),
\end{equation}
where $\theta$ are the network parameters. The network receives an observation or state representation and outputs one Q-value for each discrete action.

\begin{figure}[htbp]
    \centering
    \resizebox{0.95\textwidth}{!}{%
        \begin{tikzpicture}[
            box/.style={draw, rounded corners, thick, minimum width=3.0cm, minimum height=0.9cm, align=center, font=\small},
            tablecell/.style={draw, minimum width=0.75cm, minimum height=0.45cm, align=center},
            arrow/.style={-{Latex[length=2.5mm]}, thick}
        ]
            \node[box, draw=blue!70, fill=blue!10, font=\small\bfseries] (tab) at (0,3.2) {Tabular Q-learning};
            \foreach \i in {0,...,3} {
                \foreach \j in {0,...,3} {
                    \pgfmathsetmacro{\shade}{15 + \i*8 + \j*6}
                    \node[tablecell, fill=blue!\shade!white, draw=blue!40] at (-1.1+0.75*\j, 1.8-0.45*\i) {};
                }
            }
            \node[font=\small, color=blue!70] at (0,-0.2) {$Q[s,a]$ stored explicitly};

            \node[box, draw=orange!80, fill=orange!15, font=\small\bfseries] (dqn) at (8,3.2) {Deep Q-Network};
            \node[box, draw=green!60!black, fill=green!10, font=\small] (state) at (3.7,1.2) {state or pixels\\$s$};
            \node[box, draw=purple!70, fill=purple!10, font=\small] (net) at (8,1.2) {neural network\\$Q(\cdot;\theta)$};
            \node[box, draw=red!70!black, fill=red!8, font=\small] (out) at (11.8,1.2) {$Q(s,a_1)$\\$Q(s,a_2)$\\$\cdots$};

            \draw[arrow, color=green!60!black] (state) -- (net);
            \draw[arrow, color=purple!70] (net) -- (out);
            \draw[arrow, dashed, color=gray!60] (tab.east) -- node[above, align=center, font=\scriptsize, color=gray!70!black] {replace table by\\function approximator} (dqn.west);
        \end{tikzpicture}%
    }
    \caption{DQN replaces the Q-table with a neural network. Instead of storing one value per state-action pair, the network maps an input state to a vector of action values. This allows generalization across similar observations, but also creates new instability because the function approximator is updated by bootstrapped targets.}
    \label{fig:ch5_table_to_dqn}
\end{figure}
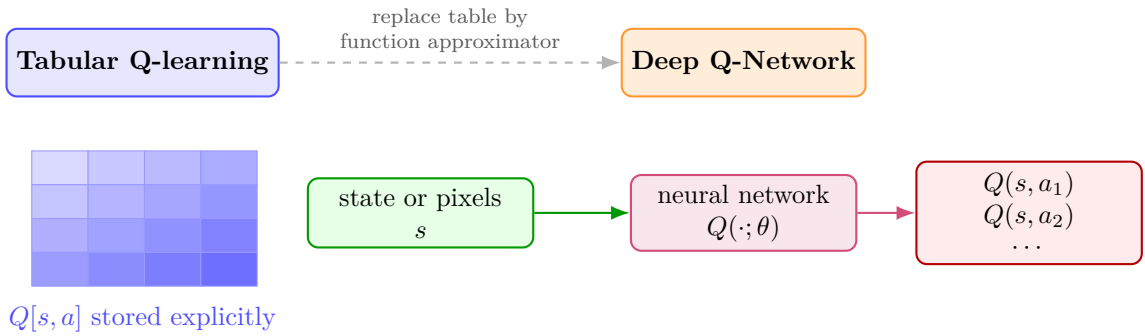

For a transition $(s_t,a_t,r_{t+1},s_{t+1},d_t)$, where $d_t$ indicates whether the next state is terminal, DQN uses the target
\begin{equation}
    y_t = r_{t+1} + \gamma(1-d_t)\max_{a'} Q(s_{t+1},a';\theta^-),
    \label{eq:dqn_target}
\end{equation}
where $\theta^-$ are the parameters of the target network. The online network parameters $\theta$ are updated to reduce the discrepancy between $Q(s_t,a_t;\theta)$ and $y_t$.

The simplest squared-error loss is
\begin{equation}
    L(\theta) = \E_{(s,a,r,s',d)\sim \D}\left[\left(y - Q(s,a;\theta)\right)^2\right].
    \label{eq:dqn_loss}
\end{equation}
In practice, the \emph{Nature} DQN used a clipped error closely related to the Huber loss to reduce sensitivity to large temporal-difference errors \citep{mnih2015human}. Modern implementations usually use the Huber loss:
\begin{equation}
    \ell_\kappa(\delta)=
    \begin{cases}
        \frac{1}{2}\delta^2, & |\delta| \leq \kappa,\\
        \kappa(|\delta| - \frac{1}{2}\kappa), & |\delta| > \kappa,
    \end{cases}
    \quad \text{where } \delta = y - Q(s,a;\theta).
\end{equation}

\section{The central difficulty: bootstrapping with a moving neural approximator}

Naive neural Q-learning is unstable because the target depends on the model being trained. The target is not a fixed label like in supervised learning. It is computed from a changing value function. This creates a moving-target problem:
\begin{equation}
    y_t(\theta) = r_{t+1} + \gamma\max_{a'}Q(s_{t+1},a';\theta).
\end{equation}
If the same network computes both the prediction and the target, then the target moves every time the network updates. Training can chase its own predictions.

This is part of the broader instability known as the deadly triad: function approximation, bootstrapping, and off-policy learning \citep{sutton2018reinforcement,tsitsiklis1997analysis,baird1995residual}. DQN contains all three:
\begin{itemize}[leftmargin=*]
    \item \textbf{function approximation}: a neural network approximates $Q$;
    \item \textbf{bootstrapping}: targets use $\max_{a'}Q(s',a')$;
    \item \textbf{off-policy learning}: the Q-learning target learns a greedy policy while behavior is epsilon-greedy.
\end{itemize}

\begin{figure}[htbp]
    \centering
    \begin{tikzpicture}[
        circ/.style={circle, draw, thick, minimum size=2.7cm, align=center, font=\small\bfseries},
        note/.style={draw, rounded corners, thick, minimum width=3.2cm, minimum height=0.8cm, align=center, font=\small\bfseries},
        arrow/.style={-{Latex[length=2.5mm]}, thick}
    ]
        \node[circ, draw=blue!70, fill=blue!10] (fa) at (0,2.6) {Function\\approximation};
        \node[circ, draw=green!60!black, fill=green!10] (boot) at (-2.2,-1.3) {Bootstrapping};
        \node[circ, draw=orange!80, fill=orange!15] (off) at (2.2,-1.3) {Off-policy\\learning};
        \node[note, draw=red!70!black, fill=red!10] (unstable) at (0,-3.8) {Potential instability\\{\normalfont\small the deadly triad}};

        \draw[arrow, color=blue!60] (fa) -- (unstable);
        \draw[arrow, color=green!60!black] (boot) -- (unstable);
        \draw[arrow, color=orange!70] (off) -- (unstable);

        \node[align=center, font=\small, color=purple!70, fill=white, inner sep=3pt, rounded corners] at (0,0.6) {DQN uses all three};
    \end{tikzpicture}
    \caption{DQN combines function approximation, bootstrapping, and off-policy learning. This combination can be unstable in general. Experience replay and target networks are practical stabilization mechanisms, not mathematical magic.}
    \label{fig:ch5_deadly_triad}
\end{figure}
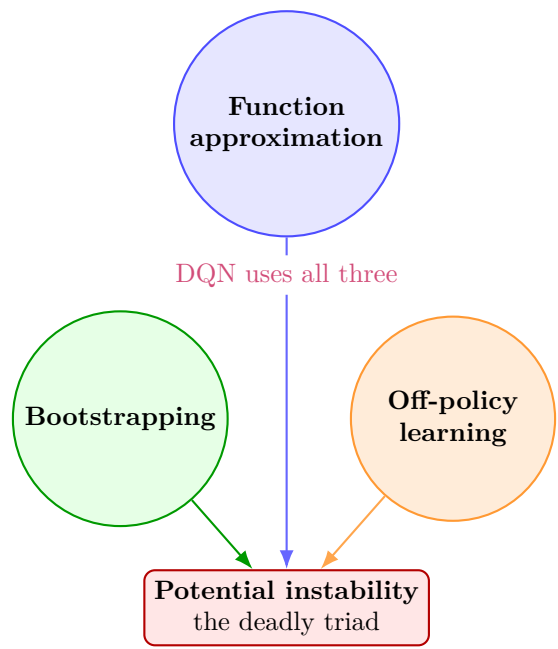

\subsection{A short sketch of Baird's counterexample}

The deadly triad is not only a theoretical warning. Baird's counterexample is a classic demonstration that temporal-difference learning with function approximation can diverge even in a small, carefully constructed Markov process \citep{baird1995residual,sutton2018reinforcement}. The example is often called \emph{Baird's star} because its states are commonly drawn as several upper states connected to a special lower state.

The exact feature vectors are not necessary for our purpose here. What matters is the interaction between three ingredients: function approximation, bootstrapping, and off-policy learning. The value function is represented by parameters rather than a table; the temporal-difference target depends on the current value estimate; and the data are generated by a behavior policy that differs from the policy being evaluated.

The striking point is that the value function in Baird's example is only linear. No deep neural network is involved. Nevertheless, the TD updates can drive the parameters away from the correct solution instead of toward it. This shows that instability is not merely a problem caused by large neural networks. It can already appear in classical reinforcement learning when approximation, bootstrapping, and off-policy learning are combined.

This observation is important for understanding DQN. Deep Q-learning combines all three elements of the deadly triad: it uses a neural network as a function approximator, it bootstraps through the Bellman target, and it learns from replayed experience that may not match the current policy exactly. DQN became successful not because this instability disappeared, but because experience replay and target networks made the learning process stable enough in practice.

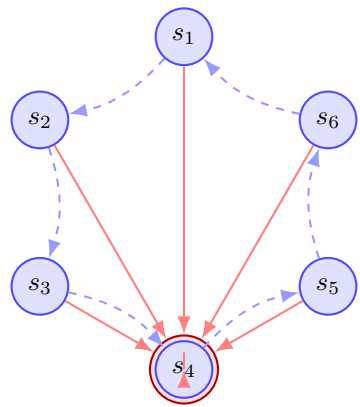
\begin{figure}[htbp]
    \centering
    \vspace{0.3cm}
    \begin{tikzpicture}[
        state/.style={circle, draw=blue!70, thick, minimum size=0.75cm, align=center, fill=blue!12, font=\small\bfseries},
        lower/.style={circle, draw=red!70!black, thick, minimum size=0.9cm, align=center, fill=red!12, font=\small\bfseries},
        arrow/.style={-{Latex[length=2.4mm]}, thick},
        dashedarrow/.style={-{Latex[length=2.4mm]}, thick, dashed}
    ]
        \node[lower] (l) at (0,-1.8) {$s_7$};
        \foreach \i/\ang in {1/90,2/150,3/210,4/270,5/330,6/30} {
            \node[state] (s\i) at ({2.2*cos(\ang)},{2.2*sin(\ang)+0.4}) {$s_\i$};
            \draw[arrow, color=red!50] (s\i) -- (l);
        }
        \foreach \i/\j in {1/2,2/3,3/4,4/5,5/6,6/1} {
            \draw[dashedarrow, color=blue!40, bend left=18] (s\i) to (s\j);
        }
        \node[align=center, font=\small, color=purple!70] at (0,3.6) {Small state space + linear approximation\\can still diverge under off-policy bootstrapping};
    \end{tikzpicture}
    \caption{A simplified sketch of the intuition behind Baird's counterexample. The exact feature construction is omitted, but the message is central: off-policy temporal-difference learning with function approximation and bootstrapping can diverge even in a small problem. DQN inherits this danger because it uses a nonlinear approximator, bootstrapped targets, and off-policy replay.}
    \label{fig:ch5_baird_sketch}
\end{figure}

\begin{warningbox}{Why this matters for DQN}
DQN did not make the deadly triad disappear. It made one particular instance of neural Q-learning work well enough by adding stabilizers. Experience replay changes the sampling process, target networks slow the target motion, reward clipping controls target scale, and gradient clipping limits rare large updates. These are practical controls on instability, not a general convergence proof.
\end{warningbox}

DQN's contribution was not to prove that this combination is always safe. It was to discover an engineering recipe that worked surprisingly well on a difficult benchmark.

\section{DQN as an architecture and as a training system}

A common misunderstanding is to define DQN as just “a neural network for Q-learning.” This is incomplete. DQN is better understood as a complete training system with the following components:
\begin{enumerate}[leftmargin=*]
    \item a discrete-action environment, such as Atari;
    \item visual preprocessing;
    \item frame stacking to provide short-term motion information;
    \item a convolutional network that outputs Q-values;
    \item epsilon-greedy exploration;
    \item an experience replay buffer;
    \item minibatch stochastic gradient descent;
    \item a target network;
    \item reward clipping;
    \item periodic evaluation.
\end{enumerate}

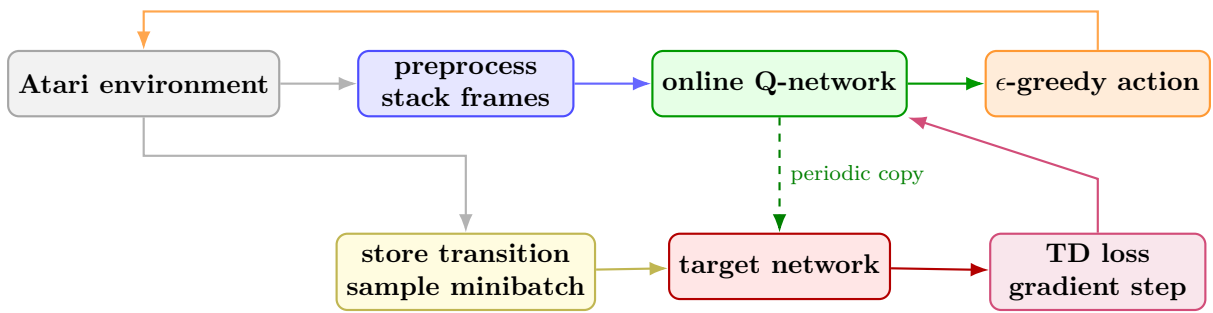
\begin{figure}[htbp]
    \centering
    \resizebox{\textwidth}{!}{%
        \begin{tikzpicture}[
            box/.style={draw, rounded corners, thick, minimum width=2.8cm, minimum height=0.85cm, align=center, font=\small},
            arrow/.style={-{Latex[length=2.4mm]}, thick},
            node distance=1.2cm and 1.0cm
        ]
            \node[box, draw=gray!70, fill=gray!10, font=\small\bfseries] (env) {Atari environment};
            \node[box, draw=blue!70, fill=blue!10, font=\small\bfseries, right=of env] (obs) {preprocess\\stack frames};
            \node[box, draw=green!60!black, fill=green!10, font=\small\bfseries, right=of obs] (qnet) {online Q-network};
            \node[box, draw=orange!80, fill=orange!15, font=\small\bfseries, right=of qnet] (action) {$\epsilon$-greedy action};

            \node[box, draw=yellow!70!black, fill=yellow!15, font=\small\bfseries, below=1.5cm of obs] (replay) {store transition\\sample minibatch};
            \node[box, draw=red!70!black, fill=red!10, font=\small\bfseries, below=1.5cm of qnet] (target) {target network};
            \node[box, draw=purple!70, fill=purple!10, font=\small\bfseries, below=1.5cm of action] (loss) {TD loss\\gradient step};

            \draw[arrow, color=gray!60] (env) -- (obs);
            \draw[arrow, color=blue!60] (obs) -- (qnet);
            \draw[arrow, color=green!60!black] (qnet) -- (action);
            \draw[arrow, color=orange!70] (action.north) -- ++(0,0.5) -| (env.north);
            \draw[arrow, color=gray!60] (env.south) -- ++(0,-0.5) -| (replay.north);
            \draw[arrow, color=yellow!70!black] (replay) -- (target);
            \draw[arrow, color=red!70!black] (target) -- (loss);
            \draw[arrow, color=purple!70] (loss.north) -- ++(0,0.7) -- (qnet.south east);
            \draw[arrow, dashed, color=green!50!black] (qnet.south) -- node[right, font=\scriptsize, color=green!50!black] {periodic copy} (target.north);
        \end{tikzpicture}%
    }
    \caption{DQN is a closed-loop training system. The agent collects transitions, stores them in replay memory, samples random minibatches, computes targets with a delayed target network, updates the online network, and periodically copies online parameters to the target network.}
    \label{fig:ch5_training_loop}
\end{figure}

Each component addresses a specific problem. Frame stacking addresses partial observability in raw frames. Experience replay addresses sample correlation and data reuse. The target network addresses moving targets. Reward clipping limits the scale of temporal-difference errors across different games. Epsilon-greedy exploration allows the agent to discover actions while still exploiting learned values.

\begin{table}[htbp]
    \centering
    \caption{DQN mechanisms, the problem each mechanism addresses, and the cost or limitation it introduces.}
    \label{tab:ch5_mechanisms}
    \small
    \begin{tabularx}{\textwidth}{p{0.20\textwidth}X X p{0.22\textwidth}}
        \toprule
        \textbf{Mechanism} & \textbf{Problem addressed} & \textbf{Core idea} & \textbf{Cost or limitation} \\
        \midrule
        Frame stacking & A single frame may not contain velocity or direction. & Use the last four preprocessed frames as the state input. & Only short memory; not a general solution to partial observability. \\
        Experience replay & Consecutive transitions are highly correlated and data is expensive. & Store transitions and train on random minibatches. & Replay data can become stale and is not truly IID. \\
        Target network & The Bellman target moves whenever the online network changes. & Compute targets using delayed parameters $\theta^-$. & Targets are only piecewise stable and can lag behind learning. \\
        Reward clipping & Atari games have very different reward scales. & Map rewards to $\{-1,0,+1\}$ before updating. & Changes the original objective and removes magnitude information. \\
        $\epsilon$-greedy exploration & A greedy policy may never discover useful actions. & Choose random actions with probability $\epsilon$. & Simple, but weak in sparse-reward or hard-exploration tasks. \\
        CNN encoder & Raw pixels are too large for tabular methods. & Learn visual features jointly with the Q-function. & Requires many samples and can overfit to benchmark-specific visual structure. \\
        Huber or clipped TD loss & Large TD errors can destabilize updates. & Use a loss less sensitive than pure squared error. & It treats symptoms of instability, not the underlying cause. \\
        \bottomrule
    \end{tabularx}
\end{table}

Table~\ref{tab:ch5_mechanisms} is one of the most useful ways to remember DQN. Every mechanism is a response to a concrete failure mode. DQN should therefore be understood less as a single trick and more as a compact set of stabilizing design choices.

\section{Input preprocessing: pixels, frame stacking, and action repeat}

Atari frames are visual images. A raw Atari frame has color channels and a relatively large resolution. DQN reduces the input complexity using preprocessing. A typical DQN preprocessing pipeline includes:
\begin{itemize}[leftmargin=*]
    \item converting the frame to grayscale;
    \item resizing to $84\times84$ pixels;
    \item optionally max-pooling over consecutive frames to reduce flickering;
    \item stacking the last four processed frames;
    \item repeating the selected action for several emulator frames.
\end{itemize}

Frame stacking is crucial because a single image often does not reveal velocity. In Pong or Breakout, a single frame shows the ball position, but not its direction of motion. By stacking several recent frames, the network can infer movement.

\begin{figure}[htbp]
    \centering
    \resizebox{0.92\textwidth}{!}{%
        \begin{tikzpicture}[
            frame/.style={draw, rounded corners, minimum width=1.4cm, minimum height=1.0cm, align=center, font=\small},
            box/.style={draw, rounded corners, thick, minimum width=2.7cm, minimum height=0.9cm, align=center, font=\small},
            arrow/.style={-{Latex[length=2.0mm]}, thick}
        ]
            \node[frame, draw=gray!60, fill=gray!10] (f1) at (0,0) {$x_{t-3}$};
            \node[frame, draw=gray!60, fill=gray!18] (f2) at (0.3,0.25) {$x_{t-2}$};
            \node[frame, draw=gray!60, fill=gray!26] (f3) at (0.6,0.5) {$x_{t-1}$};
            \node[frame, draw=gray!70, fill=gray!34] (f4) at (0.9,0.55) {$x_t$};

            \node[box, draw=blue!70, fill=blue!10, font=\small\bfseries] (stack) at (5.5,0.5) {stacked state\\{\normalfont\small $s_t \in \mathbb{R}^{4\times84\times84}$}};
            \node[box, draw=green!60!black, fill=green!10, font=\small\bfseries] (cnn) at (9.9,0.5) {CNN can infer\\motion};

            \draw[arrow, color=purple!70] (f4.east) -- node[above=4pt, font=\scriptsize, color=purple!70] {preprocess + stack} (stack.west);
            \draw[arrow, color=green!60!black] (stack) -- (cnn);

            \node[below=0.5cm of stack, align=center, font=\small, color=gray!70!black] {A single frame gives position. Multiple frames give approximate velocity.};
        \end{tikzpicture}%
    }
    \caption{Frame stacking gives the Q-network short-term memory. In many Atari games, one frame is not Markovian because velocity is hidden. Stacking recent frames helps the network infer motion without using a recurrent model.}
    \label{fig:ch5_frame_stack}
\end{figure}

Reward clipping was also important in Atari. Since different games have different score scales, DQN clipped positive rewards to $+1$, negative rewards to $-1$, and zero rewards to $0$ \citep{mnih2015human}. This made the learning problem more uniform across games, but it also changed the objective. A clipped reward does not distinguish between a small positive game score and a very large positive game score. This is often acceptable for learning robust behavior, but it is not the same as optimizing the original game score.

\begin{warningbox}{Reward clipping changes the objective}
Reward clipping stabilizes learning by controlling the scale of targets, but it also removes information about reward magnitude. This is acceptable in many Atari experiments, but in real applications such as healthcare, UAV control, or network optimization, clipping rewards without understanding the mission objective can hide important trade-offs.
\end{warningbox}

\section{The convolutional Q-network}

DQN uses a convolutional neural network because the input is an image-like tensor. The network maps a stack of frames to a vector of Q-values. If the environment has $|\A|$ discrete actions, the output layer has $|\A|$ units:
\begin{equation}
    f_\theta(s) = \left(Q(s,a_1;\theta), Q(s,a_2;\theta),\ldots,Q(s,a_{|\A|};\theta)\right).
\end{equation}
The greedy action is
\begin{equation}
    a^*(s)=\argmax_{a\in\A} Q(s,a;\theta).
\end{equation}

\begin{figure}[htbp]
    \centering
    \resizebox{\textwidth}{!}{%
        \begin{tikzpicture}[
            layer/.style={draw, rounded corners, thick, minimum width=2.3cm, minimum height=0.9cm, align=center},
            arrow/.style={-{Latex[length=2.5mm]}, thick},
            node distance=0.55cm
        ]
            \node[layer, fill=gray!10] (input) {Input\\$4\times84\times84$};
            \node[layer, right=of input, fill=blue!5] (conv1) {Conv 1\\filters};
            \node[layer, right=of conv1, fill=blue!8] (conv2) {Conv 2\\filters};
            \node[layer, right=of conv2, fill=blue!10] (conv3) {Conv 3\\filters};
            \node[layer, right=of conv3, fill=green!6] (fc) {Fully connected\\features};
            \node[layer, right=of fc, fill=orange!8] (out) {Output\\$|\A|$ Q-values};
            \draw[arrow] (input) -- (conv1);
            \draw[arrow] (conv1) -- (conv2);
            \draw[arrow] (conv2) -- (conv3);
            \draw[arrow] (conv3) -- (fc);
            \draw[arrow] (fc) -- (out);
            \node[below=1.0cm of conv2, align=center] {Convolutions learn visual features. The final layer predicts one Q-value per action.};
        \end{tikzpicture}%
    }
    \caption{The DQN architecture maps stacked preprocessed frames to a vector of action values. The convolutional layers learn visual features, and the final layer predicts one Q-value for each discrete action.}
    \label{fig:ch5_cnn_architecture}
\end{figure}

The important design choice is that the network outputs all action values in one forward pass. This is efficient for discrete action spaces. For continuous actions, however, maximizing $Q(s,a)$ over $a$ is no longer simple. This is one reason why later chapters introduce policy-gradient and actor-critic methods for continuous control.

\section{Experience replay: learning from remembered transitions}

In online RL, consecutive samples are strongly correlated. If an agent moves right in a game, the next observation is very similar to the previous one. Training a neural network on highly correlated samples can be inefficient and unstable. Experience replay addresses this problem by storing transitions in a memory buffer and sampling random minibatches for training \citep{lin1992self,mnih2013atari,mnih2015human}.

A replay buffer stores tuples
\begin{equation}
    (s_t,a_t,r_{t+1},s_{t+1},d_t).
\end{equation}
At each learning step, the agent samples a minibatch
\begin{equation}
    \B = \{(s_i,a_i,r_i,s'_i,d_i)\}_{i=1}^{B}
\end{equation}
from the replay buffer and computes a stochastic estimate of the DQN loss.

\begin{figure}[htbp]
    \centering
    \resizebox{0.95\textwidth}{!}{%
        \begin{tikzpicture}[
            trans/.style={draw, rounded corners, minimum width=1.25cm, minimum height=0.55cm, align=center, font=\small},
            box/.style={draw, rounded corners, thick, minimum width=3.2cm, minimum height=0.9cm, align=center, font=\small},
            arrow/.style={-{Latex[length=2.5mm]}, thick}
        ]
            \foreach \i in {0,...,7} {
                \pgfmathsetmacro{\shade}{10 + \i*5}
                \node[trans, draw=red!60, fill=red!\shade!white] (t\i) at (1.35*\i,1.5) {$\tau_{\i}$};
            }
            \node[box, draw=yellow!70!black, fill=yellow!15, font=\small\bfseries] (buffer) at (4.7,0) {Replay buffer\\{\normalfont\small many past transitions}};
            \node[box, draw=blue!70, fill=blue!10, font=\small\bfseries] (batch) at (4.7,-1.8) {Random minibatch\\{\normalfont\small less correlated updates}};

            \foreach \i in {0,...,7} {\draw[arrow, color=red!40] (t\i) -- (buffer.north);}
            \draw[arrow, color=blue!60] (buffer) -- (batch);

            \node[below=0.7cm of batch, align=center, font=\small, color=gray!70!black] {Replay improves data reuse and reduces the correlation between consecutive training samples.};
        \end{tikzpicture}%
    }
    \caption{Experience replay converts an online stream of correlated transitions into randomly sampled minibatches. This improves data reuse and makes neural-network training closer to supervised minibatch learning.}
    \label{fig:ch5_replay}
\end{figure}

Replay has several benefits:
\begin{itemize}[leftmargin=*]
    \item it reduces temporal correlation between samples;
    \item it allows each transition to be used multiple times;
    \item it smooths the training distribution over recent experience;
    \item it makes GPU minibatch training natural.
\end{itemize}

Replay also has limitations. Very old transitions may come from a policy very different from the current policy. This makes DQN off-policy. Q-learning can in principle learn off-policy, but with nonlinear function approximation this is not guaranteed to be harmless. Later methods such as prioritized replay sample more informative transitions more often \citep{schaul2016prioritized}.

\subsection{Replay and the approximate IID assumption}

In supervised learning, minibatch stochastic gradient descent is often motivated by the assumption that training examples are independent and identically distributed. Reinforcement learning violates this assumption in two ways. First, consecutive transitions are temporally correlated because they come from the same trajectory. Second, the data distribution changes because the behavior policy changes during learning.

Let the online stream of transitions be
\begin{equation}
    \tau_t=(s_t,a_t,r_{t+1},s_{t+1},d_t).
\end{equation}
Without replay, a minibatch built from adjacent samples looks like
\begin{equation}
    \{\tau_t,\tau_{t+1},\ldots,\tau_{t+B-1}\},
\end{equation}
where nearby samples are strongly correlated. With replay, DQN samples from the empirical buffer distribution
\begin{equation}
    \hat{\mu}_{\D}(\tau)=\frac{1}{|\D|}\sum_{j=1}^{|\D|}\mathbf{1}\{\tau=\tau_j\}.
\end{equation}
This makes the update closer to minibatch supervised learning, but it does not make samples truly IID. The buffer itself was generated by a sequence of changing behavior policies $\{\mu_0,\mu_1,\ldots\}$, and the environment dynamics still couple samples through trajectories. Replay is therefore best described as a \emph{decorrelation and data-reuse mechanism}, not as a perfect IID generator \citep{lin1992self,zhang2017deeper,fedus2020revisiting}.

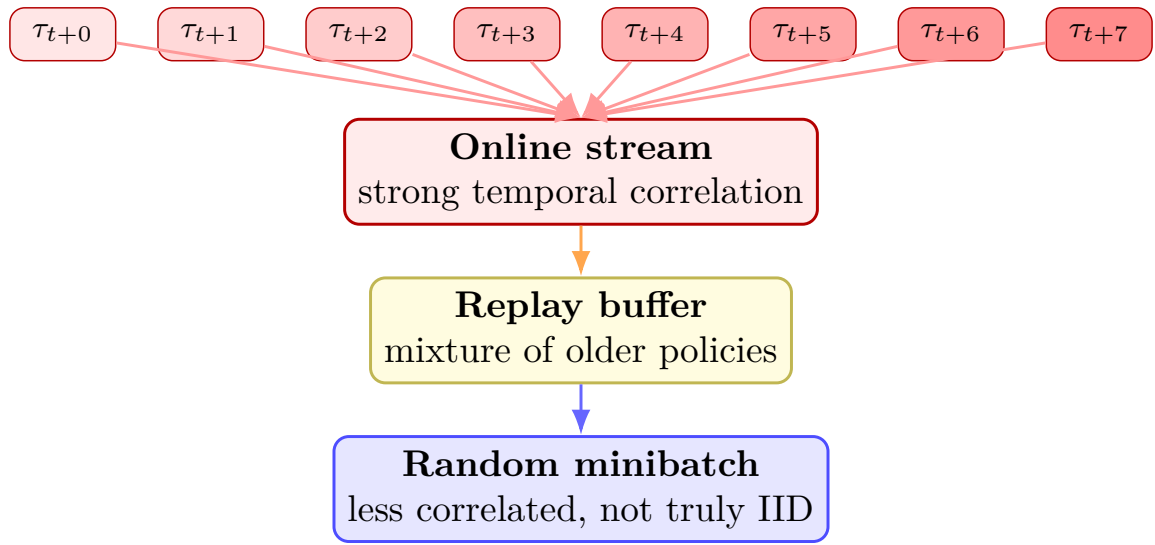
\begin{figure}[htbp]
    \centering
    \resizebox{0.95\textwidth}{!}{%
        \begin{tikzpicture}[
            trans/.style={draw, rounded corners, minimum width=1.0cm, minimum height=0.5cm, align=center, font=\scriptsize},
            box/.style={draw, rounded corners, thick, minimum width=3.3cm, minimum height=0.8cm, align=center, font=\small},
            arrow/.style={-{Latex[length=2.4mm]}, thick}
        ]
            \foreach \i in {0,...,7} {
                \pgfmathsetmacro{\shade}{10 + \i*5}
                \node[trans, draw=red!70!black, fill=red!\shade!white] (o\i) at (1.4*\i,1.8) {$\tau_{t+\i}$};
            }
            \node[box, draw=red!70!black, fill=red!8, font=\small\bfseries] (online) at (4.9,0.5) {Online stream\\{\normalfont\small strong temporal correlation}};
            \node[box, draw=yellow!70!black, fill=yellow!15, font=\small\bfseries] (buffer) at (4.9,-1.0) {Replay buffer\\{\normalfont\small mixture of older policies}};
            \node[box, draw=blue!70, fill=blue!10, font=\small\bfseries] (batch) at (4.9,-2.5) {Random minibatch\\{\normalfont\small less correlated, not truly IID}};

            \foreach \i in {0,...,7} {\draw[arrow, color=red!40] (o\i) -- (online.north);}
            \draw[arrow, color=orange!70] (online) -- (buffer);
            \draw[arrow, color=blue!60] (buffer) -- (batch);
        \end{tikzpicture}%
    }
    \caption{Experience replay reduces local temporal correlation by sampling from a buffer of past transitions. However, the buffer is still produced by changing policies and environment dynamics. Replay is therefore an approximation to IID minibatch training, not a guarantee of IID data.}
    \label{fig:ch5_replay_iid}
\end{figure}

This distinction matters in real systems. In a non-stationary network, transitions from yesterday's traffic pattern may not represent today's traffic pattern. In a multi-UAV system, replay data may become stale when other UAVs change their policies. Replay-buffer size, sampling strategy, and data freshness are therefore algorithmic choices, not only memory-management details.

\section{Target networks: slowing down the moving target}

The target network is a delayed copy of the online Q-network. The online network has parameters $\theta$. The target network has parameters $\theta^-$. The target is computed using $\theta^-$:
\begin{equation}
    y = r + \gamma(1-d)\max_{a'}Q(s',a';\theta^-).
\end{equation}
Every fixed number of environment steps or gradient steps, the target parameters are updated:
\begin{equation}
    \theta^- \leftarrow \theta.
\end{equation}
This makes the target piecewise stationary: it still changes, but less frequently.

\begin{figure}[htbp]
    \centering
    \resizebox{0.95\textwidth}{!}{%
        \begin{tikzpicture}[
            box/.style={draw, rounded corners, thick, minimum width=3.1cm, minimum height=0.9cm, align=center},
            arrow/.style={-{Latex[length=2.5mm]}, thick},
            node distance=1.1cm
        ]
            \node[box, fill=green!5] (online) {Online network\\$Q(s,a;\theta)$};
            \node[box, right=of online, fill=orange!8] (pred) {Prediction\\$Q(s_t,a_t;\theta)$};
            \node[box, below=of online, fill=blue!5] (targetnet) {Target network\\$Q(s,a;\theta^-)$};
            \node[box, right=of targetnet, fill=red!5] (target) {Target\\$r+\gamma\max_{a'}Q(s',a';\theta^-)$};
            \node[box, right=1.2cm of pred, fill=purple!6] (loss) {Loss\\prediction vs target};
            \draw[arrow] (online) -- (pred);
            \draw[arrow] (targetnet) -- (target);
            \draw[arrow] (pred) -- (loss);
            \draw[arrow] (target) -- (loss);
            \draw[arrow, dashed] (online.south) -- node[right, font=\scriptsize, align=center] {copy every\\$C$ steps} (targetnet.north);
        \end{tikzpicture}%
    }
    \caption{The target network computes the bootstrap target using delayed parameters $\theta^-$. This reduces the moving-target instability caused by using the same network to compute both predictions and targets.}
    \label{fig:ch5_target_network}
\end{figure}
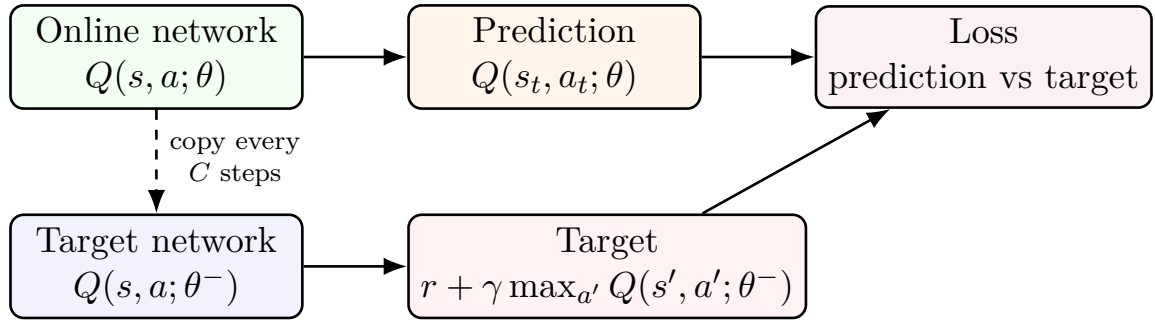

The target network can be understood as a practical compromise. If the target changes every step, learning may be unstable. If the target never changes, learning cannot progress. Periodic copying gives the optimizer a temporarily stable objective.

\section{The DQN loss and semi-gradient update}

For a minibatch $\B$, DQN computes
\begin{equation}
    y_i = r_i + \gamma(1-d_i)\max_{a'}Q(s'_i,a';\theta^-),
\end{equation}
then minimizes
\begin{equation}
    L(\theta) = \frac{1}{|\B|}\sum_{i\in \B}\ell\left(y_i - Q(s_i,a_i;\theta)\right),
\end{equation}
where $\ell$ is often squared error or Huber loss.

This is a semi-gradient method. During the online network update, the target $y_i$ is treated as constant with respect to $\theta$. In code, this means the target is computed under a no-gradient context:
\begin{equation}
    \nabla_\theta L(\theta) \approx -\frac{1}{|\B|}\sum_{i\in\B} \ell'(\delta_i)\nabla_\theta Q(s_i,a_i;\theta),
\end{equation}
where
\begin{equation}
    \delta_i = y_i - Q(s_i,a_i;\theta).
\end{equation}

\begin{warningbox}{Do not backpropagate through the target}
In DQN, the target is treated as a fixed label during the gradient step. In PyTorch this usually means computing the target inside \texttt{torch.no\_grad()}. Accidentally allowing gradients through the target network changes the algorithm and can produce unstable or incorrect updates.
\end{warningbox}

\section{Epsilon-greedy exploration and evaluation}

DQN uses epsilon-greedy exploration. At training time,
\begin{equation}
    a_t =
    \begin{cases}
        \text{random action}, & \text{with probability } \epsilon,\\
        \argmax_a Q(s_t,a;\theta), & \text{with probability } 1-\epsilon.
    \end{cases}
\end{equation}
The exploration rate $\epsilon$ is often annealed from a high value to a lower value during training.

Exploration during training and evaluation should be distinguished. A training policy may use significant random exploration. An evaluation policy usually uses a small epsilon or a greedy policy, depending on the benchmark protocol. In Atari, evaluation details matter: no-op starts, human starts, sticky actions, episode length, and frame skip can change reported scores \citep{machado2018revisiting}.

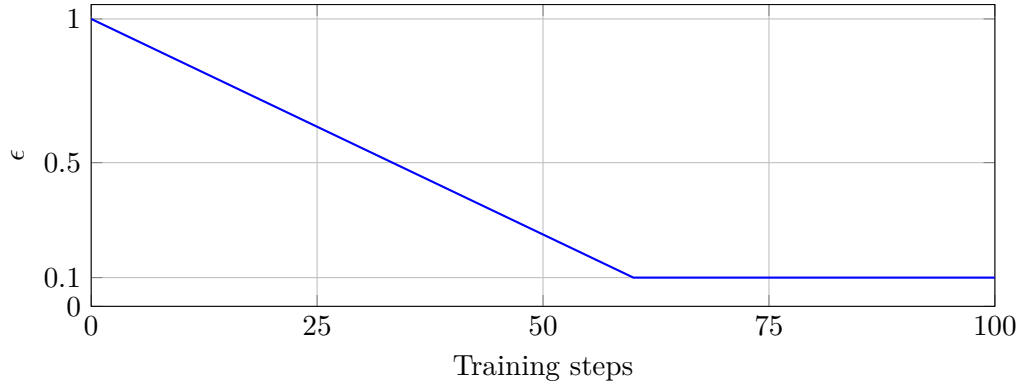
\begin{figure}[htbp]
    \centering
    \begin{tikzpicture}
        \begin{axis}[
            width=0.85\textwidth,
            height=0.35\textwidth,
            xlabel={Training steps},
            ylabel={$\epsilon$},
            xmin=0, xmax=100,
            ymin=0, ymax=1.05,
            grid=both,
            xtick={0,25,50,75,100},
            ytick={0,0.1,0.5,1.0}
        ]
            \addplot[thick, blue] coordinates {(0,1.0) (60,0.1) (100,0.1)};
        \end{axis}
    \end{tikzpicture}
    \caption{A typical epsilon schedule decreases exploration over training. Early exploration helps discover useful behavior; later exploitation allows the agent to refine value estimates around better policies. The exact schedule is an important hyperparameter.}
    \label{fig:ch5_epsilon_schedule}
\end{figure}

\section{Algorithm box: Deep Q-learning with experience replay}

\begin{tcolorbox}[
    title={Algorithm 5.1: Deep Q-learning with Experience Replay},
    colback=gray!3,
    colframe=black!70,
    fonttitle=\bfseries,
    arc=2mm,
    boxrule=0.8pt
]
    \begin{enumerate}[leftmargin=*]
        \item Initialize replay memory $\D$ with capacity $N$.
        \item Initialize online Q-network $Q(s,a;\theta)$ randomly.
        \item Initialize target Q-network $Q(s,a;\theta^-)$ with $\theta^- \leftarrow \theta$.
        \item For each episode:
        \begin{enumerate}
            \item Initialize the environment and construct the initial state $s_0$.
            \item For each time step $t$:
            \begin{enumerate}
                \item Select action $a_t$ using epsilon-greedy exploration from $Q(s_t,\cdot;\theta)$.
                \item Execute $a_t$, observe reward $r_{t+1}$, next observation $s_{t+1}$, and terminal flag $d_t$.
                \item Store $(s_t,a_t,r_{t+1},s_{t+1},d_t)$ in $\D$.
                \item Sample a random minibatch from $\D$.
                \item Compute targets
                \[
                y_i = r_i + \gamma(1-d_i)\max_{a'}Q(s'_i,a';\theta^-).
                \]
                \item Update $\theta$ by minimizing
                \[
                \frac{1}{B}\sum_i \ell\left(y_i-Q(s_i,a_i;\theta)\right).
                \]
                \item Every $C$ steps, update the target network: $\theta^- \leftarrow \theta$.
            \end{enumerate}
        \end{enumerate}
    \end{enumerate}
\end{tcolorbox}

This algorithm looks like supervised learning, but the labels are generated by the agent itself through the Bellman target. This is why DQN is sometimes described as supervised regression on temporal-difference targets, but that description must be used carefully. Unlike ordinary supervised learning, the dataset is generated by the current behavior policy, and the targets depend on the learned value function.

\section{Python implementation}

This section gives a compact but realistic PyTorch implementation of the essential DQN components. It is not intended to be a fully optimized Atari implementation. It is written to expose the logic clearly.

\subsection{Replay buffer}

\Needspace{18\baselineskip}
\begin{lstlisting}[style=pythonstyle,caption={A replay buffer for DQN transitions.},label={lst:ch5_replay_buffer}]
from dataclasses import dataclass
import random
import numpy as np
import torch

@dataclass
class Transition:
    state: np.ndarray
    action: int
    reward: float
    next_state: np.ndarray
    done: bool

class ReplayBuffer:
    def __init__(self, capacity: int):
        self.capacity = int(capacity)
        self.storage = []
        self.position = 0

    def __len__(self):
        return len(self.storage)

    def push(self, state, action, reward, next_state, done):
        item = Transition(state, int(action), float(reward), next_state, bool(done))
        if len(self.storage) < self.capacity:
            self.storage.append(item)
        else:
            self.storage[self.position] = item
        self.position = (self.position + 1) % self.capacity

    def sample(self, batch_size: int, device: torch.device):
        batch = random.sample(self.storage, batch_size)

        states = torch.as_tensor(
            np.stack([b.state for b in batch]),
            dtype=torch.float32,
            device=device,
        )
        actions = torch.as_tensor(
            [b.action for b in batch],
            dtype=torch.long,
            device=device,
        ).unsqueeze(1)
        rewards = torch.as_tensor(
            [b.reward for b in batch],
            dtype=torch.float32,
            device=device,
        ).unsqueeze(1)
        next_states = torch.as_tensor(
            np.stack([b.next_state for b in batch]),
            dtype=torch.float32,
            device=device,
        )
        dones = torch.as_tensor(
            [b.done for b in batch],
            dtype=torch.float32,
            device=device,
        ).unsqueeze(1)
        return states, actions, rewards, next_states, dones
\end{lstlisting}

For Atari-like inputs, states should typically be normalized to $[0,1]$ before entering the network. A common choice is to store frames as \texttt{uint8} to save memory and convert to \texttt{float32} only when sampling. Listing~\ref{lst:ch5_replay_buffer} uses float conversion directly for simplicity.

\subsection{Atari-style Q-network}

\Needspace{18\baselineskip}
\begin{lstlisting}[style=pythonstyle,caption={A PyTorch convolutional Q-network for stacked 84-by-84 frames.},label={lst:ch5_q_network}]
import torch.nn as nn
import torch.nn.functional as F

class AtariDQN(nn.Module):
    """Convolutional Q-network for input shape (B, 4, 84, 84)."""

    def __init__(self, num_actions: int):
        super().__init__()
        self.conv1 = nn.Conv2d(4, 32, kernel_size=8, stride=4)
        self.conv2 = nn.Conv2d(32, 64, kernel_size=4, stride=2)
        self.conv3 = nn.Conv2d(64, 64, kernel_size=3, stride=1)
        self.fc1 = nn.Linear(64 * 7 * 7, 512)
        self.out = nn.Linear(512, num_actions)

    def forward(self, x):
        # x is expected in [0, 1], shape (batch, 4, 84, 84)
        x = F.relu(self.conv1(x))
        x = F.relu(self.conv2(x))
        x = F.relu(self.conv3(x))
        x = torch.flatten(x, start_dim=1)
        x = F.relu(self.fc1(x))
        return self.out(x)
\end{lstlisting}

The output dimension equals the number of discrete actions. If the environment has six actions, the network returns six Q-values. The selected action is the index of the largest Q-value, unless epsilon-greedy exploration chooses a random action.

\subsection{Epsilon-greedy action selection}

\Needspace{14\baselineskip}
\begin{lstlisting}[style=pythonstyle,caption={Epsilon-greedy action selection.},label={lst:ch5_epsilon_greedy}]
def select_action(q_net, state, epsilon: float, num_actions: int, device):
    if random.random() < epsilon:
        return random.randrange(num_actions)

    state_t = torch.as_tensor(state, dtype=torch.float32, device=device).unsqueeze(0)
    with torch.no_grad():
        q_values = q_net(state_t)
    action = int(torch.argmax(q_values, dim=1).item())
    return action

class LinearSchedule:
    def __init__(self, start: float, end: float, duration: int):
        self.start = float(start)
        self.end = float(end)
        self.duration = int(duration)

    def value(self, step: int) -> float:
        frac = min(step / self.duration, 1.0)
        return self.start + frac * (self.end - self.start)
\end{lstlisting}

\subsection{One DQN gradient update}

\Needspace{22\baselineskip}
\begin{lstlisting}[style=pythonstyle,caption={A single DQN minibatch update using a target network.},label={lst:ch5_dqn_update}]
def dqn_update(q_net, target_net, optimizer, replay, batch_size,
               gamma: float, device, grad_clip: float = 10.0):
    if len(replay) < batch_size:
        return None

    states, actions, rewards, next_states, dones = replay.sample(batch_size, device)

    # Q(s, a; theta) for the actions actually taken.
    q_values = q_net(states).gather(1, actions)

    # Bellman targets. Do not backpropagate through the target network.
    with torch.no_grad():
        next_q_values = target_net(next_states).max(dim=1, keepdim=True).values
        targets = rewards + gamma * (1.0 - dones) * next_q_values

    loss = F.smooth_l1_loss(q_values, targets)

    optimizer.zero_grad(set_to_none=True)
    loss.backward()
    nn.utils.clip_grad_norm_(q_net.parameters(), grad_clip)
    optimizer.step()

    with torch.no_grad():
        td_error = (targets - q_values).abs().mean().item()
    return {"loss": float(loss.item()), "td_error": td_error}
\end{lstlisting}

Several details are important here:
\begin{itemize}[leftmargin=*]
    \item \texttt{gather} selects the Q-value corresponding to the action taken in each transition.
    \item \texttt{torch.no\_grad()} prevents gradients from flowing through the target.
    \item the target network is separate from the online network.
    \item Huber loss is used through \texttt{smooth\_l1\_loss}.
    \item gradient clipping can prevent rare unstable updates.
\end{itemize}

\begin{warningbox}{Three common implementation bugs}
\begin{enumerate}[leftmargin=*]
    \item \textbf{Wrong terminal handling.} If a transition is terminal, the future value term must be removed: $y=r$, not $r+\gamma\max_{a'}Q(s',a')$.
    \item \textbf{Gradient through the target.} The bootstrap target should be computed under \texttt{torch.no\_grad()}. Otherwise the optimizer may alter both sides of the regression target.
    \item \textbf{Evaluation with training epsilon.} During evaluation, use a fixed evaluation protocol. Accidentally evaluating with a high training epsilon makes a good policy look bad; evaluating with a different environment wrapper can make results incomparable.
\end{enumerate}
\end{warningbox}

\subsection{Training loop skeleton}

\Needspace{24\baselineskip}
\begin{lstlisting}[style=pythonstyle,caption={A high-level DQN training loop skeleton. Environment preprocessing is assumed to produce states of shape (4,84,84).},label={lst:ch5_training_loop}]
def train_dqn(env, num_actions: int, total_steps: int = 1_000_000):
    device = torch.device("cuda" if torch.cuda.is_available() else "cpu")

    q_net = AtariDQN(num_actions).to(device)
    target_net = AtariDQN(num_actions).to(device)
    target_net.load_state_dict(q_net.state_dict())
    target_net.eval()

    optimizer = torch.optim.Adam(q_net.parameters(), lr=1e-4)
    replay = ReplayBuffer(capacity=100_000)
    eps_schedule = LinearSchedule(start=1.0, end=0.1, duration=100_000)

    gamma = 0.99
    batch_size = 32
    learning_starts = 10_000
    train_freq = 4
    target_update_freq = 10_000

    state, info = env.reset()
    episode_return = 0.0

    for step in range(1, total_steps + 1):
        epsilon = eps_schedule.value(step)
        action = select_action(q_net, state, epsilon, num_actions, device)

        next_state, reward, terminated, truncated, info = env.step(action)
        done = terminated or truncated

        # Atari DQN often clips rewards to {-1, 0, +1}.
        clipped_reward = float(np.sign(reward))

        replay.push(state, action, clipped_reward, next_state, done)
        state = next_state
        episode_return += reward

        if done:
            state, info = env.reset()
            episode_return = 0.0

        if step > learning_starts and step % train_freq == 0:
            metrics = dqn_update(
                q_net, target_net, optimizer, replay,
                batch_size, gamma, device
            )

        if step % target_update_freq == 0:
            target_net.load_state_dict(q_net.state_dict())

    return q_net
\end{lstlisting}

In a production implementation, the environment wrapper must handle Atari-specific preprocessing carefully: action repeat, max-pooling over frames, life-loss handling if desired, frame stacking, normalization, and evaluation protocol. Libraries such as ALE, Gymnasium, and mature RL codebases implement these details, but the conceptual logic remains the same \citep{bellemare2013ale,machado2018revisiting,brockman2016openai}.

\subsection{A minimal vector-state DQN for debugging}

Before training on Atari, it is often better to debug DQN on a small vector-observation environment. This does not test visual representation learning, but it tests replay, target networks, epsilon-greedy exploration, and TD loss.

\Needspace{16\baselineskip}
\begin{lstlisting}[style=pythonstyle,caption={A small MLP Q-network useful for debugging DQN logic before using pixels.},label={lst:ch5_mlp_dqn}]
class MLPDQN(nn.Module):
    def __init__(self, obs_dim: int, num_actions: int):
        super().__init__()
        self.net = nn.Sequential(
            nn.Linear(obs_dim, 128), nn.ReLU(),
            nn.Linear(128, 128), nn.ReLU(),
            nn.Linear(128, num_actions),
        )

    def forward(self, x):
        return self.net(x)
\end{lstlisting}

\begin{practicebox}{Practical advice}
Debug your DQN implementation in stages. First test the replay buffer. Then test action selection. Then test whether the loss decreases on a fixed batch. Then train on a simple environment. Only after that move to Atari-style image input. Many failed DQN experiments come from implementation mistakes, not from the algorithm itself.
\end{practicebox}

\section{Practical engineering details and debugging checklist}

DQN is sensitive to implementation choices. The following checklist is useful when debugging:
\begin{enumerate}[leftmargin=*]
    \item \textbf{Observation shape}: verify that states have shape $(4,84,84)$ or the expected format.
    \item \textbf{Normalization}: make sure pixel values are scaled consistently, often to $[0,1]$.
    \item \textbf{Action indexing}: confirm that the network output dimension matches the action space.
    \item \textbf{Replay warm-up}: do not train before the replay buffer contains enough diverse transitions.
    \item \textbf{Target network}: make sure the target network is periodically updated and not optimized directly.
    \item \textbf{Terminal handling}: multiply future value by $(1-d)$ for terminal transitions.
    \item \textbf{No gradient through target}: compute targets inside \texttt{torch.no\_grad()}.
    \item \textbf{Reward scale}: understand whether reward clipping is appropriate.
    \item \textbf{Exploration schedule}: ensure epsilon does not decay too quickly.
    \item \textbf{Evaluation protocol}: evaluate with a consistent policy and environment setting.
\end{enumerate}

\begin{figure}[htbp]
    \centering
    \resizebox{0.95\textwidth}{!}{%
        \begin{tikzpicture}[
            box/.style={draw, rounded corners, thick, minimum width=3.0cm, minimum height=0.85cm, align=center},
            arrow/.style={-{Latex[length=2.3mm]}, thick},
            node distance=0.7cm
        ]
            \node[box, fill=red!5] (bug) {Common failure\\unstable or flat reward};
            \node[box, below left=of bug, fill=gray!8] (obs) {check observations};
            \node[box, below=of bug, fill=gray!8] (target) {check target network};
            \node[box, below right=of bug, fill=gray!8] (done) {check terminal flags};
            \node[box, below=of target, fill=gray!8] (loss) {check loss and gradients};
            \node[box, below left=of loss, fill=gray!8] (eps) {check exploration};
            \node[box, below right=of loss, fill=gray!8] (eval) {check evaluation};
            \draw[arrow] (bug) -- (obs);
            \draw[arrow] (bug) -- (target);
            \draw[arrow] (bug) -- (done);
            \draw[arrow] (target) -- (loss);
            \draw[arrow] (loss) -- (eps);
            \draw[arrow] (loss) -- (eval);
        \end{tikzpicture}%
    }
    \caption{A debugging map for DQN. Training failure can come from many sources: incorrect preprocessing, missing terminal handling, gradients through the target, insufficient exploration, or inconsistent evaluation.}
    \label{fig:ch5_debugging_map}
\end{figure}
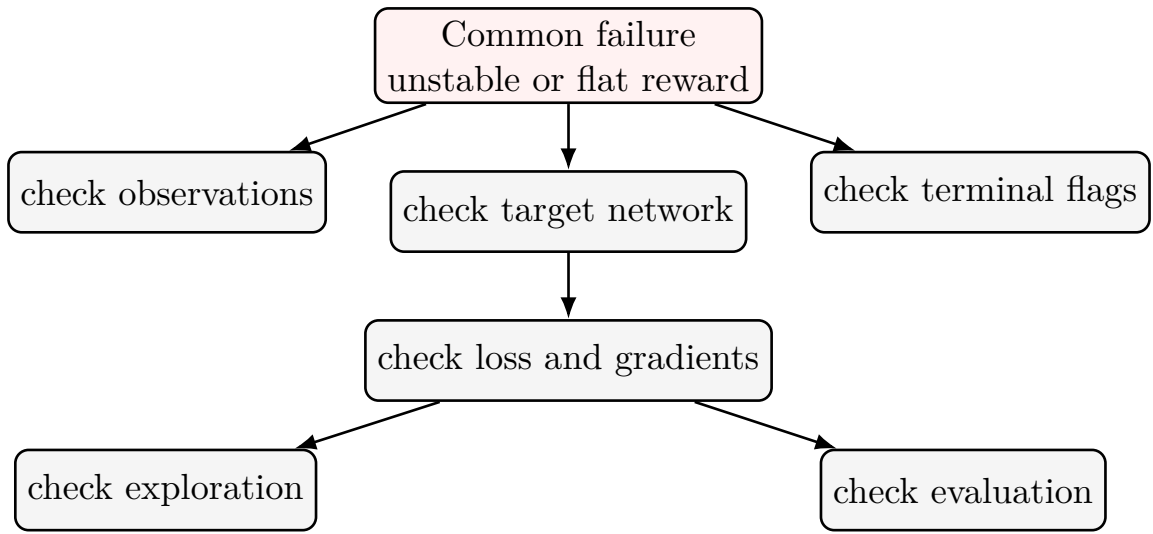

\section{What DQN achieved and what it did not solve}

DQN achieved several things:
\begin{itemize}[leftmargin=*]
    \item it learned from high-dimensional pixels;
    \item it used the same general architecture across many games;
    \item it combined deep learning with temporal-difference control;
    \item it made experience replay and target networks standard tools;
    \item it launched the modern wave of deep reinforcement learning.
\end{itemize}

But DQN also had important limitations:
\begin{itemize}[leftmargin=*]
    \item it was designed for discrete action spaces;
    \item it could be sample-inefficient;
    \item it could overestimate action values because of the max operator;
    \item it used relatively simple epsilon-greedy exploration;
    \item it struggled with sparse rewards and long-horizon exploration;
    \item it did not solve transfer across tasks;
    \item it did not provide safety guarantees;
    \item it was sensitive to implementation and evaluation protocols.
\end{itemize}

\subsection{Explicit limitations: what DQN cannot do well}

The discrete-action limitation is especially important. DQN works naturally when the network can output one Q-value per action. Atari has a small discrete action set. A UAV, however, may need continuous velocity commands, continuous transmit power, continuous beam direction, or continuous bandwidth allocation. In such settings, computing
\begin{equation}
    \argmax_{a\in\A} Q(s,a;\theta)
\end{equation}
may require solving a continuous optimization problem at every decision step. This is one reason why deterministic policy-gradient, actor-critic, and maximum-entropy methods became important for continuous control.

DQN also struggles when the discrete action space is extremely large. For example, if an SDN controller must jointly choose paths for hundreds of flows, the number of discrete joint actions may be combinatorial. A network that outputs one Q-value per joint action becomes impractical. This motivates action factorization, hierarchical control, policy-gradient methods, or multi-agent decomposition.

Finally, vanilla DQN is not a safety algorithm. It may learn unsafe actions during exploration, and its Q-values do not provide formal constraint satisfaction. For safety-critical domains such as UAV control, SD-WAN traffic engineering, autonomous driving, or healthcare, DQN must be combined with constraints, shields, control barrier functions, offline validation, or conservative deployment methods.

These limitations directly motivated later algorithms. Double DQN addressed overestimation bias \citep{vanhasselt2010double,vanhasselt2016deep}. Dueling networks separated state-value and advantage estimation \citep{wang2016dueling}. Prioritized replay changed how transitions are sampled \citep{schaul2016prioritized}. Distributional RL learned a distribution over returns rather than only the mean \citep{bellemare2017distributional}. Rainbow combined several improvements into one strong agent \citep{hessel2018rainbow}.

\begin{figure}[htbp]
    \centering
    \resizebox{\textwidth}{!}{%
        \begin{tikzpicture}[
            event/.style={draw, rounded corners, thick, minimum width=2.6cm, minimum height=0.85cm, align=center},
            arrow/.style={-{Latex[length=2.5mm]}, thick},
            node distance=0.55cm
        ]
            \node[event, fill=gray!8] (dqn13) {2013\\Atari DQN};
            \node[event, right=of dqn13, fill=blue!5] (nature15) {2015\\Nature DQN};
            \node[event, right=of nature15, fill=green!5] (double) {2016\\Double DQN};
            \node[event, right=of double, fill=orange!8] (duel) {2016\\Dueling + PER};
            \node[event, right=of duel, fill=purple!6] (dist) {2017\\Distributional RL};
            \node[event, right=of dist, fill=red!5] (rainbow) {2018\\Rainbow};
            \draw[arrow] (dqn13) -- (nature15);
            \draw[arrow] (nature15) -- (double);
            \draw[arrow] (double) -- (duel);
            \draw[arrow] (duel) -- (dist);
            \draw[arrow] (dist) -- (rainbow);
        \end{tikzpicture}%
    }
    \caption{DQN opened a family of value-based deep RL algorithms. Many later methods can be understood as addressing specific limitations of the original DQN system. Chapter~6 studies these improvements in detail.}
    \label{fig:ch5_dqn_family_timeline}
\end{figure}
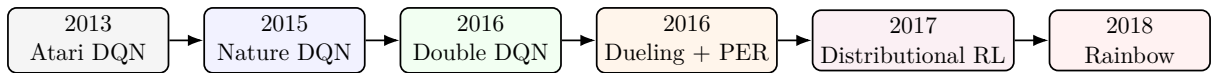

\section{Running example: DQN ideas for UAV and network control}

The Atari setting may seem far from UAV or communication-network control, but the core DQN ideas transfer conceptually.

Suppose a UAV-assisted network has a discrete action space:
\begin{equation}
    \A = \{\text{move north},\text{move south},\text{move east},\text{move west},\text{hover},\text{go recharge}\}.
\end{equation}
The observation might be a vector of battery level, user density, QoS metrics, and channel conditions. Or it might be a spatial heatmap showing user demand and signal quality. A DQN-like agent could learn
\begin{equation}
    Q(s,a;\theta)
\end{equation}
for each discrete action.

The analogy is shown in Table~\ref{tab:ch5_atari_uav}.

\begin{table}[htbp]
    \centering
    \caption{Mapping DQN concepts from Atari to UAV/network control.}
    \label{tab:ch5_atari_uav}
    \begin{tabular}{p{0.28\textwidth}p{0.30\textwidth}p{0.32\textwidth}}
        \toprule
        \textbf{DQN component} & \textbf{Atari example} & \textbf{UAV/network example} \\
        \midrule
        State/observation & stacked game frames & QoS vector, user heatmap, channel map \\
        Action & joystick/fire action & movement, serving mode, routing choice \\
        Reward & clipped game score & QoS gain minus energy and violations \\
        Replay buffer & past game transitions & past network states and decisions \\
        Target network & stable Bellman target & stable target for delayed QoS effects \\
        CNN representation & visual game features & spatial demand/interference features \\
        Discrete Q-values & one value per game action & one value per control action \\
        \bottomrule
    \end{tabular}
\end{table}

\subsection{Worked mini-example: DQN for UAV service-control}

Consider a simplified single-UAV system. The UAV flies at a fixed altitude and must choose one of six discrete actions every second:
\begin{equation}
    \A=\{\text{north},\text{south},\text{east},\text{west},\text{hover},\text{recharge}\}.
\end{equation}
The observation is a vector
\begin{equation}
    s_t = [x_t,y_t,b_t,u_t^{A},u_t^{B},u_t^{C},\overline{\mathrm{SINR}}_t,\overline{L}_t,\rho_t],
\end{equation}
where $(x_t,y_t)$ is UAV position, $b_t$ is battery level, $u_t^A,u_t^B,u_t^C$ are nearby user counts by priority class, $\overline{\mathrm{SINR}}_t$ is average SINR, $\overline{L}_t$ is average latency, and $\rho_t$ is traffic load. A simple discrete-action DQN can output six values:
\begin{equation}
    Q(s_t,\cdot;\theta) = [Q_N,Q_S,Q_E,Q_W,Q_H,Q_R].
\end{equation}
The selected action is the largest Q-value except during exploration.

A possible reward is
\begin{equation}
    r_t = 2.0\,\mathrm{QoS}_A(t)+1.0\,\mathrm{QoS}_B(t)+0.5\,\mathrm{QoS}_C(t)
    -0.05\,\Delta E_t - 5.0\,\mathbf{1}\{b_t < b_{\min}\},
\end{equation}
where high-priority users receive stronger weight, energy usage is penalized, and critically low battery receives a large penalty. A replay transition has the form
\begin{equation}
    (s_t, a_t, r_t, s_{t+1}, d_t).
\end{equation}
The DQN target becomes
\begin{equation}
    y_t = r_t + \gamma(1-d_t)\max_{a'}Q(s_{t+1},a';\theta^-).
\end{equation}

This toy formulation is useful because it shows where DQN fits and where it breaks. It fits when movement decisions are discretized and the system can tolerate exploration in simulation. It breaks when control must be continuous, when hard safety constraints must never be violated, when many UAVs act simultaneously, or when resource allocation creates a combinatorial action space. In those cases, the DQN view is still conceptually useful, but the algorithm must be extended or replaced.

\begin{table}[htbp]
    \centering
    \caption{A small UAV service-control DQN example.}
    \label{tab:ch5_uav_worked_example}
    \small
    \begin{tabularx}{\textwidth}{p{0.22\textwidth}X X}
        \toprule
        \textbf{Element} & \textbf{Concrete design} & \textbf{Why it matters} \\
        \midrule
        State & Position, battery, user counts, SINR, latency, traffic load & Gives the Q-network the information needed to compare movement and recharge decisions. \\
        Action & North, south, east, west, hover, recharge & Keeps the action space small enough for DQN output heads. \\
        Reward & Weighted QoS minus energy and battery-risk penalty & Encourages service quality while discouraging energy-unsafe behavior. \\
        Replay & Store recent traffic and mobility transitions & Reuses simulation data, but must be refreshed if traffic patterns change. \\
        Target network & Delayed Q-network for next-state values & Stabilizes delayed QoS and battery consequences. \\
        Failure mode & Continuous trajectory and multi-UAV coordination & Indicates when actor-critic, MARL, or safe RL is needed. \\
        \bottomrule
    \end{tabularx}
\end{table}

However, the limitations also transfer. If UAV control requires continuous movement, DQN is not enough. If safety constraints must never be violated, vanilla DQN is not enough. If multiple UAVs coordinate, independent DQN may be unstable. If the environment changes due to other learning agents or SDN control, the replay data may become stale. These problems motivate later chapters on actor-critic methods, multi-agent RL, safe RL, and model-based methods.

\section{Key takeaways}

\begin{itemize}[leftmargin=*]
    \item DQN was the first widely recognized deep RL breakthrough because it learned control policies directly from high-dimensional visual input.
    \item DQN is a deep version of Q-learning, but it is not just Q-learning with a neural network. It is a complete training system with replay memory, target networks, preprocessing, exploration, and careful engineering.
    \item Experience replay reduces sample correlation and improves data reuse.
    \item Target networks reduce moving-target instability.
    \item Frame stacking gives the network short-term motion information.
    \item Reward clipping stabilizes Atari training but changes the optimization objective.
    \item DQN works naturally for discrete action spaces because the network outputs one Q-value per action.
    \item DQN remains vulnerable to overestimation, sample inefficiency, sparse rewards, weak exploration, and lack of safety guarantees.
    \item Many later value-based DRL algorithms can be understood as improvements to DQN.
\end{itemize}

\section{Exercises}

\subsection*{Conceptual exercises}

\begin{enumerate}[leftmargin=*]
    \item Explain why a single Atari frame may not be Markovian. Why does stacking four frames help?
    \item Why is DQN not merely a neural network version of Q-learning?
    \item What problem does experience replay solve? What problem can replay introduce?
    \item What problem does the target network solve?
    \item Why does reward clipping make Atari training easier? What information does it remove?
    \item Why is DQN naturally suited to discrete action spaces but not continuous action spaces?
    \item Explain the deadly triad in the context of DQN.
    \item Explain why the target network reduces but does not eliminate the moving-target problem.
\end{enumerate}

\subsection*{Mathematical exercises}

\begin{enumerate}[leftmargin=*]
    \item Given $r=1$, $\gamma=0.99$, $d=0$, and $\max_{a'}Q(s',a';\theta^-)=5$, compute the DQN target.
    \item Given $Q(s,a;\theta)=4.2$ and the target from the previous question, compute the temporal-difference error.
    \item Write the DQN loss for a minibatch of three transitions.
    \item Show why the target becomes $y=r$ when $d=1$.
    \item Compare the DQN target with the tabular Q-learning target. What changes and what stays the same?
\end{enumerate}

\subsection*{Implementation exercises}

\begin{enumerate}[leftmargin=*]
    \item Modify Listing~\ref{lst:ch5_dqn_update} to implement Double DQN. Use the online network to select the next action and the target network to evaluate it.
    \item Add soft target updates of the form $\theta^- \leftarrow \tau\theta + (1-\tau)\theta^-$.
    \item Modify the replay buffer to store observations as \texttt{uint8} and convert to \texttt{float32} only when sampling.
    \item Add logging for mean Q-value, mean target value, TD error, loss, epsilon, and episode return.
    \item Test the MLP DQN on a simple vector-observation environment before using Atari frames.
\end{enumerate}

\subsection*{Research thinking exercises}

\begin{enumerate}[leftmargin=*]
    \item In a UAV network, when would a DQN-style discrete action formulation be reasonable?
    \item When would DQN be the wrong choice for UAV control?
    \item How would you design a replay buffer for a non-stationary network environment?
    \item Should rewards be clipped in a QoS control problem? Why or why not?
    \item How could target networks be useful beyond DQN, for example in actor-critic algorithms?
\end{enumerate}

\section*{Looking Ahead to Chapter 6: Food for Thought}
\addcontentsline{toc}{section}{Looking Ahead to Chapter 6: Food for Thought}

DQN was a breakthrough, but it was not the final form of value-based deep reinforcement learning. Once researchers understood that deep Q-learning could work, the next question became: how can it work better?

Before moving to Chapter~6, consider the following questions.

\subsection*{Questions to Ponder}

\begin{enumerate}[leftmargin=*]
    \item The DQN target uses $\max_{a'}Q(s',a';\theta^-)$. What happens if the Q-values are noisy?
    \item Can the max operator systematically overestimate action values?
    \item Does every transition in replay memory deserve the same probability of being sampled?
    \item Should the network estimate only the expected return, or the full distribution of possible returns?
    \item Is it useful to separate the value of being in a state from the advantage of taking a specific action?
    \item Can several improvements be combined into one stronger agent?
    \item What should a better DQN preserve from the original algorithm, and what should it change?
\end{enumerate}

\subsection*{Main idea for the next chapter}

Chapter~6 studies the DQN family. The central idea is that the original DQN revealed both a solution and a list of weaknesses. Double DQN, dueling networks, prioritized replay, noisy networks, distributional RL, and Rainbow each address a different weakness. Understanding these improvements is not only useful for Atari; it teaches a general research pattern in DRL:

\begin{quote}
A breakthrough algorithm is rarely the end of a field. It is often the beginning of a family of methods that identify, isolate, and repair its limitations.
\end{quote}
	\chapter{From DQN to Better Value-Based DRL}
\label{ch:dqn_family}
\chaptermark{Better Value-Based DRL}

\begin{keybox}{Chapter goal}
	Chapter 5 explained why Deep Q-Networks (DQN) were historically important: DQN made Q-learning work with high-dimensional visual input by combining a convolutional network, experience replay, and a target network. Chapter 6 studies what happened next. DQN was powerful, but it was not the end of value-based deep reinforcement learning. It overestimated action values, explored crudely, learned inefficiently from replay, ignored uncertainty in returns, and treated many actions as unrelated even when their values were similar. The DQN family can be understood as a sequence of targeted repairs to these weaknesses.
\end{keybox}

\section*{Chapter Overview}
\addcontentsline{toc}{section}{Chapter Overview}

\begin{enumerate}[leftmargin=*]
	\item Why DQN needed successors
	\item The diagnostic view: symptoms, mechanisms, and costs
	\item Double DQN: reducing overestimation bias
	\item Dueling networks: separating state value from action advantage
	\item Prioritized experience replay: learning more from surprising transitions
	\item Multi-step targets: bringing reward information backward faster
	\item Noisy networks: replacing blind epsilon exploration
	\item Distributional RL: learning the return distribution, not only the mean
	\item C51: categorical distributional DQN
	\item Quantile methods: QR-DQN and IQN
	\item Rainbow DQN: combining complementary improvements
	\item Beyond Rainbow: distributed, recurrent, Munchausen, and fully parameterized variants
	\item Engineering value-based DRL: what usually breaks
	\item Research example: UAV/SDN control with value-based DRL
	\item Python implementation blocks
	\item Limitations of improved value-based methods
	\item Key takeaways
	\item Exercises
	\item Looking Ahead to Chapter 7
	\item Chapter references
\end{enumerate}

% ============================================================
\section{Why DQN needed successors}
% ============================================================

DQN solved a major representation problem. It showed that a neural network could approximate a state-action value function directly from pixels and could be trained with a temporal-difference target using replay and a slowly updated target network \citep{mnih2013atari,mnih2015human}. But DQN also exposed several weaknesses that became research opportunities.

The main issue is that DQN inherits the structure of Q-learning. It estimates
\begin{equation}
	Q(s,a;\theta) \approx Q^*(s,a),
\end{equation}
and chooses actions greedily or nearly greedily:
\begin{equation}
	a_t = \arg\max_a Q(s_t,a;\theta).
\end{equation}
The one-step DQN target is
\begin{equation}
	y_t^{\mathrm{DQN}} = r_t + \gamma (1-d_t) \max_{a'} Q(s_{t+1},a';\theta^-),
\end{equation}
where $d_t$ indicates termination and $\theta^-$ denotes target-network parameters.

This target is simple, but it hides several problems:

\begin{itemize}[leftmargin=*]
	\item The same max operation both selects and evaluates the next action, which can create overestimation bias.
	\item Uniform replay treats all transitions as equally useful, even though some transitions have much larger learning signal.
	\item One-step targets propagate delayed rewards slowly.
	\item Epsilon-greedy exploration is crude, especially in sparse-reward or structured-control problems.
	\item A single expected value hides the variability and risk of future returns.
	\item A conventional Q-network does not explicitly distinguish ``this state is good'' from ``this action is better than other actions in this state.''
\end{itemize}

The algorithms in this chapter are not random additions. Each one answers a specific diagnostic question:

\begin{quote}
	What exactly is wrong with DQN, and which mechanism repairs that failure mode?
\end{quote}

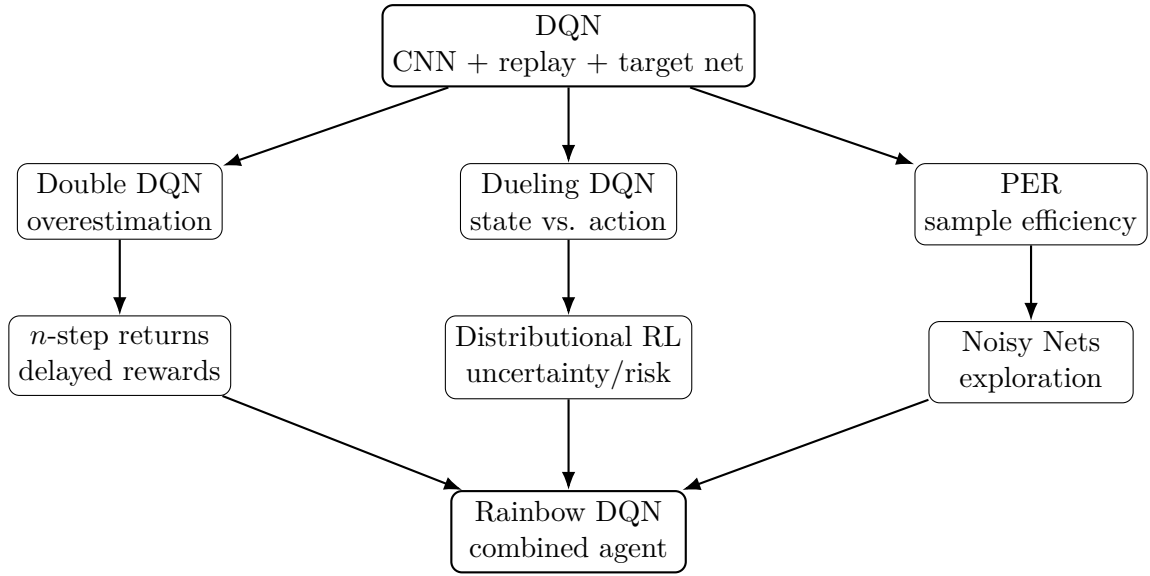
\begin{figure}[t]
	\centering
	\begin{tikzpicture}[
		box/.style={draw, rounded corners, thick, minimum width=3.1cm, minimum height=0.9cm, align=center},
		small/.style={draw, rounded corners, minimum width=2.7cm, minimum height=0.75cm, align=center},
		arrow/.style={-{Latex[length=2.4mm]}, thick},
		node distance=0.85cm
	]
		\node[box] (dqn) {DQN\\CNN + replay + target net};
		\node[small, below left=1.0cm and 2.1cm of dqn] (double) {Double DQN\\overestimation};
		\node[small, below=1.0cm of dqn] (duel) {Dueling DQN\\state vs. action};
		\node[small, below right=1.0cm and 2.1cm of dqn] (per) {PER\\sample efficiency};
		\node[small, below=1.0cm of double] (nstep) {$n$-step returns\\delayed rewards};
		\node[small, below=1.0cm of duel] (dist) {Distributional RL\\uncertainty/risk};
		\node[small, below=1.0cm of per] (noisy) {Noisy Nets\\exploration};
		\node[box, below=1.2cm of dist] (rainbow) {Rainbow DQN\\combined agent};

		\draw[arrow] (dqn) -- (double);
		\draw[arrow] (dqn) -- (duel);
		\draw[arrow] (dqn) -- (per);
		\draw[arrow] (double) -- (nstep);
		\draw[arrow] (duel) -- (dist);
		\draw[arrow] (per) -- (noisy);
		\draw[arrow] (nstep) -- (rainbow);
		\draw[arrow] (dist) -- (rainbow);
		\draw[arrow] (noisy) -- (rainbow);
	\end{tikzpicture}
	\caption{The DQN family can be understood as a set of targeted improvements. Each mechanism repairs a specific limitation of the original DQN algorithm. Rainbow later combined several of these improvements in a single agent \citep{hessel2018rainbow}.}
	\label{fig:dqn_family_overview}
\end{figure}

% ============================================================
\section{The diagnostic view: symptoms, mechanisms, and costs}
% ============================================================

A useful way to study value-based DRL is not to memorize algorithm names, but to identify symptoms. A symptom is a visible failure mode in learning: unstable value estimates, poor exploration, slow reward propagation, poor data efficiency, or risky behavior hidden by an average return. Table~\ref{tab:diagnostic_view} summarizes the main mechanisms in this chapter.

\begin{table*}[t]
	\centering
	\small
	\caption{A diagnostic view of major DQN improvements. This table is intended as a reader-facing reference: each method should be understood as a repair to a particular failure mode.}
	\label{tab:diagnostic_view}
	\begin{tabularx}{\textwidth}{p{2.4cm}YYY}
		\toprule
		\textbf{Method} & \textbf{Problem addressed} & \textbf{Core mechanism} & \textbf{Main cost or caution} \\
		\midrule
		Double DQN & Overestimation from max operator & Select action with online network, evaluate it with target network & Does not remove all estimation error; still depends on target quality \\
		\addlinespace
		Dueling DQN & Many actions have similar value; state quality and action advantage are mixed & Separate streams for $V(s)$ and $A(s,a)$, then combine into $Q(s,a)$ & Must resolve identifiability issue by subtracting mean or max advantage \\
		\addlinespace
		Prioritized replay & Uniform replay wastes updates on uninformative transitions & Sample transitions according to TD-error magnitude & Bias correction needed; can overfocus on noisy or outlier transitions \\
		\addlinespace
		$n$-step returns & One-step targets propagate delayed rewards slowly & Use several future rewards before bootstrapping & Higher variance; off-policy correction may be needed \\
		\addlinespace
		Noisy networks & Epsilon-greedy exploration is unstructured & Learn parameter-space noise inside network layers & Noise scale must be learned stably; less transparent than epsilon schedules \\
		\addlinespace
		Distributional RL & Expected value hides return uncertainty & Learn distribution of return $Z(s,a)$ rather than only $Q(s,a)$ & More complex loss and projection; risk interpretation requires care \\
		\addlinespace
		Rainbow & Single improvements are complementary & Combine Double, dueling, PER, $n$-step, NoisyNet, and distributional RL & More moving parts; harder to debug and ablate \\
		\bottomrule
	\end{tabularx}
\end{table*}

\begin{researchbox}{A research habit}
	When reading a new DRL paper, ask four questions before looking at the results: (1) What failure mode is being repaired? (2) Which part of the Bellman update, network architecture, replay distribution, or exploration rule is changed? (3) What new bias, variance, or implementation risk is introduced? (4) Is the improvement still meaningful outside Atari-style discrete-action benchmarks?
\end{researchbox}

% ============================================================
\section{Double DQN: reducing overestimation bias}
% ============================================================

\subsection{Where overestimation comes from}

The max operator in Q-learning is optimistic when value estimates contain noise. Suppose the true next-state action values are equal:
\begin{equation}
	Q^*(s',a_1)=Q^*(s',a_2)=\cdots=Q^*(s',a_m)=q.
\end{equation}
If the learned estimates are noisy,
\begin{equation}
	\hat Q(s',a_i)=q+\epsilon_i,
\end{equation}
then
\begin{equation}
	\mathbb{E}\left[\max_i \hat Q(s',a_i)\right]
	\geq \max_i \mathbb{E}[\hat Q(s',a_i)] = q.
\end{equation}
The maximum tends to select a positive noise error. This phenomenon already existed in tabular Q-learning and was studied by van Hasselt in Double Q-learning \citep{vanhasselt2010double}. In deep RL, van Hasselt, Guez, and Silver showed that DQN can suffer from substantial overestimation on Atari and proposed Double DQN as a deep version of this idea \citep{vanhasselt2016deep}.

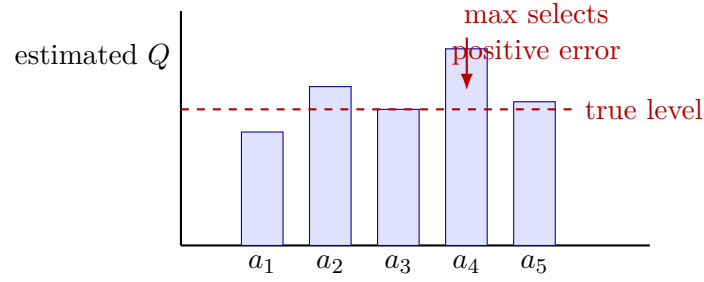
\begin{figure}[t]
	\centering
	\begin{tikzpicture}[
		bar/.style={draw, fill=gray!20, minimum width=0.55cm},
		axis/.style={thick},
		arrow/.style={-{Latex[length=2.2mm]}, thick}
	]
		\draw[axis] (0,0) -- (6.2,0);
		\draw[axis] (0,0) -- (0,3.1);
		\node[left] at (0,2.5) {estimated $Q$};
		\foreach \x/\h/\lab in {0.8/1.5/$a_1$,1.7/2.1/$a_2$,2.6/1.8/$a_3$,3.5/2.6/$a_4$,4.4/1.9/$a_5$}{
			\draw[fill=blue!12,draw=blue!60!black] (\x,0) rectangle +(0.55,\h);
			\node[below] at (\x+0.275,0) {\lab};
		}
		\draw[dashed, red!70!black, thick] (0,1.8) -- (5.2,1.8) node[right] {true level};
		\draw[arrow, red!70!black] (3.78,2.75) -- (3.78,2.05);
		\node[align=center, red!70!black] at (4.7,2.8) {max selects\\positive error};
	\end{tikzpicture}
	\caption{Overestimation bias arises because the maximum over noisy estimates tends to select actions with positive estimation errors. Double DQN reduces this effect by decoupling action selection from action evaluation \citep{vanhasselt2016deep}.}
	\label{fig:overestimation_bias}
\end{figure}

\subsection{The Double DQN target}

DQN uses
\begin{equation}
	y_t^{\mathrm{DQN}} = r_t + \gamma (1-d_t) \max_{a'} Q(s_{t+1},a';\theta^-).
\end{equation}
Double DQN decouples selection and evaluation. The online network selects the greedy action,
\begin{equation}
	a^* = \arg\max_{a'} Q(s_{t+1},a';\theta),
\end{equation}
and the target network evaluates that selected action:
\begin{equation}
	y_t^{\mathrm{DDQN}} = r_t + \gamma(1-d_t) Q(s_{t+1},a^*;\theta^-).
\end{equation}
This small change is one of the most important improvements to DQN because it attacks a precise statistical bias without changing the basic algorithmic structure.

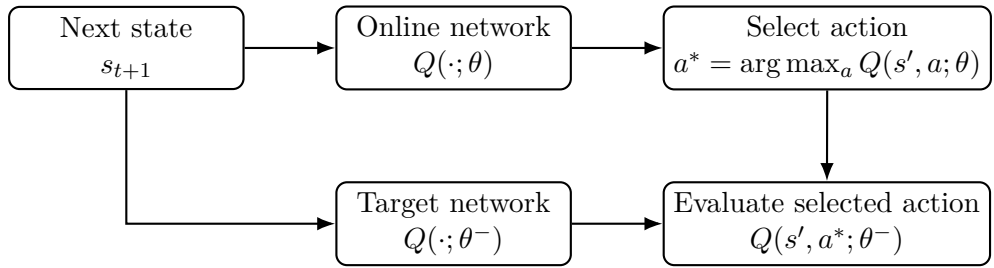
\begin{figure}[t]
	\centering
	\begin{tikzpicture}[
		box/.style={draw, rounded corners, thick, minimum width=3.1cm, minimum height=0.9cm, align=center},
		arrow/.style={-{Latex[length=2.4mm]}, thick},
		node distance=1.2cm
	]
		\node[box] (next) {Next state\\$s_{t+1}$};
		\node[box, right=of next] (online) {Online network\\$Q(\cdot;\theta)$};
		\node[box, right=of online] (select) {Select action\\$a^*=\arg\max_a Q(s',a;\theta)$};
		\node[box, below=of online] (target) {Target network\\$Q(\cdot;\theta^-)$};
		\node[box, right=of target] (eval) {Evaluate selected action\\$Q(s',a^*;\theta^-)$};
		\draw[arrow] (next) -- (online);
		\draw[arrow] (online) -- (select);
		\draw[arrow] (next) |- (target);
		\draw[arrow] (select) -- (eval);
		\draw[arrow] (target) -- (eval);
	\end{tikzpicture}
	\caption{Double DQN decouples selection and evaluation. The online network chooses the next action, while the target network evaluates it. This reduces overestimation compared with using the same target values for both operations.}
	\label{fig:double_dqn_flow}
\end{figure}

% ============================================================
\section{Dueling networks: separating state value from action advantage}
% ============================================================

In many states, the choice of action matters less than the quality of the state itself. For example, in an Atari game, the agent may be in a safe region where several actions have similar value. In a UAV network, a drone may already be positioned near a high-priority user cluster; moving slightly north or slightly east may produce similar QoS. A conventional Q-network must learn a separate output for each action even when most action differences are small.

The dueling architecture separates two concepts \citep{wang2016dueling}:
\begin{itemize}[leftmargin=*]
	\item the state value $V(s)$: how good the state is overall;
	\item the action advantage $A(s,a)$: how much better action $a$ is compared with other actions in that state.
\end{itemize}

The naive decomposition is
\begin{equation}
	Q(s,a) = V(s) + A(s,a).
\end{equation}
However, this decomposition is not identifiable: adding a constant to $V$ and subtracting it from all advantages gives the same $Q$. The dueling network resolves this by normalizing the advantage stream. A common aggregation is
\begin{equation}
	Q(s,a;\theta,\alpha,\beta) = V(s;\theta,\beta) + \left(A(s,a;\theta,\alpha) - \frac{1}{|\mathcal{A}|}\sum_{a'} A(s,a';\theta,\alpha)\right).
	\label{eq:dueling_mean}
\end{equation}

\begin{figure}[t]
	\centering
	\begin{tikzpicture}[
		box/.style={draw, rounded corners, thick, minimum width=2.8cm, minimum height=0.85cm, align=center},
		small/.style={draw, rounded corners, minimum width=2.4cm, minimum height=0.75cm, align=center},
		arrow/.style={-{Latex[length=2.3mm]}, thick},
		node distance=0.9cm
	]
		\node[box] (obs) {Observation\\$s$};
		\node[box, below=of obs] (features) {Shared encoder\\$\phi(s)$};
		\node[small, below left=1.0cm and 1.2cm of features] (value) {Value stream\\$V(s)$};
		\node[small, below right=1.0cm and 1.2cm of features] (adv) {Advantage stream\\$A(s,a)$};
		\node[box, below=2.1cm of features] (combine) {Aggregation\\$Q=V+A-\bar A$};
		\node[box, below=of combine] (qout) {Action values\\$Q(s,a_1),\ldots,Q(s,a_m)$};
		\draw[arrow] (obs) -- (features);
		\draw[arrow] (features) -- (value);
		\draw[arrow] (features) -- (adv);
		\draw[arrow] (value) -- (combine);
		\draw[arrow] (adv) -- (combine);
		\draw[arrow] (combine) -- (qout);
	\end{tikzpicture}
	\caption{The dueling architecture estimates state value and action advantage using separate streams and then combines them into action values. This helps learning when many actions have similar effects \citep{wang2016dueling}.}
	\label{fig:dueling_network}
\end{figure}
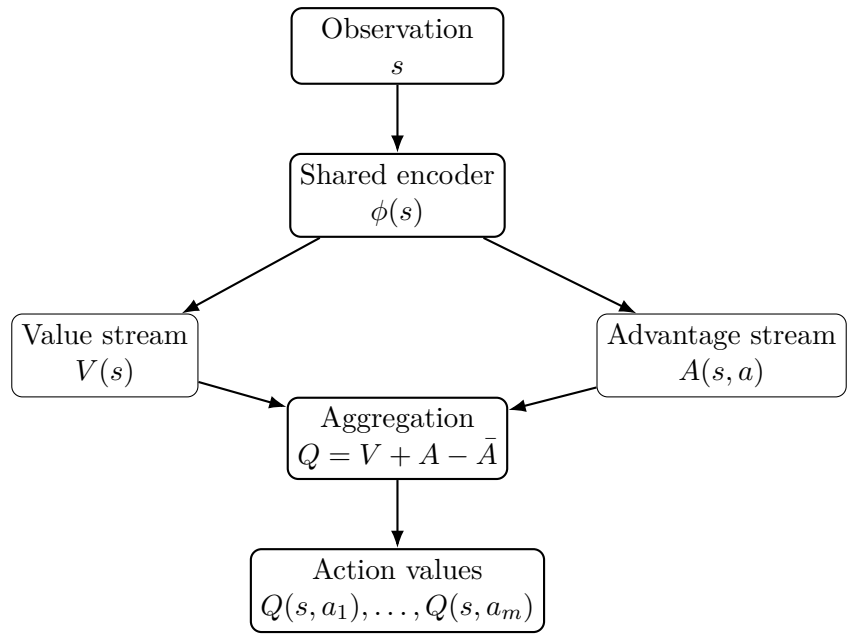

\begin{warningbox}{Implementation caution}
	Do not implement $Q=V+A$ without normalization. Without subtracting the mean or maximum advantage, $V$ and $A$ are not uniquely determined. The network can still produce Q-values, but the decomposition loses the intended meaning and can train less reliably.
\end{warningbox}

% ============================================================
\section{Prioritized experience replay}
% ============================================================

Uniform replay samples transitions randomly from the replay buffer. This reduces temporal correlation, but it ignores the fact that some transitions are more informative than others. Prioritized experience replay (PER) samples transitions with probability related to their temporal-difference error \citep{schaul2016prioritized}.

For a transition $i$, define priority
\begin{equation}
	p_i = |\delta_i| + \epsilon,
\end{equation}
where $\delta_i$ is the TD error and $\epsilon>0$ ensures every transition can be sampled. Proportional prioritization samples according to
\begin{equation}
	P(i) = \frac{p_i^\alpha}{\sum_k p_k^\alpha},
\end{equation}
where $\alpha$ controls how strongly prioritization is used. If $\alpha=0$, sampling is uniform.

Prioritized sampling changes the data distribution, so importance-sampling weights are used to reduce bias:
\begin{equation}
	w_i = \left( \frac{1}{N} \cdot \frac{1}{P(i)} \right)^\beta,
\end{equation}
usually normalized by dividing by $\max_i w_i$. The exponent $\beta$ is often annealed toward $1$ during training.

\begin{figure}[t]
	\centering
	\begin{tikzpicture}[
		cell/.style={draw, minimum width=0.9cm, minimum height=0.65cm, align=center},
		arrow/.style={-{Latex[length=2.3mm]}, thick}
	]
		\foreach \i/\p in {0/0.3,1/1.4,2/0.6,3/2.4,4/0.8,5/1.9,6/0.4}{
			\draw[cell, fill=blue!10] (\i*0.95,0) rectangle +(0.85,0.65);
			\node at (\i*0.95+0.425,0.325) {\scriptsize $i_{\i}$};
			\draw[fill=orange!45,draw=orange!80!black] (\i*0.95,0.9) rectangle +(0.85,\p);
		}
		\node[left] at (-0.25,2.7) {priority};
		\node[below] at (3.2,-0.2) {replay buffer};
		\draw[arrow, red!70!black] (3.25,3.55) -- (3.25,2.2);
		\node[red!70!black, align=center] at (5.7,3.2) {higher TD error\\sampled more often};
	\end{tikzpicture}
	\caption{Prioritized replay samples transitions with probability related to their learning signal, often measured by TD-error magnitude. Importance-sampling weights compensate for the induced sampling bias \citep{schaul2016prioritized}.}
	\label{fig:prioritized_replay}
\end{figure}
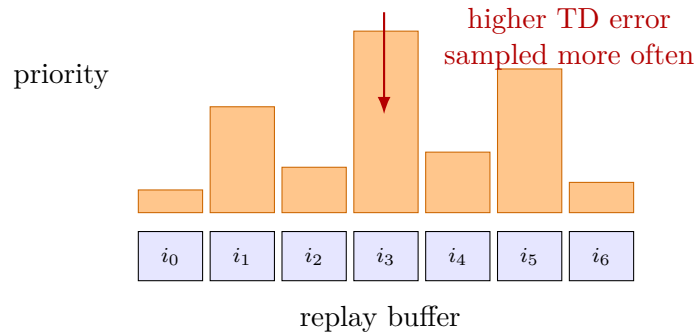

\subsection{Why PER is powerful but risky}

PER improves data efficiency because rare but informative transitions can be replayed more often. In a UAV/SDN setting, this is especially relevant for rare events: sudden congestion, handover failure, energy depletion, or QoS violation for high-priority users. Uniform replay may bury these events inside many ordinary transitions. PER can surface them.

However, TD error is not the same as usefulness. A transition may have high TD error because it is noisy, corrupted, or outside the representational capacity of the network. Over-prioritizing such transitions can destabilize learning. PER is therefore a mechanism for focusing learning, not a guarantee of better learning.

A second practical issue is the \emph{stale-priority problem}. A priority is usually computed when a transition is sampled and updated, but the network keeps changing after that update. A transition that once had a large TD error may later become easy, while another transition that once looked ordinary may become important after the representation changes. If priorities are not refreshed, the replay distribution can become a historical artifact rather than a current estimate of learning value. This is why robust PER implementations update priorities after each sampled minibatch, use a small positive floor $\epsilon$, anneal the importance-sampling correction, and avoid interpreting priority as a perfect measure of importance \citep{schaul2016prioritized}.

% ============================================================
\section{Multi-step targets}
% ============================================================

DQN uses a one-step target. One-step TD has low variance but may propagate delayed rewards slowly. Multi-step learning uses several rewards before bootstrapping:
\begin{equation}
	G_t^{(n)} = \sum_{k=0}^{n-1} \gamma^k r_{t+k} + \gamma^n Q(s_{t+n},a_{t+n};\theta^-).
\end{equation}
For control with a target network, one may use a greedy or Double-DQN style action at $s_{t+n}$.

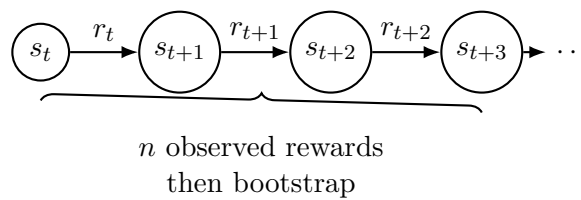
\begin{figure}[t]
	\centering
	\begin{tikzpicture}[
		state/.style={circle, draw, thick, minimum size=0.75cm},
		reward/.style={draw, rounded corners, minimum width=0.8cm, minimum height=0.55cm},
		arrow/.style={-{Latex[length=2.2mm]}, thick},
		node distance=0.9cm
	]
		\node[state] (s0) {$s_t$};
		\node[state, right=of s0] (s1) {$s_{t+1}$};
		\node[state, right=of s1] (s2) {$s_{t+2}$};
		\node[state, right=of s2] (s3) {$s_{t+3}$};
		\node[right=0.3cm of s3] (dots) {$\cdots$};
		\draw[arrow] (s0) -- node[above] {$r_t$} (s1);
		\draw[arrow] (s1) -- node[above] {$r_{t+1}$} (s2);
		\draw[arrow] (s2) -- node[above] {$r_{t+2}$} (s3);
		\draw[arrow] (s3) -- (dots);
		\draw[decorate, decoration={brace, amplitude=5pt}, thick] ($(s0.south)+(0,-0.25)$) -- node[below=0.25cm, align=center] {$n$ observed rewards\\then bootstrap} ($(s3.south)+(0,-0.25)$);
	\end{tikzpicture}
	\caption{Multi-step targets combine sampled rewards over several steps with a later bootstrap estimate. They propagate delayed rewards faster than one-step targets but usually increase variance.}
	\label{fig:n_step_return}
\end{figure}

Multi-step returns are a bridge between Monte Carlo and TD learning. When $n=1$, we recover one-step TD. As $n$ grows toward the episode length, the target becomes closer to a Monte Carlo return. Rainbow used multi-step returns as one of its components \citep{hessel2018rainbow}.

% ============================================================
\section{Noisy networks for exploration}
% ============================================================

Epsilon-greedy exploration is simple: with probability $\epsilon$, choose a random action. But random action noise is often poorly structured. In complex environments, coherent exploration may require the agent to behave differently over extended periods, not just inject independent random actions.

Noisy Networks replace some deterministic linear layers with noisy parameterized layers \citep{fortunato2018noisy}. A standard linear layer is
\begin{equation}
	y = Wx + b.
\end{equation}
A noisy linear layer samples perturbed weights and biases:
\begin{equation}
	y = (\mu_W + \sigma_W \odot \epsilon_W)x + (\mu_b + \sigma_b \odot \epsilon_b),
\end{equation}
where $\mu$ and $\sigma$ are learned parameters and $\epsilon$ is random noise. The important idea is that the exploration noise is learned and injected into the policy through parameters.

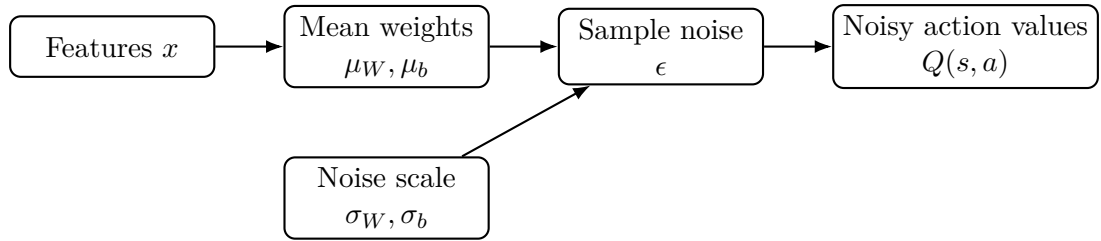
\begin{figure}[t]
	\centering
	\begin{tikzpicture}[
		box/.style={draw, rounded corners, thick, minimum width=2.7cm, minimum height=0.8cm, align=center},
		arrow/.style={-{Latex[length=2.3mm]}, thick},
		node distance=0.9cm
	]
		\node[box] (x) {Features $x$};
		\node[box, right=of x] (mu) {Mean weights\\$\mu_W,\mu_b$};
		\node[box, below=of mu] (sigma) {Noise scale\\$\sigma_W,\sigma_b$};
		\node[box, right=of mu] (noise) {Sample noise\\$\epsilon$};
		\node[box, right=of noise] (out) {Noisy action values\\$Q(s,a)$};
		\draw[arrow] (x) -- (mu);
		\draw[arrow] (mu) -- (noise);
		\draw[arrow] (sigma) -- (noise);
		\draw[arrow] (noise) -- (out);
	\end{tikzpicture}
	\caption{Noisy networks inject learned parameter-space noise into value networks. This creates structured exploration and can replace epsilon-greedy exploration in DQN-style agents \citep{fortunato2018noisy}.}
	\label{fig:noisy_networks}
\end{figure}

In UAV/network control, NoisyNet-style exploration can be interpreted as trying coherent controller variations. Instead of randomly choosing one bad routing action at a single time step, the agent may explore a slightly different value landscape over many decisions. This can be useful when good behavior requires a sequence of coordinated actions.

In practice, Noisy Networks should be presented honestly. They were an important part of the Rainbow agent and are conceptually attractive, but many modern DQN codebases still use $\epsilon$-greedy exploration because it is transparent, easy to tune, and easy to debug. NoisyNet exploration is therefore best understood as a structured exploration mechanism, not as a universal replacement for simpler exploration schedules.

% ============================================================
\section{Distributional RL: learning the return distribution}
% ============================================================

Standard Q-learning estimates the expected return:
\begin{equation}
	Q^\pi(s,a) = \mathbb{E}[G_t \mid S_t=s,A_t=a].
\end{equation}
Distributional RL instead models the random return itself:
\begin{equation}
	Z^\pi(s,a) \stackrel{D}{=} R_{t+1} + \gamma Z^\pi(S_{t+1},A_{t+1}),
\end{equation}
where $\stackrel{D}{=}$ denotes equality in distribution. The expected value is recovered by
\begin{equation}
	Q^\pi(s,a)=\mathbb{E}[Z^\pi(s,a)].
\end{equation}

Bellemare, Dabney, and Munos argued that the value distribution is fundamental, not only a tool for risk-sensitive behavior \citep{bellemare2017distributional}. This was a major conceptual shift: two actions may have the same expected return but very different risk profiles.

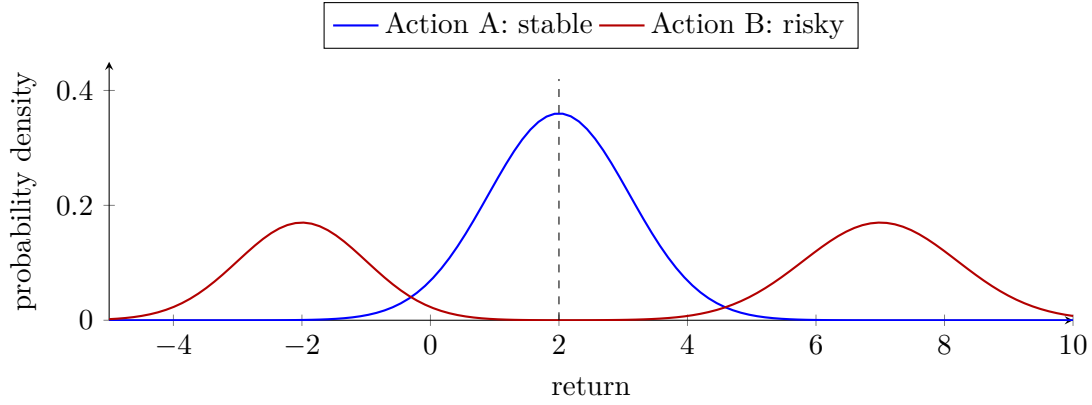
\begin{figure}[t]
	\centering
	\begin{tikzpicture}
		\begin{axis}[
			width=0.9\linewidth,
			height=5cm,
			xlabel={return},
			ylabel={probability density},
			xmin=-5,xmax=10,
			ymin=0,ymax=0.45,
			legend style={at={(0.5,1.05)},anchor=south,legend columns=2},
			axis lines=left
		]
			\addplot[blue, thick, domain=-5:10, samples=120] {0.36*exp(-0.5*((x-2)/1.1)^2)};
			\addlegendentry{Action A: stable}
			\addplot[red!70!black, thick, domain=-5:10, samples=120] {0.17*exp(-0.5*((x+2)/1.0)^2)+0.17*exp(-0.5*((x-7)/1.2)^2)};
			\addlegendentry{Action B: risky}
			\draw[dashed] (axis cs:2,0) -- (axis cs:2,0.42);
		\end{axis}
	\end{tikzpicture}
	\caption{Two actions can have similar expected return but very different return distributions. Distributional RL models the distribution of possible returns rather than only the mean. This is important for risk-sensitive domains such as robotics, finance, networking, and UAV control.}
	\label{fig:distributional_returns}
\end{figure}

\subsection{Why distribution matters for networks and UAVs}

Consider two UAV routing or movement actions. Both may have the same expected reward, but one action has low variance: it usually gives acceptable latency. The other has high variance: sometimes it gives excellent throughput, but sometimes it violates URLLC latency. For a safety-critical network, these actions are not equivalent. A distributional value function can support risk-aware policies such as choosing actions by a lower quantile or conditional value-at-risk (CVaR), although the algorithm must be designed carefully.

% ============================================================
\section{C51: categorical distributional DQN}
% ============================================================

C51 represents the return distribution using a fixed set of 51 atoms between $V_{\min}$ and $V_{\max}$ \citep{bellemare2017distributional}. Let
\begin{equation}
	z_i = V_{\min} + i\Delta z, \quad i=0,\ldots,N-1,
\end{equation}
where
\begin{equation}
	\Delta z = \frac{V_{\max}-V_{\min}}{N-1}.
\end{equation}
The network outputs probabilities over these atoms for each action:
\begin{equation}
	p_i(s,a) \approx \mathbb{P}(Z(s,a)=z_i).
\end{equation}
The expected Q-value is
\begin{equation}
	Q(s,a)=\sum_i z_i p_i(s,a).
\end{equation}

The Bellman update shifts and discounts the support:
\begin{equation}
	Tz_i = r + \gamma z_i,
\end{equation}
then projects the shifted distribution back onto the fixed atom support. This projection step is the technical heart of C51.

Projection is needed because the Bellman-shifted atoms $r+\gamma z_i$ usually do not land exactly on the original grid $\{z_0,\ldots,z_{N-1}\}$. Some shifted mass may fall between two neighboring atoms, and some may fall outside the chosen interval $[V_{\min},V_{\max}]$. C51 handles this support mismatch by clipping the shifted atoms to the fixed interval and redistributing probability mass linearly to the nearest atoms. Without this projection, the network would be asked to match a target distribution defined on a different support from the one it outputs. This is why the projection is not an implementation detail; it is what makes a fixed-support categorical distribution compatible with the Bellman update \citep{bellemare2017distributional,rowland2018analysis}.

\begin{figure}[t]
	\centering
	\begin{tikzpicture}[
		atom/.style={circle, draw, fill=blue!15, minimum size=0.45cm},
		arrow/.style={-{Latex[length=2.2mm]}, thick}
	]
		\foreach \i in {0,...,6}{
			\node[atom] (a\i) at (\i*0.8,0) {};
			\node[below] at (\i*0.8,-0.25) {\scriptsize $z_{\i}$};
		}
		\foreach \i/\h in {0/0.3,1/0.9,2/1.5,3/1.1,4/0.7,5/0.35,6/0.2}{
			\draw[fill=blue!40,draw=blue!70!black] (\i*0.8-0.18,0.25) rectangle +(0.36,\h);
		}
		\draw[arrow, red!70!black] (2.4,2.3) -- (3.3,1.6);
		\node[red!70!black, align=center] at (4.4,2.25) {Bellman shift\\and projection};
		\foreach \i in {0,...,6}{
			\node[atom, fill=orange!20] (b\i) at (\i*0.8,-2.0) {};
		}
		\foreach \i/\h in {0/0.1,1/0.25,2/0.7,3/1.25,4/1.1,5/0.55,6/0.25}{
			\draw[fill=orange!50,draw=orange!80!black] (\i*0.8-0.18,-1.75) rectangle +(0.36,\h);
		}
	\end{tikzpicture}
	\caption{C51 represents the return distribution using fixed atoms. The target distribution is shifted by reward and discounting, then projected back onto the fixed support \citep{bellemare2017distributional}.}
	\label{fig:c51_projection}
\end{figure}
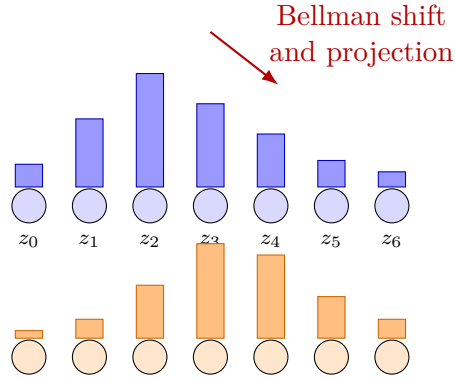

% ============================================================
\section{Quantile methods: QR-DQN and IQN}
% ============================================================

Categorical distributional RL fixes the support values and learns probabilities. Quantile methods take a different view: represent the distribution by quantiles. QR-DQN learns a fixed number of quantile values and trains them using quantile regression \citep{dabney2018qr}. IQN goes further and learns an implicit quantile function, making the quantile fraction $\tau$ an input to the network \citep{dabney2018iqn}.

For quantile regression, the quantile Huber loss for error $u$ and quantile $\tau$ is often written as
\begin{equation}
	\rho_\tau^\kappa(u)=|\tau-\mathbf{1}_{u<0}|\,\mathcal{L}_\kappa(u),
\end{equation}
where $\mathcal{L}_\kappa$ is the Huber loss. The important intuition is simple: a quantile should underpredict a specified fraction of the time and overpredict the rest.

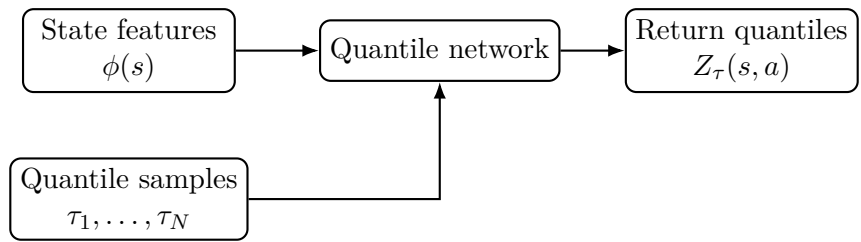
\begin{figure}[t]
	\centering
	\begin{tikzpicture}[
		box/.style={draw, rounded corners, thick, minimum width=2.8cm, minimum height=0.8cm, align=center},
		arrow/.style={-{Latex[length=2.2mm]}, thick},
		node distance=0.85cm
	]
		\node[box] (state) {State features\\$\phi(s)$};
		\node[box, below=of state] (tau) {Quantile samples\\$\tau_1,\ldots,\tau_N$};
		\node[box, right=1.1cm of state] (net) {Quantile network};
		\node[box, right=of net] (out) {Return quantiles\\$Z_{\tau}(s,a)$};
		\draw[arrow] (state) -- (net);
		\draw[arrow] (tau) -| (net);
		\draw[arrow] (net) -- (out);
	\end{tikzpicture}
	\caption{Quantile distributional methods represent return distributions through quantile values. IQN conditions the network on sampled quantile fractions, allowing a flexible implicit return distribution \citep{dabney2018iqn}.}
	\label{fig:iqn_structure}
\end{figure}

Quantile methods are particularly attractive for risk-aware control. For example, a UAV controller can choose actions using a lower quantile of the return distribution rather than the mean. This may prefer actions that avoid rare but severe QoS failures.

% ============================================================
\section{Rainbow DQN: combining complementary improvements}
% ============================================================

Rainbow combined six major DQN improvements: Double DQN, dueling architecture, prioritized replay, multi-step learning, distributional RL, and Noisy Networks \citep{hessel2018rainbow}. The important lesson is not only that the combined algorithm performed well on Atari. The deeper lesson is that different improvements repair different failure modes and are often complementary.

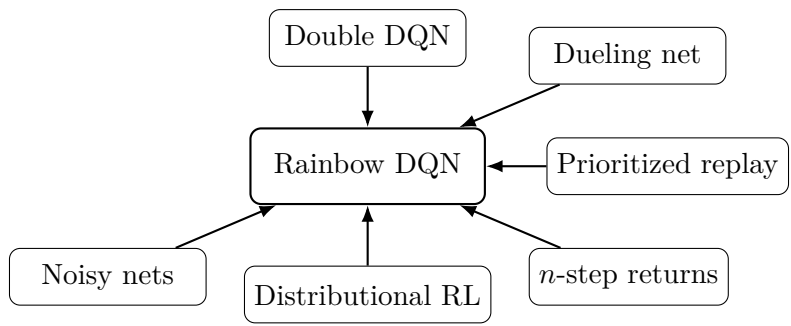
\begin{figure}[t]
	\centering
	\begin{tikzpicture}[
		component/.style={draw, rounded corners, minimum width=2.6cm, minimum height=0.75cm, align=center},
		center/.style={draw, rounded corners, thick, minimum width=3.1cm, minimum height=1.0cm, align=center},
		arrow/.style={-{Latex[length=2.2mm]}, thick},
		node distance=0.8cm
	]
		\node[center] (rainbow) {Rainbow DQN};
		\node[component, above=of rainbow] (double) {Double DQN};
		\node[component, above right=of rainbow] (dueling) {Dueling net};
		\node[component, right=of rainbow] (per) {Prioritized replay};
		\node[component, below right=of rainbow] (nstep) {$n$-step returns};
		\node[component, below=of rainbow] (dist) {Distributional RL};
		\node[component, below left=of rainbow] (noisy) {Noisy nets};
		\draw[arrow] (double) -- (rainbow);
		\draw[arrow] (dueling) -- (rainbow);
		\draw[arrow] (per) -- (rainbow);
		\draw[arrow] (nstep) -- (rainbow);
		\draw[arrow] (dist) -- (rainbow);
		\draw[arrow] (noisy) -- (rainbow);
	\end{tikzpicture}
	\caption{Rainbow DQN combines several complementary improvements. Its value is not only empirical performance but also a design lesson: value-based agents can be improved by separately addressing bias, representation, sampling, exploration, temporal credit assignment, and distributional information \citep{hessel2018rainbow}.}
	\label{fig:rainbow_stack}
\end{figure}

\subsection{What Rainbow teaches}

Rainbow is sometimes treated as a checklist of tricks. A more useful interpretation is architectural diagnosis. Double DQN repairs overestimation. Dueling networks improve value representation. PER improves sampling efficiency. Multi-step returns improve reward propagation. Noisy networks improve exploration. Distributional RL enriches the learning target.

A book chapter should also state the empirical nuance: the Rainbow components did not contribute equally. In the original Rainbow ablation study, prioritized replay and multi-step learning were among the most important contributors; removing either caused a large loss in performance on Atari. Other components were still useful, but their marginal value depended more strongly on the environment, implementation, and evaluation protocol \citep{hessel2018rainbow}. Later work revisiting Rainbow argued that smaller-scale benchmarks can reveal different component interactions and can make ablation studies more accessible, which is important for practitioners who cannot always run large Atari-scale experiments \citep{obando2021revisiting}.

For a researcher, Rainbow suggests a disciplined way to design new agents: do not add mechanisms randomly. Identify the bottleneck and choose a mechanism whose bias-variance trade-off matches that bottleneck. In a networking or UAV system, for example, multi-step targets may matter most when rewards are delayed, PER may matter most when rare violations are important, and distributional RL may matter most when tail risk matters more than average return.

% ============================================================
\section{Beyond Rainbow: distributed, recurrent, Munchausen, and fully parameterized variants}
% ============================================================

Rainbow is not the end of value-based DRL. Several later methods extended the same design philosophy in different directions.

\textbf{Ape-X} scaled DQN-style learning using many distributed actors and a centralized learner with prioritized replay \citep{horgan2018distributed}. Its main lesson is that value-based agents can benefit greatly from diverse, parallel experience collection, but this also increases the mismatch between the data-generating policies and the learner.

\textbf{R2D2} added recurrence and distributed replay, making DQN-style agents more suitable for partially observable environments \citep{kapturowski2019recurrent}. This matters because many real systems, including UAV networks and SDN controllers, observe only partial or delayed information.

\textbf{Munchausen DQN} modified the reward by adding a scaled log-policy term, connecting value-based learning to entropy-regularized ideas \citep{vieillard2020munchausen}. Its lesson is that value-based methods can borrow regularization ideas usually associated with policy-gradient and maximum-entropy methods.

\textbf{FQF}, or Fully Parameterized Quantile Function, extended quantile distributional RL by learning the quantile fractions themselves rather than fixing them in advance \citep{yang2019fqf}. This can provide a more adaptive representation of the return distribution, especially when the distribution has important structure in the tails.

These methods are important not because every practitioner should use all of them, but because they show the continuing evolution of value-based DRL: better data pipelines, memory, regularization, and distributional representations.

% ============================================================
\section{Engineering value-based DRL: what usually breaks}
% ============================================================

Improved DQN variants are powerful but fragile. Many failed experiments are caused by implementation details rather than theoretical flaws.

\begin{table*}[t]
	\centering
	\small
	\caption{Common implementation mistakes in DQN-family algorithms.}
	\label{tab:dqn_family_bugs}
	\begin{tabularx}{\textwidth}{p{3.1cm}YY}
		\toprule
		\textbf{Bug} & \textbf{Symptom} & \textbf{Fix} \\
		\midrule
		Terminal state not masked & Values explode near episode endings & Multiply bootstrap term by $(1-d_t)$ \\
		\addlinespace
		Target network receives gradients & Training unstable; memory grows unexpectedly & Compute targets under \texttt{torch.no\_grad()} \\
		\addlinespace
		Double DQN implemented as DQN & Overestimation remains & Use online network for argmax and target network for evaluation \\
		\addlinespace
		PER priorities not updated & Replay distribution becomes stale & Update priorities after every learning step \\
		\addlinespace
		Importance weights ignored in PER & Biased optimization and unstable late training & Multiply loss by normalized IS weights \\
		\addlinespace
		Dueling aggregation wrong & Value/advantage streams become ambiguous & Use $Q=V+A-\mathrm{mean}(A)$ or a similar normalization \\
		\addlinespace
		Distributional support poorly chosen & Mass collapses at $V_{\min}$ or $V_{\max}$ & Choose support consistent with reward scale and clipping \\
		\addlinespace
		Noisy layers not resampled & Exploration becomes deterministic too early & Resample noise at appropriate action-selection/update points \\
		\bottomrule
	\end{tabularx}
\end{table*}

% ============================================================
\section{Research example: UAV/SDN control with value-based DRL}
% ============================================================

Most books explain the DQN family only through Atari. That is historically correct but scientifically limiting. The same mechanisms become even more meaningful in communication-network control, SDN-assisted UAV systems, and QoS-aware resource allocation.

Consider a UAV acting as an aerial base station. The state includes position, battery level, user density, traffic demand, SINR, latency, throughput, and SDN-controller suggestions. The action set is discrete:
\begin{equation}
	\mathcal{A}=\{\mathrm{north},\mathrm{south},\mathrm{east},\mathrm{west},\mathrm{up},\mathrm{down},\mathrm{hover},\mathrm{recharge},\mathrm{follow\_SDN}\}.
\end{equation}
The reward combines QoS and penalties:
\begin{equation}
	r_t = w_q R_t^{\mathrm{QoS}} - w_e C_t^{\mathrm{energy}} - w_v C_t^{\mathrm{violation}} - w_c C_t^{\mathrm{collision}}.
\end{equation}

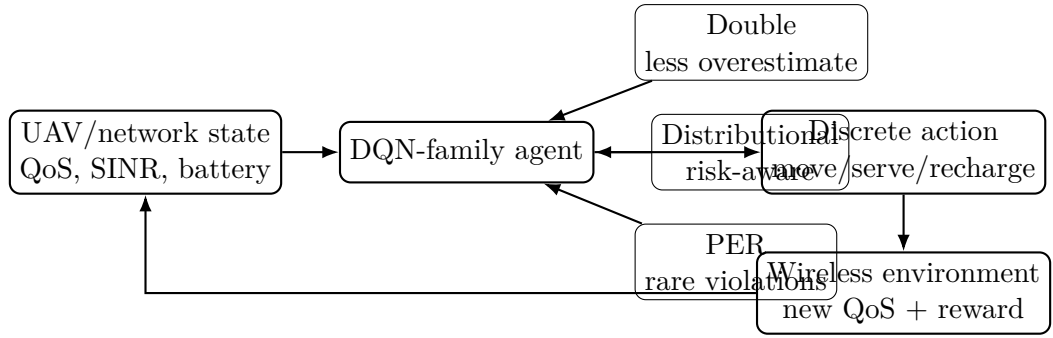
\begin{figure}[t]
	\centering
	\begin{tikzpicture}[
		box/.style={draw, rounded corners, thick, minimum width=2.9cm, minimum height=0.8cm, align=center},
		small/.style={draw, rounded corners, minimum width=2.5cm, minimum height=0.65cm, align=center},
		arrow/.style={-{Latex[length=2.2mm]}, thick},
		node distance=0.75cm
	]
		\node[box] (state) {UAV/network state\\QoS, SINR, battery};
		\node[box, right=of state] (qnet) {DQN-family agent};
		\node[small, above right=of qnet] (ddqn) {Double\\less overestimate};
		\node[small, right=of qnet] (dist) {Distributional\\risk-aware};
		\node[small, below right=of qnet] (per) {PER\\rare violations};
		\node[box, right=2.2cm of qnet] (action) {Discrete action\\move/serve/recharge};
		\node[box, below=of action] (env) {Wireless environment\\new QoS + reward};
		\draw[arrow] (state) -- (qnet);
		\draw[arrow] (qnet) -- (action);
		\draw[arrow] (action) -- (env);
		\draw[arrow] (env.west) -| (state.south);
		\draw[arrow] (ddqn) -- (qnet);
		\draw[arrow] (dist) -- (qnet);
		\draw[arrow] (per) -- (qnet);
	\end{tikzpicture}
	\caption{In UAV/SDN control, the DQN-family mechanisms have concrete interpretations. Double DQN controls optimistic QoS estimates, PER emphasizes rare violations, dueling networks help when many movements are similar, and distributional RL exposes risk hidden by average reward.}
	\label{fig:uav_dqn_family}
\end{figure}

\subsection{A worked mini-example}

Suppose two actions have the following empirical return distributions over repeated simulation episodes:
\begin{align}
	Z(s,\mathrm{hover}) &\in \{8,9,9,10,10,11,11,12\},\\
	Z(s,\mathrm{move\_toward\_URLLC}) &\in \{-10,6,8,12,14,16,18,20\}.
\end{align}
The hover action has mean return $10.0$. The movement action has mean return $10.5$, so a purely expectation-based DQN may prefer moving toward the URLLC cluster. But the lower tail tells a different story: the worst observed hover return is $8$, while the worst observed movement return is $-10$. A simple empirical summary is shown in Table~\ref{tab:uav_tail_risk_example}.

\begin{table}[t]
	\centering
	\small
	\caption{A small distributional example for UAV/SDN control. Mean return favors the risky action, while tail-aware evaluation favors the safer action.}
	\label{tab:uav_tail_risk_example}
	\begin{tabular}{lccc}
		\toprule
		\textbf{Action} & \textbf{Mean return} & \textbf{Worst observed return} & \textbf{Interpretation} \\
		\midrule
		Hover & $10.0$ & $8$ & Stable service, low tail risk \\
		Move toward URLLC & $10.5$ & $-10$ & Higher upside, severe rare failure \\
		\bottomrule
	\end{tabular}
\end{table}

The second action may have a higher mean because it sometimes gives very high QoS reward. But it also has a severe negative tail: sometimes it causes interference, energy depletion, or a latency violation. A standard DQN that learns only expected values may prefer the risky action. A distributional agent can expose the tail. A risk-sensitive controller may choose the safer action by using a lower quantile:
\begin{equation}
	a^*_{\mathrm{risk}} = \arg\max_a \mathrm{Quantile}_{0.1}\left(Z(s,a)\right).
\end{equation}
This is not a minor detail in networking. For URLLC-style users, tail behavior matters more than average behavior. A policy with excellent mean throughput but rare severe latency violations may be unacceptable.

\begin{researchbox}{What is original in this book's running example?}
	For Atari, a bad action usually means losing game score. For UAV/SDN control, a bad action may mean battery failure, QoS violation for high-priority users, safety violation, or unstable resource allocation. This changes how we interpret value-based DRL. Distributional learning is not only a performance trick; it can become a risk-analysis tool. Prioritized replay is not only a sample-efficiency trick; it can emphasize rare safety-critical events. Dueling networks are not only an architecture trick; they help distinguish globally good network states from locally useful control actions.
\end{researchbox}

% ============================================================
\section{Python implementation blocks}
% ============================================================

The following code snippets are not a complete production library. They are compact, readable implementations of the most important mechanisms. They are designed to be studied, modified, and tested.

\subsection{Double DQN target computation}

\Needspace{16\baselineskip}
\begin{lstlisting}[style=pythonstyle,caption={Double DQN target computation in PyTorch.},label={lst:double_dqn_target}]
import torch
import torch.nn.functional as F

@torch.no_grad()
def double_dqn_target(online_net, target_net, next_obs, rewards, dones, gamma):
    """Compute Double DQN targets.

    online_net selects the greedy action.
    target_net evaluates the selected action.
    dones must be 1.0 for terminal transitions and 0.0 otherwise.
    """
    # Action selection by online network
    next_q_online = online_net(next_obs)  # [B, A]
    next_actions = next_q_online.argmax(dim=1, keepdim=True)  # [B, 1]

    # Action evaluation by target network
    next_q_target = target_net(next_obs)  # [B, A]
    selected_next_q = next_q_target.gather(1, next_actions).squeeze(1)

    # Terminal masking is essential.
    targets = rewards + gamma * (1.0 - dones) * selected_next_q
    return targets

def dqn_loss(q_net, obs, actions, targets):
    q_values = q_net(obs).gather(1, actions.long().unsqueeze(1)).squeeze(1)
    return F.smooth_l1_loss(q_values, targets)
\end{lstlisting}

\subsection{Dueling Q-network}

\Needspace{18\baselineskip}
\begin{lstlisting}[style=pythonstyle,caption={A minimal dueling Q-network for vector observations.},label={lst:dueling_q_network}]
import torch
import torch.nn as nn

class DuelingQNetwork(nn.Module):
    def __init__(self, obs_dim, action_dim, hidden_dim=256):
        super().__init__()
        self.encoder = nn.Sequential(
            nn.Linear(obs_dim, hidden_dim),
            nn.ReLU(),
            nn.Linear(hidden_dim, hidden_dim),
            nn.ReLU(),
        )
        self.value_stream = nn.Sequential(
            nn.Linear(hidden_dim, hidden_dim),
            nn.ReLU(),
            nn.Linear(hidden_dim, 1),
        )
        self.advantage_stream = nn.Sequential(
            nn.Linear(hidden_dim, hidden_dim),
            nn.ReLU(),
            nn.Linear(hidden_dim, action_dim),
        )

    def forward(self, obs):
        z = self.encoder(obs)
        value = self.value_stream(z)          # [B, 1]
        advantage = self.advantage_stream(z)  # [B, A]

        # Identifiability correction: subtract mean advantage.
        q = value + advantage - advantage.mean(dim=1, keepdim=True)
        return q
\end{lstlisting}

\subsection{A simple prioritized replay buffer}

\Needspace{20\baselineskip}
\begin{lstlisting}[style=pythonstyle,caption={Readable prioritized replay buffer using NumPy probabilities. For large-scale training, a sum-tree is more efficient.},label={lst:per_simple}]
import numpy as np

class SimplePrioritizedReplay:
    def __init__(self, capacity, alpha=0.6, beta=0.4, eps=1e-6):
        self.capacity = capacity
        self.alpha = alpha
        self.beta = beta
        self.eps = eps
        self.storage = []
        self.priorities = np.zeros(capacity, dtype=np.float32)
        self.pos = 0

    def add(self, transition):
        max_prio = self.priorities.max() if self.storage else 1.0
        if len(self.storage) < self.capacity:
            self.storage.append(transition)
        else:
            self.storage[self.pos] = transition
        self.priorities[self.pos] = max_prio
        self.pos = (self.pos + 1) % self.capacity

    def sample(self, batch_size):
        n = len(self.storage)
        scaled = self.priorities[:n] ** self.alpha
        probs = scaled / scaled.sum()

        idxs = np.random.choice(n, batch_size, p=probs)
        samples = [self.storage[i] for i in idxs]

        # Importance-sampling weights
        weights = (n * probs[idxs]) ** (-self.beta)
        weights = weights / weights.max()
        return samples, idxs, weights.astype(np.float32)

    def update_priorities(self, idxs, td_errors):
        for idx, err in zip(idxs, td_errors):
            self.priorities[idx] = abs(float(err)) + self.eps
\end{lstlisting}

\subsection{Noisy linear layer}

\Needspace{20\baselineskip}
\begin{lstlisting}[style=pythonstyle,caption={A factorized NoisyLinear layer similar in spirit to NoisyNet.},label={lst:noisy_linear}]
import math
import torch
import torch.nn as nn
import torch.nn.functional as F

class NoisyLinear(nn.Module):
    def __init__(self, in_features, out_features, sigma0=0.5):
        super().__init__()
        self.in_features = in_features
        self.out_features = out_features

        self.mu_w = nn.Parameter(torch.empty(out_features, in_features))
        self.sigma_w = nn.Parameter(torch.empty(out_features, in_features))
        self.mu_b = nn.Parameter(torch.empty(out_features))
        self.sigma_b = nn.Parameter(torch.empty(out_features))

        self.register_buffer("eps_w", torch.empty(out_features, in_features))
        self.register_buffer("eps_b", torch.empty(out_features))
        self.reset_parameters(sigma0)
        self.reset_noise()

    def reset_parameters(self, sigma0):
        bound = 1.0 / math.sqrt(self.in_features)
        self.mu_w.data.uniform_(-bound, bound)
        self.mu_b.data.uniform_(-bound, bound)
        self.sigma_w.data.fill_(sigma0 / math.sqrt(self.in_features))
        self.sigma_b.data.fill_(sigma0 / math.sqrt(self.out_features))

    def _scale_noise(self, size):
        x = torch.randn(size, device=self.mu_w.device)
        return x.sign() * x.abs().sqrt()

    def reset_noise(self):
        eps_in = self._scale_noise(self.in_features)
        eps_out = self._scale_noise(self.out_features)
        self.eps_w.copy_(eps_out.outer(eps_in))
        self.eps_b.copy_(eps_out)

    def forward(self, x):
        if self.training:
            w = self.mu_w + self.sigma_w * self.eps_w
            b = self.mu_b + self.sigma_b * self.eps_b
        else:
            w = self.mu_w
            b = self.mu_b
        return F.linear(x, w, b)
\end{lstlisting}

\subsection{C51 projection skeleton}

\Needspace{20\baselineskip}
\begin{lstlisting}[style=pythonstyle,caption={C51 target projection skeleton. This function projects shifted target atoms back to the fixed support.},label={lst:c51_projection}]
import torch

def c51_projection(next_dist, rewards, dones, gamma, v_min, v_max, num_atoms):
    """Project target distribution onto fixed C51 support.

    next_dist: [B, num_atoms], probabilities for chosen next actions.
    rewards, dones: [B]
    returns: [B, num_atoms], projected target distribution.
    """
    batch_size = rewards.shape[0]
    device = rewards.device
    support = torch.linspace(v_min, v_max, num_atoms, device=device)
    delta_z = (v_max - v_min) / (num_atoms - 1)

    tz = rewards.unsqueeze(1) + gamma * (1.0 - dones.unsqueeze(1)) * support.unsqueeze(0)
    tz = tz.clamp(v_min, v_max)

    b = (tz - v_min) / delta_z
    lower = b.floor().long()
    upper = b.ceil().long()

    projected = torch.zeros_like(next_dist)
    offset = torch.arange(batch_size, device=device).unsqueeze(1) * num_atoms

    # Distribute probability mass to neighboring atoms.
    projected.view(-1).index_add_(
        0,
        (lower + offset).view(-1),
        (next_dist * (upper.float() - b)).view(-1),
    )
    projected.view(-1).index_add_(
        0,
        (upper + offset).view(-1),
        (next_dist * (b - lower.float())).view(-1),
    )

    # Handle exact integer projection where lower == upper.
    same = (upper == lower)
    if same.any():
        projected.view(-1).index_add_(
            0,
            (lower[same] + offset.expand_as(lower)[same]).view(-1),
            next_dist[same].view(-1),
        )
    return projected
\end{lstlisting}

\begin{warningbox}{Code caution for C51}
	The projection code is easy to get wrong. The most common mistakes are forgetting terminal masking, using a support range inconsistent with reward scaling, and mishandling the case where the projected atom lands exactly on a support atom. Always test the projection separately before training the full agent.
\end{warningbox}

\subsection{Quantile Huber loss for QR-DQN}

\Needspace{18\baselineskip}
\begin{lstlisting}[style=pythonstyle,caption={Quantile Huber loss used in QR-DQN style distributional learning.},label={lst:quantile_huber}]
import torch
import torch.nn.functional as F

def quantile_huber_loss(pred_quantiles, target_quantiles, taus, kappa=1.0):
    """Compute quantile Huber loss.

    pred_quantiles:   [B, N] predicted quantiles for chosen actions
    target_quantiles: [B, N] target quantiles
    taus:             [N] quantile fractions in (0, 1)
    """
    # Pairwise Bellman errors: target_j - pred_i
    diff = target_quantiles.unsqueeze(1) - pred_quantiles.unsqueeze(2)  # [B, N, N]

    abs_diff = diff.abs()
    huber = torch.where(
        abs_diff <= kappa,
        0.5 * diff.pow(2),
        kappa * (abs_diff - 0.5 * kappa),
    )

    taus = taus.view(1, -1, 1)
    weight = (taus - (diff.detach() < 0).float()).abs()
    loss = (weight * huber / kappa).sum(dim=1).mean(dim=1).mean()
    return loss
\end{lstlisting}

% ============================================================
\section{Limitations of improved value-based methods}
% ============================================================

The methods in this chapter improve DQN, but they do not make value-based DRL universal. Several limitations remain.

\begin{enumerate}[leftmargin=*]
	\item \textbf{Discrete action dependence.} DQN-style methods naturally output one value per discrete action. They are not the natural first choice for high-dimensional continuous actions such as UAV acceleration vectors, power-control levels, or continuous routing ratios.
	\item \textbf{Large discrete action spaces.} Even if actions are discrete, a very large action set can make the max operation expensive and generalization across actions weak.
	\item \textbf{Sample inefficiency.} Atari-style DQN agents often require millions of environment frames. In real systems, collecting that much online data may be impossible.
	\item \textbf{Safety.} Better value estimates do not guarantee safe exploration. Distributional values can expose risk, but they do not enforce constraints by themselves.
	\item \textbf{Representation limits.} CNNs and MLPs are not enough for all domains. Graph neural networks, recurrent networks, transformers, or world models may be needed for structured and partially observable systems.
	\item \textbf{Evaluation sensitivity.} Performance depends heavily on preprocessing, reward scaling, seeds, evaluation protocol, and environment stochasticity.
\end{enumerate}

These limitations motivate the next part of the book: policy gradients and actor-critic methods. When actions become continuous, stochastic, or high-dimensional, learning a policy directly often becomes more natural than estimating a value for every possible action.

% ============================================================
\section{Key takeaways}
% ============================================================

\begin{itemize}[leftmargin=*]
	\item DQN was a breakthrough, but its limitations created a family of improved value-based algorithms.
	\item Double DQN reduces overestimation by decoupling action selection from action evaluation.
	\item Dueling networks separate state value from action advantage and help when many actions are similarly valued.
	\item Prioritized replay samples transitions according to learning signal but requires bias correction.
	\item Multi-step returns propagate rewards faster but can increase variance.
	\item Noisy networks learn parameter-space exploration instead of relying only on epsilon-greedy randomness.
	\item Distributional RL models the full return distribution and can reveal risk hidden by expected values.
	\item C51 uses fixed categorical atoms; QR-DQN and IQN use quantile-based representations.
	\item Rainbow shows that complementary improvements can be combined, but the resulting agent is harder to debug.
	\item For UAV/SDN and communication networks, these methods are not just Atari tricks: they map naturally to overestimated QoS, rare violations, delayed effects, structured exploration, and risk-sensitive control.
\end{itemize}

% ============================================================
\section{Exercises}
% ============================================================

\subsection*{Conceptual exercises}
\begin{enumerate}[leftmargin=*]
	\item Explain why the max operator in Q-learning can cause overestimation even when individual value estimates are unbiased.
	\item Why does Double DQN reduce overestimation but not eliminate all value-estimation error?
	\item Explain why Double DQN can sometimes underestimate action values, and why mild underestimation is often preferable to systematic overestimation in control.
	\item In your own words, explain why the dueling architecture needs an identifiability correction such as subtracting the mean advantage.
	\item Why can prioritized replay be harmful if high TD error is caused by noise rather than useful learning signal?
	\item Explain why two actions with the same expected return may not be equally desirable in a safety-critical network.
\end{enumerate}

\subsection*{Mathematical exercises}
\begin{enumerate}[leftmargin=*]
	\item Derive the Double DQN target from the DQN target by separating action selection and action evaluation.
	\item Suppose $p_i=|\delta_i|+\epsilon$ and $\alpha=0$. Show that PER becomes uniform replay.
	\item For a three-step target, write $G_t^{(3)}$ explicitly.
	\item Given atoms $z=[0,1,2]$ and probabilities $p=[0.2,0.5,0.3]$, compute the expected value.
	\item For a dueling network with $V(s)=4$ and advantages $A(s,\cdot)=[1,2,3]$, compute the Q-values using mean-normalized aggregation.
\end{enumerate}

\subsection*{Coding exercises}
\begin{enumerate}[leftmargin=*]
	\item Modify a DQN implementation so that it uses the Double DQN target.
	\item Implement the dueling architecture for an Atari-style CNN.
	\item Add priority updates to a replay buffer using TD-error magnitude.
	\item Implement a small test to verify that C51 projection preserves total probability mass.
	\item Add a risk-sensitive action rule to a distributional agent by selecting actions according to a lower quantile rather than the mean.
\end{enumerate}

\subsection*{Research thinking exercises}
\begin{enumerate}[leftmargin=*]
	\item In a UAV network, which transitions should be prioritized: high reward, high TD error, safety violations, or rare high-priority user events? Defend your answer.
	\item Design a discrete action space for UAV/SDN control and explain where DQN-family methods may break down.
	\item How would you evaluate whether distributional RL improves tail latency rather than only average reward?
	\item Which Rainbow components would you keep for a safety-critical network-control task, and which would you remove or modify?
\end{enumerate}

% ============================================================
\section*{Looking Ahead to Chapter 7: Food for Thought}
\addcontentsline{toc}{section}{Looking Ahead to Chapter 7: Food for Thought}

This chapter showed how far value-based DRL can be pushed. Starting from DQN, researchers reduced overestimation, improved replay, changed architectures, learned return distributions, and combined mechanisms into Rainbow. But one major limitation remains: value-based methods are most natural when the action set is discrete and not too large.

Before moving to policy-gradient methods, pause and consider the following questions:

\begin{enumerate}[leftmargin=*]
	\item What should an agent do when actions are continuous, such as steering angle, motor torque, bandwidth fraction, transmission power, or UAV velocity?
	\item If the policy itself is stochastic, why might it be better to optimize the policy directly rather than infer it from a value function?
	\item Can we estimate how changing policy parameters changes expected return?
	\item Why might a value function still be useful even when the main object being optimized is the policy?
	\item What does it mean to follow the gradient of performance?
\end{enumerate}

\begin{quote}
	Chapter 6 improved value-based deep reinforcement learning. Chapter 7 begins a new family: learning the policy directly.
\end{quote}

% ============================================================
%\section{Chapter references}
% ============================================================
	\part{Policy Gradients and Actor-Critic Methods}
%\addcontentsline{toc}{part}{Part III --- Policy Gradients and Actor-Critic Methods}

\chapter{Why Learn a Policy Directly?}
\label{ch:policy_gradients}
\chaptermark{Why Learn a Policy Directly?}

\begin{keybox}{Chapter goal}
	Chapters 5 and 6 studied value-based deep reinforcement learning. The agent learned a value function and selected actions by maximizing that value. This is powerful for discrete action problems, but many real systems require continuous, structured, stochastic, or constrained actions. Chapter 7 introduces the central idea of policy-gradient methods: represent the policy itself as a differentiable model and optimize its parameters directly. This shift is one of the most important transitions in deep reinforcement learning.
\end{keybox}

\section*{Chapter Overview}
\addcontentsline{toc}{section}{Chapter Overview}

\begin{enumerate}[leftmargin=*]
	\item Why value-based control is not enough
	\item The policy as the primary object
	\item Continuous, structured, and stochastic action spaces
	\item The objective: expected return as a function of policy parameters
	\item Trajectory distributions and why the environment need not be differentiable
	\item The likelihood-ratio trick
	\item The policy-gradient theorem
	\item What the gradient really means
	\item Stochastic policy gradients versus deterministic policy gradients
	\item Policy parameterizations: categorical, Gaussian, squashed Gaussian, beta, and structured policies
	\item From policy gradients to modern algorithms
	\item Research example: UAV/SDN control with continuous actions
	\item Python implementation blocks
	\item Practical failure modes and debugging rules
	\item Limitations of direct policy optimization
	\item Key takeaways
	\item Exercises
	\item Looking Ahead to Chapter 8
	\item Chapter references
\end{enumerate}

% ============================================================
\section{Why value-based control is not enough}
% ============================================================

Value-based reinforcement learning begins with a natural idea: estimate how good each action is, then choose the action with the largest value. In tabular Q-learning and DQN-style methods, the policy is usually implicit:
\begin{equation}
	\pi(s) = \arg\max_a Q(s,a).
\end{equation}
This worked well in small grid worlds and many Atari games because the action space was discrete and modest. In Atari, the agent might choose among actions such as moving left, moving right, firing, or doing nothing \citep{mnih2015human,bellemare2013ale}. The network can output one number per action.

But many real systems are not like this. A robot arm does not choose among five joystick buttons. It outputs torques or target velocities. A UAV may choose a heading angle, altitude change, speed, transmit power, and bandwidth allocation. A wireless controller may choose routing fractions or resource shares. A language model chooses a distribution over tokens at each step, where the policy itself is the object being shaped by preference or verifiable rewards \citep{ouyang2022training,jaech2024openai}.

The key limitation is not only that actions may be continuous. It is that actions may be \emph{structured}: an action can contain multiple coupled decisions, constraints, bounds, safety requirements, and stochastic behavior. A value network with one output per action becomes awkward or impossible when the number of possible actions is infinite or combinatorial.

\begin{figure}[t]
	\centering
	\resizebox{0.95\textwidth}{!}{%
		\begin{tikzpicture}[
			box/.style={draw, rounded corners, thick, minimum width=3.4cm, minimum height=0.95cm, align=center, font=\small},
			small/.style={draw, rounded corners, minimum width=2.5cm, minimum height=0.75cm, align=center, font=\small},
			arrow/.style={-{Latex[length=2.5mm]}, thick}
			]
			% Value side
			\node[box, draw=blue!70, fill=blue!10, font=\small\bfseries] (value) at (0,0) {Value-Based Control\\{\normalfont\small Learn $Q(s,a)$}};
			\node[small, draw=blue!50, fill=blue!5] (disc) at (-2,-2) {Best for\\small discrete actions};
			\node[small, draw=blue!50, fill=blue!5] (max) at (2,-2) {Requires\\$\arg\max_a Q(s,a)$};
			
			% Policy side
			\node[box, draw=green!60!black, fill=green!10, font=\small\bfseries] (policy) at (8,0) {Policy-Based Control\\{\normalfont\small Learn $\pi_\theta(a|s)$}};
			\node[small, draw=green!40!black, fill=green!5] (cont) at (6,-2) {Works with\\continuous actions};
			\node[small, draw=green!40!black, fill=green!5] (prob) at (10,-2) {Can represent\\stochastic behavior};
			
			\draw[arrow, color=blue!60] (value) -- (disc);
			\draw[arrow, color=blue!60] (value) -- (max);
			\draw[arrow, color=green!60!black] (policy) -- (cont);
			\draw[arrow, color=green!60!black] (policy) -- (prob);
			\draw[arrow, color=red!70!black] (value) -- node[above=3pt, align=center, font=\scriptsize, color=red!70!black] {directly learn\\behavior} (policy);
		\end{tikzpicture}%
	}
	\caption{Value-based methods learn action values and derive behavior via $\arg\max$. Policy-based methods learn the policy directly. Actor-critic methods combine both.}
	\label{fig:value_vs_policy}
\end{figure}
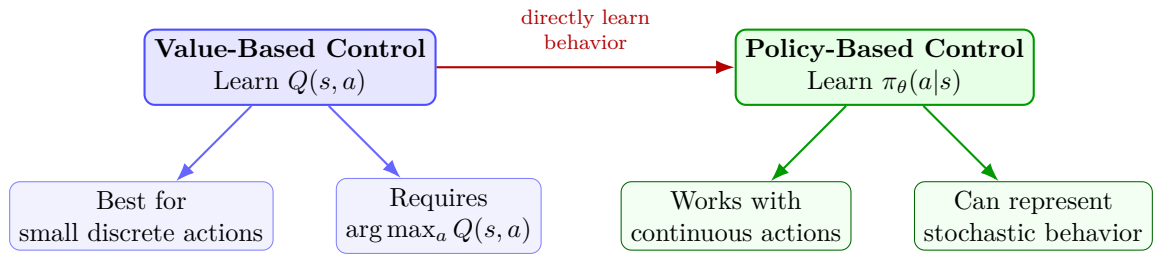

\begin{table*}[t]
	\centering
	\caption{Why direct policy learning becomes necessary.}
	\label{tab:value_vs_policy_need}
	\begin{tabularx}{\textwidth}{p{3.4cm}YY}
		\toprule
		Issue & Value-based view & Direct policy view \\
		\midrule
		Continuous actions & Requires difficult maximization over $a$ or discretization. & Policy outputs action distribution parameters directly. \\
		Large discrete actions & One Q-value per action may be expensive. & Policy can factorize or generate actions sequentially. \\
		Stochastic behavior & Usually added externally through $\epsilon$-greedy or noise. & Stochasticity is part of the policy. \\
		Safety and constraints & Argmax may select unsafe actions unless filtered. & Policy can be parameterized or projected to respect constraints. \\
		Smooth control & Greedy value maximization may be discontinuous. & Policy outputs smooth actions under continuous distributions. \\
		Language-model alignment & Token probabilities are already a policy. & Preference/reward optimization directly changes token distributions. \\
		\bottomrule
	\end{tabularx}
\end{table*}

% ============================================================
\section{The policy as the primary object}
% ============================================================

A policy is a rule for choosing actions. In a stochastic policy, the agent samples actions from a distribution:
\begin{equation}
	\pi_\theta(a|s) = \Pr(A_t=a \mid S_t=s),
\end{equation}
where $\theta$ denotes the policy parameters. In deep reinforcement learning, $\theta$ is usually the set of neural-network weights.

The philosophical shift of policy-gradient methods is simple:
\begin{quote}
	Instead of learning values and deriving behavior indirectly, learn behavior itself.
\end{quote}
This makes policy-gradient methods conceptually close to supervised learning: the model takes a state as input and outputs action probabilities or action-distribution parameters. The difference is that the training signal is not a labeled correct action. The training signal is the long-term return produced by sampled actions.

\begin{mathbox}{Two ways to represent control}
	\textbf{Value-based control:}
	\begin{equation}
		\theta_Q \leftarrow \theta_Q - \alpha \nabla_{\theta_Q} \left(y - Q(s,a;\theta_Q)\right)^2,
		\qquad
		a = \arg\max_a Q(s,a;\theta_Q).
	\end{equation}
	\textbf{Policy-gradient control:}
	\begin{equation}
		\theta_\pi \leftarrow \theta_\pi + \alpha \widehat{\nabla_{\theta_\pi} J(\theta_\pi)},
		\qquad
		a \sim \pi_{\theta_\pi}(\cdot|s).
	\end{equation}
\end{mathbox}

A direct policy can be deterministic or stochastic. A deterministic policy maps states to actions:
\begin{equation}
	a = \mu_\theta(s).
\end{equation}
A stochastic policy maps states to distributions:
\begin{equation}
	a \sim \pi_\theta(\cdot|s).
\end{equation}
Stochastic policies are central in the classical policy-gradient theorem \citep{sutton1999policy,williams1992simple}. Deterministic policies become especially important in high-dimensional continuous-control algorithms such as deterministic policy gradient (DPG), DDPG, and TD3 \citep{silver2014deterministic,lillicrap2016continuous,fujimoto2018addressing}.

% ============================================================
\section{Continuous, structured, and stochastic action spaces}
% ============================================================

The most visible reason to learn a policy directly is continuous control. Suppose a UAV chooses two continuous actions:
\begin{equation}
	a_t = (\Delta x_t, \Delta y_t),
\end{equation}
where each component lies in $[-1,1]$. A value-based method could discretize each dimension into $K$ bins, producing $K^2$ actions. If the action contains five dimensions, the same discretization produces $K^5$ actions. This is another curse of dimensionality, now in action space rather than state space.

\begin{figure}[t]
	\centering
	\begin{tikzpicture}
		\begin{axis}[
			width=0.88\textwidth,
			height=5.5cm,
			xlabel={Number of bins per action dimension $K$},
			ylabel={Number of discrete actions $K^d$},
			ymode=log,
			xmin=2, xmax=20,
			grid=major,
			legend style={at={(0.03,0.97)},anchor=north west},
			]
			\addplot+[mark=*] coordinates {(2,4) (5,25) (10,100) (20,400)};
			\addlegendentry{$d=2$}
			\addplot+[mark=square*] coordinates {(2,8) (5,125) (10,1000) (20,8000)};
			\addlegendentry{$d=3$}
			\addplot+[mark=triangle*] coordinates {(2,32) (5,3125) (10,100000) (20,3200000)};
			\addlegendentry{$d=5$}
		\end{axis}
	\end{tikzpicture}
	\caption{Discretizing continuous action spaces creates exponential growth in the number of actions. If each of $d$ action dimensions is discretized into $K$ bins, the discrete action set has $K^d$ actions. Direct policy learning avoids this enumeration by outputting action parameters directly.}
	\label{fig:action_discretization}
\end{figure}
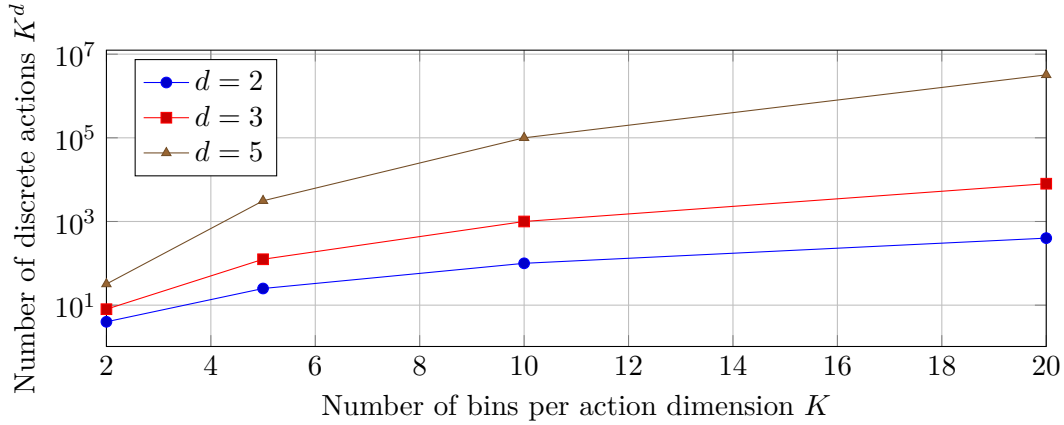

The second reason is stochasticity. Some tasks require stochastic policies, not merely for exploration but for optimality. In partially observable environments, the same observation may correspond to different hidden states; randomized behavior can be useful. In competitive multi-agent settings, deterministic behavior may be exploitable. In language models, the policy is naturally a probability distribution over tokens.

The third reason is structure. A policy can be designed to respect action constraints. For example, a Gaussian policy can be squashed by a hyperbolic tangent to enforce bounded actions, or a softmax distribution can enforce that bandwidth-allocation fractions sum to one. These design choices are difficult to express through a simple Q-table.

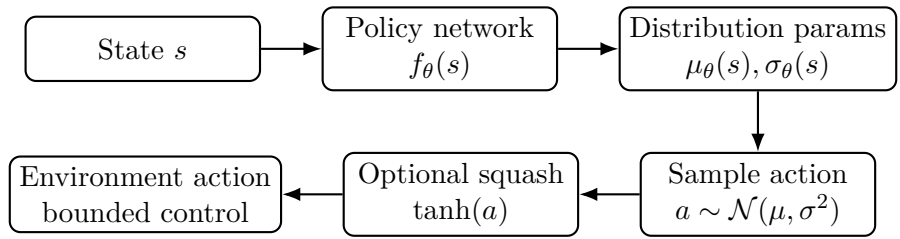
\begin{figure}[t]
	\centering
	\begin{tikzpicture}[
		box/.style={draw, rounded corners, thick, minimum width=3.1cm, minimum height=0.85cm, align=center},
		arrow/.style={-{Latex[length=2.4mm]}, thick},
		node distance=0.8cm
		]
		\node[box] (state) {State $s$};
		\node[box, right=of state] (nn) {Policy network\\$f_\theta(s)$};
		\node[box, right=of nn] (params) {Distribution params\\$\mu_\theta(s),\sigma_\theta(s)$};
		\node[box, below=of params] (sample) {Sample action\\$a \sim \mathcal{N}(\mu,\sigma^2)$};
		\node[box, left=of sample] (squash) {Optional squash\\$\tanh(a)$};
		\node[box, left=of squash] (env) {Environment action\\bounded control};
		\draw[arrow] (state) -- (nn);
		\draw[arrow] (nn) -- (params);
		\draw[arrow] (params) -- (sample);
		\draw[arrow] (sample) -- (squash);
		\draw[arrow] (squash) -- (env);
	\end{tikzpicture}
	\caption{A neural policy can output the parameters of an action distribution. For continuous bounded control, a Gaussian sample is often squashed through $\tanh$ before being sent to the environment.}
	\label{fig:gaussian_policy_pipeline}
\end{figure}

% ============================================================
\section{The objective: expected return as a function of policy parameters}
% ============================================================

Policy-gradient methods optimize an objective of the form
\begin{equation}
	J(\theta) = \mathbb{E}_{\tau \sim p_\theta(\tau)}\left[ G(\tau) \right],
\end{equation}
where $\tau$ denotes a trajectory and $G(\tau)$ denotes its return. In an episodic MDP,
\begin{equation}
	\tau = (s_0,a_0,r_1,s_1,a_1,r_2,\ldots,s_T),
\end{equation}
with return
\begin{equation}
	G(\tau) = \sum_{t=0}^{T-1} \gamma^t r_{t+1}.
\end{equation}

The trajectory distribution induced by a stochastic policy is
\begin{equation}
	p_\theta(\tau)
	= \rho_0(s_0) \prod_{t=0}^{T-1}
	\pi_\theta(a_t|s_t) p(s_{t+1}|s_t,a_t),
\end{equation}
where $\rho_0$ is the initial-state distribution and $p(s_{t+1}|s_t,a_t)$ is the environment transition model.

The objective is therefore not an ordinary supervised loss. The policy parameters affect the distribution of future data. Changing the policy changes the states that will be visited, the actions that will be sampled, and the returns that will be observed.

\begin{figure}[t]
	\centering
	\begin{tikzpicture}[
		circ/.style={circle, draw, thick, minimum size=0.8cm, align=center},
		reward/.style={rectangle, draw, rounded corners, minimum width=0.9cm, minimum height=0.6cm, align=center},
		arrow/.style={-{Latex[length=2.2mm]}, thick},
		node distance=0.95cm
		]
		\node[circ] (s0) {$s_0$};
		\node[circ, right=of s0] (a0) {$a_0$};
		\node[circ, right=of a0] (s1) {$s_1$};
		\node[circ, right=of s1] (a1) {$a_1$};
		\node[circ, right=of a1] (s2) {$s_2$};
		\node[circ, right=of s2] (a2) {$a_2$};
		\node[right=0.5cm of a2] (dots) {$\cdots$};
		\node[reward, above=0.7cm of $(a0)!0.5!(s1)$] (r1) {$r_1$};
		\node[reward, above=0.7cm of $(a1)!0.5!(s2)$] (r2) {$r_2$};
		\draw[arrow] (s0) -- node[below, font=\scriptsize] {$\pi_\theta$} (a0);
		\draw[arrow] (a0) -- node[below, font=\scriptsize] {$p$} (s1);
		\draw[arrow] (s1) -- node[below, font=\scriptsize] {$\pi_\theta$} (a1);
		\draw[arrow] (a1) -- node[below, font=\scriptsize] {$p$} (s2);
		\draw[arrow] (s2) -- node[below, font=\scriptsize] {$\pi_\theta$} (a2);
		\draw[arrow] (a2) -- (dots);
		\draw[arrow] (r1) -- (s1);
		\draw[arrow] (r2) -- (s2);
		\node[below=1.0cm of s1, align=center] {$p_\theta(\tau) = \rho_0(s_0) \prod_t \pi_\theta(a_t|s_t)p(s_{t+1}|s_t,a_t)$};
	\end{tikzpicture}
	\caption{A trajectory distribution depends on both the policy and the environment dynamics. Policy-gradient methods differentiate the expected return with respect to policy parameters, even when the environment dynamics are unknown or non-differentiable.}
	\label{fig:trajectory_distribution}
\end{figure}

% ============================================================
\section{Trajectory distributions and why the environment need not be differentiable}
% ============================================================

A common misunderstanding is that policy-gradient methods require differentiating through the environment. They do not. The key trick is that the policy is differentiable, but the environment transition probability does not need to be.

Starting from
\begin{equation}
	J(\theta) = \int p_\theta(\tau) G(\tau) d\tau,
\end{equation}
we differentiate:
\begin{align}
	\nabla_\theta J(\theta)
	&= \int \nabla_\theta p_\theta(\tau) G(\tau) d\tau.
\end{align}
Using the identity
\begin{equation}
	\nabla_\theta p_\theta(\tau)
	= p_\theta(\tau) \nabla_\theta \log p_\theta(\tau),
\end{equation}
we obtain
\begin{equation}
	\nabla_\theta J(\theta)
	= \mathbb{E}_{\tau \sim p_\theta}\left[ G(\tau) \nabla_\theta \log p_\theta(\tau) \right].
\end{equation}
Now observe that
\begin{align}
	\log p_\theta(\tau)
	&= \log \rho_0(s_0)
	+ \sum_{t=0}^{T-1} \log \pi_\theta(a_t|s_t)
	+ \sum_{t=0}^{T-1} \log p(s_{t+1}|s_t,a_t).
\end{align}
The environment terms do not depend on $\theta$, so they vanish:
\begin{equation}
	\nabla_\theta \log p_\theta(\tau)
	= \sum_{t=0}^{T-1} \nabla_\theta \log \pi_\theta(a_t|s_t).
\end{equation}
Therefore,
\begin{equation}
	\nabla_\theta J(\theta)
	= \mathbb{E}_{\tau \sim p_\theta}\left[\sum_{t=0}^{T-1} \nabla_\theta \log \pi_\theta(a_t|s_t) G(\tau)\right].
\end{equation}
This is the likelihood-ratio or score-function estimator. It is the mathematical basis of REINFORCE \citep{williams1992simple}.

A subtle but important point is that the return $G(\tau)$ is treated as a sampled scalar weight, not as a quantity through which we backpropagate. The gradient is taken through the policy log-probabilities, not through the reward function or the environment. In other words, the differentiable object is $\log \pi_\theta(a_t|s_t)$; the sampled return only tells the optimizer how strongly to increase or decrease the probability of the sampled action. This is why the same estimator can be used when the reward comes from a black-box simulator, a packet-level network emulator, a physical robot, a human-preference model, or a non-differentiable verification rule.

\begin{researchbox}{Why this matters in real systems}
	Policy gradients can be used even when the environment is a simulator, a robot, a network emulator, a traffic system, or a black-box reward function. We do not need a differentiable model of packet loss, wireless fading, collision dynamics, human preference, or game physics. We only need to evaluate sampled actions and compute gradients through the policy distribution.
\end{researchbox}

% ============================================================
\section{The likelihood-ratio trick}
% ============================================================

The identity
\begin{equation}
	\nabla_\theta p_\theta(x) = p_\theta(x) \nabla_\theta \log p_\theta(x)
\end{equation}
looks small, but it is one of the most useful tools in reinforcement learning. It turns a gradient of a probability distribution into an expectation over samples from that distribution.

For a single action, suppose the policy sampled $a$ in state $s$ and later received return $G$. The update direction is proportional to
\begin{equation}
	G \nabla_\theta \log \pi_\theta(a|s).
\end{equation}
If $G$ is high, the log-probability of the sampled action is increased. If $G$ is low or negative, the log-probability is decreased. This gives policy gradients a simple behavioral interpretation:
\begin{quote}
	Make good sampled actions more likely and bad sampled actions less likely.
\end{quote}

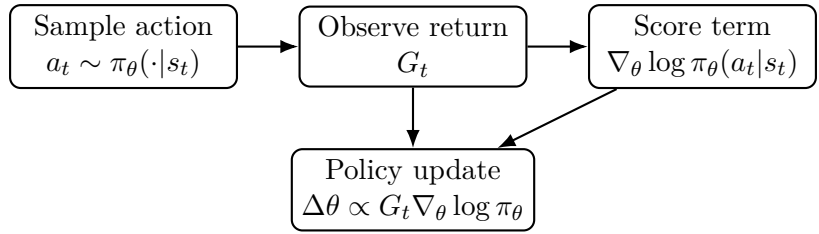
\begin{figure}[t]
	\centering
	\begin{tikzpicture}[
		box/.style={draw, rounded corners, thick, minimum width=3.0cm, minimum height=0.85cm, align=center},
		arrow/.style={-{Latex[length=2.4mm]}, thick},
		node distance=0.8cm
		]
		\node[box] (sample) {Sample action\\$a_t \sim \pi_\theta(\cdot|s_t)$};
		\node[box, right=of sample] (return) {Observe return\\$G_t$};
		\node[box, right=of return] (score) {Score term\\$\nabla_\theta \log \pi_\theta(a_t|s_t)$};
		\node[box, below=of return] (update) {Policy update\\$\Delta\theta \propto G_t \nabla_\theta \log \pi_\theta$};
		\draw[arrow] (sample) -- (return);
		\draw[arrow] (return) -- (score);
		\draw[arrow] (return) -- (update);
		\draw[arrow] (score) -- (update);
	\end{tikzpicture}
	\caption{The likelihood-ratio estimator multiplies a score term by a return estimate. The score term tells how to change the policy to make the sampled action more or less likely; the return determines the direction's weight.}
	\label{fig:score_function_update}
\end{figure}

The estimator is very general, but it can have high variance. The same return may be used to reinforce many actions in the same trajectory, so one lucky or unlucky episode can move many action probabilities in the same direction. This is the practical weakness of the simplest REINFORCE estimator. Chapter 8 begins from this problem and introduces baselines, value functions, and advantage estimates as variance-reduction tools \citep{greensmith2004variance}. For now, the important point is that direct policy optimization is possible even without differentiating through the environment.

% ============================================================
\section{The policy-gradient theorem}
% ============================================================

The trajectory-based expression is correct but inefficient because it appears to use the full trajectory return everywhere. The policy-gradient theorem gives a more useful form. For a stochastic policy in an MDP,
\begin{equation}
	\nabla_\theta J(\theta)
	\propto
	\sum_s d^{\pi_\theta}(s) \sum_a \nabla_\theta \pi_\theta(a|s) Q^{\pi_\theta}(s,a),
\end{equation}
where $d^{\pi_\theta}(s)$ is the discounted state visitation distribution under the current policy. Equivalently,
\begin{equation}
	\nabla_\theta J(\theta)
	= \mathbb{E}_{s \sim d^{\pi_\theta}, a \sim \pi_\theta}\left[
	\nabla_\theta \log \pi_\theta(a|s) Q^{\pi_\theta}(s,a)
	\right].
\end{equation}
This theorem is fundamental because it says that the policy can be improved using an action-value function, without explicitly differentiating the state distribution. It was formalized for function approximation by Sutton, McAllester, Singh, and Mansour \citep{sutton1999policy}.

The key mathematical point is that the gradient of the environment dynamics does not appear. The transition model $p(s_{t+1}|s_t,a_t)$ strongly affects which trajectories are sampled, but it is not parameterized by the policy parameters $\theta$. When differentiating the log-probability of a trajectory, the environment terms therefore drop out, and only the policy terms remain. This is the ``magic'' of the policy-gradient theorem: the agent can improve a differentiable policy while treating the world itself as an unknown, non-differentiable source of samples.

\begin{mathbox}{Policy-gradient theorem}
	\begin{equation}
		\boxed{
			\nabla_\theta J(\theta)
			= \mathbb{E}_{s,a}\left[
			\nabla_\theta \log \pi_\theta(a|s) Q^{\pi_\theta}(s,a)
			\right]
		}
	\end{equation}
	The expectation is taken over states visited by the current policy and actions sampled from that policy.
\end{mathbox}

In practice, $Q^{\pi_\theta}(s,a)$ is usually replaced by one of several estimators:
\begin{equation}
	G_t, \qquad Q_w(s_t,a_t), \qquad A_w(s_t,a_t), \qquad \hat{A}_t.
\end{equation}
These choices lead to REINFORCE, actor-critic methods, generalized advantage estimation, TRPO, PPO, and many modern policy-optimization algorithms \citep{williams1992simple,konda2000actor,schulman2015gae,schulman2015trpo,schulman2017ppo}.

\begin{figure}[t]
	\centering
	\begin{tikzpicture}[
		term/.style={draw, rounded corners, thick, minimum width=3.0cm, minimum height=0.8cm, align=center},
		arrow/.style={-{Latex[length=2.3mm]}, thick},
		node distance=0.9cm
		]
		\node[term] (score) {Score\\$\nabla_\theta \log \pi_\theta(a|s)$};
		\node[term, right=of score] (quality) {Quality signal\\$Q^\pi$, $G_t$, or $A_t$};
		\node[term, right=of quality] (grad) {Policy gradient\\increase/decrease probability};
		\draw[arrow] (score) -- (quality);
		\draw[arrow] (quality) -- (grad);
		\node[below=0.9cm of quality, align=center] {Policy gradient = direction of probability change $\times$ how good the action was};
	\end{tikzpicture}
	\caption{The policy-gradient theorem separates the gradient into two conceptual parts: a score term that changes action probabilities and a quality signal that measures whether the action was useful.}
	\label{fig:policy_gradient_theorem_intuition}
\end{figure}

% ============================================================
\section{What the gradient really means}
% ============================================================

For a categorical policy, the score term has a concrete meaning. Suppose the policy outputs probabilities over three actions:
\begin{equation}
	\pi_\theta(\cdot|s) = (0.2, 0.5, 0.3).
\end{equation}
If action 2 is sampled and receives a high return, the gradient increases the log-probability of action 2 in similar states. Because probabilities must sum to one, increasing one probability tends to decrease others.

For a Gaussian policy,
\begin{equation}
	a \sim \mathcal{N}(\mu_\theta(s), \sigma^2_\theta(s)),
\end{equation}
the gradient changes the mean and variance. If sampled actions above the current mean produce higher return, the mean may shift upward. If broader exploration helps, the variance may increase; if precise control helps, the variance may decrease.

This is why policy gradients are not merely an algorithmic trick. They provide a way to shape behavior at the distribution level.

The price is variance. A raw Monte Carlo return may reflect many sources of randomness: the selected action, earlier actions, later actions, stochastic transitions, exploration noise, and random initial states. The gradient estimator may therefore point in a useful direction only on average, while individual updates are noisy. This is the reason baselines and advantage functions are not cosmetic improvements; they are central to making policy-gradient methods usable.

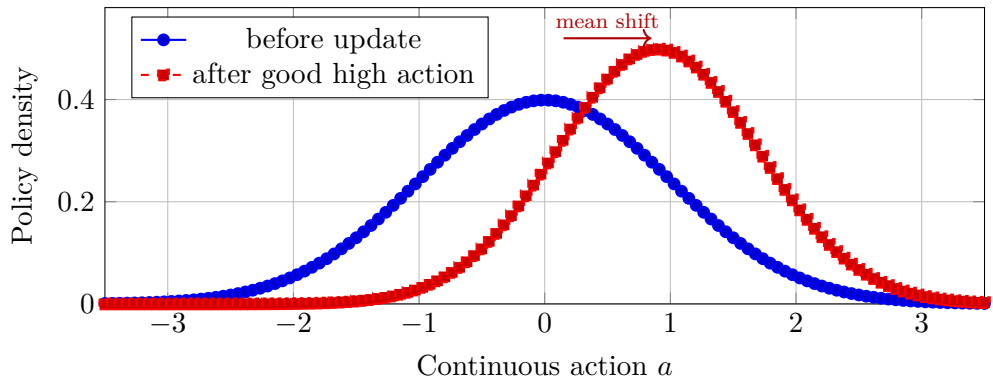
\begin{figure}[t]
	\centering
	\begin{tikzpicture}
		\begin{axis}[
			width=0.83\textwidth,
			height=5.5cm,
			xlabel={Continuous action $a$},
			ylabel={Policy density},
			xmin=-3.5, xmax=3.5,
			ymin=0, ymax=0.58,
			grid=major,
			legend style={at={(0.03,0.97)},anchor=north west},
			]
			\addplot+[domain=-3.5:3.5,samples=120,thick] {1/sqrt(2*pi)*exp(-x^2/2)};
			\addlegendentry{before update}
			\addplot+[domain=-3.5:3.5,samples=120,thick,dashed] {1/(0.8*sqrt(2*pi))*exp(-(x-0.9)^2/(2*0.8^2))};
			\addlegendentry{after good high action}
			\draw[->, thick, red!70!black] (axis cs:0.15,0.52) -- (axis cs:0.85,0.52) node[midway, above, font=\scriptsize] {mean shift};
		\end{axis}
	\end{tikzpicture}
	\caption{For a continuous Gaussian policy, a policy-gradient update can shift the mean and change the variance of the action distribution. This is more natural than discretizing the action space and learning one Q-value per discrete bin.}
	\label{fig:gaussian_policy_shift}
\end{figure}

% ============================================================
\section{Stochastic policy gradients versus deterministic policy gradients}
% ============================================================

Stochastic policy gradients optimize a distribution $\pi_\theta(a|s)$. Deterministic policy gradients optimize a deterministic mapping $\mu_\theta(s)$. The deterministic policy-gradient theorem states that, under suitable assumptions,
\begin{equation}
	\nabla_\theta J(\theta)
	= \mathbb{E}_{s \sim \rho^{\mu}}\left[
	\nabla_\theta \mu_\theta(s) \nabla_a Q^\mu(s,a)\big|_{a=\mu_\theta(s)}
	\right].
\end{equation}
This result is important for continuous control because the gradient no longer integrates over all actions. Instead, it evaluates how the critic changes with respect to the action at the current actor output \citep{silver2014deterministic}.

\begin{table*}[t]
	\centering
	\caption{Stochastic versus deterministic policy gradients.}
	\label{tab:stochastic_vs_deterministic_pg}
	\begin{tabularx}{\textwidth}{p{3.1cm}YY}
		\toprule
		Aspect & Stochastic policy gradient & Deterministic policy gradient \\
		\midrule
		Policy form & $a \sim \pi_\theta(\cdot|s)$ & $a = \mu_\theta(s)$ \\
		Gradient form & $\nabla \log \pi_\theta(a|s) Q^\pi(s,a)$ & $\nabla_\theta \mu_\theta(s) \nabla_a Q(s,a)$ \\
		Exploration & Built into stochastic policy & Usually external noise or stochastic behavior policy \\
		Common algorithms & REINFORCE, A2C/A3C, TRPO, PPO, SAC-style stochastic actors & DPG, DDPG, TD3 \\
		Strength & Natural stochasticity; supports discrete and continuous actions & Efficient for high-dimensional continuous actions \\
		Weakness & High variance; often on-policy & Requires critic quality; exploration not automatic \\
		\bottomrule
	\end{tabularx}
\end{table*}

Deterministic policy gradients are not a replacement for stochastic policy gradients. They are a different tool. In robotics and continuous-control problems, deterministic actors can be efficient. In alignment, preference learning, language modeling, and safety-sensitive exploration, stochastic policies remain central because they represent uncertainty and allow controlled sampling.

% ============================================================
\section{Policy parameterizations}
% ============================================================

The policy distribution is not a minor implementation detail. It defines the geometry of exploration, the support of possible actions, and the kind of behavior the agent can represent. Table~\ref{tab:policy_parameterizations} summarizes common choices.

\begin{table*}[t]
	\centering
	\caption{Common policy parameterizations and where they are useful.}
	\label{tab:policy_parameterizations}
	\begin{tabularx}{\textwidth}{p{3.1cm}p{3.7cm}YY}
		\toprule
		Policy type & Output & Useful for & Main caution \\
		\midrule
		Categorical softmax & Action probabilities & Discrete actions, token policies, small action sets & Large action sets can be expensive. \\
		Diagonal Gaussian & Mean and log standard deviation & Continuous control, locomotion, UAV movement & Unbounded unless squashed or clipped. \\
		Squashed Gaussian & Gaussian sample passed through $\tanh$ & Bounded continuous control & Log-probability needs change-of-variables correction. \\
		Beta policy & Shape parameters on bounded interval & Naturally bounded scalar actions & Can be numerically delicate near boundaries. \\
		Dirichlet policy & Positive vector summing to one & Allocation fractions, bandwidth shares & Parameters must remain positive; correlations can be limited. \\
		Autoregressive policy & Sequential conditional choices & Structured actions, language, combinatorial control & Sampling and training can be slower. \\
		Hybrid policy & Discrete mode plus continuous parameters & UAV mode + movement, network path + split ratio & Requires careful factorization. \\
		\bottomrule
	\end{tabularx}
\end{table*}

\subsection{Categorical policies}

For discrete actions, a neural network produces logits $z_\theta(s)$ and the policy is
\begin{equation}
	\pi_\theta(a|s) = \frac{\exp(z_a)}{\sum_b \exp(z_b)}.
\end{equation}
This is the natural policy for small action spaces and for token-level language-model policies.

\subsection{Gaussian policies}

For continuous actions, a common choice is a diagonal Gaussian:
\begin{equation}
	\pi_\theta(a|s) = \mathcal{N}\left(a; \mu_\theta(s), \mathrm{diag}(\sigma^2_\theta(s))\right).
\end{equation}
The mean controls the center of behavior; the standard deviation controls exploration.

\subsection{Squashed Gaussian policies}

If the action must lie in $[-1,1]^d$, one can sample $u$ from a Gaussian and set
\begin{equation}
	a = \tanh(u).
\end{equation}
This is common in maximum-entropy actor-critic methods such as SAC \citep{haarnoja2018soft}. However, the log-probability must account for the change of variables:
\begin{equation}
	\log \pi(a|s) = \log \mathcal{N}(u;\mu,\sigma^2) - \sum_i \log(1-\tanh^2(u_i)).
\end{equation}
Forgetting this correction is a common bug.

\subsection{Beta policies for bounded actions}

A Beta policy is another way to represent bounded continuous actions. Instead of sampling from an unbounded Gaussian and then applying $\tanh$, a Beta distribution directly lives on a bounded interval after a simple affine transformation. This can be attractive for actions such as normalized power, duty cycle, or allocation ratios. The trade-off is numerical care: when the shape parameters become too small or actions approach the boundary, gradients and log-probabilities can become delicate. For this reason, squashed Gaussians remain common in deep actor-critic implementations, while Beta policies are useful when bounded support is central to the problem design.

\subsection{Structured and hybrid policies}

Many systems require multiple decisions at once. In UAV network control, a high-level policy may choose a mode such as \emph{serve}, \emph{recharge}, or \emph{avoid congestion}, while a continuous policy chooses movement and bandwidth allocation. This gives a hybrid policy:
\begin{equation}
	\pi(a|s) = \pi(m|s) \pi(u|s,m),
\end{equation}
where $m$ is a discrete mode and $u$ is a continuous control vector.

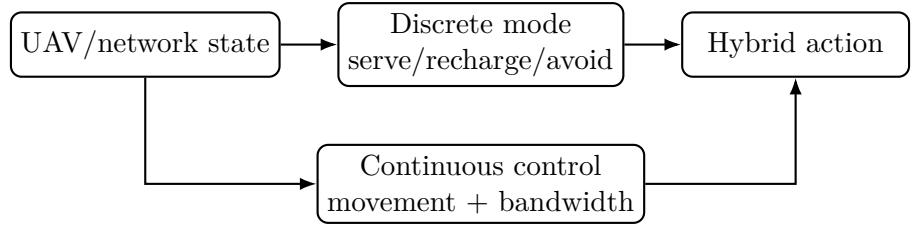
\begin{figure}[t]
	\centering
	\begin{tikzpicture}[
		box/.style={draw, rounded corners, thick, minimum width=3.0cm, minimum height=0.85cm, align=center},
		arrow/.style={-{Latex[length=2.2mm]}, thick},
		node distance=0.75cm
		]
		\node[box] (state) {UAV/network state};
		\node[box, right=of state] (mode) {Discrete mode\\serve/recharge/avoid};
		\node[box, below=of mode] (cont) {Continuous control\\movement + bandwidth};
		\node[box, right=of mode] (action) {Hybrid action};
		\draw[arrow] (state) -- (mode);
		\draw[arrow] (state) |- (cont);
		\draw[arrow] (mode) -- (action);
		\draw[arrow] (cont) -| (action);
	\end{tikzpicture}
	\caption{A structured policy can factorize a complex decision into a discrete mode and continuous parameters. This is often more natural for UAV/SDN systems than a flat discrete action space.}
	\label{fig:hybrid_policy}
\end{figure}

% ============================================================
\section{From policy gradients to modern algorithms}
% ============================================================

The policy-gradient theorem is the starting point, not the end. Modern algorithms add mechanisms to address high variance, unstable updates, sample inefficiency, and unsafe exploration.

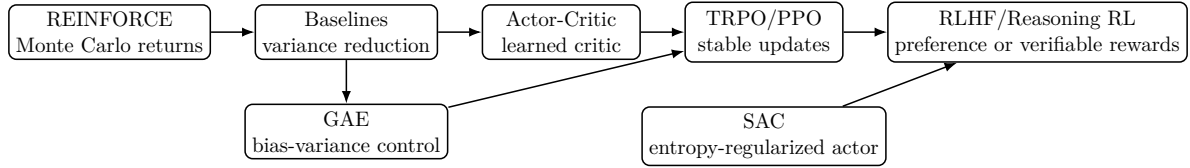
\begin{figure}[t]
	\centering
	\resizebox{0.98\textwidth}{!}{%
		\begin{tikzpicture}[
			box/.style={draw, rounded corners, thick, minimum width=2.9cm, minimum height=0.8cm, align=center},
			arrow/.style={-{Latex[length=2.3mm]}, thick},
			node distance=0.8cm
			]
			\node[box] (reinforce) {REINFORCE\\Monte Carlo returns};
			\node[box, right=of reinforce] (baseline) {Baselines\\variance reduction};
			\node[box, right=of baseline] (ac) {Actor-Critic\\learned critic};
			\node[box, below=of baseline] (gae) {GAE\\bias-variance control};
			\node[box, right=of ac] (trpo) {TRPO/PPO\\stable updates};
			\node[box, below=of trpo] (sac) {SAC\\entropy-regularized actor};
			\node[box, right=of trpo] (modern) {RLHF/Reasoning RL\\preference or verifiable rewards};
			
			\draw[arrow] (reinforce) -- (baseline);
			\draw[arrow] (baseline) -- (ac);
			\draw[arrow] (ac) -- (trpo);
			\draw[arrow] (baseline) -- (gae);
			\draw[arrow] (gae) -- (trpo);
			\draw[arrow] (trpo) -- (modern);
			\draw[arrow] (sac) -- (modern);
		\end{tikzpicture}%
	}
	\caption{Policy gradients form the foundation for many later algorithms. REINFORCE leads to baselines and actor-critic methods; trust-region and proximal methods stabilize policy updates; entropy-regularized actors support robust exploration; large-scale RLHF and reasoning RL optimize policies over token distributions.}
	\label{fig:policy_gradient_lineage}
\end{figure}

\subsection{REINFORCE}

REINFORCE uses Monte Carlo returns in the score-function estimator \citep{williams1992simple}. It is simple and unbiased, but high variance. Chapter 8 studies it in detail.

\subsection{Actor-critic methods}

Actor-critic methods combine a policy (actor) with a learned value estimator (critic). The critic reduces variance by estimating $V(s)$, $Q(s,a)$, or the advantage $A(s,a)$ \citep{konda2000actor,peters2008natural}.

\subsection{Trust-region and proximal methods}

Large policy updates can destroy performance. Natural-gradient policy optimization views the update through the geometry of the policy distribution rather than ordinary Euclidean parameter space, an idea developed in early policy-gradient work and later connected to trust-region methods \citep{kakade2001natural,schulman2015trpo}. TRPO constrains updates using a KL-divergence trust region \citep{schulman2015trpo}. PPO simplifies this idea using a clipped surrogate objective, becoming one of the most widely used policy-gradient algorithms \citep{schulman2017ppo}.

\subsection{Maximum-entropy actor learning}

SAC adds an entropy bonus to the objective, encouraging policies that achieve high reward while maintaining useful stochasticity \citep{haarnoja2018soft}. This is especially important in continuous control. Conceptually, maximum-entropy actor learning is the bridge between policy gradients and entropy-regularized control: the policy is not only optimized for reward, but also shaped to remain robust, exploratory, and less brittle under uncertainty. This idea is useful when connecting DRL to safety-aware control, including constrained UAV and network-control settings.

\subsection{Policy optimization in language models}

Modern language models are naturally policies over tokens. RLHF fine-tunes a language-model policy using a learned reward model and policy optimization, often based on PPO-style updates \citep{ouyang2022training}. PPO became attractive in this setting largely for practical reasons: it gives a simple way to limit policy drift from a supervised model while optimizing a learned reward, and it is easier to implement at scale than full trust-region optimization. Recent reasoning-oriented systems use reinforcement learning to improve multi-step problem solving and verifiable reasoning behavior \citep{jaech2024openai,shao2024deepseekmath}. Direct Preference Optimization (DPO) later showed that some preference-optimization objectives can bypass explicit RL while remaining connected to the same constrained policy-optimization view \citep{rafailov2023dpo}.

% ============================================================
\section{Research example: UAV/SDN control with continuous actions}
% ============================================================

To make the motivation concrete, consider a UAV-assisted SDN network. A UAV must choose movement and resource-control actions at each time step. A discrete DQN-style policy may choose among actions such as north, south, east, west, up, down, hover, or recharge. This is useful for early experiments, but it becomes limiting when the UAV must control continuous heading, speed, altitude, and bandwidth shares.

A direct policy can instead output a continuous action:
\begin{equation}
	a_t = [\Delta x_t, \Delta y_t, \Delta z_t, p_t, b^{\mathrm{URLLC}}_t, b^{\mathrm{eMBB}}_t, b^{\mathrm{mMTC}}_t],
\end{equation}
where $p_t$ is transmit power and $b_t$ terms are bandwidth shares. The policy can be factorized as
\begin{equation}
	\pi_\theta(a_t|s_t)
	= \pi_\theta(\Delta x_t,\Delta y_t,\Delta z_t,p_t|s_t)
	\pi_\theta(b_t|s_t).
\end{equation}
The movement/power part can be modeled by a squashed Gaussian, while the bandwidth vector can be modeled by a softmax or Dirichlet-like allocation.

The action-count difference becomes visible quickly. If heading is discretized into 5 bins, vertical motion into 3 bins, speed into 4 bins, and bandwidth policy into 4 templates, a DQN-style controller already has
\begin{equation}
	5 \times 3 \times 4 \times 4 = 240
\end{equation}
discrete action templates. Adding transmit power with only 5 levels increases this to $1200$ templates. A direct policy, by contrast, can output a compact continuous vector such as heading, climb rate, speed, bandwidth share, and transmit power.

\begin{table*}[t]
	\centering
	\caption{DQN-style discretization versus direct policy learning in UAV/SDN control.}
	\label{tab:uav_policy_vs_dqn}
	\begin{tabularx}{\textwidth}{p{3.2cm}YY}
		\toprule
		Decision & DQN-style value-based approach & Direct policy-gradient approach \\
		\midrule
		Movement & Choose among fixed directions. & Output continuous $\Delta x,\Delta y,\Delta z$. \\
		Power control & Discretize into power levels. & Output bounded continuous power. \\
		Bandwidth allocation & Enumerate allocation templates. & Output allocation fractions that sum to one. \\
		Exploration & Epsilon-greedy over templates. & Stochastic action distribution with learned variance. \\
		Safety & Add external safety filter after argmax. & Combine constrained parameterization with safety filter. \\
		Scalability & Action count grows combinatorially. & Action dimension grows linearly. \\
		\bottomrule
	\end{tabularx}
\end{table*}

\subsection{Worked numerical mini-example}

Suppose the UAV observes a 64-dimensional network state containing position, battery, traffic load, channel quality, and per-class QoS. A DQN-style policy with 1200 templates must output 1200 Q-values for every state, even though many templates differ only slightly or violate physical constraints. Adding one more control variable multiplies the number of templates again.

A policy-gradient actor can instead output seven interpretable numbers: three movement parameters, one power parameter, and three bandwidth logits. The output size grows with the number of control variables, not with the product of discretization bins. More importantly, the policy can be designed so that movement is bounded, power stays in a valid interval, and bandwidth shares sum to one. The network learns a distribution over feasible controls rather than scoring thousands of hand-designed templates.

\begin{researchbox}{Book-level lesson from the UAV example}
	The advantage of policy gradients is not only ``continuous actions.'' The deeper advantage is that the policy can encode structure: bounded movement, normalized bandwidth, stochastic exploration, safety-aware projection, and task modes. For real networked autonomy, this structure is often more important than the algorithm name.
\end{researchbox}

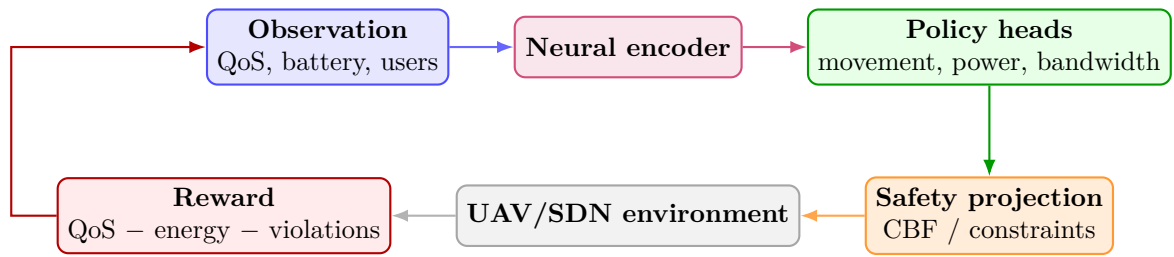
\begin{figure}[t]
	\centering
	\begin{tikzpicture}[
		box/.style={draw, rounded corners, thick, minimum width=3.0cm, minimum height=0.8cm, align=center, font=\small},
		arrow/.style={-{Latex[length=2.2mm]}, thick},
		node distance=0.85cm
		]
		\node[box, draw=blue!70, fill=blue!10, font=\small\bfseries] (obs) {Observation\\{\normalfont\small QoS, battery, users}};
		\node[box, draw=purple!70, fill=purple!10, font=\small\bfseries, right=of obs] (enc) {Neural encoder};
		\node[box, draw=green!60!black, fill=green!10, font=\small\bfseries, right=of enc] (heads) {Policy heads\\{\normalfont\small movement, power, bandwidth}};
		\node[box, draw=orange!80, fill=orange!15, font=\small\bfseries, below=1.2cm of heads] (safe) {Safety projection\\{\normalfont\small CBF / constraints}};
		\node[box, draw=gray!70, fill=gray!10, font=\small\bfseries, left=of safe] (env) {UAV/SDN environment};
		\node[box, draw=red!70!black, fill=red!8, font=\small\bfseries, left=of env] (reward) {Reward\\{\normalfont\small QoS $-$ energy $-$ violations}};
		
		\draw[arrow, color=blue!60] (obs) -- (enc);
		\draw[arrow, color=purple!70] (enc) -- (heads);
		\draw[arrow, color=green!60!black] (heads) -- (safe);
		\draw[arrow, color=orange!70] (safe) -- (env);
		\draw[arrow, color=gray!60] (env) -- (reward);
		\draw[arrow, color=red!70!black] (reward.west) -- ++(-0.6,0) |- (obs.west);
	\end{tikzpicture}
	\caption{A direct policy-gradient architecture for UAV/SDN control can use multiple policy heads and an explicit safety projection. This is a more natural design than enumerating a large discrete action set.}
	\label{fig:uav_policy_gradient_architecture}
\end{figure}

% ============================================================
\section{Python implementation blocks}
% ============================================================

The code in this section is not intended to replace a complete training library. Its purpose is to make the mathematical ideas executable and visible.

\subsection{Categorical policy for discrete actions}

\Needspace{16\baselineskip}
\begin{lstlisting}[style=pythonstyle,caption={A categorical neural policy for discrete actions.},label={lst:categorical_policy}]
import torch
import torch.nn as nn
from torch.distributions import Categorical

class CategoricalPolicy(nn.Module):
    def __init__(self, obs_dim: int, act_dim: int, hidden: int = 128):
        super().__init__()
        self.net = nn.Sequential(
            nn.Linear(obs_dim, hidden),
            nn.Tanh(),
            nn.Linear(hidden, hidden),
            nn.Tanh(),
            nn.Linear(hidden, act_dim),
        )

    def forward(self, obs: torch.Tensor) -> Categorical:
        logits = self.net(obs)
        return Categorical(logits=logits)

    def act(self, obs: torch.Tensor):
        dist = self.forward(obs)
        action = dist.sample()
        logp = dist.log_prob(action)
        entropy = dist.entropy()
        return action, logp, entropy

def policy_gradient_loss(log_probs, returns):
    # Gradient ascent on expected return is implemented as
    # gradient descent on the negative objective.
    return -(log_probs * returns).mean()
\end{lstlisting}

\subsection{Diagonal Gaussian policy for continuous control}

\Needspace{18\baselineskip}
\begin{lstlisting}[style=pythonstyle,caption={A diagonal Gaussian policy for continuous actions.},label={lst:gaussian_policy}]
import torch
import torch.nn as nn
from torch.distributions import Normal, Independent

class GaussianPolicy(nn.Module):
    def __init__(self, obs_dim: int, act_dim: int, hidden: int = 256):
        super().__init__()
        self.body = nn.Sequential(
            nn.Linear(obs_dim, hidden),
            nn.Tanh(),
            nn.Linear(hidden, hidden),
            nn.Tanh(),
        )
        self.mean_head = nn.Linear(hidden, act_dim)
        self.log_std = nn.Parameter(torch.zeros(act_dim))

    def forward(self, obs: torch.Tensor) -> Independent:
        h = self.body(obs)
        mean = self.mean_head(h)
        log_std = self.log_std.clamp(-20.0, 2.0)
        std = log_std.exp().expand_as(mean)
        return Independent(Normal(mean, std), 1)

    def act(self, obs: torch.Tensor):
        dist = self.forward(obs)
        action = dist.sample()
        logp = dist.log_prob(action)
        return action, logp
\end{lstlisting}

\subsection{Squashed Gaussian policy with log-probability correction}

\Needspace{22\baselineskip}
\begin{lstlisting}[style=pythonstyle,caption={Squashed Gaussian sampling with the tanh log-probability correction.},label={lst:squashed_gaussian}]
import torch
from torch.distributions import Normal, Independent

LOG_STD_MIN = -20.0
LOG_STD_MAX = 2.0
EPS = 1e-6

def sample_squashed_gaussian(mean, log_std):
    """Sample bounded actions in [-1, 1] and compute corrected log-prob.

    Common bug: computing log_prob after tanh without the correction.
    The correction is required because tanh changes the density.
    """
    log_std = torch.clamp(log_std, LOG_STD_MIN, LOG_STD_MAX)
    std = torch.exp(log_std)
    base = Independent(Normal(mean, std), 1)

    # Reparameterized sample: u = mean + std * noise.
    u = base.rsample()
    a = torch.tanh(u)

    # Change-of-variables correction.
    logp_u = base.log_prob(u)
    correction = torch.log(1.0 - a.pow(2) + EPS).sum(dim=-1)
    logp_a = logp_u - correction
    return a, logp_a
\end{lstlisting}

\subsection{Trajectory-based policy-gradient update}

\Needspace{20\baselineskip}
\begin{lstlisting}[style=pythonstyle,caption={Minimal trajectory-based policy-gradient update.},label={lst:trajectory_pg}]
def compute_discounted_returns(rewards, gamma: float):
    returns = []
    G = 0.0
    for r in reversed(rewards):
        G = r + gamma * G
        returns.append(G)
    returns.reverse()
    return torch.tensor(returns, dtype=torch.float32)

def reinforce_update(policy, optimizer, log_probs, rewards, gamma=0.99):
    returns = compute_discounted_returns(rewards, gamma)

    # Standardize returns to reduce scale-related instability.
    # This does not change the conceptual estimator, but it affects optimization.
    returns = (returns - returns.mean()) / (returns.std() + 1e-8)

    log_probs = torch.stack(log_probs)
    loss = -(log_probs * returns).sum()

    optimizer.zero_grad()
    loss.backward()
    torch.nn.utils.clip_grad_norm_(policy.parameters(), max_norm=1.0)
    optimizer.step()
    return float(loss.item())
\end{lstlisting}

\subsection{Actor update using a learned critic}

\Needspace{18\baselineskip}
\begin{lstlisting}[style=pythonstyle,caption={Actor update when an advantage estimate is available.},label={lst:advantage_actor_update}]
def actor_loss_from_advantages(dist, actions, advantages, entropy_coef=0.0):
    """Policy-gradient loss using advantage estimates.

    advantages should usually be detached from the actor update.
    Common bug: letting gradients flow into the critic through advantages.
    """
    logp = dist.log_prob(actions)
    entropy = dist.entropy().mean()
    pg_loss = -(logp * advantages.detach()).mean()
    return pg_loss - entropy_coef * entropy
\end{lstlisting}

\subsection{Deterministic actor update}

\Needspace{16\baselineskip}
\begin{lstlisting}[style=pythonstyle,caption={Deterministic policy-gradient actor update using a critic.},label={lst:dpg_actor_update}]
def deterministic_actor_loss(actor, critic, obs_batch):
    """Maximize Q(s, mu(s)) by minimizing its negative.

    critic(obs, action) should return Q-values. During the actor update,
    gradients flow through the action into the actor, but critic parameters
    are usually not updated by this loss.
    """
    action = actor(obs_batch)
    q_value = critic(obs_batch, action)
    return -q_value.mean()
\end{lstlisting}

\subsection{A UAV policy head with bounded movement and bandwidth allocation}

\Needspace{24\baselineskip}
\begin{lstlisting}[style=pythonstyle,caption={A structured UAV policy with bounded movement and normalized bandwidth allocation.},label={lst:uav_policy_head}]
class UAVPolicy(nn.Module):
    def __init__(self, obs_dim: int, hidden: int = 256):
        super().__init__()
        self.encoder = nn.Sequential(
            nn.Linear(obs_dim, hidden),
            nn.ReLU(),
            nn.Linear(hidden, hidden),
            nn.ReLU(),
        )

        # Movement: dx, dy, dz and power, bounded by tanh.
        self.move_mean = nn.Linear(hidden, 4)
        self.move_log_std = nn.Parameter(torch.full((4,), -0.5))

        # Bandwidth allocation among URLLC, eMBB, and mMTC.
        self.bandwidth_logits = nn.Linear(hidden, 3)

    def forward(self, obs):
        h = self.encoder(obs)
        mean = self.move_mean(h)
        log_std = self.move_log_std.expand_as(mean)

        # See Listing \ref{lst:squashed_gaussian}.
        bounded_move, logp_move = sample_squashed_gaussian(mean, log_std)

        # Softmax ensures bandwidth fractions sum to one.
        bandwidth = torch.softmax(self.bandwidth_logits(h), dim=-1)

        action = torch.cat([bounded_move, bandwidth], dim=-1)
        return action, logp_move
\end{lstlisting}

% ============================================================
\section{Practical failure modes and debugging rules}
% ============================================================

Policy-gradient code often runs without errors while learning nothing. Table~\ref{tab:pg_debugging} lists common issues.

\begin{table*}[t]
	\centering
	\caption{Common policy-gradient bugs and how to detect them.}
	\label{tab:pg_debugging}
	\begin{tabularx}{\textwidth}{p{3.4cm}YY}
		\toprule
		Bug & Symptom & Fix \\
		\midrule
		Wrong sign of loss & Policy gets worse while loss decreases. & Remember: gradient ascent on return is negative loss in PyTorch. \\
		Forgetting to detach advantages & Critic receives unintended actor gradients. & Use \texttt{advantages.detach()} in actor loss. \\
		No log-prob correction after $\tanh$ & Continuous policy unstable or entropy incorrect. & Apply change-of-variables correction. \\
		Unbounded standard deviation & Actions explode or entropy collapses. & Clamp log standard deviation. \\
		No entropy monitoring & Exploration silently disappears. & Track policy entropy during training. \\
		High-variance returns & Learning is noisy and seed-dependent. & Normalize returns; use baselines or advantages. \\
		Bad action scaling & Policy outputs valid numbers but environment interprets them wrongly. & Explicitly map $[-1,1]$ to physical bounds. \\
		Ignoring evaluation stochasticity & Evaluation results vary wildly. & Evaluate both stochastic and mean-action policies. \\
		\bottomrule
	\end{tabularx}
\end{table*}

\begin{warningbox}{Do not confuse differentiable policy with differentiable environment}
	Policy-gradient methods require gradients through the policy log-probability, not through the environment transition. This is why they can optimize policies in simulators, physical systems, network emulators, and language-model reward settings where the reward process is not differentiable.
\end{warningbox}

% ============================================================
\section{Limitations of direct policy optimization}
% ============================================================

Policy-gradient methods solve important problems, but they introduce new ones.

\begin{enumerate}[leftmargin=*]
	\item \textbf{High variance.} Monte Carlo policy-gradient estimators can require many samples.
	\item \textbf{On-policy data.} Many policy-gradient methods require data from the current policy, reducing sample efficiency.
	\item \textbf{Sensitivity to step size.} A policy update that is too large can destroy behavior.
	\item \textbf{Exploration collapse.} The policy can become deterministic too early.
	\item \textbf{Reward hacking.} Directly optimizing a policy can exploit misspecified rewards.
	\item \textbf{Safety constraints.} A policy distribution can sample unsafe actions unless constrained or filtered.
	\item \textbf{Credit assignment.} Long-horizon tasks still require good return or advantage estimation.
	\end{enumerate}

These limitations explain the evolution of later chapters. Chapter 8 introduces baselines and advantages to reduce variance. Chapter 9 introduces actor-critic learning. Chapter 10 studies PPO, which stabilizes policy updates. Chapter 11 studies SAC, which uses entropy regularization for robust continuous control.

% ============================================================
\section{Key takeaways}
% ============================================================

\begin{itemize}[leftmargin=*]
	\item Value-based methods derive behavior from a value function; policy-gradient methods optimize behavior directly.
	\item Direct policy learning is especially important for continuous, structured, stochastic, or constrained action spaces.
	\item The policy objective is the expected return under the trajectory distribution induced by the policy.
	\item The likelihood-ratio trick allows policy gradients without differentiating through the environment.
	\item The policy-gradient theorem expresses the gradient using a score term and an action-quality signal.
	\item Stochastic policies are natural for exploration, uncertainty, and language-model token distributions.
	\item Deterministic policy gradients are useful for high-dimensional continuous control but rely heavily on critic quality.
	\item Policy parameterization matters: categorical, Gaussian, squashed Gaussian, beta, Dirichlet, autoregressive, and hybrid policies encode different assumptions.
	\item In UAV/SDN systems, direct policies can output continuous movement, power, and bandwidth-allocation controls more naturally than DQN-style discrete action templates.
\end{itemize}

% ============================================================
\section{Exercises}
% ============================================================

\subsection*{Conceptual exercises}
\begin{enumerate}[leftmargin=*]
	\item Explain why a DQN-style Q-network becomes awkward for a five-dimensional continuous action space.
	\item What does it mean to learn the policy directly rather than derive it from a value function?
	\item Why can policy-gradient methods be used when the environment is not differentiable?
	\item Give one reason why stochastic policies can be useful even after training.
	\item Explain why a policy-gradient update can be interpreted as increasing the probability of good sampled actions.
\end{enumerate}

\subsection*{Mathematical exercises}
\begin{enumerate}[leftmargin=*]
	\item Derive $\nabla_\theta p_\theta(x)=p_\theta(x)\nabla_\theta\log p_\theta(x)$.
	\item Starting from a trajectory distribution, show why the environment transition terms disappear from $\nabla_\theta\log p_\theta(\tau)$.
	\item For a Gaussian policy $a \sim \mathcal{N}(\mu_\theta(s),\sigma^2)$, explain qualitatively how the gradient changes the mean when high actions receive high return.
	\item Suppose an action space has $d=6$ dimensions and each dimension is discretized into $K=10$ bins. How many discrete actions are produced?
	\item Write the deterministic policy-gradient update and explain the role of $\nabla_a Q(s,a)$.
	\item Show that subtracting a state-dependent baseline $b(s_t)$ from the return does not change the expected policy-gradient estimator:
	\[
	\mathbb{E}_{a_t\sim \pi_\theta(\cdot|s_t)}
	\left[\nabla_\theta \log \pi_\theta(a_t|s_t)b(s_t)\right] = 0.
	\]
	Why can this reduce variance?
\end{enumerate}

\subsection*{Coding exercises}
\begin{enumerate}[leftmargin=*]
	\item Implement a categorical policy and train it on a small bandit problem.
	\item Modify Listing~\ref{lst:gaussian_policy} so that the standard deviation is state-dependent.
	\item Implement the squashed Gaussian correction and verify that removing it changes the log-probability.
	\item Build a toy continuous-control environment where the best action is $a^*(s)=\sin(s)$ and train a Gaussian policy to imitate reward feedback.
	\item Extend Listing~\ref{lst:uav_policy_head} by adding a discrete mission-mode head.
\end{enumerate}

\subsection*{Research thinking exercises}
\begin{enumerate}[leftmargin=*]
	\item In a UAV network, when would a Gaussian movement policy be better than a discrete movement policy?
	\item How would you parameterize a policy whose bandwidth-allocation vector must sum to one?
	\item What safety constraints should be added before deploying a stochastic policy in a real UAV system?
	\item In language-model RLHF, why is the policy naturally a probability distribution?
	\item Explain one way policy optimization can cause reward hacking in networking or language-model settings.
\end{enumerate}

% ============================================================
\section*{Looking Ahead to Chapter 8: Food for Thought}
\addcontentsline{toc}{section}{Looking Ahead to Chapter 8: Food for Thought}
% ============================================================

Chapter 7 introduced the idea of optimizing the policy directly. The central equation was
\begin{equation}
	\nabla_\theta J(\theta)
	= \mathbb{E}\left[\nabla_\theta \log \pi_\theta(a|s)Q^\pi(s,a)\right].
\end{equation}
This equation is elegant, but it hides a practical problem: how do we estimate the quality term?

If we use the full Monte Carlo return, the estimate is unbiased but often very noisy. If we use a learned value function, the estimate may have lower variance but can introduce bias. If we subtract a baseline, the expected gradient remains correct, but the variance can drop dramatically.

\subsection*{Questions to ponder}
\begin{enumerate}[leftmargin=*]
	\item Why is the return from one sampled trajectory such a noisy training signal?
	\item How can subtracting a baseline reduce variance without changing the expected gradient?
	\item What is the difference between a return, a value, a Q-value, and an advantage?
	\item Why does the advantage $A(s,a)=Q(s,a)-V(s)$ better capture whether an action was better than expected?
	\item How does REINFORCE connect the likelihood-ratio trick to a complete algorithm?
\end{enumerate}

\begin{quote}
	Chapter 7 explained why learning the policy directly is necessary. Chapter 8 explains how to make the simplest policy-gradient estimator usable in practice.
\end{quote}

% ============================================================
% \section{Chapter references}
% ============================================================
	\chapter{REINFORCE, Baselines, and Advantage Functions}
\label{ch:reinforce_baselines}
\chaptermark{REINFORCE, Baselines, and Advantages}

\begin{keybox}{Chapter goal}
	Chapter 7 explained why one might learn a policy directly. This chapter explains the first practical answer: the REINFORCE estimator, together with the variance-reduction ideas that make policy-gradient learning usable. The main story is not that REINFORCE is the best modern algorithm. It is that REINFORCE exposes the central difficulty of policy gradients: the gradient is easy to write down, but hard to estimate with low variance. Baselines, reward-to-go, advantage functions, and generalized advantage estimation are the conceptual bridge from simple policy-gradient learning to actor-critic methods, PPO, SAC, RLHF, and modern reasoning-oriented reinforcement learning.
\end{keybox}

\section*{Chapter Overview}
\addcontentsline{toc}{section}{Chapter Overview}

\begin{enumerate}[leftmargin=*]
	\item The problem left open by Chapter 7
	\item REINFORCE as Monte Carlo policy gradient
	\item Full-return REINFORCE and the credit-assignment problem
	\item The causality trick and reward-to-go
	\item Baselines as control variates
	\item Why a state-dependent baseline is unbiased
	\item Advantage functions
	\item TD errors, n-step advantages, and GAE
	\item Normalization, whitening, and practical scaling
	\item Entropy regularization and exploration
	\item Python implementation: from REINFORCE to GAE
	\item UAV/SDN worked example: advantage as service-quality surprise
	\item Modern extension: REINFORCE-style learning for LLMs
	\item Common implementation bugs
	\item Exercises
	\item Looking Ahead to Chapter 9
\end{enumerate}

\section{The problem left open by Chapter 7}

Chapter 7 introduced the policy-gradient theorem. In its most recognizable form, it states that a policy can be improved by moving its parameters in the direction
\begin{equation}
	\grad_\theta J(\theta)
	= \E_{s \sim d^{\pi_\theta},\, a \sim \pi_\theta}
	\left[\grad_\theta \log \pi_\theta(a \given s) Q^{\pi_\theta}(s,a)\right].
	\label{eq:pg-theorem-ch8}
\end{equation}
This formula is powerful because it does not require differentiating through the environment dynamics. The transition probabilities determine which trajectories are sampled, but the gradient with respect to policy parameters appears only through the log probability of the policy. This is the mathematical reason policy-gradient methods can be used with black-box simulators, physical robots, packet-level network simulators, wireless channels, and human-preference reward models \citep{williams1992simple,sutton1999policy,sutton2018reinforcement}.

However, Equation~\eqref{eq:pg-theorem-ch8} hides an immediate practical problem. The true action-value function $Q^{\pi_\theta}(s,a)$ is not known. The agent must estimate it from sampled experience. The simplest possible estimate is the return observed after taking an action. This gives REINFORCE \citep{williams1992simple}.

\begin{figure}[t]
	\centering
	\begin{tikzpicture}[
		box/.style={draw, rounded corners, thick, minimum width=2.6cm, minimum height=0.85cm, align=center},
		arrow/.style={-{Latex[length=2.5mm]}, thick},
		node distance=0.65cm
		]
		\node[box] (rollout) {Collect\\trajectory};
		\node[box, right=of rollout] (return) {Compute\\returns};
		\node[box, right=of return] (baseline) {Subtract\\baseline};
		\node[box, right=of baseline] (loss) {Policy-gradient\\loss};
		\node[box, right=of loss] (update) {Update\\policy};
		\draw[arrow] (rollout) -- (return);
		\draw[arrow] (return) -- node[above,font=\scriptsize] {high variance} (baseline);
		\draw[arrow] (baseline) -- node[above,font=\scriptsize] {advantage} (loss);
		\draw[arrow] (loss) -- (update);
		\node[below=0.65cm of baseline, align=center, font=\small] {Chapter 8 explains the middle of this pipeline:\\how to transform raw returns into useful learning signals.};
	\end{tikzpicture}
	\caption{The policy-gradient pipeline. REINFORCE begins with sampled trajectories and uses returns as learning signals. Baselines and advantage estimators transform those raw returns into lower-variance signals.}
	\label{fig:ch8_pipeline}
\end{figure}

The central question of this chapter is therefore:
\begin{quote}
	How can we estimate the policy gradient from trajectories without drowning the learning signal in variance?
\end{quote}

This chapter answers that question in stages. First, we derive REINFORCE. Then we show why using the full episode return gives a noisy credit-assignment signal. Next, we introduce reward-to-go and baselines. Finally, we define advantage functions and generalized advantage estimation, the estimator that became one of the practical foundations of modern actor-critic algorithms and PPO \citep{schulman2015gae,schulman2017ppo}.

\section{REINFORCE as Monte Carlo policy gradient}

Consider an episodic task. A trajectory is
\begin{equation}
	\tau = (S_0,A_0,R_1,S_1,A_1,R_2,\ldots,S_T).
\end{equation}
The trajectory probability under policy $\pi_\theta$ is
\begin{equation}
	p_\theta(\tau)
	= p(S_0)\prod_{t=0}^{T-1}
	\pi_\theta(A_t \given S_t)
	p(S_{t+1},R_{t+1} \given S_t,A_t).
\end{equation}
The objective is the expected return
\begin{equation}
	J(\theta) = \E_{\tau \sim p_\theta}[G_0].
\end{equation}
Using the likelihood-ratio identity,
\begin{equation}
	\grad_\theta p_\theta(\tau)
	= p_\theta(\tau)\grad_\theta \log p_\theta(\tau),
\end{equation}
we obtain
\begin{align}
	\grad_\theta J(\theta)
	&= \grad_\theta \int p_\theta(\tau) G_0(\tau)\,d\tau \\
	&= \int p_\theta(\tau) \grad_\theta \log p_\theta(\tau) G_0(\tau)\,d\tau \\
	&= \E_{\tau \sim p_\theta}
	\left[\grad_\theta \log p_\theta(\tau) G_0\right].
\end{align}
Because the environment dynamics do not depend on $\theta$, the log trajectory probability contains policy terms plus environment terms, but only the policy terms have nonzero gradient:
\begin{equation}
	\grad_\theta \log p_\theta(\tau)
	= \sum_{t=0}^{T-1} \grad_\theta \log \pi_\theta(A_t \given S_t).
\end{equation}
Therefore,
\begin{equation}
	\grad_\theta J(\theta)
	= \E_{\tau \sim p_\theta}
	\left[\sum_{t=0}^{T-1}
	\grad_\theta \log \pi_\theta(A_t \given S_t)G_0\right].
	\label{eq:reinforce-full-return}
\end{equation}
Equation~\eqref{eq:reinforce-full-return} is the full-return REINFORCE estimator. It says that every action in the trajectory is reinforced in proportion to the same episode return.

\begin{keybox}{Key idea: score-function estimation}
	The return $G_0$ is treated as a sampled scalar weight. The gradient is taken through $\log \pi_\theta(A_t\given S_t)$, not through the environment or the reward. This is why REINFORCE can work when the reward is non-differentiable, delayed, or produced by a simulator.
\end{keybox}

\section{Full-return REINFORCE and the credit-assignment problem}

Full-return REINFORCE is unbiased under standard sampling assumptions, but it can have very high variance. The reason is easy to see: every action in the episode is multiplied by the same return $G_0$.

Suppose a robot takes ten actions before reaching a goal. If the final outcome is good, full-return REINFORCE pushes up the probability of all ten actions. But not all ten actions were necessarily useful. Some may have been irrelevant, and one may even have been harmful but compensated for later. The return tells us that the trajectory was good, but not precisely which action was responsible.

This is a credit-assignment problem. The agent must decide which earlier decisions deserve credit or blame for later rewards.

\begin{figure}[t]
	\centering
	\begin{tikzpicture}[
		state/.style={circle,draw,thick,minimum size=0.72cm},
		reward/.style={rectangle,draw,rounded corners,minimum width=0.9cm,minimum height=0.55cm,align=center},
		arrow/.style={-{Latex[length=2.2mm]},thick},
		node distance=0.9cm
		]
		\node[state] (s0) {$S_0$};
		\node[state, right=of s0] (s1) {$S_1$};
		\node[state, right=of s1] (s2) {$S_2$};
		\node[state, right=of s2] (s3) {$S_3$};
		\node[state, right=of s3] (s4) {$S_T$};
		\draw[arrow] (s0) -- node[below] {$A_0$} (s1);
		\draw[arrow] (s1) -- node[below] {$A_1$} (s2);
		\draw[arrow] (s2) -- node[below] {$A_2$} (s3);
		\draw[arrow] (s3) -- node[below] {$\cdots$} (s4);
		\node[reward, above=0.75cm of s4] (r) {$G_0$};
		\draw[arrow] (r) -- (s0);
		\draw[arrow] (r) -- (s1);
		\draw[arrow] (r) -- (s2);
		\draw[arrow] (r) -- (s3);
		\node[below=1.0cm of s2, align=center,font=\small] {Full-return REINFORCE assigns the same episode-level signal to every action.\\This is unbiased, but often noisy.};
	\end{tikzpicture}
	\caption{Full-return credit assignment. A single trajectory return gives a coarse learning signal to all actions in the episode, even though some actions may be more responsible than others.}
	\label{fig:full-return-credit}
\end{figure}

The variance of REINFORCE is not a minor implementation inconvenience. It is the reason simple policy-gradient methods often need many samples. Variance reduction is therefore not an optional detail; it is the main bridge from the elegant policy-gradient theorem to practical deep reinforcement learning \citep{greensmith2004variance,schulman2015gae}.

\section{The causality trick and reward-to-go}

The first variance-reduction improvement is based on causality. An action at time \(t\)
cannot influence rewards that occurred before time \(t\). Therefore, when updating the
policy for action \(A_t\), we do not need the full episode return \(G_0\). We can use the
reward-to-go:
\begin{equation}
	G_t
	=
	\sum_{k=t}^{T-1}
	\gamma^{k-t} R_{k+1}.
\end{equation}
The reward-to-go estimator is
\begin{equation}
	\nabla_\theta J(\theta)
	=
	\mathbb{E}
	\left[
	\sum_{t=0}^{T-1}
	\nabla_\theta \log \pi_\theta(A_t \mid S_t)G_t
	\right].
	\label{eq:reward-to-go-pg}
\end{equation}
This estimator keeps the gradient unbiased while removing irrelevant past rewards from
the learning signal.

This argument is closely related to the baseline identity introduced in the next section.
In both cases, the key fact is that the expected score function is zero:
\begin{equation}
	\mathbb{E}_{a\sim \pi_\theta(\cdot\mid s)}
	\left[
	\nabla_\theta \log \pi_\theta(a\mid s)
	\right]
	=
	0.
	\label{eq:score_function_zero}
\end{equation}
Therefore, terms that do not depend on the sampled action at the current decision point
can be added or removed without changing the expected policy gradient. The causality
trick removes irrelevant past rewards; a baseline removes a reference value. Both reduce
variance while preserving the expected gradient.

\begin{figure}[t]
	\centering
	\begin{tikzpicture}[
		time/.style={circle,draw,thick,minimum size=0.7cm},
		reward/.style={rectangle,draw,rounded corners,minimum width=0.7cm,minimum height=0.5cm},
		arrow/.style={-{Latex[length=2mm]},thick},
		node distance=0.85cm
		]
		\node[time] (s0) {$S_0$};
		\node[time, right=of s0] (s1) {$S_1$};
		\node[time, right=of s1] (s2) {$S_2$};
		\node[time, right=of s2] (s3) {$S_3$};
		\node[time, right=of s3] (s4) {$S_4$};
		\foreach \i/\j in {s0/s1,s1/s2,s2/s3,s3/s4}{\draw[arrow] (\i) -- (\j);}
		\node[reward, above=0.7cm of $(s0)!0.5!(s1)$] (r1) {$R_1$};
		\node[reward, above=0.7cm of $(s1)!0.5!(s2)$] (r2) {$R_2$};
		\node[reward, above=0.7cm of $(s2)!0.5!(s3)$] (r3) {$R_3$};
		\node[reward, above=0.7cm of $(s3)!0.5!(s4)$] (r4) {$R_4$};
		\draw[decorate,decoration={brace,amplitude=5pt},thick] ($(r3.north west)+(0,0.25)$) -- node[above=5pt] {$G_2=R_3+\gamma R_4+\cdots$} ($(r4.north east)+(0,0.25)$);
		\node[below=0.9cm of s2, align=center,font=\small] {For action $A_2$, rewards before $S_2$ are removed.\\Only future rewards remain in the learning signal.};
	\end{tikzpicture}
	\caption{The causality trick. The action at time $t$ can affect future rewards but not past rewards. Reward-to-go therefore reduces variance without changing the expected gradient.}
	\label{fig:reward-to-go}
\end{figure}
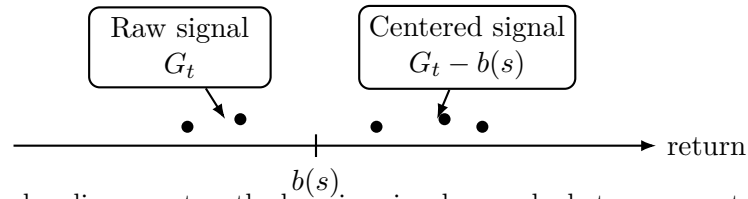

Reward-to-go is often the first practical improvement over the simplest REINFORCE estimator. However, it does not solve the main variance problem. Future returns can still be large, noisy, and state-dependent. A good outcome from a difficult state and a good outcome from an easy state should not produce the same policy update. This motivates baselines.

\section{Baselines as control variates}

A baseline is a quantity subtracted from the return before multiplying by the score function. The common baseline is a state-value estimate $b(S_t)$:
\begin{equation}
	\grad_\theta J(\theta)
	= \E\left[\sum_{t=0}^{T-1}
	\grad_\theta \log \pi_\theta(A_t\given S_t)
	\left(G_t - b(S_t)\right)\right].
	\label{eq:pg-baseline}
\end{equation}
The baseline does not change the expected gradient if it does not depend on the sampled action. But it can greatly reduce variance.

The intuitive reason is simple. The raw return answers: ``How good was the outcome?'' The baseline-adjusted return answers: ``How good was the outcome compared with what was expected from this state?''

\begin{figure}[t]
	\centering
	\begin{tikzpicture}[
		axisline/.style={thick,-{Latex[length=2mm]}},
		dot/.style={circle,fill,inner sep=1.6pt},
		box/.style={draw,rounded corners,thick,minimum width=2.4cm,minimum height=0.7cm,align=center}
		]
		\draw[axisline] (0,0) -- (8.5,0) node[right] {return};
		\draw[thick] (4,0.15) -- (4,-0.15) node[below] {$b(s)$};
		\node[dot] at (2.3,0.25) {};
		\node[dot] at (3.0,0.35) {};
		\node[dot] at (4.8,0.25) {};
		\node[dot] at (5.7,0.35) {};
		\node[dot] at (6.2,0.25) {};
		\node[box] (raw) at (2.2,1.3) {Raw signal\\$G_t$};
		\node[box] (centered) at (6.0,1.3) {Centered signal\\$G_t-b(s)$};
		\draw[-{Latex[length=2mm]},thick] (raw) -- (2.8,0.35);
		\draw[-{Latex[length=2mm]},thick] (centered) -- (5.6,0.35);
		\node[align=center,font=\small] at (4,-1.0) {Subtracting a baseline recenters the learning signal around what was expected.\\Positive values mean better than expected; negative values mean worse than expected.};
	\end{tikzpicture}
	\caption{A baseline acts like a control variate. It can center the return distribution and reduce gradient variance without changing the expected gradient.}
	\label{fig:baseline-centering}
\end{figure}

This is the same conceptual move used in statistics under the name \emph{control variates}. Greensmith, Bartlett, and Baxter analyzed variance-reduction techniques for policy-gradient estimates and showed that the choice of baseline can strongly affect estimator variance \citep{greensmith2004variance}. In deep RL practice, the baseline is usually learned as a value function.

\section{Why a state-dependent baseline is unbiased}

The key identity is
\begin{equation}
	\E_{a \sim \pi_\theta(\cdot\given s)}
	\left[\grad_\theta \log \pi_\theta(a\given s)b(s)\right] = 0.
\end{equation}
To see why, move the baseline outside the expectation over actions, because it does not depend on $a$:
\begin{align}
	\sum_a \pi_\theta(a\given s)\grad_\theta \log \pi_\theta(a\given s)b(s)
	&= b(s) \sum_a \pi_\theta(a\given s)
	\frac{\grad_\theta \pi_\theta(a\given s)}{\pi_\theta(a\given s)} \\
	&= b(s) \sum_a \grad_\theta \pi_\theta(a\given s) \\
	&= b(s) \grad_\theta \sum_a \pi_\theta(a\given s) \\
	&= b(s) \grad_\theta 1 \\
	&= 0.
\end{align}
For continuous actions, the sum becomes an integral and the same argument holds under standard regularity assumptions.

This is the mathematical justification for subtracting a value function from the return. The expected gradient is unchanged, but the variance can be much lower. The baseline is therefore not a reward-shaping term. It should not change what the agent is optimizing. It changes how noisy the gradient estimate is.

\begin{warningbox}{Common confusion}
	A baseline is not a penalty and not an extra reward. If it is action-independent, it does not change the expected policy gradient. It only changes the variance of the estimator. This distinction is crucial in implementation: the policy loss uses the advantage as a weight, while the value loss trains the baseline separately.
\end{warningbox}

\section{Advantage functions}

The most important baseline is the state-value function:
\begin{equation}
	b(s) = V^\pi(s).
\end{equation}
Subtracting this baseline from the action value gives the advantage function:
\begin{equation}
	A^\pi(s,a) = Q^\pi(s,a) - V^\pi(s).
\end{equation}
The advantage measures whether action $a$ was better or worse than the typical action chosen by the policy in state $s$.

If $A^\pi(s,a)>0$, the action was better than expected. The policy should increase its probability.

If $A^\pi(s,a)<0$, the action was worse than expected. The policy should decrease its probability.

If $A^\pi(s,a)\approx 0$, the action was about average. The policy update should be small.

\begin{figure}[t]
	\centering
	\begin{tikzpicture}[
		bar/.style={draw,thick,minimum width=0.8cm},
		axis/.style={thick,-{Latex[length=2mm]}},
		lab/.style={font=\small,align=center}
		]
		\draw[axis] (0,0) -- (8,0) node[right] {actions};
		\draw[axis] (0,-1.8) -- (0,2.2) node[above] {value};
		\draw[thick,dashed] (0,0.7) -- (7.4,0.7) node[right] {$V^\pi(s)$};
		\draw[fill=gray!20] (1,0) rectangle (1.8,1.5);
		\draw[fill=gray!20] (2.4,0) rectangle (3.2,0.2);
		\draw[fill=gray!20] (3.8,0) rectangle (4.6,1.0);
		\draw[fill=gray!20] (5.2,0) rectangle (6.0,-0.6);
		\node[lab] at (1.4,-0.35) {$a_1$};
		\node[lab] at (2.8,-0.35) {$a_2$};
		\node[lab] at (4.2,-0.35) {$a_3$};
		\node[lab] at (5.6,-0.35) {$a_4$};
		\draw[{Latex[length=2mm]}-{Latex[length=2mm]},thick,blue!70!black] (1.95,0.7) -- (1.95,1.5) node[midway,right,font=\scriptsize] {$A>0$};
		\draw[{Latex[length=2mm]}-{Latex[length=2mm]},thick,red!70!black] (6.15,-0.6) -- (6.15,0.7) node[midway,right,font=\scriptsize] {$A<0$};
		\node[align=center,font=\small] at (4,-1.35) {Advantage is not absolute goodness. It is goodness relative to the state baseline.};
	\end{tikzpicture}
	\caption{Advantage as relative action quality. The baseline $V^\pi(s)$ represents the expected value of the state. The advantage tells whether a particular action was better or worse than that expectation.}
	\label{fig:advantage-relative}
\end{figure}
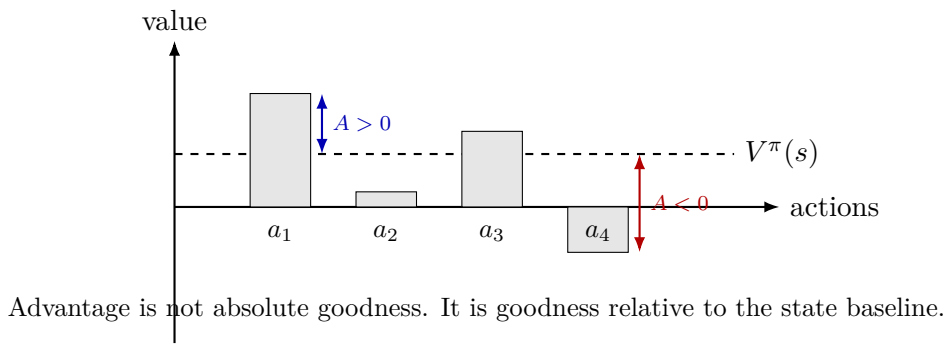

Using advantages, the policy-gradient theorem becomes
\begin{equation}
	\grad_\theta J(\theta)
	= \E\left[\grad_\theta \log \pi_\theta(A_t\given S_t) A^{\pi_\theta}(S_t,A_t)\right].
	\label{eq:pg-advantage}
\end{equation}
This is the form used in most modern policy-gradient and actor-critic methods.

\section{TD errors, n-step advantages, and GAE}

The advantage $A^\pi(s,a)$ is not known exactly. We must estimate it. Several estimators are common.

\subsection{Monte Carlo advantage}

The simplest advantage estimate is
\begin{equation}
	\hat{A}_t^{\text{MC}} = G_t - V_\phi(S_t),
\end{equation}
where $V_\phi$ is a learned value baseline. This estimator has low bias if the return is sampled correctly, but high variance.

\subsection{One-step TD advantage}

A lower-variance estimator uses a one-step temporal-difference error:
\begin{equation}
	\delta_t = R_{t+1} + \gamma V_\phi(S_{t+1}) - V_\phi(S_t).
\end{equation}
This can be interpreted as a one-step estimate of the advantage. It has lower variance but can be biased if the value function is inaccurate.

\subsection{n-step advantage}

An $n$-step estimator uses several real rewards before bootstrapping:
\begin{equation}
	\hat{A}_t^{(n)} =
	\sum_{l=0}^{n-1}\gamma^l R_{t+l+1}
	+ \gamma^n V_\phi(S_{t+n}) - V_\phi(S_t).
\end{equation}
Larger $n$ reduces bootstrap bias but increases variance. Smaller $n$ increases bootstrap dependence but usually reduces variance.

\subsection{Generalized Advantage Estimation}

Generalized Advantage Estimation (GAE) combines multi-step TD errors using an
exponentially weighted sum \citep{schulman2015gae}. Recall that the one-step TD
residual is
\begin{equation}
	\delta_t
	=
	r_t
	+
	\gamma (1-d_{t+1}) V_\phi(s_{t+1})
	-
	V_\phi(s_t),
	\label{eq:td_residual_gae}
\end{equation}
where \(d_{t+1}=1\) if \(s_{t+1}\) is terminal and \(d_{t+1}=0\) otherwise. GAE is then
defined as
\begin{equation}
	\hat{A}_t^{\mathrm{GAE}(\gamma,\lambda)}
	=
	\sum_{l=0}^{\infty}(\gamma\lambda)^l \delta_{t+l}.
	\label{eq:gae}
\end{equation}

The parameter \(\lambda \in [0,1]\) controls the bias--variance trade-off. If
\(\lambda\) is close to \(0\), the estimator behaves like a one-step TD advantage.
If \(\lambda\) is close to \(1\), it behaves more like a Monte Carlo estimator.

In implementation, GAE is usually not computed by explicitly evaluating the full
sum in Eq.~\eqref{eq:gae}. Instead, it is computed by a single backward recursion
through the trajectory:
\begin{equation}
	\hat{A}_t
	=
	\delta_t
	+
	\gamma\lambda (1-d_{t+1}) \hat{A}_{t+1}.
	\label{eq:gae_recursion}
\end{equation}
If the next state is terminal, then \(d_{t+1}=1\), so the recursion stops and no
value estimate is bootstrapped across the episode boundary. This recursive form is
one reason GAE is computationally cheap: after the value predictions are available,
all advantages can be computed in one backward pass.

\begin{table}[t]
	\centering
	\caption{Qualitative bias--variance behavior of GAE as \(\lambda\) changes.}
	\label{tab:gae_bias_variance}
	\begin{tabular}{lll}
		\toprule
		\(\lambda\) & Variance & Bias \\
		\midrule
		\(0.0\) & Low & High; close to one-step TD \\
		\(0.5\) & Medium & Medium \\
		\(0.95\) & Medium-high & Low; common practical choice \\
		\(1.0\) & High & Low; close to Monte Carlo \\
		\bottomrule
	\end{tabular}
\end{table}

Table~\ref{tab:gae_bias_variance} should be read qualitatively. The exact bias and
variance depend on the environment, the quality of the value function, the trajectory
length, and the reward scale.

\begin{figure}[t]
	\centering
	\begin{tikzpicture}[
		est/.style={draw,rounded corners,thick,minimum width=2.3cm,minimum height=0.75cm,align=center},
		arrow/.style={-{Latex[length=2.3mm]},thick},
		node distance=0.6cm
		]
		\node[est] (td) {TD(0)\\low variance\\more bias};
		\node[est, right=of td] (nstep) {$n$-step\\middle ground};
		\node[est, right=of nstep] (gae) {GAE\\weighted mixture};
		\node[est, right=of gae] (mc) {Monte Carlo\\low bias\\high variance};
		\draw[arrow] (td) -- (nstep);
		\draw[arrow] (nstep) -- (gae);
		\draw[arrow] (gae) -- (mc);
		\node[below=0.8cm of gae, align=center,font=\small] {$\lambda$ moves GAE along the bias-variance spectrum.\\In practice, values such as $\lambda=0.95$ are common starting points.};
	\end{tikzpicture}
	\caption{Advantage estimators trade bias against variance. GAE provides a smooth interpolation between one-step TD and Monte Carlo style estimation.}
	\label{fig:gae-bias-variance}
\end{figure}

GAE is one of the most important estimators in modern DRL. It is used in many actor-critic and PPO implementations because it often gives a strong practical balance between stability and sample efficiency \citep{schulman2015gae,schulman2017ppo}.

\section{Normalization, whitening, and practical scaling}

In theory, subtracting a baseline is enough to preserve the expected gradient. In practice, policy-gradient training is sensitive to the scale of the advantage. If advantages are extremely large, the policy update can become too aggressive. If they are tiny, the update can vanish.

A common practical trick is advantage normalization:
\begin{equation}
	\tilde{A}_t
	=
	\frac{\hat{A}_t - \mu_A}{\sigma_A + \epsilon},
	\label{eq:advantage_normalization}
\end{equation}
where \(\mu_A\) and \(\sigma_A\) are computed over the batch. This operation changes the exact gradient direction, so it is not a purely theoretical identity. But it often improves optimization in deep learning implementations.

A practical warning is that advantage normalization depends on the batch used to compute the mean and standard deviation. When advantage normalization is applied per minibatch, as in many PPO-style implementations, the minibatch statistics can themselves be noisy for small minibatches. This is one reason large-rollout or large-batch PPO often trains more stably: the advantage statistics are estimated from a broader sample of trajectories before the data are split into optimization minibatches.

For language-model RL, group-normalized rewards or advantages are common. In GRPO-style methods, multiple completions are sampled for the same prompt, and rewards are normalized within the group \citep{shao2024deepseekmath,deepseek2025r1}. This creates a prompt-local baseline: a completion is rewarded if it is better than other completions for the same prompt, not merely if its absolute reward is high.

\begin{warningbox}{Practical warning}
	Advantage normalization is useful, but it can hide reward-scale problems. If rewards are badly designed, normalizing the advantages may make training appear stable while the learned behavior optimizes the wrong objective. Always inspect raw rewards, raw returns, value predictions, advantages, and task metrics separately.
\end{warningbox}

\section{Entropy regularization and exploration}

REINFORCE can prematurely collapse to nearly deterministic behavior. This means that the policy may assign very high probability to a small set of actions before it has explored enough alternatives. Once this happens, learning can become brittle: the agent keeps reinforcing early choices, even if better actions exist.

To encourage exploration, many policy-gradient implementations add an entropy bonus:
\begin{equation}
	J_{\mathrm{entropy}}(\theta)
	=
	J(\theta)
	+
	\beta
	\mathbb{E}_{s\sim d^\pi}
	\left[
	\mathcal{H}(\pi_\theta(\cdot \mid s))
	\right],
	\label{eq:entropy_regularized_objective}
\end{equation}
where the entropy of a discrete policy is
\begin{equation}
	\mathcal{H}(\pi_\theta(\cdot \mid s))
	=
	-
	\sum_a
	\pi_\theta(a \mid s)
	\log \pi_\theta(a \mid s).
	\label{eq:discrete_policy_entropy}
\end{equation}

The entropy coefficient \(\beta\) controls how strongly stochasticity is encouraged. A larger \(\beta\) keeps the policy more exploratory, while a smaller \(\beta\) allows the policy to become more deterministic. In practice, \(\beta\) must be tuned carefully. If it is too large, the policy may remain too random and fail to exploit good actions. If it is too small, the policy may collapse early and stop exploring.

For continuous policies, the same idea applies, but the entropy is defined over a probability density rather than a probability mass function. For example, Gaussian policies have analytic entropy, which makes entropy regularization easy to implement. This is one reason Gaussian policies are widely used in continuous-control policy-gradient methods.

Entropy regularization is a bridge to later algorithms. In PPO, entropy is often used as an auxiliary exploration bonus. In maximum-entropy RL and SAC, entropy becomes part of the main objective rather than only a helper term. This changes the interpretation of the optimal policy: the agent is encouraged not only to achieve high reward, but also to preserve useful uncertainty in its actions \citep{haarnoja2018soft}. This idea is important in continuous control and can also be connected to robust or safety-aware control, where overly deterministic behavior may be undesirable under uncertainty.

In UAV or SDN control, this point is practical. If a policy becomes deterministic too early, it may always choose the same routing split, UAV direction, or bandwidth allocation, even when the network state changes. Entropy regularization keeps alternative actions alive long enough for the agent to discover better strategies under varying traffic load, channel quality, and safety constraints.

\section{Python implementation: from REINFORCE to GAE}

This section gives compact PyTorch code for the main ideas. The code is intentionally minimal. Production training loops require logging, seeding, vectorized environments, checkpointing, numerical checks, and evaluation without exploration noise.

\subsection{Reward-to-go and normalization}

\Needspace{16\baselineskip}
\begin{lstlisting}[style=pythonstyle,caption={Computing reward-to-go and normalized advantages.},label={lst:returns}]
import torch


def reward_to_go(rewards, gamma=0.99):
    """Compute discounted reward-to-go for one episode.

    Args:
        rewards: list or 1D tensor [T]
        gamma: discount factor

    Returns:
        Tensor [T], where out[t] = r_t + gamma r_{t+1} + ...
    """
    returns = []
    running = 0.0
    for r in reversed(rewards):
        running = float(r) + gamma * running
        returns.append(running)
    returns.reverse()
    return torch.tensor(returns, dtype=torch.float32)


def normalize(x, eps=1e-8):
    """Normalize a batch of returns or advantages."""
    return (x - x.mean()) / (x.std(unbiased=False) + eps)


# Example
rewards = [0.0, 0.0, 1.0, 2.0]
G = reward_to_go(rewards, gamma=0.9)
A = normalize(G)
print(G)  # tensor([2.2680, 2.5200, 2.8000, 2.0000])
print(A)
\end{lstlisting}

\subsection{Categorical policy for discrete actions}

\Needspace{18\baselineskip}
\begin{lstlisting}[style=pythonstyle,caption={A minimal categorical policy network.},label={lst:categorical-policy}]
import torch
import torch.nn as nn
from torch.distributions import Categorical


class CategoricalPolicy(nn.Module):
    def __init__(self, obs_dim, act_dim, hidden=128):
        super().__init__()
        self.net = nn.Sequential(
            nn.Linear(obs_dim, hidden),
            nn.Tanh(),
            nn.Linear(hidden, hidden),
            nn.Tanh(),
            nn.Linear(hidden, act_dim),
        )

    def forward(self, obs):
        logits = self.net(obs)
        return Categorical(logits=logits)

    def act(self, obs):
        dist = self.forward(obs)
        action = dist.sample()
        logp = dist.log_prob(action)
        entropy = dist.entropy()
        return action, logp, entropy
\end{lstlisting}

\subsection{REINFORCE update without a baseline}

\Needspace{18\baselineskip}
\begin{lstlisting}[style=pythonstyle,caption={REINFORCE policy update using reward-to-go.},label={lst:reinforce-update}]
def reinforce_update(policy, optimizer, log_probs, rewards, gamma=0.99):
    """One REINFORCE update for a single episode.

    log_probs: list of log pi(a_t|s_t), each a scalar tensor
    rewards: list of rewards collected after each action
    """
    returns = reward_to_go(rewards, gamma=gamma)
    returns = normalize(returns)

    log_probs = torch.stack(log_probs)
    loss = -(log_probs * returns).sum()

    optimizer.zero_grad(set_to_none=True)
    loss.backward()
    torch.nn.utils.clip_grad_norm_(policy.parameters(), max_norm=1.0)
    optimizer.step()

    return float(loss.detach())
\end{lstlisting}

This is the cleanest implementation of REINFORCE. It is also very noisy. The next step is to learn a baseline.

\subsection{Learning a value baseline}

\Needspace{20\baselineskip}
\begin{lstlisting}[style=pythonstyle,caption={Policy update with a learned value baseline.},label={lst:baseline-update}]
class ValueNet(nn.Module):
    def __init__(self, obs_dim, hidden=128):
        super().__init__()
        self.net = nn.Sequential(
            nn.Linear(obs_dim, hidden),
            nn.Tanh(),
            nn.Linear(hidden, hidden),
            nn.Tanh(),
            nn.Linear(hidden, 1),
        )

    def forward(self, obs):
        return self.net(obs).squeeze(-1)


def reinforce_with_baseline_update(
    policy, value_fn, policy_opt, value_opt,
    observations, log_probs, rewards, gamma=0.99,
):
    obs = torch.stack(observations)       # [T, obs_dim]
    log_probs = torch.stack(log_probs)    # [T]
    returns = reward_to_go(rewards, gamma=gamma)

    values = value_fn(obs)                # [T]
    advantages = returns - values.detach()
    advantages = normalize(advantages)

    # Policy loss: do not backpropagate through the advantage.
    policy_loss = -(log_probs * advantages).mean()

    # Value loss: train baseline to predict returns.
    value_loss = 0.5 * (values - returns).pow(2).mean()

    policy_opt.zero_grad(set_to_none=True)
    policy_loss.backward()
    torch.nn.utils.clip_grad_norm_(policy.parameters(), max_norm=1.0)
    policy_opt.step()

    value_opt.zero_grad(set_to_none=True)
    value_loss.backward()
    torch.nn.utils.clip_grad_norm_(value_fn.parameters(), max_norm=1.0)
    value_opt.step()

    return float(policy_loss.detach()), float(value_loss.detach())
\end{lstlisting}

The line `values.detach()` is important. The policy update should treat the advantage as a scalar weight. The value network is trained by its own value loss. Accidentally allowing gradients from the policy loss into the value network is a common implementation bug.

\subsection{Generalized Advantage Estimation}

The theory of GAE was introduced earlier in Eqs.~\eqref{eq:td_residual_gae}--\eqref{eq:gae_recursion} and Table~\ref{tab:gae_bias_variance}. The implementation below computes the same backward recursion directly.

\Needspace{18\baselineskip}
\begin{lstlisting}[style=pythonstyle,caption={Generalized Advantage Estimation.},label={lst:gae}]
def compute_gae(rewards, values, dones, gamma=0.99, lam=0.95):
    """Compute GAE advantages and lambda returns.

    Args:
        rewards: tensor [T]
        values: tensor [T + 1], includes bootstrap value V(s_T)
        dones: tensor [T], 1 if transition ended episode else 0
    """
    T = len(rewards)
    advantages = torch.zeros(T, dtype=torch.float32)
    last_gae = 0.0

    for t in reversed(range(T)):
        nonterminal = 1.0 - float(dones[t])
        delta = rewards[t] + gamma * values[t + 1] * nonterminal - values[t]
        last_gae = delta + gamma * lam * nonterminal * last_gae
        advantages[t] = last_gae

    returns = advantages + values[:-1]
    return advantages, returns
\end{lstlisting}

The terminal mask is not optional. If \texttt{dones[t]} is ignored, the algorithm may
bootstrap across episode boundaries. This creates fake value information and can seriously
corrupt training.

\begin{table}[t]
	\centering
	\caption{Advantage values in a UAV/SDN decision state.}
	\label{tab:uav-advantage-example}
	\begin{tabular}{lccc}
		\toprule
		Action & Reward-to-go $G_t$ & Baseline $V(s_t)$ & Advantage $G_t - V(s_t)$ \\
		\midrule
		Move to URLLC hotspot & $6.8$ & $4.2$ & $+2.6$ \\
		Stay and serve current users & $4.0$ & $4.2$ & $-0.2$ \\
		Move toward charging station & $2.7$ & $4.2$ & $-1.5$ \\
		\bottomrule
	\end{tabular}
\end{table}

The raw returns say that all three actions achieved some positive outcome. The advantages tell a sharper story: moving toward the URLLC hotspot was much better than expected, staying was slightly worse than expected, and moving toward the charging station was poor for this particular state.

\begin{figure}[t]
	\centering
	\begin{tikzpicture}[
		act/.style={draw,rounded corners,thick,minimum width=2.5cm,minimum height=0.7cm,align=center},
		arrow/.style={-{Latex[length=2.2mm]},thick},
		positive/.style={draw=blue!60!black,fill=blue!8},
		negative/.style={draw=red!70!black,fill=red!8}
		]
		\node[act,positive] (a1) at (0,1.5) {URLLC hotspot\\$A=+2.6$};
		\node[act] (a2) at (0,0) {Stay\\$A=-0.2$};
		\node[act,negative] (a3) at (0,-1.5) {Recharge direction\\$A=-1.5$};
		\node[act] (policy) at (5,0) {Policy update\\$\grad\log\pi(a|s)A$};
		\draw[arrow,blue!70!black] (a1) -- node[above,font=\scriptsize] {increase probability} (policy);
		\draw[arrow] (a2) -- node[above,font=\scriptsize] {small decrease} (policy);
		\draw[arrow,red!70!black] (a3) -- node[below,font=\scriptsize] {decrease probability} (policy);
	\end{tikzpicture}
	\caption{Advantage-based policy update in a UAV/SDN control state. The policy is not simply reinforced by positive reward; it is reinforced by better-than-expected performance.}
	\label{fig:uav-advantage-update}
\end{figure}
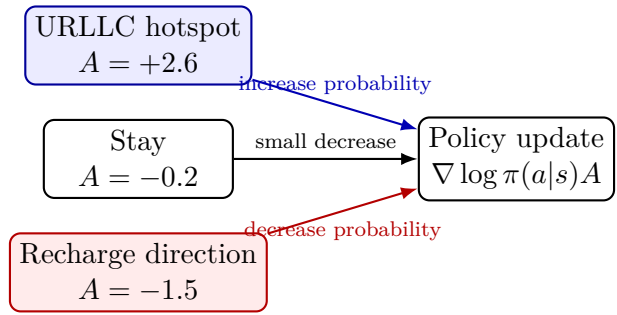

This example also explains why a learned baseline can be dangerous if it is inaccurate. If the value network consistently underestimates congested states, many actions will appear artificially good. If it overestimates recovery states, the agent may fail to learn useful emergency behaviors. In safety-critical networking, value diagnostics are therefore not optional.

\Needspace{18\baselineskip}
\begin{lstlisting}[style=pythonstyle,caption={A small UAV/SDN advantage computation example.},label={lst:uav-advantage}]
import torch

# Example reward-to-go from three sampled decisions in the same state.
returns = torch.tensor([6.8, 4.0, 2.7])
value_baseline = torch.tensor(4.2)
advantages = returns - value_baseline

for name, G, A in zip(
    ["move_to_urllc", "stay", "move_to_charge"],
    returns, advantages,
):
    direction = "increase" if A > 0 else "decrease"
    print(f"{name:16s} return={G:.1f} advantage={A:+.1f} -> {direction}")
\end{lstlisting}

\section{Modern extension: REINFORCE-style learning for LLMs}

REINFORCE is old, but it is not obsolete. In fact, recent language-model alignment and reasoning work revived interest in REINFORCE-style estimators because a completion can be treated as a trajectory and a reward model or verifier can score the final output \citep{ouyang2022training,ahmadian2024back,shao2024deepseekmath,deepseek2025r1}.

For a prompt \(x\), an LLM samples a completion \(y \sim \pi_\theta(\cdot \mid x)\). A reward function gives \(R(x,y)\). A sequence-level REINFORCE estimator is
\begin{equation}
	\nabla_\theta J(\theta)
	\approx
	\left(R(x,y)-b(x)\right)
	\nabla_\theta \log \pi_\theta(y \mid x),
\end{equation}
where
\begin{equation}
	\log \pi_\theta(y \mid x)
	=
	\sum_{t=1}^{|y|}
	\log \pi_\theta(y_t \mid x,y_{<t}).
\end{equation}
This is mathematically close to episodic REINFORCE, except that the trajectory is a token sequence and the reward may come from a human-preference model, a unit test, a mathematical verifier, or a rule-based score.

This sequence-level formulation is simple and scalable, but it does not fully solve token-level credit assignment. When a long completion receives one final reward, it remains difficult to know which tokens were responsible for success or failure. This is one reason credit assignment in reasoning-oriented language-model reinforcement learning remains an active research problem.

\subsection{RLOO and group-relative advantages}

In REINFORCE Leave-One-Out (RLOO), multiple completions are sampled for the same prompt. For each completion, the baseline is computed from the rewards of the other completions. This creates a critic-free baseline: the algorithm reduces variance without learning a separate value function. The idea is closely related to leave-one-out baselines studied by Kool, van Hoof, and Welling, where multiple sampled solutions are used to construct a baseline from the batch itself \citep{kool2019buy}.

Suppose a prompt produces \(K\) sampled completions:
\[
y_1, y_2, \ldots, y_K.
\]
Each completion receives a reward:
\[
R(x,y_1), R(x,y_2), \ldots, R(x,y_K).
\]
For completion \(i\), the leave-one-out baseline is
\begin{equation}
	b_i(x)
	=
	\frac{1}{K-1}
	\sum_{j\neq i}
	R(x,y_j).
\end{equation}
The corresponding advantage is
\begin{equation}
	\hat{A}_i
	=
	R(x,y_i)-b_i(x).
\end{equation}

This means that each completion is judged relative to other completions sampled for the same prompt. A high reward is useful only if it is high compared with the alternatives. This is the same principle introduced earlier in the chapter: raw reward is often less useful than reward relative to a baseline.

GRPO-style methods use group-relative normalization to avoid training a separate critic in some large-language-model reinforcement learning settings \citep{shao2024deepseekmath,deepseek2025r1}. The core idea is again the same: replace raw reward by performance relative to a group baseline. This makes the policy-gradient update more stable while avoiding the cost and difficulty of training a large value model.

\begin{figure}[t]
	\centering
	\begin{tikzpicture}[
		box/.style={draw,rounded corners,thick,minimum width=2.5cm,minimum height=0.75cm,align=center},
		small/.style={draw,rounded corners,minimum width=1.6cm,minimum height=0.6cm,align=center},
		arrow/.style={-{Latex[length=2.2mm]},thick},
		node distance=0.65cm
		]
		\node[box] (prompt) {Prompt $x$};
		\node[small, right=1.0cm of prompt, yshift=0.9cm] (y1) {$y_1$, $R_1$};
		\node[small, right=1.0cm of prompt] (y2) {$y_2$, $R_2$};
		\node[small, right=1.0cm of prompt, yshift=-0.9cm] (y3) {$y_3$, $R_3$};
		\node[box, right=1.2cm of y2] (group) {Group baseline\\mean reward};
		\node[box, right=1.2cm of group] (adv) {Relative\\advantages};
		\draw[arrow] (prompt) -- (y1);
		\draw[arrow] (prompt) -- (y2);
		\draw[arrow] (prompt) -- (y3);
		\draw[arrow] (y1) -- (group);
		\draw[arrow] (y2) -- (group);
		\draw[arrow] (y3) -- (group);
		\draw[arrow] (group) -- (adv);
		\node[below=0.9cm of group, align=center,font=\small] {Multiple completions for the same prompt create a local baseline.\\The policy learns which completion was better than its siblings.};
	\end{tikzpicture}
	\caption{REINFORCE-style advantage estimation for language-model RL. RLOO and GRPO-style methods use multiple completions for the same prompt to construct a prompt-local baseline or normalized advantage.}
	\label{fig:llm-group-advantage}
\end{figure}

\Needspace{18\baselineskip}
\begin{lstlisting}[style=pythonstyle,caption={RLOO and GRPO-style group advantages.},label={lst:rloo-grpo}]
def rloo_advantages(rewards):
    """Leave-one-out advantages for K completions of one prompt."""
    rewards = torch.as_tensor(rewards, dtype=torch.float32)
    K = rewards.numel()
    total = rewards.sum()
    loo_baseline = (total - rewards) / max(K - 1, 1)
    return rewards - loo_baseline


def group_normalized_advantages(rewards, eps=1e-8):
    """GRPO-style normalized group advantages."""
    rewards = torch.as_tensor(rewards, dtype=torch.float32)
    return (rewards - rewards.mean()) / (rewards.std(unbiased=False) + eps)


rewards = torch.tensor([0.0, 1.0, 1.0, 0.5])
print(rloo_advantages(rewards))
print(group_normalized_advantages(rewards))
\end{lstlisting}

This modern connection is useful for understanding why Chapter 8 matters beyond classical control. Baselines and advantages are no longer only tools for robots or games. They are also central to how large models are optimized from human preferences, verifiable rewards, and reasoning outcomes.

\section{Common implementation bugs}

Table~\ref{tab:ch8-bugs} summarizes common bugs in REINFORCE and advantage-based training.

\begin{table}[t]
	\centering
	\caption{Common implementation bugs in REINFORCE, baselines, and advantage estimation.}
	\label{tab:ch8-bugs}
	\begin{tabularx}{\textwidth}{p{3.3cm}X X}
		\toprule
		Bug & Symptom & Fix \\
		\midrule
		Forgetting to detach advantages & Value network receives gradients from policy loss & Use `advantages.detach()` in policy loss \\
		Bootstrapping across terminal states & Sudden value spikes near episode ends & Use done masks in TD and GAE \\
		Using full return when reward-to-go is needed & Very noisy updates and slow learning & Use $G_t$, not always $G_0$ \\
		No value-loss scale control & Critic loss dominates policy learning & Track policy loss, value loss, entropy separately \\
		Normalizing across wrong dimension & Group or episode structure is destroyed & Normalize over the intended batch/group \\
		Evaluating with sampling accidentally on & Noisy evaluation curves & Use deterministic or fixed evaluation protocol \\
		Ignoring raw metrics & Reward improves but task gets worse & Log reward components and real task metrics \\
		Wrong sign in policy loss & Policy learns to avoid good actions & Minimize $-\log\pi(a|s)\hat{A}$ \\
		\bottomrule
	\end{tabularx}
\end{table}

\section{Exercises}

\subsection*{Conceptual exercises}

\begin{enumerate}[leftmargin=*]
	\item Explain why REINFORCE can optimize a policy without differentiating through the environment.
	\item Why does full-return REINFORCE have high variance?
	\item Explain the difference between return, value, and advantage.
	\item Why does subtracting a state-dependent baseline not change the expected policy gradient?
	\item In a UAV network, give an example where a positive reward could still correspond to a negative advantage.
	\item Why can advantage normalization be useful in practice but not a purely theoretical identity?
	\item Explain why RLOO and GRPO-style methods can be viewed as modern baseline methods for language-model RL.
\end{enumerate}

\subsection*{Mathematical exercises}

\begin{enumerate}[leftmargin=*]
	\item Prove that
	\[
	\E_{a\sim\pi_\theta(\cdot|s)}[\grad_\theta\log\pi_\theta(a|s)b(s)] = 0.
	\]
	\item Derive the reward-to-go policy-gradient estimator from the full-return estimator using causality.
	\item Given $V(s_t)=3.0$, $r_{t+1}=1.0$, $V(s_{t+1})=4.0$, and $\gamma=0.9$, compute the TD error.
	\item Compute GAE for a three-step trajectory with rewards $(1,0,2)$, values $(0.5,0.7,0.2,0)$, $\gamma=0.9$, and $\lambda=0.95$.
	\item Show how the $n$-step advantage becomes closer to Monte Carlo as $n$ increases.
\end{enumerate}

\subsection*{Coding exercises}

\begin{enumerate}[leftmargin=*]
	\item Implement REINFORCE with reward-to-go on a simple Gymnasium environment.
	\item Add a value baseline and compare learning variance across five random seeds.
	\item Implement GAE and verify that terminal masks prevent bootstrapping across episode boundaries.
	\item Modify the RLOO code to handle a batch of prompts, each with $K$ completions.
	\item For a UAV control problem, log raw reward, return, value prediction, advantage, and policy entropy. Plot all five over training.
\end{enumerate}

\section*{Looking Ahead to Chapter 9: Food for Thought}
\addcontentsline{toc}{section}{Looking Ahead to Chapter 9: Food for Thought}

This chapter introduced REINFORCE, baselines, and advantage functions. The pattern is now clear: the policy needs a learning signal, and the quality of that signal determines the stability of learning. REINFORCE gives an unbiased but noisy signal. Baselines reduce variance. Advantage functions tell the policy whether an action was better or worse than expected.

However, a new question appears:
\begin{quote}
	If a value function is so useful as a baseline, why not learn it continuously and use it as a critic for every policy update?
\end{quote}
This question leads directly to actor-critic learning. In Chapter 9, the policy becomes the actor and the value estimator becomes the critic. The actor chooses actions. The critic evaluates them. Together, they form the core architecture behind A2C, A3C, DDPG, TD3, SAC, PPO, and many modern policy-optimization systems.

Before moving on, keep the following questions in mind:
\begin{enumerate}[leftmargin=*]
	\item When does a baseline become a critic?
	\item How accurate must the critic be for the actor to improve?
	\item Can a bad critic make the actor worse?
	\item How should we balance policy loss, value loss, and entropy?
	\item In safety-critical systems, should the critic estimate only reward, or also risk and constraint violation?
\end{enumerate}

% \chapter*{Chapter 8 References}
% \addcontentsline{toc}{chapter}{Chapter 8 References}
	\chapter{Actor-Critic Learning}
\label{ch:actor_critic}
\chaptermark{Actor-Critic Learning}

\begin{keybox}{Chapter goal}
	This chapter explains how policy learning and value learning are combined in actor-critic methods. The actor chooses actions. The critic evaluates states, actions, or advantages. The resulting feedback is used to update the actor with much lower variance than pure Monte Carlo policy gradients, while remaining more flexible than value-based control. By the end of the chapter, the reader should understand A2C, A3C, value critics, Q critics, advantage critics, off-policy corrections, continuous-control actor-critics, and the practical engineering choices that make actor-critic methods work.
\end{keybox}

\section*{Chapter Overview}
\addcontentsline{toc}{section}{Chapter Overview}

\begin{enumerate}[leftmargin=*]
	\item Why actor-critic learning exists
	\item The actor and the critic
	\item From REINFORCE to actor-critic
	\item What can the critic estimate?
	\item One-step, n-step, and GAE actor-critic targets
	\item The actor-critic loss as an engineering object
	\item A2C and A3C
	\item Actor-critic in language-model RL
	\item Shared networks, gradient interference, and representation learning
	\item Off-policy actor-critic and importance sampling
	\item Continuous-control actor-critics: DDPG, TD3, and SAC preview
	\item Practical failure modes and debugging
	\item UAV/SDN worked example
	\item Python implementation
	\item Algorithm box: on-policy advantage actor-critic
	\item What actor-critic teaches us
	\item Exercises
	\item Looking Ahead to Chapter 10
	\item Chapter references
\end{enumerate}

\section{Why actor-critic learning exists}

Chapter~7 introduced the policy-gradient idea: learn a policy directly by following an estimate of the gradient of expected return. Chapter~8 showed that REINFORCE can be improved by reward-to-go, baselines, advantages, and generalized advantage estimation. These tools reduce variance, but a central tension remains.

Pure Monte Carlo policy gradients can learn without a value function, but they often have high variance. Value-based methods can use bootstrapping and learn from local temporal-difference errors, but they are awkward for continuous actions and do not directly optimize a parameterized stochastic policy. Actor-critic learning combines the two ideas.

The \emph{actor} is the policy. It decides what to do. The \emph{critic} is a learned evaluator. It estimates how good the actor's choices are. The actor uses the critic's estimate as a lower-variance learning signal.

Historically, actor-critic methods grew from the same policy-gradient tradition as REINFORCE, but they introduced a learned value function into the policy update. Konda and Tsitsiklis gave a foundational analysis of actor-critic algorithms, and Sutton et al. connected policy-gradient methods with value-function approximation in a general framework \citep{konda2000actor,sutton1999policy}. Standard textbook treatments and later surveys place actor-critic learning at the intersection of policy search, value-function approximation, and temporal-difference learning \citep{sutton2018reinforcement,grondman2012survey}. Modern deep reinforcement learning then turned this idea into scalable neural algorithms such as A3C, A2C, DDPG, TD3, PPO, and SAC \citep{mnih2016asynchronous,lillicrap2016continuous,fujimoto2018addressing,schulman2017ppo,haarnoja2018soft}.

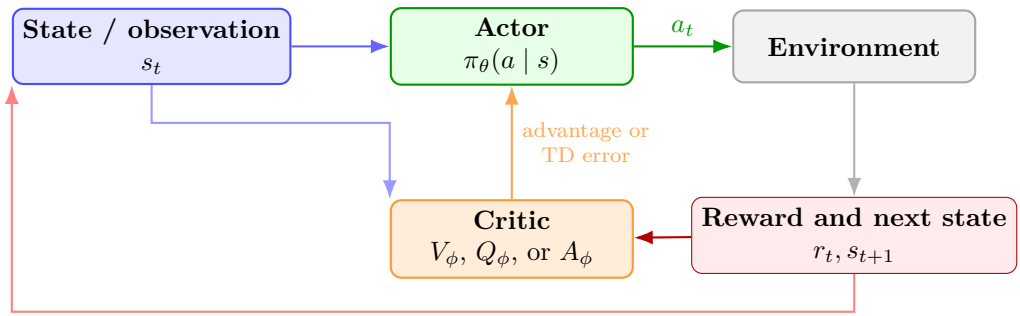
\begin{figure}[t]
	\centering
	\begin{tikzpicture}[
		box/.style={draw, rounded corners, thick, minimum width=3.2cm, minimum height=0.95cm, align=center, font=\small},
		small/.style={draw, rounded corners, minimum width=2.7cm, minimum height=0.75cm, align=center, font=\small},
		arrow/.style={-{Latex[length=2.5mm]}, thick},
		node distance=1.3cm
		]
		\node[box, draw=blue!70, fill=blue!10, font=\small\bfseries] (state) {State / observation\\{\normalfont\small $s_t$}};
		\node[box, draw=green!60!black, fill=green!10, font=\small\bfseries, right=of state] (actor) {Actor\\{\normalfont\small $\pi_\theta(a\mid s)$}};
		\node[box, draw=gray!70, fill=gray!10, font=\small\bfseries, right=of actor] (env) {Environment};
		\node[box, draw=orange!80, fill=orange!15, font=\small\bfseries, below=1.5cm of actor] (critic) {Critic\\{\normalfont\small $V_\phi$, $Q_\phi$, or $A_\phi$}};
		\node[small, draw=red!70!black, fill=red!8, font=\small\bfseries, below=1.5cm of env] (reward) {Reward and next state\\{\normalfont\small $r_t, s_{t+1}$}};

		\draw[arrow, color=blue!60] (state) -- (actor);
		\draw[arrow, color=green!60!black] (actor) -- node[above, font=\small, color=green!60!black] {$a_t$} (env);
		\draw[arrow, color=gray!60] (env) -- (reward);
		\draw[arrow, color=red!70!black] (reward) -- (critic);
		\draw[arrow, color=blue!40] (state.south) -- ++(0,-0.5) -| (critic.north west);
		\draw[arrow, color=orange!70] (critic) -- node[right, align=center, font=\scriptsize, color=orange!70] {advantage or\\TD error} (actor);
		\draw[arrow, color=red!50] (reward.south) -- ++(0,-0.5) -| (state.south west);
	\end{tikzpicture}
	\caption{The actor-critic idea. The actor chooses actions. The critic evaluates the consequences and produces a learning signal for the actor. The critic usually learns by temporal-difference bootstrapping, while the actor learns by a policy-gradient update.}
	\label{fig:actor_critic_loop}
\end{figure}

Actor-critic methods are central because they form the backbone of many modern DRL algorithms. PPO is an actor-critic method with a clipped policy update. SAC is an entropy-regularized off-policy actor-critic method. DDPG and TD3 are deterministic actor-critic methods for continuous control. IMPALA uses distributed actors with a value-learning correction. Even many RLHF systems for language models are conceptually actor-critic systems when a policy model is optimized using either a learned reward, a baseline, or a value estimate.

\section{The actor and the critic}

An actor-critic method has two learned objects:
\begin{align}
	\text{actor:} \quad & \pi_\theta(a\mid s), \\
	\text{critic:} \quad & V_\phi(s), \; Q_\phi(s,a), \; \text{or} \; A_\phi(s,a).
\end{align}
The parameters \(\theta\) control the policy, while the parameters \(\phi\) control the value estimator. In some implementations, the actor and critic have separate networks. In others, they share a representation trunk and use separate output heads.

The critic is not a second agent. It is a learned measuring instrument. It tells the actor whether an action was better or worse than expected. If the critic is accurate, the actor receives a cleaner gradient. If the critic is wrong, the actor may be pushed in the wrong direction.

\begin{keybox}{Core intuition}
	REINFORCE says: ``increase the probability of actions that led to high return.'' Actor-critic says: ``increase the probability of actions that were better than the critic expected.''
\end{keybox}

This difference matters because raw returns can vary wildly across states. A reward of 5 may be excellent in a congested network and terrible in an easy network. The critic provides context by estimating the expected future return from the current state. The actor then learns from a relative signal, often an advantage:
\begin{equation}
	A^\pi(s,a) = Q^\pi(s,a) - V^\pi(s).
\end{equation}

\section{From REINFORCE to actor-critic}

The REINFORCE update with a baseline is the Monte Carlo starting point for this transition \citep{williams1992simple}:
\begin{equation}
	\nabla_\theta J(\theta)
	=
	\E_\pi\left[
	\sum_{t}
	\nabla_\theta \log \pi_\theta(a_t\mid s_t)
	\left(G_t - b(s_t)\right)
	\right].
	\label{eq:reinforce_baseline_ac}
\end{equation}
If the baseline is a learned state-value function, \(b(s_t)=V_\phi(s_t)\), then
\begin{equation}
	\hat{A}_t = G_t - V_\phi(s_t).
\end{equation}
This is already a simple actor-critic method: the critic learns \(V_\phi\), and the actor uses the estimated advantage.

However, a full Monte Carlo return \(G_t\) may still be high variance. Actor-critic methods usually replace it with a bootstrapped target. The one-step TD advantage is
\begin{equation}
	\hat{A}_t^{\mathrm{TD}}
	=
	r_t + \gamma V_\phi(s_{t+1}) - V_\phi(s_t),
	\label{eq:td_advantage_ch9}
\end{equation}
with terminal masking when \(s_{t+1}\) is terminal. This quantity is also the TD error:
\begin{equation}
	\delta_t = r_t + \gamma V_\phi(s_{t+1}) - V_\phi(s_t).
\end{equation}
The actor update becomes
\begin{equation}
	\nabla_\theta J(\theta)
	\approx
	\E_\pi\left[
	\nabla_\theta \log \pi_\theta(a_t\mid s_t)\, \delta_t
	\right].
	\label{eq:actor_critic_td_update}
\end{equation}

\begin{figure}[t]
	\centering
	\begin{tikzpicture}[
		box/.style={draw, rounded corners, thick, minimum width=3.1cm, minimum height=0.8cm, align=center, font=\small},
		arrow/.style={-{Latex[length=2.3mm]}, thick},
		node distance=2.0cm
		]
		\node[box, draw=blue!70, fill=blue!10, font=\small\bfseries] (mc) {REINFORCE\\{\normalfont\small $G_t$}};
		\node[box, draw=orange!80, fill=orange!15, font=\small\bfseries, right=of mc] (baseline) {Baseline\\{\normalfont\small $G_t - V_\phi(s_t)$}};
		\node[box, draw=green!60!black, fill=green!10, font=\small\bfseries, right=of baseline] (td) {Actor-Critic\\{\normalfont\small $r_t+\gamma V_\phi(s_{t+1})-V_\phi(s_t)$}};
		\node[box, draw=purple!70, fill=purple!10, font=\small\bfseries, below=1.2cm of baseline] (gae) {GAE\\{\normalfont\small weighted TD errors}};

		\draw[arrow, color=red!70!black] (mc) -- node[above=5pt, font=\scriptsize, color=red!70!black] {lower variance} (baseline);
		\draw[arrow, color=red!70!black] (baseline) -- node[above=5pt, font=\scriptsize, color=red!70!black] {bootstrap} (td);
		\draw[arrow, color=orange!70] (baseline) -- (gae);
		\draw[arrow, color=green!60!black] (td) -- (gae);
	\end{tikzpicture}
	\caption{From REINFORCE to actor-critic. Actor-critic methods replace raw Monte Carlo returns by learned baselines, TD errors, n-step targets, or GAE advantages. This reduces variance but introduces bias through bootstrapping and critic approximation.}
	\label{fig:reinforce_to_ac}
\end{figure}
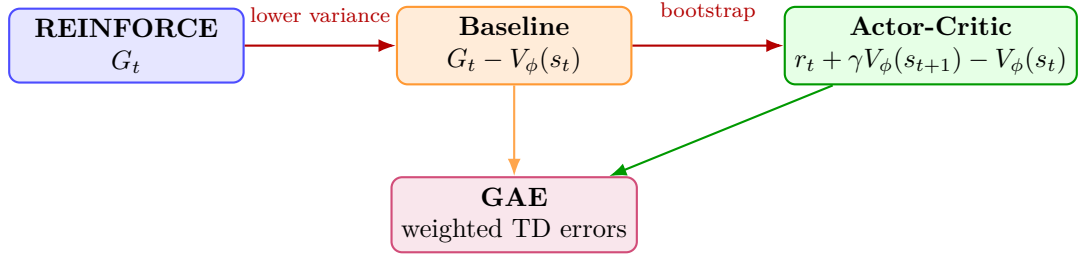

The price of this improvement is bias. A bootstrapped target depends on the critic. If the critic is inaccurate, the actor update is biased. Thus, actor-critic methods trade variance for bias. Much of modern actor-critic design is about managing this trade-off.

\section{What can the critic estimate?}

There is no single critic. Different actor-critic algorithms use different value objects.

\begin{table}[t]
	\centering
	\caption{Common critic choices in actor-critic algorithms.}
	\label{tab:critic_types}
\begin{tabular}{p{2.6cm}p{3.7cm}p{5.2cm}}
		\toprule
		Critic type & Estimate & Typical use \\
		\midrule
		State-value critic & \(V_\phi(s)\) & On-policy actor-critic, A2C, A3C, PPO \\
		Action-value critic & \(Q_\phi(s,a)\) & DDPG, TD3, SAC, deterministic actor updates \\
		Advantage critic & \(A_\phi(s,a)\) & Advantage actor-critic, compatible approximation \\
		Distributional critic & Distribution of \(Z(s,a)\) & Risk-sensitive control, distributional actor-critic variants \\
		Soft critic & \(Q_\phi(s,a)\) with entropy-regularized target & SAC and maximum-entropy RL \\
		Safety critic & Cost value \(V_c(s)\) or \(Q_c(s,a)\) & Constrained or safe actor-critic \\
		\bottomrule
	\end{tabular}
\end{table}

A state-value critic is natural for stochastic on-policy methods. The actor samples actions, and the critic estimates whether the resulting return was better than expected for that state. A Q critic is more natural when the actor needs a differentiable objective through the action, as in deterministic continuous-control methods. A safety critic estimates costs or constraint violations rather than reward.

\begin{figure}[t]
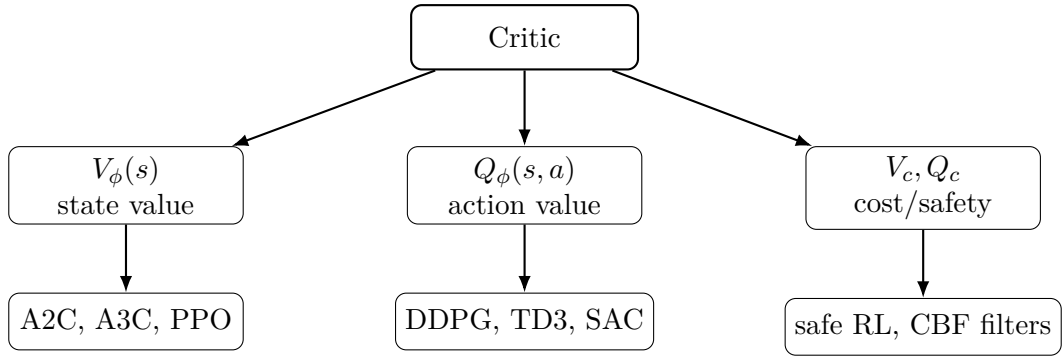

	\centering
	%
	\caption{Different critics support different actor-critic families. On-policy methods often use a state-value critic, continuous-control methods often use an action-value critic, and safe actor-critic methods may add a cost or safety critic.}
	\label{fig:critic_types}
\end{figure}

\section{One-step, n-step, and GAE actor-critic targets}

The critic needs targets. The simplest target is one-step TD:
\begin{equation}
	y_t^{(1)} = r_t + \gamma (1-d_{t+1})V_\phi(s_{t+1}).
\end{equation}
The critic minimizes
\begin{equation}
	L_V(\phi) = \frac{1}{2}\left(V_\phi(s_t)-y_t^{(1)}\right)^2.
\end{equation}
The actor uses the corresponding advantage
\begin{equation}
	\hat{A}_t = y_t^{(1)} - V_\phi(s_t).
\end{equation}

An n-step target uses more real rewards before bootstrapping:
\begin{equation}
	y_t^{(n)}
	=
	\sum_{k=0}^{n-1}\gamma^k r_{t+k}
	+
	\gamma^n V_\phi(s_{t+n}).
\end{equation}
In practice, terminal masks are required so that the target does not bootstrap beyond an episode boundary.

GAE, introduced in Chapter~8, combines a whole spectrum of n-step TD errors \citep{schulman2015gae}. In implementation, this is usually computed with the backward recursion discussed in Chapter~8, rather than by explicitly summing all future residuals. It is often the default advantage estimator in PPO-style actor-critic implementations because it provides a practical bias-variance trade-off.

\begin{figure}[t]
	\centering
	\begin{tikzpicture}[
		box/.style={draw, rounded corners, thick, minimum width=2.5cm, minimum height=0.8cm, align=center, font=\small},
		arrow/.style={-{Latex[length=2.3mm]}, thick},
		node distance=0.9cm
		]
		\node[box, draw=blue!70, fill=blue!10, font=\small\bfseries] (one) {1-step TD\\{\normalfont\scriptsize low variance, more bias}};
		\node[box, draw=orange!80, fill=orange!15, font=\small\bfseries, right=of one] (nstep) {n-step\\{\normalfont\scriptsize middle ground}};
		\node[box, draw=purple!70, fill=purple!10, font=\small\bfseries, right=of nstep] (gae) {GAE\\{\normalfont\scriptsize weighted mixture}};
		\node[box, draw=green!60!black, fill=green!10, font=\small\bfseries, right=of gae] (mc) {Monte Carlo\\{\normalfont\scriptsize low bias, high variance}};

		\draw[arrow, color=red!70!black] (one) -- (nstep);
		\draw[arrow, color=red!70!black] (nstep) -- (gae);
		\draw[arrow, color=red!70!black] (gae) -- (mc);

		\node[below=0.8cm of $(nstep)!0.5!(gae)$, align=center, font=\small, color=gray!70!black] {Moving right uses more sampled rewards before bootstrapping.\\Actor-critic methods choose where to sit on this spectrum.};
	\end{tikzpicture}
	\caption{Actor-critic target choices. One-step TD has low variance but can be biased by the critic. Monte Carlo returns are less biased but high variance. n-step returns and GAE interpolate between these extremes.}
	\label{fig:target_spectrum_ch9}
\end{figure}

\section{The actor-critic loss as an engineering object}

A modern on-policy actor-critic implementation often optimizes a combined loss:
\begin{equation}
	L(\theta,\phi)
	=
	L_{\mathrm{policy}}(\theta)
	+
	c_v L_{\mathrm{value}}(\phi)
	-
	c_e L_{\mathrm{entropy}}(\theta).
	\label{eq:combined_ac_loss}
\end{equation}
A common policy loss is
\begin{equation}
	L_{\mathrm{policy}}(\theta)
	=
	-\E_t\left[\log \pi_\theta(a_t\mid s_t)\,\sg(\hat{A}_t)\right],
	\label{eq:ac_policy_loss}
\end{equation}
where \(\sg(\cdot)\) denotes stop-gradient. The value loss is
\begin{equation}
	L_{\mathrm{value}}(\phi)
	=
	\frac{1}{2}\E_t\left[\left(V_\phi(s_t)-\hat{R}_t\right)^2\right],
\end{equation}
and the entropy term encourages exploration:
\begin{equation}
	L_{\mathrm{entropy}}(\theta)
	=
	\E_t\left[\mathcal{H}(\pi_\theta(\cdot\mid s_t))\right].
\end{equation}

The actor and critic losses use related quantities, but their gradients should be kept conceptually separate. The policy loss treats the advantage estimate as a fixed learning signal for the actor. The value loss trains the critic toward a target return. In code, this means that advantages in the actor loss and returns in the critic loss are usually detached in the appropriate places. A common implementation bug is to let the actor gradient flow through the value estimate used to construct the advantage, which turns a variance-reduction baseline into an unintended part of the actor objective.

\begin{pitfallbox}{Stop-gradient is not cosmetic}
	The actor should not backpropagate through the advantage target as if the advantage were a differentiable objective for the critic. In most actor-critic code, advantages used in the policy loss are detached. Without this stop-gradient, actor and critic gradients can mix in unintended ways.
\end{pitfallbox}

The combined loss is an engineering object, not a single clean theoretical objective. The coefficients \(c_v\) and \(c_e\), advantage normalization, rollout length, critic target, optimizer, and gradient clipping all matter. Many practical actor-critic failures are not caused by the policy-gradient theorem; they are caused by the details of this loss implementation.

\section{A2C and A3C}

A3C, or Asynchronous Advantage Actor-Critic, was one of the first deep actor-critic methods to become widely influential \citep{mnih2016asynchronous}. It used many parallel workers, each interacting with its own copy of the environment and asynchronously updating shared global parameters. The parallelism helped decorrelate data and stabilized training without the replay buffer used by DQN.

A2C, or Advantage Actor-Critic, is the synchronous version. Instead of asynchronous workers updating shared parameters at different times, A2C collects rollouts from multiple environments, aggregates them into a batch, and performs a synchronized update. A2C is easier to implement reproducibly on modern hardware.

\begin{figure}[t]
	\centering
	\begin{tikzpicture}[
		worker/.style={draw, rounded corners, minimum width=2.6cm, minimum height=0.75cm, align=center, font=\small},
		box/.style={draw, rounded corners, thick, minimum width=3.0cm, minimum height=0.85cm, align=center, font=\small},
		arrow/.style={-{Latex[length=2.2mm]}, thick},
		node distance=0.75cm
		]
		\node[worker, draw=blue!70, fill=blue!10, font=\small\bfseries] (w1) {Worker 1\\{\normalfont\small rollout}};
		\node[worker, draw=green!60!black, fill=green!10, font=\small\bfseries, below=of w1] (w2) {Worker 2\\{\normalfont\small rollout}};
		\node[worker, draw=orange!80, fill=orange!15, font=\small\bfseries, below=of w2] (w3) {Worker 3\\{\normalfont\small rollout}};
		\node[box, draw=purple!70, fill=purple!10, font=\small\bfseries, right=1.5cm of w2] (batch) {Batch of trajectories};
		\node[box, draw=red!70!black, fill=red!8, font=\small\bfseries, right=1.5cm of batch] (update) {Synchronous\\A2C update};

		\draw[arrow, color=blue!60] (w1) -- (batch);
		\draw[arrow, color=green!60!black] (w2) -- (batch);
		\draw[arrow, color=orange!70] (w3) -- (batch);
		\draw[arrow, color=red!70!black] (batch) -- (update);

		\node[below=2.9cm of batch, align=center, font=\small, color=gray!70!black] {A3C updates asynchronously. A2C synchronizes rollouts before the update.};
	\end{tikzpicture}
	\caption{A2C and A3C use multiple actor-learners. A3C performs asynchronous updates, while A2C typically collects parallel rollouts and updates synchronously.}
	\label{fig:a2c_a3c_parallel}
\end{figure}

\begin{table}[t]
	\centering
	\caption{A2C versus A3C.}
	\label{tab:a2c_a3c}
	\begin{tabular}{p{2.5cm}p{4.6cm}p{4.6cm}}
		\toprule
		Aspect & A3C & A2C \\
		\midrule
		Update style & Asynchronous workers update shared parameters & Synchronous batch update from multiple environments \\
		Hardware fit & Strong historical fit for multi-core CPU & Strong fit for GPU/accelerator batching \\
		Reproducibility & More nondeterminism due to async updates & Easier to reproduce \\
		Stability source & Parallel decorrelated experience & Parallel decorrelated experience plus synchronized optimization \\
		Modern use & Historically important, less common in clean baselines & Common as a simple on-policy baseline \\
		\bottomrule
	\end{tabular}
\end{table}

\section{Actor-critic in language-model RL}

The actor-critic view also helps interpret reinforcement learning from human feedback. In PPO-style RLHF, the language model is the actor: it produces a token sequence conditioned on a prompt. A reward model, verifier, or human-preference signal provides a scalar evaluation of the generated response. A value head attached to the policy model often serves as the critic by predicting the expected future reward, including penalties such as the KL term that keeps the policy close to a reference model \citep{ouyang2022training,schulman2017ppo}.

This connection is useful pedagogically. The actor is not a robot motor controller; it is a probability distribution over tokens. The critic is not estimating physical return; it is estimating expected sequence quality under the reward model and regularization terms. But the structure is the same: the actor changes the probability of actions, while the critic reduces variance by estimating what outcome was expected for the current state or prefix.

\section{Shared networks, gradient interference, and representation learning}

Actor-critic models often share a neural representation. For example, a convolutional or MLP trunk processes the observation, then separate actor and critic heads produce action logits and value predictions. Sharing saves computation and can improve representation learning, but it also creates interference: the actor and critic may prefer different features.

Recent work has studied how actor and critic representations interact in deep RL, including the gap between actor-critic and pure policy-gradient methods and the specialization of actor and critic representations \citep{wen2021gap,garcin2025interplay}. One key point is that the actor needs features useful for choosing actions, while the critic needs features useful for predicting returns. These are related but not identical objectives. When the critic loss dominates, the shared representation may become good for value prediction but poor for policy improvement. When the actor dominates, the critic may become inaccurate and destabilize training.

\begin{figure}[t]
	\centering
	\begin{tikzpicture}[
		box/.style={draw,rounded corners,thick,minimum width=3.0cm,minimum height=0.85cm,align=center},
		head/.style={draw,rounded corners,minimum width=2.6cm,minimum height=0.75cm,align=center},
		arrow/.style={-{Latex[length=2.2mm]},thick},
		node distance=0.85cm
		]
		\node[box] (obs) {Observation};
		\node[box, right=of obs] (trunk) {Shared representation\\$h_\psi(s)$};
		\node[head, above right=0.7cm and 1.2cm of trunk] (actor) {Actor head\\$\pi_\theta$};
		\node[head, below right=0.7cm and 1.2cm of trunk] (critic) {Critic head\\$V_\phi$};
		\draw[arrow] (obs) -- (trunk);
		\draw[arrow] (trunk) -- (actor);
		\draw[arrow] (trunk) -- (critic);
		\node[below=2.15cm of trunk, align=center,font=\small] {Shared features reduce cost, but actor and critic gradients can compete.};
	\end{tikzpicture}
	\caption{Shared actor-critic networks. A shared trunk can be efficient, but actor and critic losses may push the representation in different directions. Separate networks reduce interference but increase computation.}
	\label{fig:shared_actor_critic}
\end{figure}

\begin{warningbox}{Practical rule}
	If training is unstable, inspect actor loss, value loss, entropy, explained variance, gradient norms, and advantage scale separately. A single total loss is rarely enough to diagnose actor-critic training.
\end{warningbox}

\section{Off-policy actor-critic and importance sampling}

On-policy actor-critic uses data generated by the current policy. This is simple but sample-inefficient. Off-policy actor-critic tries to learn from data generated by another policy, a replay buffer, older policies, or distributed actors.

If actions are sampled from a behavior policy \(\mu\) but we want to update a target policy \(\pi_\theta\), importance sampling can correct the mismatch:
\begin{equation}
	\rho_t = \frac{\pi_\theta(a_t\mid s_t)}{\mu(a_t\mid s_t)}.
\end{equation}
The policy-gradient term becomes weighted by \(\rho_t\). However, importance ratios can have high variance, especially over long trajectories. Practical methods clip, truncate, or otherwise regularize these ratios.

IMPALA introduced V-trace targets for scalable distributed actor-critic learning with off-policy corrections \citep{espeholt2018impala}. Degris et al. studied off-policy actor-critic updates and showed how off-policy policy gradients can be corrected under suitable assumptions \citep{degris2012offpolicy}. These ideas matter because modern RL systems often collect data asynchronously, from multiple actors, or from replay buffers.

Off-policy actor-critic also re-enters the deadly-triad regime discussed in Chapter~5: it combines function approximation, bootstrapping, and off-policy data. Importance sampling, target networks, clipped corrections, and conservative updates can mitigate this instability, but they do not make it disappear. This is why off-policy actor-critic methods are often more sample-efficient than on-policy methods, but also more delicate to tune.

\begin{figure}[t]
	\centering
	\begin{tikzpicture}[
		box/.style={draw,rounded corners,thick,minimum width=3.0cm,minimum height=0.85cm,align=center},
		arrow/.style={-{Latex[length=2.2mm]},thick},
		node distance=0.8cm
		]
		\node[box] (beh) {Behavior policy\\$\mu$};
		\node[box, right=of beh] (data) {Collected data\\$(s,a,r,s')$};
		\node[box, right=of data] (target) {Target policy\\$\pi_\theta$};
		\node[box, below=of data] (ratio) {Correction\\$\rho=\pi/\mu$};
		\draw[arrow] (beh) -- (data);
		\draw[arrow] (data) -- (target);
		\draw[arrow] (data) -- (ratio);
		\draw[arrow] (ratio) -| (target);
		\node[below=0.8cm of ratio, align=center,font=\small] {Off-policy actor-critic must correct or control the mismatch between data collection and policy update.};
	\end{tikzpicture}
	\caption{Off-policy actor-critic. Data may be generated by a behavior policy different from the policy being optimized. Importance ratios or specialized targets such as V-trace can reduce the mismatch, but they introduce variance and implementation complexity.}
	\label{fig:offpolicy_ac}
\end{figure}

\section{Continuous-control actor-critics: DDPG, TD3, and SAC preview}

For continuous actions, it is often natural for the actor to output actions directly. Natural actor-critic methods introduced an earlier line of work that uses the geometry of the policy space to precondition the actor update \citep{peters2008natural,bhatnagar2009natural}. Deterministic policy-gradient methods use an actor \(a=\mu_\theta(s)\) and a Q critic \(Q_\phi(s,a)\). The deterministic actor update is
\begin{equation}
	\nabla_\theta J(\theta)
	\approx
	\E_s\left[
	\nabla_a Q_\phi(s,a)\vert_{a=\mu_\theta(s)}
	\nabla_\theta \mu_\theta(s)
	\right].
	\label{eq:dpg_actor_update_ch9}
\end{equation}
This is the foundation of DDPG \citep{silver2014deterministic,lillicrap2016continuous}. TD3 improves DDPG by using clipped double critics, delayed policy updates, and target policy smoothing \citep{fujimoto2018addressing}. Clipped double critics reduce overestimation by taking the smaller of two target Q estimates, echoing the overestimation problem discussed for Double DQN. Delayed policy updates let the critic move closer to a useful estimate before the actor follows its gradient. Target policy smoothing adds small noise to the target action so that the critic learns a smoother Q surface around the action selected by the actor. SAC uses a stochastic actor and an entropy-regularized critic, making exploration part of the objective \citep{haarnoja2018soft}.

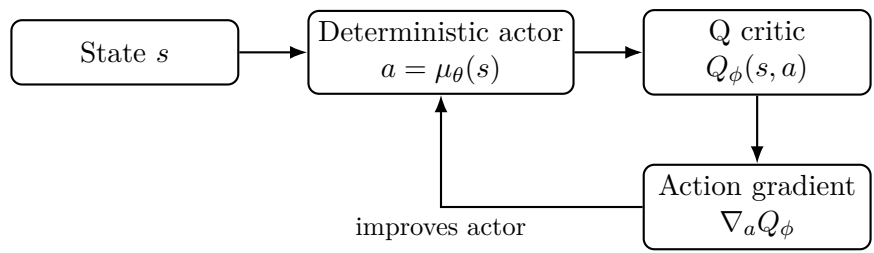
\begin{figure}[t]
	\centering
	\begin{tikzpicture}[
		box/.style={draw,rounded corners,thick,minimum width=3.0cm,minimum height=0.85cm,align=center},
		arrow/.style={-{Latex[length=2.3mm]},thick},
		node distance=0.9cm
		]
		\node[box] (state) {State $s$};
		\node[box, right=of state] (actor) {Deterministic actor\\$a=\mu_\theta(s)$};
		\node[box, right=of actor] (critic) {Q critic\\$Q_\phi(s,a)$};
		\node[box, below=of critic] (grad) {Action gradient\\$\nabla_a Q_\phi$};
		\draw[arrow] (state) -- (actor);
		\draw[arrow] (actor) -- (critic);
		\draw[arrow] (critic) -- (grad);
		\draw[arrow] (grad) -| node[below,font=\small] {improves actor} (actor);
	\end{tikzpicture}
	\caption{Deterministic actor-critic. The critic is differentiable with respect to the action, so the actor can be updated by following the action-gradient of the critic. This is the core idea behind DDPG and TD3.}
	\label{fig:deterministic_ac}
\end{figure}

\begin{table}[t]
	\centering
	\caption{Continuous-control actor-critic algorithms previewed in this chapter.}
	\label{tab:continuous_ac_preview}
	\begin{tabular}{p{2.2cm}p{3.6cm}p{5.7cm}}
		\toprule
		Algorithm & Actor type & Key critic idea \\
		\midrule
		DDPG & Deterministic & Single Q critic, replay buffer, target networks \\
		TD3 & Deterministic & Two Q critics, clipped target, delayed actor update \\
		SAC & Stochastic & Soft Q critic with entropy-regularized Bellman target \\
		PPO & Stochastic & Usually state-value critic with clipped on-policy actor update \\
		\bottomrule
	\end{tabular}
\end{table}

Chapter~10 studies PPO as the practical on-policy workhorse. Chapter~11 studies SAC and maximum-entropy RL in depth. Here, the key message is that these are not separate species: they are all actor-critic methods with different choices of actor, critic, data regime, and stabilization mechanism.

\section{Practical failure modes and debugging}

Actor-critic methods are powerful because they combine policy optimization and value learning. They are fragile for the same reason. There are two learning systems coupled together. If the critic is poor, the actor receives bad gradients. If the actor changes too fast, the critic's data distribution shifts. If entropy collapses early, exploration dies. If advantages are badly scaled, the actor update becomes too large or too small.

\begin{table}[t]
	\centering
	\caption{Common actor-critic failure modes and diagnostics.}
	\label{tab:ac_failures}
	\begin{tabular}{p{3.0cm}p{4.0cm}p{4.6cm}}
		\toprule
		Symptom & Likely cause & What to inspect \\
		\midrule
		Value loss explodes & Bad target scale, missing terminal mask, too high learning rate & Rewards, dones, value targets, gradient norms \\
		Policy entropy collapses early & Too little exploration, entropy coefficient too small & Entropy curve, action distribution \\
		Actor improves then crashes & Critic bias, too-large policy steps & KL, advantage scale, value error \\
		Returns high variance & Poor baseline or unstable critic & Explained variance, GAE lambda, rollout length \\
		No learning & Rewards too sparse, advantages near zero, wrong log-probs & Reward scale, log-prob computation, action sampling \\
		Continuous actor saturates & Tanh actions at bounds, poor action scaling & Pre-tanh means, action histogram \\
		Unsafe actions persist & Reward does not encode constraints, no safety critic/filter & Cost violations, constraint metrics, safety layer \\
		\bottomrule
	\end{tabular}
\end{table}

\begin{pitfallbox}{The critic can make the actor confidently wrong}
	A bad critic is worse than no critic when it produces systematic wrong advantages. The actor will not simply learn slowly; it may learn the wrong behavior efficiently. This is why value diagnostics are not optional in actor-critic experiments.
\end{pitfallbox}

\section[UAV/SDN actor-critic example]{UAV/SDN worked example: actor-critic control with QoS and safety}
\sectionmark{UAV/SDN actor-critic example}

Consider an SDN-assisted UAV network. The actor outputs a continuous action:
\begin{equation}
	a_t = [\Delta x, \Delta y, \Delta z, b_A, b_B, b_C, p_{tx}],
\end{equation}
where \(\Delta x, \Delta y, \Delta z\) are movement commands, \(b_A, b_B, b_C\) are bandwidth shares for user classes A/B/C, and \(p_{tx}\) is transmit power. The critic estimates the value of the current network state.

A state-value critic might estimate
\begin{equation}
	V_\phi(s_t) \approx \E[\text{future QoS reward} - \text{energy cost} - \text{safety penalties}].
\end{equation}
A safety critic might estimate expected future constraint cost:
\begin{equation}
	V^c_\psi(s_t) \approx \E\left[\sum_{k\geq t}\gamma^{k-t} c_k\right],
\end{equation}
where \(c_k\) counts latency violations, collision risk, or battery-safety violations.

A concrete numerical interpretation helps. Suppose that in a congested cell the reward critic predicts \(V_r(s_t)=18.0\), meaning that the current state is expected to yield about 18 units of future QoS-adjusted reward. Suppose the safety critic predicts \(V_c(s_t)=0.30\), meaning that the expected discounted constraint cost is 0.30. If one sampled action produces a realized reward-to-go of 22.5 and a realized cost-to-go of 0.55, then its reward advantage is positive, \(A_r=22.5-18.0=4.5\), but its cost advantage is also positive, \(A_c=0.55-0.30=0.25\). In a Lagrangian actor update, the action is attractive for QoS but unattractive for safety. The actor should therefore increase this action only if the reward gain justifies the additional constraint risk after applying the current safety multiplier.

\begin{table}[t]
	\centering
	\caption{Concrete actor-critic interpretation for UAV/SDN control.}
	\label{tab:uav_ac_example}
\begin{tabular}{p{2.9cm}p{8.6cm}}
		\toprule
		Component & UAV/SDN interpretation \\
		\midrule
		Actor & Maps network state to UAV movement, bandwidth shares, and power control \\
		Reward critic & Predicts long-term QoS, throughput, latency, and energy trade-off \\
		Safety critic & Predicts future constraint risk such as collision, battery depletion, or URLLC violations \\
		Advantage & Measures whether the chosen action performed better than expected in that traffic state \\
		Entropy & Keeps alternative routing/movement decisions alive during uncertain traffic changes \\
		Terminal mask & Prevents bootstrapping after drone failure, episode end, or mission reset \\
		\bottomrule
	\end{tabular}
\end{table}

\begin{researchbox}{Research-grade extension}
	A distinctive UAV/SDN actor-critic system can use multiple critics: one reward critic, one cost critic, and one risk critic. The actor is then updated using a Lagrangian or safety-filtered objective. This creates a bridge between actor-critic learning, constrained RL, and control barrier functions.
\end{researchbox}

\begin{figure}[t]
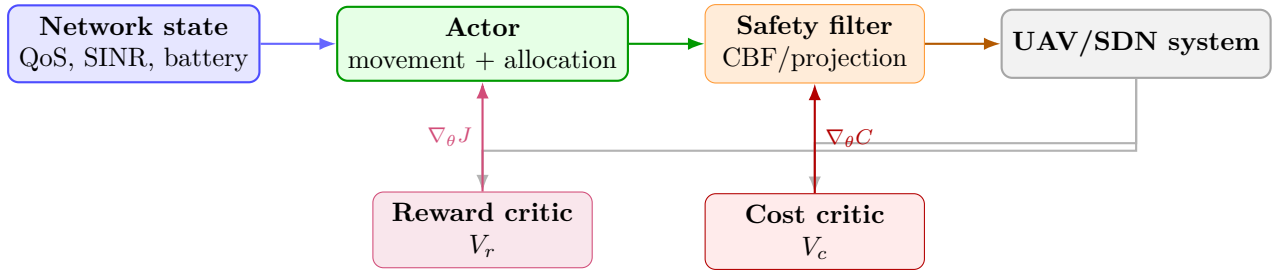

	\centering
	%
	\caption{A UAV/SDN actor--critic architecture. The actor proposes continuous control actions; a safety filter can modify actions before execution; reward and cost critics provide learning signals. This architecture connects actor--critic learning to safe control and constrained network optimization.}
	\label{fig:uav_sdn_ac_architecture}
\end{figure}

\section{Python implementation}

This section gives compact PyTorch code for a practical on-policy actor-critic. The code is intentionally minimal but includes the implementation details that most often cause silent bugs: terminal masks, log-probabilities, detached advantages, entropy, and value loss.

\subsection{Actor-critic network for discrete actions}

\Needspace{18\baselineskip}
\begin{lstlisting}[style=pythonstyle,caption={A shared-trunk actor-critic network for discrete actions.},label={lst:ac_network_discrete}]
import torch
import torch.nn as nn
import torch.nn.functional as F
from torch.distributions import Categorical

class ActorCriticDiscrete(nn.Module):
    def __init__(self, obs_dim, act_dim, hidden=128):
        super().__init__()
        self.trunk = nn.Sequential(
            nn.Linear(obs_dim, hidden), nn.Tanh(),
            nn.Linear(hidden, hidden), nn.Tanh(),
        )
        self.policy_head = nn.Linear(hidden, act_dim)
        self.value_head = nn.Linear(hidden, 1)

    def forward(self, obs):
        h = self.trunk(obs)
        logits = self.policy_head(h)
        value = self.value_head(h).squeeze(-1)
        return logits, value

    def act(self, obs):
        logits, value = self.forward(obs)
        dist = Categorical(logits=logits)
        action = dist.sample()
        log_prob = dist.log_prob(action)
        entropy = dist.entropy()
        return action, log_prob, entropy, value

    def evaluate_actions(self, obs, actions):
        logits, value = self.forward(obs)
        dist = Categorical(logits=logits)
        log_probs = dist.log_prob(actions)
        entropy = dist.entropy()
        return log_probs, entropy, value
\end{lstlisting}

\subsection{Rollout storage and GAE}

\Needspace{18\baselineskip}
\begin{lstlisting}[style=pythonstyle,caption={GAE computation for actor-critic rollouts.},label={lst:ac_gae}]
def compute_gae(rewards, values, dones, gamma=0.99, lam=0.95):
    """Compute GAE advantages and value targets.

    rewards: tensor [T]
    values: tensor [T + 1], including bootstrap value at final state
    dones: tensor [T], 1 if transition ended an episode
    """
    T = len(rewards)
    advantages = torch.zeros(T, dtype=torch.float32)
    last_gae = 0.0

    for t in reversed(range(T)):
        nonterminal = 1.0 - float(dones[t])
        delta = rewards[t] + gamma * values[t + 1] * nonterminal - values[t]
        last_gae = delta + gamma * lam * nonterminal * last_gae
        advantages[t] = last_gae

    returns = advantages + values[:-1]
    return advantages, returns
\end{lstlisting}

\subsection{A2C-style update}

\Needspace{18\baselineskip}
\begin{lstlisting}[style=pythonstyle,caption={A2C-style actor-critic update.},label={lst:a2c_update}]
def a2c_update(model, optimizer, obs, actions, old_log_probs,
               returns, advantages, value_coef=0.5, entropy_coef=0.01,
               max_grad_norm=0.5):
    """One actor-critic update from a rollout batch."""
    log_probs, entropy, values = model.evaluate_actions(obs, actions)

    # Important: advantages are treated as targets for the actor.
    adv = advantages.detach()
    adv = (adv - adv.mean()) / (adv.std(unbiased=False) + 1e-8)

    policy_loss = -(log_probs * adv).mean()
    value_loss = 0.5 * (returns.detach() - values).pow(2).mean()
    entropy_bonus = entropy.mean()

    loss = policy_loss + value_coef * value_loss - entropy_coef * entropy_bonus

    optimizer.zero_grad(set_to_none=True)
    loss.backward()
    nn.utils.clip_grad_norm_(model.parameters(), max_grad_norm)
    optimizer.step()

    stats = {
        "loss": float(loss.detach()),
        "policy_loss": float(policy_loss.detach()),
        "value_loss": float(value_loss.detach()),
        "entropy": float(entropy_bonus.detach()),
        "adv_mean": float(advantages.mean()),
        "adv_std": float(advantages.std(unbiased=False)),
    }
    return stats
\end{lstlisting}

\subsection{Continuous actor and value critic for UAV control}

\Needspace{20\baselineskip}
\begin{lstlisting}[style=pythonstyle,caption={Continuous actor-critic policy head for UAV/SDN control.},label={lst:uav_actor_critic}]
from torch.distributions import Normal

class UAVActorCritic(nn.Module):
    """Continuous actor-critic for UAV movement and resource allocation."""
    def __init__(self, obs_dim, hidden=256):
        super().__init__()
        self.trunk = nn.Sequential(
            nn.Linear(obs_dim, hidden), nn.ReLU(),
            nn.Linear(hidden, hidden), nn.ReLU(),
        )
        # Action: dx, dy, dz, raw bandwidth logits for A/B/C, transmit power
        self.mean_head = nn.Linear(hidden, 7)
        self.log_std = nn.Parameter(torch.ones(7) * -0.5)
        self.value_head = nn.Linear(hidden, 1)

    def forward(self, obs):
        h = self.trunk(obs)
        mean = self.mean_head(h)
        std = self.log_std.exp().expand_as(mean)
        value = self.value_head(h).squeeze(-1)
        return mean, std, value

    def sample_action(self, obs):
        mean, std, value = self.forward(obs)
        dist = Normal(mean, std)
        raw = dist.rsample()
        log_prob = dist.log_prob(raw).sum(dim=-1)

        # Movement commands bounded to [-1, 1]
        move = torch.tanh(raw[..., 0:3])

        # Bandwidth shares normalized with softmax
        bw = torch.softmax(raw[..., 3:6], dim=-1)

        # Power bounded to [0, 1]
        power = torch.sigmoid(raw[..., 6:7])
        action = torch.cat([move, bw, power], dim=-1)
        entropy = dist.entropy().sum(dim=-1)
        return action, log_prob, entropy, value
\end{lstlisting}

This UAV actor is deliberately structured. Movement is bounded with \(\tanh\), bandwidth shares sum to one through a softmax, and transmit power is bounded by a sigmoid. This is better than asking a generic Gaussian to learn physical constraints from reward penalties alone.

\subsection{Soft target updates for off-policy actor-critic}

\Needspace{14\baselineskip}
\begin{lstlisting}[style=pythonstyle,caption={Soft target-network update used in DDPG, TD3, and SAC.},label={lst:soft_update}]
def soft_update(source, target, tau=0.005):
    """Polyak averaging: target <- tau*source + (1-tau)*target."""
    with torch.no_grad():
        for p, p_targ in zip(source.parameters(), target.parameters()):
            p_targ.data.mul_(1.0 - tau)
            p_targ.data.add_(tau * p.data)
\end{lstlisting}

\section{Algorithm box: on-policy advantage actor-critic}

\begin{tcolorbox}[
	title={Algorithm 9.1: On-policy advantage actor-critic},
	colback=blue!15,
	colframe=blue!60!black,
	colupper=black,
	fonttitle=\bfseries
]
	\begin{enumerate}[leftmargin=*]
		\item Initialize actor parameters \(\theta\) and critic parameters \(\phi\).
		\item For each iteration:
		\begin{enumerate}[leftmargin=*]
			\item Collect a rollout using \(a_t\sim \pi_\theta(\cdot\mid s_t)\).
			\item Store \((s_t,a_t,r_t,d_t,\log\pi_\theta(a_t\mid s_t),V_\phi(s_t))\).
			\item Bootstrap the final value \(V_\phi(s_T)\), unless the episode ended.
			\item Compute advantages using TD, n-step returns, or GAE.
			\item Update the critic by minimizing value prediction error.
			\item Update the actor using \(-\log\pi_\theta(a_t\mid s_t)\hat{A}_t\).
			\item Optionally add entropy bonus and gradient clipping.
		\end{enumerate}
	\end{enumerate}
\end{tcolorbox}

\section{What actor-critic teaches us}

Actor-critic learning is not merely a compromise between policy gradients and value learning. It is a modular design pattern. Once the actor and critic are separated, we can change the critic target, add entropy, introduce replay, add double critics, add cost critics, share or separate representations, use recurrent memory, or insert safety filters.

This modularity explains why actor-critic algorithms dominate modern continuous-control and policy-optimization research. PPO, SAC, TD3, IMPALA, and many safe RL methods are all actor-critic systems with different answers to the same questions:
\begin{enumerate}[leftmargin=*]
	\item What does the actor output?
	\item What does the critic estimate?
	\item Is the data on-policy or off-policy?
	\item How is the critic target constructed?
	\item How large can the actor update be?
	\item How is exploration maintained?
	\item How are safety constraints represented?
\end{enumerate}

\section{Exercises}

\subsection*{Conceptual exercises}
\begin{enumerate}[leftmargin=*]
	\item Explain in your own words why actor-critic methods usually have lower variance than REINFORCE.
	\item Why can a bad critic make policy learning worse rather than merely slower?
	\item Compare A2C and A3C. Why is A2C often easier to reproduce?
	\item Why is a state-value critic common in PPO, while a Q critic is common in DDPG, TD3, and SAC?
	\item In a UAV/SDN problem, what would a safety critic estimate? Give three possible safety costs.
	\item When would you prefer separate actor and critic networks instead of a shared trunk? Discuss computation, representation interference, and critic accuracy.
\end{enumerate}

\subsection*{Mathematical exercises}
\begin{enumerate}[leftmargin=*]
	\item Starting from the baseline policy-gradient estimator, show how replacing \(G_t-b(s_t)\) by a TD error leads to the one-step actor-critic update.
	\item Derive the deterministic policy-gradient update in Eq.~\eqref{eq:dpg_actor_update_ch9} using the chain rule.
	\item For a two-step trajectory, write the one-step TD target and the two-step target for \(V(s_0)\). Which one bootstraps sooner?
	\item Suppose \(V_\phi(s_t)=5.0\), \(r_t=1.2\), \(V_\phi(s_{t+1})=4.5\), \(\gamma=0.99\), and the next state is nonterminal. Compute the TD advantage.
	\item Suppose the same transition is terminal. Recompute the TD advantage and explain the difference.
\end{enumerate}

\subsection*{Implementation exercises}
\begin{enumerate}[leftmargin=*]
	\item Modify Listing~\ref{lst:a2c_update} to log the approximate KL divergence between old and new policies.
	\item Add value clipping to the critic loss, similar to PPO implementations.
	\item Implement a shared-trunk A2C agent and log actor loss, critic loss, entropy, and explained variance separately.
	\item Extend Listing~\ref{lst:uav_actor_critic} with a separate safety critic head.
	\item Implement n-step return computation with terminal masks.
	\item Plot entropy, value loss, policy loss, and average return for a simple actor-critic agent. Which curve would you inspect first if learning collapses?
\end{enumerate}

\section*{Looking Ahead to Chapter 10: Food for Thought}
\addcontentsline{toc}{section}{Looking Ahead to Chapter 10: Food for Thought}

Actor-critic methods combine policy learning and value learning, but they introduce a new danger: the actor can change too much after one update. If the policy changes too aggressively, the data collected under the old policy may no longer be useful, the critic may become inaccurate, and performance may collapse.

Chapter~10 studies PPO, one of the most widely used practical actor-critic algorithms. PPO asks a simple question:
\begin{quote}
	How can we update the policy enough to improve it, but not so much that learning becomes unstable?
\end{quote}

Before moving on, consider:
\begin{enumerate}[leftmargin=*]
	\item How large should a policy update be?
	\item How can we measure whether a new policy is too different from the old one?
	\item Why is a clipped objective easier to implement than a constrained trust-region method?
	\item Why did PPO become especially important for RLHF and large-scale policy optimization?
\end{enumerate}

% \section{Chapter references}
	\chapter[PPO: The Practical Workhorse]{Proximal Policy Optimization: The Practical Workhorse}
\label{ch:ppo}
\chaptermark{PPO: The Practical Workhorse}

\begin{keybox}{Chapter goal}
	This chapter explains why Proximal Policy Optimization (PPO) became one of the most widely used deep reinforcement learning algorithms. PPO is not the most theoretically elegant policy-gradient method, nor is it always the most sample-efficient. Its importance comes from a rare combination: it is simple enough to implement, stable enough to train, flexible across discrete and continuous actions, and adaptable to robotics, games, network control, and language-model alignment. By the end of the chapter, the reader should understand the clipped surrogate objective, KL control, GAE-based PPO, value clipping, entropy bonuses, implementation details, diagnostics, PPO for UAV/SDN control, and the connection between PPO, RLHF, GRPO, and modern reasoning-oriented RL.
\end{keybox}

\section*{Chapter Overview}
\addcontentsline{toc}{section}{Chapter Overview}

\begin{enumerate}[leftmargin=*]
	\item Why PPO became the practical workhorse
	\item The problem PPO tries to solve: destructive policy updates
	\item From TRPO to PPO
	\item The probability ratio
	\item The clipped surrogate objective
	\item PPO as actor-critic with GAE
	\item Value loss, entropy bonus, and the full PPO objective
	\item PPO-Clip, PPO-Penalty, and KL control
	\item Minibatches, epochs, and the danger of stale on-policy data
	\item Practical implementation details that matter
	\item Python implementation
	\item PPO diagnostics and debugging
	\item PPO for UAV/SDN control
	\item PPO in RLHF, GRPO, and reasoning models
	\item Limitations of PPO
	\item Exercises
	\item Looking Ahead to Chapter 11
	\item Chapter references
\end{enumerate}

\section{Why PPO became the practical workhorse}

The previous chapters built the path to PPO. Chapter~7 introduced direct policy learning. Chapter~8 showed how baselines, reward-to-go, and advantage functions reduce policy-gradient variance. Chapter~9 combined policy learning and value learning into actor-critic methods. PPO is one of the most important practical outcomes of this progression.

PPO was introduced by Schulman et al. as a family of policy-gradient methods that alternate between collecting data with the current policy and optimizing a surrogate objective for several epochs using minibatches. The design goal was pragmatic: obtain many of the practical stability benefits of Trust Region Policy Optimization (TRPO) while avoiding the complexity of second-order constrained optimization \citep{schulman2017ppo,schulman2015trpo}. In the original PPO work, the clipped objective gave a simple first-order method that performed well across simulated robotic control and Atari environments \citep{schulman2017ppo}.

PPO became a workhorse because it occupies a useful middle ground. REINFORCE is simple but too noisy. TRPO is stable but more complex. DDPG and TD3 are more sample-efficient in continuous control but more sensitive to replay-buffer and critic errors. SAC is powerful and sample-efficient, but introduces entropy temperature tuning and off-policy design choices. PPO is on-policy, conceptually clean, and often robust enough to be a strong default baseline.

This comparison also previews why PPO is not the only actor-critic design. TD3, for example, improves DDPG through three stabilizers: clipped double critics reduce overestimation in the same spirit as Double DQN; delayed actor updates give the critic more time to stabilize before the actor follows its gradients; and target policy smoothing makes the learned Q-surface less brittle around a chosen continuous action. PPO chooses a different trade-off: it avoids replay-buffer instability by staying close to on-policy data, at the cost of lower sample efficiency.

\begin{keybox}{Core intuition}
	A vanilla policy gradient says: ``move the policy in the direction that improves the advantage estimate.'' PPO adds: ``but do not move it too far in one update.''
\end{keybox}

This simple idea matters because neural policies can change dramatically after one gradient step. A policy update that looks beneficial on sampled data can destroy performance by making the new policy too different from the policy that generated the data. PPO controls this by comparing the new policy to the old policy through a probability ratio.

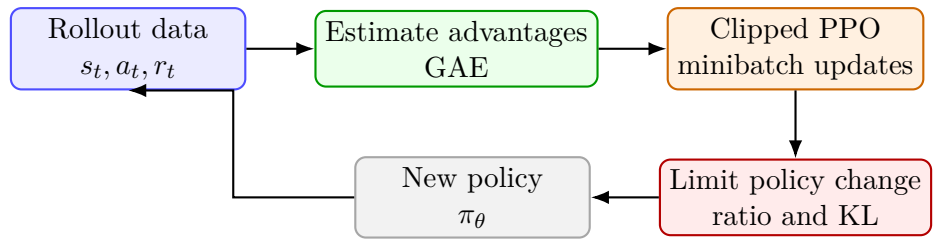
\begin{figure}[t]
	\centering
	\begin{tikzpicture}[
		box/.style={draw,rounded corners,thick,minimum width=3.1cm,minimum height=0.85cm,align=center},
		small/.style={draw,rounded corners,minimum width=2.5cm,minimum height=0.75cm,align=center},
		arrow/.style={-{Latex[length=2.4mm]},thick},
		node distance=0.9cm
		]
		\node[box, fill=blue!8, draw=blue!70] (rollout) {Rollout data\\$s_t,a_t,r_t$};
		\node[box, right=of rollout, fill=green!8, draw=green!60!black] (adv) {Estimate advantages\\GAE};
		\node[box, right=of adv, fill=orange!12, draw=orange!80!black] (update) {Clipped PPO\\minibatch updates};
		\node[box, below=of update, fill=red!8, draw=red!70!black] (limit) {Limit policy change\\ratio and KL};
		\node[box, left=of limit, fill=gray!10, draw=gray!70] (newpol) {New policy\\$\pi_\theta$};

		\draw[arrow] (rollout) -- (adv);
		\draw[arrow] (adv) -- (update);
		\draw[arrow] (update) -- (limit);
		\draw[arrow] (limit) -- (newpol);
		\draw[arrow] (newpol.west) -- ++(-1.6,0) |- (rollout.south);
	\end{tikzpicture}
	\caption{PPO alternates between collecting on-policy data, estimating advantages, and performing several clipped minibatch updates. The clipping mechanism discourages policy updates that move too far from the behavior policy that collected the data.}
	\label{fig:ppo_loop_overview}
\end{figure}

\section{The problem PPO tries to solve: destructive policy updates}

Policy-gradient methods update parameters using sampled trajectories. But sampled trajectories are local evidence. They tell us how the old policy behaved, not necessarily how a very different new policy will behave. If the update is too large, the new policy may assign much higher probability to actions that looked good only because of noise, estimation error, or a lucky trajectory.

This problem is especially severe in deep RL because policy networks are nonlinear. A moderate parameter update can sharply change action probabilities. In continuous control, a Gaussian mean may move too far. In discrete control, action probabilities may collapse. In language-model RL, a token distribution may shift so far that generated completions move outside the region where the reward model or verifier is reliable.

\begin{pitfallbox}{A central PPO warning}
	PPO does not make the policy update mathematically safe. It only discourages large likelihood-ratio changes on sampled actions. If advantages are wrong, value estimates are poor, rewards are mis-scaled, or too many epochs are used on stale data, PPO can still fail.
\end{pitfallbox}

The destructive-update problem is the policy-gradient analogue of instability in value-based learning. In DQN, target networks and replay stabilize bootstrapped value learning. In PPO, clipping and KL diagnostics stabilize actor-critic policy optimization. These mechanisms do not prove optimality; they make training behave well enough in many practical settings.

\section{From TRPO to PPO}

TRPO was motivated by a theoretical performance-improvement bound. The idea is to improve the policy while constraining how far the new policy moves from the old policy, often using a KL-divergence trust region \citep{schulman2015trpo}. A simplified TRPO-style optimization problem is
\begin{align}
	\max_\theta \quad & \E_t\left[ r_t(\theta) \hat{A}_t \right], \\
	\text{subject to} \quad & \E_t\left[ \KL\left(\pi_{\theta_{\mathrm{old}}}(\cdot\mid s_t) \;\|\; \pi_\theta(\cdot\mid s_t)\right) \right] \leq \delta.
\end{align}
TRPO is powerful but more complex than ordinary gradient descent because it involves a constrained optimization problem and approximate second-order information. In practice, this connects to the natural-gradient view of policy optimization: the update direction is shaped by the geometry of the policy distribution, often through an approximation to a Fisher-information matrix inverse rather than a plain Euclidean gradient.

PPO keeps the trust-region spirit but replaces the hard constraint with a simpler objective. Instead of solving a constrained problem, PPO modifies the surrogate objective so that policy-ratio changes beyond a threshold stop providing additional improvement. This gives a first-order method that is much easier to implement with ordinary deep-learning optimizers.

\begin{figure}[t]
	\centering
	\begin{tikzpicture}[
		box/.style={draw,rounded corners,thick,minimum width=3.5cm,minimum height=0.9cm,align=center},
		arrow/.style={-{Latex[length=2.4mm]},thick},
		node distance=1.5cm
		]
		\node[box, fill=blue!8, draw=blue!70] (pg) {Vanilla policy gradient\\large update possible};
		\node[box, right=of pg, fill=green!8, draw=green!60!black] (trpo) {TRPO\\explicit KL trust region};
		\node[box, right=of trpo, fill=orange!12, draw=orange!80!black] (ppo) {PPO\\clipped surrogate objective};

		\draw[arrow] (pg) -- node[above,align=center,font=\small] {stability\\problem} (trpo);
		\draw[arrow] (trpo) -- node[above,align=center,font=\small] {simpler\\first-order\\ method} (ppo);

		\node[below=0.8cm of pg, align=center,font=\small] {Simple but fragile};
		\node[below=0.8cm of trpo, align=center,font=\small] {Stable but complex};
		\node[below=0.8cm of ppo, align=center,font=\small] {Stable enough and easy};
	\end{tikzpicture}
	\caption{PPO can be understood as a practical approximation to the trust-region idea. TRPO constrains the KL divergence explicitly; PPO uses clipping or KL penalties to discourage excessive policy movement while remaining easy to implement with first-order optimization.}
	\label{fig:trpo_to_ppo}
\end{figure}
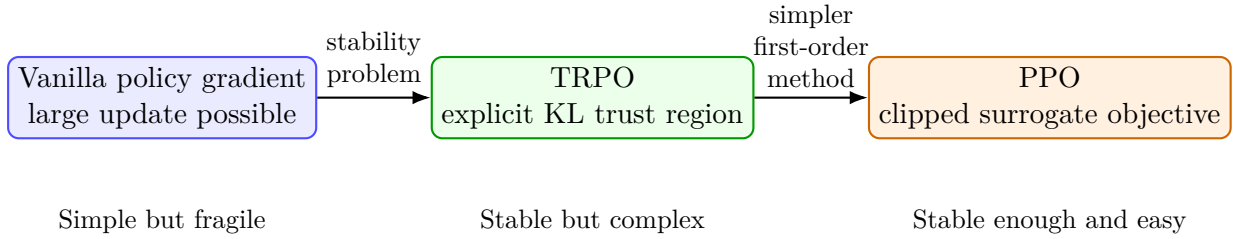

\section{The probability ratio}

The central quantity in PPO is the probability ratio
\begin{equation}
	r_t(\theta)
	=
	\frac{\pi_\theta(a_t\mid s_t)}{\pi_{\theta_{\mathrm{old}}}(a_t\mid s_t)}.
	\label{eq:ppo_ratio}
\end{equation}
This ratio compares how likely the new policy is to choose the sampled action relative to the old policy that generated the data. In implementation, the ratio is usually computed in log-space for numerical stability:
\begin{equation}
	r_t(\theta)
	=
	\exp\left(
	\log \pi_\theta(a_t\mid s_t)
	-
	\log \pi_{\theta_{\mathrm{old}}}(a_t\mid s_t)
	\right).
	\label{eq:ppo_log_ratio}
\end{equation}
This avoids explicitly dividing very small probabilities, which is especially important for Gaussian policies and language-model policies whose token probabilities can be tiny.

If \(r_t(\theta)=1\), the new policy assigns the same probability to the sampled action as the old policy. If \(r_t(\theta)>1\), the new policy has increased the probability of that action. If \(r_t(\theta)<1\), the new policy has decreased it.

The unclipped policy-gradient surrogate is
\begin{equation}
	L^{\mathrm{PG}}(\theta)
	=
	\E_t\left[r_t(\theta)\hat{A}_t\right].
\end{equation}
This is the same importance-ratio form used in policy gradients: increase the likelihood of actions with positive advantage and decrease the likelihood of actions with negative advantage.

The problem is that \(r_t(\theta)\) can become too large or too small. PPO modifies this objective by clipping the ratio.

\section{The clipped surrogate objective}

PPO-Clip uses the objective
\begin{equation}
	L^{\mathrm{CLIP}}(\theta)
	=
	\E_t
	\left[
	\min\left(
	r_t(\theta)\hat{A}_t,
	\clip(r_t(\theta),1-\epsilon,1+\epsilon)\hat{A}_t
	\right)
	\right].
	\label{eq:ppo_clip_objective}
\end{equation}
The hyperparameter \(\epsilon\), often around \(0.1\) to \(0.3\), controls the clipping range.

The minimum is important. PPO does not simply clip the ratio everywhere. It chooses the more conservative of the unclipped and clipped objectives. This creates a pessimistic surrogate: once the policy update improves the objective too much by moving the ratio outside the allowed range, the clipped term prevents further gain from that sample.

For a positive advantage, the action was better than expected. PPO wants to increase its probability, but not beyond \(1+\epsilon\). For a negative advantage, the action was worse than expected. PPO wants to decrease its probability, but not below \(1-\epsilon\). Equivalently, the gradient is allowed to correct harmful policy movement, but it stops rewarding further movement once the policy has already changed too far in the apparently beneficial direction. This asymmetry is the reason the clipped objective acts like a conservative lower-bound surrogate rather than a simple symmetric clipping operation.

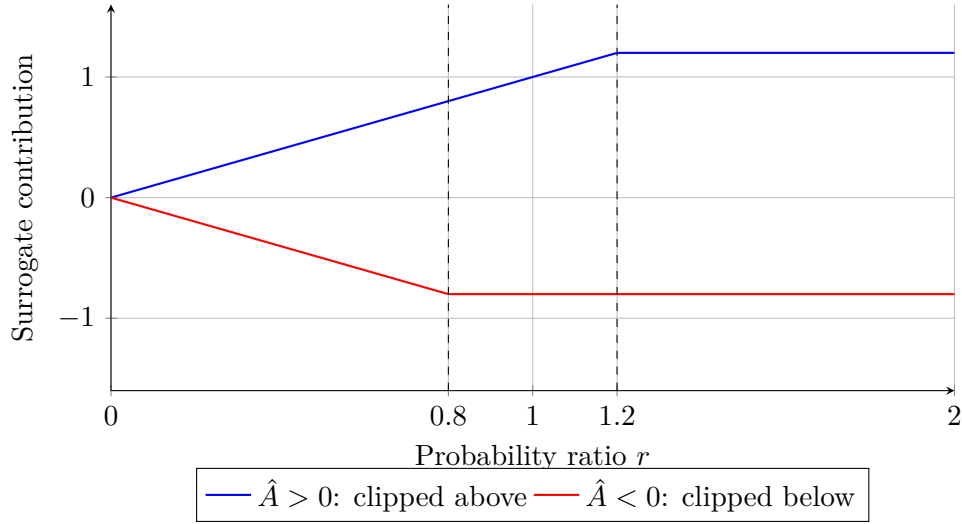
\begin{figure}[t]
	\centering
	\begin{tikzpicture}
		\begin{axis}[
			width=0.80\textwidth,
			height=0.42\textwidth,
			xlabel={Probability ratio $r$},
			ylabel={Surrogate contribution},
			xmin=0.0, xmax=2.0,
			ymin=-1.6, ymax=1.6,
			axis lines=left,
			grid=both,
			legend style={at={(0.5,-0.20)},anchor=north,legend columns=2},
			xtick={0,0.8,1.0,1.2,2.0},
			ytick={-1,0,1}
			]
			\addplot[blue, thick, domain=0:2, samples=200] {min(x,1.2)};
			\addlegendentry{$\hat{A}>0$: clipped above}
			\addplot[red, thick, domain=0:2, samples=200] {max(-x,-0.8)};
			\addlegendentry{$\hat{A}<0$: clipped below}
			\draw[dashed] (axis cs:0.8,-1.6) -- (axis cs:0.8,1.6);
			\draw[dashed] (axis cs:1.2,-1.6) -- (axis cs:1.2,1.6);
		\end{axis}
	\end{tikzpicture}
	\caption{The PPO clipped surrogate for positive and negative advantages when \(\epsilon=0.2\). Note the asymmetry: for positive advantages, the upper side is clipped, so increasing the ratio above \(1+\epsilon\) gives no further benefit; for negative advantages, the lower side is clipped, so decreasing the ratio below \(1-\epsilon\) gives no further benefit.}
	\label{fig:ppo_clip_curve}
\end{figure}

\begin{warningbox}{What clipping does and does not do}
	Clipping is not a hard KL constraint. The policy can still move substantially, especially in states or actions not represented in the minibatch. For this reason, practical PPO implementations monitor approximate KL divergence, clip fraction, entropy, value loss, and explained variance.
\end{warningbox}

\section{PPO as actor-critic with GAE}

In most deep-RL implementations, PPO is an actor-critic algorithm. The policy network is the actor. The value network is the critic. Rollouts are collected using the current policy, then advantages are estimated using generalized advantage estimation, introduced in Chapter~8:
\begin{equation}
	\hat{A}_t^{\mathrm{GAE}(\gamma,\lambda)}
	=
	\sum_{l=0}^{\infty}(\gamma\lambda)^l \delta_{t+l}.
\end{equation}
In practice this is computed by the backward recursion
\begin{equation}
	\hat{A}_t
	=
	\delta_t
	+
	\gamma\lambda(1-d_{t+1})\hat{A}_{t+1}.
\end{equation}

The value target is usually
\begin{equation}
	\hat{R}_t
	=
	\hat{A}_t + V_{\phi_{\mathrm{old}}}(s_t).
\end{equation}
The actor is updated by the PPO clipped objective. The critic is updated by regression to \(\hat{R}_t\). The actor and critic losses are conceptually separate even if they share a neural-network trunk.

\begin{figure}[t]
	\centering
	\begin{tikzpicture}[
		box/.style={draw,rounded corners,thick,minimum width=3.0cm,minimum height=0.85cm,align=center,font=\small},
		arrow/.style={-{Latex[length=2.3mm]},thick},
		node distance=0.85cm
		]
		\node[box, draw=blue!70, fill=blue!8] (roll) {Rollout with old policy\\$\pi_{\theta_{old}}$};
		\node[box, right=of roll, draw=green!60!black, fill=green!8] (gae) {GAE advantages\\$\hat{A}_t$};
		\node[box, right=of gae, draw=orange!80!black, fill=orange!12] (ratio) {Policy ratio\\$r_t(\theta)$};
		\node[box, below=of ratio, draw=red!70!black, fill=red!8] (actorloss) {Actor loss\\clipped surrogate};
		\node[box, below=of gae, draw=purple!70, fill=purple!8] (value) {Value target\\$\hat{R}_t$};
		\node[box, below=of roll, draw=gray!70, fill=gray!10] (criticloss) {Critic loss\\MSE};

		\draw[arrow] (roll) -- (gae);
		\draw[arrow] (gae) -- (ratio);
		\draw[arrow] (ratio) -- (actorloss);
		\draw[arrow] (gae) -- (value);
		\draw[arrow] (value) -- (criticloss);
		\draw[arrow] (roll) -- (criticloss);
	\end{tikzpicture}
	\caption{PPO is usually implemented as an actor-critic method. The actor uses the clipped surrogate objective, while the critic learns a value function from GAE-based returns. The actor loss should not backpropagate through the advantage computation.}
	\label{fig:ppo_actor_critic_losses}
\end{figure}
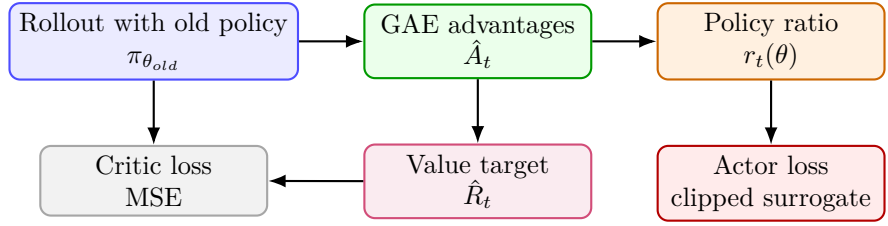

\section{Value loss, entropy bonus, and the full PPO objective}

A common PPO implementation minimizes a total loss of the form
\begin{equation}
	\mathcal{L}(\theta,\phi)
	=
	\mathcal{L}_{\mathrm{policy}}(\theta)
	+
	c_v \mathcal{L}_{\mathrm{value}}(\phi)
	-
	c_e \E_t\left[\mathcal{H}(\pi_\theta(\cdot\mid s_t))\right],
	\label{eq:ppo_total_loss}
\end{equation}
where \(c_v\) weights the critic loss and \(c_e\) weights the entropy bonus. The entropy term plays the same exploration role discussed in Chapter~8, but in PPO it is usually an auxiliary bonus rather than the central maximum-entropy objective used later in SAC. Because most deep-learning libraries minimize losses, the policy loss is usually the negative clipped objective:
\begin{equation}
	\mathcal{L}_{\mathrm{policy}}(\theta)
	=
	-\E_t\left[
	\min\left(
	r_t(\theta)\hat{A}_t,
	\clip(r_t(\theta),1-\epsilon,1+\epsilon)\hat{A}_t
	\right)
	\right].
\end{equation}
The critic loss is often
\begin{equation}
	\mathcal{L}_{\mathrm{value}}(\phi)
	=
	\frac{1}{2}\E_t\left[(V_\phi(s_t)-\hat{R}_t)^2\right].
\end{equation}
Some implementations also use value clipping, analogous to policy-ratio clipping:
\begin{align}
	V^{\mathrm{clip}}_\phi(s_t)
	&=
	V_{\phi_{\mathrm{old}}}(s_t)
	+
	\clip\left(V_\phi(s_t)-V_{\phi_{\mathrm{old}}}(s_t),-\epsilon_v,\epsilon_v\right), \\
	\mathcal{L}_{\mathrm{value}}^{\mathrm{clip}}
	&=
	\frac{1}{2}\E_t\left[
	\max\left(
	(V_\phi(s_t)-\hat{R}_t)^2,
	(V^{\mathrm{clip}}_\phi(s_t)-\hat{R}_t)^2
	\right)
	\right].
\end{align}

\begin{pitfallbox}{Stop-gradient is not optional}
	The advantages and value targets should be treated as fixed targets during the PPO update. In code, they should be detached from the computation graph. If the policy loss backpropagates through the advantage or return computation, the actor can corrupt the critic target and the update no longer matches the PPO objective.
\end{pitfallbox}

\section{PPO-Clip, PPO-Penalty, and KL control}

The most widely used version is PPO-Clip, but the original PPO paper also considered a KL-penalty form \citep{schulman2017ppo}. A KL-penalty objective can be written as
\begin{equation}
	L^{\mathrm{KLPEN}}(\theta)
	=
	\E_t\left[r_t(\theta)\hat{A}_t
	-
	\beta\KL\left(\pi_{\theta_{\mathrm{old}}}(\cdot\mid s_t)\;\|\;\pi_\theta(\cdot\mid s_t)\right)
	\right].
\end{equation}
Here \(\beta\) controls how strongly policy movement is penalized. Some implementations adapt \(\beta\) to keep the measured KL near a target.

Even PPO-Clip implementations often monitor KL and stop updates early if the KL becomes too high. A common approximate KL for sampled actions is
\begin{equation}
	\widehat{\KL}
	\approx
	\E_t\left[\log \pi_{\theta_{\mathrm{old}}}(a_t\mid s_t)-\log \pi_\theta(a_t\mid s_t)\right].
\end{equation}
Another useful diagnostic is the clip fraction: the fraction of samples for which \(r_t(\theta)\) lies outside \([1-\epsilon,1+\epsilon]\). A very high clip fraction means many samples are saturated and the update may no longer be extracting useful gradient information.

\begin{table}[t]
	\centering
	\caption{PPO mechanisms and what they control.}
	\label{tab:ppo_mechanisms}
	\begin{tabular}{p{0.25\textwidth}p{0.38\textwidth}p{0.25\textwidth}}
		\toprule
		Mechanism & Problem controlled & Cost or caveat \\
		\midrule
		Ratio clipping & Prevents large likelihood-ratio improvement on sampled actions & Does not enforce a hard KL bound \\
		KL early stopping & Detects excessive policy movement & Requires choosing a target KL \\
		GAE & Reduces policy-gradient variance & Introduces bias controlled by \(\lambda\) \\
		Entropy bonus & Prevents premature deterministic collapse & Too much entropy slows exploitation \\
		Advantage normalization & Stabilizes optimization scale & Small minibatches give noisy statistics \\
		Value clipping & Prevents abrupt critic changes & Can slow value learning \\
		Multiple epochs & Improves data reuse & Too many epochs make data stale \\
		\bottomrule
	\end{tabular}
\end{table}

\section{Minibatches, epochs, and stale on-policy data}

PPO is often described as an on-policy algorithm, but it does not use each sample only once. It collects a rollout with \(\pi_{\theta_{\mathrm{old}}}\), then performs multiple epochs of minibatch optimization on that rollout. This is one reason PPO is more sample-efficient than one-update REINFORCE or A2C.

However, this reuse has a danger. After several epochs, the current policy may become meaningfully different from the behavior policy that generated the data. This is the K-epoch dilemma: using the same rollout for multiple optimization epochs improves data efficiency, but it also makes the data increasingly stale. PPO's reuse of each batch across \(K\) epochs is therefore partially off-policy within a single update; the clipped objective is what prevents this stale data from destabilizing the policy too quickly, but it does not remove the mismatch completely. 

\begin{warningbox}{PPO is only approximately on-policy during optimization}
	A rollout is on-policy when collected. After several optimization epochs, the policy has changed. PPO is best understood as controlled reuse of recently on-policy data, not as fully off-policy learning.
\end{warningbox}

This is why PPO has important engineering hyperparameters: rollout length, number of environments, minibatch size, number of epochs, clip range, target KL, learning rate, entropy coefficient, value coefficient, and advantage normalization. Empirical studies have shown that such implementation choices can substantially affect performance, sometimes as much as the high-level algorithm itself \citep{engstrom2020implementation,andrychowicz2021what,huang2022details}.

\section{Practical implementation details that matter}

PPO is famous for being simple, but the simple mathematical objective hides many implementation choices. This is one reason different PPO implementations can behave differently even when they use the same clipped objective.

\begin{table}[t]
	\centering
	\caption{Common PPO implementation details and why they matter.}
	\label{tab:ppo_implementation_details}
	\begin{tabular}{p{0.28\textwidth}p{0.57\textwidth}}
		\toprule
		Detail & Why it matters \\
		\midrule
		Store old log-probabilities & The ratio must compare the new policy against the policy that generated the data. \\
		Detach advantages and returns & Prevents policy loss from backpropagating through critic targets. \\
		Normalize advantages over rollout batch & Stabilizes policy-gradient scale. \\
		Use terminal masks in GAE & Prevents bootstrapping across episode boundaries. \\
		Track approximate KL & Detects destructive policy movement even with clipping. \\
		Track clip fraction & Reveals whether most samples are saturated. \\
		Separate actor and critic learning rates if needed & Critic instability can poison the policy update. \\
		Initialize log standard deviation carefully & Too small causes deterministic collapse; too large causes noisy behavior. \\
		Use observation/reward normalization when appropriate & Reduces scale sensitivity in continuous control. \\
		Evaluate without exploration noise only when policy semantics allow it & Stochastic policies may need mean-action evaluation for continuous control. \\
		\bottomrule
	\end{tabular}
\end{table}

The key lesson from modern PPO studies is that the implementation is part of the algorithm. Engstrom et al. showed that code-level choices can explain a large part of PPO's empirical behavior relative to TRPO \citep{engstrom2020implementation}. Andrychowicz et al. studied many on-policy actor-critic choices at scale and found that practical design decisions strongly affect results \citep{andrychowicz2021what}. Huang et al. documented a large checklist of PPO implementation details, emphasizing that faithful reproduction requires attention to these details \citep{huang2022details}.

\section{Python implementation}

This section gives compact PyTorch-style code for PPO. The code is not a full training framework, but it shows the core computations that must be correct.

\subsection{Categorical and Gaussian policies}

\Needspace{14\baselineskip}
\begin{lstlisting}[style=pythonstyle,caption={Categorical and diagonal-Gaussian policy helpers.},label={lst:ppo_policy_helpers}]
import torch
import torch.nn as nn
import torch.nn.functional as F
from torch.distributions import Categorical, Normal, Independent

class ActorCritic(nn.Module):
    def __init__(self, obs_dim, action_dim, discrete=True):
        super().__init__()
        self.discrete = discrete
        self.trunk = nn.Sequential(
            nn.Linear(obs_dim, 128), nn.Tanh(),
            nn.Linear(128, 128), nn.Tanh(),
        )
        if discrete:
            self.pi = nn.Linear(128, action_dim)
        else:
            self.mu = nn.Linear(128, action_dim)
            self.log_std = nn.Parameter(torch.zeros(action_dim))
        self.v = nn.Linear(128, 1)

    def dist_and_value(self, obs):
        h = self.trunk(obs)
        value = self.v(h).squeeze(-1)
        if self.discrete:
            logits = self.pi(h)
            dist = Categorical(logits=logits)
        else:
            mu = self.mu(h)
            std = self.log_std.exp().expand_as(mu)
            dist = Independent(Normal(mu, std), 1)
        return dist, value

    def act(self, obs):
        dist, value = self.dist_and_value(obs)
        action = dist.sample()
        logp = dist.log_prob(action)
        entropy = dist.entropy()
        return action, logp, entropy, value
\end{lstlisting}

For continuous control, this simple Gaussian policy assumes unconstrained actions. If the environment uses bounded actions, the policy output should be scaled, clipped, or transformed carefully. Squashed Gaussian policies require a log-probability correction as discussed in Chapter~7.

\subsection{GAE and PPO minibatch loss}

\Needspace{16\baselineskip}
\begin{lstlisting}[style=pythonstyle,caption={GAE computation for PPO rollouts.},label={lst:ppo_gae}]
def compute_gae(rewards, values, dones, gamma=0.99, lam=0.95):
    """Compute GAE advantages and returns.

    rewards: [T, N]
    values:  [T + 1, N]
    dones:   [T, N], 1 if terminal else 0
    """
    T = rewards.shape[0]
    advantages = torch.zeros_like(rewards)
    last_gae = torch.zeros_like(rewards[0])

    for t in reversed(range(T)):
        nonterminal = 1.0 - dones[t].float()
        delta = rewards[t] + gamma * values[t + 1] * nonterminal - values[t]
        last_gae = delta + gamma * lam * nonterminal * last_gae
        advantages[t] = last_gae

    returns = advantages + values[:-1]
    return advantages, returns
\end{lstlisting}

\Needspace{18\baselineskip}
\begin{lstlisting}[style=pythonstyle,caption={Core PPO clipped loss with value loss and entropy bonus.},label={lst:ppo_loss}]
def ppo_loss(model, obs, actions, old_logp, advantages, returns,
             clip_eps=0.2, vf_coef=0.5, ent_coef=0.01,
             value_clip_eps=None, old_values=None):
    dist, values = model.dist_and_value(obs)
    new_logp = dist.log_prob(actions)
    entropy = dist.entropy().mean()

    # Probability ratio r_t(theta). Old log-probs must be detached.
    log_ratio = new_logp - old_logp.detach()
    ratio = torch.exp(log_ratio)

    # Advantages are fixed targets for the actor.
    adv = advantages.detach()
    pg_unclipped = ratio * adv
    pg_clipped = torch.clamp(ratio, 1.0 - clip_eps, 1.0 + clip_eps) * adv
    policy_loss = -torch.min(pg_unclipped, pg_clipped).mean()

    target_returns = returns.detach()
    if value_clip_eps is not None and old_values is not None:
        v_clipped = old_values + torch.clamp(
            values - old_values, -value_clip_eps, value_clip_eps
        )
        v_loss_unclipped = (values - target_returns).pow(2)
        v_loss_clipped = (v_clipped - target_returns).pow(2)
        value_loss = 0.5 * torch.max(v_loss_unclipped, v_loss_clipped).mean()
    else:
        value_loss = 0.5 * (values - target_returns).pow(2).mean()

    total_loss = policy_loss + vf_coef * value_loss - ent_coef * entropy

    with torch.no_grad():
        approx_kl = (old_logp.detach() - new_logp).mean()
        clip_frac = ((ratio < 1.0 - clip_eps) | (ratio > 1.0 + clip_eps)).float().mean()

    stats = {
        "policy_loss": float(policy_loss.detach()),
        "value_loss": float(value_loss.detach()),
        "entropy": float(entropy.detach()),
        "approx_kl": float(approx_kl.detach()),
        "clip_frac": float(clip_frac.detach()),
    }
    return total_loss, stats
\end{lstlisting}

\subsection{A compact PPO update loop}

\Needspace{20\baselineskip}
\begin{lstlisting}[style=pythonstyle,caption={PPO minibatch update with KL early stopping.},label={lst:ppo_update}]
def normalize(x, eps=1e-8):
    return (x - x.mean()) / (x.std(unbiased=False) + eps)

def ppo_update(model, optimizer, batch, epochs=10, minibatch_size=256,
               clip_eps=0.2, target_kl=0.02):
    obs = batch["obs"]
    actions = batch["actions"]
    old_logp = batch["logp"]
    old_values = batch["values"]
    returns = batch["returns"]
    advantages = normalize(batch["advantages"])

    n = obs.shape[0]
    inds = torch.arange(n)
    last_stats = {}

    for epoch in range(epochs):
        perm = inds[torch.randperm(n)]
        for start in range(0, n, minibatch_size):
            mb = perm[start:start + minibatch_size]
            loss, stats = ppo_loss(
                model,
                obs[mb], actions[mb], old_logp[mb],
                advantages[mb], returns[mb],
                clip_eps=clip_eps,
                value_clip_eps=clip_eps,
                old_values=old_values[mb],
            )
            optimizer.zero_grad(set_to_none=True)
            loss.backward()
            nn.utils.clip_grad_norm_(model.parameters(), max_norm=0.5)
            optimizer.step()
            last_stats = stats

        # Stop if the whole epoch has already moved the policy too far.
        if last_stats.get("approx_kl", 0.0) > 1.5 * target_kl:
            break

    return last_stats
\end{lstlisting}

\begin{keybox}{Why this code matters}
	The most important lines are not the neural-network layers. They are the old log-probabilities, detached advantages, clipped ratio, terminal masks in GAE, approximate KL, and clip fraction. These are the parts that make PPO different from a generic actor-critic update.
\end{keybox}

\section{PPO diagnostics and debugging}

PPO is popular partly because it exposes useful diagnostics. A serious implementation should log more than episode return.

\begin{table}[t]
	\centering
	\caption{PPO diagnostics and how to interpret them.}
	\label{tab:ppo_diagnostics}
	\begin{tabular}{p{0.23\textwidth}p{0.35\textwidth}p{0.30\textwidth}}
		\toprule
		Diagnostic & Suspicious pattern & Possible cause \\
		\midrule
		Approximate KL & Jumps above target & Learning rate too high; too many epochs; advantage scale too large \\
		Clip fraction & Near zero & Updates too small; learning rate too low \\
		Clip fraction & Very high & Most samples saturated; stale data; large policy step \\
		Entropy & Collapses early & Exploration loss too weak; reward scale too strong \\
		Value loss & Explodes & Bad returns; missing terminal masks; critic learning rate too high \\
		Explained variance & Near zero or negative & Critic not learning useful values \\
		Gradient norm & Explodes or becomes zero & Learning rate too high; bad advantage scale; saturated network \\
		Advantage mean/std & Extreme scale & Reward normalization or value function problem \\
		Constraint violations & Improve reward but violate safety & Reward objective and safety objective misaligned \\
		\bottomrule
	\end{tabular}
\end{table}

\begin{figure}[t]
	\centering
	\begin{tikzpicture}
		\begin{axis}[
			width=0.82\textwidth,
			height=0.42\textwidth,
			xlabel={Training update},
			ylabel={Diagnostic value},
			xmin=0,xmax=100,
			ymin=0,ymax=1.2,
			grid=both,
			legend style={at={(0.5,-0.20)},anchor=north,legend columns=3}
			]
			\addplot[blue, thick, domain=0:100, samples=100] {0.15 + 0.04*sin(deg(x/6))};
			\addlegendentry{approx. KL}
			\addplot[orange!80!black, thick, domain=0:100, samples=100] {0.25 + 0.12*sin(deg(x/8+1))};
			\addlegendentry{clip fraction}
			\addplot[green!60!black, thick, domain=0:100, samples=100] {0.85*exp(-x/90)+0.15};
			\addlegendentry{entropy}
		\end{axis}
	\end{tikzpicture}
	\caption{PPO should be monitored through multiple diagnostics, not only reward. Approximate KL, clip fraction, and entropy reveal whether the policy update is too weak, too strong, or collapsing prematurely.}
	\label{fig:ppo_diagnostics}
\end{figure}
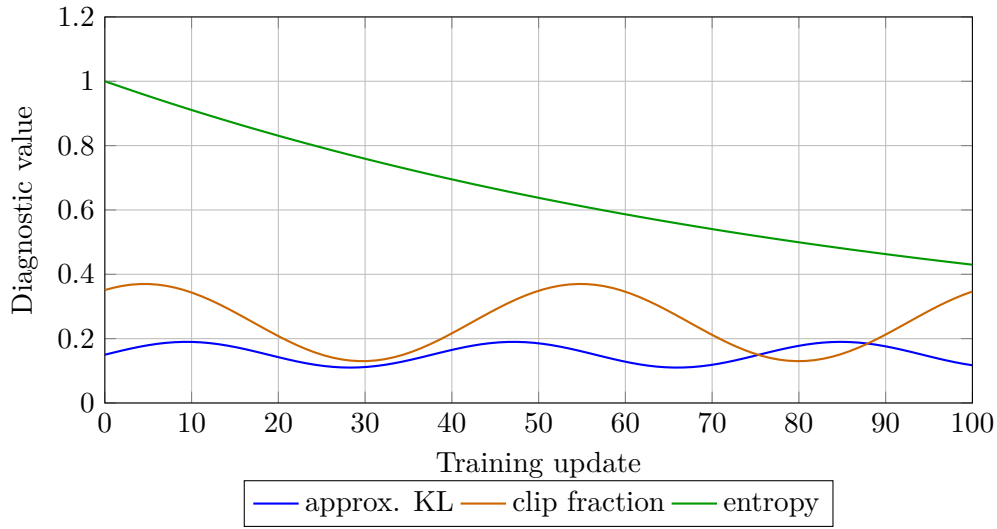

\section{PPO for UAV/SDN control}

PPO is a natural candidate for UAV/SDN control when the action space is continuous or hybrid. A UAV may output continuous movement commands, transmission power, service allocation weights, or routing proportions. DQN-style discretization can become unnatural, while PPO can directly optimize a stochastic policy over continuous actions.

Consider a UAV-assisted SDN controller with an action vector
\begin{equation}
	a_t
	=
	[\Delta x,\Delta y,\Delta z, p_{tx}, b_A,b_B,b_C],
\end{equation}
where \(\Delta x,\Delta y,\Delta z\) are movement commands, \(p_{tx}\) is transmit power, and \(b_A,b_B,b_C\) are bandwidth shares for three service classes. A PPO actor can output means and standard deviations for movement and power, plus a normalized allocation head for bandwidth.

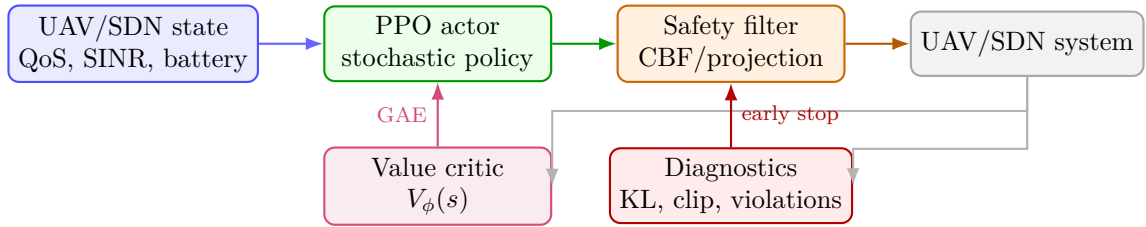
\begin{figure}[t]
	\centering
	\begin{tikzpicture}[
		box/.style={draw,rounded corners,thick,minimum width=3.0cm,minimum height=0.85cm,align=center,font=\small},
		arrow/.style={-{Latex[length=2.3mm]},thick},
		node distance=0.85cm
		]
		\node[box, draw=blue!70, fill=blue!8] (state) {UAV/SDN state\\QoS, SINR, battery};
		\node[box, right=of state, draw=green!60!black, fill=green!8] (ppo) {PPO actor\\stochastic policy};
		\node[box, right=of ppo, draw=orange!80!black, fill=orange!12] (filter) {Safety filter\\CBF/projection};
		\node[box, right=of filter, draw=gray!70, fill=gray!10] (env) {UAV/SDN system};
		\node[box, below=of ppo, draw=purple!70, fill=purple!8] (critic) {Value critic\\$V_\phi(s)$};
		\node[box, below=of filter, draw=red!70!black, fill=red!8] (diag) {Diagnostics\\KL, clip, violations};

		\draw[arrow, draw=blue!60] (state) -- (ppo);
		\draw[arrow, draw=green!60!black] (ppo) -- (filter);
		\draw[arrow, draw=orange!70!black] (filter) -- (env);
		\draw[arrow, draw=gray!60] (env.south) -- ++(0,-0.45) -| (critic.east);
		\draw[arrow, draw=gray!60] (env.south) -- ++(0,-0.95) -| (diag.east);
		\draw[arrow, draw=purple!70] (critic.north) -- node[left,font=\scriptsize,text=purple!70] {GAE} (ppo.south);
		\draw[arrow, draw=red!70!black] (diag.north) -- node[right,font=\scriptsize,text=red!70!black] {early stop} (filter.south);
	\end{tikzpicture}
	\caption{A PPO-style UAV/SDN control architecture. PPO proposes stochastic continuous actions, a safety filter can project unsafe actions before execution, and diagnostics monitor both optimization stability and safety.}
	\label{fig:ppo_uav_sdn_architecture}
\end{figure}

\begin{researchbox}{PPO plus safety filters}
	In safety-critical UAV/SDN systems, PPO should not be the only safety mechanism. A practical architecture can use PPO for learning a high-performance stochastic policy, while a control barrier function or projection layer modifies actions before execution. PPO then learns from the consequences of safety-filtered actions. This connects policy optimization with safe control, but it also creates a subtle learning issue: the action sampled by the actor may differ from the action executed by the system. The logged action, log-probability, and reward attribution must be handled carefully. 
\end{researchbox}

\subsection{Concrete PPO reward example}

Suppose the UAV reward at time \(t\) is
\begin{equation}
	r_t
	=
	2.0\,\mathrm{QoS}_t
	-
	0.5\,\mathrm{Energy}_t
	-
	3.0\,\mathrm{Violation}_t.
\end{equation}
If the UAV achieves \(\mathrm{QoS}_t=0.85\), uses \(\mathrm{Energy}_t=0.40\), and produces no safety violation, then
\begin{equation}
	r_t = 2.0(0.85)-0.5(0.40)-3.0(0)=1.50.
\end{equation}
If the value critic expected \(V(s_t)=1.10\), the immediate advantage-like surprise is positive. PPO increases the likelihood of the movement and allocation pattern, but clipping prevents the policy from overcommitting to it after a single rollout. 

\Needspace{15\baselineskip}
\begin{lstlisting}[style=pythonstyle,caption={A structured PPO actor head for UAV/SDN continuous and allocation actions.},label={lst:ppo_uav_actor}]
class UAVPPOActor(nn.Module):
    def __init__(self, obs_dim):
        super().__init__()
        self.trunk = nn.Sequential(
            nn.Linear(obs_dim, 128), nn.Tanh(),
            nn.Linear(128, 128), nn.Tanh(),
        )
        self.move_mu = nn.Linear(128, 3)      # dx, dy, dz
        self.power_mu = nn.Linear(128, 1)     # transmit power command
        self.alloc_logits = nn.Linear(128, 3) # class A/B/C bandwidth shares
        self.log_std = nn.Parameter(torch.zeros(4))
        self.value = nn.Linear(128, 1)

    def forward(self, obs):
        h = self.trunk(obs)
        mu = torch.cat([self.move_mu(h), self.power_mu(h)], dim=-1)
        std = self.log_std.exp().expand_as(mu)
        cont_dist = Independent(Normal(mu, std), 1)
        alloc = torch.softmax(self.alloc_logits(h), dim=-1)
        v = self.value(h).squeeze(-1)
        return cont_dist, alloc, v
\end{lstlisting}

The allocation head is deterministic in this simple sketch. A fully stochastic hybrid actor would also define a distribution over allocations, such as a Dirichlet policy or a logistic-normal distribution. The correct choice depends on whether the bandwidth shares should be explored as random actions or computed as deterministic functions of the sampled movement and power.

\section{PPO in RLHF, GRPO, and reasoning models}

PPO became especially visible outside robotics because it was used in reinforcement learning from human feedback. In the InstructGPT pipeline, a language model was first supervised-fine-tuned, then a reward model was trained from human preferences, and finally the policy was optimized using PPO-style RLHF \citep{ouyang2022training}. In this setting, the policy is the language model, the critic is often a value head, and the reward combines a learned preference reward with a KL penalty to a reference model.

A simplified RLHF-style objective is
\begin{equation}
	R_{\mathrm{RLHF}}(x,y)
	=
	R_{\mathrm{RM}}(x,y)
	-
	\beta
	\KL\left(\pi_\theta(\cdot\mid x)\;\|\;\pi_{\mathrm{ref}}(\cdot\mid x)\right),
\end{equation}
where \(R_{\mathrm{RM}}\) is the reward-model score and \(\pi_{\mathrm{ref}}\) is a reference policy. This KL penalty plays a role similar in spirit to PPO's trust-region idea: it discourages the optimized policy from moving too far from a known reference distribution.

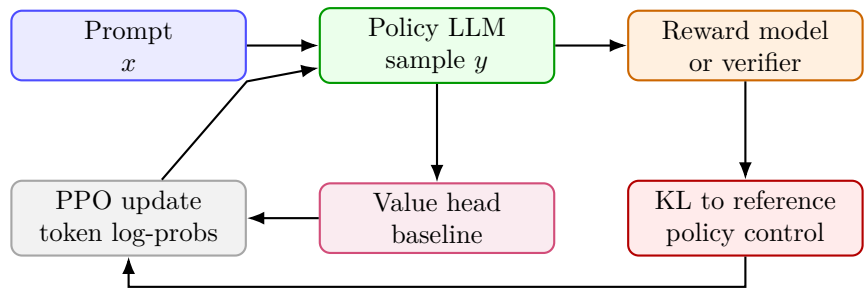
\begin{figure}[t]
	\centering
	\begin{tikzpicture}[
		box/.style={draw,rounded corners,thick,minimum width=3.1cm,minimum height=0.85cm,align=center,font=\small},
		arrow/.style={-{Latex[length=2.3mm]},thick},
		node distance=0.95cm
		]
		\node[box, draw=blue!70, fill=blue!8] (prompt) {Prompt\\$x$};
		\node[box, right=of prompt, draw=green!60!black, fill=green!8] (policy) {Policy LLM\\sample $y$};
		\node[box, right=of policy, draw=orange!80!black, fill=orange!12] (reward) {Reward model\\or verifier};
		\node[box, below=1.3cm of policy, draw=purple!70, fill=purple!8] (value) {Value head\\baseline};
		\node[box, below=1.3cm of reward, draw=red!70!black, fill=red!8] (kl) {KL to reference\\policy control};
		\node[box, below=1.3cm of prompt, draw=gray!70, fill=gray!10] (ppo) {PPO update\\token log-probs};

		\draw[arrow] (prompt) -- (policy);
		\draw[arrow] (policy) -- (reward);
		\draw[arrow] (policy) -- (value);
		\draw[arrow] (reward) -- (kl);
		\draw[arrow] (kl.south) -- ++(0,-0.4) -| (ppo.south);
		\draw[arrow] (value) -- (ppo);
		\draw[arrow] (ppo) -- (prompt.south east) -- (policy);
	\end{tikzpicture}
	\caption{PPO-style RLHF for language models. A policy model samples completions, a reward model or verifier scores them, a value head estimates a baseline, and a KL term discourages drifting too far from a reference model.}
	\label{fig:ppo_rlhf_pipeline}
\end{figure}

Recent reasoning-oriented language-model work extended this policy-optimization thread. DeepSeekMath introduced Group Relative Policy Optimization (GRPO), a PPO variant that avoids a separate critic by using group-relative rewards for multiple completions of the same prompt \citep{shao2024deepseekmath}. DeepSeek-R1 and related reasoning models showed that reinforcement learning can incentivize reasoning behaviors such as verification, self-reflection, and strategy adaptation \citep{deepseek2025r1}. OpenAI's o1 system card similarly describes large-scale reinforcement learning for reasoning-oriented models \citep{jaech2024openai}.

A simplified group-relative advantage for completion \(y_i\) from prompt \(x\) can be written as
\begin{equation}
	\hat{A}_i
	=
	\frac{R(x,y_i)-\mu_g}{\sigma_g+\varepsilon},
	\qquad
	\mu_g=\frac{1}{K}\sum_{j=1}^K R(x,y_j),
	\label{eq:grpo_group_advantage}
\end{equation}
where \(\mu_g\) and \(\sigma_g\) are the mean and standard deviation of rewards within the group of completions sampled for the same prompt. This makes the baseline prompt-local: a completion is reinforced because it is better than its siblings, not merely because its absolute reward is high. 

\begin{table}[t]
	\centering
	\caption{PPO, RLHF-PPO, and GRPO-style optimization.}
	\label{tab:ppo_grpo_comparison}
	\small
	\begin{tabular}{p{0.17\textwidth}p{0.25\textwidth}p{0.24\textwidth}p{0.18\textwidth}}
		\toprule
		Method & Policy update control & Baseline / critic & Typical domain \\
		\midrule
		PPO-Clip & Ratio clipping & Value critic & Robotics, games, control \\
		RLHF-PPO & PPO plus KL to reference & Value head & LLM alignment \\
		GRPO-style & Group-relative norm. plus PPO-like ratios & Group baseline; often no critic & Reasoning RL \\
		\bottomrule
	\end{tabular}
\end{table}

\begin{warningbox}{Modern LLM note}
	PPO and GRPO are not magic alignment methods. They optimize the reward signal they are given. If the reward model, verifier, or rule-based score is incomplete, the policy may exploit it. This is the language-model version of reward hacking.
\end{warningbox}

\section{Limitations of PPO}

PPO is popular, but it is not universally best. Its main limitations are:
\begin{enumerate}[leftmargin=*]
	\item \textbf{On-policy sample cost.} PPO discards old data after limited reuse. Off-policy methods such as SAC can be much more sample-efficient.
	\item \textbf{Sensitivity to implementation details.} Normalization, initialization, advantage computation, KL monitoring, and rollout structure matter greatly \citep{engstrom2020implementation,huang2022details}.
	\item \textbf{No hard safety guarantee.} Clipping controls policy ratios on sampled actions, not physical safety or constraint satisfaction.
	\item \textbf{Clip saturation.} If many ratios are clipped, the gradient can become uninformative.
	\item \textbf{Weak exploration in sparse rewards.} PPO's entropy bonus may not be enough for hard-exploration tasks.
	\item \textbf{Critic dependence.} Bad value estimates produce bad advantages, which can mislead the actor.
	\item \textbf{Reward hacking.} In RLHF or network-control settings, PPO may exploit imperfect reward functions.
\end{enumerate}

\begin{keybox}{What PPO really teaches}
	PPO is not important because clipping is a perfect theoretical solution. PPO is important because it shows how much practical stability can be obtained by combining policy ratios, conservative updates, actor-critic learning, GAE, normalization, entropy bonuses, and careful diagnostics.
\end{keybox}

\section{Exercises}

\subsection*{Conceptual exercises}
\begin{enumerate}[leftmargin=*]
	\item Explain why PPO is often described as a practical approximation to TRPO.
	\item What does the probability ratio \(r_t(\theta)\) measure? Why is it central to PPO?
	\item Why does PPO clip the objective differently for positive and negative advantages?
	\item Explain why PPO is on-policy but still performs multiple epochs of minibatch optimization on the same rollout.
	\item Why does PPO not provide a hard safety guarantee for UAV control?
	\item In RLHF, why is a KL penalty to a reference policy useful?
\end{enumerate}

\subsection*{Mathematical exercises}
\begin{enumerate}[leftmargin=*]
	\item For \(\epsilon=0.2\), compute the clipped ratio for \(r=0.6,0.9,1.1,1.5\).
	\item Suppose \(\hat{A}_t=2.0\), \(r_t=1.4\), and \(\epsilon=0.2\). Compute the unclipped and clipped PPO objective contributions.
	\item Suppose \(\hat{A}_t=-3.0\), \(r_t=0.5\), and \(\epsilon=0.2\). Compute the unclipped and clipped contributions. Which one does the minimum choose?
	\item Show why the clipped surrogate is conservative for positive advantages when \(r_t>1+\epsilon\).
	\item Derive the PPO policy loss used in code when the optimizer minimizes rather than maximizes.
\end{enumerate}

\subsection*{Coding exercises}
\begin{enumerate}[leftmargin=*]
	\item Implement the PPO clipped loss for a categorical policy.
	\item Add value clipping to an existing PPO implementation.
	\item Log approximate KL, clip fraction, entropy, and explained variance during training.
	\item Modify the UAV actor in Listing~\ref{lst:ppo_uav_actor} to make bandwidth allocation stochastic using a Dirichlet distribution.
	\item Add KL early stopping to a PPO update loop and test how it changes training stability.
\end{enumerate}

\subsection*{Research thinking exercises}
\begin{enumerate}[leftmargin=*]
	\item In UAV/SDN control, what should be handled by the PPO reward and what should be handled by a safety filter?
	\item How can PPO fail if the safety filter changes the executed action but the log-probability corresponds to the unfiltered action?
	\item Compare PPO and SAC for a continuous-control networking task. Which would you choose and why?
	\item For language-model RL, compare PPO with GRPO-style group-relative optimization. What is gained and what is lost by removing the critic?
\end{enumerate}

\section*{Looking Ahead to Chapter 11: Food for Thought}
\addcontentsline{toc}{section}{Looking Ahead to Chapter 11: Food for Thought}

PPO controls policy updates by clipping probability ratios and monitoring KL divergence. It is stable and flexible, but it remains on-policy and can require many environment interactions. Chapter~11 turns to Soft Actor-Critic (SAC), a different actor-critic philosophy.

The central questions for the next chapter are:
\begin{enumerate}[leftmargin=*]
	\item What if exploration is not just an auxiliary entropy bonus, but part of the main objective?
	\item Can an actor-critic method be both stochastic and off-policy?
	\item How does maximum-entropy RL connect reward maximization to robustness and uncertainty?
	\item Why is SAC often more sample-efficient than PPO in continuous-control tasks?
	\item What new difficulties appear when learning from a replay buffer with stochastic actors and critics?
\end{enumerate}

\begin{quote}
	Chapter~10 showed how PPO makes policy-gradient learning stable enough to become a practical default. Chapter~11 shows how maximum-entropy actor-critic learning makes stochastic exploration a central part of the objective itself.
\end{quote}

% \section{Chapter references}
	\chapter[SAC and Maximum-Entropy RL]{Soft Actor-Critic and Maximum-Entropy Reinforcement Learning}
\label{ch:sac}
\chaptermark{SAC and Maximum-Entropy RL}

\begin{keybox}{Chapter thesis}
	Soft Actor-Critic (SAC) is not merely ``actor-critic with entropy.'' It is the point where three ideas meet: off-policy learning, stochastic continuous-control policies, and maximum-entropy decision-making. SAC became important because it solved a practical tension: PPO is robust but sample-hungry, while deterministic off-policy methods such as DDPG and TD3 can be sample-efficient but brittle. SAC offers a stochastic, entropy-regularized, off-policy alternative that is often stable enough for real continuous-control systems.
\end{keybox}

\section*{Chapter Overview}
\addcontentsline{toc}{section}{Chapter Overview}

\begin{enumerate}[leftmargin=*]
	\item Why Chapter 11 matters
	\item From ordinary return to maximum-entropy return
	\item Entropy as exploration, robustness, and regularization
	\item Soft value functions and the soft Bellman equations
	\item From soft policy iteration to SAC
	\item The SAC architecture
	\item The squashed Gaussian policy
	\item Critic learning and clipped double Q targets
	\item Actor learning through the reparameterization trick
	\item Automatic entropy-temperature tuning
	\item The complete SAC algorithm
	\item Concrete PyTorch implementation
	\item Diagnostics, debugging, and implementation traps
	\item SAC for UAV/SDN continuous control
	\item Beyond vanilla SAC: TQC, REDQ, DroQ, discrete SAC, and 2026 directions
	\item Limitations and when not to use SAC
	\item Exercises
	\item Looking Ahead to Chapter 12
\end{enumerate}

\section{Why Chapter 11 matters}

The previous chapters developed the policy-gradient and actor-critic view of reinforcement learning. Chapter~7 explained why one may learn a policy directly. Chapter~8 introduced REINFORCE, baselines, advantages, and Generalized Advantage Estimation. Chapter~9 showed how actors and critics cooperate. Chapter~10 then explained why PPO became the practical workhorse: it is easy to implement, robust across many tasks, and well-suited to on-policy optimization.

However, PPO has a major weakness: it is typically sample-inefficient. It collects fresh data, performs several epochs of updates, and then discards or heavily devalues those trajectories. In robotics, wireless control, SDN, UAV control, and physical systems, collecting interaction data can be expensive. This motivates off-policy actor-critic algorithms, which can reuse past experience from a replay buffer.

Before SAC, popular off-policy continuous-control methods included DDPG and TD3 \citep{lillicrap2016continuous,fujimoto2018addressing}. They are sample-efficient because they reuse replay data, but they rely on deterministic policies and can be sensitive to exploration noise, Q-function error, and hyperparameters. SAC was introduced to combine off-policy sample efficiency with stochastic policy learning under the maximum-entropy framework \citep{haarnoja2018soft,haarnoja2018apps}.

\begin{figure}[t]
	\centering
	\resizebox{\textwidth}{!}{%
		\begin{tikzpicture}[
			box/.style={draw,rounded corners,thick,minimum width=3.2cm,minimum height=0.9cm,align=center,font=\small},
			arrow/.style={-{Latex[length=2.3mm]},thick},
			node distance=1.9cm
			]
			\node[box,fill=blue!8,draw=blue!70] (ppo) {PPO\\{\normalfont on-policy, robust, sample-hungry}};
			\node[box,fill=orange!10,draw=orange!80!black,right=of ppo] (ddpg) {DDPG/TD3\\{\normalfont off-policy, deterministic, brittle}};
			\node[box,fill=green!10,draw=green!60!black,right=of ddpg] (sac) {SAC\\{\normalfont off-policy, stochastic, entropy-regularized}};
			\draw[arrow] (ppo) -- node[above,font=\scriptsize,align=center] {need reuse\\of data} (ddpg);
			\draw[arrow] (ddpg) -- node[above,font=\scriptsize,align=center] {need robust\\exploration} (sac);
		\end{tikzpicture}%
	}
	\caption{Why SAC appears after PPO, DDPG, and TD3. PPO is robust but often sample-inefficient; DDPG and TD3 reuse data but rely on deterministic policies and external exploration noise. SAC combines off-policy actor-critic learning with entropy-regularized stochastic control.}
	\label{fig:ppo_td3_sac_motivation}
\end{figure}
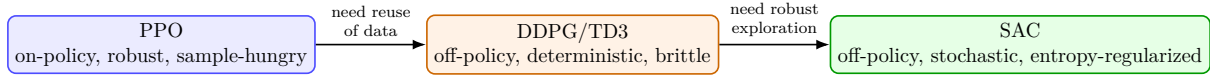

\begin{table}[t]
	\centering
	\caption{Where the main SAC components first appeared in the book. SAC is best understood as a synthesis of ideas introduced across earlier chapters.}
	\label{tab:sac_component_map}
	\begin{tabular}{p{0.25\textwidth}p{0.42\textwidth}p{0.23\textwidth}}
		\toprule
		SAC component & Earlier idea & Where it appeared \\
		\midrule
		Replay buffer & Reusing off-policy experience to stabilize and improve sample efficiency & Chapter~5 \\
		Twin critics / clipped double Q & Reducing overestimation by comparing multiple value estimates & Chapters~6 and~9 \\
		Entropy term & Keeping policies stochastic and preventing premature collapse & Chapter~8 \\
		Actor--critic structure & Coupling a policy learner with a value estimator & Chapter~9 \\
		Reparameterized continuous policy & Differentiating through sampled continuous actions & Chapter~7 \\
		Safety layer or constrained critic & Separating reward optimization from constraint satisfaction & Chapter~9 and Chapter~12 preview \\
		\bottomrule
	\end{tabular}
\end{table}

\section{From ordinary return to maximum-entropy return}

The standard reinforcement-learning objective maximizes expected discounted reward:
\begin{equation}
	J(\pi)
	=
	\E_{\tau\sim\pi}
	\left[
	\sum_{t=0}^{\infty}\gamma^t r(s_t,a_t)
	\right].
	\label{eq:standard_rl_objective_ch11}
\end{equation}
Maximum-entropy reinforcement learning modifies this objective by rewarding both task performance and policy entropy:
\begin{equation}
	J_{\mathrm{MaxEnt}}(\pi)
	=
	\E_{\tau\sim\pi}
	\left[
	\sum_{t=0}^{\infty}\gamma^t
	\left(
	r(s_t,a_t)
	+
	\alpha\mathcal{H}(\pi(\cdot\given s_t))
	\right)
	\right].
	\label{eq:maxent_objective_ch11}
\end{equation}
Equivalently, if an action is sampled from the policy, the entropy term can be written using the log-probability of that action:
\begin{equation}
	\mathcal{H}(\pi(\cdot\given s_t))
	=
	\E_{a_t\sim\pi}
	[-\log \pi(a_t\given s_t)].
\end{equation}
Thus the objective can also be viewed as maximizing
\begin{equation}
	\E
	\left[
	\sum_{t=0}^{\infty}\gamma^t
	\left(
	r(s_t,a_t)
	-
	\alpha \log \pi(a_t\given s_t)
	\right)
	\right].
	\label{eq:maxent_sampled_objective}
\end{equation}
The temperature parameter \(\alpha>0\) controls the trade-off between reward and entropy. A large \(\alpha\) encourages a highly stochastic policy. A small \(\alpha\) makes the objective closer to ordinary reward maximization.

\begin{keybox}{The SAC intuition}
	Ordinary RL asks: ``Which action gives high return?'' Maximum-entropy RL asks: ``Which policy gives high return while staying as random as possible among similarly good actions?'' SAC turns this principle into an off-policy actor-critic algorithm.
\end{keybox}

\section{Entropy as exploration, robustness, and regularization}

Entropy is often introduced as an exploration bonus, but in SAC it has a deeper role. It regularizes the policy toward uncertainty unless the Q-function provides strong evidence that one action is better. This has three consequences.

First, entropy improves exploration. The policy does not collapse too quickly to a narrow action distribution. Second, entropy improves robustness. If several actions have similar expected return, the policy can keep probability mass over all of them rather than committing prematurely. Third, entropy smooths optimization. The policy loss contains both a value-seeking term and a log-probability term, which prevents extremely sharp policies early in training.

This is especially useful in continuous control. A UAV does not simply choose ``left'' or ``right.'' It may choose a movement vector, a climb rate, a transmit-power adjustment, and a bandwidth allocation. A deterministic policy can become brittle when the state estimate is noisy. A stochastic policy can maintain controlled variability.

\begin{warningbox}{Entropy is not random behavior for its own sake}
	A common misunderstanding is that SAC wants the agent to be random forever. This is not correct. SAC rewards entropy only insofar as the temperature \(\alpha\) and target entropy make it useful. If an action is clearly better, the Q-function can dominate the entropy term. SAC encourages useful stochasticity, not permanent indecision.
\end{warningbox}

\section{Soft value functions and the soft Bellman equations}

In ordinary RL, the value of a state under policy \(\pi\) is the expected discounted sum of rewards. In maximum-entropy RL, the value includes entropy. The soft action-value function is
\begin{equation}
	Q^{\pi}_{\mathrm{soft}}(s_t,a_t)
	=
	\E_{\pi}
	\left[
	\sum_{k=t}^{\infty}\gamma^{k-t}
	\left(
	r(s_k,a_k)-\alpha\log\pi(a_k\given s_k)
	\right)
	\given s_t,a_t
	\right].
\end{equation}
The corresponding soft state-value function is
\begin{equation}
	V^{\pi}_{\mathrm{soft}}(s)
	=
	\E_{a\sim\pi(\cdot\given s)}
	\left[
	Q^{\pi}_{\mathrm{soft}}(s,a)
	-
	\alpha\log\pi(a\given s)
	\right].
	\label{eq:soft_v_function}
\end{equation}
The soft Bellman backup is therefore
\begin{equation}
	Q^{\pi}_{\mathrm{soft}}(s,a)
	=
	r(s,a)
	+
	\gamma
	\E_{s'\sim p(\cdot\given s,a)}
	\left[
	V^{\pi}_{\mathrm{soft}}(s')
	\right].
	\label{eq:soft_bellman_policy}
\end{equation}
Like the standard Bellman backup, the soft Bellman backup is a $\gamma$-contraction in the sup-norm, so soft policy evaluation converges to a unique fixed point for any bounded reward. This gives the soft Bellman equations the same fixed-point interpretation as the classical Bellman equations, but with entropy included in the value of future behavior.

For a learned critic, this structure leads directly to the SAC target:
\begin{equation}
	y
	=
	r
	+
	\gamma(1-d)
	\left(
	\min_{i=1,2}Q_{\bar\theta_i}(s',a')
	-
	\alpha\log\pi_{\phi}(a'\given s')
	\right),
	\label{eq:sac_target}
\end{equation}
where \(a'\sim\pi_{\phi}(\cdot\given s')\), \(d\) is a terminal flag, and \(\bar\theta_i\) are target critic parameters.

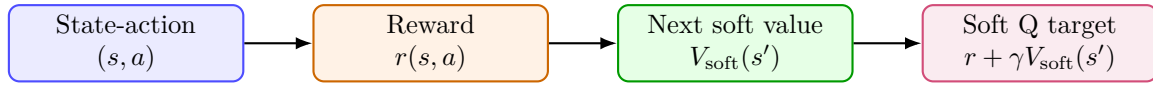
\begin{figure}[t]
	\centering
	\begin{tikzpicture}[
		box/.style={draw,rounded corners,thick,minimum width=3.1cm,minimum height=0.85cm,align=center,font=\small},
		arrow/.style={-{Latex[length=2.2mm]},thick},
		node distance=0.9cm
		]
		\node[box,fill=blue!8,draw=blue!70] (s) {State-action\\$(s,a)$};
		\node[box,fill=orange!10,draw=orange!80!black,right=of s] (r) {Reward\\$r(s,a)$};
		\node[box,fill=green!10,draw=green!60!black,right=of r] (next) {Next soft value\\$V_{\mathrm{soft}}(s')$};
		\node[box,fill=purple!8,draw=purple!70,right=of next] (q) {Soft Q target\\$r+\gamma V_{\mathrm{soft}}(s')$};
		\draw[arrow] (s) -- (r);
		\draw[arrow] (r) -- (next);
		\draw[arrow] (next) -- (q);
		\node[below=0.8cm of next,align=center,font=\small] {Soft value includes both expected return and entropy.};
	\end{tikzpicture}
	\caption{The soft Bellman backup. SAC changes the value target by subtracting the log-probability term, which rewards entropy in the policy.}
	\label{fig:soft_bellman_backup}
\end{figure}

\section{From soft policy iteration to SAC}

Soft policy iteration alternates between two conceptual steps \citep{haarnoja2018soft,haarnoja2018apps}. In soft policy evaluation, the algorithm estimates the soft Q-function for the current policy. In soft policy improvement, it updates the policy toward actions with high soft value.

For discrete actions, the optimal maximum-entropy policy has a Boltzmann form:
\begin{equation}
	\pi^*(a\given s)
	\propto
	\exp\left(\frac{1}{\alpha}Q^*(s,a)\right).
	\label{eq:boltzmann_policy}
\end{equation}
This equation is useful conceptually. It says that better actions receive exponentially larger probability, but suboptimal actions are not assigned zero probability unless \(\alpha\to0\). In continuous action spaces, normalizing this distribution exactly is usually intractable. SAC therefore learns a parameterized stochastic policy and optimizes it by gradient descent.

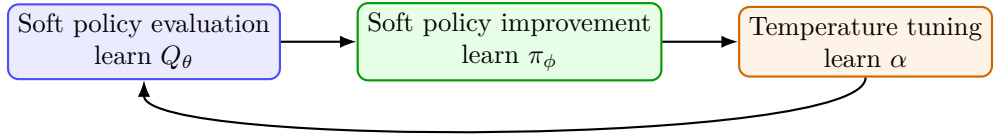
\begin{figure}[t]
	\centering
	\begin{tikzpicture}[
		box/.style={draw,rounded corners,thick,minimum width=3.2cm,minimum height=0.85cm,align=center,font=\small},
		arrow/.style={-{Latex[length=2.2mm]},thick},
		node distance=1.0cm
		]
		\node[box,fill=blue!8,draw=blue!70] (eval) {Soft policy evaluation\\learn $Q_{\theta}$};
		\node[box,fill=green!10,draw=green!60!black,right=of eval] (imp) {Soft policy improvement\\learn $\pi_{\phi}$};
		\node[box,fill=orange!10,draw=orange!80!black,right=of imp] (temp) {Temperature tuning\\learn $\alpha$};
		\draw[arrow] (eval) -- (imp);
		\draw[arrow] (imp) -- (temp);
		\draw[arrow] (temp.south) .. controls +(0,-0.9) and +(0,-0.9) .. (eval.south);
	\end{tikzpicture}
	\caption{The conceptual loop behind SAC. SAC learns critics, a stochastic actor, and often the entropy temperature.}
	\label{fig:soft_policy_iteration_sac}
\end{figure}

\section{The SAC architecture}

Modern SAC typically uses the following components:
\begin{itemize}
	\item a replay buffer storing off-policy transitions;
	\item two Q-critics \(Q_{\theta_1}\) and \(Q_{\theta_2}\);
	\item two slowly updated target critics \(Q_{\bar\theta_1}\) and \(Q_{\bar\theta_2}\);
	\item a stochastic actor \(\pi_{\phi}\), usually a squashed Gaussian policy;
	\item an entropy temperature \(\alpha\), often learned automatically.
\end{itemize}

The two critics reduce overestimation bias using the clipped double-Q idea introduced in TD3 and adopted by SAC \citep{fujimoto2018addressing,haarnoja2018apps}. The replay buffer makes SAC off-policy and sample-efficient. The stochastic actor makes exploration part of the policy itself rather than an external noise process.

\begin{figure}[t]
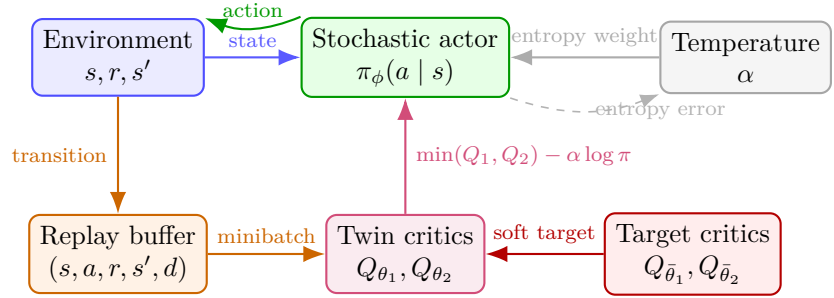

	\centering

	\caption{The SAC architecture. SAC stores environment transitions in a replay buffer, trains twin critics using soft Bellman targets, updates a stochastic actor using critic values and the entropy term, and often learns the entropy temperature automatically.}
	\label{fig:sac_architecture}
\end{figure}

\section{The squashed Gaussian policy}

For continuous control, SAC often uses a Gaussian policy followed by a tanh squashing function:
\begin{align}
	u &= \mu_{\phi}(s)+\sigma_{\phi}(s)\odot \epsilon,
	\qquad \epsilon\sim\Normal(0,I),\\
	a &= \tanh(u).
\end{align}
The tanh transform keeps each action dimension in \((-1,1)\). The action can then be rescaled to the physical action bounds of the environment.

The log-probability must include the change-of-variables correction:
\begin{equation}
	\log\pi_{\phi}(a\given s)
	=
	\log\Normal(u;\mu_{\phi}(s),\sigma_{\phi}(s))
	-
	\sum_i \log\left(1-\tanh^2(u_i)+\epsilon\right).
	\label{eq:tanh_logprob_correction_ch11}
\end{equation}
Forgetting this correction is one of the most common SAC implementation bugs.
A related practical detail is that implementations clip $\log\sigma_{\phi}(s)$ to a range such as $[-20,2]$ to prevent the policy from becoming numerically singular or collapsing to determinism. The constants \texttt{LOG\_STD\_MIN} and \texttt{LOG\_STD\_MAX} in Listing~\ref{lst:sac_actor} implement exactly this clamp.

\begin{figure}[t]
	\centering
	\begin{tikzpicture}[
		box/.style={draw,rounded corners,thick,minimum width=3.1cm,minimum height=0.8cm,align=center,font=\small},
		arrow/.style={-{Latex[length=2.2mm]},thick},
		node distance=0.9cm
		]
		\node[box,fill=blue!8,draw=blue!70] (s) {State $s$};
		\node[box,fill=green!10,draw=green!60!black,right=of s] (params) {Network outputs\\$\mu_\phi(s),\log\sigma_\phi(s)$};
		\node[box,fill=orange!10,draw=orange!80!black,right=of params] (sample) {Reparameterize\\$u=\mu+\sigma\epsilon$};
		\node[box,fill=purple!8,draw=purple!70,right=of sample] (tanh) {Squash\\$a=\tanh(u)$};
		\draw[arrow] (s) -- (params);
		\draw[arrow] (params) -- (sample);
		\draw[arrow] (sample) -- (tanh);
		\node[below=0.7cm of sample,align=center,font=\small] {The log-probability must include the tanh Jacobian correction.};
	\end{tikzpicture}
	\caption{The squashed Gaussian policy used in SAC. The Gaussian sample is differentiable through the reparameterization trick, and tanh bounds the action.}
	\label{fig:squashed_gaussian_policy_ch11}
\end{figure}

\begin{pitfallbox}{The tanh correction is not optional}
	If the log-probability is computed before tanh and used directly, the entropy term is wrong. The actor then optimizes the wrong objective. This bug may not crash training, but it can silently produce unstable policies or poor exploration.
\end{pitfallbox}

\section{Critic learning and clipped double-Q targets}

SAC uses two critics and trains each by minimizing a Bellman error:
\begin{equation}
	J_Q(\theta_i)
	=
	\E_{(s,a,r,s',d)\sim\mathcal{D}}
	\left[
	\left(
	Q_{\theta_i}(s,a)-y
	\right)^2
	\right],
	\qquad i\in\{1,2\},
	\label{eq:sac_critic_loss}
\end{equation}
where the target \(y\) is defined in Eq.~\eqref{eq:sac_target}. The minimum of the two target critics reduces overestimation bias:
\begin{equation}
	\min_{i=1,2}Q_{\bar\theta_i}(s',a').
\end{equation}
This is the continuous-control counterpart of a lesson already seen in Double DQN and TD3: when the critic is noisy, maximizing or trusting a single critic can produce overly optimistic value estimates. Taking the minimum yields a deliberately pessimistic estimate: it tends to underestimate rather than overestimate. In actor-critic learning, underestimation is usually less dangerous than overestimation, because it may slow policy improvement, whereas overestimation can actively drive the actor toward illusory high-value actions. 

\begin{warningbox}{SAC is still inside the deadly-triad regime}
	SAC is off-policy, bootstraps through a learned critic, and uses nonlinear function approximation. Therefore it still lives inside the deadly-triad regime discussed in Chapter~5. Replay, target networks, clipped double critics, entropy regularization, and conservative learning rates mitigate instability; they do not make instability impossible.
\end{warningbox}

\section{Actor learning through the reparameterization trick}

The SAC actor is trained to choose actions that have high Q-value and high entropy. The common actor loss is
\begin{equation}
	J_{\pi}(\phi)
	=
	\E_{s\sim\mathcal{D},\epsilon\sim\Normal}
	\left[
	\alpha\log\pi_{\phi}(a_{\phi}(s,\epsilon)\given s)
	-
	\min_{i=1,2}Q_{\theta_i}(s,a_{\phi}(s,\epsilon))
	\right].
	\label{eq:sac_actor_loss}
\end{equation}
Minimizing this loss increases Q-values while also encouraging entropy. The reparameterization trick makes the sampled action differentiable with respect to the actor parameters:
\begin{equation}
	a_{\phi}(s,\epsilon)=\tanh(\mu_{\phi}(s)+\sigma_{\phi}(s)\odot\epsilon).
\end{equation}
This is a key difference from REINFORCE-style score-function estimators. SAC does not need to estimate the actor gradient only through \(\grad_{\phi}\log\pi_{\phi}\). It can backpropagate through the sampled action into the critic.

\begin{keybox}{What the actor loss says}
	The SAC actor loss says: choose actions that the critic believes are good, but pay a penalty if the policy becomes too certain. The actor is pulled toward high-value actions and pushed away from premature determinism.
\end{keybox}

\section{Automatic entropy-temperature tuning}

The temperature \(\alpha\) is one of SAC's most important hyperparameters. If it is too large, the policy remains too random. If it is too small, the policy becomes nearly deterministic and loses the benefit of maximum-entropy control.

Modern SAC usually tunes \(\alpha\) automatically by matching the policy entropy to a target entropy \(\mathcal{H}_{\mathrm{target}}\) \citep{haarnoja2018apps}. A common loss for \(\alpha\) is
\begin{equation}
	J(\alpha)
	=
	\E_{a_t\sim\pi_{\phi}}
	\left[
	-\alpha
	\left(
	\log\pi_{\phi}(a_t\given s_t)
	+
	\mathcal{H}_{\mathrm{target}}
	\right)
	\right].
	\label{eq:alpha_loss_ch11}
\end{equation}
This formulation is the dual of a constrained optimization problem: $\alpha$ acts as the Lagrange multiplier for the entropy constraint $\mathbb{E}[\mathcal{H}(\pi(\cdot\given s))] \geq \mathcal{H}_{\mathrm{target}}$, increasing when the policy is too deterministic and decreasing when it is too random. This Lagrangian view will return in Chapter~12 in a different role: as the multiplier on safety constraints rather than entropy constraints. 

In practice, implementations often optimize \(\log\alpha\) rather than \(\alpha\) directly to keep \(\alpha>0\).

A common default for continuous actions is
\begin{equation}
	\mathcal{H}_{\mathrm{target}}=-\dim(\mathcal{A}).
\end{equation}
This default is not a law. It is a useful starting point. In safety-critical control, one may need to tune the target entropy more conservatively or combine SAC with a safety layer.

\begin{figure}[t]
	\centering
	\begin{tikzpicture}[
		box/.style={draw,rounded corners,thick,minimum width=3.2cm,minimum height=0.85cm,align=center,font=\small},
		arrow/.style={-{Latex[length=2.2mm]},thick},
		node distance=1.0cm
		]
		\node[box,fill=blue!8,draw=blue!70] (entropy) {Observed entropy\\$-\log\pi(a\given s)$};
		\node[box,fill=orange!10,draw=orange!80!black,right=of entropy] (compare) {Compare with\\target entropy};
		\node[box,fill=green!10,draw=green!60!black,right=of compare] (alpha) {Update\\$\alpha$};
		\node[box,fill=purple!8,draw=purple!70,right=of alpha] (actor) {Actor loss\\$\alpha\log\pi-Q$};
		\draw[arrow] (entropy) -- (compare);
		\draw[arrow] (compare) -- (alpha);
		\draw[arrow] (alpha) -- (actor);
	\end{tikzpicture}
	\caption{Automatic entropy-temperature tuning. SAC adapts \(\alpha\) so the policy entropy stays near a target value.}
	\label{fig:alpha_tuning_ch11}
\end{figure}
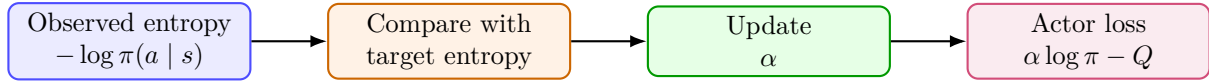

\section{The complete SAC algorithm}

\begin{center}
	\begin{minipage}{0.94\textwidth}
		\begin{tcolorbox}[title={Algorithm 11.1: Soft Actor-Critic},colback=gray!4,colframe=black!70,fonttitle=\bfseries]
			\begin{enumerate}[leftmargin=*]
				\item Initialize actor \(\pi_{\phi}\), critics \(Q_{\theta_1},Q_{\theta_2}\), target critics \(Q_{\bar\theta_1},Q_{\bar\theta_2}\), replay buffer \(\mathcal{D}\), and temperature \(\alpha\).
				\item For each environment step:
				\begin{enumerate}[leftmargin=*]
					\item sample action \(a_t\sim\pi_{\phi}(\cdot\given s_t)\);
					\item execute action and observe \((r_t,s_{t+1},d_t)\);
					\item store \((s_t,a_t,r_t,s_{t+1},d_t)\) in \(\mathcal{D}\).
				\end{enumerate}
				\item For each gradient update:
				\begin{enumerate}[leftmargin=*]
					\item sample a minibatch from \(\mathcal{D}\);
					\item sample next actions \(a'\sim\pi_{\phi}(\cdot\given s')\);
					\item build the soft Bellman target using Eq.~\eqref{eq:sac_target};
					\item update both critics using Eq.~\eqref{eq:sac_critic_loss};
					\item sample current actions \(a\sim\pi_{\phi}(\cdot\given s)\);
					\item update the actor using Eq.~\eqref{eq:sac_actor_loss};
					\item update \(\alpha\) using Eq.~\eqref{eq:alpha_loss_ch11};
					\item softly update target critics.
				\end{enumerate}
			\end{enumerate}
		\end{tcolorbox}
	\end{minipage}
\end{center}

\section{Concrete PyTorch implementation}

This section gives compact code fragments. They are intentionally explicit rather than maximally optimized.

\subsection{Replay buffer}

\Needspace{18\baselineskip}
\begin{lstlisting}[style=pythonstyle,caption={Minimal replay buffer for SAC.},label={lst:sac_replay_buffer}]
import random
from collections import deque
import numpy as np
import torch
import torch.nn as nn
import torch.nn.functional as F

class ReplayBuffer:
    def __init__(self, capacity):
        self.buffer = deque(maxlen=capacity)

    def push(self, state, action, reward, next_state, done):
        self.buffer.append((state, action, reward, next_state, done))

    def sample(self, batch_size, device):
        batch = random.sample(self.buffer, batch_size)
        s, a, r, sp, d = map(np.array, zip(*batch))
        return (
            torch.as_tensor(s, dtype=torch.float32, device=device),
            torch.as_tensor(a, dtype=torch.float32, device=device),
            torch.as_tensor(r, dtype=torch.float32, device=device).unsqueeze(-1),
            torch.as_tensor(sp, dtype=torch.float32, device=device),
            torch.as_tensor(d, dtype=torch.float32, device=device).unsqueeze(-1),
        )

    def __len__(self):
        return len(self.buffer)
\end{lstlisting}

\subsection{Squashed Gaussian actor}

\Needspace{24\baselineskip}
\begin{lstlisting}[style=pythonstyle,caption={Squashed Gaussian actor with tanh log-probability correction.},label={lst:sac_actor}]
LOG_STD_MIN = -20
LOG_STD_MAX = 2
EPS = 1e-6

class SquashedGaussianActor(nn.Module):
    def __init__(self, obs_dim, act_dim, hidden=256, action_scale=1.0, action_bias=0.0):
        super().__init__()
        self.net = nn.Sequential(
            nn.Linear(obs_dim, hidden), nn.ReLU(),
            nn.Linear(hidden, hidden), nn.ReLU(),
        )
        self.mu = nn.Linear(hidden, act_dim)
        self.log_std = nn.Linear(hidden, act_dim)
        self.register_buffer("action_scale", torch.as_tensor(action_scale))
        self.register_buffer("action_bias", torch.as_tensor(action_bias))

    def forward(self, obs):
        h = self.net(obs)
        mu = self.mu(h)
        log_std = self.log_std(h).clamp(LOG_STD_MIN, LOG_STD_MAX)
        std = log_std.exp()
        return mu, std

    def sample(self, obs):
        mu, std = self(obs)
        dist = torch.distributions.Normal(mu, std)
        u = dist.rsample()                 # reparameterized sample
        tanh_u = torch.tanh(u)
        action = tanh_u * self.action_scale + self.action_bias

        log_prob = dist.log_prob(u)
        log_prob -= torch.log(1.0 - tanh_u.pow(2) + EPS)
        log_prob = log_prob.sum(dim=-1, keepdim=True)
        return action, log_prob, torch.tanh(mu) * self.action_scale + self.action_bias
\end{lstlisting}

\subsection{Critic networks}

\Needspace{18\baselineskip}
\begin{lstlisting}[style=pythonstyle,caption={Q-critic for continuous actions.},label={lst:sac_critic}]
class QCritic(nn.Module):
    def __init__(self, obs_dim, act_dim, hidden=256):
        super().__init__()
        self.q = nn.Sequential(
            nn.Linear(obs_dim + act_dim, hidden), nn.ReLU(),
            nn.Linear(hidden, hidden), nn.ReLU(),
            nn.Linear(hidden, 1),
        )

    def forward(self, obs, act):
        x = torch.cat([obs, act], dim=-1)
        return self.q(x)
\end{lstlisting}

\subsection{SAC update}

\Needspace{28\baselineskip}
\begin{lstlisting}[style=pythonstyle,caption={One SAC update step with automatic entropy tuning.},label={lst:sac_update}]
def soft_update(target, source, tau):
    with torch.no_grad():
        for p_targ, p in zip(target.parameters(), source.parameters()):
            p_targ.data.mul_(1.0 - tau).add_(tau * p.data)

def sac_update(batch, actor, q1, q2, q1_targ, q2_targ,
               actor_opt, q_opt, log_alpha, alpha_opt,
               gamma=0.99, tau=0.005, target_entropy=-1.0):
    obs, act, rew, next_obs, done = batch
    alpha = log_alpha.exp()

    # Critic target: no gradients through target calculation.
    with torch.no_grad():
        next_action, next_logp, _ = actor.sample(next_obs)
        q1_next = q1_targ(next_obs, next_action)
        q2_next = q2_targ(next_obs, next_action)
        q_next = torch.min(q1_next, q2_next) - alpha * next_logp
        target_q = rew + gamma * (1.0 - done) * q_next

    q1_pred = q1(obs, act)
    q2_pred = q2(obs, act)
    q_loss = F.mse_loss(q1_pred, target_q) + F.mse_loss(q2_pred, target_q)

    q_opt.zero_grad(set_to_none=True)
    q_loss.backward()
    q_opt.step()

    # Actor update: freeze critics conceptually; gradients flow through action into actor.
    new_action, logp, _ = actor.sample(obs)
    q_pi = torch.min(q1(obs, new_action), q2(obs, new_action))
    actor_loss = (alpha.detach() * logp - q_pi).mean()

    actor_opt.zero_grad(set_to_none=True)
    actor_loss.backward()
    actor_opt.step()

    # Temperature update: optimize log_alpha for positivity.
    alpha_loss = -(log_alpha * (logp.detach() + target_entropy)).mean()
    alpha_opt.zero_grad(set_to_none=True)
    alpha_loss.backward()
    alpha_opt.step()

    soft_update(q1_targ, q1, tau)
    soft_update(q2_targ, q2, tau)

    return {
        "q_loss": float(q_loss.detach()),
        "actor_loss": float(actor_loss.detach()),
        "alpha_loss": float(alpha_loss.detach()),
        "alpha": float(alpha.detach()),
        "entropy_est": float((-logp).mean().detach()),
    }
\end{lstlisting}

\begin{warningbox}{Stop-gradients in SAC}
	The critic target must be computed under \texttt{torch.no\_grad()}. The actor update should not update critic parameters, even though it uses the critic to evaluate sampled actions. The temperature loss should detach the log-probability. These detachments are not cosmetic; they define the algorithm. 
\end{warningbox}

\section{Diagnostics, debugging, and implementation traps}

SAC is often easier to tune than DDPG, but it is not automatic. Table~\ref{tab:sac_debugging} summarizes common failure modes.

\begin{table}[t]
	\centering
	\caption{SAC debugging checklist.}
	\label{tab:sac_debugging}
	\begin{tabular}{p{0.25\textwidth}p{0.35\textwidth}p{0.30\textwidth}}
		\toprule
		Symptom & Likely cause & What to inspect \\
		\midrule
		Q-values explode & Missing terminal mask, high reward scale, target update bug & Bellman target, \texttt{done}, reward normalization \\
		Entropy collapses early & \(\alpha\) too small or target entropy too low & \(\alpha\), entropy, log standard deviation \\
		Policy remains random & \(\alpha\) too large or critic not learning & actor loss, Q-loss, target entropy \\
		NaNs in log-probability & Missing tanh correction epsilon or extreme log standard deviation & Eq.~\eqref{eq:tanh_logprob_correction_ch11}, log-std clamp \\
		Actor improves then collapses & Critic overestimation or stale replay distribution & min-Q target, replay age, learning rates \\
		Good simulation, bad deployment & Policy exploits simulator noise or lacks safety constraints & domain randomization, safety filter, real constraints \\
		Unsafe UAV actions & Entropy explores unsafe regions & CBF shield, action projection, constrained critic \\
		\bottomrule
	\end{tabular}
\end{table}

\subsection{Practical metrics to log}

A SAC implementation should log more than episodic return. At minimum, track critic loss, actor loss, \(\alpha\), estimated entropy, mean log standard deviation, Q-value scale, target Q scale, replay-buffer size, episode length, and domain-specific safety violations. For UAV/SDN control, also log latency, SINR, throughput, handover rate, energy consumption, CBF corrections, and constraint violations.

\section{SAC for UAV/SDN continuous control}

SAC is especially natural for UAV/SDN control because many actions are continuous or hybrid. A UAV may need to output movement direction, speed, altitude change, power level, and bandwidth shares. Discretizing all these quantities leads to combinatorial explosion. SAC can instead output a continuous vector directly.

\subsection{A structured UAV action vector}

Consider a UAV controller with action
\begin{equation}
	a_t
	=
	[\Delta x,\Delta y,\Delta z, p_{\mathrm{tx}}, b_A,b_B,b_C],
\end{equation}
where \(\Delta x,\Delta y,\Delta z\) control movement, \(p_{\mathrm{tx}}\) controls transmit power, and \(b_A,b_B,b_C\) represent bandwidth shares for user classes. The movement and power components can be squashed with tanh and rescaled to physical bounds. The bandwidth shares can be normalized so they are nonnegative and sum to one.

\Needspace{22\baselineskip}
\begin{lstlisting}[style=pythonstyle,caption={Structured SAC actor head for UAV/SDN control.},label={lst:uav_sac_actor}]
class UAVSACActor(nn.Module):
    def __init__(self, obs_dim, hidden=256):
        super().__init__()
        self.backbone = nn.Sequential(
            nn.Linear(obs_dim, hidden), nn.ReLU(),
            nn.Linear(hidden, hidden), nn.ReLU(),
        )
        self.move_mu = nn.Linear(hidden, 3)       # dx, dy, dz
        self.move_log_std = nn.Linear(hidden, 3)
        self.power_mu = nn.Linear(hidden, 1)
        self.power_log_std = nn.Linear(hidden, 1)
        self.bandwidth_logits = nn.Linear(hidden, 3)

    def forward(self, obs):
        h = self.backbone(obs)
        move_mu = self.move_mu(h)
        move_std = self.move_log_std(h).clamp(-20, 2).exp()
        power_mu = self.power_mu(h)
        power_std = self.power_log_std(h).clamp(-20, 2).exp()
        bandwidth = torch.softmax(self.bandwidth_logits(h), dim=-1)
        return move_mu, move_std, power_mu, power_std, bandwidth
\end{lstlisting}

This policy is not a full SAC actor by itself because it mixes sampled continuous components with deterministic normalized bandwidth. But it shows how SAC ideas can be adapted to structured control. A more complete implementation would sample bandwidth shares from a logistic-normal or Dirichlet distribution and include the correct log-probability term.

\begin{researchbox}{SAC plus safety filters}
	For UAV/SDN systems, SAC should not be deployed as an unconstrained controller. Vanilla SAC optimizes a reward-plus-entropy objective; it does not by itself guarantee collision avoidance, battery safety, latency bounds, interference limits, or regulatory constraints. A practical UAV/SDN architecture should therefore combine SAC with a control barrier function, action projection layer, constrained-RL objective, safety critic, or a supervisory SDN policy. The actor proposes a continuous action; the safety layer modifies it if necessary; the critic should learn from the executed action, not merely from the unsafe pre-filter proposal. This creates a bridge between maximum-entropy exploration, safe control, and constrained network optimization. 
\end{researchbox}

\begin{figure}[t]
	\centering
	\begin{tikzpicture}[
		box/.style={draw,rounded corners,thick,minimum width=3.0cm,minimum height=0.85cm,align=center,font=\small},
		small/.style={draw,rounded corners,minimum width=2.8cm,minimum height=0.75cm,align=center,font=\small},
		arrow/.style={-{Latex[length=2.2mm]},thick},
		node distance=0.95cm
		]
		\node[box,fill=blue!8,draw=blue!70,font=\small\bfseries] (state) {Network state\\{\normalfont\small QoS, SINR, battery}};
		\node[box,fill=green!10,draw=green!60!black,font=\small\bfseries,right=of state] (actor) {SAC actor\\{\normalfont\small stochastic action}};
		\node[box,fill=orange!10,draw=orange!80!black,font=\small\bfseries,right=of actor] (shield) {Safety layer\\{\normalfont\small CBF/projection}};
		\node[box,fill=gray!10,draw=gray!70,font=\small\bfseries,right=of shield] (env) {UAV/SDN system};

		\node[small,fill=purple!8,draw=purple!70,font=\small\bfseries,below=1.3cm of actor] (critics) {Twin critics\\{\normalfont\small $Q_1,Q_2$}};
		\node[small,fill=red!6,draw=red!70!black,font=\small\bfseries,below=1.3cm of shield] (alpha) {Temperature\\{\normalfont\small $\alpha$}};

		% main flow
		\draw[arrow,draw=blue!70] (state) -- (actor);
		\draw[arrow,draw=green!60!black] (actor) -- (shield);
		\draw[arrow,draw=orange!80!black] (shield) -- (env);

		% feedback from environment
		\draw[arrow,draw=gray!60] (env.south) -- ++(0,-0.5) -| (critics.east);
		\draw[arrow,draw=gray!60] (env.south) -- ++(0,-0.5) -| (alpha.east);

		% learning signals - separated anchors
		\draw[arrow,draw=purple!70] (critics.north) -- node[left,font=\scriptsize,color=purple!70] {$\nabla_\phi J$} (actor.south west);
		\draw[arrow,draw=red!70!black] (alpha.north) -- node[right,font=\scriptsize,color=red!70!black] {entropy} (shield.south);

	\end{tikzpicture}
	\caption{SAC for UAV/SDN control. The actor proposes stochastic continuous actions, a safety layer can project actions before execution, and twin critics plus entropy-temperature tuning provide the learning signal.}
	\label{fig:sac_uav_sdn}
\end{figure}
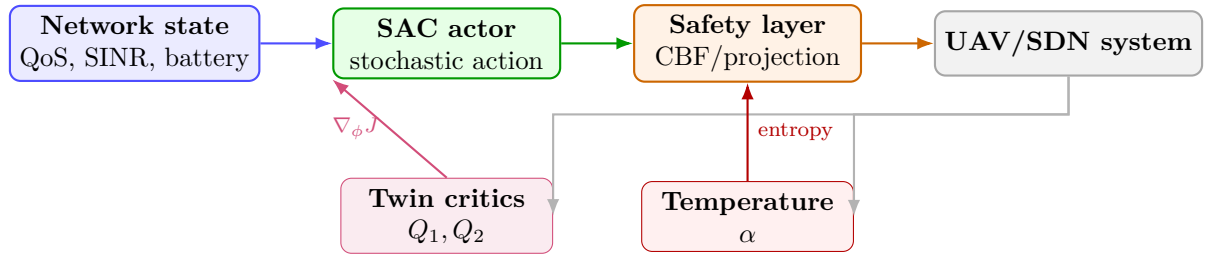

\section{Beyond vanilla SAC: variants and 2026 directions}

SAC is a foundation, not the end of off-policy continuous control. Several later methods modify its critic, replay ratio, distributional representation, or action space.

\subsection{TQC: distributional critics for overestimation control}

Truncated Quantile Critics (TQC) combine SAC-style actor learning with distributional critics and truncation of high quantiles to reduce overestimation bias \citep{kuznetsov2020tqc}. This is particularly relevant when critic overestimation leads the actor toward unsafe or unrealistic actions. In UAV/SDN control, a distributional critic can also support tail-risk reasoning, such as avoiding actions with rare but severe latency violations. This connects back to the tail-risk reasoning illustrated in Chapter~6, where distributional value estimates can distinguish mean-good but tail-bad actions from mean-modest but tail-safe ones. 

\subsection{REDQ and DroQ: higher update-to-data ratios}

REDQ improves sample efficiency by using an ensemble of critics and a high update-to-data ratio \citep{chen2021redq}. DroQ uses dropout Q-functions to obtain similar benefits with lower computational cost \citep{hiraoka2021droq}. These methods are important when real environment interaction is expensive but gradient computation is cheap.

\subsection{Discrete SAC and hybrid action spaces}

SAC was originally most successful for continuous actions. Discrete SAC variants adapt the maximum-entropy actor-critic idea to discrete actions \citep{christodoulou2019discrete}. Later work revisited discrete SAC and found underestimation and instability issues in challenging discrete-action domains \citep{zhou2022revisiting}. This matters for hybrid systems such as UAV/SDN control, where some decisions are discrete, such as handover or routing mode, while others are continuous, such as power or movement.

\subsection{2025--2026 frontier: guidance, structure, and entropy design}

Recent work increasingly treats SAC not only as an algorithm but as a component inside structured control systems. Examples include discrete-action redesign, parameterized-action variants, and guided SAC systems that use external policies or language-model guidance to improve sample efficiency \citep{ma2026guidedsac}. The 2026 GuidedSAC line should be read as an emerging research direction rather than a settled replacement for SAC: the external guidance can shape exploration, but the underlying stability still depends on entropy, critic accuracy, replay, target networks, and off-policy learning discipline. 

\begin{table}[t]
	\centering
	\caption{SAC and related off-policy continuous-control methods.}
	\label{tab:sac_variants}
	\begin{tabular}{p{0.20\textwidth}p{0.34\textwidth}p{0.36\textwidth}}
		\toprule
		Method & Core idea & Main caution \\
		\midrule
		SAC & maximum-entropy off-policy actor-critic & sensitive to log-probability and temperature details \\
		TD3 & deterministic actor with clipped double critics & exploration noise is external and can be brittle \\
		TQC & distributional critics with truncated high quantiles & more complex critic implementation \\
		REDQ & critic ensembles and high update-to-data ratio & high compute and careful tuning needed \\
		DroQ & dropout critics for efficient high replay ratio & regularization settings matter \\
		Discrete SAC & entropy-regularized actor-critic for discrete actions & can suffer from underestimation/instability \\
		\bottomrule
	\end{tabular}
\end{table}

\section{Limitations and when not to use SAC}

SAC is powerful, but it is not universally best.
\begin{enumerate}[leftmargin=*]
	\item \textbf{Discrete actions:} SAC is strongest in continuous control. Discrete SAC exists, but DQN-family methods or PPO may be better depending on the problem.
	\item \textbf{Very sparse rewards:} entropy alone may not solve exploration if reward is extremely rare.
	\item \textbf{Safety-critical exploration:} stochastic actions can visit unsafe regions unless a shield, CBF, or constrained method is used.
	\item \textbf{High-dimensional action spaces:} Gaussian policies may struggle when actions have complex constraints such as simplex or hybrid structures.
	\item \textbf{Offline-only data:} vanilla SAC can extrapolate beyond the dataset. Offline RL methods such as CQL or IQL are usually safer.
	\item \textbf{Implementation sensitivity:} the tanh correction, terminal mask, target updates, and entropy tuning are easy to get wrong.
\end{enumerate}

\section{Exercises}

\subsection*{Conceptual exercises}
\begin{enumerate}[leftmargin=*]
	\item Explain the difference between ordinary reward maximization and maximum-entropy reward maximization.
	\item Why does SAC use a stochastic policy even in continuous-control tasks where deterministic actions are possible?
	\item Why is the tanh log-probability correction necessary?
	\item Explain why automatic entropy tuning is useful. What can go wrong if \(\alpha\) is too large or too small?
	\item Why should SAC not be deployed directly in a safety-critical UAV system without a safety layer?
\end{enumerate}

\subsection*{Mathematical exercises}
\begin{enumerate}[leftmargin=*]
	\item Starting from the maximum-entropy objective, derive the sampled entropy form involving \(-\alpha\log\pi(a_t\given s_t)\).
	\item Derive the SAC critic target in Eq.~\eqref{eq:sac_target} from the soft Bellman equation.
	\item Show how the actor loss in Eq.~\eqref{eq:sac_actor_loss} trades off Q-value and entropy.
	\item For a two-dimensional continuous action space, what is the default target entropy \(-\dim(\mathcal{A})\)? Why is this only a heuristic?
	\item Explain how the clipped double-Q target reduces overestimation bias.
\end{enumerate}

\subsection*{Coding exercises}
\begin{enumerate}[leftmargin=*]
	\item Implement the squashed Gaussian log-probability correction and test it on random Gaussian samples.
	\item Add gradient clipping to the SAC update in Listing~\ref{lst:sac_update}.
	\item Modify the actor to support asymmetric action bounds.
	\item Implement a deterministic evaluation function that uses the mean action instead of sampling.
	\item Extend the UAV actor in Listing~\ref{lst:uav_sac_actor} so bandwidth shares are sampled from a differentiable distribution rather than produced deterministically.
\end{enumerate}

\subsection*{Research-thinking exercises}
\begin{enumerate}[leftmargin=*]
	\item Design a SAC-based controller for UAV trajectory and bandwidth allocation. What are the state, action, reward, and safety constraints?
	\item How would you combine SAC with a control barrier function? Should the critic learn from proposed actions or executed actions after projection?
	\item Compare PPO and SAC for SD-WAN traffic engineering. Which one is easier to deploy safely, and which one is more sample-efficient?
	\item Propose a tail-risk version of SAC using distributional critics for latency-sensitive networking.
\end{enumerate}

\section*{Looking Ahead to Chapter 12: Food for Thought}
\addcontentsline{toc}{section}{Looking Ahead to Chapter 12: Food for Thought}

SAC completes the core actor-critic arc of Part~III. We have now seen on-policy policy optimization, variance reduction, actor-critic learning, PPO, and maximum-entropy off-policy control. Yet all these methods still share a limitation: they learn mainly from real or simulated interaction. They do not explicitly learn a model of how the world changes.

Chapter~12 begins Part~IV by asking a new question:
\begin{quote}
	Can an agent improve sample efficiency by learning a model of the environment and using it to imagine or plan future experience?
\end{quote}

Before moving on, consider the following questions.
\begin{enumerate}[leftmargin=*]
	\item SAC reuses past data, but can it reason about actions it has never tried?
	\item What would it mean for a UAV to learn a predictive model of user mobility, channel quality, or battery dynamics?
	\item When is model learning helpful, and when can model error make control worse?
	\item Can world models reduce the need for unsafe real-world exploration?
\end{enumerate}

% \chapter*{Chapter 11 references}
% \addcontentsline{toc}{chapter}{Chapter 11 references}
	\part{Planning, Models, and Offline Learning}
% \addcontentsline{toc}{part}{Part IV -- Deep Reinforcement Learning: From First Principles to 2026}

\chapter[Model-Based Deep Reinforcement Learning]{Model-Based Deep Reinforcement Learning}
\label{ch:model_based_rl}
\chaptermark{Model-Based Deep RL}

\begin{keybox}{Chapter thesis}
	Model-based deep reinforcement learning is not simply ``learning a simulator.'' It is the study of how an agent can learn a predictive model of the world and use that model for planning, policy improvement, data augmentation, uncertainty estimation, or imagination. A useful model need not reconstruct every detail of the environment. It must predict the information that matters for decisions: rewards, values, constraints, future observations, or latent state transitions.
\end{keybox}

\section*{Chapter Overview}
\addcontentsline{toc}{section}{Chapter Overview}

\begin{enumerate}[leftmargin=*]
	\item Why Chapter 12 matters
	\item What is a model?
	\item From model-free learning to model-based learning
	\item The Dyna idea: learning, acting, and planning in one loop
	\item Model errors and the compounding-error problem
	\item Uncertainty-aware models and ensembles
	\item Short-horizon model rollouts and MBPO
	\item Planning with learned models: MPC and CEM
	\item Latent world models from pixels
	\item Dreamer-style imagination learning
	\item MuZero-style value-equivalent models
	\item TD-MPC and TD-MPC2: latent planning for continuous control
	\item Model-based RL for UAV/SDN control
	\item Concrete PyTorch building blocks
	\item Diagnostics, debugging, and failure modes
	\item Frontier note: diffusion and foundation world models
	\item Limitations and when not to use model-based RL
	\item Exercises
	\item Looking Ahead to Chapter 13
\end{enumerate}

\section{Why Chapter 12 matters}

Part~III developed policy-gradient and actor-critic methods. PPO learned from fresh on-policy rollouts. SAC reused data through a replay buffer and entropy-regularized off-policy learning. These methods learn policies or values from interaction, but they do not explicitly ask a deeper question:

\begin{quote}
	Can the agent learn how the world changes, and then use that knowledge to improve future decisions?
\end{quote}

This is the central question of model-based reinforcement learning. A model-based agent learns a model of the environment and uses it to plan, imagine future trajectories, generate synthetic experience, or improve value estimates. This idea is old: dynamic programming used known transition models; Dyna integrated real experience and model-generated updates \citep{sutton1990integrated}; and optimal control has long relied on predictive models. What changed in deep reinforcement learning is that the model can now be learned from high-dimensional observations, including images, network telemetry, robot sensors, and latent representations.

Model-based deep RL matters because interaction is expensive. A robot cannot safely fall millions of times. A UAV cannot freely crash into obstacles during training. A network controller cannot intentionally overload a production SD-WAN or wireless slice. If a learned model can reduce the number of real interactions needed, model-based RL becomes a practical tool rather than only a theoretical idea.

Constrained and safety-aware RL methods using Lagrangian formulations are developed in a later chapter. This chapter takes the complementary perspective that better models can also reduce safety risk by predicting consequences before acting. A world model does not replace a safety layer, but it can help the agent see danger earlier.

\begin{figure}[t]
	\centering
	\begin{tikzpicture}[
		box/.style={draw,rounded corners,thick,minimum width=3.1cm,minimum height=0.9cm,align=center,font=\small},
		arrow/.style={-{Latex[length=2.2mm]},thick},
		node distance=1.5cm
		]
		\node[box,fill=blue!8,draw=blue!70] (real) {Real environment\\{\normalfont\small expensive interaction}};
		\node[box,fill=orange!10,draw=orange!80!black,right=of real] (model) {Learned model\\{\normalfont\small cheap imagination}};
		\node[box,fill=green!10,draw=green!60!black,right=of model] (policy) {Policy / planner\\{\normalfont\small improved decisions}};
		\draw[arrow,draw=blue!70] (real) -- node[above,font=\scriptsize] {data} (model);
		\draw[arrow,draw=orange!80!black] (model) -- node[above,font=\scriptsize] {predictions} (policy);
		\draw[arrow,draw=green!60!black] (policy.south) to[bend left=12] node[below,font=\scriptsize] {actions} (real.south);
	\end{tikzpicture}
	\caption{The central loop of model-based reinforcement learning. Real interaction trains a model; the learned model supports planning or imagined policy improvement; the improved policy acts again in the real environment.}
	\label{fig:mbrl_central_loop}
\end{figure}
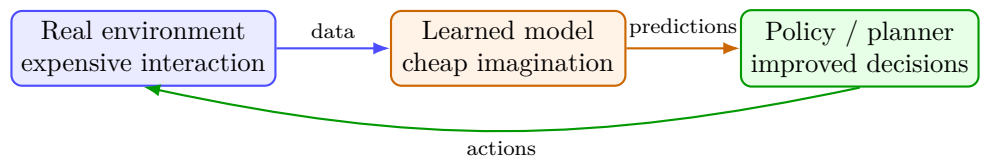

Recent model-based methods show why this topic is central to modern RL. DreamerV3 learns a world model and improves behavior by imagining future scenarios, reporting strong results across more than 150 tasks with a single configuration \citep{hafner2023dreamerv3,hafner2025dreamerv3nature}. MuZero learns a model sufficient for planning without being given the rules and achieved superhuman performance across Go, chess, shogi, and Atari \citep{schrittwieser2020muzero}. TD-MPC2 scales latent model-predictive control across more than 100 continuous-control tasks and studies scaling to large multitask agents \citep{hansen2024tdmpc2}.

\section{What is a model?}

In an MDP, a complete model consists of transition and reward functions:
\begin{equation}
	p(s_{t+1},r_t\given s_t,a_t).
	\label{eq:full_mdp_model}
\end{equation}
A deterministic model may be written as
\begin{equation}
	s_{t+1} = f(s_t,a_t), \qquad r_t = r(s_t,a_t).
\end{equation}
A stochastic model predicts a distribution:
\begin{equation}
	\hat{p}_\psi(s_{t+1}\given s_t,a_t),
	\qquad
	\hat{r}_\psi(s_t,a_t),
\end{equation}
where \(\psi\) are learned parameters.

However, modern model-based RL often uses models that are not complete simulators. A model may predict only rewards and values. A model may operate in a latent space instead of pixel space. A model may be value-equivalent: it may ignore details that are irrelevant for planning. A model may predict safety costs or constraint violations rather than reconstructing all observations.

\begin{table}[t]
	\centering
	\caption{Different meanings of ``model'' in model-based deep RL. A model is useful when it predicts what the decision algorithm needs, not necessarily every detail of the world.}
	\label{tab:model_types}
	\begin{tabular}{p{0.23\textwidth}p{0.36\textwidth}p{0.30\textwidth}}
		\toprule
		Model type & What it predicts & Typical use \\
		\midrule
		State-transition model & Next state or observation & Planning, Dyna updates, MPC \\
		Reward model & Immediate reward & Model-based value targets \\
		Cost/safety model & Constraint cost, collision, violation risk & Safe planning, shielding, CBF support \\
		Latent dynamics model & Next latent state & World models from pixels or high-dimensional telemetry \\
		Value-equivalent model & Reward, value, and policy-relevant predictions & MuZero-style planning \\
		Uncertainty model & Distribution or ensemble disagreement & Risk-aware planning and rollout truncation \\
		\bottomrule
	\end{tabular}
\end{table}

\begin{warningbox}{A model need not be visually accurate}
	A common mistake is to believe that model-based RL requires a perfect video predictor. For control, visual realism is not the goal. A useful model should preserve decision-relevant structure. In a UAV/SDN setting, predicting future latency, SINR, load, battery, and violation risk may be more important than reconstructing a photorealistic scene.
\end{warningbox}

\section{From model-free learning to model-based learning}

A model-free method learns values or policies directly from real experience:
\begin{equation}
	(s_t,a_t,r_t,s_{t+1}) \longrightarrow \text{update } Q_\theta \text{ or } \pi_\phi.
\end{equation}
A model-based method adds an intermediate learned predictor:
\begin{equation}
	(s_t,a_t) \longrightarrow \hat{s}_{t+1},\hat{r}_t,\hat{c}_t,
\end{equation}
and then uses the prediction for planning or learning.

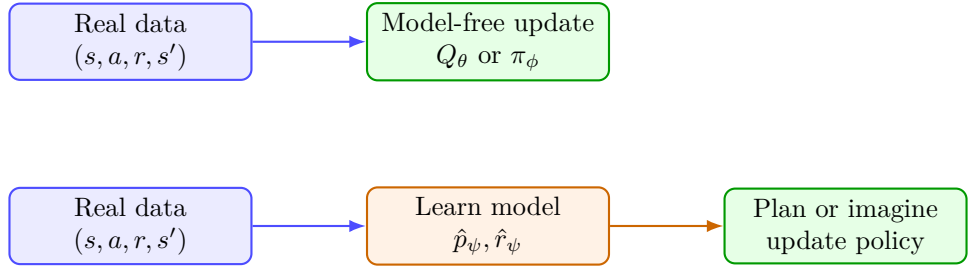
\begin{figure}[t]
	\centering
	\begin{tikzpicture}[
		box/.style={draw,rounded corners,thick,minimum width=3.2cm,minimum height=0.9cm,align=center,font=\small},
		arrow/.style={-{Latex[length=2.2mm]},thick},
		node distance=1.5cm
		]
		\node[box,fill=blue!8,draw=blue!70] (mfdata) {Real data\\$(s,a,r,s')$};
		\node[box,fill=green!10,draw=green!60!black,right=of mfdata] (mfagent) {Model-free update\\$Q_\theta$ or $\pi_\phi$};
		\draw[arrow,draw=blue!70] (mfdata) -- (mfagent);

		\node[box,fill=blue!8,draw=blue!70,below=1.4cm of mfdata] (mbdata) {Real data\\$(s,a,r,s')$};
		\node[box,fill=orange!10,draw=orange!80!black,right=of mbdata] (mbmodel) {Learn model\\$\hat{p}_\psi,\hat{r}_\psi$};
		\node[box,fill=green!10,draw=green!60!black,right=of mbmodel] (mbagent) {Plan or imagine\\update policy};
		\draw[arrow,draw=blue!70] (mbdata) -- (mbmodel);
		\draw[arrow,draw=orange!80!black] (mbmodel) -- (mbagent);
	\end{tikzpicture}
	\caption{Model-free and model-based learning. Model-free methods update the policy or value function directly from real transitions. Model-based methods first learn a predictive model and then use it for planning, synthetic rollouts, or imagined actor-critic learning.}
	\label{fig:model_free_vs_model_based}
\end{figure}

The advantage of model-based RL is sample efficiency. The disadvantage is model bias. If the model is wrong, the agent can exploit prediction errors and learn behavior that works inside imagination but fails in reality. This is the central tension of the chapter:

\begin{quote}
	Model-based RL trades real-environment cost for model-bias risk.
\end{quote}

\section{The Dyna idea: learning, acting, and planning in one loop}

Dyna is the classical template behind many modern model-based methods \citep{sutton1990integrated}. It interleaves three processes:
\begin{enumerate}[leftmargin=*]
	\item act in the real environment;
	\item learn a model from real transitions;
	\item use the model to generate simulated updates.
\end{enumerate}

For tabular Q-learning, a Dyna update may use real and model-generated transitions:
\begin{equation}
	Q(s,a)
	\leftarrow
	Q(s,a)
	+
	\alpha
	\left[ r + \gamma \max_{a'} Q(s',a') - Q(s,a) \right].
\end{equation}
The same update can be applied to real transitions and imagined transitions.

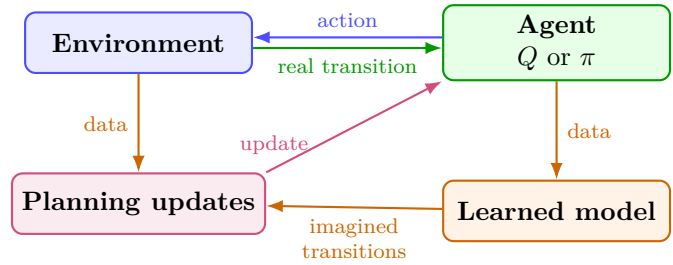
\begin{figure}[t]
	\centering
	\begin{tikzpicture}[
		box/.style={draw,rounded corners,thick,minimum width=3.0cm,minimum height=0.8cm,align=center,font=\small},
		arrow/.style={-{Latex[length=2.2mm]},thick},
		node distance=2.5cm
		]
		\node[box,fill=blue!8,draw=blue!70,font=\small\bfseries] (env) {Environment};
		\node[box,fill=green!10,draw=green!60!black,font=\small\bfseries,right=of env] (agent) {Agent\\{\normalfont\small $Q$ or $\pi$}};
		\node[box,fill=purple!8,draw=purple!70,font=\small\bfseries,below=1.3cm of env] (planning) {Planning updates};
		\node[box,fill=orange!10,draw=orange!80!black,font=\small\bfseries,below=1.3cm of agent] (model) {Learned model};

		% Real interaction
		\draw[arrow,draw=blue!70] ([yshift=2pt]agent.west) -- node[above,font=\scriptsize,color=blue!70] {action} ([yshift=2pt]env.east);
		\draw[arrow,draw=green!60!black] ([yshift=-2pt]env.east) -- node[below,font=\scriptsize,color=green!60!black] {real transition} ([yshift=-2pt]agent.west);

		% Data to model
		\draw[arrow,draw=orange!80!black] (env.south) -- node[left,font=\scriptsize,color=orange!80!black] {data} (planning.north);
		\draw[arrow,draw=orange!80!black] (agent.south) -- node[right,font=\scriptsize,color=orange!80!black] {data} (model.north);

		% Model to planning
		\draw[arrow,draw=orange!80!black] (model) -- node[below,font=\scriptsize,color=orange!80!black,align=center] {imagined\\transitions} (planning);

		% Planning back to agent
		\draw[arrow,draw=purple!70] (planning.north east) -- node[left,font=\scriptsize,color=purple!70,pos=0.3] {update} (agent.south west);
	\end{tikzpicture}
	\caption{The Dyna architecture. The agent learns from real experience and from model-generated planning updates. Modern model-based deep RL can be viewed as a deep, high-dimensional version of this idea.}
	\label{fig:dyna_loop}
\end{figure}

\Needspace{16\baselineskip}
\begin{lstlisting}[style=pythonstyle,caption={A minimal tabular Dyna-Q update.},label={lst:dyna_q}]
import random
from collections import defaultdict

def dyna_q_step(Q, model, s, a, r, sp, gamma=0.99, alpha=0.1,
                planning_steps=10, actions=None):
    """One real Q-learning update plus model-based planning updates.

    Q: defaultdict mapping (state, action) -> value
    model: dictionary mapping (state, action) -> (reward, next_state)
    """
    # Real update
    best_next = max(Q[(sp, ap)] for ap in actions)
    td_target = r + gamma * best_next
    Q[(s, a)] += alpha * (td_target - Q[(s, a)])

    # Update learned one-step model
    model[(s, a)] = (r, sp)

    # Planning updates from simulated experience
    keys = list(model.keys())
    for _ in range(planning_steps):
        ss, aa = random.choice(keys)
        rr, ssp = model[(ss, aa)]
        best_next = max(Q[(ssp, ap)] for ap in actions)
        target = rr + gamma * best_next
        Q[(ss, aa)] += alpha * (target - Q[(ss, aa)])

    return Q, model
\end{lstlisting}

Dyna shows the basic promise of model-based learning: one real transition can support many planning updates. But it also shows the danger. If the learned model stores an incorrect transition, the agent may reinforce a wrong value many times.

\section{Model errors and the compounding-error problem}

Suppose a learned model has one-step prediction error:
\begin{equation}
	\epsilon_t
	=
	\lVert \hat{s}_{t+1} - s_{t+1} \rVert.
\end{equation}
If the model is rolled forward for many steps, small errors can accumulate. A state prediction error at time \(t+1\) becomes the input for the next prediction, which may produce a larger error at time \(t+2\), and so on. This is the compounding-error problem.

Theoretical analyses make this intuition precise. For Lipschitz dynamics, multi-step prediction error can grow roughly like \(O(\epsilon L^H/(L-1))\) when the one-step error is bounded by \(\epsilon\), the effective Lipschitz constant is \(L\), and the rollout horizon is \(H\) \citep{asadi2018lipschitz}. The exact constants depend on the metric and assumptions, but the lesson is simple: if the learned dynamics are not contractive, long model rollouts can amplify small one-step errors very quickly. 

\begin{figure}[t]
	\centering
	\begin{tikzpicture}[
		state/.style={circle,draw,thick,minimum size=0.7cm,align=center,font=\small},
		pred/.style={circle,draw,thick,dashed,minimum size=0.7cm,align=center,font=\small},
		arrow/.style={-{Latex[length=2.1mm]},thick},
		node distance=1.2cm
		]
		\node[state,fill=blue!8,draw=blue!70] (s0) {$s_t$};
		\node[state,fill=blue!8,draw=blue!70,right=of s0] (s1) {$s_{t+1}$};
		\node[state,fill=blue!8,draw=blue!70,right=of s1] (s2) {$s_{t+2}$};
		\node[state,fill=blue!8,draw=blue!70,right=of s2] (s3) {$s_{t+3}$};
		\node[pred,fill=orange!10,draw=orange!80!black,below=0.8cm of s1] (h1) {$\hat{s}_{t+1}$};
		\node[pred,fill=orange!10,draw=orange!80!black,below=1.0cm of s2] (h2) {$\hat{s}_{t+2}$};
		\node[pred,fill=orange!10,draw=orange!80!black,below=1.2cm of s3] (h3) {$\hat{s}_{t+3}$};
		\draw[arrow,draw=blue!70] (s0) -- (s1);
		\draw[arrow,draw=blue!70] (s1) -- (s2);
		\draw[arrow,draw=blue!70] (s2) -- (s3);
		\draw[arrow,draw=orange!80!black] (s0) -- (h1);
		\draw[arrow,draw=orange!80!black] (h1) -- (h2);
		\draw[arrow,draw=orange!80!black] (h2) -- (h3);
		\draw[<->,draw=red!70!black,thick] (s1) -- node[right,font=\scriptsize] {$\epsilon_1$} (h1);
		\draw[<->,draw=red!70!black,thick] (s2) -- node[right,font=\scriptsize] {$\epsilon_2$} (h2);
		\draw[<->,draw=red!70!black,thick] (s3) -- node[right,font=\scriptsize] {$\epsilon_3$} (h3);
	\end{tikzpicture}
	\caption{Compounding model error. A small one-step prediction error can grow when model predictions are recursively fed back into the model. This is why many practical methods use short model rollouts, uncertainty estimates, or latent representations.}
	\label{fig:compounding_error}
\end{figure}
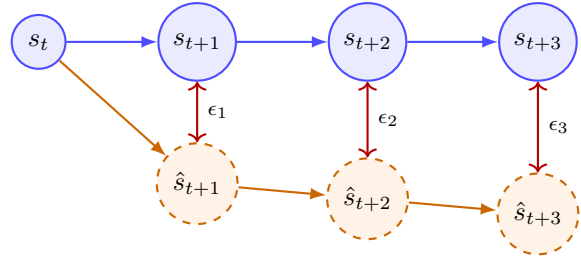

The more an agent plans inside a flawed model, the more it may exploit model errors. This is sometimes called model exploitation.

\begin{pitfallbox}{A model can be wrong in the direction the policy wants}
	In supervised learning, a prediction error is often just an error. In model-based RL, the policy actively searches for actions that maximize predicted return. This means it can find and exploit the model's blind spots. A slightly wrong model can become dangerously wrong when optimized against. 
\end{pitfallbox}

\section{Uncertainty-aware models and ensembles}

A practical response to model bias is to estimate uncertainty. If a model is uncertain about a region of state-action space, the agent can plan conservatively, shorten rollouts, or collect more real data.

One common method is to train an ensemble of dynamics models \citep{chua2018pets,janner2019mbpo}:
\begin{equation}
	\left\{ \hat{f}_{\psi_1},\hat{f}_{\psi_2},\ldots,\hat{f}_{\psi_M} \right\}.
\end{equation}
The ensemble mean gives a prediction, while disagreement provides an uncertainty estimate:
\begin{equation}
	\mu(s,a)=\frac{1}{M}\sum_{m=1}^{M}\hat{f}_{\psi_m}(s,a),
\end{equation}
\begin{equation}
	u(s,a)=\frac{1}{M}\sum_{m=1}^{M}\left\lVert \hat{f}_{\psi_m}(s,a)-\mu(s,a)\right\rVert^2.
\end{equation}

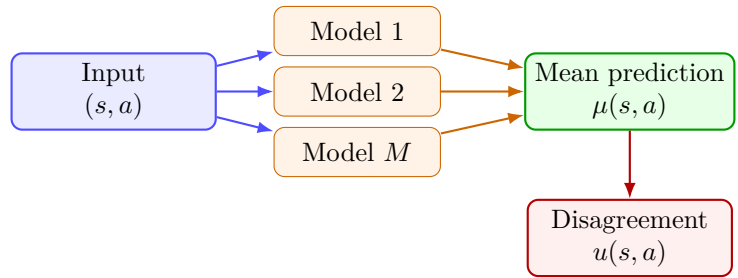
\begin{figure}[t]
	\centering
	\begin{tikzpicture}[
		box/.style={draw,rounded corners,thick,minimum width=2.7cm,minimum height=0.8cm,align=center,font=\small},
		small/.style={draw,rounded corners,minimum width=2.2cm,minimum height=0.65cm,align=center,font=\small},
		arrow/.style={-{Latex[length=2.1mm]},thick},
		node distance=0.75cm
		]
		\node[box,fill=blue!8,draw=blue!70] (input) {Input\\$(s,a)$};
		\node[small,fill=orange!10,draw=orange!80!black,right=of input,yshift=0.8cm] (m1) {Model 1};
		\node[small,fill=orange!10,draw=orange!80!black,right=of input] (m2) {Model 2};
		\node[small,fill=orange!10,draw=orange!80!black,right=of input,yshift=-0.8cm] (m3) {Model $M$};
		\node[box,fill=green!10,draw=green!60!black,right=1.1cm of m2] (mean) {Mean prediction\\$\mu(s,a)$};
		\node[box,fill=red!6,draw=red!70!black,below=0.9cm of mean] (unc) {Disagreement\\$u(s,a)$};
		\draw[arrow,draw=blue!70] (input) -- (m1);
		\draw[arrow,draw=blue!70] (input) -- (m2);
		\draw[arrow,draw=blue!70] (input) -- (m3);
		\draw[arrow,draw=orange!80!black] (m1) -- (mean);
		\draw[arrow,draw=orange!80!black] (m2) -- (mean);
		\draw[arrow,draw=orange!80!black] (m3) -- (mean);
		\draw[arrow,draw=red!70!black] (mean) -- (unc);
	\end{tikzpicture}
	\caption{Ensemble uncertainty. Multiple learned models predict the next state or latent state. Agreement gives confidence; disagreement warns that the model may be extrapolating beyond the data distribution.}
	\label{fig:ensemble_uncertainty}
\end{figure}

\Needspace{18\baselineskip}
\begin{lstlisting}[style=pythonstyle,caption={A compact probabilistic ensemble dynamics model.},label={lst:ensemble_model}]
import torch
import torch.nn as nn
import torch.nn.functional as F

class DynamicsModel(nn.Module):
    def __init__(self, state_dim, action_dim, hidden=256):
        super().__init__()
        self.net = nn.Sequential(
            nn.Linear(state_dim + action_dim, hidden), nn.ReLU(),
            nn.Linear(hidden, hidden), nn.ReLU(),
            nn.Linear(hidden, 2 * state_dim)  # mean and log variance
        )
        self.state_dim = state_dim

    def forward(self, s, a):
        x = torch.cat([s, a], dim=-1)
        mean, logvar = self.net(x).chunk(2, dim=-1)
        logvar = torch.clamp(logvar, -10.0, 2.0)
        return mean, logvar

    def loss(self, s, a, sp):
        mean, logvar = self.forward(s, a)
        inv_var = torch.exp(-logvar)
        # Gaussian negative log likelihood, up to constants
        return ((mean - sp).pow(2) * inv_var + logvar).mean()

def ensemble_predict(models, s, a):
    means = []
    for model in models:
        mean, _ = model(s, a)
        means.append(mean)
    preds = torch.stack(means, dim=0)       # [M, B, state_dim]
    mean = preds.mean(dim=0)
    disagreement = preds.var(dim=0).mean(dim=-1)
    return mean, disagreement
\end{lstlisting}

\section{Short-horizon model rollouts and MBPO}

Model-Based Policy Optimization (MBPO) uses a practical insight: learned models are often accurate for short horizons even when long-horizon rollouts are unreliable \citep{janner2019mbpo}. Instead of replacing the real environment with the model, MBPO uses the model to generate short synthetic rollouts starting from real states in the replay buffer.

Let \(H\) be the rollout horizon. A short model rollout generates
\begin{equation}
	s_0 \sim \mathcal{D}_{\mathrm{real}},
	\quad
	a_t \sim \pi(\cdot\given s_t),
	\quad
	\hat{s}_{t+1}\sim \hat{p}_\psi(\cdot\given s_t,a_t),
	\quad
	t=0,\ldots,H-1.
\end{equation}
The synthetic transitions are stored in a model buffer and used to train a model-free algorithm such as SAC, introduced in Chapter~11. Practical MBPO implementations train the policy on a mixture of real and synthetic transitions; once the model is sufficiently trusted, synthetic transitions often dominate the update data by a factor of roughly \(5\)--\(20\), while real transitions remain essential anchors against model bias. 

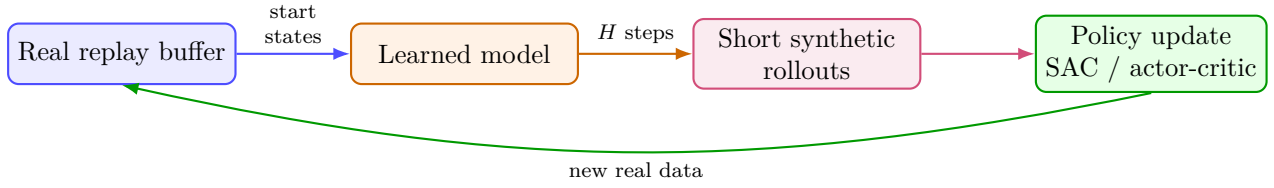
\begin{figure}[t]
	\centering
	\begin{tikzpicture}[
		box/.style={draw,rounded corners,thick,minimum width=3.0cm,minimum height=0.8cm,align=center,font=\small},
		arrow/.style={-{Latex[length=2.2mm]},thick},
		node distance=1.5cm
		]
		\node[box,fill=blue!8,draw=blue!70] (real) {Real replay buffer};
		\node[box,fill=orange!10,draw=orange!80!black,right=of real] (model) {Learned model};
		\node[box,fill=purple!8,draw=purple!70,right=of model] (synthetic) {Short synthetic\\rollouts};
		\node[box,fill=green!10,draw=green!60!black,right=of synthetic] (policy) {Policy update\\SAC / actor-critic};
		\draw[arrow,draw=blue!70] (real) -- node[above,font=\scriptsize,align=center] {start\\states} (model);
		\draw[arrow,draw=orange!80!black] (model) -- node[above,font=\scriptsize] {$H$ steps} (synthetic);
		\draw[arrow,draw=purple!70] (synthetic) -- (policy);
		\draw[arrow,draw=green!60!black] (policy.south) to[bend left=12] node[below,font=\scriptsize] {new real data} (real.south);
	\end{tikzpicture}
	\caption{MBPO-style short-horizon rollouts. The learned model is used only for short synthetic trajectories starting from real replay states. This reduces compounding error while improving sample efficiency.}
	\label{fig:mbpo_loop}
\end{figure}

\Needspace{18\baselineskip}
\begin{lstlisting}[style=pythonstyle,caption={Short model rollouts for MBPO-style data augmentation.},label={lst:short_rollouts}]
@torch.no_grad()
def generate_model_rollouts(policy, model, start_states, horizon=5):
    """Generate short synthetic rollouts from real replay states."""
    synthetic = []
    s = start_states

    for _ in range(horizon):
        a = policy.sample_action(s)
        sp_mean, sp_logvar = model(s, a)
        noise = torch.randn_like(sp_mean)
        sp = sp_mean + noise * torch.exp(0.5 * sp_logvar)

        # In practice reward/done can be learned or computed from a task model.
        r = learned_reward(s, a, sp)
        d = learned_done(s, a, sp)

        synthetic.append((s, a, r, sp, d))
        s = sp

    return synthetic
\end{lstlisting}

\begin{warningbox}{Do not roll out too far too early}
	Long imagined rollouts can be harmful when the model is immature. Many practical systems increase the rollout horizon slowly, use uncertainty thresholds, or terminate rollouts when ensemble disagreement becomes too large.
\end{warningbox}

\section{Planning with learned models: MPC and CEM}

A different use of a learned model is model-predictive control (MPC). Instead of training a policy by imagined rollouts, the agent plans online. At each decision step, it searches for an action sequence that maximizes predicted return under the model, executes only the first action, then replans at the next state.

Given horizon \(H\), an action sequence \(a_{0:H-1}\), and learned model \(\hat{p}_\psi\), the planner solves
\begin{equation}
	a^*_{0:H-1}
	=
	\argmax_{a_{0:H-1}}
	\E_{\hat{p}_\psi}
	\left[
	\sum_{t=0}^{H-1}\gamma^t \hat{r}(\hat{s}_t,a_t)
	\right].
\end{equation}
Only \(a^*_0\) is executed. This receding-horizon structure is crucial: it allows the controller to correct planning errors using new real observations.

One common black-box optimizer for MPC is the Cross-Entropy Method (CEM). CEM samples action sequences, evaluates them under the model, keeps elite sequences, refits a Gaussian distribution to the elites, and repeats. In continuous-control systems, practical MPC implementations often add action-smoothness penalties or temporally correlated sampling noise so that planned actions do not jitter from one step to the next \citep{chua2018pets}.

MPC and imagined-policy learning use models in different ways. MPC plans online at every decision step; it does not require a learned policy, but it can be expensive at runtime. Dreamer-style agents train a policy offline or asynchronously using imagined rollouts; execution is cheap because the actor produces an action directly, but the learned behavior depends strongly on the quality of the world model. In short, MPC spends computation at decision time, while Dreamer spends computation during training. 

\begin{figure}[t]
	\centering
	\begin{tikzpicture}[
		box/.style={draw,rounded corners,thick,minimum width=2.9cm,minimum height=0.8cm,align=center,font=\small},
		arrow/.style={-{Latex[length=2.2mm]},thick},
		node distance=0.75cm
		]
		\node[box,fill=blue!8,draw=blue!70] (sample) {Sample action\\sequences};
		\node[box,fill=orange!10,draw=orange!80!black,right=of sample] (eval) {Evaluate in\\learned model};
		\node[box,fill=purple!8,draw=purple!70,right=of eval] (elite) {Select elite\\sequences};
		\node[box,fill=green!10,draw=green!60!black,right=of elite] (refit) {Refit sampling\\distribution};
		\node[box,fill=gray!10,draw=gray!70,below=1.1cm of eval] (execute) {Execute first\\action only};
		\draw[arrow,draw=blue!70] (sample) -- (eval);
		\draw[arrow,draw=orange!80!black] (eval) -- (elite);
		\draw[arrow,draw=purple!70] (elite) -- (refit);
		\draw[arrow,draw=green!60!black] (refit.south) to[bend left=18] (sample.south);
		\draw[arrow,draw=gray!70] (elite.south) |- (execute.east);
	\end{tikzpicture}
	\caption{CEM planning for model-predictive control. The planner samples action sequences, evaluates them in the model, refits the sampling distribution to elite candidates, and executes only the first action before replanning.}
	\label{fig:cem_mpc}
\end{figure}
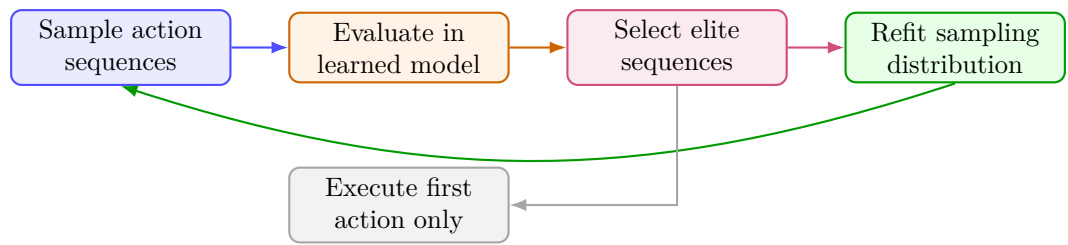

\Needspace{20\baselineskip}
\begin{lstlisting}[style=pythonstyle,caption={CEM planner for continuous-action model-predictive control.},label={lst:cem_planner}]
@torch.no_grad()
def cem_plan(model, reward_fn, state, action_dim, horizon=15,
             population=512, elite_frac=0.1, iterations=5,
             action_low=-1.0, action_high=1.0, gamma=0.99):
    """Plan an action sequence and return the first action."""
    device = state.device
    mean = torch.zeros(horizon, action_dim, device=device)
    std = torch.ones(horizon, action_dim, device=device)
    n_elite = max(1, int(population * elite_frac))

    for _ in range(iterations):
        eps = torch.randn(population, horizon, action_dim, device=device)
        actions = mean.unsqueeze(0) + std.unsqueeze(0) * eps
        actions = torch.clamp(actions, action_low, action_high)

        s = state.repeat(population, 1)
        returns = torch.zeros(population, device=device)
        discount = 1.0

        for t in range(horizon):
            a = actions[:, t]
            sp_mean, _ = model(s, a)
            r = reward_fn(s, a, sp_mean)
            returns += discount * r.squeeze(-1)
            s = sp_mean
            discount *= gamma

        elite_idx = returns.topk(n_elite).indices
        elite_actions = actions[elite_idx]
        mean = elite_actions.mean(dim=0)
        std = elite_actions.std(dim=0).clamp_min(1e-3)

    return torch.clamp(mean[0], action_low, action_high)
\end{lstlisting}

\section{Latent world models from pixels}

For high-dimensional observations such as images or dense network heatmaps, predicting directly in observation space can be expensive and unnecessary. Latent world models learn a compact state representation \(z_t\) and predict dynamics in that latent space:
\begin{equation}
	z_t = e_\psi(o_t),
	\qquad
	z_{t+1}\sim \hat{p}_\psi(\cdot\given z_t,a_t).
\end{equation}
World Models by Ha and Schmidhuber popularized the view that an agent can learn a compressed model of the environment and train a controller inside it \citep{ha2018worldmodels}. PlaNet and Dreamer developed latent dynamics for planning and actor-critic learning from pixels \citep{hafner2019planet,hafner2020dreamer,hafner2021dreamerv2,hafner2023dreamerv3}.

A common latent structure is a recurrent state-space model (RSSM), combining deterministic memory and stochastic latent variables:
\begin{align}
	h_t &= f_\psi(h_{t-1}, z_{t-1}, a_{t-1}), \\
	z_t &\sim p_\psi(z_t\given h_t), \\
	o_t &\sim p_\psi(o_t\given h_t,z_t), \\
	r_t &\sim p_\psi(r_t\given h_t,z_t).
\end{align}
The deterministic state \(h_t\) carries memory; the stochastic latent \(z_t\) captures uncertainty. 

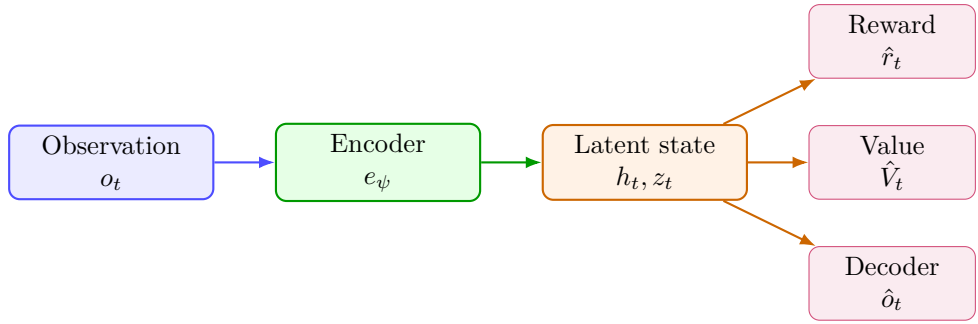
\begin{figure}[t]
	\centering
	\begin{tikzpicture}[
		box/.style={draw,rounded corners,thick,minimum width=2.7cm,minimum height=0.8cm,align=center,font=\small},
		small/.style={draw,rounded corners,minimum width=2.2cm,minimum height=0.65cm,align=center,font=\small},
		arrow/.style={-{Latex[length=2.1mm]},thick},
		node distance=0.8cm
		]
		\node[box,fill=blue!8,draw=blue!70] (obs) {Observation\\$o_t$};
		\node[box,fill=green!10,draw=green!60!black,right=of obs] (enc) {Encoder\\$e_\psi$};
		\node[box,fill=orange!10,draw=orange!80!black,right=of enc] (latent) {Latent state\\$h_t,z_t$};
		\node[small,fill=purple!8,draw=purple!70,right=of latent,yshift=1.6cm] (rew) {Reward\\$\hat r_t$};
		\node[small,fill=purple!8,draw=purple!70,right=of latent] (val) {Value\\$\hat V_t$};
		\node[small,fill=purple!8,draw=purple!70,right=of latent,yshift=-1.6cm] (dec) {Decoder\\$\hat o_t$};
		\draw[arrow,draw=blue!70] (obs) -- (enc);
		\draw[arrow,draw=green!60!black] (enc) -- (latent);
		\draw[arrow,draw=orange!80!black] (latent) -- (rew);
		\draw[arrow,draw=orange!80!black] (latent) -- (val);
		\draw[arrow,draw=orange!80!black] (latent) -- (dec);
	\end{tikzpicture}
	\caption{A latent world model. Observations are encoded into a latent state. The model predicts rewards, values, future latents, and sometimes reconstructions. Modern agents often plan or learn policies in this latent space rather than directly in pixel space.}
	\label{fig:latent_world_model}
\end{figure}

\section{Dreamer-style imagination learning}

Dreamer uses a learned latent world model to improve an actor and critic from imagined trajectories \citep{hafner2020dreamer,hafner2021dreamerv2,hafner2023dreamerv3,hafner2025dreamerv3nature}. The key idea is that once the model can roll forward in latent space, the policy can be optimized using imagined rewards and values without querying the real environment for every update.

For imagined latent states \(z_t\), actions \(a_t\sim\pi_\phi(\cdot\given z_t)\), and predicted rewards \(\hat{r}_t\), a lambda-return target is
\begin{equation}
	V_t^\lambda
	=
	\hat{r}_t
	+
	\gamma
	\left((1-\lambda)V_\xi(z_{t+1}) + \lambda V_{t+1}^\lambda\right).
\end{equation}
The actor is trained to choose actions that lead to high imagined returns.

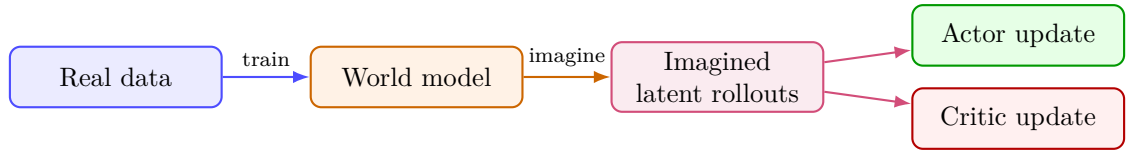
\begin{figure}[t]
	\centering
	\begin{tikzpicture}[
		box/.style={draw,rounded corners,thick,minimum width=2.8cm,minimum height=0.8cm,align=center,font=\small},
		arrow/.style={-{Latex[length=2.2mm]},thick},
		node distance=1.15cm
		]
		\node[box,fill=blue!8,draw=blue!70] (data) {Real data};
		\node[box,fill=orange!10,draw=orange!80!black,right=of data] (world) {World model};
		\node[box,fill=purple!8,draw=purple!70,right=of world] (imag) {Imagined\\latent rollouts};
		\node[box,fill=green!10,draw=green!60!black,right=of imag,yshift=0.55cm] (actor) {Actor update};
		\node[box,fill=red!6,draw=red!70!black,right=of imag,yshift=-0.55cm] (critic) {Critic update};
		\draw[arrow,draw=blue!70] (data) -- node[above,font=\scriptsize] {train} (world);
		\draw[arrow,draw=orange!80!black] (world) -- node[above,font=\scriptsize] {imagine} (imag);
		\draw[arrow,draw=purple!70] (imag) -- (actor);
		\draw[arrow,draw=purple!70] (imag) -- (critic);
	\end{tikzpicture}
	\caption{Dreamer-style imagination learning. The world model is trained from real data. The actor and critic are then updated on imagined latent trajectories.}
	\label{fig:dreamer_imagination}
\end{figure}

\Needspace{18\baselineskip}
\begin{lstlisting}[style=pythonstyle,caption={Lambda returns for imagined trajectories.},label={lst:lambda_returns}]
def lambda_returns(rewards, values, discounts, lam=0.95):
    """Compute lambda returns for imagined trajectories.

    rewards:   tensor [T, B]
    values:    tensor [T + 1, B]
    discounts: tensor [T, B], usually gamma * (1 - done)
    """
    T = rewards.shape[0]
    returns = torch.zeros_like(rewards)
    next_return = values[-1]

    for t in reversed(range(T)):
        bootstrap = (1.0 - lam) * values[t + 1] + lam * next_return
        next_return = rewards[t] + discounts[t] * bootstrap
        returns[t] = next_return

    return returns
\end{lstlisting}

\begin{researchbox}{World models for UAV/SDN telemetry}
	In UAV/SDN control, a world model need not generate images. It can learn latent dynamics of traffic load, SINR, latency, UAV battery, user mobility, and queue state. The imagined rollouts can then estimate future QoS degradation or battery risk before executing a real action.
\end{researchbox}

\section{MuZero-style value-equivalent models}

MuZero is model-based, but not because it learns a visually faithful simulator. It learns a model that supports planning by predicting reward, value, and policy quantities \citep{schrittwieser2020muzero}. The model has three parts:
\begin{align}
	h_0 &= \mathrm{representation}_\psi(o_{1:t}), \\
	h_{k+1}, \hat{r}_{k+1} &= \mathrm{dynamics}_\psi(h_k,a_k), \\
	\hat{p}_k, \hat{v}_k &= \mathrm{prediction}_\psi(h_k).
\end{align}
The model is not trained to reconstruct the observation. It is trained to be useful for planning. MuZero uses Monte Carlo Tree Search inside the learned latent model to generate planning targets for the policy and value heads. In other words, the learned model is coupled to search: its purpose is not visual prediction, but value-equivalent planning.

\begin{figure}[t]
	\centering
	\begin{tikzpicture}[
		box/.style={draw,rounded corners,thick,minimum width=2.9cm,minimum height=0.8cm,align=center,font=\small},
		arrow/.style={-{Latex[length=2.2mm]},thick},
		node distance=0.9cm
		]
		\node[box,fill=blue!8,draw=blue!70,font=\small\bfseries] (obs) {Observation\\{\normalfont\small history}};
		\node[box,fill=green!10,draw=green!60!black,font=\small\bfseries,right=of obs] (repr) {Representation\\{\normalfont\small $h_0$}};
		\node[box,fill=orange!10,draw=orange!80!black,font=\small\bfseries,right=of repr] (dyn) {Recurrent\\{\normalfont\small dynamics}};
		\node[box,fill=purple!8,draw=purple!70,font=\small\bfseries,right=1.2cm of dyn,yshift=0.6cm] (pred) {Policy + value};
		\node[box,fill=red!6,draw=red!70!black,font=\small\bfseries,right=1.2cm of dyn,yshift=-0.6cm] (rew) {Reward};

		\draw[arrow,draw=blue!70] (obs) -- (repr);
		\draw[arrow,draw=green!60!black] (repr) -- (dyn);
		\draw[arrow,draw=orange!80!black] (dyn) -- (pred);
		\draw[arrow,draw=orange!80!black] (dyn) -- (rew);
		\draw[arrow,draw=orange!80!black]
		([xshift=-4pt]dyn.north) to[out=100,in=80,looseness=4.5]
		node[above=4pt,font=\scriptsize,color=orange!80!black] {unroll}
		([xshift=4pt]dyn.north);
	\end{tikzpicture}
	\caption{MuZero-style value-equivalent modeling. The model predicts the quantities needed for search: reward, policy, and value. It does not need to reconstruct the full observation.}
	\label{fig:muzero_model}
\end{figure}
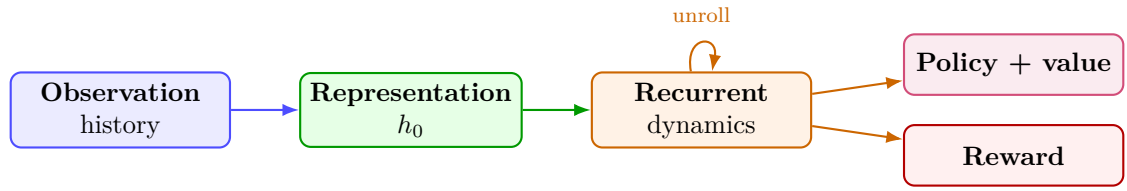

EfficientZero extended this line toward sample-efficient visual control, reporting strong performance on Atari 100k with limited data by combining MuZero-style planning with self-supervised consistency and value prefix prediction \citep{ye2021efficientzero}. EfficientZero V2 extended the approach to broader discrete and continuous control settings \citep{wang2024efficientzerov2}.

\section{TD-MPC and TD-MPC2: latent planning for continuous control}

TD-MPC and TD-MPC2 combine learned latent dynamics, temporal-difference learning, and model-predictive control \citep{hansen2022tdmpc,hansen2024tdmpc2}. Unlike pixel reconstruction world models, TD-MPC is decoder-free: it learns a latent model optimized for control-relevant predictions.

A simplified TD-MPC-style objective combines:
\begin{align}
	\mathcal{L}_{\mathrm{reward}} &= \lVert \hat{r}_t-r_t\rVert^2, \\
	\mathcal{L}_{\mathrm{value}} &= \lVert \hat{Q}(z_t,a_t)-y_t\rVert^2, \\
	\mathcal{L}_{\mathrm{consistency}} &= \lVert \hat{z}_{t+1}-\mathrm{sg}(z_{t+1})\rVert^2,
\end{align}
where \(\mathrm{sg}\) denotes stop-gradient. The planner then uses the latent model to search for actions. 

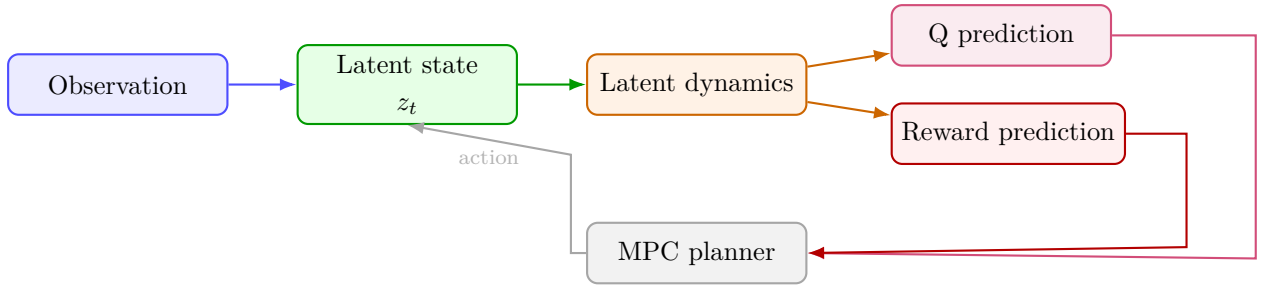
\begin{figure}[t]
	\centering
	\begin{tikzpicture}[
		box/.style={draw,rounded corners,thick,minimum width=2.9cm,minimum height=0.8cm,
			align=center,font=\small},
		arrow/.style={-{Latex[length=2.2mm]},thick},
		node distance=0.9cm
		]

		% Main pipeline nodes
		\node[box,fill=blue!8,draw=blue!70]                           (obs)    {Observation};
		\node[box,fill=green!10,draw=green!60!black,right=of obs]     (latent) {Latent state\\[2pt]$z_t$};
		\node[box,fill=orange!10,draw=orange!80!black,right=of latent](model)  {Latent dynamics};

		% Output nodes
		\node[box,fill=purple!8,draw=purple!70,
		right=1.1cm of model, yshift= 0.65cm] (q)   {Q prediction};
		\node[box,fill=red!6,draw=red!70!black,
		right=1.1cm of model, yshift=-0.65cm] (r)   {Reward prediction};

		% MPC planner below latent dynamics
		\node[box,fill=gray!10,draw=gray!70,
		below=1.4cm of model]                 (mpc) {MPC planner};

		% Forward arrows
		\draw[arrow,draw=blue!70]         (obs)    -- (latent);
		\draw[arrow,draw=green!60!black]  (latent) -- (model);
		\draw[arrow,draw=orange!80!black] (model)  -- (q);
		\draw[arrow,draw=orange!80!black] (model)  -- (r);

		% Q -> MPC: go right past the box, then down, then left into mpc.east
		\draw[arrow,draw=purple!70]
		(q.east)
		-- ++(1.9,0)
		-- ++(0,-2.95)
		-- (mpc.east);

		% Reward -> MPC: go right past the box, then down, then left into mpc.east
		\draw[arrow,draw=red!70!black]
		(r.east)
		-- ++(0.8,0)
		-- ++(0.0,-1.5)
		-- ([yshift=0pt]mpc.east);

		% MPC feedback: left, up, into latent.west
		\draw[arrow,draw=gray!70]
		(mpc.west)
		-- ++(-0.2,0)
		-- ++(-0.0,1.3)
		-- node[below,font=\scriptsize,color=gray!60]{action}
		(latent.south);

	\end{tikzpicture}
	\caption{TD-MPC-style latent planning. The model predicts control-relevant
		quantities in latent space and uses MPC to search for high-value action sequences.}
	\label{fig:tdmpc_latent_planning}
\end{figure}

TD-MPC2 is important for the 2026 view of model-based RL because it emphasizes scalability and robustness across many continuous-control tasks with a single hyperparameter setting, including large multitask agents \citep{hansen2024tdmpc2}. It is one of the strongest model-based agents for continuous control as of the mid-2020s, but it is not the final word: robustness, deployment reliability, and multi-task transfer remain open problems.

\section{Model-based RL for UAV/SDN control}

For UAV-assisted wireless networks and SDN control, model-based RL is especially attractive. The environment is dynamic, partially observable, and constrained. Real exploration can be costly or unsafe. A learned model can forecast whether an action will degrade latency, SINR, energy, or safety.

A useful UAV/SDN predictive model may estimate:
\begin{itemize}
	\item next UAV position and battery;
	\item future user density and mobility clusters;
	\item SINR and throughput under candidate movement and bandwidth actions;
	\item queueing delay and latency risk;
	\item collision or separation-margin violations;
	\item CBF slack or safety-filter intervention probability.
\end{itemize}

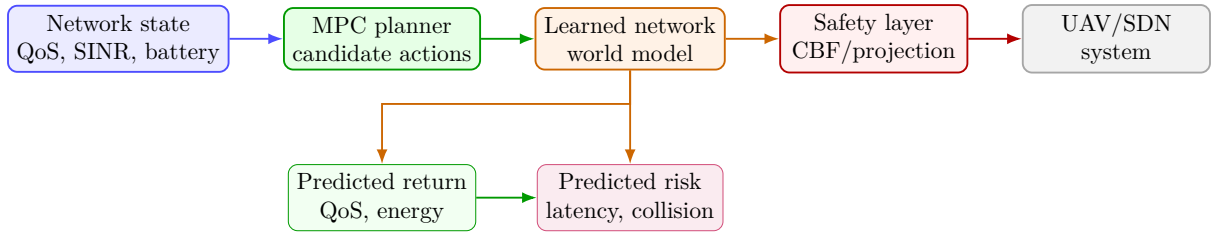
\begin{figure}[t]
	\centering
	\resizebox{\columnwidth}{!}{%
		\begin{tikzpicture}[
			box/.style={draw,rounded corners,thick,minimum width=2.8cm,minimum height=0.8cm,
				align=center,font=\small},
			small/.style={draw,rounded corners,minimum width=2.5cm,minimum height=0.7cm,
				align=center,font=\small},
			arrow/.style={-{Latex[length=2.2mm]},thick},
			node distance=0.8cm
			]

			% Top row
			\node[box,fill=blue!8,draw=blue!70]                             (state)   {Network state\\QoS, SINR, battery};
			\node[box,fill=green!10,draw=green!60!black,right=of state]     (planner) {MPC planner\\candidate actions};
			\node[box,fill=orange!10,draw=orange!80!black,right=of planner] (model)   {Learned network\\world model};
			\node[box,fill=red!6,draw=red!70!black,right=of model]          (safety)  {Safety layer\\CBF/projection};
			\node[box,fill=gray!10,draw=gray!70,right=of safety]            (env)     {UAV/SDN\\system};

			% Bottom row
			\node[small,fill=green!5,draw=green!60!black,
			below=1.4cm of planner]                                   (score)   {Predicted return\\QoS, energy};
			\node[small,fill=purple!8,draw=purple!70,
			below=1.4cm of model]                                     (risk)    {Predicted risk\\latency, collision};

			% Top row arrows
			\draw[arrow,draw=blue!70]         (state)   -- (planner);
			\draw[arrow,draw=green!60!black]  (planner) -- (model);
			\draw[arrow,draw=orange!80!black] (model)   -- (safety);
			\draw[arrow,draw=red!70!black]    (safety)  -- (env);

			% Model -> score: go down then left
			\draw[arrow,draw=orange!80!black]
			(model.south) -- ++(0,-0.5) -| (score.north);

			% Model -> risk: straight down
			\draw[arrow,draw=orange!80!black]
			(model.south) -- (risk.north);

			% Score -> risk
			\draw[arrow,draw=green!60!black] (score.east) -- (risk.west);

		\end{tikzpicture}%
	}
	\caption{Model-based UAV/SDN control. A learned network model predicts future QoS
		and risk for candidate actions. An MPC planner can choose actions that improve service
		while a safety layer enforces hard constraints before execution.}
	\label{fig:uav_sdn_mbrl}
\end{figure}

\begin{researchbox}{Research signature: model-based safe DRL for UAV/SDN}
	A distinctive research direction is to combine learned world models with safety filters. The world model predicts future QoS and risk; the planner selects actions that optimize reward; a CBF or projection layer enforces hard safety constraints; and the critic or model is trained on the executed action after projection. This avoids teaching the model that unsafe pre-filter actions were actually executed.
\end{researchbox}

\Needspace{20\baselineskip}
\begin{lstlisting}[style=pythonstyle,caption={A UAV/SDN predictive model with reward and cost heads.},label={lst:uav_sdn_model}]
class UAVSDNWorldModel(nn.Module):
    """Predict next latent state, QoS reward, and safety cost."""
    def __init__(self, state_dim, action_dim, latent_dim=128, hidden=256):
        super().__init__()
        self.encoder = nn.Sequential(
            nn.Linear(state_dim, hidden), nn.ReLU(),
            nn.Linear(hidden, latent_dim), nn.ReLU()
        )
        self.dynamics = nn.Sequential(
            nn.Linear(latent_dim + action_dim, hidden), nn.ReLU(),
            nn.Linear(hidden, latent_dim)
        )
        self.reward_head = nn.Linear(latent_dim, 1)
        self.cost_head = nn.Linear(latent_dim, 1)
        self.latency_head = nn.Linear(latent_dim, 1)
        self.sinr_head = nn.Linear(latent_dim, 1)

    def forward(self, state, action):
        z = self.encoder(state)
        zp = self.dynamics(torch.cat([z, action], dim=-1))
        pred = {
            "next_latent": zp,
            "reward": self.reward_head(zp),
            "cost": torch.nn.functional.softplus(self.cost_head(zp)),
            "latency": torch.nn.functional.softplus(self.latency_head(zp)),
            "sinr": self.sinr_head(zp),
        }
        return pred
\end{lstlisting}

\section{Concrete PyTorch building blocks}

This section collects minimal components that can be reused in research code. They are intentionally compact, but they expose the main engineering decisions: model loss, uncertainty penalty, planning, and safe execution.

\subsection{Training a one-step model}

\Needspace{18\baselineskip}
\begin{lstlisting}[style=pythonstyle,caption={Training a one-step dynamics model.},label={lst:model_train_step}]
def train_world_model(model, optimizer, batch):
    s, a, r, sp, done = batch
    pred = model(s, a)

    # If the model predicts latent state, compare to encoded next state.
    with torch.no_grad():
        target_zp = model.encoder(sp)

    dyn_loss = (pred["next_latent"] - target_zp).pow(2).mean()
    reward_loss = (pred["reward"] - r).pow(2).mean()

    # Example safety labels: positive cost for unsafe transitions.
    cost_target = batch_cost_label(s, a, sp, done)
    cost_loss = (pred["cost"] - cost_target).pow(2).mean()

    loss = dyn_loss + reward_loss + 0.5 * cost_loss
    optimizer.zero_grad()
    loss.backward()
    torch.nn.utils.clip_grad_norm_(model.parameters(), 10.0)
    optimizer.step()

    return {"model_loss": float(loss.detach())}
\end{lstlisting}

\subsection{Uncertainty-penalized planning}

\Needspace{20\baselineskip}
\begin{lstlisting}[style=pythonstyle,caption={Uncertainty-penalized score for safe planning.},label={lst:uncertainty_penalty}]
def score_action_sequence(models, reward_fn, cost_fn, state, actions, gamma=0.99, uncertainty_coef=1.0,
                          cost_coef=5.0):
    """Evaluate one candidate action sequence with an ensemble."""
    total = 0.0
    discount = 1.0
    s = state

    for a in actions:
        preds = []
        for model in models:
            pred = model(s, a)
            preds.append(pred["next_latent"])
        stacked = torch.stack(preds, dim=0)
        zp = stacked.mean(dim=0)
        uncertainty = stacked.var(dim=0).mean()

        reward = reward_fn(s, a, zp)
        cost = cost_fn(s, a, zp)
        total = total + discount * (reward - cost_coef * cost
                                    - uncertainty_coef * uncertainty)
        s = zp
        discount *= gamma

    return total
\end{lstlisting}

\subsection{Safe MPC wrapper}

\Needspace{20\baselineskip}
\begin{lstlisting}[style=pythonstyle,caption={Safe MPC wrapper: plan, project, execute.},label={lst:safe_mpc_wrapper}]
def safe_model_based_action(state, planner, safety_filter, env):
    """Research pattern for UAV/SDN model-based safe control.

    1. Plan an action using the learned model.
    2. Project or filter the action using hard safety constraints.
    3. Execute the filtered action.
    4. Store the executed action, not only the proposed action.
    """
    proposed_action = planner.plan(state)
    executed_action, safety_info = safety_filter.project(state, proposed_action)
    next_state, reward, done, info = env.step(executed_action)

    transition = {
        "state": state,
        "proposed_action": proposed_action,
        "executed_action": executed_action,
        "reward": reward,
        "next_state": next_state,
        "done": done,
        "safety_info": safety_info,
    }
    return next_state, transition
\end{lstlisting}

\section{Diagnostics, debugging, and failure modes}

Model-based RL has different failure modes from model-free RL. A high policy reward is not enough. The model itself must be audited.

\begin{table}[t]
	\centering
	\caption{Common model-based RL failure modes and diagnostics.}
	\label{tab:mbrl_debugging}
	\begin{tabular}{p{0.23\textwidth}p{0.33\textwidth}p{0.32\textwidth}}
		\toprule
		Symptom & Likely cause & What to inspect \\
		\midrule
		Great imagined return, poor real return & Model exploitation & Real-vs-model rollout error, uncertainty, horizon \\
		Good one-step prediction, bad planning & Compounding error & Multi-step rollout accuracy \\
		Unstable policy updates & Model buffer too large or stale & Real-to-synthetic ratio, model age \\
		Planner chooses extreme actions & Model extrapolation & Ensemble disagreement, action bounds \\
		Unsafe behavior after filtering & Training on proposed instead of executed action & Replay storage and critic targets \\
		World model ignores rare events & Imbalanced data & Safety-cost labels, rare-event replay \\
		High reconstruction quality, poor control & Wrong objective & Reward/value prediction, latent controllability \\
		\bottomrule
	\end{tabular}
\end{table}

\begin{warningbox}{Prediction accuracy is not the same as control usefulness}
	A model with low pixel reconstruction loss may still be poor for control if it fails to predict rewards, values, constraints, or action-sensitive latent changes. Always evaluate a model on decision-relevant metrics, not only reconstruction error.
\end{warningbox}

\section{Frontier note: diffusion and foundation world models}

A 2026-facing view of model-based RL should also mention a parallel line of work: large generative world models. DIAMOND trains an RL agent inside a diffusion world model for Atari, showing that diffusion models can be adapted to long-horizon environment simulation when efficiency and temporal stability are handled carefully \citep{alonso2024diamond}. Genie studies generative interactive environments trained from unlabelled videos, introducing the idea of a large foundation world model with a learned latent action interface \citep{bruce2024genie}. These systems are not drop-in replacements for MBPO, Dreamer, or MuZero, but they point toward a future where world models may be pre-trained at scale and then adapted for control, robotics, games, or network digital twins.

For UAV/SDN research, the lesson is not that one should generate photorealistic video of the network. The lesson is that scalable generative models may provide richer simulated futures, rare-event augmentation, and scenario generation. A practical communication-network world model still needs to predict decision-relevant variables: load, SINR, latency, queueing, battery, mobility, interference, and constraint risk.

\section{Limitations and when not to use model-based RL}

Model-based RL is powerful, but it is not always the right tool.

\begin{enumerate}[leftmargin=*]
	\item \textbf{Model bias can dominate.} If the learned model is wrong in important regions, planning can make the policy worse.
	\item \textbf{Long-horizon prediction is hard.} Small errors compound over time, especially in chaotic or partially observed environments.
	\item \textbf{High-dimensional models are expensive.} World models can be computationally heavy, especially from pixels.
	\item \textbf{Uncertainty estimates can be misleading.} Ensembles capture some epistemic uncertainty but not all forms of distribution shift.
	\item \textbf{Planning costs time.} MPC can be too slow for high-frequency control unless the planner is carefully optimized.
	\item \textbf{Reward models can be exploited.} Learned rewards may be easier to exploit than real rewards.
	\item \textbf{Safety is not automatic.} A model-based planner can still propose unsafe actions unless constraints, shields, CBFs, or conservative penalties are used.
\end{enumerate}

\section{Exercises}

\subsection*{Conceptual exercises}
\begin{enumerate}[leftmargin=*]
	\item Explain the difference between a complete environment model and a value-equivalent model.
	\item Why can a model with good one-step prediction still fail during multi-step planning?
	\item Why does MBPO use short model rollouts rather than replacing the environment completely?
	\item Why is uncertainty estimation especially important in model-based RL?
	\item In a UAV/SDN system, which variables should a world model predict for safe control?
\end{enumerate}

\subsection*{Mathematical exercises}
\begin{enumerate}[leftmargin=*]
	\item Derive the Dyna-Q update from the ordinary Q-learning update by replacing a real transition with a model-generated transition.
	\item Suppose a model has one-step error bounded by \(\epsilon\) and a Lipschitz constant \(L\) for the dynamics. Sketch why multi-step error can grow with horizon.
	\item Starting from the CEM objective, explain why executing only the first planned action reduces the effect of model error.
	\item Derive the lambda-return recursion used in Dreamer-style imagined actor-critic learning.
	\item For an ensemble of \(M\) models, derive the sample variance estimate used as a disagreement penalty.
\end{enumerate}

\subsection*{Coding exercises}
\begin{enumerate}[leftmargin=*]
	\item Implement a one-step dynamics model for a simple continuous-control environment.
	\item Add an ensemble of five models and compute disagreement-based uncertainty.
	\item Implement CEM planning using the learned model.
	\item Modify the CEM planner to penalize predicted safety cost.
	\item Implement MBPO-style short rollouts and compare different rollout horizons.
\end{enumerate}

\subsection*{Research thinking exercises}
\begin{enumerate}[leftmargin=*]
	\item Design a model-based controller for UAV placement under latency and battery constraints. What should the model predict?
	\item Should a safety filter be placed before or after the learned model in training? Explain the difference between proposed and executed actions.
	\item How would you evaluate whether a learned network world model is useful for QoS control?
	\item Compare Dreamer-style imagination and MPC-style online planning for SDN traffic engineering.
	\item Propose an uncertainty-aware model-based safe-RL architecture that combines ensembles, CBF projection, and actor-critic learning.
\end{enumerate}

\section*{Looking Ahead to Chapter 13: Food for Thought}
\addcontentsline{toc}{section}{Looking Ahead to Chapter 13: Food for Thought}

Chapter~12 showed that a model can be used in many ways: Dyna-style planning updates, MBPO-style short rollouts, MPC, latent imagination, and value-equivalent planning. But one method deserves its own chapter because it changed the way researchers think about models and planning: MuZero.

MuZero asks a subtle question:
\begin{quote}
	If the goal is planning, does the model need to predict the real next observation, or only the quantities needed by search?
\end{quote}

Before moving to Chapter~13, consider these questions:
\begin{enumerate}[leftmargin=*]
	\item What information is necessary for planning?
	\item Can a model be useful even if it cannot reconstruct observations?
	\item Why might reward, value, and policy predictions be more important than pixel prediction?
	\item How does tree search use a learned model differently from Dreamer-style actor-critic imagination?
	\item In UAV/SDN control, what would a value-equivalent model predict?
\end{enumerate}

\begin{quote}
	Chapter~12 explained how models support planning and imagination. Chapter~13 asks how little a model needs to know in order to plan effectively.
\end{quote}
	\chapter[MuZero and Learning Without Known Rules]{MuZero and Learning Without Known Rules}
\label{ch:muzero}
\chaptermark{MuZero and Learned-Rule Planning}

\begin{keybox}{Chapter thesis}
	MuZero is not just a stronger DQN, and it is not just a world model. It is a decision-making system that learns an abstract, value-equivalent model and uses search inside that model to improve action selection. The learned model is not trained to reconstruct observations perfectly. It is trained to predict the quantities that matter for planning: rewards, values, and improved policies. This distinction is why MuZero is one of the most important bridges between model-based RL, planning, and deep value learning.
\end{keybox}

\section*{Chapter Overview}
\addcontentsline{toc}{section}{Chapter Overview}

\begin{enumerate}[leftmargin=*]
	\item Why Chapter 13 matters
	\item From AlphaGo and AlphaZero to MuZero
	\item The central idea: planning without known rules
	\item What MuZero learns: representation, dynamics, and prediction
	\item Value-equivalent models: not all models need pixels
	\item Monte Carlo Tree Search inside a learned model
	\item The PUCT selection rule
	\item MuZero training targets and unrolled loss
	\item Root exploration, visit counts, and action selection
	\item Reanalyse and MuZero Unplugged
	\item EfficientZero, Gumbel MuZero, Sampled MuZero, and Stochastic MuZero
	\item Concrete PyTorch building blocks
	\item UAV/SDN scenario: MuZero-style planning for safe network control
	\item Debugging, diagnostics, and failure modes
	\item Limitations and when not to use MuZero
	\item Exercises
	\item Looking Ahead to Chapter 14
\end{enumerate}

\section{Why Chapter 13 matters}

Chapter~12 introduced model-based deep reinforcement learning. We saw that learned models can be used for Dyna-style updates, short model rollouts, model predictive control, latent imagination, and world-model policy learning. This chapter studies a different and historically important branch of model-based RL: \emph{search with a learned model}.

The motivating question is:
\begin{quote}
	Can an agent plan effectively even when it does not know the rules of the environment?
\end{quote}

In classical planning, the transition model is given. In chess, Go, or a deterministic simulator, a planner can enumerate legal actions and compute the next state. AlphaZero used this idea powerfully: it combined a policy-value network with Monte Carlo Tree Search, but still relied on a perfect simulator of the game rules \citep{silver2018alphazero}. MuZero removed that assumption. It learned its own abstract model and used that learned model for search \citep{schrittwieser2020muzero}.

This chapter is important for three reasons. First, MuZero shows that a model does not need to predict observations in pixel space to be useful. Second, MuZero shows that planning and learning can be tightly coupled: search improves training targets, and training improves future search. Third, MuZero-style algorithms have influenced later work on sample efficiency, offline learning, stochastic environments, complex action spaces, and general MCTS-based RL systems \citep{schrittwieser2021unplugged,hubert2021sampled,antonoglou2022stochastic,niu2023lightzero,wang2024efficientzerov2}.

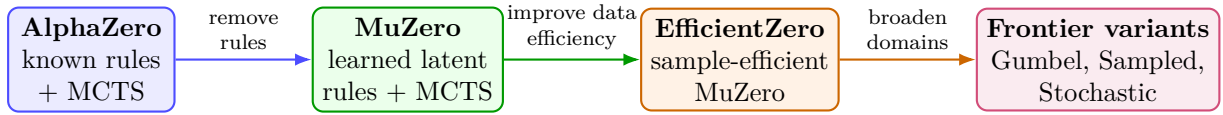
\begin{figure}[t]
	\centering
	\begin{tikzpicture}[
		box/.style={draw,rounded corners,thick,minimum width=2.2cm,minimum height=1.1cm,
			align=center,font=\small},
		arrow/.style={-{Latex[length=2.2mm]},thick},
		node distance=1.8cm
		]
		\node[box,fill=blue!8,draw=blue!70] (az)
		{\textbf{AlphaZero}\\\small known rules\\\small + MCTS};

		\node[box,fill=green!8,draw=green!60!black,right=of az] (muz)
		{\textbf{MuZero}\\\small learned latent\\\small rules + MCTS};

		\node[box,fill=orange!10,draw=orange!80!black,right=of muz] (ez)
		{\textbf{EfficientZero}\\\small sample-efficient\\\small MuZero};

		\node[box,fill=purple!8,draw=purple!70,right=of ez] (front)
		{\textbf{Frontier variants}\\\small Gumbel, Sampled,\\\small Stochastic};

		\draw[arrow,draw=blue!70]
		(az) -- node[above,font=\scriptsize,align=center]{remove\\rules} (muz);
		\draw[arrow,draw=green!60!black]
		(muz) -- node[above,font=\scriptsize,align=center]{improve data\\efficiency} (ez);
		\draw[arrow,draw=orange!80!black]
		(ez) -- node[above,font=\scriptsize,align=center]{broaden\\domains} (front);

	\end{tikzpicture}
	\caption{The historical arc from AlphaZero to MuZero-style agents. AlphaZero uses
		search with known rules. MuZero learns an abstract model and searches inside it.
		Later variants improve sample efficiency, action-space handling, stochasticity,
		and general deployment.}
	\label{fig:alphazero_muzero_arc}
\end{figure}

\section{From AlphaGo and AlphaZero to MuZero}

AlphaGo combined supervised learning, reinforcement learning, value networks, policy networks, and Monte Carlo Tree Search to defeat expert Go players \citep{silver2016mastering}. AlphaZero simplified and generalized the system by learning from self-play and using a single policy-value network for Go, chess, and shogi \citep{silver2018alphazero}. But AlphaZero still required known environment rules. Given a board position and a legal move, the game engine could compute the next board position exactly.

MuZero keeps the planning idea but removes the need for known rules. Instead of using an external simulator inside MCTS, MuZero learns a model with three neural functions:
\begin{align}
	s^0 &= h_\theta(o_{1:t}), \\
	(r^k, s^k) &= g_\theta(s^{k-1}, a^k), \\
	(p^k, v^k) &= f_\theta(s^k).
\end{align}
Here $h_\theta$ maps observations or histories into a latent state, $g_\theta$ predicts the next latent state and reward after an action, and $f_\theta$ predicts a policy prior and value from the latent state.

\begin{table}[t]
	\centering
	\caption{AlphaZero versus MuZero. The key difference is not search itself, but where the search model comes from.}
	\label{tab:alphazero_vs_muzero}
	\begin{tabular}{p{3.0cm}p{5.2cm}p{5.2cm}}
		\toprule
		Aspect & AlphaZero & MuZero \\
		\midrule
		Environment rules & Known simulator/rules are used in search & Learned latent dynamics are used in search \\
		State used in search & True game state & Learned hidden state \\
		Model prediction & Exact legal transition & Reward, next latent state, policy, value \\
		Observation reconstruction & Not needed & Not needed \\
		Planning method & MCTS & MCTS inside learned model \\
		Key risk & Requires known rules & Learned model can be biased or opaque \\
		\bottomrule
	\end{tabular}
\end{table}

\section{The central idea: planning without known rules}

A naive model-based agent tries to learn the environment transition $p(s'\given s,a)$ and reward $r(s,a)$. MuZero does something subtler. It learns a latent state transition model that is useful for planning, even if the latent state is not interpretable as a physical state.

The system is trained so that, after unrolling the model through a sequence of actions, its predictions match observed rewards, improved policies from search, and value targets. The model is therefore \emph{value-equivalent}: it preserves the quantities needed for control, not necessarily every detail needed for reconstruction.

\begin{warningbox}{Important distinction}
	MuZero does not learn a video predictor. It does not need to generate the next image. It learns an abstract model that supports value prediction and policy improvement. A model can be bad at reconstructing pixels but still good for decision-making; conversely, a visually accurate model can be poor for control if it misses reward-relevant structure.
\end{warningbox}

\section{What MuZero learns: representation, dynamics, and prediction}

As previewed in Chapter~12, MuZero learns three functions rather than a conventional pixel-predictive simulator. This chapter develops those functions in depth, emphasizing how they support search and value-equivalent planning.

MuZero uses three neural functions.

\subsection{Representation function}

The representation function maps observations to the root latent state:
\begin{equation}
	s^0 = h_\theta(o_{1:t}).
\end{equation}
In Atari, $o_{1:t}$ may be a stack of frames. In a UAV/SDN controller, it may include recent telemetry, UAV positions, queue states, SINR maps, packet loss, latency, and battery levels. If the environment is partially observable, the representation may need recurrent or history-based structure.

\subsection{Dynamics function}

The dynamics function predicts the next latent state and immediate reward:
\begin{equation}
	(r^k, s^k) = g_\theta(s^{k-1}, a^k).
\end{equation}
The reward prediction matters because MCTS accumulates rewards along imagined paths. The latent transition matters because search must evaluate deeper action sequences.

\subsection{Prediction function}

The prediction function outputs a policy prior and value:
\begin{equation}
	(p^k, v^k) = f_\theta(s^k).
\end{equation}
The policy prior guides exploration in the tree. The value prediction evaluates leaf nodes. Together they make search much more efficient than blind rollout.

\begin{figure}[t]
	\centering
	\resizebox{0.98\textwidth}{!}{%
		\begin{tikzpicture}[
			box/.style={draw,rounded corners,thick,minimum width=3.1cm,minimum height=0.85cm,
				align=center,font=\small},
			arrow/.style={-{Latex[length=2.2mm]},thick},
			node distance=0.9cm
			]

			% Top row
			\node[box,fill=blue!8,draw=blue!70]                           (obs) {Observation history\\$o_{1:t}$};
			\node[box,fill=green!10,draw=green!60!black,right=of obs]     (h)   {Representation\\$h_\theta$};
			\node[box,fill=gray!8,draw=gray!70,right=of h]                (s0)  {Latent state\\$s^0$};
			\node[box,fill=purple!8,draw=purple!70,right=of s0]           (f)   {Prediction\\$f_\theta(s)$};
			\node[box,fill=red!7,draw=red!70!black,right=of f]            (out) {Policy prior + value\\$p,v$};

			% Dynamics below latent state
			\node[box,fill=orange!10,draw=orange!80!black,below=1.2cm of s0] (g) {Dynamics\\$g_\theta(s,a)$};

			% Top row arrows
			\draw[arrow,draw=blue!70]        (obs) -- (h);
			\draw[arrow,draw=green!60!black] (h)   -- (s0);
			\draw[arrow,draw=purple!70]      (s0)  -- (f);
			\draw[arrow,draw=purple!70]      (f)   -- (out);

			% s0 -> g (down), labelled a^1 on the right
			\draw[arrow,draw=orange!80!black]
			(s0.south) -- node[right,font=\scriptsize]{$a^1$} (g.north);

			% g -> s0: exit right, go up, enter s0 from right side as s^1
			\draw[arrow,draw=orange!80!black]
			(g.east)
			-- ++(0.9,0.0)
			-- ++(0,0.4)
			-- node[above,font=\scriptsize]{$r^1,s^1$}
			(s0.south);

		\end{tikzpicture}%
	}
	\caption{The MuZero network decomposition. The representation function initializes a
		latent state, the dynamics function unrolls imagined actions, and the prediction
		function gives policy priors and values for search.}
	\label{fig:muzero_networks}
\end{figure}
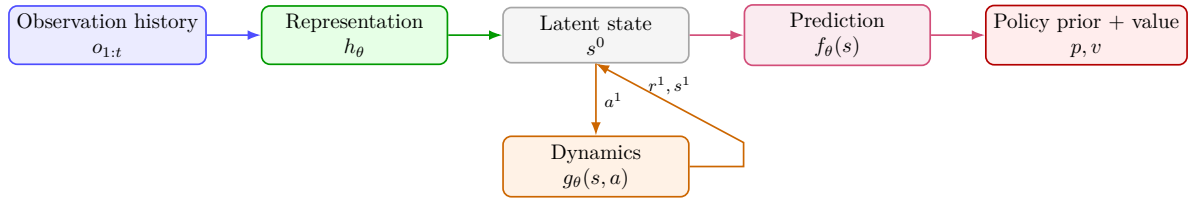

\section{Value-equivalent models: not all models need pixels}

A standard world model may be trained to predict future observations. MuZero is different. It is trained to predict rewards, values, and policies. This is why MuZero is often described as learning a value-equivalent model.

For a model to be useful for search, it should preserve the consequences of actions for return. It does not need to preserve irrelevant details. In a wireless control problem, for example, the agent may not need to reconstruct the exact user interface, terrain texture, or packet trace. It needs to predict how actions affect latency, throughput, battery, interference, and future value.

\begin{researchbox}{Research view: value equivalence for UAV/SDN}
	A UAV/SDN MuZero model should not be judged only by next-state prediction error. A model that predicts user coordinates accurately but fails to predict latency violations is not useful. A model that predicts QoS risk, energy cost, and safety-relevant future value may be useful even if it is not a full simulator. This is the core reason value-equivalent modeling is attractive for network control.
\end{researchbox}

A recent line of work asks what MuZero actually learns and how well its latent model generalizes beyond the policy distribution that trained it \citep{he2024whatmodel}. This is important because MuZero's model may be excellent for the current policy but less reliable for evaluating very different policies.

\section{Monte Carlo Tree Search inside a learned model}

MuZero uses MCTS to improve action selection. Each MCTS simulation starts from the root latent state $s^0$ and follows a tree of imagined actions. MCTS uses a tree because each path represents a different decision sequence. In deterministic domains, transposition tables can sometimes merge equivalent states into a graph, but MuZero-style implementations usually keep the tree structure because it simplifies backup rules and preserves consistent visit-count statistics at each node. A simulation has four conceptual phases:

\begin{enumerate}[leftmargin=*]
	\item \textbf{Selection}: choose actions recursively using a search rule that balances value and exploration.
	\item \textbf{Expansion}: when a new node is reached, use the dynamics and prediction networks to expand it.
	\item \textbf{Evaluation}: use the predicted value at the new node.
	\item \textbf{Backup}: propagate the value estimate back along the visited path.
\end{enumerate}

\begin{figure}[t]
	\centering
	\begin{tikzpicture}[
		node/.style={circle,draw,thick,minimum size=0.75cm,align=center,font=\small},
		leaf/.style={circle,draw,dashed,minimum size=0.72cm,align=center,font=\small},
		arrow/.style={-{Latex[length=2.1mm]},thick}
		]
		\node[node,fill=blue!8,draw=blue!70] (root) at (0,0) {$s^0$};

		% Level 1 — spread wider
		\node[node,fill=green!8,draw=green!60!black] (a) at (-2.5,-1.6) {$s^1_a$};
		\node[node,fill=green!8,draw=green!60!black] (b) at ( 0.0,-1.6) {$s^1_b$};
		\node[node,fill=green!8,draw=green!60!black] (c) at ( 2.5,-1.6) {$s^1_c$};

		% Level 2 — each pair centred under its parent, no overlap
		\node[leaf] (aa) at (-3.3,-3.1) {$s^2$};
		\node[leaf] (ab) at (-1.7,-3.1) {$s^2$};
		\node[leaf] (ba) at (-0.6,-3.1) {$s^2$};
		\node[leaf] (bb) at ( 0.6,-3.1) {$s^2$};
		\node[leaf] (ca) at ( 1.7,-3.1) {$s^2$};
		\node[leaf] (cb) at ( 3.3,-3.1) {$s^2$};

		\draw[arrow] (root) -- node[left, font=\scriptsize]{$a$} (a);
		\draw[arrow] (root) -- node[right,font=\scriptsize]{$b$} (b);
		\draw[arrow] (root) -- node[right,font=\scriptsize]{$c$} (c);
		\draw[arrow] (a) -- (aa);
		\draw[arrow] (a) -- (ab);
		\draw[arrow] (b) -- (ba);
		\draw[arrow] (b) -- (bb);
		\draw[arrow] (c) -- (ca);
		\draw[arrow] (c) -- (cb);

	\end{tikzpicture}
	\caption{MuZero plans in a learned latent state space. Search does not require a
		perfect environment simulator; it uses the learned dynamics and prediction functions.
		MCTS expands a tree in latent space; search statistics become improved policy targets.}
	\label{fig:muzero_mcts_tree}
\end{figure}
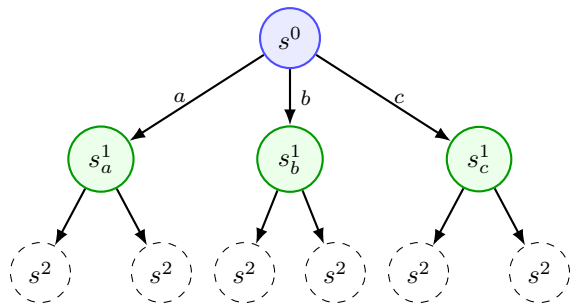

\section{The PUCT selection rule}

At an internal node, MuZero chooses an action by maximizing a score that combines value estimates and exploration guided by the policy prior. A common form is:
\begin{equation}
	a^* = \argmax_a \left[ Q(s,a) + U(s,a) \right],
\end{equation}
where
\begin{equation}
	U(s,a) = c_{\mathrm{puct}} P(s,a) \frac{\sqrt{\sum_b N(s,b)}}{1 + N(s,a)}.
	\label{eq:puct}
\end{equation}
Here $Q(s,a)$ is the search value estimate, $P(s,a)$ is the policy prior, and $N(s,a)$ is the visit count. The prior pulls search toward promising actions; the visit-count term pushes search toward actions that have not yet been explored.

\begin{pitfallbox}{PUCT is not just epsilon-greedy search}
	PUCT is structured exploration inside a tree. The policy prior biases the tree toward plausible actions, while visit counts prevent the search from collapsing too early. If the prior is poor, search can still recover by exploring alternatives. If the value estimates are poor, search can be misled even when the prior is reasonable.
\end{pitfallbox}

\section{MuZero training targets and unrolled loss}

MuZero trains on stored trajectories. At a real time step $t$, the representation network produces $s_t^0$. The dynamics network is unrolled through the sequence of real actions $a_{t+1},\ldots,a_{t+K}$. At each unroll step $k$, the model predicts reward $r_t^k$, value $v_t^k$, and policy $p_t^k$.

The training targets are:
\begin{itemize}
	\item observed rewards $u_{t+k}$;
	\item value targets $z_{t+k}$, often n-step bootstrapped returns;
	\item policy targets $\pi_{t+k}$ derived from MCTS visit counts.
\end{itemize}

A simplified MuZero loss is:
\begin{equation}
	\mathcal{L}_t(\theta)
	=
	\sum_{k=0}^{K}
	\left[
	\ell^v(z_{t+k}, v_t^k)
	+
	\ell^p(\pi_{t+k}, p_t^k)
	+
	\ell^r(u_{t+k}, r_t^k)
	\right]
	+ c\|\theta\|_2^2.
	\label{eq:muzero_loss}
\end{equation}

For scalar values and rewards, $\ell^v$ and $\ell^r$ may be mean-squared error. In large-scale MuZero implementations, value and reward are often represented using a categorical support transform, making these losses cross-entropies over discrete support bins. This categorical support transform is the same basic idea developed in Chapter~6 for distributional value learning such as C51 and QR-DQN: representing scalar quantities as distributions over a discrete support enables cross-entropy losses and often gives more stable learning than direct regression.

\begin{figure}[t]
	\centering
	\begin{tikzpicture}[
		box/.style={draw,rounded corners,thick,minimum width=2.6cm,minimum height=0.8cm,align=center,font=\small},
		target/.style={draw,rounded corners,minimum width=2.5cm,minimum height=0.7cm,align=center,font=\small},
		arrow/.style={-{Latex[length=2.2mm]},thick},
		node distance=0.7cm
		]
		\node[box,fill=blue!8,draw=blue!70] (s0) {$s^0$};
		\node[box,fill=orange!10,draw=orange!80!black,right=of s0] (s1) {$s^1$};
		\node[box,fill=orange!10,draw=orange!80!black,right=of s1] (s2) {$s^2$};
		\node[box,fill=orange!10,draw=orange!80!black,right=of s2] (s3) {$s^3$};
		\draw[arrow] (s0) -- node[above,font=\scriptsize] {$a_{t+1}$} (s1);
		\draw[arrow] (s1) -- node[above,font=\scriptsize] {$a_{t+2}$} (s2);
		\draw[arrow] (s2) -- node[above,font=\scriptsize] {$a_{t+3}$} (s3);
		\node[target,fill=green!8,draw=green!60!black,below=0.9cm of s0] (t0) {$z_t,\pi_t$};
		\node[target,fill=green!8,draw=green!60!black,below=0.9cm of s1] (t1) {$u_{t+1},z_{t+1},\pi_{t+1}$};
		\node[target,fill=green!8,draw=green!60!black,below=0.9cm of s2] (t2) {$u_{t+2},z_{t+2},\pi_{t+2}$};
		\node[target,fill=green!8,draw=green!60!black,below=0.9cm of s3] (t3) {$u_{t+3},z_{t+3},\pi_{t+3}$};
		\draw[arrow,draw=green!50!black] (s0) -- (t0);
		\draw[arrow,draw=green!50!black] (s1) -- (t1);
		\draw[arrow,draw=green!50!black] (s2) -- (t2);
		\draw[arrow,draw=green!50!black] (s3) -- (t3);
	\end{tikzpicture}
	\caption{MuZero training unroll. The learned dynamics model is unrolled through real actions, and each unroll step is trained against reward, value, and search-policy targets.}
	\label{fig:muzero_unroll_loss}
\end{figure}
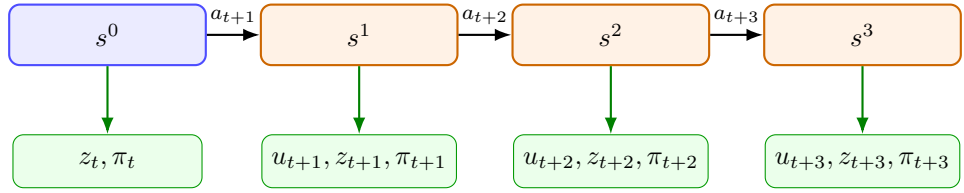

\section{Root exploration, visit counts, and action selection}

After many simulations, MCTS produces visit counts $N(s^0,a)$ at the root. These counts define an improved policy target:
\begin{equation}
	\pi(a\given s^0) = \frac{N(s^0,a)^{1/\tau}}{\sum_b N(s^0,b)^{1/\tau}},
\end{equation}
where $\tau$ is a temperature. During training, the agent may sample from this distribution. A common schedule keeps $\tau>0$ early in training to encourage exploration and then anneals toward $\tau\to 0$ so that the policy becomes increasingly greedy with respect to visit counts. During evaluation, it often chooses the most visited action.

The policy head is trained to match these visit-count targets. This is why search improves the network: the policy network learns to imitate improved decisions produced by planning. Later, the improved policy network guides future search more effectively.

\begin{keybox}{The MuZero learning loop}
	Search improves the policy target. Training improves the policy, value, and model. The improved network improves the next search. MuZero is therefore a loop of planning-improves-learning and learning-improves-planning.
\end{keybox}

\section{Reanalyse and MuZero Unplugged}

MuZero is expensive because search is performed during data generation. Reanalyse reuses stored trajectories by recomputing improved policy and value targets using the current network. MuZero Unplugged combines this idea with MuZero to support online and offline reinforcement learning from different data budgets \citep{schrittwieser2021unplugged}.

The key idea is that stored data are not frozen labels. A trajectory collected by an older policy can be revisited. The current network can run search from old states and produce better targets. This makes MuZero-style learning attractive when interaction is expensive or when only logged data are available.

More concretely, Reanalyse takes old trajectories from replay, selects previously observed states, and reruns MCTS using the current representation, dynamics, prediction, and value networks. The resulting fresh visit-count policy targets and value targets can be better than the original targets because the network has improved since the data were collected. In this sense, old experience becomes useful again: the raw observations and actions are reused, but the supervision attached to them is refreshed by current search. This is one of the main mechanisms behind MuZero-style data efficiency.

\begin{warningbox}{Offline caveat}
	Reanalyse does not magically solve offline RL. If the dataset lacks coverage of important actions or states, the learned model and search may extrapolate poorly. Planning with a biased model can confidently improve in the wrong direction. Offline MuZero-style learning therefore requires careful regularization, validation, and uncertainty awareness.
\end{warningbox}

\section[MuZero-style variants]{EfficientZero, Gumbel MuZero, Sampled MuZero, and Stochastic MuZero}

MuZero inspired several important variants.

\subsection{EfficientZero and EfficientZero V2}

EfficientZero improves sample efficiency by strengthening self-supervised consistency, value prefix prediction, and model learning components \citep{ye2021efficientzero}. EfficientZero V2 extends the line toward both discrete and continuous control with limited data \citep{wang2024efficientzerov2}. The practical lesson is that MuZero's architecture is powerful but data efficiency depends heavily on representation learning, target construction, and model regularization.

\subsection{Gumbel MuZero}

Gumbel MuZero improves planning by using Gumbel-based policy improvement. It was designed to improve action selection especially under small search budgets \citep{danihelka2022gumbel}. A distinctive feature is that the search-improved policy can be shown to improve over the prior under the Gumbel policy-improvement construction, which is unusual among MCTS variants that are often used mainly as heuristics. Empirically, this makes Gumbel MuZero especially attractive when only a small number of simulations is affordable. The lesson is that the search procedure itself is an algorithmic object, not a fixed black box.

\subsection{Sampled MuZero}

Standard MuZero assumes that actions can be enumerated during tree search. This is hard in continuous or very large action spaces. Sampled MuZero plans over sampled action subsets, allowing MuZero-style ideas to extend to complex action spaces \citep{hubert2021sampled}.

\subsection{Stochastic MuZero}

Standard MuZero is strongest in deterministic or effectively deterministic settings. Stochastic MuZero extends the framework to stochastic environments by introducing stochastic transitions through afterstates and chance outcomes \citep{antonoglou2022stochastic}. This matters for real systems where uncertainty is intrinsic, including wireless networks, user mobility, and packet-level dynamics.

\subsection{LightZero and general MCTS systems}

LightZero is a NeurIPS 2023 benchmark and toolkit that decomposes and benchmarks MCTS/MuZero-style systems across more general sequential decision scenarios, including domains beyond board games \citep{niu2023lightzero}. This reflects a broader 2026-facing trend: MuZero is no longer only a game-playing algorithm. It is part of a larger family of planning-with-learned-model systems.

\begin{table}[t]
	\centering
	\caption{MuZero-style variants and what problem they address.}
	\label{tab:muzero_variants}
	\begin{tabular}{p{3.2cm}p{5.1cm}p{5.3cm}}
		\toprule
		Method & Main problem addressed & Core idea \\
		\midrule
		MuZero & Planning without known rules & Learn latent dynamics for MCTS \\
		MuZero Unplugged & Data reuse and offline/low-data learning & Reanalyse old trajectories with current search \\
		EfficientZero & Sample efficiency & Stronger self-supervised and value-prefix learning \\
		Gumbel MuZero & Better search with small budgets & Gumbel-based policy improvement \\
		Sampled MuZero & Large or continuous action spaces & Plan over sampled action subsets \\
		Stochastic MuZero & Inherent stochasticity & Afterstates and stochastic tree search \\
		LightZero & General deployment and benchmarking & Modular MCTS/MuZero toolkit and benchmark \\
		\bottomrule
	\end{tabular}
\end{table}

\section{Concrete PyTorch building blocks}

The following code is not a production MuZero implementation. A full system requires parallel actors, replay storage, target support transforms, root exploration, reanalyse, and efficient batched MCTS. The goal here is to make the core machinery understandable.

\subsection{A minimal MuZero-style network}

\Needspace{18\baselineskip}
\begin{lstlisting}[style=pythonstyle,caption={Minimal MuZero-style network skeleton.},label={lst:muzero_network}]
import torch
import torch.nn as nn
import torch.nn.functional as F

class MuZeroNet(nn.Module):
    def __init__(self, obs_dim, latent_dim, action_dim):
        super().__init__()
        self.action_dim = action_dim
        self.representation = nn.Sequential(
            nn.Linear(obs_dim, 256), nn.ReLU(),
            nn.Linear(256, latent_dim), nn.Tanh()
        )
        self.action_embed = nn.Embedding(action_dim, latent_dim)
        self.dynamics_body = nn.Sequential(
            nn.Linear(2 * latent_dim, 256), nn.ReLU(),
            nn.Linear(256, latent_dim), nn.Tanh()
        )
        self.reward_head = nn.Linear(latent_dim, 1)
        self.policy_head = nn.Linear(latent_dim, action_dim)
        self.value_head = nn.Linear(latent_dim, 1)

    def initial_inference(self, obs):
        latent = self.representation(obs)
        policy_logits = self.policy_head(latent)
        value = self.value_head(latent)
        reward = torch.zeros_like(value)
        return latent, reward, value, policy_logits

    def recurrent_inference(self, latent, action):
        a = self.action_embed(action)
        x = torch.cat([latent, a], dim=-1)
        next_latent = self.dynamics_body(x)
        reward = self.reward_head(next_latent)
        policy_logits = self.policy_head(next_latent)
        value = self.value_head(next_latent)
        return next_latent, reward, value, policy_logits
\end{lstlisting}

\subsection{Search node and PUCT selection}

\Needspace{18\baselineskip}
\begin{lstlisting}[style=pythonstyle,caption={MCTS node and PUCT selection.},label={lst:puct_node}]
import math

class Node:
    def __init__(self, prior):
        self.prior = float(prior)
        self.visit_count = 0
        self.value_sum = 0.0
        self.reward = 0.0
        self.hidden_state = None
        self.children = {}

    def expanded(self):
        return len(self.children) > 0

    def value(self):
        if self.visit_count == 0:
            return 0.0
        return self.value_sum / self.visit_count

def puct_score(parent, child, pb_c_base=19652, pb_c_init=1.25):
    pb_c = math.log((parent.visit_count + pb_c_base + 1) / pb_c_base) + pb_c_init
    pb_c *= math.sqrt(parent.visit_count + 1e-8) / (child.visit_count + 1)
    prior_score = pb_c * child.prior
    value_score = child.value()
    return value_score + prior_score

def select_child(node):
    return max(node.children.items(), key=lambda item: puct_score(node, item[1]))
\end{lstlisting}

\subsection{Expansion and backup}

\Needspace{20\baselineskip}
\begin{lstlisting}[style=pythonstyle,caption={Expansion and backup for a simplified MuZero MCTS.},label={lst:mcts_expand_backup}]
def expand_node(node, hidden_state, reward, policy_logits):
    node.hidden_state = hidden_state
    node.reward = float(reward)
    probs = torch.softmax(policy_logits, dim=-1).detach().cpu().numpy()
    for action, prior in enumerate(probs):
        node.children[action] = Node(prior)

def backpropagate(search_path, value, discount):
    # search_path contains nodes from root to leaf.
    for node in reversed(search_path):
        node.value_sum += float(value)
        node.visit_count += 1
        value = node.reward + discount * value

def run_mcts(root, network, num_simulations, discount=0.997):
    for _ in range(num_simulations):
        node = root
        search_path = [node]
        action_history = []

        while node.expanded():
            action, node = select_child(node)
            search_path.append(node)
            action_history.append(action)

        parent = search_path[-2]
        action = torch.tensor([action_history[-1]], dtype=torch.long)
        latent = parent.hidden_state
        next_latent, reward, value, policy_logits = network.recurrent_inference(latent, action)
        expand_node(node, next_latent, reward.item(), policy_logits[0])
        backpropagate(search_path, value.item(), discount)
\end{lstlisting}

This code is simplified. Production implementations batch neural network inference, normalize values in the tree, add Dirichlet noise at the root, mask illegal actions, transform scalar values onto categorical supports, and store search statistics for training.

\subsection{MuZero training loss}

\Needspace{20\baselineskip}
\begin{lstlisting}[style=pythonstyle,caption={Unrolled MuZero loss for scalar reward and value targets.},label={lst:muzero_loss}]
def muzero_unroll_loss(network, obs, actions, target_rewards,
                       target_values, target_policies, value_coef=1.0,
                       reward_coef=1.0, policy_coef=1.0):
    """Simplified unrolled MuZero training loss.

    obs: [B, obs_dim]
    actions: [B, K] integer actions used for unroll
    target_rewards: [B, K]
    target_values: [B, K + 1]
    target_policies: [B, K + 1, action_dim]
    """
    latent, reward, value, policy_logits = network.initial_inference(obs)
    losses = []

    value_loss = F.mse_loss(value.squeeze(-1), target_values[:, 0])
    policy_loss = F.cross_entropy(policy_logits, target_policies[:, 0].argmax(dim=-1))
    losses.append(value_coef * value_loss + policy_coef * policy_loss)

    K = actions.shape[1]
    for k in range(K):
        latent, reward, value, policy_logits = network.recurrent_inference(
            latent, actions[:, k]
        )
        r_loss = F.mse_loss(reward.squeeze(-1), target_rewards[:, k])
        v_loss = F.mse_loss(value.squeeze(-1), target_values[:, k + 1])
        # If target_policies are distributions, use cross-entropy with soft labels.
        logp = F.log_softmax(policy_logits, dim=-1)
        p_loss = -(target_policies[:, k + 1] * logp).sum(dim=-1).mean()
        losses.append(reward_coef * r_loss + value_coef * v_loss + policy_coef * p_loss)

        # Stop gradient scaling is often used in real implementations.
        latent = latent.detach() + (latent - latent.detach()) * 0.5

    return sum(losses) / len(losses)
\end{lstlisting}

\section[UAV/SDN MuZero scenario]{UAV/SDN scenario: MuZero-style planning for safe network control}

A UAV/SDN controller is not a board game. Actions may involve movement, bandwidth allocation, transmission power, handover decisions, and SDN routing. The transition dynamics depend on mobility, wireless channels, queueing, interference, and battery. A perfect simulator is rarely available. This is exactly the kind of setting where MuZero's value-equivalent idea is conceptually useful.

\begin{figure}[t]
	\centering
	\resizebox{0.98\textwidth}{!}{%
		\begin{tikzpicture}[
			box/.style={draw,rounded corners,thick,minimum width=3.0cm,minimum height=0.85cm,align=center,font=\small},
			small/.style={draw,rounded corners,minimum width=2.6cm,minimum height=0.72cm,align=center,font=\small},
			arrow/.style={-{Latex[length=2.2mm]},thick},
			node distance=1.3cm
			]
			\node[box,fill=blue!8,draw=blue!70] (state) {Network telemetry\\QoS, SINR, battery};
			\node[box,fill=green!10,draw=green!60!black,right=of state] (repr) {Representation\\$h_\theta$};
			\node[box,fill=gray!8,draw=gray!70,right=of repr] (latent) {Latent network state\\$s^0$};
			\node[box,fill=orange!10,draw=orange!80!black,right=of latent] (mcts) {MCTS planning\\learned model};
			\node[box,fill=red!7,draw=red!70!black,right=of mcts] (shield) {Safety layer\\CBF/projection};
			\node[box,fill=purple!8,draw=purple!70,below=1.3cm of mcts] (targets) {Search targets\\policy, value, risk};
			\node[box,fill=gray!10,draw=gray!70,right=of shield] (env) {UAV/SDN system};
			\draw[arrow,draw=blue!70] (state) -- (repr);
			\draw[arrow,draw=green!60!black] (repr) -- (latent);
			\draw[arrow,draw=gray!70] (latent) -- (mcts);
			\draw[arrow,draw=orange!80!black] (mcts) -- node[above,font=\scriptsize,align=center] {planned\\ action} (shield);
			\draw[arrow,draw=red!70!black] (shield) -- node[above,font=\scriptsize,align=center] {executed\\action} (env);
			\draw[arrow,draw=purple!70] (mcts) -- (targets);
			\draw[arrow,draw=gray!60] (env.south) |- (targets.east);
		\end{tikzpicture}%
	}
	\caption{A MuZero-style UAV/SDN control scenario. The agent maps telemetry into a latent state, searches with a learned model, applies a safety layer before execution, and trains from both environment outcomes and search-improved targets.}
	\label{fig:uav_sdn_muzero_scenario}
\end{figure}
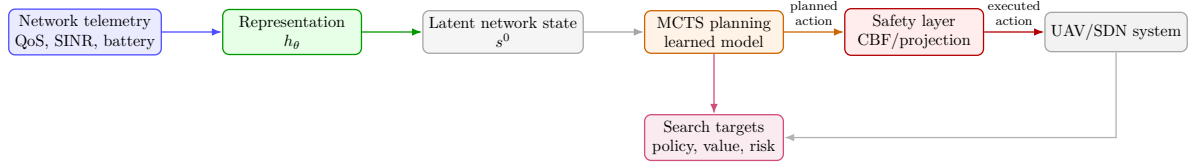

\subsection{A concrete numerical planning example}

Suppose the root network state has three candidate high-level actions:
\begin{enumerate}[leftmargin=*]
	\item $a_1$: move toward URLLC hotspot;
	\item $a_2$: stay and rebalance bandwidth;
	\item $a_3$: move toward charging station.
\end{enumerate}
After MCTS, the visit counts and predicted values are:

\begin{table}[t]
	\centering
	\caption{Example MuZero-style root search statistics for UAV/SDN control.}
	\label{tab:uav_muzero_root_stats}
	\begin{tabular}{lccc}
		\toprule
		Action & Visit count & Mean value & Safety note \\
		\midrule
		Move to URLLC hotspot & 62 & 18.5 & high QoS, moderate battery cost \\
		Stay and rebalance & 31 & 16.8 & stable, lower risk \\
		Move to charger & 7 & 12.2 & safe energy, poor immediate QoS \\
		\bottomrule
	\end{tabular}
\end{table}

With temperature $\tau=1$, the search-policy target assigns most probability to the hotspot action. The visit counts can be interpreted through the PUCT rule in Eq.~\eqref{eq:puct}. Early in search, the exploration term $U(s,a)$ gives all three actions some chance to be examined. As simulations accumulate, the action with consistently higher $Q(s,a)+U(s,a)$ receives more visits, so search concentrates on $a_1$ while still allocating some simulations to $a_2$ and $a_3$. In this example, $a_1$ wins because its high predicted QoS return dominates its moderate battery cost. 

However, a safety layer may still modify the exact movement vector if the planned trajectory violates collision or battery constraints. The critic and dynamics model should be trained using the executed action and observed outcome, not only the pre-filter proposal. This point connects MuZero-style planning to the safety-filter discussion in Chapters~11 and later safe-RL chapters.

\subsection{Sampled actions for continuous UAV control}

For continuous UAV control, enumerating all actions is impossible. A Sampled MuZero-style controller can propose a finite candidate set from a policy distribution, then search only over those candidates.

\Needspace{18\baselineskip}
\begin{lstlisting}[style=pythonstyle,caption={Sampling candidate UAV actions for Sampled MuZero-style planning.},label={lst:sampled_actions_uav}]
def sample_uav_actions(policy, state, num_actions=16):
    """Return candidate continuous actions for sampled planning.

    Example action: [dx, dy, dz, bandwidth_urllc, bandwidth_embb, power]
    The policy proposes candidates; the search evaluates them.
    """
    with torch.no_grad():
        dist = policy(state)
        actions = dist.sample((num_actions,))

        # Project bandwidth components onto a simplex-like feasible range.
        move = actions[..., :3].clamp(-1.0, 1.0)
        bw_raw = actions[..., 3:5]
        bw = torch.softmax(bw_raw, dim=-1)
        power = actions[..., 5:6].clamp(0.0, 1.0)
        return torch.cat([move, bw, power], dim=-1)
\end{lstlisting}

\begin{researchbox}{Research direction: risk-aware MuZero for UAV/SDN}
	A distinctive research-grade extension is to augment MuZero with reward, cost, and uncertainty heads. Search can then optimize not only expected return, but also tail latency, battery risk, collision risk, and model uncertainty. This creates a bridge between MuZero, distributional RL, safe model-based RL, and CBF-style action projection.
\end{researchbox}

\section{Debugging, diagnostics, and failure modes}

MuZero-style agents are powerful but hard to debug. The model, search, value targets, policy targets, replay, and environment interaction can fail in different ways.

\begin{table}[t]
	\centering
	\caption{Common MuZero failure modes and diagnostics.}
	\label{tab:muzero_debugging}
	\begin{tabular}{p{3.5cm}p{5.0cm}p{5.0cm}}
		\toprule
		Symptom & Likely cause & What to inspect \\
		\midrule
		Search policy collapses early & Poor priors, too little exploration, bad value scale & Root visit counts, Dirichlet noise, PUCT constant \\
		Value diverges & Unstable targets or support transform bug & Value target range, reward scaling, bootstrapping horizon \\
		Model predicts rewards but planning fails & Latent dynamics not useful for deeper search & Unroll consistency, model value accuracy by depth \\
		Training improves but evaluation does not & Search/training mismatch or overfitting to targets & Evaluation temperature, reanalyse targets, replay freshness \\
		MCTS too slow & Too many simulations or unbatched inference & Batch inference, simulation budget, action pruning \\
		Unsafe planned actions & Search objective lacks constraints & Safety layer, cost head, constraint-aware search \\
		Offline performance collapses & Dataset coverage too narrow & State-action coverage, uncertainty, conservative regularization \\
		\bottomrule
	\end{tabular}
\end{table}

\begin{pitfallbox}{A learned model can make search confidently wrong}
	Planning amplifies model errors. If the learned dynamics or value head is biased, MCTS may repeatedly choose branches where the model is optimistic but the real environment is dangerous or low-value. MuZero-style agents therefore need validation of search depth, value calibration, uncertainty, and safety constraints.
\end{pitfallbox}

\section{Limitations and when not to use MuZero}

MuZero is impressive, but it is not always the right tool.

\begin{enumerate}[leftmargin=*]
	\item \textbf{Compute cost}: MCTS requires many neural network evaluations per environment step.
	\item \textbf{Implementation complexity}: a correct MuZero implementation is much harder than DQN, PPO, or SAC.
	\item \textbf{Action-space difficulty}: standard tree search works best when actions can be enumerated.
	\item \textbf{Model opacity}: the latent model may be difficult to interpret.
	\item \textbf{Stochastic environments}: standard deterministic MuZero may be insufficient without stochastic extensions.
	\item \textbf{Offline risk}: search may exploit model errors under poor data coverage.
	\item \textbf{Safety}: MuZero does not automatically guarantee constraints. Safe deployment still requires constrained objectives, shields, CBFs, or supervisory control.
\end{enumerate}

For many continuous-control problems, SAC or TD-MPC-style methods may be simpler and more practical. In latency-critical deployments, even 50 MCTS simulations per decision may be too expensive; some systems may use only a tiny search budget, a single network evaluation with action ranking, or reserve full MCTS for training-time policy improvement rather than runtime control. For large language models, search-style planning may be useful in verification or tree-of-thought settings, but it is not a direct replacement for policy optimization. The right lesson is not that MuZero should be used everywhere. The right lesson is that learning a model for search can be extraordinarily powerful when the search budget, action space, and model accuracy align. 

\section{Exercises}

\subsection*{Conceptual exercises}
\begin{enumerate}[leftmargin=*]
	\item Explain the difference between a pixel-predictive world model and a value-equivalent MuZero model.
	\item Why can MuZero plan without knowing the true environment rules?
	\item Why is MCTS more useful when guided by both a policy prior and a value estimate?
	\item Explain why MuZero may fail in stochastic environments without a stochastic-model extension.
	\item Why can a safety layer be necessary even if MuZero performs planning?
\end{enumerate}

\subsection*{Mathematical exercises}
\begin{enumerate}[leftmargin=*]
	\item Starting from the PUCT rule in Eq.~\eqref{eq:puct}, explain how increasing $N(s,a)$ affects the exploration bonus.
	\item Given root visit counts $N=(62,31,7)$ and temperature $\tau=1$, compute the normalized search policy target.
	\item Suppose a three-step MuZero unroll predicts rewards $1.0,0.5,-0.2$ and final value $5.0$ with discount $0.99$. Compute the bootstrapped value target at the root.
	\item Show how the loss in Eq.~\eqref{eq:muzero_loss} changes if reward and value are represented as categorical supports rather than scalars.
\end{enumerate}

\subsection*{Implementation exercises}
\begin{enumerate}[leftmargin=*]
	\item Extend Listing~\ref{lst:puct_node} to mask illegal actions.
	\item Add Dirichlet noise to the root priors before MCTS.
	\item Modify Listing~\ref{lst:mcts_expand_backup} to batch recurrent inference for multiple leaves.
	\item Implement categorical value support encoding and decoding.
	\item Implement a small MuZero-style agent for a toy grid-world with unknown transition rules.
\end{enumerate}

\subsection*{Research thinking exercises}
\begin{enumerate}[leftmargin=*]
	\item Design a MuZero-style latent model for UAV/SDN control. What should the representation, dynamics, and prediction heads output?
	\item How would you add a cost critic or safety head to the MuZero prediction function?
	\item In a wireless network, what errors would be more dangerous: reward prediction error, latency prediction error, or value prediction error? Why?
	\item Compare Sampled MuZero and SAC for continuous UAV movement. Which would be easier to deploy and why?
	\item Propose an experiment to test whether a MuZero latent model is value-equivalent but not observation-equivalent.
\end{enumerate}

\section*{Looking Ahead to Chapter 14: Food for Thought}
\addcontentsline{toc}{section}{Looking Ahead to Chapter 14: Food for Thought}

MuZero learns from interaction and search. But many real domains do not allow additional online interaction. Hospitals, autonomous driving fleets, production networks, and industrial control systems often have logged datasets but cannot safely run arbitrary exploratory policies. This leads to the next major topic: offline reinforcement learning.

Before moving on, consider the following questions.

\begin{enumerate}[leftmargin=*]
	\item What if the agent cannot collect new data?
	\item What if the dataset was generated by a weak or biased policy?
	\item How can a learned policy avoid choosing actions that are absent from the dataset?
	\item Can model-based planning help offline RL, or does it amplify model errors?
	\item How should offline RL balance improvement against conservatism?
\end{enumerate}

Chapter~13 showed how search can improve policies when a learned model is available. Chapter~14 asks what happens when the only available experience is a fixed dataset.
	\chapter[Offline Reinforcement Learning]{Offline Reinforcement Learning}
\chaptermark{Offline RL}
\label{ch:offline_rl}

\begin{keybox}{Chapter goal}
	Offline reinforcement learning asks whether an agent can learn a useful policy from a fixed dataset without further interaction with the environment. This chapter explains why the problem is fundamentally harder than ordinary supervised learning, how extrapolation error appears, how major algorithms such as BCQ, BEAR, CQL, TD3+BC, and IQL control out-of-distribution actions, and how offline RL connects to robotics, communication networks, medical decision-making, recommender systems, autonomous driving, and modern sequence-modeling approaches.
\end{keybox}

\section*{Chapter Overview}
\addcontentsline{toc}{section}{Chapter Overview}
\begin{enumerate}[leftmargin=*]
	\item Why offline reinforcement learning matters
	\item Offline RL as a different problem, not just ``RL without exploration''
	\item Formal setup and dataset notation
	\item Why behavior cloning is not enough
	\item Distribution shift and extrapolation error
	\item The three design principles of offline RL
	\item Behavior-constrained methods: BCQ, BEAR, and policy regularization
	\item Conservative value learning: CQL
	\item In-sample learning and IQL
	\item TD3+BC and simple actor regularization
	\item Sequence-modeling view: return-conditioned policies and Decision Transformer
	\item Generative and diffusion policies for offline RL
	\item Model-based offline RL and pessimistic planning
	\item Dataset quality, coverage, corruption, and evaluation
	\item Offline-to-online fine-tuning
	\item UAV/SDN worked scenario: learning from network logs
	\item Practical implementation code
	\item Failure modes and debugging checklist
	\item Limitations and frontiers toward 2026
	\item Exercises
	\item Looking ahead
\end{enumerate}

% ============================================================
\section{Why offline reinforcement learning matters}

Online reinforcement learning assumes that the agent can act in the environment, observe consequences, and keep improving through trial and error. This assumption is powerful, but it is often unrealistic. A robot cannot safely crash thousands of times. A medical system cannot try arbitrary treatments. A network controller cannot intentionally overload a production backbone to learn a better routing policy. A UAV cannot repeatedly collide with obstacles, violate airspace constraints, or drain its battery during exploration.

Offline reinforcement learning, also called batch reinforcement learning, addresses this setting. The agent receives a fixed dataset of past experience and must learn a policy without additional environment interaction. The dataset may contain demonstrations, expert trajectories, random behavior, logs from a production controller, mixed-quality human decisions, or historical telemetry from an autonomous system \citep{levine2020offline,fu2020d4rl,gulcehre2020rlunplugged}.

\begin{figure}[t]
	\centering
	\begin{tikzpicture}[
		box/.style={draw,rounded corners,thick,minimum width=3.2cm,minimum height=0.9cm,align=center,font=\small},
		arrow/.style={-{Latex[length=2.2mm]},thick},
		node distance=1.5cm
		]

		\node[box,fill=blue!8,draw=blue!70] (online_env) {Environment};
		\node[box,fill=green!8,draw=green!60!black,right=of online_env] (online_agent) {Online RL agent};
		\draw[arrow,draw=blue!70] (online_env) to[bend left=15] node[above,font=\scriptsize] {state, reward} (online_agent);
		\draw[arrow,draw=green!60!black] (online_agent) to[bend left=15] node[below,font=\scriptsize] {action} (online_env);
		\node[below=0.65cm of online_agent,align=center,font=\small] {Learns by interacting.\\Exploration is possible but risky.};

		\node[box,fill=orange!10,draw=orange!80!black,below=2.4cm of online_env] (data) {Fixed dataset $\D$};
		\node[box,fill=purple!8,draw=purple!70,right=of data] (offline_agent) {Offline RL learner};
		\node[box,fill=gray!8,draw=gray!70,right=of offline_agent] (policy) {Deployed policy};
		\draw[arrow,draw=orange!80!black] (data) -- node[above,font=\scriptsize,align=center] {logged\\transitions} (offline_agent);
		\draw[arrow,draw=purple!70] (offline_agent) -- node[above,font=\scriptsize,align=center] {train\\once} (policy);
		\node[below=0.65cm of offline_agent,align=center,font=\small] {Learns without new interaction.\\Safety depends on dataset coverage.};

	\end{tikzpicture}
	\caption{Online RL learns through interaction; offline RL learns from a fixed dataset. Offline RL is attractive when exploration is expensive, dangerous, slow, or ethically impossible.}
	\label{fig:online_vs_offline_rl}
\end{figure}
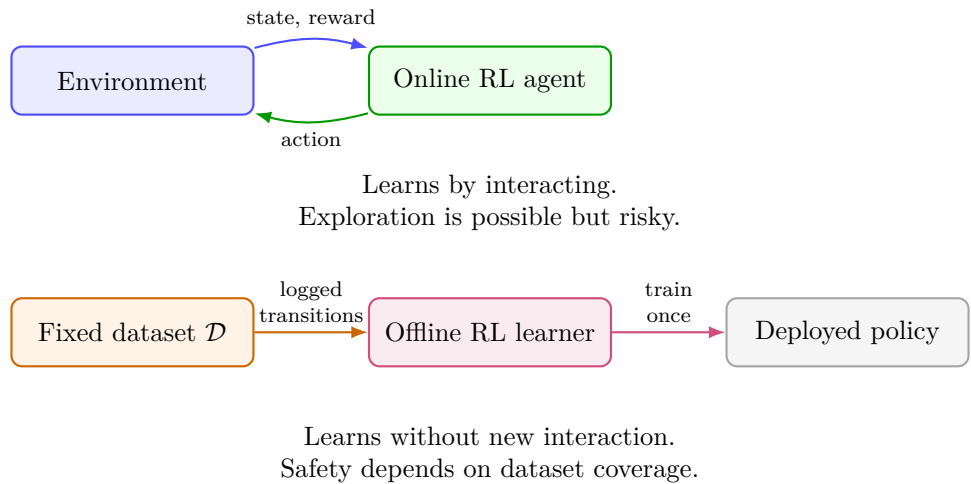

The promise is enormous: if we can learn from existing data, then reinforcement learning becomes compatible with many real-world domains where online exploration is unacceptable. The difficulty is equally serious: the learned policy may choose actions that are not well represented in the dataset. The value function must then estimate consequences for actions it has never seen. This is the central danger of offline RL.

\section{Offline RL is not simply online RL without exploration}

It is tempting to think of offline RL as ordinary RL with exploration disabled. This is misleading. In online RL, if the agent is uncertain about an action, it can try it and observe the outcome. In offline RL, the dataset is all the agent has. If the dataset does not contain enough examples of a state-action region, the agent cannot query the environment to fill the gap.

This creates a strict asymmetry:
\begin{itemize}
	\item online RL can be unsafe because it explores;
	\item offline RL can be unsafe because it cannot verify extrapolations.
\end{itemize}

\begin{warningbox}{The core offline RL danger}
	Offline RL agents may optimize value estimates for actions that are outside the support of the dataset. These actions can receive unrealistically high predicted values because no data exists to correct the estimate. This is known as extrapolation error or out-of-distribution value error \citep{fujimoto2019offpolicy,levine2020offline}.
\end{warningbox}

This is why offline RL is not just supervised learning either. Behavior cloning can imitate the dataset, but it cannot easily improve beyond the behavior policy. Offline RL tries to improve, but improvement requires reasoning about alternative actions. That reasoning is exactly where extrapolation error appears.

% ============================================================
\section{Formal setup}

Assume an MDP with states $s\in\Sspace$, actions $a\in\A$, reward $r$, transition kernel $p(s'\mid s,a)$, discount factor $\gamma$, and policy $\pi(a\mid s)$. Offline RL receives a static dataset
\begin{equation}
	\D = \{(s_i,a_i,r_i,s'_i,d_i)\}_{i=1}^{N},
\end{equation}
where $d_i$ is a terminal indicator. The data are generated by an unknown behavior policy or mixture of policies, often denoted $\pi_\beta$.

The offline RL objective is to learn a policy $\pi_\theta$ that maximizes return:
\begin{equation}
	J(\pi_\theta)
	=
	\E_{\pi_\theta}\left[\sum_{t=0}^{\infty}\gamma^t r_t\right],
\end{equation}
using only $\D$, without collecting new transitions.

The key mismatch is that the training distribution is governed by the dataset distribution $d^{\pi_\beta}(s,a)$, while the learned policy induces $d^{\pi_\theta}(s,a)$. If $\pi_\theta$ moves far from $\pi_\beta$, the agent enters unsupported regions.

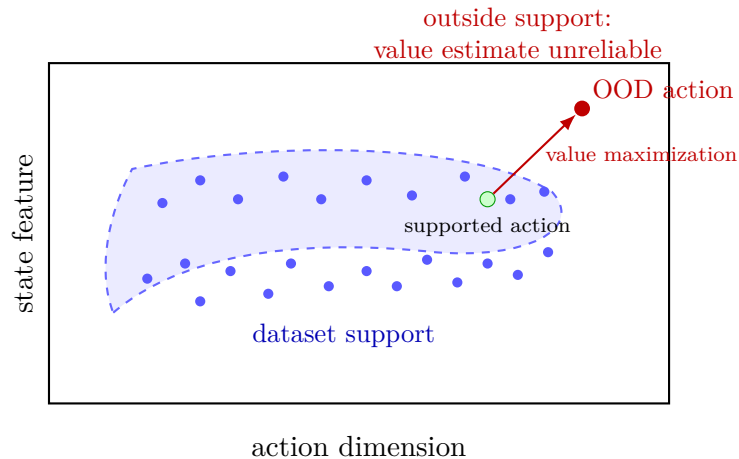
\begin{figure}[t]
	\centering
	\begin{tikzpicture}[
		point/.style={circle,fill=blue!65,inner sep=1.35pt},
		query/.style={circle,fill=red!75!black,inner sep=2.1pt},
		arrow/.style={-{Latex[length=2.2mm]},thick}
		]

		% Plot frame
		\draw[thick] (-4.1,-2.25) rectangle (4.1,2.25);

		% Axis labels
		\node[below] at (0,-2.55) {action dimension};
		\node[rotate=90] at (-4.45,0) {state feature};

		% Dataset support region
		\fill[blue!8]
		(-3.25,-1.05)
		.. controls (-2.2,-0.15) and (-0.4,-0.10) .. (1.0,-0.25)
		.. controls (2.2,-0.35) and (3.0,0.05) .. (2.55,0.55)
		.. controls (1.4,1.25) and (-1.4,1.20) .. (-3.0,0.85)
		.. controls (-3.35,0.20) and (-3.45,-0.55) .. (-3.25,-1.05);

		\draw[thick,dashed,blue!60]
		(-3.25,-1.05)
		.. controls (-2.2,-0.15) and (-0.4,-0.10) .. (1.0,-0.25)
		.. controls (2.2,-0.35) and (3.0,0.05) .. (2.55,0.55)
		.. controls (1.4,1.25) and (-1.4,1.20) .. (-3.0,0.85)
		.. controls (-3.35,0.20) and (-3.45,-0.55) .. (-3.25,-1.05);

		\node[font=\small,blue!70!black] at (-0.2,-1.35) {dataset support};

		% Logged dataset points
		\foreach \x/\y in {-2.8/-0.6,-2.3/-0.4,-2.1/-0.9,-1.7/-0.5,-1.2/-0.8,-0.9/-0.4,-0.4/-0.7,0.1/-0.5,0.5/-0.7,0.9/-0.35,1.3/-0.65,1.7/-0.4,2.1/-0.55,2.5/-0.25}
		\node[point] at (\x,\y) {};

		\foreach \x/\y in {-2.6/0.4,-2.1/0.7,-1.6/0.45,-1.0/0.75,-0.5/0.45,0.1/0.7,0.7/0.5,1.4/0.75,2.0/0.45,2.45/0.55}
		\node[point] at (\x,\y) {};

		% Supported candidate action
		\node[circle,draw=green!60!black,fill=green!20,inner sep=2.0pt,
		label=below:{\scriptsize supported action}] (safe) at (1.7,0.45) {};

		% OOD learned action
		\node[query] (ood) at (2.95,1.65) {};
		\node[align=center,font=\small,text=red!75!black] at (2.1,2.65)
		{outside support:\\value estimate unreliable};

		\draw[arrow,draw=red!75!black]
		(safe) -- node[right,font=\scriptsize,text=red!75!black] {value maximization} (ood);

		\node[above right,font=\small,text=red!75!black] at (ood)
		{OOD action};

	\end{tikzpicture}
	\caption{Offline RL must avoid unsupported state--action regions. The blue points represent logged data, and the dashed region shows a two-dimensional projection of dataset support. Value maximization can push a learned policy toward out-of-distribution actions, where the learned value estimate may be unreliable.}
	\label{fig:offline_support_mismatch}
\end{figure}

% ============================================================
\section{Behavior cloning: the supervised baseline}

The simplest offline method is behavior cloning (BC). It treats the dataset as supervised learning and fits
\begin{equation}
	\pi_\theta(a\mid s) \approx \pi_\beta(a\mid s).
\end{equation}
For continuous actions, a Gaussian policy can be trained by minimizing mean squared error or negative log-likelihood:
\begin{equation}
	\mathcal{L}_{\mathrm{BC}}(\theta)
	=
	\E_{(s,a)\sim\D}
	\left[\|\pi_\theta(s)-a\|_2^2\right].
\end{equation}

BC is stable because it never asks the policy to choose unsupported actions during training. But it has two major limitations:
\begin{enumerate}[leftmargin=*]
	\item it cannot easily improve beyond the behavior policy;
	\item it suffers from compounding errors when small mistakes move the agent into states not well represented in the data.
\end{enumerate}

Modern imitation-learning variants improve the basic BC template. Implicit Behavioral Cloning uses an energy-based model to represent multi-modal action distributions, while Action Chunking Transformers predict short action chunks to reduce compounding error in fine-grained robot manipulation \citep{florence2022ibc,zhao2023act}. These methods are not offline RL in the value-optimization sense, but they are important baselines whenever the dataset contains strong demonstrations.

\Needspace{16\baselineskip}
\begin{lstlisting}[style=pythonstyle,caption={Behavior cloning for continuous actions.},label={lst:bc}]
import torch
import torch.nn.functional as F

def behavior_cloning_update(policy, optimizer, batch):
    """Supervised imitation from an offline dataset.

    batch["obs"]     : tensor [B, obs_dim]
    batch["actions"] : tensor [B, act_dim]
    """
    obs = batch["obs"]
    actions = batch["actions"]

    pred_actions = policy(obs)
    loss = F.mse_loss(pred_actions, actions)

    optimizer.zero_grad()
    loss.backward()
    optimizer.step()
    return {"bc_loss": float(loss.detach())}
\end{lstlisting}

BC should be the first baseline in any offline RL study. If a sophisticated offline RL algorithm cannot beat BC, the dataset may not contain enough information for reliable improvement, or the offline RL algorithm may be overfitting value estimates.

% ============================================================
\section{Distribution shift and extrapolation error}

Offline RL usually uses value functions to ask a counterfactual question: what would happen if the agent took action $a$ in state $s$? In the dataset, only some actions were actually taken. A learned critic may assign high value to unsupported actions because the Bellman target does not contain enough corrective data.

For Q-learning, the target is
\begin{equation}
	y = r + \gamma \max_{a'} Q_{\bar\theta}(s',a').
\end{equation}
The maximization over $a'$ is dangerous offline. It can select an action that the dataset never took at $s'$. If $Q$ overestimates that action, the error bootstraps backward and contaminates other estimates. This is the offline analogue of the overestimation problem discussed in Chapter~6 for DQN-style value learning: maximization selects the optimistic error, and bootstrapping then propagates it. Offline RL adds a second difficulty: the selected action may also be outside the dataset distribution.

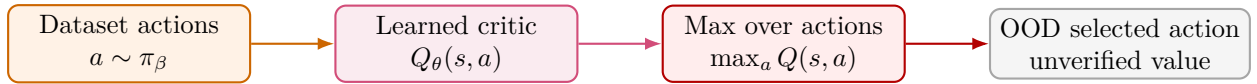
\begin{figure}[t]
	\centering
	\begin{tikzpicture}[
		box/.style={draw,rounded corners,thick,minimum width=3.2cm,minimum height=0.85cm,align=center,font=\small},
		arrow/.style={-{Latex[length=2.2mm]},thick},
		node distance=1.1cm
		]

		\node[box,fill=orange!10,draw=orange!80!black] (data) {Dataset actions\\$a\sim \pi_\beta$};
		\node[box,fill=purple!8,draw=purple!70,right=of data] (critic) {Learned critic\\$Q_\theta(s,a)$};
		\node[box,fill=red!7,draw=red!70!black,right=of critic] (max) {Max over actions\\$\max_a Q(s,a)$};
		\node[box,fill=gray!8,draw=gray!70,right=of max] (ood) {OOD selected action\\unverified value};
		\draw[arrow,draw=orange!80!black] (data) -- (critic);
		\draw[arrow,draw=purple!70] (critic) -- (max);
		\draw[arrow,draw=red!70!black] (max) -- (ood);
		\node[below=0.85cm of max,align=center,font=\small] {Bellman backup can amplify unsupported high Q-values.};

	\end{tikzpicture}
	\caption{Extrapolation error in offline Q-learning. The value maximization step may select out-of-distribution actions, and bootstrapping can propagate the resulting error.}
	\label{fig:extrapolation_error}
\end{figure}

This problem was studied explicitly in batch/offline RL work before the modern deep RL wave and later became central in algorithms such as BCQ, BEAR, and CQL \citep{fujimoto2019offpolicy,kumar2019bear,kumar2020cql}.

% ============================================================
\section{Three design principles of offline RL}

Most offline RL algorithms combine one or more of three principles.

\begin{table}[t]
	\centering
	\caption{Three major design principles in offline reinforcement learning.}
	\label{tab:offline_principles}
	\begin{tabularx}{\textwidth}{p{3.2cm}p{5.5cm}X}
		\toprule
		Principle & Main idea & Representative methods \\
		\midrule
		Behavior constraint & Keep the learned policy close to the data-generating behavior policy & BC, BCQ, BEAR, BRAC, TD3+BC \\
		Conservatism / pessimism & Penalize high values for unsupported actions so the critic underestimates OOD actions & CQL, MOPO, MOReL-style pessimistic models \\
		In-sample learning & Avoid evaluating arbitrary actions; learn values and policies using actions present in the dataset & IQL, expectile value learning, advantage-weighted regression \\
		\bottomrule
	\end{tabularx}
\end{table}

The best method depends heavily on dataset quality. Expert demonstrations may favor behavior cloning or advantage-weighted imitation. Mixed datasets often benefit from IQL or CQL. Narrow datasets require strong constraints. Large, diverse datasets may support more aggressive policy improvement.

% ============================================================
\section{Behavior-constrained offline RL: BCQ, BEAR, and regularization}

Behavior-constrained methods try to prevent the learned policy from selecting actions that are far from the dataset support.

\subsection{BCQ}

Batch-Constrained deep Q-learning (BCQ) learns a generative model of dataset actions and restricts action selection to actions likely under the behavior distribution \citep{fujimoto2019offpolicy}. For continuous control, BCQ samples candidate actions from a learned behavior model, perturbs them slightly, and selects the candidate with highest Q-value.

\begin{equation}
	a^*(s) = \argmax_{a\in\{a_1,\ldots,a_K\}} Q_\theta(s,a),
	\quad a_k \sim G_\omega(s).
\end{equation}

The important idea is not the exact architecture. The important idea is \emph{batch constraint}: do not maximize over arbitrary actions; maximize over plausible dataset-like actions.

\subsection{BEAR and distribution matching}

BEAR constrains the learned policy so that it remains close to the behavior policy under a distributional distance such as maximum mean discrepancy (MMD) \citep{kumar2019bear}. The policy objective becomes
\begin{equation}
	\max_\pi \; \E_{s\sim\D,a\sim\pi}[Q(s,a)]
	\quad \text{subject to} \quad
	D(\pi(\cdot\mid s),\pi_\beta(\cdot\mid s)) \leq \epsilon.
\end{equation}

This approach makes the trade-off explicit: improve the policy, but only inside a trust region around the dataset behavior.

\subsection{BRAC, AWR, and AWAC}

Behavior Regularized Actor Critic (BRAC) makes policy regularization a first-class objective, while AWR and AWAC use advantage-weighted regression to imitate good actions more strongly than bad actions \citep{wu2019brac,peng2019awr,nair2020awac}.

A generic advantage-weighted actor update is
\begin{equation}
	\mathcal{L}_{\mathrm{AWR}}(\theta)
	=
	-\E_{(s,a)\sim\D}
	\left[
	\exp\left(\frac{\hat A(s,a)}{\lambda}\right)
	\log \pi_\theta(a\mid s)
	\right].
\end{equation}

This keeps the policy close to dataset actions, but gives more weight to actions estimated to be better than expected.

% ============================================================
\section{Conservative value learning: CQL}

Conservative Q-Learning (CQL) addresses extrapolation error by explicitly lowering Q-values for actions outside the dataset while maintaining high values for dataset actions \citep{kumar2020cql}. A simplified CQL objective is
\begin{equation}
	\begin{aligned}
		\mathcal{L}_{\mathrm{CQL}}(Q)
		=
		&\;\mathcal{L}_{\mathrm{TD}}(Q)
		+
		\alpha
		\left(
		\E_{s\sim\D,a\sim\mu(\cdot\mid s)}[Q(s,a)]
		-
		\E_{(s,a)\sim\D}[Q(s,a)]
		\right),
	\end{aligned}
\end{equation}
where $\mu$ samples actions from a broad proposal distribution, often including policy actions, random actions, and current action samples.

Practical CQL implementations use either a fixed conservatism weight $\alpha$, often in the range $1$--$10$, or an adaptive Lagrangian formulation that ties $\alpha$ to the gap between out-of-distribution and in-distribution Q-values \citep{kumar2020cql}. This mirrors the Lagrangian-dual idea seen in automatic entropy-temperature tuning: a multiplier is adjusted so that the learned critic remains conservative enough without becoming uselessly pessimistic. 

CQL is also part of a broader pessimism pattern that appeared earlier in the book. Double DQN reduced optimistic value errors in Chapter~6, while TD3 and SAC used clipped double critics in Chapters~9 and~11 to prefer the smaller of two Q-estimates. CQL pushes this idea further for the offline setting: it does not merely choose the smaller critic estimate; it explicitly penalizes high values for actions that are poorly supported by the dataset. 

\begin{figure}[t]
	\centering
	\begin{tikzpicture}[
		box/.style={draw,rounded corners,thick,minimum width=3.3cm,minimum height=0.9cm,
			align=center,font=\small},
		arrow/.style={-{Latex[length=2.2mm]},thick},
		node distance=1.0cm
		]

		\node[box,fill=orange!10,draw=orange!80!black] (dataset)
		{Dataset action\\$a\sim\mathcal{D}$};

		\node[box,fill=red!7,draw=red!70!black,below=0.9cm of dataset] (ood)
		{Sampled OOD action\\$a\sim\mu$};

		\node[box,fill=purple!8,draw=purple!70,
		right=1.5cm of dataset,yshift=-0.45cm] (critic)
		{Critic\\$Q_\theta(s,a)$};

		\node[box,fill=blue!8,draw=blue!70,right=1.5cm of critic] (loss)
		{CQL regularizer\\push OOD values down};

		\draw[arrow,draw=orange!80!black] (dataset) -- (critic);
		\draw[arrow,draw=red!70!black]    (ood)     -- (critic);
		\draw[arrow,draw=purple!70]       (critic)  -- (loss);

		% Label moved well below both left-column boxes
		\node[below=1.4cm of ood, align=center, font=\small]
		{CQL prefers high values on data actions\\and low values elsewhere.};

	\end{tikzpicture}
	\caption{CQL adds pessimism to the critic. Unsupported actions are penalized so that
		value maximization is less likely to exploit overestimated out-of-distribution actions.}
	\label{fig:cql_pessimism}
\end{figure}

CQL is influential because it turns offline RL into a conservative value-estimation problem. Instead of trusting Q-values everywhere, it intentionally learns lower values for actions that are not supported by the dataset.

% ============================================================
\section{In-sample learning and Implicit Q-Learning}

Implicit Q-Learning (IQL) avoids querying the critic on arbitrary out-of-dataset actions during value learning \citep{kostrikov2021iql}. It uses three components:
\begin{enumerate}[leftmargin=*]
	\item learn a state value $V_\psi(s)$ using expectile regression from dataset actions;
	\item learn $Q_\theta(s,a)$ using TD targets based on $V_\psi(s')$;
	\item extract a policy by advantage-weighted regression on dataset actions.
\end{enumerate}

The expectile value loss is
\begin{equation}
	\mathcal{L}_V(\psi)
	=
	\E_{(s,a)\sim\D}
	\left[
	|\tau - \one_{Q_\theta(s,a)-V_\psi(s)<0}|
	\left(Q_\theta(s,a)-V_\psi(s)\right)^2
	\right].
\end{equation}
For $\tau>0.5$, the value function estimates an upper expectile of dataset action values, approximating the value of better-than-average actions while still remaining inside the dataset support.

\Needspace{18\baselineskip}
\begin{lstlisting}[style=pythonstyle,caption={Expectile loss used in IQL-style value learning.},label={lst:expectile}]
import torch
import torch.nn.functional as F

def expectile_loss(diff, tau=0.7):
    """IQL expectile regression loss.

    diff = q_values - v_values
    tau > 0.5 fits an upper expectile.
    """
    weight = torch.where(diff > 0, tau, 1.0 - tau)
    return (weight * diff.pow(2)).mean()

def iql_value_update(value_net, q_net, optimizer, batch, tau=0.7):
    obs = batch["obs"]
    actions = batch["actions"]

    with torch.no_grad():
        q = q_net(obs, actions)
    v = value_net(obs)
    loss = expectile_loss(q - v, tau=tau)

    optimizer.zero_grad()
    loss.backward()
    optimizer.step()
    return {"value_loss": float(loss.detach())}
\end{lstlisting}

IQL is popular because it is comparatively simple, stable, and effective on many D4RL-style continuous-control tasks. Its key lesson is that offline RL can improve without explicitly maximizing over unsupported actions.

% ============================================================
\section{TD3+BC: a simple and strong baseline}

TD3+BC modifies TD3 by adding a behavior-cloning term to the actor objective \citep{fujimoto2021td3bc}. If the actor is $\pi_\theta$, the objective can be written as minimizing
\begin{equation}
	\mathcal{L}_{\pi}(\theta)
	=
	-\lambda \E_{s\sim\D}[Q(s,\pi_\theta(s))]
	+
	\E_{(s,a)\sim\D}\left[\|\pi_\theta(s)-a\|_2^2\right].
\end{equation}
The first term improves the policy according to the critic. The second term keeps the policy close to the dataset.

\Needspace{18\baselineskip}
\begin{lstlisting}[style=pythonstyle,caption={TD3+BC actor loss.},label={lst:td3bc}]
def td3_bc_actor_loss(actor, critic, batch, alpha=2.5):
    obs = batch["obs"]
    data_actions = batch["actions"]

    policy_actions = actor(obs)
    q_values = critic(obs, policy_actions)

    # Normalize Q scale as in many TD3+BC implementations.
    lam = alpha / q_values.abs().mean().detach().clamp(min=1e-6)

    q_loss = -lam * q_values.mean()
    bc_loss = F.mse_loss(policy_actions, data_actions)
    return q_loss + bc_loss, {"q_loss": float(q_loss.detach()),
                              "bc_loss": float(bc_loss.detach())}
\end{lstlisting}

TD3+BC is valuable pedagogically because it shows the core offline RL trade-off in one line: improve according to $Q$, but do not move too far from the data. The TD3 part also inherits the clipped-double-critic idea introduced in actor-critic methods: using the smaller of two critic estimates reduces overestimation before the behavior-cloning term restricts the actor to dataset-supported actions. 

% ============================================================
\section{Sequence modeling view and Decision Transformer}

Decision Transformer reframes offline RL as conditional sequence modeling \citep{chen2021decisiontransformer}. Instead of learning a Bellman backup, it trains a transformer to predict actions conditioned on past states, actions, and desired return-to-go:
\begin{equation}
	(\hat R_1,s_1,a_1,\hat R_2,s_2,a_2,\ldots).
\end{equation}

The policy is queried by choosing a target return and autoregressively predicting actions. This approach avoids explicit Q-learning and therefore avoids some extrapolation-error mechanisms. However, it inherits sequence-modeling challenges: context length, return conditioning, dataset quality, and sensitivity to the target return.

\begin{figure}[t]
	\centering
	\begin{tikzpicture}[
		token/.style={draw,rounded corners,minimum width=1.2cm,minimum height=0.65cm,align=center,font=\small},
		box/.style={draw,rounded corners,thick,minimum width=2.6cm,minimum height=0.8cm,align=center,font=\small},
		arrow/.style={-{Latex[length=2mm]},thick}
		]
		\node[token,fill=green!8,draw=green!60!black] (r1) at (0,0) {$\hat R_1$};
		\node[token,fill=blue!8,draw=blue!70] (s1) at (1.35,0) {$s_1$};
		\node[token,fill=orange!10,draw=orange!80!black] (a1) at (2.7,0) {$a_1$};
		\node[token,fill=green!8,draw=green!60!black] (r2) at (4.05,0) {$\hat R_2$};
		\node[token,fill=blue!8,draw=blue!70] (s2) at (5.4,0) {$s_2$};
		\node[token,fill=orange!10,draw=orange!80!black] (a2) at (6.9,0) {$a_2$};
		\node[box,fill=purple!8,draw=purple!70] (tr) at (3.4,-1.6) {Transformer\\sequence model};
		\draw[arrow,draw=purple!70] (r1) -- (tr);
		\draw[arrow,draw=purple!70] (s1) -- (tr);
		\draw[arrow,draw=purple!70] (a1) -- (tr);
		\draw[arrow,draw=purple!70] (r2) -- (tr);
		\draw[arrow,draw=purple!70] (s2) -- (tr);
		\draw[arrow,draw=purple!70] (tr) -- node[right,font=\scriptsize] {predict next action} (a2);
	\end{tikzpicture}
	\caption{Decision Transformer treats offline RL as return-conditioned sequence modeling. The model predicts actions from a context of return-to-go, states, and previous actions.}
	\label{fig:decision_transformer_sequence}
\end{figure}
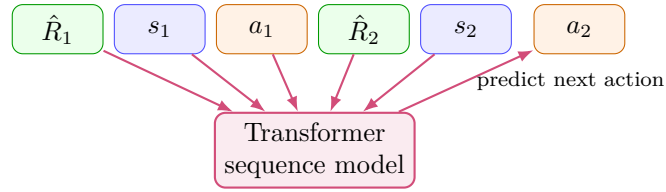

A later chapter develops Decision Transformers in depth. In this chapter, the important point is that offline RL has two broad families: value-conservative methods and sequence-modeling methods.

A subtle limitation is trajectory stitching. Value-based offline RL can, in principle, combine good local transitions from different trajectories through Bellman backups. Return-conditioned sequence models instead imitate trajectory prefixes and may struggle to assemble a new high-return behavior from good fragments that never appeared together in one logged trajectory. Later work on Q-learning Decision Transformers and Elastic Decision Transformers explicitly targets this stitching limitation \citep{yamagata2023qdt,wu2023elastic}. 

% ============================================================
\section{Generative and diffusion policies for offline RL}

Offline datasets often contain multi-modal behavior. In a robotics dataset, the same state may allow several valid grasps. In a UAV dataset, a drone may move left or right around an obstacle depending on traffic demand. A simple Gaussian policy may average these behaviors and produce an invalid action.

Generative policies address this by modeling complex action distributions. Diffusion-QL, Diffuser, Diffusion Policy, and related diffusion-policy methods use diffusion models to represent expressive behavior priors and then bias generation toward high-value actions or stable imitation policies \citep{janner2022diffuser,wang2022diffusionql,chi2023diffusionpolicy}. This line has become increasingly important through 2024--2026 because it connects offline RL to modern generative modeling and to high-dimensional visuomotor control.

\begin{researchbox}{Why diffusion matters for offline RL}
	A diffusion policy can represent multiple plausible actions for the same state. This is useful when the offline dataset is multi-modal. The challenge is not only to imitate the data, but to select high-value actions while staying near the dataset support.
\end{researchbox}

Diffusion policies are not a replacement for all offline RL methods. They are more computationally expensive, harder to tune, and may require many denoising steps at inference time. Recent work studies distillation and faster sampling to reduce this cost.

% ============================================================
\section{Model-based offline RL}

Model-based offline RL learns a model from the dataset and uses it for policy learning or planning. The danger is that the learned model may be wrong outside the dataset support. Therefore, model-based offline RL often uses pessimism or uncertainty penalties.

MOPO penalizes rewards using model uncertainty, while MOReL-style approaches construct pessimistic MDPs that discourage leaving known regions \citep{yu2020mopo,kidambi2020morel}. A generic uncertainty-penalized reward is
\begin{equation}
	\tilde r(s,a) = r(s,a) - \lambda \, u(s,a),
\end{equation}
where $u(s,a)$ measures model uncertainty. In MOPO-style methods, this uncertainty is typically epistemic uncertainty estimated by ensemble disagreement: it measures what the learned model does not know because of limited data, rather than irreducible aleatoric stochasticity in the environment. This connects back to model-based RL, but with a stricter offline safety requirement: the model should not hallucinate profitable transitions in unknown regions.

% ============================================================
\section{Dataset quality, coverage, and corruption}

Dataset quality determines what offline RL can achieve. A dataset can be expert, medium, random, mixed, narrow, broad, corrupted, nonstationary, or biased. The same algorithm may work well on one dataset and fail on another.

\begin{table}[t]
	\centering
	\caption{Offline dataset types and likely algorithmic behavior.}
	\label{tab:dataset_types}
	\begin{tabularx}{\textwidth}{p{2.6cm}p{5.2cm}X}
		\toprule
		Dataset type & Typical property & Useful first baselines \\
		\midrule
		Expert & High-quality, narrow support & BC, IQL, AWR/AWAC \\
		Medium & Mixed competent and poor actions & IQL, CQL, TD3+BC \\
		Random & Broad but low-reward actions & CQL, model-based pessimism, maybe poor policy quality \\
		Medium-expert & Multi-modal quality & IQL, CQL, Decision Transformer \\
		Corrupted logs & Sensor/action/reward errors & robust sequence modeling, filtering, uncertainty diagnostics \\
		Production network logs & Safe but conservative behavior & TD3+BC, IQL, offline-to-online fine-tuning \\
		\bottomrule
	\end{tabularx}
\end{table}

D4RL helped standardize offline RL evaluation by providing datasets of varying quality for locomotion, navigation, manipulation, and other tasks \citep{fu2020d4rl}. RL Unplugged provided another benchmark suite with logged data from several domains \citep{gulcehre2020rlunplugged}. In the 2024--2026 ecosystem, actively maintained dataset libraries such as Minari provide Gymnasium-compatible offline RL datasets and data-handling utilities, including reproductions of D4RL-style datasets \citep{minari2024}.

However, benchmark performance should not be confused with deployment readiness. Production network logs have additional structure that generic benchmarks rarely capture. Most actions are conservative because they were produced by rule-based controllers or human operators; rewards may be delayed because QoS is measured over windows rather than instantaneously; reward definitions may combine telemetry from multiple systems; and the logging policy itself may have changed during data collection. For UAV/SDN control, these details are not bookkeeping issues—they determine whether an offline policy is learning a stable control rule or exploiting artifacts of old operational procedures. 

\begin{warningbox}{Offline evaluation is hard}
	In supervised learning, test accuracy can be measured on held-out labels. In offline RL, the true value of a learned policy usually requires environment interaction, which the offline setting forbids. Off-policy evaluation is possible but difficult and often high-variance or biased.
\end{warningbox}

% ============================================================
\section{Offline-to-online fine-tuning}

Offline RL is often used as a safe initialization stage. The agent first learns from existing data, then cautiously fine-tunes online with constraints, safety filters, or human supervision. This is attractive in robotics and networks: the offline policy avoids random exploration, while online fine-tuning adapts to deployment conditions.

A practical offline-to-online pipeline is:
\begin{enumerate}[leftmargin=*]
	\item collect or curate a dataset;
	\item train BC and offline RL baselines;
	\item evaluate in simulation or conservative off-policy evaluation;
	\item deploy with a safety layer or limited action authority;
	\item fine-tune online with strict monitoring.
\end{enumerate}

Offline-to-online RL is not a loophole around safety. The moment the policy starts collecting new data, the system becomes an online control system and must satisfy online safety requirements.

To make the trade-offs concrete, suppose the dataset contains 100K UAV/SDN transitions, of which only about 3\% involve aggressive bandwidth reallocation with a mean logged reward of 21, while the dominant 97\% follow a conservative rule-based policy with mean reward 16. The four main algorithm families would treat this dataset as follows:

\begin{table}[t]
	\centering
	\caption{Worked example: how four offline methods treat the same UAV/SDN dataset.}
	\label{tab:uav_offline_worked_example}
	\begin{tabular}{lp{5.8cm}p{4.0cm}}
		\toprule
		Method & Treatment of the 3\% aggressive actions & Expected outcome \\
		\midrule
		BC & Mostly ignored; imitate dominant conservative policy & Reward $\approx 16$, very safe \\
		TD3+BC & BC term anchors near logged behavior; $Q$ pulls cautiously toward aggressive actions & Reward $\approx 17$--$18$ \\
		CQL & Aggressive-action $Q$-values are penalized as poorly supported & Reward $\approx 17$, conservative \\
		IQL & Upper-expectile $V$ never queries OOD actions; advantage-weighted actor leans on better-than-average in-sample actions & Reward $\approx 17.5$, stable \\
		\bottomrule
	\end{tabular}
\end{table}

The numbers are illustrative in Table~\ref{tab:uav_offline_worked_example}, not measured, but the comparison captures the central trade-off. BC is safe but cannot improve; CQL and IQL improve cautiously by keeping the policy near the data; TD3+BC trades a slightly higher reward ceiling for a slightly weaker safety anchor. The right choice depends on whether the aggressive 3\% are genuinely high-value or, as the production-log discussion warned, artifacts of an old controller. 

% ============================================================
\section{UAV/SDN worked scenario: learning from network logs}

Consider a UAV-assisted SDN wireless network. The operator has months of logs generated by a rule-based controller and occasional human interventions. Each transition contains:
\begin{equation}
	(s_t,a_t,r_t,s_{t+1},d_t),
\end{equation}
where $s_t$ includes UAV location, battery, traffic load, SINR, latency, and user priority; $a_t$ includes movement direction, bandwidth allocation, and power level; $r_t$ combines QoS reward, energy cost, and safety penalty.

\begin{figure}[t]
	\centering
	\begin{tikzpicture}[
		box/.style={draw,rounded corners,thick,minimum width=3.0cm,minimum height=0.85cm,align=center,font=\small},
		small/.style={draw,rounded corners,minimum width=2.7cm,minimum height=0.75cm,align=center,font=\small},
		arrow/.style={-{Latex[length=2.2mm]},thick},
		node distance=0.9cm
		]
		\node[box,fill=blue!8,draw=blue!70] (logs) {UAV/SDN logs\\QoS, action, outcome};
		\node[box,fill=orange!10,draw=orange!80!black,right=of logs] (dataset) {Offline dataset\\$\D$};
		\node[box,fill=purple!8,draw=purple!70,right=of dataset] (learner) {Offline RL learner\\IQL / CQL / TD3+BC};
		\node[box,fill=green!10,draw=green!60!black,right=of learner] (policy) {Candidate policy\\movement + allocation};
		\node[small,fill=red!7,draw=red!70!black,below=1.1cm of policy] (safety) {Safety gate\\CBF / rules / SDN};
		\node[small,fill=gray!8,draw=gray!70,below=1.1cm of learner] (eval) {Offline diagnostics\\coverage + OPE};

		\draw[arrow,draw=blue!70] (logs) -- (dataset);
		\draw[arrow,draw=orange!80!black] (dataset) -- (learner);
		\draw[arrow,draw=purple!70] (learner) -- (policy);
		\draw[arrow,draw=purple!70] (learner) -- (eval);
		\draw[arrow,draw=green!60!black] (policy) -- (safety);
		\draw[arrow,draw=red!70!black] (safety.west) -- ++(-1.4,0) |- (eval.east);
	\end{tikzpicture}
	\caption{Offline RL for UAV/SDN control. Historical network logs are converted into an offline dataset, used to train a policy, checked with coverage diagnostics and off-policy evaluation, and deployed only behind a safety gate.}
	\label{fig:uav_offline_rl_scenario}
\end{figure}
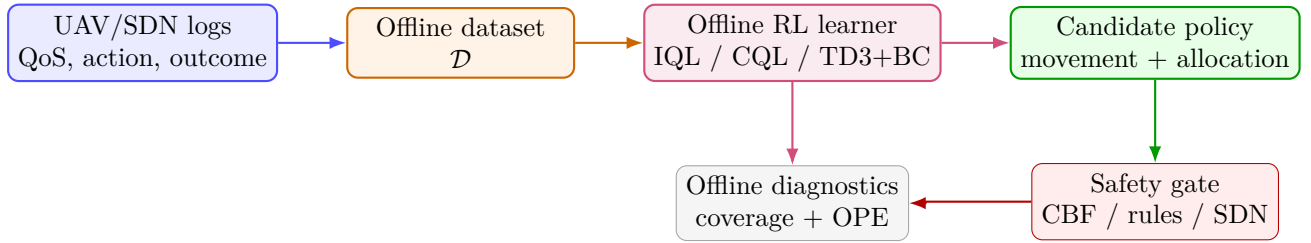

A naive offline RL system might learn that aggressive bandwidth allocation improves throughput. But if such actions are rare in the logs, their Q-values may be unreliable. A safer system uses one or more of the following:
\begin{itemize}
	\item TD3+BC or IQL to stay close to logged behavior;
	\item CQL to penalize unsupported high-Q actions;
	\item uncertainty diagnostics to detect low-coverage states;
	\item a CBF or projection layer to enforce battery, collision, and latency constraints at deployment.
\end{itemize}

\begin{researchbox}{Research signature: offline safe UAV/SDN learning}
	A distinctive research direction is to combine offline RL with safety filters. The offline learner proposes a policy from logs; a CBF or SDN safety layer prevents unsafe execution; and the critic is trained to evaluate the executed safe action rather than only the pre-filter action. This links offline RL, safe RL, and network control.
\end{researchbox}

% ============================================================
\section{Practical implementation code}

\subsection{Offline dataset loader}

\Needspace{18\baselineskip}
\begin{lstlisting}[style=pythonstyle,caption={A minimal offline transition dataset.},label={lst:offline_dataset}]
from torch.utils.data import Dataset

class OfflineRLDataset(Dataset):
    def __init__(self, obs, actions, rewards, next_obs, dones):
        self.obs = torch.as_tensor(obs, dtype=torch.float32)
        self.actions = torch.as_tensor(actions, dtype=torch.float32)
        self.rewards = torch.as_tensor(rewards, dtype=torch.float32).unsqueeze(-1)
        self.next_obs = torch.as_tensor(next_obs, dtype=torch.float32)
        self.dones = torch.as_tensor(dones, dtype=torch.float32).unsqueeze(-1)

    def __len__(self):
        return self.obs.shape[0]

    def __getitem__(self, idx):
        return {
            "obs": self.obs[idx],
            "actions": self.actions[idx],
            "rewards": self.rewards[idx],
            "next_obs": self.next_obs[idx],
            "dones": self.dones[idx],
        }
\end{lstlisting}

\subsection{IQL actor extraction}

\Needspace{18\baselineskip}
\begin{lstlisting}[style=pythonstyle,caption={IQL-style advantage-weighted actor extraction.},label={lst:iql_actor}]
def iql_actor_loss(policy, value_net, q_net, batch, beta=3.0, max_weight=100.0):
    obs = batch["obs"]
    actions = batch["actions"]

    with torch.no_grad():
        q = q_net(obs, actions)
        v = value_net(obs)
        adv = q - v
        weights = torch.exp(beta * adv).clamp(max=max_weight)

    log_prob = policy.log_prob(obs, actions)
    loss = -(weights * log_prob).mean()
    return loss, {"adv_mean": float(adv.mean()),
                  "weight_mean": float(weights.mean())}
\end{lstlisting}

\subsection{CQL penalty sketch}

\Needspace{18\baselineskip}
\begin{lstlisting}[style=pythonstyle,caption={Simplified continuous-action CQL penalty.},label={lst:cql_penalty}]
def cql_penalty(q_net, policy, obs, data_actions, num_random=10):
    """Sketch of a CQL-style conservative regularizer.

    This is not a full production CQL implementation. It shows the idea:
    compare Q-values on broad candidate actions against Q-values on data actions.
    """
    B, act_dim = data_actions.shape

    # Random actions in normalized action range [-1, 1].
    random_actions = torch.empty(B, num_random, act_dim, device=obs.device).uniform_(-1, 1)

    # Current policy actions.
    with torch.no_grad():
        policy_actions = policy.sample(obs).unsqueeze(1)  # [B, 1, act_dim]

    candidate_actions = torch.cat([random_actions, policy_actions], dim=1)
    obs_rep = obs.unsqueeze(1).expand(-1, candidate_actions.shape[1], -1)

    q_candidates = q_net(obs_rep.reshape(-1, obs.shape[-1]),
                         candidate_actions.reshape(-1, act_dim))
    q_candidates = q_candidates.view(B, -1)

    q_data = q_net(obs, data_actions)
    conservative_gap = torch.logsumexp(q_candidates, dim=1, keepdim=True) - q_data
    return conservative_gap.mean()
\end{lstlisting}

\subsection{Coverage diagnostic}

\Needspace{16\baselineskip}
\begin{lstlisting}[style=pythonstyle,caption={Simple nearest-neighbor coverage score for candidate actions.},label={lst:coverage}]
def action_coverage_score(candidate_actions, dataset_actions, k=5):
    """Return mean distance to k nearest dataset actions.

    Larger score means the candidate actions are farther from the dataset support.
    For real systems, use state-conditioned coverage, not only action distance.
    """
    dist = torch.cdist(candidate_actions, dataset_actions)
    knn_dist, _ = torch.topk(dist, k=k, largest=False, dim=1)
    return knn_dist.mean(dim=1)
\end{lstlisting}

% ============================================================
\section{Failure modes and debugging checklist}

\begin{table}[t]
	\centering
	\caption{Common offline RL failure modes.}
	\label{tab:offline_failure_modes}
	\begin{tabularx}{\textwidth}{p{3.2cm}p{4.2cm}X}
		\toprule
		Symptom & Likely cause & What to inspect \\
		\midrule
		Policy worse than BC & Value overfitting or poor dataset quality & Compare against BC; inspect Q-values and coverage \\
		Very high predicted Q, poor rollout & Extrapolation error & Evaluate OOD action distance; use CQL or stronger BC regularization \\
		Actor copies bad actions & Dataset has mixed or poor behavior & Use advantage weighting, IQL, filtering, or return-conditioned training \\
		No improvement over behavior & Constraint too strong & Relax BC weight or conservative penalty carefully \\
		Unsafe deployment actions & Offline objective ignores constraints & Add safety gate, CBF, constrained RL, or action projection \\
		Works on benchmark, fails in system & Dataset shift or hidden confounders & Validate coverage, nonstationarity, logging policy, and reward definition \\
		Policy improves on dataset but fails after deployment & Training/deployment distribution drift & Validate dataset recency, stationarity assumptions, and drift in users, traffic, channels, or logging policy \\
		\bottomrule
	\end{tabularx}
\end{table}

A serious offline RL project should report BC, a simple offline RL baseline, dataset coverage diagnostics, held-out behavior prediction, reward distribution, and deployment safety checks. Reward alone is not sufficient.

% ============================================================
\section{Limitations and frontiers toward 2026}

Offline RL has made major progress, but several limitations remain.

\begin{enumerate}[leftmargin=*]
	\item \textbf{Evaluation remains difficult.} Without online testing, it is hard to know whether a learned policy is actually better.
	\item \textbf{Dataset coverage limits improvement.} No algorithm can reliably infer consequences for actions completely absent from the data without additional assumptions.
	\item \textbf{Conservatism can be too strong.} Pessimistic algorithms may avoid useful improvement if uncertainty is overestimated.
	\item \textbf{Sequence models need large data and may struggle with stitching.} Decision Transformer-style approaches can be powerful but may require broad, high-quality datasets and may not naturally recombine useful sub-trajectories unless additional value, model-based, or adaptive-context mechanisms are added.
	\item \textbf{Generative policies are expressive but expensive.} Diffusion policies can represent multi-modal behavior, but inference cost and stability remain concerns.
	\item \textbf{Safety is not automatic.} Offline training does not guarantee safe deployment; safety constraints need explicit treatment.
\end{enumerate}

Toward 2026, the frontier includes diffusion policies, transformer-based offline RL, robust learning under corrupted data, offline-to-online adaptation, foundation-model-guided control, and constrained offline RL for safety-critical systems. Recent ICLR 2025--2026 work studies adaptive feature fusion, in-sample learning, diffusion regularization, and generative trajectory modeling for offline RL. These directions are promising, but they do not remove the core lesson of the chapter: \emph{offline RL is only as reliable as its dataset support, value estimation discipline, and deployment safeguards}. 

% ============================================================
\section{Exercises}

\subsection*{Conceptual exercises}
\begin{enumerate}[leftmargin=*]
	\item Explain why offline RL is not just supervised learning.
	\item Why can Q-learning fail when trained only on a fixed dataset?
	\item Compare BC, TD3+BC, CQL, and IQL in terms of how they avoid unsupported actions.
	\item Why might an expert-only dataset be less useful for policy improvement than a medium-expert dataset?
	\item Why is offline evaluation harder in RL than in supervised learning?
\end{enumerate}

\subsection*{Mathematical exercises}
\begin{enumerate}[leftmargin=*]
	\item Show how the TD target $r+\gamma\max_a Q(s',a)$ can select an action not present in the dataset.
	\item Derive the gradient of the TD3+BC actor loss with respect to actor parameters.
	\item For IQL, explain how expectile regression changes as $\tau$ moves from $0.5$ to $0.9$.
	\item For CQL, explain why the term $\E_{a\sim\mu}[Q(s,a)]-\E_{a\sim\D}[Q(s,a)]$ encourages conservative values.
\end{enumerate}

\subsection*{Coding exercises}
\begin{enumerate}[leftmargin=*]
	\item Implement behavior cloning and evaluate it as a baseline before any offline RL algorithm.
	\item Implement the TD3+BC actor loss in Listing~\ref{lst:td3bc} and test different BC weights.
	\item Add the CQL penalty in Listing~\ref{lst:cql_penalty} to a Q-learning update.
	\item Compute a simple state-conditioned coverage score for candidate actions using nearest neighbors.
\end{enumerate}

\subsection*{Research thinking exercises}
\begin{enumerate}[leftmargin=*]
	\item Design an offline RL pipeline for UAV/SDN control using production network logs. What data would you need?
	\item How would you detect whether a learned action is outside the dataset support?
	\item In safe offline RL, should the safety layer be included during training, deployment, or both?
	\item For a medical dataset, when would you prefer BC over offline RL?
	\item For a language-model reasoning dataset, what is the offline RL analogue of behavior support?
\end{enumerate}

% ============================================================
\section*{Looking Ahead to Chapter 15: Food for Thought}
\addcontentsline{toc}{section}{Looking Ahead to Chapter 15: Decision Transformers and Sequence Modeling}

This chapter introduced Decision Transformer as one branch of offline RL. The next chapter develops that idea deeply. It asks a different question from CQL or IQL:
\begin{quote}
	Can reinforcement learning be reformulated as conditional sequence modeling, where the agent predicts actions from past context and a desired return?
\end{quote}

This shift connects offline RL to transformers, trajectory modeling, large-scale pretraining, return conditioning, and modern foundation-model thinking. It does not remove the offline RL challenges of dataset quality and support, but it offers a new representation of the problem.

\begin{quote}
	Chapter 14 explained how to learn safely from fixed datasets.\\
	Chapter 15 asks whether trajectories themselves can be treated as language-like sequences.
\end{quote}

% ============================================================
	\chapter[Decision Transformers and Sequence Modeling]{Decision Transformers and Sequence Modeling}
\label{ch:decision_transformers}
\chaptermark{Decision Transformers}

\begin{keybox}{Chapter goal}
	Decision Transformer shows that offline RL can be reformulated as return-conditioned sequence modeling. Instead of learning a Bellman backup, a causal transformer predicts actions from a history of returns-to-go, states, and previous actions. This chapter develops the full Decision Transformer family: token sequences, training, inference, trajectory stitching, QDT and Elastic DT variants, UAV/SDN applications, and the connections to broader sequence-model control systems.
\end{keybox}

\section*{Chapter Overview}
\addcontentsline{toc}{section}{Chapter Overview}

\begin{enumerate}[leftmargin=*]
	\item Why sequence modeling changes the offline RL perspective
	\item Trajectories as token sequences
	\item Decision Transformer: return-conditioned behavior modeling
	\item Training objective and data construction
	\item Architecture and causal masking
	\item Inference by desired return
	\item PyTorch implementation: data, model, training, and rollout
	\item Trajectory Transformer and planning as sequence generation
	\item Reinforcement learning via supervised learning
	\item Trajectory stitching: the central limitation
	\item QDT, Elastic DT, and stitching-aware variants
	\item UAV/SDN trajectory modeling scenario
	\item Practical implementation details
	\item Frontiers: online DTs, multi-game DTs, prompts, and 2026 directions
	\item Failure modes and debugging
	\item Exercises and looking ahead
\end{enumerate}

\section{Why sequence modeling changes the offline RL perspective}

Offline reinforcement learning in Chapter~14 was presented as a value-learning and conservatism problem: we have a fixed dataset, and the learned policy must avoid unsupported actions. Conservative Q-learning, IQL, TD3+BC, and model-based offline RL all retain something close to the classical RL view: learn values, improve a policy, and be careful about distribution shift.

Decision Transformer (DT) proposes a different view: maybe a policy can be learned as a conditional sequence model. Instead of learning a Bellman backup, the model learns to predict the next action from a history of returns-to-go, states, and previous actions \citep{chen2021decisiontransformer}. In this view, offline RL resembles language modeling: a trajectory is a sentence, states and actions are tokens, and a desired return is a prompt.

Chapter~14 introduced Decision Transformer briefly as one of the major sequence-modeling alternatives to value-based offline RL. This chapter develops that family in depth: how trajectories become token sequences, how return-to-go conditioning works, why inference is closed-loop, and where sequence modeling succeeds or fails compared with Bellman-style methods.

\begin{keybox}{Core idea}
	A Decision Transformer does not ask: ``What is the value of this action?'' It asks: ``Given this history and a desired future return, what action would likely appear next in a successful trajectory?''
\end{keybox}

This shift is powerful because it imports the machinery of modern sequence modeling into reinforcement learning: causal attention, long context windows, supervised learning losses, pretraining, prompting, multitask conditioning, and scaling. But it also changes the failure modes. A Decision Transformer can imitate and condition on high-return behavior, but it may struggle to stitch together pieces of different trajectories in ways that value-based dynamic programming can naturally do \citep{brandfonbrener2022when,yamagata2023qdt,wu2023elastic}.

\begin{researchbox}{A running interpretation for this book}
	For UAV/SDN control, a Decision Transformer can be trained on logs of network states, actions, QoS rewards, and safety outcomes. At deployment, the desired return can encode an operational target such as high QoS satisfaction, low energy usage, or low violation rate. This makes sequence modeling attractive for operator-guided network control, but it does not remove the need for safety filtering or dataset coverage checks.
\end{researchbox}

\section{Trajectories as token sequences}

A standard offline RL dataset contains trajectories
\begin{equation}
	\tau = (s_0,a_0,r_1,s_1,a_1,r_2,\ldots,s_T).
\end{equation}
Decision Transformer reorganizes each trajectory into a sequence of triplets:
\begin{equation}
	(\hat{R}_0,s_0,a_0,\hat{R}_1,s_1,a_1,\ldots,\hat{R}_{T-1},s_{T-1},a_{T-1}),
\end{equation}
where \(\hat{R}_t\) is the return-to-go:
\begin{equation}
	\hat{R}_t = \sum_{k=t}^{T-1} r_{k+1}.
	\label{eq:rtg}
\end{equation}
Some implementations use discounted returns-to-go, but the original Decision Transformer commonly used undiscounted returns-to-go for benchmark tasks \citep{chen2021decisiontransformer}.

The interleaving order is a convention rather than a new RL theorem. The original Decision Transformer uses the order \((\hat R_t,s_t,a_t)\), repeated over time, so the action token is predicted after the return and state tokens for that timestep. Other implementations may use slightly different orderings, but they must preserve causal masking: the predicted action must not attend to future rewards, future states, or future actions.

\begin{figure}[t]
	\centering
	\begin{tikzpicture}[
		token/.style={draw,rounded corners,thick,minimum width=1.35cm,minimum height=0.62cm,align=center,font=\small},
		arrow/.style={-{Latex[length=2.2mm]},thick},
		node distance=0.10cm
		]
		\node[token,fill=blue!8,draw=blue!70] (r0) {$\hat R_0$};
		\node[token,fill=green!8,draw=green!60!black,right=of r0] (s0) {$s_0$};
		\node[token,fill=orange!12,draw=orange!80!black,right=of s0] (a0) {$a_0$};
		\node[token,fill=blue!8,draw=blue!70,right=of a0] (r1) {$\hat R_1$};
		\node[token,fill=green!8,draw=green!60!black,right=of r1] (s1) {$s_1$};
		\node[token,fill=orange!12,draw=orange!80!black,right=of s1] (a1) {$a_1$};
		\node[right=0.10cm of a1] (dots) {$\cdots$};
		\node[token,fill=blue!8,draw=blue!70,right=0.10cm of dots] (rt) {$\hat R_t$};
		\node[token,fill=green!8,draw=green!60!black,right=of rt] (st) {$s_t$};
		\node[token,fill=orange!12,draw=orange!80!black,right=of st] (at) {$a_t$};

		\draw[decorate,decoration={brace,amplitude=5pt},thick] ($(r0.north west)+(0,0.15)$) -- ($(a0.north east)+(0,0.15)$) node[midway,above=0.25cm,font=\small] {one decision step};
		\draw[arrow,draw=black!55] (r0.south) -- ++(0,-0.55) -| node[pos=0.25,below,font=\scriptsize] {causal context predicts next action} (a0.south);
	\end{tikzpicture}
	\caption{Decision Transformer views a trajectory as a token sequence of return-to-go, state, and action. The model is trained autoregressively: past returns, states, and actions provide context for predicting the next action.}
	\label{fig:dt_tokens}
\end{figure}
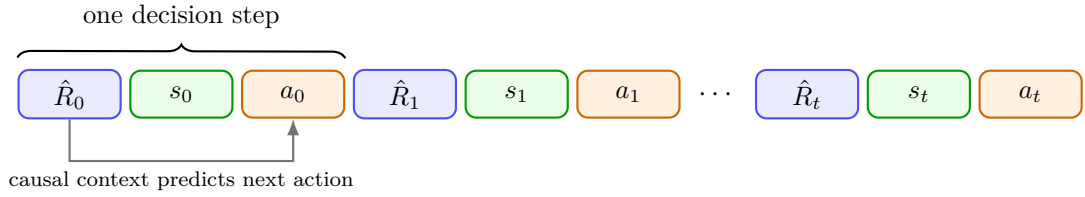

This representation is the bridge between offline RL and sequence modeling. The desired return acts like a command. The history of states and actions acts like context. The next action is the supervised target.

\section[Return-conditioned behavior modeling]{Decision Transformer as return-conditioned behavior modeling}

Let \(K\) be the context length. The model receives the most recent \(K\) triplets and predicts the next action:
\begin{equation}
	\pi_\theta(a_t\mid \hat{R}_{t-K:t},s_{t-K:t},a_{t-K:t-1}).
	\label{eq:dt_policy}
\end{equation}
For continuous actions, the model can regress directly to the dataset action using a mean-squared error loss. For discrete actions, it predicts a categorical distribution using cross-entropy.

The supervised loss is
\begin{equation}
	\mathcal{L}_{\mathrm{DT}}(\theta)
	=
	\E_{\tau\sim \D}\left[
	\sum_{t=0}^{T-1}
	\ell\left(
	f_\theta(\hat{R}_{\leq t},s_{\leq t},a_{<t}), a_t
	\right)
	\right],
	\label{eq:dt_loss}
\end{equation}
where \(\ell\) is MSE for continuous actions or cross-entropy for discrete actions.

\begin{warningbox}{Decision Transformer is not Bellman learning}
	Decision Transformer does not fit a value function and does not perform temporal-difference backups. It is closer to conditional behavior cloning: it clones actions from trajectories while conditioning on a desired future return. This makes it simple and stable, but it also means that the model inherits the coverage and compositional limits of the dataset.
\end{warningbox}

\section{Training objective and data construction}

The training procedure has four practical steps.

\begin{enumerate}[leftmargin=*]
	\item Split offline data into trajectories.
	\item Compute return-to-go \(\hat R_t\) for each timestep.
	\item Sample fixed-length context windows.
	\item Train a causal transformer to predict actions at the state/action positions.
\end{enumerate}

The target return should be scaled. In D4RL-style continuous-control tasks, returns may have a very different magnitude from states and actions, so a return scale is usually used:
\begin{equation}
	\tilde R_t = \frac{\hat R_t}{c_R}.
\end{equation}
If \(c_R\) is too small, the return token dominates the embedding. If it is too large, the model may ignore the return prompt.

\begin{table}[t]
	\centering
	\caption{Key objects in Decision Transformer training.}
	\label{tab:dt_objects}
	\begin{tabularx}{\textwidth}{p{2.6cm}p{4.0cm}X}
		\toprule
		Object & Shape / example & Purpose \\
		\midrule
		State \(s_t\) & vector, image embedding, network state & describes current condition \\
		Action \(a_t\) & continuous vector or discrete id & supervised prediction target \\
		Return-to-go \(\hat R_t\) & scalar & desired future performance prompt \\
		Timestep \(t\) & integer & positional/temporal context \\
		Attention mask & binary mask & prevents padding from affecting training \\
		Context length \(K\) & e.g., 20, 30, 100 & controls historical memory \\
		\bottomrule
	\end{tabularx}
\end{table}

For variable-length trajectories, padding direction is a practical detail. Many Decision Transformer implementations left-pad short context windows so that the most recent real timesteps remain at the right edge of the context, closest to the action being predicted. Whatever convention is used, the attention mask must hide padded tokens; otherwise the model can learn spurious patterns from padding rather than from trajectory history.

\section{Architecture and causal masking}

A Decision Transformer embeds three streams: returns-to-go, states, and actions. The embeddings are interleaved and passed through a causal transformer. The output at the state position is used to predict the next action.

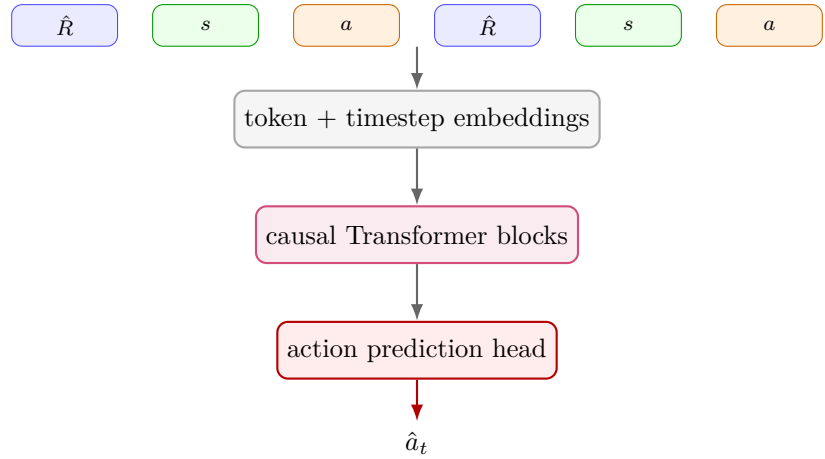
\begin{figure}[t]
	\centering
	\begin{tikzpicture}[
		box/.style={draw,rounded corners,thick,minimum width=2.7cm,minimum height=0.75cm,align=center,font=\small},
		small/.style={draw,rounded corners,minimum width=1.4cm,minimum height=0.55cm,align=center,font=\scriptsize},
		arrow/.style={-{Latex[length=2.2mm]},thick},
		node distance=0.45cm
		]
		\node[small,fill=blue!8,draw=blue!70] (r) {$\hat R$};
		\node[small,fill=green!8,draw=green!60!black,right=of r] (s) {$s$};
		\node[small,fill=orange!12,draw=orange!80!black,right=of s] (a) {$a$};
		\node[small,fill=blue!8,draw=blue!70,right=of a] (r2) {$\hat R$};
		\node[small,fill=green!8,draw=green!60!black,right=of r2] (s2) {$s$};
		\node[small,fill=orange!12,draw=orange!80!black,right=of s2] (a2) {$a$};

		\node[box,below=0.85cm of $(s)!0.5!(s2)$,fill=gray!8,draw=gray!70] (embed) {token + timestep embeddings};
		\node[box,below=0.75cm of embed,fill=purple!8,draw=purple!70] (trans) {causal Transformer blocks};
		\node[box,below=0.75cm of trans,fill=red!7,draw=red!70!black] (head) {action prediction head};
		\node[below=0.55cm of head,font=\small] (out) {$\hat a_t$};

		\draw[arrow,draw=black!60] ($(r.south)!0.5!(a2.south)$) -- (embed.north);
		\draw[arrow,draw=black!60] (embed) -- (trans);
		\draw[arrow,draw=black!60] (trans) -- (head);
		\draw[arrow,draw=red!70!black] (head) -- (out);
	\end{tikzpicture}
	\caption{Decision Transformer architecture. Return-to-go, state, and action tokens are embedded with timestep information and processed by a causal transformer. The action head predicts the next action from the available context.}
	\label{fig:dt_architecture}
\end{figure}

Causal masking is essential. At training time, the model must not see future actions. It should predict \(a_t\) using only the return prompt and trajectory history available up to time \(t\). This is the same autoregressive discipline that makes language models predictive rather than simply reconstructive \citep{vaswani2017attention,chen2021decisiontransformer}.

\section{Inference by desired return}

At deployment, the user chooses an initial target return \(\hat R_0^{\mathrm{target}}\). The model observes \(s_0\) and predicts an action. After executing the action and receiving reward \(r_1\), the target return is updated:
\begin{equation}
	\hat R_{t+1}^{\mathrm{target}}
	=
	\hat R_t^{\mathrm{target}} - r_{t+1}.
	\label{eq:dt_rtg_update}
\end{equation}
The target is therefore a remaining-budget signal: if the desired total return was high and the agent already received reward, the remaining desired return decreases.

\begin{keybox}{Decision Transformer inference loop}
	\begin{enumerate}[leftmargin=*]
		\item Set the initial desired return \(\hat R_0^{\mathrm{target}}\).
		\item At timestep \(t\), feed the context \((\hat R_{\leq t},s_{\leq t},a_{<t})\) to the transformer.
		\item Predict and execute \(a_t\).
		\item Observe \(r_{t+1}\) and \(s_{t+1}\).
		\item Update the remaining return using \(\hat R_{t+1}^{\mathrm{target}} = \hat R_t^{\mathrm{target}} - r_{t+1}\).
		\item Slide the context window forward and repeat.
	\end{enumerate}
\end{keybox}

This update rule is the closed-loop control mechanism of Decision Transformer. The transformer does not simply output a complete plan once; it is repeatedly reconditioned on the actual rewards and states observed during execution.

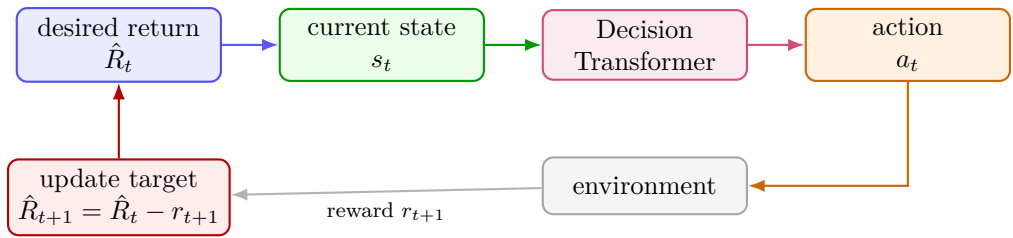
\begin{figure}[t]
	\centering
	\begin{tikzpicture}[
		box/.style={draw,rounded corners,thick,minimum width=2.7cm,minimum height=0.75cm,align=center,font=\small},
		arrow/.style={-{Latex[length=2.2mm]},thick},
		node distance=0.75cm
		]
		\node[box,fill=blue!8,draw=blue!70] (target) {desired return\\$\hat R_t$};
		\node[box,fill=green!8,draw=green!60!black,right=of target] (state) {current state\\$s_t$};
		\node[box,fill=purple!8,draw=purple!70,right=of state] (dt) {Decision\\Transformer};
		\node[box,fill=orange!12,draw=orange!80!black,right=of dt] (action) {action\\$a_t$};
		\node[box,fill=gray!8,draw=gray!70,below=1.0cm of dt] (env) {environment};
		\node[box,fill=red!7,draw=red!70!black,below=1.0cm of target] (update) {update target\\$\hat R_{t+1}=\hat R_t-r_{t+1}$};

		\draw[arrow,draw=blue!65] (target) -- (state);
		\draw[arrow,draw=green!60!black] (state) -- (dt);
		\draw[arrow,draw=purple!70] (dt) -- (action);
		\draw[arrow,draw=orange!80!black] (action) |- (env);
		\draw[arrow,draw=gray!60] (env) -- node[below,font=\scriptsize] {reward $r_{t+1}$} (update);
		\draw[arrow,draw=red!70!black] (update) -- (target);
	\end{tikzpicture}
	\caption{Decision Transformer inference. A target return conditions the policy. After each reward is observed, the remaining desired return is updated and fed back into the next context window.}
	\label{fig:dt_inference}
\end{figure}

Choosing the target return is not trivial. If the target is too low, the agent may imitate mediocre trajectories. If it is too high, the model may be prompted outside the dataset distribution. Multi-game and target-return optimization work studies how to set or adapt desired returns rather than choosing them manually \citep{lee2022multigame,tatematsu2025mtro}.

\section{PyTorch implementation: trajectory dataset}

The following code constructs context windows for a Decision Transformer. It handles return-to-go, padding, and masks. The exact state/action normalization should be computed from the training dataset only.

\Needspace{18\baselineskip}
\begin{lstlisting}[style=pythonstyle,caption={Trajectory dataset for Decision Transformer.},label={lst:dt_dataset}]
import torch
from torch.utils.data import Dataset

class DecisionTransformerDataset(Dataset):
    def __init__(self, trajectories, context_len, state_mean, state_std,
                 rtg_scale=1000.0):
        self.trajs = trajectories
        self.K = context_len
        self.state_mean = torch.as_tensor(state_mean, dtype=torch.float32)
        self.state_std = torch.as_tensor(state_std, dtype=torch.float32)
        self.rtg_scale = rtg_scale
        self.index = []
        for ti, traj in enumerate(self.trajs):
            T = len(traj["rewards"])
            for start in range(T):
                self.index.append((ti, start))

    def __len__(self):
        return len(self.index)

    def __getitem__(self, idx):
        ti, start = self.index[idx]
        traj = self.trajs[ti]
        end = min(start + self.K, len(traj["rewards"]))

        states = torch.as_tensor(traj["states"][start:end], dtype=torch.float32)
        actions = torch.as_tensor(traj["actions"][start:end], dtype=torch.float32)
        rewards = torch.as_tensor(traj["rewards"][start:end], dtype=torch.float32)
        timesteps = torch.arange(start, end, dtype=torch.long)

        # Return-to-go for the full original trajectory.
        full_rewards = torch.as_tensor(traj["rewards"], dtype=torch.float32)
        rtg_full = torch.flip(torch.cumsum(torch.flip(full_rewards, dims=[0]), dim=0), dims=[0])
        rtg = rtg_full[start:end] / self.rtg_scale

        L = end - start
        pad = self.K - L
        states = (states - self.state_mean) / (self.state_std + 1e-6)

        if pad > 0:
            states = torch.cat([states, torch.zeros(pad, states.shape[-1])], dim=0)
            actions = torch.cat([actions, torch.zeros(pad, actions.shape[-1])], dim=0)
            rtg = torch.cat([rtg, torch.zeros(pad)], dim=0)
            timesteps = torch.cat([timesteps, torch.zeros(pad, dtype=torch.long)], dim=0)

        mask = torch.zeros(self.K, dtype=torch.float32)
        mask[:L] = 1.0
        return states, actions, rtg.unsqueeze(-1), timesteps, mask
\end{lstlisting}

\section{PyTorch implementation: Decision Transformer model}

Listing~\ref{lst:dt_model} gives a compact continuous-action Decision Transformer. It uses a standard transformer encoder with a causal mask. In production implementations, one often uses GPT-style blocks, layer normalization choices, dropout, and careful batching, but the mathematical structure is the same.

\Needspace{20\baselineskip}
\begin{lstlisting}[style=pythonstyle,caption={A compact continuous-action Decision Transformer.},label={lst:dt_model}]
import torch
import torch.nn as nn
import torch.nn.functional as F

class MiniDecisionTransformer(nn.Module):
    def __init__(self, state_dim, action_dim, hidden=128,
                 max_timestep=4096, n_layers=3, n_heads=4):
        super().__init__()
        self.hidden = hidden
        self.state_embed = nn.Linear(state_dim, hidden)
        self.action_embed = nn.Linear(action_dim, hidden)
        self.rtg_embed = nn.Linear(1, hidden)
        self.time_embed = nn.Embedding(max_timestep, hidden)

        layer = nn.TransformerEncoderLayer(
            d_model=hidden,
            nhead=n_heads,
            dim_feedforward=4 * hidden,
            dropout=0.1,
            batch_first=True,
            activation="gelu",
        )
        self.transformer = nn.TransformerEncoder(layer, num_layers=n_layers)
        self.action_head = nn.Sequential(
            nn.LayerNorm(hidden),
            nn.Linear(hidden, action_dim),
            nn.Tanh(),  # assumes actions are normalized to [-1, 1]
        )

    def forward(self, states, actions, rtg, timesteps, attention_mask=None):
        # states/actions: [B, K, dim], rtg: [B, K, 1], timesteps: [B, K]
        B, K, _ = states.shape
        time_emb = self.time_embed(timesteps)

        r_tok = self.rtg_embed(rtg) + time_emb
        s_tok = self.state_embed(states) + time_emb
        a_tok = self.action_embed(actions) + time_emb

        # Interleave tokens: (R_0, s_0, a_0, R_1, s_1, a_1, ...)
        tokens = torch.stack([r_tok, s_tok, a_tok], dim=2).reshape(B, 3 * K, self.hidden)

        causal_mask = torch.triu(torch.ones(3 * K, 3 * K, device=states.device), diagonal=1)
        causal_mask = causal_mask.masked_fill(causal_mask == 1, float("-inf"))

        if attention_mask is not None:
            key_padding = (attention_mask.repeat_interleave(3, dim=1) == 0)
        else:
            key_padding = None

        h = self.transformer(tokens, mask=causal_mask, src_key_padding_mask=key_padding)

        # Predict action from state-token positions.
        state_positions = h[:, 1::3, :]
        pred_actions = self.action_head(state_positions)
        return pred_actions
\end{lstlisting}

\Needspace{15\baselineskip}
\begin{lstlisting}[style=pythonstyle,caption={Decision Transformer training step for continuous actions.},label={lst:dt_training}]
def dt_training_step(model, batch, optimizer):
    states, actions, rtg, timesteps, mask = batch
    pred_actions = model(states, actions, rtg, timesteps, attention_mask=mask)

    # Only train on real, non-padding positions.
    loss_per_step = (pred_actions - actions).pow(2).mean(dim=-1)
    loss = (loss_per_step * mask).sum() / (mask.sum() + 1e-6)

    optimizer.zero_grad(set_to_none=True)
    loss.backward()
    torch.nn.utils.clip_grad_norm_(model.parameters(), 1.0)
    optimizer.step()
    return float(loss.detach())
\end{lstlisting}

\section{PyTorch implementation: autoregressive rollout}

At evaluation time, the policy must maintain a rolling context. The target return is updated after each reward. For UAV/SDN applications, a safety layer may be applied after the DT proposes an action.

\Needspace{18\baselineskip}
\begin{lstlisting}[style=pythonstyle,caption={Autoregressive Decision Transformer rollout.},label={lst:dt_rollout}]
@torch.no_grad()
def rollout_dt(env, model, target_return, state_mean, state_std,
               rtg_scale=1000.0, context_len=20, safety_filter=None):
    states, actions, rtgs, timesteps = [], [], [], []
    state, done = env.reset(), False
    total_return = 0.0
    t = 0

    while not done:
        states.append(torch.as_tensor(state, dtype=torch.float32))
        rtgs.append(torch.tensor([target_return / rtg_scale], dtype=torch.float32))
        timesteps.append(torch.tensor(t, dtype=torch.long))

        if len(actions) < len(states):
            # Placeholder previous action for current position.
            action_dim = env.action_space.shape[0]
            actions.append(torch.zeros(action_dim))

        s = torch.stack(states[-context_len:])
        a = torch.stack(actions[-context_len:])
        r = torch.stack(rtgs[-context_len:])
        ts = torch.stack(timesteps[-context_len:])

        # Left pad if context is short.
        pad = context_len - len(s)
        if pad > 0:
            s = torch.cat([torch.zeros(pad, s.shape[-1]), s], dim=0)
            a = torch.cat([torch.zeros(pad, a.shape[-1]), a], dim=0)
            r = torch.cat([torch.zeros(pad, 1), r], dim=0)
            ts = torch.cat([torch.zeros(pad, dtype=torch.long), ts], dim=0)
            mask = torch.cat([torch.zeros(pad), torch.ones(context_len - pad)]).unsqueeze(0)
        else:
            mask = torch.ones(1, context_len)

        s = ((s - state_mean) / (state_std + 1e-6)).unsqueeze(0)
        a = a.unsqueeze(0)
        r = r.unsqueeze(0)
        ts = ts.unsqueeze(0)

        pred = model(s, a, r, ts, attention_mask=mask)[0, -1]
        action = pred.cpu().numpy()
        if safety_filter is not None:
            action = safety_filter(state, action)

        next_state, reward, done, info = env.step(action)
        actions[-1] = torch.as_tensor(action, dtype=torch.float32)
        total_return += reward
        target_return -= reward
        state = next_state
        t += 1

    return total_return
\end{lstlisting}

\section{Trajectory Transformer and planning by sequence generation}

Decision Transformer predicts actions conditioned on desired return. Trajectory Transformer instead models trajectories more directly as a joint sequence over states, actions, and rewards, then uses search or planning in the learned sequence model \citep{janner2021trajectory}. The difference is subtle but important:

\begin{itemize}
	\item Decision Transformer is primarily a conditional policy model.
	\item Trajectory Transformer is closer to a generative model over future trajectories.
\end{itemize}

\begin{figure}[t]
	\centering
	\begin{tikzpicture}[
		box/.style={draw,rounded corners,thick,minimum width=3.2cm,minimum height=0.8cm,
			align=center,font=\small},
		arrow/.style={-{Latex[length=2.2mm]},thick},
		node distance=1.9cm
		]

		\node[box,fill=blue!8,draw=blue!70] (dt)
		{Decision Transformer\\conditional policy};

		\node[box,fill=green!8,draw=green!60!black,below=of dt] (tt)
		{Trajectory Transformer\\trajectory generator};

		\node[box,fill=orange!10,draw=orange!80!black,right=2.8cm of dt] (action)
		{predict next action};

		\node[box,fill=purple!8,draw=purple!70,right=2.8cm of tt] (plan)
		{search over futures};

		% Arrows with labels above/below the arrow, not overlapping boxes
		\draw[arrow,draw=blue!65]
		(dt) -- node[above,font=\scriptsize]{condition on return} (action);

		\draw[arrow,draw=green!60!black]
		(tt) -- node[below,font=\scriptsize,align=center]{model\\future sequence} (plan);

		% Vertical brace or note — placed below both right boxes
		\node[align=center,font=\small,below=0.5cm of plan]
		{Both use sequence modeling,\\but their control mechanisms differ.};

	\end{tikzpicture}
	\caption{Decision Transformer and Trajectory Transformer both use sequence modeling,
		but DT is used as a return-conditioned policy while Trajectory Transformer supports
		planning by generating or searching over future trajectories.}
	\label{fig:dt_vs_tt}
\end{figure}
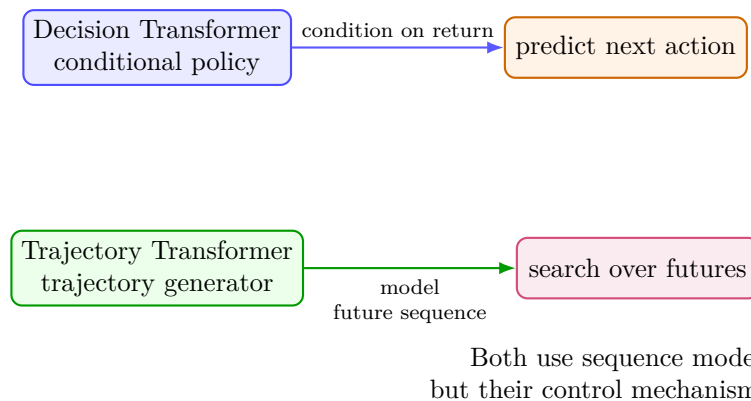

Unlike Decision Transformer, which conditions on a desired return and predicts only actions, Trajectory Transformer also predicts future states and rewards. This allows beam-search planning over candidate future trajectories at inference time, blurring the boundary between sequence modeling and model-based planning. In that sense, Trajectory Transformer is closer to a learned simulator plus planner, while Decision Transformer is closer to a conditional policy model.

This distinction connects Chapter~15 back to Chapter~12 and Chapter~13. Trajectory Transformer is a sequence-modeling cousin of model-based planning; Decision Transformer is a sequence-modeling cousin of conditional imitation learning.

\section{Reinforcement learning via supervised learning}

Decision Transformer belongs to a broader family sometimes called reinforcement learning via supervised learning (RvS) \citep{emmons2022rvs}. These methods ask when RL can be reduced to conditional supervised learning. Instead of fitting a Bellman equation, they train a policy conditioned on information such as goals, rewards, or returns.

\begin{table}[t]
	\centering
	\caption{Several ways to use supervised learning for offline control.}
	\label{tab:rvs_family}
	\begin{tabularx}{\textwidth}{p{3.0cm}p{4.1cm}X}
		\toprule
		Method family & Conditioning signal & Main idea \\
		\midrule
		Behavior cloning & none & imitate dataset actions \\
		RvS & goal or reward & condition policy on desired outcome \citep{emmons2022rvs} \\
		Decision Transformer & return-to-go and history & autoregressively predict actions \citep{chen2021decisiontransformer} \\
		Prompt-DT & trajectory prompt & adapt to new tasks from demonstrations \citep{xu2022prompting} \\
		GDT & generalized hindsight information & match future-dependent statistics \citep{furuta2022generalized} \\
		\bottomrule
	\end{tabularx}
\end{table}

The strength of RvS-style methods is simplicity. They use stable supervised objectives and avoid the deadly triad of off-policy bootstrapping. Their weakness is that they depend heavily on dataset quality. If the dataset does not contain high-return behavior, or if high-return behavior requires combining fragments across trajectories, pure sequence modeling may be insufficient.

The RvS perspective also connects naturally to language-model post-training. Techniques such as RLHF and DPO can be read as large-scale forms of conditioning on preferred outcomes, where the conditioning signal is a learned preference or reward rather than a scalar environment return \citep{ouyang2022training,rafailov2023dpo}. This connection is one reason Decision Transformer is not merely an offline-RL curiosity: it is part of a broader shift toward treating decision-making traces as sequences that can be modeled, prompted, and refined. 

\section{Trajectory stitching: the central limitation}

Trajectory stitching means combining good parts of different trajectories to produce a better policy than any single demonstrated trajectory. Value-based RL can sometimes do this naturally: if a transition leads to a valuable state, dynamic programming can propagate that value backward even if the full trajectory was not optimal.

Decision Transformer is weaker at stitching because it is trained to imitate sequences that actually occurred in the dataset. If no trajectory contains the complete desired combination, a return-conditioned model may not invent it reliably \citep{brandfonbrener2022when,yamagata2023qdt,wu2023elastic}.

Kumar et al. showed empirically on standard offline-RL benchmarks that conventional value-based offline RL can outperform Decision Transformer when trajectory stitching is required \citep{kumar2022when}. This does not mean Decision Transformer is weak in general; rather, it clarifies the regime where Bellman-style value propagation remains especially useful.

\begin{figure}[t]
	\centering
	\begin{tikzpicture}[
		state/.style={circle,draw,thick,minimum size=0.75cm,align=center,font=\small},
		arrow/.style={-{Latex[length=2.2mm]},thick},
		dashedarrow/.style={-{Latex[length=2.2mm]},thick,dashed},
		node distance=0.75cm
		]
		\node[state,fill=blue!8,draw=blue!70] (a0) at (0,1.2) {$A$};
		\node[state,fill=blue!8,draw=blue!70] (b0) at (1.5,1.2) {$B$};
		\node[state,fill=blue!8,draw=blue!70] (c0) at (3.0,1.2) {$C$};
		\node[state,fill=green!8,draw=green!60!black] (a1) at (0,-1.0) {$A$};
		\node[state,fill=green!8,draw=green!60!black] (d1) at (1.5,-1.0) {$D$};
		\node[state,fill=green!8,draw=green!60!black] (e1) at (3.0,-1.0) {$E$};
		\node[state,fill=red!8,draw=red!70!black] (goal) at (4.8,0.1) {$G$};

		\draw[arrow,draw=blue!65] (a0) -- (b0);
		\draw[arrow,draw=blue!65] (b0) -- (c0);
		\draw[arrow,draw=green!60!black] (a1) -- (d1);
		\draw[arrow,draw=green!60!black] (d1) -- (e1);
		\draw[arrow,draw=gray!60] (c0) -- (goal);
		\draw[arrow,draw=gray!60] (e1) -- (goal);
		\draw[dashedarrow,draw=red!70!black] (b0) -- node[right,font=\scriptsize,text=red!70!black] {useful stitched path} (d1);
		\node[align=center,font=\small] at (2.4,-2.0) {Value methods may stitch good transitions.\\Pure imitation-style sequence models may not.};
	\end{tikzpicture}
	\caption{Trajectory stitching. Two suboptimal dataset trajectories may contain useful segments. Dynamic-programming methods can propagate value across segments, while pure return-conditioned imitation may struggle if the stitched path never appeared.}
	\label{fig:trajectory_stitching}
\end{figure}
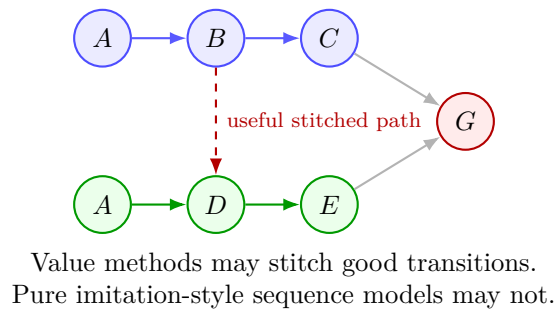

\section{QDT, Elastic DT, and stitching-aware variants}

Q-learning Decision Transformer (QDT) addresses stitching by using dynamic-programming information to relabel returns-to-go before training a DT-style model \citep{yamagata2023qdt}. The key idea is to let a value-learning method infer better future values, then use those values as improved conditioning targets for sequence modeling.

This return-relabeling step is structurally analogous to the reanalyse mechanism in MuZero Unplugged from Chapter~13: both reuse stored trajectories by attaching fresh, network-improved supervision targets. In QDT, the improved target is a relabeled return-to-go; in MuZero-style reanalyse, it is often a search-improved policy or value target. 

Elastic Decision Transformer (EDT) attacks the stitching problem from another angle. It adjusts the effective history length during action inference, allowing the model to use different amounts of context depending on the decision situation \citep{wu2023elastic}. The motivation is that too much trajectory history can anchor the model to a suboptimal past, while too little history may remove useful context.

\begin{table}[t]
	\centering
	\caption{Decision-Transformer variants and what they try to fix.}
	\label{tab:dt_variants}
	\begin{tabularx}{\textwidth}{p{3.0cm}p{4.2cm}X}
		\toprule
		Variant & Main change & Problem addressed \\
		\midrule
		DT & condition on return-to-go & simple offline sequence policy \citep{chen2021decisiontransformer} \\
		Online DT & offline pretraining plus online finetuning & adapts after deployment \citep{zheng2022online_dt} \\
		QDT & relabel returns using Q-learning & improves stitching \citep{yamagata2023qdt} \\
		EDT & elastic history length & reduces harmful context anchoring \citep{wu2023elastic} \\
		Prompt-DT & demonstration prompt & few-shot policy generalization \citep{xu2022prompting} \\
		Multi-game DT & one model across games & scaling and multitask transfer \citep{lee2022multigame} \\
		\bottomrule
	\end{tabularx}
\end{table}

A useful way to read this line of work is: Decision Transformer made offline RL look like sequence modeling; later methods reintroduced pieces of RL that sequence modeling alone was missing—online adaptation, value relabeling, trajectory stitching, prompting, and target-return selection.

\section{UAV/SDN scenario: sequence modeling for network control}

For this book, the most important nonstandard example is UAV/SDN control. Suppose we have historical logs from an SDN-assisted UAV network. Each timestep contains a state vector:
\begin{equation}
	s_t = [x_t,y_t,z_t,b_t,\mathrm{SINR}_t,\mathrm{latency}_t,\mathrm{load}_t,\mathrm{violations}_t,\ldots],
\end{equation}
and an action vector:
\begin{equation}
	a_t = [\Delta x_t,\Delta y_t,\Delta z_t,\rho_t^{\mathrm{URLLC}},\rho_t^{\mathrm{eMBB}},\rho_t^{\mathrm{mMTC}},p_t].
\end{equation}
The reward may combine QoS, energy, and safety:
\begin{equation}
	r_t = w_q R_t^{\mathrm{QoS}} - w_e C_t^{\mathrm{energy}} - w_s C_t^{\mathrm{safety}}.
\end{equation}
A Decision Transformer can condition on a desired network objective, such as high cumulative QoS reward under low safety violations.

\begin{figure}[t]
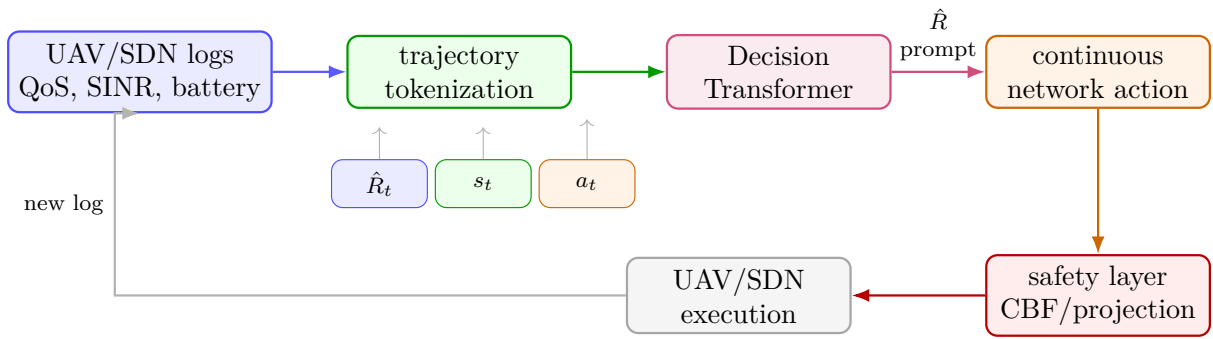

	\centering
	\resizebox{\columnwidth}{!}{%
%
	}
	\caption{A Decision Transformer scenario for UAV/SDN control. Historical network
		logs are converted into return-conditioned trajectory sequences. The model proposes
		continuous actions, but a safety layer should still check or project actions before
		execution.}
	\label{fig:uav_sdn_dt_scenario}
\end{figure}

\subsection{Concrete numerical example}

Suppose three historical UAV trajectory windows have the following normalized return-to-go values and outcomes:

\begin{table}[t]
	\centering
	\caption{Illustrative UAV/SDN return-conditioned sequence examples.}
	\label{tab:uav_dt_numerical}
	\small
	\begin{tabularx}{\textwidth}{p{1.2cm}p{1.8cm}p{2.3cm}p{2.2cm}p{1.7cm}X}
		\toprule
		Window & Target RTG & QoS satisfaction & Safety violations & Energy cost & Action pattern \\
		\midrule
		A & 0.35 & 61\% & 0.02 & low & conservative hover + serve \\
		B & 0.72 & 82\% & 0.05 & medium & move to hotspot + allocate URLLC \\
		C & 0.90 & 88\% & 0.22 & high & aggressive movement + high power \\
		\bottomrule
	\end{tabularx}
\end{table}

A naive deployment might prompt the model with RTG \(0.90\) and obtain aggressive behavior. A safer operator may prompt with RTG \(0.72\), then rely on a CBF/projection layer to prevent constraint violations. This example shows why target return is a control knob, not a magic guarantee. A high target return can also push the model toward rare or unsafe regions of the dataset.

\section{Practical implementation details}

Decision Transformers are easy to describe but sensitive to details.

\begin{table}[t]
	\centering
	\caption{Implementation details that matter for Decision Transformers.}
	\label{tab:dt_practical_details}
	\begin{tabularx}{\textwidth}{p{3.3cm}p{4.8cm}X}
		\toprule
		Detail & Why it matters & Practical rule \\
		\midrule
		State normalization & stabilizes transformer input scale & compute mean/std from training data \\
		Return scale & prevents return token domination & tune \(c_R\) per domain \\
		Context length & controls memory and compute & longer is not always better \\
		Padding mask & prevents fake tokens from training model & always mask padded timesteps \\
		Action normalization & makes MSE meaningful & scale continuous actions to \([-1,1]\) \\
		Target return & controls deployment behavior & avoid prompts far beyond dataset support \\
		Safety layer & DT has no hard safety guarantee & use projection/CBF when constraints matter \\
		\bottomrule
	\end{tabularx}
\end{table}

\begin{warningbox}{High target return is not a safety certificate}
	Conditioning on a high return does not prove that the selected action is safe or even feasible. If high-return trajectories are rare, noisy, or unsafe, the model may imitate undesirable behavior. In safety-critical networked systems, target-return conditioning should be combined with validation, uncertainty checks, and safety filters.
\end{warningbox}

\Needspace{16\baselineskip}
\begin{lstlisting}[style=pythonstyle,caption={UAV/SDN state and action sequence construction.},label={lst:uav_dt_builder}]
def build_uav_dt_trajectory(log_rows, reward_fn):
    """Convert network-control logs into a DT trajectory.

    Each row may contain UAV position, battery, QoS, SINR, latency,
    load, previous action, and safety indicators.
    """
    states, actions, rewards = [], [], []
    for row in log_rows:
        state = [
            row["x"], row["y"], row["z"], row["battery"],
            row["sinr"], row["latency_ms"], row["traffic_load"],
            row["qos_satisfaction"], row["safety_violation"],
        ]
        action = [
            row["dx"], row["dy"], row["dz"],
            row["bw_urlcc"], row["bw_embb"], row["bw_mmtc"],
            row["tx_power"],
        ]
        reward = reward_fn(row)
        states.append(state)
        actions.append(action)
        rewards.append(reward)
    return {"states": states, "actions": actions, "rewards": rewards}
\end{lstlisting}

\section[Frontiers in Decision Transformers]{Frontiers: online DTs, multi-game DTs, prompts, and 2026 directions}

The first Decision Transformer results were offline. But the field quickly moved toward online finetuning, multitask policies, prompt-based adaptation, and target-return selection.

Online Decision Transformer blends offline sequence-model pretraining with online finetuning \citep{zheng2022online_dt}. Multi-Game Decision Transformer shows that one transformer can be trained across many Atari games, connecting RL sequence models to the scaling behavior seen in language and vision \citep{lee2022multigame}. Prompt-DT uses demonstration prompts for few-shot policy generalization \citep{xu2022prompting}. More recent work studies target-return optimization and pure RL finetuning of Decision Transformers, reflecting the open question of how sequence-model policies should be adapted after deployment \citep{tatematsu2025mtro,luo2026dtfinetuning}.

A broader 2025--2026 perspective is that modern reasoning models can be viewed as massive sequence-decision systems. Models such as OpenAI o1 and DeepSeek-R1 operate over token trajectories, receive outcome-based or verifier-based supervision, and improve through a mixture of supervised sequence modeling and reinforcement-learning post-training \citep{jaech2024openai,deepseek2025r1}. The Decision Transformer and RvS lens helps explain this hybrid structure: supervised pretraining teaches the model to imitate useful traces, while RL-style post-training changes which traces are preferred.

\begin{table}[t]
	\centering
	\caption{Frontier directions for Decision Transformers and sequence-model control.}
	\label{tab:dt_frontiers}
	\begin{tabularx}{\textwidth}{p{3.4cm}p{4.8cm}X}
		\toprule
		Direction & Motivation & Open issue \\
		\midrule
		Online DT & adapt after offline pretraining & balancing supervised relabeling and RL gradients \\
		Multi-game DT & scale across tasks & target return calibration per task \\
		Prompt-DT & few-shot adaptation & prompt selection and OOD generalization \\
		QDT / EDT & improve stitching & combine sequence modeling with value reasoning \\
		Trajectory generators & model behaviors or futures & planning cost and safety validation \\
		UAV/SDN sequence policies & use real operator logs & nonstationary networks and safety constraints \\
		\bottomrule
	\end{tabularx}
\end{table}

\section{Failure modes and debugging}

\begin{table}[t]
	\centering
	\caption{Common Decision Transformer failure modes.}
	\label{tab:dt_debugging}
	\begin{tabularx}{\textwidth}{p{3.2cm}p{4.6cm}X}
		\toprule
		Symptom & Likely cause & What to inspect \\
		\midrule
		Ignores target return & return scale too large or dataset has weak return diversity & RTG normalization, return histogram \\
		Performs like BC & target return not informative & compare to behavior cloning baseline \\
		Unstable outputs & action normalization missing & action mean/std, tanh scaling \\
		Bad long-horizon behavior & context too short or poor dataset coverage & context length, trajectory quality \\
		Unsafe high-return actions & high-return logs include violations & reward design, safety labels, CBF layer \\
		Poor deployment despite low training loss & covariate shift after first mistakes & closed-loop evaluation, not just action MSE \\
		Cannot stitch skills & pure imitation objective & QDT, EDT, or value-based offline RL \\
		\bottomrule
	\end{tabularx}
\end{table}

\section{Limitations}

Decision Transformers are not a replacement for all of offline RL. Their limitations are important.

\begin{enumerate}[leftmargin=*]
	\item They depend heavily on dataset coverage and quality.
	\item Target returns must be chosen carefully.
	\item They may struggle with trajectory stitching.
	\item They can be expensive for long contexts because transformer attention scales quadratically with context length. For long UAV or network-telemetry logs, such as 24-hour operational windows, context length and compute become first-order design constraints.
	\item They optimize a supervised prediction loss, not a Bellman consistency objective.
	\item Closed-loop errors can accumulate even when offline action prediction loss is low.
	\item They do not provide hard safety guarantees.
	\item They may require large datasets and careful normalization to exploit transformer capacity.
\end{enumerate}

\section{Exercises}

\subsection*{Conceptual exercises}
\begin{enumerate}[leftmargin=*]
	\item Explain why Decision Transformer can be viewed as conditional behavior cloning rather than value-based RL.
	\item Why can a target return that is too high produce poor or unsafe behavior?
	\item Explain the trajectory stitching limitation. Why can value-based methods sometimes stitch better than sequence models?
	\item Compare Decision Transformer and Trajectory Transformer. Which one is closer to a policy model, and which one is closer to a planning model?
	\item In UAV/SDN control, what operational meaning could the desired return prompt have?
\end{enumerate}

\subsection*{Mathematical exercises}
\begin{enumerate}[leftmargin=*]
	\item Given rewards \([2, -1, 3, 4]\), compute the undiscounted return-to-go at each timestep.
	\item Write the supervised MSE loss for a continuous-action Decision Transformer over a batch with padding masks.
	\item Suppose a target return is updated by \(\hat R_{t+1}=\hat R_t-r_{t+1}\). If \(\hat R_0=100\) and rewards are \([10,5,-2]\), compute the remaining target returns.
	\item Show how a discrete-action Decision Transformer loss becomes cross-entropy over action tokens.
\end{enumerate}

\subsection*{Coding exercises}
\begin{enumerate}[leftmargin=*]
	\item Modify Listing~\ref{lst:dt_dataset} to use discounted returns-to-go.
	\item Add a categorical action head to Listing~\ref{lst:dt_model} for discrete action spaces.
	\item Implement an evaluation routine that tests three target returns and plots achieved return versus target return.
	\item Add a safety filter to Listing~\ref{lst:dt_rollout} that clips UAV movement and bandwidth allocation to feasible ranges.
\end{enumerate}

\subsection*{Research exercises}
\begin{enumerate}[leftmargin=*]
	\item Design a Decision Transformer dataset for SD-WAN routing logs. What are the state tokens, action tokens, and reward signal?
	\item Propose a method for choosing target return automatically in a nonstationary UAV network.
	\item Compare QDT and CQL for a dataset containing mostly suboptimal network-controller trajectories.
	\item How could Prompt-DT be used to adapt a UAV controller to a new city using only a few demonstration trajectories?
\end{enumerate}

\section*{Looking Ahead to Chapter 16: Food for Thought}
\addcontentsline{toc}{section}{Looking Ahead to Chapter 16: Food for Thought}

This chapter treated a trajectory as a sequence generated by one agent. The next part of the book changes the setting: what happens when many agents act at once?

In multi-agent reinforcement learning, each agent's experience depends on the policies of other agents. For UAV networks, this means that one UAV's movement affects interference, coverage, user assignment, and the rewards of other UAVs. The trajectory is no longer only a sequence; it is a coupled set of interacting sequences.

Questions to carry forward:
\begin{enumerate}[leftmargin=*]
	\item If each agent has its own trajectory, how should a centralized learner represent the joint history?
	\item Can transformers help model multi-agent interaction graphs?
	\item When should agents share policies, critics, or communication messages?
	\item How does credit assignment change when a team reward depends on many agents?
	\item How can safety filters remain decentralized while training uses centralized information?
\end{enumerate}

\begin{quote}
	Chapter~15 showed how a single trajectory can become a sequence model. Chapter~16 asks how learning changes when many trajectories interact.
\end{quote}
	\part{Multi-Agent, Hierarchical, and Safe DRL}
\chapter[Multi-Agent Reinforcement Learning]{Multi-Agent Reinforcement Learning}
\label{ch:marl}
\chaptermark{Multi-Agent RL}

\begin{keybox}{Chapter goal}
	Multi-agent reinforcement learning extends the single-agent framework to systems where many agents learn, coordinate, and act simultaneously. This chapter develops the formal foundations of Markov games and Dec-POMDPs, the main design principles of cooperative MARL—centralized training with decentralized execution, value decomposition, and multi-agent actor-critic methods—graph-based communication, and the UAV/SDN application of cooperative multi-UAV control.
\end{keybox}

\section*{Chapter Overview}
\addcontentsline{toc}{section}{Chapter Overview}

\begin{enumerate}[leftmargin=*]
	\item Why multi-agent reinforcement learning matters
	\item From MDPs to Markov games and Dec-POMDPs
	\item What makes MARL hard: non-stationarity, credit, coordination, and scale
	\item Independent learning and why it is both useful and dangerous
	\item Centralized training with decentralized execution
	\item Value decomposition: VDN, QMIX, QTRAN, and QPLEX
	\item Multi-agent actor-critic: MADDPG, COMA, MAPPO, and HAPPO
	\item Communication, attention, and graph-based coordination
	\item Sequence modeling and transformer-based MARL
	\item Mean-field, population, and heterogeneous-agent MARL
	\item Multi-agent replay, fingerprints, and recurrent policies
	\item UAV/SDN worked scenario: safe cooperative multi-UAV control
	\item Python implementation patterns
	\item Evaluation, benchmarks, and reproducibility
	\item Failure modes and debugging checklist
	\item Limitations and when not to use MARL
	\item Exercises
	\item Looking Ahead to Chapter 17
\end{enumerate}

\section{Why multi-agent reinforcement learning matters}

Most chapters so far studied a single agent. The agent observed a state, chose an action, received a reward, and learned from the consequences. That picture is already difficult, but many real systems are not single-agent systems. A wireless network contains multiple UAVs, base stations, users, controllers, and edge servers. A traffic system contains many vehicles and intersections. A swarm-robotics system contains many robots with partial observations. A market contains many adaptive participants. A language-agent workflow may contain multiple specialized agents that plan, critique, verify, and execute.

Multi-agent reinforcement learning (MARL) studies learning when several agents act in the same environment. This simple change makes the problem qualitatively different. The environment is no longer merely stochastic; it is partly shaped by other learning agents. If one agent changes its policy, the effective transition and reward distributions seen by the others also change. This creates moving targets, coordination failures, credit-assignment problems, communication bottlenecks, and new safety risks.

\begin{keybox}{Core idea}
	Single-agent RL asks: \emph{what should one agent do?} Multi-agent RL asks: \emph{how should several agents learn, coordinate, and remain safe while each agent's behavior changes the learning problem faced by the others?}
\end{keybox}

MARL is not only about making many copies of a single-agent algorithm. Independent DQN or independent PPO can be surprisingly strong baselines, especially when agents share parameters and operate in symmetric roles. But as soon as agents have different observations, different roles, different constraints, or coupled rewards, more structure is needed. The central design question becomes: how much global information should be used during training, and how much information can each agent rely on during execution?

This question leads to one of the main ideas of the chapter: centralized training with decentralized execution (CTDE). During training, a critic, mixer, or coordinator may access global state, joint actions, or other agents' trajectories. During execution, each agent must act using only its local observation, local memory, and possibly limited communication. CTDE is one of the most important principles in cooperative MARL and appears in MADDPG, COMA, VDN, QMIX, MAPPO, MAT, and many modern variants \citep{lowe2017maddpg,foerster2018coma,sunehag2018vdn,rashid2018qmix,yu2022mappo,wen2022mat}.

\begin{figure}[t]
	\centering
	\begin{tikzpicture}[
		box/.style={draw,rounded corners,thick,minimum width=3.0cm,minimum height=0.85cm,align=center,font=\small},
		arrow/.style={-{Latex[length=2.3mm]},thick},
		node distance=1.5cm
		]
		\node[box,fill=blue!8,draw=blue!70] (single) {Single-agent RL\\one policy, one learner};
		\node[box,fill=orange!10,draw=orange!80!black,right=of single] (many) {Multi-agent RL\\many changing policies};
		\node[box,fill=green!8,draw=green!60!black,right=of many] (ctde) {CTDE\\global training, local execution};
		\draw[arrow,draw=blue!70] (single) -- node[above,font=\scriptsize] {add agents} (many);
		\draw[arrow,draw=orange!80!black] (many) -- node[above,font=\scriptsize,align=center] {stabilize\\learning} (ctde);
		\node[below=0.8cm of many,align=center,font=\small] {new difficulties: non-stationarity, coordination, credit assignment, communication, safety};
	\end{tikzpicture}
	\caption{The transition from single-agent RL to multi-agent RL. Adding agents creates new learning difficulties because each agent's policy changes the environment faced by the others. CTDE is a common response: use global information while training, but preserve decentralized execution.}
	\label{fig:marl_transition}
\end{figure}

\section{From MDPs to Markov games and Dec-POMDPs}

A single-agent MDP contains states, actions, rewards, transition probabilities, and a discount factor. MARL generalizes this setting by giving the environment multiple agents. A common formal model is the Markov game, also called a stochastic game, whose roots go back to Shapley's stochastic games and Littman's Markov-game formulation for RL \citep{shapley1953stochastic,littman1994markov}. For $N$ agents, a Markov game can be written as
\begin{equation}
	\mathcal{G}
	=
	\left(
	\mathcal{S},
	\{\mathcal{A}_i\}_{i=1}^{N},
	P,
	\{r_i\}_{i=1}^{N},
	\gamma
	\right),
\end{equation}
where $\mathcal{S}$ is the state space, $\mathcal{A}_i$ is the action space of agent $i$, $P(s'\given s,a_1,\ldots,a_N)$ is the transition function, $r_i$ is the reward of agent $i$, and $\gamma$ is the discount factor.

The joint action is
\begin{equation}
	\mathbf{a}
	=
	(a_1,a_2,\ldots,a_N),
\end{equation}
and the joint policy may factorize as
\begin{equation}
	\boldsymbol{\pi}(\mathbf{a}\given s)
	=
	\prod_{i=1}^{N}\pi_i(a_i\given o_i),
\end{equation}
where $o_i$ is agent $i$'s local observation. This factorization is central to decentralized execution: each agent acts locally, even if the joint behavior is evaluated globally.

In fully cooperative settings, all agents share the same team reward:
\begin{equation}
	r_1(s,\mathbf{a}) = r_2(s,\mathbf{a}) = \cdots = r_N(s,\mathbf{a}) = r(s,\mathbf{a}).
\end{equation}
In mixed or competitive settings, rewards differ. Competitive MARL includes zero-sum games, while mixed settings include cooperation and competition simultaneously.

Many real systems are partially observable. A decentralized partially observable Markov decision process (Dec-POMDP) models a cooperative multi-agent task where agents receive local observations and must coordinate without seeing the full state \citep{oliehoek2016dec}. In a Dec-POMDP, the global state may be available to a simulator or centralized trainer, but not to individual agents at execution time.

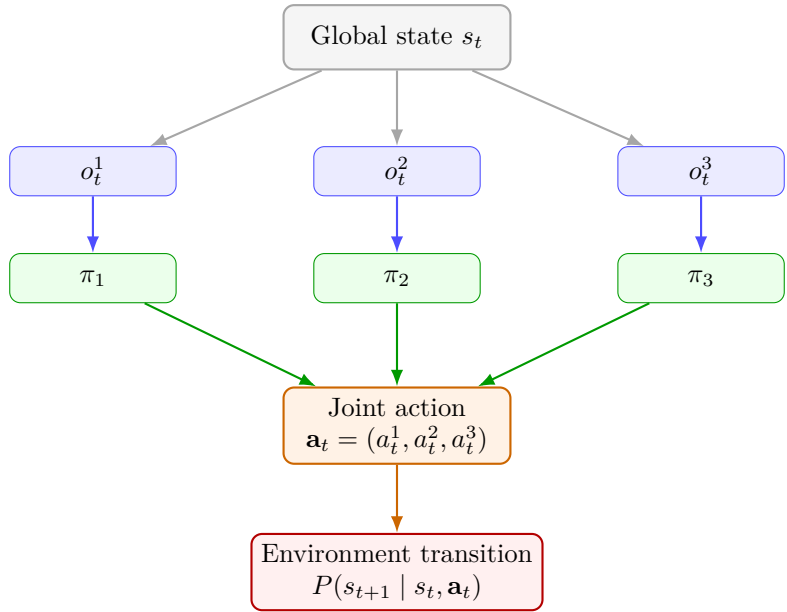
\begin{figure}[t]
	\centering
	\begin{tikzpicture}[
		box/.style={draw,rounded corners,thick,minimum width=3.0cm,minimum height=0.85cm,align=center,font=\small},
		small/.style={draw,rounded corners,minimum width=2.2cm,minimum height=0.65cm,align=center,font=\small},
		arrow/.style={-{Latex[length=2.2mm]},thick},
		node distance=0.8cm
		]
		\node[box,fill=gray!8,draw=gray!70] (state) {Global state $s_t$};
		\node[small,fill=blue!8,draw=blue!70,below left=1.0cm and 1.4cm of state] (o1) {$o_t^1$};
		\node[small,fill=blue!8,draw=blue!70,below=1.0cm of state] (o2) {$o_t^2$};
		\node[small,fill=blue!8,draw=blue!70,below right=1.0cm and 1.4cm of state] (o3) {$o_t^3$};
		\node[small,fill=green!8,draw=green!60!black,below=0.75cm of o1] (p1) {$\pi_1$};
		\node[small,fill=green!8,draw=green!60!black,below=0.75cm of o2] (p2) {$\pi_2$};
		\node[small,fill=green!8,draw=green!60!black,below=0.75cm of o3] (p3) {$\pi_3$};
		\node[box,fill=orange!10,draw=orange!80!black,below=1.1cm of p2] (joint) {Joint action\\$\mathbf{a}_t=(a_t^1,a_t^2,a_t^3)$};
		\node[box,fill=red!6,draw=red!70!black,below=0.9cm of joint] (env) {Environment transition\\$P(s_{t+1}\given s_t,\mathbf{a}_t)$};
		\draw[arrow,draw=gray!70] (state) -- (o1);
		\draw[arrow,draw=gray!70] (state) -- (o2);
		\draw[arrow,draw=gray!70] (state) -- (o3);
		\draw[arrow,draw=blue!70] (o1) -- (p1);
		\draw[arrow,draw=blue!70] (o2) -- (p2);
		\draw[arrow,draw=blue!70] (o3) -- (p3);
		\draw[arrow,draw=green!60!black] (p1) -- (joint);
		\draw[arrow,draw=green!60!black] (p2) -- (joint);
		\draw[arrow,draw=green!60!black] (p3) -- (joint);
		\draw[arrow,draw=orange!80!black] (joint) -- (env);
	\end{tikzpicture}
	\caption{A Dec-POMDP view of cooperative MARL. The environment has a global state, but each agent acts using only its local observation. The joint action determines the next state and team reward.}
	\label{fig:decpomdp_structure}
\end{figure}

\section{What makes MARL hard}

MARL is difficult because the learning problem combines all the difficulties of single-agent RL with several new ones.

\subsection{Non-stationarity}

In single-agent RL, a fixed environment transition model $P(s'\given s,a)$ is usually assumed. In MARL, from the perspective of agent $i$, the transition distribution depends on the policies of the other agents:
\begin{equation}
	P_i(s'\given s,a_i)
	=
	\sum_{a_{-i}} P(s'\given s,a_i,a_{-i})\prod_{j\neq i}\pi_j(a_j\given o_j).
\end{equation}
If the other agents update their policies, then $P_i$ changes. Thus, each agent sees a non-stationary learning problem.

\subsection{Credit assignment}

In cooperative MARL, the team may receive a single reward, but many agents contributed to it. If three UAVs jointly improve QoS, which UAV deserves credit? If one UAV causes interference, which policy update should be penalized? This is the multi-agent credit-assignment problem.

\subsection{Coordination and relative overgeneralization}

Some tasks require agents to coordinate on compatible actions. Independent learners may find locally safe but globally poor behavior. For example, two UAVs may both choose to serve the same hotspot, causing congestion, while another cell remains uncovered. This is not necessarily because either UAV is irrational; it may be because the reward signal is insufficiently informative about the joint action.

\subsection{Scalability}

The joint action space grows exponentially:
\begin{equation}
	|\mathcal{A}_{\mathrm{joint}}|
	=
	\prod_{i=1}^{N}|\mathcal{A}_i|.
\end{equation}
If each of $N$ agents has $m$ actions, the joint action space has $m^N$ actions. With 10 agents and 5 actions each, this is $5^{10}=9{,}765{,}625$ joint actions. This is why naive joint-action Q-learning is rarely practical.

\begin{warningbox}{Why MARL is not just single-agent RL with a bigger action space}
	Treating all agents as one giant centralized agent gives a clean mathematical model, but it destroys scalability and decentralized execution. The action space grows exponentially, observations become high-dimensional, and the learned policy may require information that individual agents cannot access at runtime.
\end{warningbox}

\section{Independent learning: useful but dangerous}

The simplest MARL method is independent learning. Each agent runs a single-agent RL algorithm using its own observations and rewards, while treating other agents as part of the environment. Independent Q-learning (IQL) and independent PPO (IPPO) are examples.

Independent learning is attractive because it is simple, scalable, and easy to implement. It often works surprisingly well when agents are homogeneous, share parameters, and receive dense rewards. Empirical studies have shown that independent learners can achieve strong behavior in some cooperative and competitive domains, especially when stabilizing devices such as parameter sharing, fingerprints, or recurrent policies are used \citep{tampuu2017multiagent,foerster2017stabilising}. But it ignores non-stationarity and coordination structure.

For agent $i$, independent Q-learning updates
\begin{equation}
	Q_i(o_i,a_i)
	\leftarrow
	Q_i(o_i,a_i)
	+
	\alpha
	\left[
	r_i
	+
	\gamma\max_{a_i'}Q_i(o_i',a_i')
	-
	Q_i(o_i,a_i)
	\right].
\end{equation}
The update looks familiar, but the hidden assumption is false: the environment is not stationary from the agent's perspective.

\begin{figure}[t]
	\centering
	\begin{tikzpicture}[
		box/.style={draw,rounded corners,thick,minimum width=3.1cm,minimum height=0.85cm,align=center,font=\small},
		arrow/.style={-{Latex[length=2.2mm]},thick},
		node distance=0.9cm
		]
		\node[box,fill=green!8,draw=green!60!black] (agent1) {Agent 1\\updates $\pi_1$};
		\node[box,fill=blue!8,draw=blue!70,right=of agent1] (agent2) {Agent 2\\updates $\pi_2$};
		\node[box,fill=orange!10,draw=orange!80!black,right=of agent2] (agent3) {Agent 3\\updates $\pi_3$};
		\node[box,fill=red!6,draw=red!70!black,below=1.2cm of agent2] (effective) {Each agent sees a moving environment};
		\draw[arrow,draw=green!60!black] (agent1) -- (effective);
		\draw[arrow,draw=blue!70] (agent2) -- (effective);
		\draw[arrow,draw=orange!80!black] (agent3) -- (effective);
		\node[below=0.8cm of effective,align=center,font=\small] {The learning target changes because the other agents' policies change.};
	\end{tikzpicture}
	\caption{Independent learners face non-stationarity. Each agent updates its own policy, thereby changing the effective environment experienced by the others.}
	\label{fig:independent_nonstationarity}
\end{figure}

\begin{table}[t]
	\centering
	\caption{Independent learning in MARL: why it is still used and why it fails.}
	\label{tab:independent_learning}
	\begin{tabular}{p{0.25\textwidth}p{0.32\textwidth}p{0.32\textwidth}}
		\toprule
		Property & Advantage & Risk \\
		\midrule
		Simple implementation & Reuses single-agent DQN/PPO/SAC code & Ignores non-stationarity \\
		Scalable with agent count & No centralized critic needed & Poor coordination in tightly coupled tasks \\
		Local observations & Matches decentralized execution & Learns from incomplete information \\
		Parameter sharing & Efficient for homogeneous agents & Can fail for heterogeneous roles \\
		\bottomrule
	\end{tabular}
\end{table}

\section{Centralized training with decentralized execution}

CTDE is the dominant cooperative MARL paradigm. It uses more information during training than during execution. The centralized trainer may observe global state, joint actions, or other agents' hidden states. But during execution, each agent must use only its local observation and possibly allowed communication.

Formally, a decentralized policy has the form
\begin{equation}
	a_i \sim \pi_i(\cdot\given o_i),
\end{equation}
while a centralized critic may evaluate
\begin{equation}
	Q_i(s,a_1,\ldots,a_N)
	\quad\text{or}\quad
	V(s).
\end{equation}
The critic is removed or ignored at execution time.

\begin{figure}[t]
	\centering
	\resizebox{\columnwidth}{!}{%
		\begin{tikzpicture}[
			box/.style={draw,rounded corners,thick,minimum width=3.2cm,minimum height=0.85cm,
				align=center,font=\small},
			small/.style={draw,rounded corners,minimum width=2.4cm,minimum height=0.65cm,
				align=center,font=\small},
			arrow/.style={-{Latex[length=2.2mm]},thick},
			node distance=1.9cm
			]

			% Top row
			\node[box,fill=blue!8,draw=blue!70]                          (global)
			{Training data\\$s,o_1,\ldots,o_N,\mathbf{a},r$};
			\node[box,fill=purple!8,draw=purple!70,right=1.6cm of global](critic)
			{Centralized critic\\or mixer};
			\node[box,fill=green!8,draw=green!60!black,right=1.6cm of critic](policies)
			{Decentralized policies\\$\pi_i(a_i\mid o_i)$};

			% Execution box below policies
			\node[small,fill=orange!10,draw=orange!80!black,
			below=1.3cm of policies]                               (exec)
			{Execution\\local observations only};

			% Arrows
			\draw[arrow,draw=blue!70]        (global)   -- (critic);
			\draw[arrow,draw=purple!70]      (critic)   --
			node[above,font=\scriptsize,align=center]{learning\\signal} (policies);
			\draw[arrow,draw=green!60!black] (policies) -- (exec);

			% Note — below critic, well clear of exec box
			\node[align=center,font=\small,below=1.3cm of critic]
			{The critic can use information\\unavailable to agents at runtime.};

		\end{tikzpicture}%
	}
	\caption{Centralized training with decentralized execution. The training process may
		use global state and joint actions, while each deployed agent acts using only local
		information.}
	\label{fig:ctde}
\end{figure}
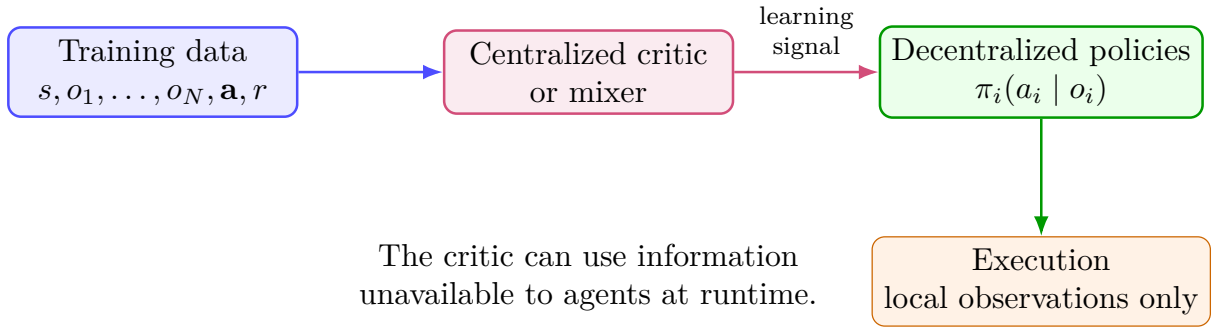

CTDE is not one algorithm. It is a design principle. MADDPG uses centralized critics and decentralized deterministic actors \citep{lowe2017maddpg}. COMA uses a centralized critic with a counterfactual baseline \citep{foerster2018coma}. VDN and QMIX use centralized value mixing for decentralized action-value functions \citep{sunehag2018vdn,rashid2018qmix}. MAPPO uses PPO-style decentralized actors with centralized value functions and implementation details that make PPO surprisingly competitive in cooperative MARL \citep{yu2022mappo}.
\begin{researchbox}{Comparative MARL evidence from UAV-assisted 5G slicing}
A concrete example of CTDE in UAV-assisted communication systems is our comparative study of MAPPO, MADDPG, and MADQN for multi-UAV 5G network slicing. The study compares an on-policy centralized-critic method (MAPPO), a deterministic multi-agent actor-critic method (MADDPG), and a value-based baseline (MADQN) under heterogeneous QoS and energy objectives. Its main lesson is not that one algorithm dominates in every scenario, but that algorithm suitability depends on action structure, environment openness, and the trade-off between QoS, stability, and energy consumption. This kind of evidence is especially useful because it grounds the CTDE idea in a realistic cyber-physical networking system rather than only in game-style benchmarks.
\end{researchbox}

\section{Value decomposition: VDN, QMIX, QTRAN, and QPLEX}

Value-decomposition methods are designed for cooperative tasks with a shared team reward. The goal is to learn decentralized action-value functions $Q_i(o_i,a_i)$ while using a centralized training signal for the team value $Q_{\mathrm{tot}}$.

\subsection{VDN}

Value-Decomposition Networks (VDN) assume that the team action-value function decomposes additively \citep{sunehag2018vdn}:
\begin{equation}
	Q_{\mathrm{tot}}(s,\mathbf{a})
	=
	\sum_{i=1}^{N} Q_i(o_i,a_i).
\end{equation}
This makes decentralized execution simple: each agent greedily chooses
\begin{equation}
	a_i^* = \argmax_{a_i} Q_i(o_i,a_i),
\end{equation}
and the joint greedy action is consistent with maximizing the sum. But the additive assumption is restrictive. It cannot represent many coordination tasks where the value of one agent's action depends nonlinearly on another agent's action.

\subsection{QMIX}

QMIX relaxes VDN by using a monotonic mixing network \citep{rashid2018qmix}:
\begin{equation}
	Q_{\mathrm{tot}}
	=
	f_{\mathrm{mix}}\left(Q_1,\ldots,Q_N,s\right),
\end{equation}
with the monotonicity constraint
\begin{equation}
	\frac{\partial Q_{\mathrm{tot}}}{\partial Q_i} \geq 0
	\quad \forall i.
\end{equation}
This constraint ensures that decentralized greedy action selection remains consistent with centralized greedy action selection:
\begin{equation}
	\argmax_{\mathbf{a}} Q_{\mathrm{tot}}(s,\mathbf{a})
	=
	\left(
	\argmax_{a_1}Q_1(o_1,a_1),
	\ldots,
	\argmax_{a_N}Q_N(o_N,a_N)
	\right).
\end{equation}
This property is known as Individual-Global-Max (IGM) consistency. QMIX uses hypernetworks conditioned on the global state to generate nonnegative mixing weights.

\begin{figure}[t]
	\centering
	\begin{tikzpicture}[
		box/.style={draw,rounded corners,thick,minimum width=2.7cm,minimum height=0.75cm,align=center,font=\small},
		arrow/.style={-{Latex[length=2.2mm]},thick},
		node distance=0.8cm
		]
		\node[box,fill=blue!8,draw=blue!70] (q1) {$Q_1(o_1,a_1)$};
		\node[box,fill=blue!8,draw=blue!70,below=0.5cm of q1] (q2) {$Q_2(o_2,a_2)$};
		\node[box,fill=blue!8,draw=blue!70,below=0.5cm of q2] (q3) {$Q_N(o_N,a_N)$};
		\node[box,fill=orange!10,draw=orange!80!black,right=1.4cm of q2] (mix) {Monotonic mixer\\$\partial Q_{tot}/\partial Q_i\geq 0$};
		\node[box,fill=green!8,draw=green!60!black,right=1.4cm of mix] (qt) {$Q_{\mathrm{tot}}(s,\mathbf{a})$};
		\node[box,fill=purple!8,draw=purple!70,above=0.9cm of mix] (state) {Global state $s$\\hypernetwork input};
		\draw[arrow,draw=blue!70] (q1) -- (mix);
		\draw[arrow,draw=blue!70] (q2) -- (mix);
		\draw[arrow,draw=blue!70] (q3) -- (mix);
		\draw[arrow,draw=purple!70] (state) -- (mix);
		\draw[arrow,draw=orange!80!black] (mix) -- (qt);
		\node[below=1.9cm of mix,align=center,font=\small] {Centralized training learns $Q_{tot}$; decentralized execution uses each $Q_i$.};
	\end{tikzpicture}
	\caption{The QMIX idea. Per-agent utilities are combined by a monotonic mixer conditioned on the global state. The monotonic constraint preserves decentralized greedy action selection.}
	\label{fig:qmix_mixer}
\end{figure}

\subsection{Beyond QMIX}

QTRAN tries to relax the monotonic restriction by transforming the joint action-value function while preserving decentralized optimality \citep{son2019qtran}. QPLEX uses a duplex dueling architecture to represent richer value factorization while maintaining IGM consistency \citep{wang2021qplex}. Weighted QMIX and related variants attempt to correct representational or optimization limitations of monotonic mixing. The main lesson is not that one mixer solves all tasks. The lesson is that value decomposition is a trade-off between expressiveness, stability, and decentralized execution.

\begin{table}[t]
	\centering
	\caption{Value-decomposition methods.}
	\label{tab:value_decomposition}
	\begin{tabular}{p{0.18\textwidth}p{0.30\textwidth}p{0.37\textwidth}}
		\toprule
		Method & Main assumption & Strength and limitation \\
		\midrule
		VDN & Additive team value & Simple and stable, but limited expressiveness \\
		QMIX & Monotonic mixing & Preserves decentralized greedy execution, but cannot represent non-monotonic coordination \\
		QTRAN & Transformation with optimality constraints & More expressive, but harder to optimize \\
		QPLEX & Duplex dueling with IGM & More expressive while preserving IGM, but more complex \\
		\bottomrule
	\end{tabular}
\end{table}

\section{Multi-agent actor-critic methods}

Value decomposition is powerful for discrete cooperative tasks. Actor-critic methods are more flexible for continuous control, mixed action spaces, and heterogeneous agents.

\subsection{MADDPG}

MADDPG extends deterministic actor-critic learning to multiple agents \citep{lowe2017maddpg}. Each agent has a decentralized actor
\begin{equation}
	a_i = \mu_i(o_i),
\end{equation}
while a centralized critic may condition on the full state and joint action:
\begin{equation}
	Q_i(s,a_1,\ldots,a_N).
\end{equation}
The actor gradient is
\begin{equation}
	\grad_{\theta_i}J_i
	\approx
	\E\left[
	\grad_{\theta_i}\mu_i(o_i)
	\grad_{a_i}Q_i(s,a_1,\ldots,a_N)
	\right].
\end{equation}
MADDPG is natural for continuous multi-agent control, but it can be sensitive to replay staleness, critic overfitting, and non-stationarity.

\subsection{COMA}

COMA addresses credit assignment using a counterfactual baseline \citep{foerster2018coma}. The critic estimates $Q(s,\mathbf{a})$, and agent $i$'s advantage is computed by comparing the actual joint action with alternatives for agent $i$ while holding other agents fixed:
\begin{equation}
	A_i^{\mathrm{COMA}}(s,\mathbf{a})
	=
	Q(s,\mathbf{a})
	-
	\sum_{a_i'} \pi_i(a_i'\given o_i)
	Q\bigl(s,(\mathbf{a}_{-i},a_i')\bigr).
	\label{eq:coma_advantage}
\end{equation}
The second term is the counterfactual baseline: it averages over what agent $i$ would have done under its current policy while all other agents' actions are fixed. This asks: how much did agent $i$'s chosen action improve the joint value compared with what that agent would have done on average?

\subsection{MAPPO and HAPPO}

MAPPO applies PPO-style policy optimization to cooperative MARL with centralized value functions and practical implementation choices \citep{yu2022mappo}. Its success is important because it showed that strong on-policy baselines can compete with more specialized off-policy MARL algorithms when implementation details are tuned carefully. The lesson is similar to single-agent PPO in Chapter~10: value normalization, advantage normalization, rollout structure, and critic design can matter as much as the nominal algorithm.

Recent applications of value-decomposition and graph-attention MARL to UAV-assisted network slicing have shown that variants such as GAD-QMIX can balance per-tier QoS with interference, energy, and safety constraints in 5G/6G scenarios \citep{bista2024gadqmix}.

HATRPO and HAPPO extend trust-region and PPO-style ideas to heterogeneous-agent learning with sequential policy updates and monotonic improvement arguments \citep{kuba2021happo}. These methods are useful reminders that multi-agent policy optimization is not simply PPO applied independently to every agent. The order and coupling of policy updates can matter. A useful systems-oriented comparison comes from UAV-assisted 5G network slicing, where MAPPO, MADDPG, and MADQN can be interpreted as representatives of three major MARL families: on-policy centralized-critic policy optimization, deterministic centralized-critic continuous control, and value-based multi-agent control. In such settings, MAPPO often provides the most balanced performance when stability, coordination quality, and QoS fairness all matter; MADDPG can be attractive in continuous-control settings but may be more sensitive to critic instability and replay issues; MADQN remains a simpler baseline when the action space is discretized and computational efficiency matters. This comparison is valuable pedagogically because it connects algorithm families to practical deployment trade-offs.

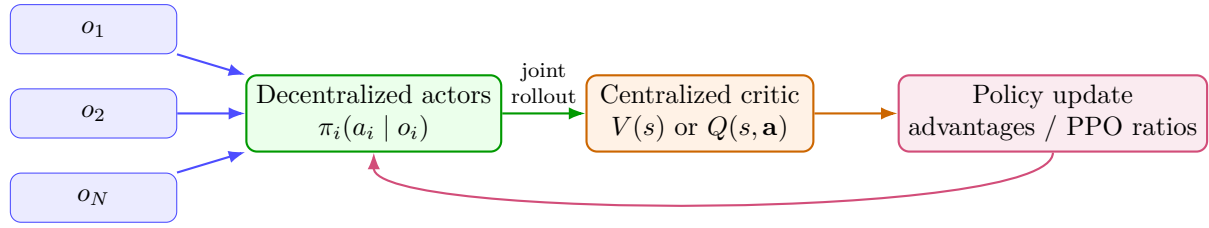
\begin{figure}[t]
	\centering
	\begin{tikzpicture}[
		box/.style={draw,rounded corners,thick,minimum width=3.0cm,minimum height=0.78cm,align=center,font=\small},
		small/.style={draw,rounded corners,minimum width=2.2cm,minimum height=0.65cm,align=center,font=\small},
		arrow/.style={-{Latex[length=2.2mm]},thick},
		node distance=1.8cm
		]
		\node[small,fill=blue!8,draw=blue!70] (o1) {$o_1$};
		\node[small,fill=blue!8,draw=blue!70,below=0.45cm of o1] (o2) {$o_2$};
		\node[small,fill=blue!8,draw=blue!70,below=0.45cm of o2] (o3) {$o_N$};
		\node[box,fill=green!8,draw=green!60!black,right=0.9cm of o2] (actors) {Decentralized actors\\$\pi_i(a_i\given o_i)$};
		\node[box,fill=orange!10,draw=orange!80!black,right=1.1cm of actors] (critic) {Centralized critic\\$V(s)$ or $Q(s,\mathbf{a})$};
		\node[box,fill=purple!8,draw=purple!70,right=1.1cm of critic] (loss) {Policy update\\advantages / PPO ratios};
		\draw[arrow,draw=blue!70] (o1) -- (actors);
		\draw[arrow,draw=blue!70] (o2) -- (actors);
		\draw[arrow,draw=blue!70] (o3) -- (actors);
		\draw[arrow,draw=green!60!black] (actors) -- node[above,font=\scriptsize,align=center] {joint\\rollout} (critic);
		\draw[arrow,draw=orange!80!black] (critic) -- (loss);
		\draw[arrow,draw=purple!70] (loss.south) .. controls +(0,-0.9) and +(0,-0.9) .. (actors.south);
	\end{tikzpicture}
	\caption{Centralized-critic actor-critic MARL. Decentralized actors choose actions from local observations, while a centralized critic uses global information to compute lower-variance learning signals.}
	\label{fig:marl_actor_critic}
\end{figure}
A useful systems-oriented comparison comes from UAV-assisted 5G network slicing, where MADQN, MADDPG, and MAPPO can be interpreted as representatives of three major MARL families: value-based multi-agent control, deterministic centralized-critic continuous control, and on-policy centralized-critic policy optimization. In such settings, MADQN can remain attractive when the action space is discretized and computational efficiency matters; MADDPG is natural for continuous-control settings but can be more sensitive to critic instability and replay staleness; and MAPPO often provides the most balanced performance when stability, coordination quality, and QoS fairness all matter. This comparison is valuable because it connects abstract MARL families to practical deployment trade-offs in UAV-assisted network slicing\citep{bista2025marl_uav_slicing}.
\begin{table}[t]
\centering
\caption{A practical comparison of MARL families for UAV-assisted network slicing.}
\label{tab:marl_uav_compare}
\begin{tabular}{p{0.17\textwidth}p{0.22\textwidth}p{0.24\textwidth}p{0.24\textwidth}}
\toprule
Method & Family & Main strength & Main limitation \\
\midrule
MADQN & Value-based MARL & Simple and efficient for discretized control & Limited flexibility for continuous or hybrid actions \\
MADDPG & Centralized-critic deterministic actor-critic & Natural for continuous multi-UAV control & Sensitive to critic instability, replay staleness, and tuning \\
MAPPO & On-policy centralized-critic actor-critic & Strong stability and competitive cooperative performance & Higher rollout cost and on-policy sample demand \\
\bottomrule
\end{tabular}
\end{table}

\section{Communication, attention, and graph-based coordination}

Some tasks cannot be solved well if agents act only from local observations without communication. Communication learning methods allow agents to exchange messages. CommNet uses continuous communication channels among agents \citep{sukhbaatar2016commnet}. DIAL and RIAL study differentiable communication protocols \citep{foerster2016dial}. Attention-based critics and graph neural networks learn which agents matter for a decision \citep{iqbal2019maac,jiang2020dgn}.

Communication is not automatically helpful. Learned messages can become a non-stationary channel, overfit to a training distribution, create bandwidth or latency overhead, and expose agents to adversarial or corrupted information in mixed settings. A communication module should therefore be evaluated with ablations that remove, delay, corrupt, or restrict messages.

A graph view is natural in UAV and network control. Agents are nodes. Edges represent interference, communication links, backhaul dependencies, proximity, or shared users. A graph policy can aggregate messages from neighbors:
\begin{equation}
	h_i^{(k+1)}
	=
	\sigma\left(
	W h_i^{(k)}
	+
	\sum_{j\in\mathcal{N}(i)}\alpha_{ij}W_m h_j^{(k)}
	\right),
\end{equation}
where $\alpha_{ij}$ is an attention weight. This allows each UAV to reason about nearby UAVs without requiring a fixed number of agents.

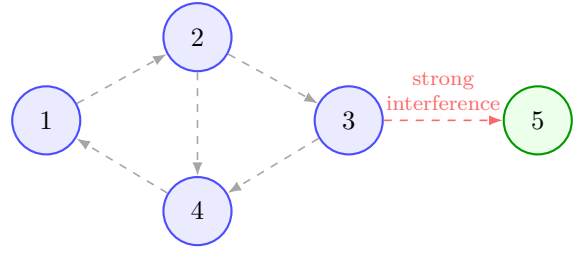
\begin{figure}[t]
	\centering
	\begin{tikzpicture}[
		agent/.style={circle,draw,thick,minimum size=0.9cm,align=center,font=\small},
		arrow/.style={-{Latex[length=2mm]},thick},
		msg/.style={-{Latex[length=1.8mm]},semithick,dashed}
		]
		\node[agent,fill=blue!8,draw=blue!70]        (a1) at (0,0)     {$1$};
		\node[agent,fill=blue!8,draw=blue!70]        (a2) at (2.0,1.1) {$2$};
		\node[agent,fill=blue!8,draw=blue!70]        (a3) at (4.0,0)   {$3$};
		\node[agent,fill=blue!8,draw=blue!70]        (a4) at (2.0,-1.2){$4$};
		\node[agent,fill=green!8,draw=green!60!black](a5) at (6.5,0)   {$5$};

		\draw[msg,draw=gray!70] (a1) -- (a2);
		\draw[msg,draw=gray!70] (a2) -- (a3);
		\draw[msg,draw=gray!70] (a3) -- (a4);
		\draw[msg,draw=gray!70] (a4) -- (a1);
		\draw[msg,draw=gray!70] (a2) -- (a4);

		% interference arrow — label placed above the midpoint, clear of node 5
		\draw[msg,draw=red!60]
		(a3) -- node[above,font=\scriptsize,text=red!60,align=center]
		{strong\\interference} (a5);

	\end{tikzpicture}
	\caption{A graph view of MARL. Each agent is a node, and edges represent
		communication, interference, proximity, or task coupling. Attention weights
		can learn which neighbors matter. Agents exchange local messages over a
		learned or physical interaction graph.}
	\label{fig:graph_marl}
\end{figure}

\section{Sequence modeling and transformer-based MARL}

Chapter~15 explained how Decision Transformers treat trajectories as sequences. Multi-agent sequence modeling applies similar ideas to joint decision-making. Multi-Agent Transformer (MAT) casts cooperative MARL as a sequence modeling problem over agents and actions \citep{wen2022mat}. Instead of choosing all agents' actions independently at the same time, the model can generate an action sequence over agents, using previous agents' choices as context.

MAT can be viewed as a multi-agent extension of the sequence-modeling perspective developed in Chapter~15: a transformer treats joint actions as a sequence and predicts them autoregressively conditioned on observations. This perspective connects MARL to the transformer-based agentic AI systems discussed in later chapters. It also exposes a key modeling choice: agent ordering. If the model generates actions in the order $(1,2,\ldots,N)$, early agents influence later ones. Recent sequence-modeling variants study how ordering, heterogeneity, and generalization across different agent counts affect performance.

\begin{keybox}{Sequence-model view of MARL}
	A joint action can be generated as an autoregressive sequence:
	\[
	\pi(\mathbf{a}\given \mathbf{o})
	=
	\prod_{i=1}^{N}\pi(a_i\given \mathbf{o},a_{<i}).
	\]
	This turns simultaneous coordination into structured conditional prediction.
\end{keybox}

The sequence view is powerful, but it should not hide the decentralization requirement. If execution is decentralized, the model must either produce local policies, compress global context into allowed communication, or use centralized planning only when communication and latency budgets permit it.

\section{Mean-field, population, and heterogeneous-agent MARL}

When the number of agents is large, pairwise modeling becomes expensive. Mean-field MARL approximates the influence of many neighbors by an average action or distribution \citep{yang2018meanfield}, with later extensions considering heterogeneous populations and richer interaction structures \citep{subramanian2020heterogeneousmf}. Instead of conditioning on every neighbor action, agent $i$ may condition on
\begin{equation}
	\bar{a}_{\mathcal{N}(i)}
	=
	\frac{1}{|\mathcal{N}(i)|}
	\sum_{j\in\mathcal{N}(i)}a_j.
\end{equation}
This is useful for large populations, though it can lose important individual interactions.

Population-based MARL studies robustness against diverse partners and opponents. In cooperative settings, a policy should coordinate with teammates. In mixed settings, it should not overfit to one opponent. Techniques such as self-play, population-based training, fictitious play, and league training are important in competitive and mixed domains.

Heterogeneous-agent MARL allows agents to have different observation spaces, action spaces, roles, or dynamics. This is essential in realistic UAV/SDN systems: one UAV may be a high-altitude relay, another may be a low-altitude access point, and the SDN controller may act at a different time scale. HAPPO/HATRPO and heterogeneous-agent learning frameworks address some of this complexity \citep{kuba2021happo,zhong2024heterogeneous}.

\section{Multi-agent replay, fingerprints, and recurrent policies}

Off-policy MARL often uses replay buffers. But replay is more dangerous in MARL than in single-agent RL because old data were generated by old combinations of policies. A transition stored at time $t$ may become misleading after several agents update their policies. One way to reduce this problem is to store a fingerprint, such as the training iteration, exploration rate, or policy version, together with each transition \citep{foerster2017stabilising}. The learner can then condition on the fingerprint to partially identify the data distribution.

Partial observability also makes recurrent policies important. RIAL/DIAL-style communication policies and recurrent multi-agent actor-critic variants use memory to stabilize decisions when each agent sees only a local fragment of the environment \citep{foerster2016dial,lowe2017maddpg}. In SMAC-style tasks, robot swarms, and UAV networks, an agent's current observation may not reveal hidden traffic demand, opponent intent, or teammate state. A recurrent policy can maintain memory:
\begin{equation}
	h_{i,t} = f_\theta(h_{i,t-1},o_{i,t}),
	\qquad
	a_{i,t}\sim \pi_i(\cdot\given h_{i,t}).
\end{equation}

\begin{warningbox}{Replay in MARL is stale faster than replay in single-agent RL}
	A replay transition is not only old because the value function changed. It is old because the behavior policies of multiple agents changed. This is why MARL replay often needs fingerprints, recurrent state, centralized critics, or short replay windows.
\end{warningbox}

\section{UAV/SDN worked scenario: cooperative multi-UAV control}

Consider a UAV-assisted wireless network with four UAVs serving three classes of users. URLLC users require low latency, eMBB users require high throughput, and mMTC users require reliable low-rate service. UAVs interfere with one another if they move too close or reuse spectrum poorly. Each UAV observes local users, local SINR, local battery, and nearby UAV positions. A centralized SDN controller observes global traffic and can provide training-time labels, rewards, constraints, or guidance.

\begin{figure}[t]
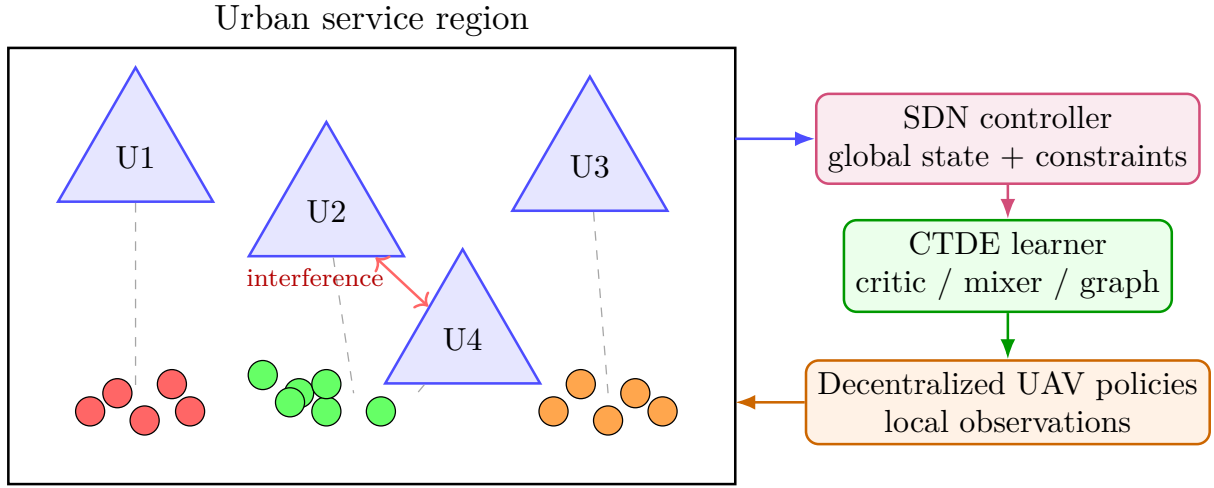

	\centering
	\resizebox{\columnwidth}{!}{%
%
	}
	\caption{A multi-agent UAV/SDN scenario. UAVs coordinate movement, spectrum
		allocation, and user association. The SDN controller can provide centralized
		training information, while deployed UAV policies must act from local observations
		and limited communication.}
	\label{fig:uav_sdn_marl_scenario}
\end{figure}

A simple cooperative reward might be
\begin{equation}
	r_t
	=
	w_q R_{\mathrm{QoS}}
	-
	w_e C_{\mathrm{energy}}
	-
	w_i C_{\mathrm{interference}}
	-
	w_s C_{\mathrm{safety}}.
\end{equation}
But this scalar reward hides multi-agent structure. One UAV may consume energy to improve global latency. Another may move away from a hotspot to reduce interference. A third may serve low-rate IoT users so that a high-capacity UAV can serve URLLC users. The team reward alone does not say which policy should change.

A CTDE design can use:
\begin{itemize}
	\item decentralized UAV actors $\pi_i(a_i\given o_i)$;
	\item a centralized reward critic $V_r(s)$;
	\item a centralized cost critic $V_c(s)$ for safety or constraint violations;
	\item a graph-attention module over UAV interference edges;
	\item an SDN guidance signal used during training or as a safety supervisor.
\end{itemize}
A comparative MARL study is especially informative in this setting because the same UAV slicing problem can favor different algorithmic families for different reasons. An on-policy method such as MAPPO may provide strong stability and balanced cooperative performance under shared QoS objectives, a deterministic actor-critic method such as MADDPG may exploit continuous control structure more directly, and a value-based baseline such as MADQN may remain attractive when the control interface is discretized. This illustrates a broader lesson of MARL systems design: algorithm selection should follow the structure of the control problem, not only benchmark popularity.
\begin{researchbox}{A distinctive safe-MARL architecture for UAV/SDN}
	A strong UAV/SDN MARL system should not only maximize team reward. It should separate service quality, energy, interference, and safety into interpretable learning signals. A practical architecture may combine CTDE, graph attention, multi-critic actor-critic learning, and a safety filter such as CBF projection. This connects MARL to the forthcoming safe-RL treatment in Chapter~18, where constrained MDPs, Lagrangian methods, shields, and control barrier functions are developed explicitly.
\end{researchbox}

\subsection{Concrete numerical example}

Suppose three UAVs choose whether to move toward a URLLC hotspot. The team reward and interference cost are estimated after one rollout window.

\begin{table}[t]
	\centering
	\caption{A numerical credit-assignment example for cooperative UAV control.}
	\label{tab:uav_credit_assignment}
	\begin{tabular}{p{0.46\textwidth}ccc}
		\toprule
		Joint behavior & QoS reward & Interference cost & Team score \\
		\midrule
		Only UAV 1 moves & 18.0 & 2.0 & 16.0 \\
		UAV 1 and UAV 2 move & 24.0 & 9.0 & 15.0 \\
		UAV 1 moves, UAV 2 relays, UAV 3 covers IoT & 26.0 & 3.5 & 22.5 \\
		All UAVs move to hotspot & 29.0 & 14.0 & 15.0 \\
		\bottomrule
	\end{tabular}
\end{table}

A greedy local policy might send all UAVs to the hotspot because each sees URLLC demand. But the best joint behavior in Table~\ref{tab:uav_credit_assignment} is role specialization: one UAV serves the hotspot, one acts as relay, and one covers IoT users. This is why MARL needs coordination, not merely independent reward maximization. %The editorial report specifically highlights this numerical example as one of the chapter's strongest concrete teaching devices.

\section{Python implementation patterns}

This section gives compact implementation patterns. The goal is not to provide a full library, but to show the design skeletons that readers can adapt.

\subsection{Multi-agent rollout batch}

\Needspace{18\baselineskip}
\begin{lstlisting}[style=pythonstyle,caption={A minimal multi-agent rollout container.},label={lst:marl_batch}]
import torch

class MultiAgentBatch:
    """Stores one rollout for N agents over T steps.

    obs:      [T, N, obs_dim]
    actions:  [T, N]
    rewards:  [T] or [T, N]
    dones:    [T]
    states:   [T, state_dim] optional global state for CTDE
    """
    def __init__(self, obs, actions, rewards, dones, states=None, masks=None):
        self.obs = obs
        self.actions = actions
        self.rewards = rewards
        self.dones = dones
        self.states = states
        self.masks = masks

    def to(self, device):
        for name, value in self.__dict__.items():
            if torch.is_tensor(value):
                setattr(self, name, value.to(device))
        return self
\end{lstlisting}

\subsection{VDN mixer}

\Needspace{14\baselineskip}
\begin{lstlisting}[style=pythonstyle,caption={VDN mixer: additive team value.},label={lst:vdn_mixer}]
import torch.nn as nn

class VDNMixer(nn.Module):
    def forward(self, agent_qs):
        """agent_qs: [batch, n_agents]
        returns:   [batch, 1]
        """
        return agent_qs.sum(dim=-1, keepdim=True)
\end{lstlisting}

\subsection{QMIX mixer}

\Needspace{22\baselineskip}
\begin{lstlisting}[style=pythonstyle,caption={Simplified QMIX monotonic mixer with hypernetworks.},label={lst:qmix_mixer}]
class QMIXMixer(nn.Module):
    def __init__(self, n_agents, state_dim, hidden_dim=32):
        super().__init__()
        self.n_agents = n_agents
        self.hidden_dim = hidden_dim

        # Hypernetworks generate non-negative mixing weights from state.
        self.hyper_w1 = nn.Linear(state_dim, n_agents * hidden_dim)
        self.hyper_b1 = nn.Linear(state_dim, hidden_dim)
        self.hyper_w2 = nn.Linear(state_dim, hidden_dim)
        self.hyper_b2 = nn.Sequential(
            nn.Linear(state_dim, hidden_dim), nn.ReLU(), nn.Linear(hidden_dim, 1)
        )

    def forward(self, agent_qs, states):
        """agent_qs: [B, N], states: [B, state_dim]."""
        B = agent_qs.shape[0]
        q = agent_qs.view(B, 1, self.n_agents)

        w1 = torch.abs(self.hyper_w1(states)).view(B, self.n_agents, self.hidden_dim)
        b1 = self.hyper_b1(states).view(B, 1, self.hidden_dim)
        hidden = torch.relu(torch.bmm(q, w1) + b1)

        w2 = torch.abs(self.hyper_w2(states)).view(B, self.hidden_dim, 1)
        b2 = self.hyper_b2(states).view(B, 1, 1)
        q_tot = torch.bmm(hidden, w2) + b2
        return q_tot.view(B, 1)
\end{lstlisting}

The absolute value in Listing~\ref{lst:qmix_mixer} enforces nonnegative mixing weights. In production implementations, softplus is sometimes used instead of absolute value to avoid nondifferentiability at zero. %The editorial report specifically calls out this practical note as accurate and worth keeping.

\subsection{Centralized critic for MAPPO}

\Needspace{22\baselineskip}
\begin{lstlisting}[style=pythonstyle,caption={A centralized value critic for MAPPO-style CTDE.},label={lst:mappo_critic}]
class CentralizedValueCritic(nn.Module):
    def __init__(self, state_dim, hidden_dim=256):
        super().__init__()
        self.net = nn.Sequential(
            nn.Linear(state_dim, hidden_dim), nn.ReLU(),
            nn.Linear(hidden_dim, hidden_dim), nn.ReLU(),
            nn.Linear(hidden_dim, 1)
        )

    def forward(self, global_state):
        return self.net(global_state).squeeze(-1)


def mappo_value_loss(critic, states, returns):
    values = critic(states)
    return 0.5 * (values - returns.detach()).pow(2).mean()
\end{lstlisting}

\subsection{MADDPG centralized critic update}

\Needspace{22\baselineskip}
\begin{lstlisting}[style=pythonstyle,caption={Centralized critic target for MADDPG-style training.},label={lst:maddpg_critic}]
def maddpg_critic_loss(critic_i, target_critic_i, target_actors,
                       batch, gamma=0.99):
    """batch contains obs, actions, rewards_i, next_obs, dones, state."""
    obs, actions = batch["obs"], batch["actions"]
    next_obs = batch["next_obs"]
    rewards_i, dones = batch["rewards_i"], batch["dones"]

    with torch.no_grad():
        next_actions = []
        for j, actor_j in enumerate(target_actors):
            next_actions.append(actor_j(next_obs[:, j]))
        next_actions = torch.cat(next_actions, dim=-1)
        next_obs_flat = next_obs.reshape(next_obs.shape[0], -1)
        target_q = target_critic_i(next_obs_flat, next_actions)
        y = rewards_i + gamma * (1.0 - dones) * target_q

    obs_flat = obs.reshape(obs.shape[0], -1)
    act_flat = actions.reshape(actions.shape[0], -1)
    q = critic_i(obs_flat, act_flat)
    return 0.5 * (q - y).pow(2).mean()
\end{lstlisting}

\subsection{Graph message passing for UAV agents}

\Needspace{22\baselineskip}
\begin{lstlisting}[style=pythonstyle,caption={Simple graph message passing for UAV MARL.},label={lst:graph_message_passing}]
class GraphMessageLayer(nn.Module):
    def __init__(self, in_dim, msg_dim):
        super().__init__()
        self.self_proj = nn.Linear(in_dim, msg_dim)
        self.msg_proj = nn.Linear(in_dim, msg_dim)
        self.attn = nn.Linear(2 * msg_dim, 1)

    def forward(self, h, adjacency):
        """h: [B, N, D], adjacency: [B, N, N] with 1 for neighbors."""
        B, N, _ = h.shape
        self_h = self.self_proj(h)
        msg_h = self.msg_proj(h)

        hi = self_h.unsqueeze(2).expand(B, N, N, -1)
        hj = msg_h.unsqueeze(1).expand(B, N, N, -1)
        score = self.attn(torch.cat([hi, hj], dim=-1)).squeeze(-1)
        score = score.masked_fill(adjacency == 0, -1e9)
        alpha = torch.softmax(score, dim=-1)
        messages = torch.bmm(alpha, msg_h)
        return torch.relu(self_h + messages)
\end{lstlisting}

\subsection{Action masking for decentralized execution}

\Needspace{18\baselineskip}
\begin{lstlisting}[style=pythonstyle,caption={Action masking for invalid decentralized actions.},label={lst:action_masking}]
def masked_categorical_logits(logits, action_mask):
    """action_mask: 1 for valid actions, 0 for invalid actions."""
    invalid = action_mask <= 0
    return logits.masked_fill(invalid, -1e9)


def sample_masked_action(logits, action_mask):
    masked_logits = masked_categorical_logits(logits, action_mask)
    dist = torch.distributions.Categorical(logits=masked_logits)
    action = dist.sample()
    logp = dist.log_prob(action)
    return action, logp
\end{lstlisting}

Action masking is essential in many MARL environments. In UAV control, a drone may not be allowed to move outside the flight region, choose a depleted battery mode, or allocate bandwidth it does not have. Masking prevents the policy from assigning probability mass to invalid actions. It is not a substitute for safety verification, but it prevents a large class of avoidable errors.

\section{Evaluation, benchmarks, and reproducibility}

MARL evaluation is fragile. A method can look strong because of a specific map, agent count, reward shaping, implementation detail, or seed. Standard benchmarks help, but they also age. The original StarCraft Multi-Agent Challenge (SMAC) became widely used, and SMACv2 introduced procedurally generated scenarios and stronger generalization tests \citep{samvelyan2019smac,ellis2023smacv2}. PettingZoo standardized many multi-agent environments behind a common API \citep{terry2021pettingzoo}. BenchMARL provides a 2024 benchmarking library for standardized MARL training and evaluation \citep{bettini2024benchmarl}.

\begin{table}[t]
	\centering
	\caption{MARL benchmark and evaluation considerations.}
	\label{tab:marl_benchmarks}
	\begin{tabular}{p{0.25\textwidth}p{0.31\textwidth}p{0.31\textwidth}}
		\toprule
		Benchmark/tool & What it is useful for & Warning \\
		\midrule
		MPE / particle worlds & Simple coordination, communication, continuous spaces & Too small for strong claims \\
		SMAC & Cooperative micromanagement and partial observability & Can overfit to fixed maps \\
		SMACv2 & Procedural generalization and stronger partial observability & Still game-like, not physical networking \\
		Hanabi & Ad hoc cooperation and hidden information & Highly domain-specific \\
		PettingZoo & Common API across many MARL tasks & API standard, not a scientific guarantee \\
		BenchMARL & Reproducible PyTorch/TorchRL benchmarking & Still requires careful task selection \\
		UAV/SDN simulator & Domain-specific QoS, energy, safety, mobility & Must report realism assumptions and constraints \\
		\bottomrule
	\end{tabular}
\end{table}

A strong MARL evaluation should report more than team return. For UAV/SDN systems, it should include QoS satisfaction, latency, SINR, throughput, packet loss, energy, collision rate, interference, fairness across user classes, communication overhead, controller latency, and safety violations.

\begin{warningbox}{Do not compare MARL algorithms only by final return}
	MARL methods may have similar return but very different safety, fairness, communication cost, and robustness. In networked systems, a method that gets higher return by violating latency constraints or draining UAV batteries is not actually better.
\end{warningbox}

\section{Failure modes and debugging checklist}

\begin{table}[t]
	\centering
	\caption{Common MARL failure modes.}
	\label{tab:marl_failures}
	\begin{tabular}{p{0.25\textwidth}p{0.30\textwidth}p{0.30\textwidth}}
		\toprule
		Symptom & Likely cause & What to inspect \\
		\midrule
		Training unstable across seeds & Non-stationarity or high-variance critic & Policy update size, replay age, centralized critic inputs \\
		Agents collapse to same role & Poor credit assignment or parameter sharing mismatch & Role features, rewards, entropy, diversity metrics \\
		High return but poor QoS fairness & Reward hides minority users & Per-class metrics and constraint violations \\
		QMIX learns but execution fails & Bad observation design or action masks & Local observations, masks, IGM assumption \\
		MAPPO underperforms IPPO & Centralized critic overfits or value normalization wrong & Value loss, explained variance, critic inputs \\
		Communication policy noisy & Message channel overfits or lacks bottleneck & Message entropy, attention maps, ablations without communication \\
		UAVs collide or cluster & Reward lacks safety/repulsion or filter not trained on executed action & Collision rate, CBF interventions, critic targets \\
		Offline replay hurts & Multi-agent replay too stale & Policy version fingerprints and replay window length \\
		\bottomrule
	\end{tabular}
\end{table}

\section{Limitations and when not to use MARL}

\begin{enumerate}[leftmargin=*]
	\item \textbf{Non-stationarity is hard to eliminate.} CTDE, fingerprints, and recurrent policies reduce it, but they do not remove the fact that other agents keep changing the learning problem.
	\item \textbf{The joint action space still grows explosively.} Value decomposition, graph structure, and mean-field approximations help, but large multi-agent systems remain difficult to scale.
	\item \textbf{Credit assignment remains unsolved in general.} Counterfactual baselines and value mixers help in structured cases, but sparse delayed team rewards can still make learning unstable or misleading.
	\item \textbf{Communication adds overhead and attack surface.} Learned messages can be delayed, corrupted, bandwidth-expensive, or strategically manipulated in mixed settings.
	\item \textbf{Evaluation is fragile.} Many MARL results depend strongly on map design, reward shaping, seed choice, parameter sharing, and implementation details.
	\item \textbf{Safety is harder than in single-agent RL.} Constraints may be individual, pairwise, or collective. A globally high reward can still hide collisions, interference spikes, unfair service, or unsafe emergent coordination.
\end{enumerate}

\section{Exercises}

\subsection*{Conceptual exercises}
\begin{enumerate}[leftmargin=*]
	\item Explain why other learning agents make the environment non-stationary from the viewpoint of one agent.
	\item Compare independent PPO, MAPPO, and QMIX for a cooperative UAV task. Which assumptions does each method make?
	\item Why does CTDE preserve decentralized execution while still using global information during training?
	\item Explain why a high team reward can still hide poor credit assignment.
	\item Why is communication not always beneficial in MARL?
\end{enumerate}

\subsection*{Mathematical exercises}
\begin{enumerate}[leftmargin=*]
	\item Suppose $N$ agents each have $m$ discrete actions. Derive the size of the joint action space and compute it for $N=8$, $m=6$.
	\item Show that VDN satisfies decentralized greedy consistency for additive $Q_{tot}$.
	\item Explain why the monotonicity condition $\partial Q_{tot}/\partial Q_i\geq 0$ is sufficient for decentralized greedy action selection in QMIX.
	\item Derive the COMA counterfactual advantage for a two-agent discrete-action case.
	\item For a graph MARL policy, write a message-passing update where attention weights depend on relative distance between UAVs.
\end{enumerate}

\subsection*{Coding exercises}
\begin{enumerate}[leftmargin=*]
	\item Implement a shared-parameter independent PPO policy for $N$ homogeneous agents.
	\item Extend Listing~\ref{lst:qmix_mixer} with a two-layer hypernetwork and softplus nonnegative weights.
	\item Implement action masking for a multi-UAV environment with invalid movement and invalid bandwidth actions.
	\item Add fingerprints to a multi-agent replay buffer: store episode index, exploration rate, and policy version.
	\item Use Listing~\ref{lst:graph_message_passing} to build a graph policy that only exchanges messages between UAVs within a fixed interference radius.
	\item Reproduce a small GAD-QMIX-style UAV slicing experiment: implement three traffic classes, compare independent DQN, QMIX, and a graph-attention QMIX variant, and report per-class throughput, latency, interference, and safety violations.
\end{enumerate}

\subsection*{Research thinking exercises}
\begin{enumerate}[leftmargin=*]
	\item Design a CTDE architecture for SDN-assisted UAV networks with reward, cost, and fairness critics.
	\item In the UAV scenario of Figure~\ref{fig:uav_sdn_marl_scenario}, decide whether you would use QMIX, MAPPO, MAT, or graph actor-critic. Justify your answer.
	\item How would you evaluate whether learned communication is actually useful rather than decorative?
	\item Propose an ablation study to separate the benefits of centralized critics, graph attention, action masking, and safety filtering.
	\item Explain how MARL safety differs from single-agent safety. Which constraints are individual, and which are collective?
\end{enumerate}

\section*{Looking Ahead to Chapter 17: Food for Thought}
\addcontentsline{toc}{section}{Looking Ahead to Chapter 17: Food for Thought}

This chapter studied several agents learning at the same time. The next chapter asks a different question: what if one agent, or a team of agents, must reason at multiple time scales? Hierarchical reinforcement learning decomposes behavior into high-level decisions and low-level control. In a UAV system, a high-level policy may choose whether to serve URLLC users, recharge, explore, relay, or avoid interference. A low-level policy then converts that option into movement, bandwidth, and power-control actions.

\begin{quote}
	Chapter~16 asked how many agents can coordinate. Chapter~17 asks how complex behavior can be decomposed into decisions at different levels of abstraction.
\end{quote}

% \section*{Chapter references}
% \addcontentsline{toc}{section}{Chapter references}
	\chapter[Hierarchical Reinforcement Learning]{Hierarchical Reinforcement Learning}
\chaptermark{Hierarchical RL}
\label{ch:hrl}

\begin{keybox}{Chapter goal}
	Hierarchical reinforcement learning formalizes long-horizon decision-making through temporal abstraction. This chapter develops the options framework and SMDP foundations, option-critic learning, goal-conditioned HRL, unsupervised skill discovery, feudal and event-conditioned architectures, and the UAV/SDN application of hierarchical cooperative control.
\end{keybox}

\section*{Chapter Overview}
\addcontentsline{toc}{section}{Chapter Overview}

\begin{enumerate}[leftmargin=*]
	\item Why hierarchy matters
	\item The core idea: temporal abstraction
	\item Semi-Markov decision processes
	\item The options framework
	\item Option-value functions and SMDP Q-learning
	\item MAXQ, HAMs, and early task-decomposition views
	\item Option-Critic and end-to-end option learning
	\item Goal-conditioned HRL and subgoal discovery
	\item Skill discovery and latent-skill policies
	\item Feudal, manager-worker, and event-conditioned HRL
	\item Offline and sequence-modeling views of HRL
	\item Hierarchical MARL and UAV/SDN control
	\item Python implementation patterns
	\item Practical failure modes
	\item Research frontiers toward 2026
	\item Limitations and when not to use HRL
	\item Exercises
	\item Looking Ahead to Chapter 18
\end{enumerate}

\section{Why hierarchy matters}

The previous chapters developed the main families of modern deep reinforcement learning: value-based learning, policy-gradient methods, actor-critic learning, PPO, SAC, model-based DRL, MuZero-style search, offline RL, Decision Transformers, and multi-agent RL. These methods can be powerful, but they still struggle with one of the oldest difficulties in reinforcement learning: long-horizon decision-making.

A flat policy chooses primitive actions at every time step. This is natural for simple environments, but it becomes inefficient when useful behavior spans many time steps. A UAV does not only choose a small velocity vector every few milliseconds. It also chooses a mission mode: serve a hotspot, recharge, explore a region, avoid interference, or coordinate with other UAVs. A robot does not only output torques; it chooses to grasp, lift, carry, open, inspect, or navigate. A language-model agent does not only emit tokens; it may first plan, then solve subproblems, then verify.

Hierarchical reinforcement learning (HRL) formalizes this idea. It introduces temporally extended decisions: actions that last for multiple low-level time steps. These temporally extended actions may be called \emph{options}, \emph{skills}, \emph{subpolicies}, \emph{subtasks}, \emph{macro-actions}, or \emph{abstract actions}, depending on the tradition. The central idea is the same: a high-level controller chooses what should be done, and a low-level controller decides how to do it.

\begin{keybox}{Core intuition}
	Flat RL asks: what primitive action should I take now? Hierarchical RL asks: what skill or subgoal should I pursue now, and how should lower-level control execute it?
\end{keybox}

Historically, HRL connects several lines of work: feudal reinforcement learning \citep{dayan1993feudal}, hierarchical abstract machines \citep{parr1998reinforcement}, MAXQ value decomposition \citep{dietterich2000hierarchical}, and the options framework \citep{sutton1999between}. Modern deep HRL extends these ideas using neural option policies, goal-conditioned policies, latent skills, representation learning, unsupervised skill discovery, language-guided planning, and multi-agent hierarchies \citep{bacon2017option,vezhnevets2017feudal,nachum2018data,pateria2021survey,zhang2025llmhrl}.

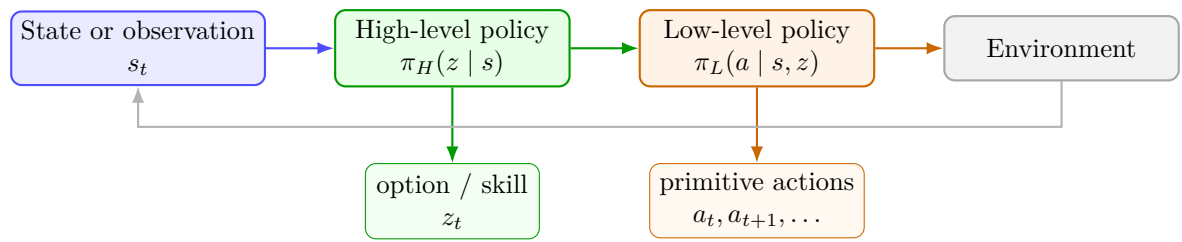
\begin{figure}[t]
	\centering
	\begin{tikzpicture}[
		box/.style={draw,rounded corners,thick,minimum width=3.1cm,minimum height=0.85cm,align=center,font=\small},
		small/.style={draw,rounded corners,minimum width=2.3cm,minimum height=0.75cm,align=center,font=\small},
		arrow/.style={-{Latex[length=2.2mm]},thick},
		node distance=0.9cm
		]
		\node[box,fill=blue!8,draw=blue!70] (state) {State or observation\\$s_t$};
		\node[box,fill=green!10,draw=green!60!black,right=of state] (manager) {High-level policy\\$\pi_H(z\mid s)$};
		\node[box,fill=orange!10,draw=orange!80!black,right=of manager] (worker) {Low-level policy\\$\pi_L(a\mid s,z)$};
		\node[box,fill=gray!10,draw=gray!70,right=of worker] (env) {Environment};
		\node[small,fill=green!5,draw=green!60!black,below=1.0cm of manager] (subgoal) {option / skill\\$z_t$};
		\node[small,fill=orange!5,draw=orange!80!black,below=1.0cm of worker] (actions) {primitive actions\\$a_t,a_{t+1},\ldots$};

		\draw[arrow,draw=blue!70] (state) -- (manager);
		\draw[arrow,draw=green!60!black] (manager) -- (worker);
		\draw[arrow,draw=orange!80!black] (worker) -- (env);
		\draw[arrow,draw=green!60!black] (manager) -- (subgoal);
		\draw[arrow,draw=orange!80!black] (worker) -- (actions);
		\draw[arrow,draw=gray!60] (env.south) -- ++(0,-0.6) -| (state.south);
	\end{tikzpicture}
	\caption{The basic two-level HRL architecture. A high-level policy chooses a temporally extended intention, option, skill, or subgoal; a low-level policy translates that abstraction into primitive actions.}
	\label{fig:hrl_basic_architecture}
\end{figure}

\section{Temporal abstraction: the central idea}

Temporal abstraction means that a decision can persist for more than one primitive time step. Instead of choosing a new primitive action every step, the agent may choose an option that executes until it terminates. This changes the effective horizon of the learning problem.

Suppose a low-level controller acts every $\Delta t=0.03$ seconds. A 30-second UAV mission contains 1000 primitive decisions. A high-level controller that changes intent every 3 seconds faces only 10 high-level decisions. The hierarchy does not remove the low-level control problem, but it decomposes the long horizon into a shorter high-level horizon and reusable low-level behaviors.

\begin{table}[t]
	\centering
	\caption{Flat control versus hierarchical control.}
	\label{tab:flat_vs_hierarchical}
	\begin{tabular}{p{3.4cm}p{4.4cm}p{4.4cm}}
		\toprule
		Aspect & Flat RL & Hierarchical RL \\
		\midrule
		Decision unit & Primitive action & Skill, option, subgoal, or mission mode \\
		Temporal scale & One environment step & Multiple environment steps \\
		Credit assignment & Every primitive action must be evaluated & High-level and low-level credit can be separated \\
		Exploration & Random primitive actions may be inefficient & Explore in skill/subgoal space \\
		Reuse & Behavior is usually tied to the learned policy & Skills can be reused across tasks \\
		Failure mode & Long-horizon sparse rewards & Bad subgoals or poorly trained low-level skills \\
		\bottomrule
	\end{tabular}
\end{table}

\begin{warningbox}{Hierarchy is not automatically better}
	Hierarchy helps only if the abstraction is useful. Bad subgoals, unstable termination conditions, or low-level skills that do not reliably execute high-level commands can make learning worse than a flat baseline.
\end{warningbox}

\section{Semi-Markov decision processes}

Options naturally lead to semi-Markov decision processes (SMDPs). In an ordinary MDP, every action consumes one time step. In an SMDP, an action may last for a random duration $k$. The option starts in state $s_t$, executes primitive actions internally, terminates after $k$ steps, and returns control to the high-level policy in state $s_{t+k}$.

The discounted reward accumulated during the option is
\begin{equation}
	R_t^{(k)}
	=
	\sum_{j=0}^{k-1} \gamma^j r_{t+j+1}.
	\label{eq:smdp_option_return}
\end{equation}
The corresponding SMDP Bellman equation for an option-value function is
\begin{equation}
	Q^\Omega(s,o)
	=
	\E\left[
	R_t^{(k)} + \gamma^k V^\Omega(s_{t+k})
	\given s_t=s, o_t=o
	\right],
	\label{eq:smdp_bellman}
\end{equation}
	where $\Omega$ denotes the high-level policy over options.

\begin{figure}[t]
	\centering
	\begin{tikzpicture}[
		tick/.style={circle,draw,minimum size=0.65cm,font=\small},
		opt/.style={draw,rounded corners,thick,minimum width=2.8cm,minimum height=0.75cm,align=center,font=\small},
		arrow/.style={-{Latex[length=2.2mm]},thick},
		node distance=0.8cm
		]
		\node[tick] (s0) {$s_t$};
		\node[tick,right=of s0] (s1) {$s_{t+1}$};
		\node[tick,right=of s1] (s2) {$s_{t+2}$};
		\node[right=0.45cm of s2] (dots) {$\cdots$};
		\node[tick,right=0.85cm of dots] (sk) {$s_{t+k}$};
		\node[opt,above=0.9cm of s1] (option) {option $o$ executes};
		\draw[arrow] (s0) -- node[below,font=\scriptsize] {$a_t$} (s1);
		\draw[arrow] (s1) -- node[below,font=\scriptsize] {$a_{t+1}$} (s2);
		\draw[arrow] (s2) -- (dots);
		\draw[arrow] (dots) -- node[below,font=\scriptsize] {$a_{t\!+\!k\!-\!1}$} (sk);
		\draw[decorate,decoration={brace,amplitude=5pt},thick] ($(s0.north west)+(0,0.35)$) -- node[above=6pt,font=\small] {duration $k$} ($(sk.north east)+(0,0.25)$);
		\draw[arrow,draw=green!60!black] (option.south) -- (s1.north);
	\end{tikzpicture}
	\caption{In an SMDP, a temporally extended option can last for multiple primitive steps. The high-level decision receives a $k$-step discounted return and then continues from the termination state.}
	\label{fig:smdp_timeline}
\end{figure}
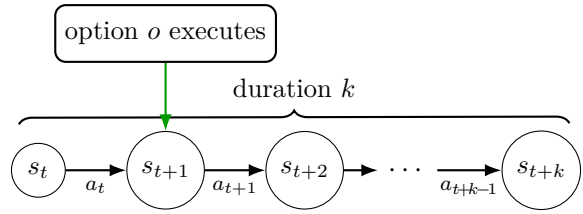

\section{The options framework}

The options framework is the most widely used formalism for temporally extended actions \citep{sutton1999between}. An option $o$ is defined by three components:
\begin{equation}
	o = (I_o, \pi_o, \beta_o),
\end{equation}
where
\begin{itemize}
	\item $I_o \subseteq \mathcal{S}$ is the initiation set, the states where option $o$ can start;
	\item $\pi_o(a\mid s)$ is the intra-option policy, the low-level policy used while the option is active;
	\item $\beta_o(s) \in [0,1]$ is the termination function, the probability that the option terminates in state $s$.
\end{itemize}

The sigmoid parameterization of $\beta_o(s)$ is common, but it is not the only possible termination model. Some systems use deterministic termination rules, learned duration distributions, threshold-based termination, or event-triggered termination conditions when the domain structure is known.

\begin{figure}[t]
	\centering
	\begin{tikzpicture}[
		comp/.style={draw,rounded corners,thick,minimum width=3.0cm,minimum height=0.9cm,align=center,font=\small},
		arrow/.style={-{Latex[length=2.2mm]},thick},
		node distance=1.9cm
		]
		\node[comp,fill=blue!8,draw=blue!70] (init) {Initiation set\\$I_o$};
		\node[comp,fill=green!10,draw=green!60!black,right=of init] (policy) {Intra-option policy\\$\pi_o(a\mid s)$};
		\node[comp,fill=red!7,draw=red!70!black,right=of policy] (term) {Termination rule\\$\beta_o(s)$};
		\draw[arrow,draw=blue!70] (init) -- node[above,font=\scriptsize] {can start} (policy);
		\draw[arrow,draw=green!60!black] (policy) -- node[above,font=\scriptsize] {executes} (term);
		\node[below=0.75cm of policy,align=center,font=\small] {An option is not just a label. It specifies where it can begin, how it acts, and when it stops.};
	\end{tikzpicture}
	\caption{The three components of an option: initiation set, intra-option policy, and termination rule.}
	\label{fig:option_components}
\end{figure}

Primitive actions are a special case of options that last for exactly one step. Therefore, HRL does not replace ordinary RL; it generalizes the action space to include temporally extended actions.

\section{Option-value functions and SMDP Q-learning}

If options are given, the high-level problem can be solved using SMDP variants of value learning. A simple off-policy SMDP Q-learning update is
\begin{equation}
	Q(s_t,o_t)
	\leftarrow
	Q(s_t,o_t)
	+
	\alpha
	\left[
	R_t^{(k)} + \gamma^k \max_{o'} Q(s_{t+k},o') - Q(s_t,o_t)
	\right].
	\label{eq:smdp_q_learning}
\end{equation}
Compared with ordinary Q-learning, the only difference is that the reward is a $k$-step option return and the discount is $\gamma^k$.

\begin{lstlisting}[style=pythonstyle,caption={SMDP Q-learning update for temporally extended options.},label={lst:smdp_q_update}]
def smdp_q_update(Q, s, option, option_rewards, s_next, gamma=0.99, lr=0.1):
    """One SMDP Q-learning update.

    option_rewards: list of primitive rewards collected while option executed.
    """
    k = len(option_rewards)
    discounted_return = 0.0
    for j, r in enumerate(option_rewards):
        discounted_return += (gamma ** j) * r

    target = discounted_return + (gamma ** k) * max(Q[s_next].values())
    td_error = target - Q[s][option]
    Q[s][option] += lr * td_error
    return td_error
\end{lstlisting}

This update is simple, but it assumes that useful options already exist. Much of modern HRL is about learning options, discovering subgoals, or learning continuous latent skills automatically. The next two sections turn to two complementary problems: how to learn options end-to-end with Option-Critic, and how to discover useful subgoals or latent skills.

\section{MAXQ, HAMs, and early decomposition views}

Before deep HRL, several influential frameworks treated hierarchy as task decomposition. Hierarchical abstract machines (HAMs) constrain the policy through a hand-designed finite-state controller \citep{parr1998reinforcement}. MAXQ decomposes the value function of a task into values of subtasks and completion functions \citep{dietterich2000hierarchical}. Feudal RL separates managers that set goals from workers that execute them \citep{dayan1993feudal}.

These methods are historically important because they introduced three ideas that remain central:
\begin{enumerate}
	\item the agent can reason at multiple temporal scales;
	\item subproblems can have their own value functions;
	\item decomposition can make long-horizon tasks easier, but only when the decomposition is meaningful.
\end{enumerate}

\begin{table}[t]
	\centering
	\caption{Early HRL formalisms and their lasting ideas.}
	\label{tab:early_hrl}
	\begin{tabular}{p{3.0cm}p{4.2cm}p{4.6cm}}
		\toprule
		Method & Core idea & Lasting lesson \\
		\midrule
		Feudal RL & Managers set commands for workers & Use different temporal scales for control \\
		HAMs & Hand-designed abstract machines constrain behavior & Structure can reduce search but may bias learning \\
		MAXQ & Decompose task value into subtask values & Value decomposition can expose reusable subtasks \\
		Options & Learn or specify temporally extended actions & Skills can be treated as actions in an SMDP \\
		\bottomrule
	\end{tabular}
\end{table}

\section{Option-Critic: learning options end-to-end}

The option-critic architecture learns intra-option policies and termination functions end-to-end \citep{bacon2017option}. Instead of assuming fixed options, the agent learns both what each option does and when it should stop.

The intra-option policy-gradient update has the form
\begin{equation}
	\nabla_\theta J
	=
	\E\left[
	\nabla_\theta \log \pi_{o,\theta}(a\mid s)\, Q_U(s,o,a)
	\right],
	\label{eq:option_critic_policy_gradient}
\end{equation}
where $Q_U(s,o,a)$ is the value of taking primitive action $a$ while executing option $o$.

The termination gradient is often written as
\begin{equation}
	\nabla_\vartheta J
	=
	-\E\left[
	\nabla_\vartheta \beta_{o,\vartheta}(s')\, A_\Omega(s',o)
	\right],
	\label{eq:option_critic_termination_gradient}
\end{equation}
where
\begin{equation}
	A_\Omega(s',o)
	=
	Q_\Omega(s',o)-V_\Omega(s')
\end{equation}
measures whether continuing option $o$ is better than switching according to the policy over options. If the current option has positive advantage, termination is discouraged. If it has negative advantage, termination is encouraged.

\begin{figure}[t]
	\centering
	\begin{tikzpicture}[
		box/.style={draw,rounded corners,thick,minimum width=3.0cm,minimum height=0.85cm,align=center,font=\small},
		arrow/.style={-{Latex[length=2.2mm]},thick},
		node distance=1.6cm
		]
		\node[box,fill=blue!8,draw=blue!70] (s) {state $s$};
		\node[box,fill=green!10,draw=green!60!black,right=of s] (po) {policy over options\\$\mu(o\mid s)$};
		\node[box,fill=orange!10,draw=orange!80!black,right=of po] (io) {intra-option policy\\$\pi_o(a\mid s)$};
		\node[box,fill=gray!10,draw=gray!70,right=of io] (env) {environment};
		\node[box,fill=red!7,draw=red!70!black,below=1.1cm of io] (term) {termination\\$\beta_o(s)$};
		\draw[arrow,draw=blue!70] (s) -- (po);
		\draw[arrow,draw=green!60!black] (po) -- node[above,font=\scriptsize] {choose $o$} (io);
		\draw[arrow,draw=orange!80!black] (io) -- node[above,font=\scriptsize] {action $a$} (env);
		\draw[arrow,draw=gray!60] (env.south) |- (term.east);
		\draw[arrow,draw=red!70!black] (term.west) -- node[above,font=\scriptsize,align=center,pos=0.25,xshift=-1.9cm] {continue or stop} (po.south);
	\end{tikzpicture}
	\caption{Option-Critic learns intra-option policies and termination functions. A termination rule decides whether control remains with the current option or returns to the policy over options.}
	\label{fig:option_critic_architecture}
\end{figure}
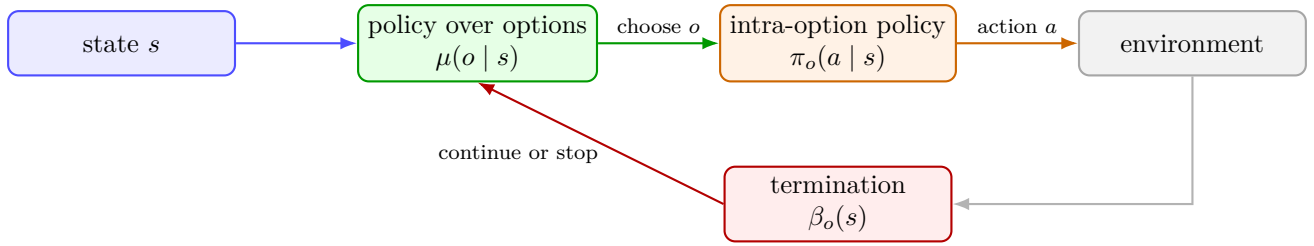

\begin{lstlisting}[style=pythonstyle,caption={A minimal option-critic style neural module.},label={lst:option_critic_module}]
import torch
import torch.nn as nn
import torch.nn.functional as F

class OptionCriticNet(nn.Module):
    def __init__(self, obs_dim, act_dim, num_options, hidden=128):
        super().__init__()
        self.trunk = nn.Sequential(
            nn.Linear(obs_dim, hidden), nn.ReLU(),
            nn.Linear(hidden, hidden), nn.ReLU()
        )
        self.q_options = nn.Linear(hidden, num_options)          # Q_Omega(s,o)
        self.terminations = nn.Linear(hidden, num_options)       # beta_o(s)
        self.intra_options = nn.Linear(hidden, num_options * act_dim)
        self.num_options = num_options
        self.act_dim = act_dim

    def forward(self, obs):
        h = self.trunk(obs)
        q_o = self.q_options(h)
        beta = torch.sigmoid(self.terminations(h))
        logits = self.intra_options(h).view(-1, self.num_options, self.act_dim)
        return q_o, beta, logits

    def option_action_dist(self, obs, option_ids):
        _, _, logits = self.forward(obs)
        chosen_logits = logits[torch.arange(obs.shape[0]), option_ids]
        return torch.distributions.Categorical(logits=chosen_logits)
\end{lstlisting}

\begin{pitfallbox}{Option collapse}
	End-to-end option learning often discovers multiple options that behave almost identically. This is called option collapse. Diversity regularization, termination penalties, information-theoretic objectives, or task-structured subgoals may be needed to make options distinct and reusable. One canonical remedy is a deliberation cost, which penalizes frequent option termination and encourages options to persist long enough to represent distinct, reusable behaviors.
	
\end{pitfallbox}

\section{Goal-conditioned HRL and subgoal discovery}

Another major view of HRL uses goals or subgoals. A high-level policy chooses a subgoal $g_t$, and a low-level goal-conditioned policy acts to reach it:
\begin{equation}
	g_t \sim \pi_H(g\mid s_t),
	\qquad
	a_t \sim \pi_L(a\mid s_t,g_t).
\end{equation}
The low-level policy often receives an intrinsic reward such as
\begin{equation}
	r_t^{\mathrm{int}}
	=
	-\left\| f(s_{t+1}) - g_t \right\|_2,
	\label{eq:goal_intrinsic_reward}
\end{equation}
where $f(s)$ maps states into a goal representation space.

Goal-conditioned HRL makes sense when the environment has meaningful intermediate states. In navigation, subgoals may be waypoints. In manipulation, they may be object poses. In UAV/SDN control, they may be target regions, QoS modes, or traffic classes.

A practical difficulty is that high-level subgoals may be impossible for the low-level policy to achieve. HIRO addresses this by relabeling high-level actions to make off-policy learning more consistent with the low-level behavior that actually occurred \citep{nachum2018data}.

\begin{lstlisting}[style=pythonstyle,caption={HIRO-style subgoal relabeling sketch.},label={lst:hiro_relabeling}]
def relabel_subgoal(states, original_goal, low_level_actions, encoder):
    """Sketch of subgoal relabeling.

    In practice, HIRO searches candidate goals and chooses the one that makes
    the observed low-level action sequence most likely under the current worker.
    This simplified version uses achieved displacement as a relabeled goal.
    """
    z_start = encoder(states[0])
    z_end = encoder(states[-1])
    achieved_goal = z_end - z_start
    return achieved_goal
\end{lstlisting}

\section{Skill discovery and latent-skill policies}

Sometimes rewards are sparse or tasks are not known in advance. In that case, the agent may learn reusable skills without external task rewards. Unsupervised skill discovery methods learn a latent variable $z$ and a policy $\pi(a\mid s,z)$ such that different values of $z$ produce distinguishable behaviors.

DIAYN maximizes the mutual information between the skill $z$ and the states visited by the skill \citep{eysenbach2018diayn}. The conceptual objective is
\begin{equation}
	I(Z;S)=\mathcal{H}(Z)-\mathcal{H}(Z\mid S),
	\label{eq:skill_mi}
\end{equation}
which encourages different skill labels to lead to distinguishable state distributions. Because the true posterior $p(z\mid s)$ is unknown, DIAYN trains a discriminator $q_\phi(z\mid s)$ and uses the variational objective
\begin{equation}
	\max_{\pi,\phi}\; \mathbb{E}\left[\log q_\phi(z\mid s)\right] + \mathcal{H}(Z) + \beta\mathcal{H}(A\mid S,Z).
	\label{eq:diayn_objective}
\end{equation}
The first term rewards states from which the skill can be identified, while the entropy term prevents the skill-conditioned policy from collapsing too early. DADS learns skills whose effects on the environment are predictable and distinguishable \citep{sharma2020dads}. Skill-prior methods such as SPiRL and OPAL learn latent action spaces from offline datasets or demonstrations, then use RL in the learned skill space \citep{pertsch2021spirl,ajay2021opal}.

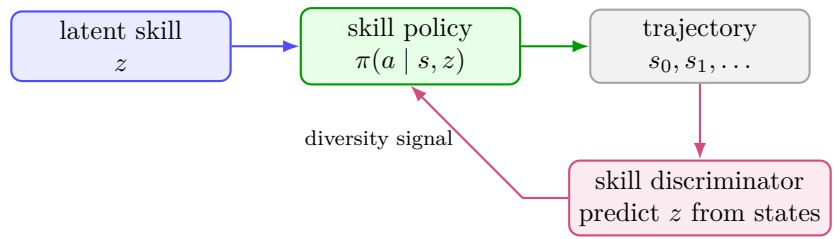
\begin{figure}[t]
	\centering
	\begin{tikzpicture}[
		box/.style={draw,rounded corners,thick,minimum width=2.9cm,minimum height=0.8cm,align=center,font=\small},
		arrow/.style={-{Latex[length=2.2mm]},thick},
		node distance=0.9cm
		]
		\node[box,fill=blue!8,draw=blue!70] (z) {latent skill\\$z$};
		\node[box,fill=green!10,draw=green!60!black,right=of z] (policy) {skill policy\\$\pi(a\mid s,z)$};
		\node[box,fill=gray!10,draw=gray!70,right=of policy] (traj) {trajectory\\$s_0,s_1,\ldots$};
		\node[box,fill=purple!8,draw=purple!70,below=1.0cm of traj] (disc) {skill discriminator\\predict $z$ from states};
		\draw[arrow,draw=blue!70] (z) -- (policy);
		\draw[arrow,draw=green!60!black] (policy) -- (traj);
		\draw[arrow,draw=purple!70] (traj) -- (disc);
		\draw[arrow,draw=purple!70]
		(disc.west)
		-- ++(-0.6,0)
		-- node[above,font=\scriptsize,xshift=-1.4cm]{diversity signal}
		++(-1.0,1.0)
		-- (policy.south);
	\end{tikzpicture}
	\caption{Unsupervised skill discovery learns a latent-conditioned policy whose behaviors are distinguishable. The discriminator provides an intrinsic reward encouraging diverse skills.}
	\label{fig:skill_discovery}
\end{figure}

\section{Feudal, manager-worker, and event-conditioned HRL}

Feudal and manager-worker methods separate high-level intent from low-level action. A manager emits a goal vector or command, and a worker receives intrinsic reward for following it \citep{dayan1993feudal,vezhnevets2017feudal}. This is different from ordinary option learning because the high-level action can be continuous and state-dependent rather than a discrete option index. Manager-worker hierarchies face two credit-assignment problems: across time, deciding which step caused which reward; and across levels, deciding whether a bad outcome was the manager's fault for choosing a poor goal or the worker's fault for failing to execute it.

For cyber-physical and networking systems, a natural extension is \emph{event-conditioned HRL}. The high-level policy does not need to act at a fixed frequency. Instead, high-level decisions can be triggered by events: congestion, battery threshold crossing, QoS violation, safety margin shrinkage, topology change, or new mission demand.

\begin{researchbox}{Event-conditioned HRL for UAV/SDN systems}
	A distinctive HRL architecture for UAV-assisted networks is to let the high-level policy choose mission modes only when important events occur. For example: \emph{serve URLLC hotspot}, \emph{return to charger}, \emph{spread out to reduce interference}, or \emph{follow SDN load-balancing guidance}. The low-level policy then executes continuous control until the next event. This reduces unnecessary high-level switching and connects HRL to safe control, CBF filters, and SDN supervision.
\end{researchbox}

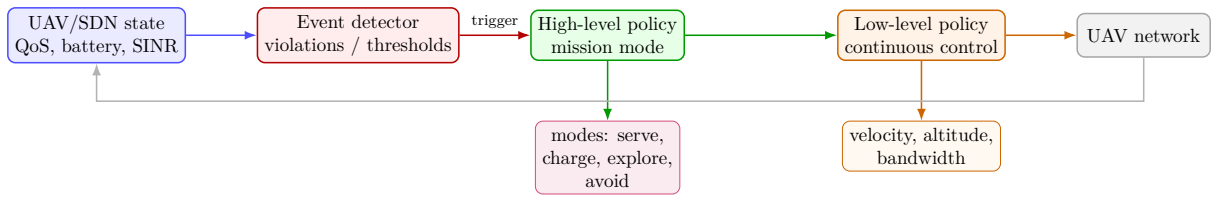
\begin{figure}[t]
	\centering
	\resizebox{\columnwidth}{!}{%
		\begin{tikzpicture}[
			box/.style={draw,rounded corners,thick,minimum width=2.45cm,minimum height=0.8cm,
				align=center,font=\small},
			small/.style={draw,rounded corners,minimum width=2.3cm,minimum height=0.75cm,
				align=center,font=\small},
			arrow/.style={-{Latex[length=2.2mm]},thick},
			node distance=1.3cm
			]

			\node[box,fill=blue!8,draw=blue!70]                        (state)
			{UAV/SDN state\\QoS, battery, SINR};
			\node[box,fill=red!7,draw=red!70!black,right=of state]     (event)
			{Event detector\\violations / thresholds};
			\node[box,fill=green!10,draw=green!60!black,right=of event](high)
			{High-level policy\\mission mode};
			\node[box,fill=orange!10,draw=orange!80!black,
			right=2.8cm of high]                                 (low)
			{Low-level policy\\continuous control};
			\node[box,fill=gray!10,draw=gray!70,right=of low]          (env)
			{UAV network};

			% Bottom boxes — wrapped text to stay narrow
			\node[small,fill=purple!8,draw=purple!70,
			below=1.1cm of high]                                 (modes)
			{modes: serve,\\charge, explore,\\avoid};
			\node[small,fill=orange!5,draw=orange!80!black,
			below=1.1cm of low]                                  (actions)
			{velocity, altitude,\\bandwidth};

			% Arrows
			\draw[arrow,draw=blue!70]         (state) -- (event);
			\draw[arrow,draw=red!70!black]    (event) --
			node[above,font=\scriptsize]{trigger} (high);
			\draw[arrow,draw=green!60!black]  (high)  -- (low);
			\draw[arrow,draw=orange!80!black] (low)   -- (env);
			\draw[arrow,draw=green!60!black]  (high)  -- (modes);
			\draw[arrow,draw=orange!80!black] (low)   -- (actions);
			\draw[arrow,draw=gray!60]
			(env.south) -- ++(0,-0.8) -| (state.south);

		\end{tikzpicture}%
	}
	\caption{Event-conditioned HRL for UAV/SDN control. High-level mission modes are
		triggered by events such as QoS violations, low battery, congestion, or
		safety-margin changes, while low-level policies execute continuous actions.}
	\label{fig:event_conditioned_uav_hrl}
\end{figure}

\section{Offline and sequence-modeling views of HRL}

Chapters 14 and 15 developed offline RL and trajectory sequence modeling. HRL can also be viewed through this lens. Offline datasets often contain repeated behavioral chunks: driving maneuvers, robot manipulation primitives, UAV routing patterns, or network load-balancing responses. Skill-prior methods learn these chunks as latent actions and then plan or optimize in the latent skill space. SPiRL and OPAL are examples of this idea: they compress offline behavior into reusable latent skills and then let downstream RL explore at the skill level rather than at every primitive action step \citep{pertsch2021spirl,ajay2021opal}.

This view is powerful because it avoids learning every primitive behavior from scratch. However, it inherits offline RL's support problem. If the high-level policy selects a skill in a context where that skill was never seen, the low-level decoder may produce unreliable behavior.

\begin{warningbox}{Offline skills inherit dataset support limits}
	Learning a skill library from logs does not remove the offline RL problem. A learned skill prior is trustworthy only where the dataset contains enough examples of similar state-skill pairs.
\end{warningbox}

\section{Hierarchical MARL and UAV/SDN control}

Chapter 16 introduced multi-agent RL. HRL becomes especially useful when many agents must coordinate across different temporal scales. In a multi-UAV network, a global or regional high-level controller may allocate mission roles, while each UAV executes a low-level policy locally. This combines CTDE, hierarchical control, and safe execution.

A concrete two-level design is:
\begin{itemize}
	\item high-level controller every $K$ steps: assign UAVs to service zones, charging, exploration, or interference-avoidance roles;
	\item low-level controller every step: choose movement, altitude, bandwidth split, or power control;
	\item safety layer every step: enforce collision, battery, latency, and interference constraints.
\end{itemize}

\begin{table}[t]
	\centering
	\caption{Example HRL decomposition for UAV-assisted network slicing.}
	\label{tab:uav_hrl_decomposition}
	\begin{tabular}{p{3.0cm}p{4.1cm}p{4.9cm}}
		\toprule
		Level & Decision & Example variables \\
		\midrule
		High level & Mission mode / subgoal & serve URLLC, recharge, explore, spread out, follow SDN guidance \\
		Middle level & Resource mode & prioritize latency, prioritize throughput, conserve energy, fairness mode \\
		Low level & Primitive control & velocity, altitude, user association, bandwidth split, transmit power \\
		Safety layer & Action correction & collision avoidance, CBF projection, battery guard, QoS constraint filter \\
		\bottomrule
	\end{tabular}
\end{table}

\subsection{Concrete numerical example}

Suppose a UAV observes the following state summary:
\begin{equation}
	\mathrm{battery}=18\%,\quad
	\mathrm{URLLC\ violation\ rate}=0.22,\quad
	\mathrm{nearest\ charger}=35\mathrm{m}.
\end{equation}
A flat reward may be ambiguous: serving the hotspot gives immediate QoS reward, but risks battery failure. A high-level critic may assign option values by combining the expected $K$-step option return with a risk or energy penalty:
\begin{equation}
	V_{\mathrm{option}}(s,o)
	=
	R_t^{(K)}(s,o)
	-
	\lambda_e\,\mathrm{EnergyRisk}(s,o),
	\label{eq:uav_option_value}
\end{equation}
where $R_t^{(K)}$ summarizes the temporally extended QoS benefit of executing option $o$ and the second term penalizes options that create battery or mission-completion risk. A high-level critic may then assign option values as follows:

\begin{table}[t]
	\centering
	\caption{Example high-level option values for a UAV.}
	\label{tab:uav_option_values}
	\begin{tabular}{lccc}
		\toprule
		Option & Immediate QoS gain & Energy risk penalty & High-level value \\
		\midrule
		Serve hotspot & $+8.0$ & $-6.5$ & $1.5$ \\
		Move to charger & $-1.5$ & $+6.0$ & $4.5$ \\
		Spread out & $+2.5$ & $-1.0$ & $1.5$ \\
		Explore uncovered cell & $+1.0$ & $-2.0$ & $-1.0$ \\
		\bottomrule
	\end{tabular}
\end{table}

The high-level policy should choose \emph{move to charger}, even though it has negative immediate QoS gain, because the temporally extended value is higher. This is precisely the kind of long-horizon trade-off HRL is designed to represent.

\section{Python implementation patterns}

\subsection{Two-level policy network}

\Needspace{18\baselineskip}
\begin{lstlisting}[style=pythonstyle,caption={A simple two-level HRL actor for discrete modes and continuous actions.},label={lst:two_level_hrl_actor}]
import torch
import torch.nn as nn
import torch.nn.functional as F

class TwoLevelHRLActor(nn.Module):
    def __init__(self, obs_dim, num_modes, cont_action_dim, hidden=128):
        super().__init__()
        self.encoder = nn.Sequential(
            nn.Linear(obs_dim, hidden), nn.ReLU(),
            nn.Linear(hidden, hidden), nn.ReLU()
        )
        self.mode_head = nn.Linear(hidden, num_modes)
        self.low_mean = nn.Linear(hidden + num_modes, cont_action_dim)
        self.low_log_std = nn.Linear(hidden + num_modes, cont_action_dim)

    def forward_high(self, obs):
        h = self.encoder(obs)
        mode_logits = self.mode_head(h)
        return torch.distributions.Categorical(logits=mode_logits), h

    def forward_low(self, obs, mode_id):
        dist_mode, h = self.forward_high(obs)
        mode_onehot = F.one_hot(mode_id, dist_mode.logits.shape[-1]).float()
        z = torch.cat([h, mode_onehot], dim=-1)
        mean = self.low_mean(z)
        log_std = self.low_log_std(z).clamp(-5.0, 2.0)
        std = log_std.exp()
        return torch.distributions.Normal(mean, std)
\end{lstlisting}

\subsection{Event-triggered high-level switching}

\Needspace{16\baselineskip}
\begin{lstlisting}[style=pythonstyle,caption={Event-triggered high-level decision logic for UAV/SDN HRL.},label={lst:event_triggered_hrl}]
def should_resample_mode(state, steps_since_mode, max_duration=50):
    """Return True if the high-level policy should choose a new mode."""
    if steps_since_mode >= max_duration:
        return True
    if state["battery"] < 0.20:
        return True
    if state["urlcc_violation_rate"] > 0.10:
        return True
    if state["collision_margin"] < 0.15:
        return True
    if state["traffic_shift_detected"]:
        return True
    return False
\end{lstlisting}

\subsection{Hierarchical rollout storage}

\Needspace{18\baselineskip}
\begin{lstlisting}[style=pythonstyle,caption={Rollout storage that separates high-level and low-level credit assignment.},label={lst:hierarchical_rollout_storage}]
class HierarchicalRollout:
    def __init__(self):
        self.low_steps = []       # every primitive step
        self.high_segments = []   # one entry per option or mode

    def add_low_step(self, obs, mode, action, reward, done, info):
        self.low_steps.append({
            "obs": obs,
            "mode": mode,
            "action": action,
            "reward": reward,
            "done": done,
            "info": info,
        })

    def close_high_segment(self, start_idx, end_idx, mode, next_obs, gamma=0.99):
        rewards = [x["reward"] for x in self.low_steps[start_idx:end_idx]]
        ret = sum((gamma ** j) * r for j, r in enumerate(rewards))
        self.high_segments.append({
            "start": start_idx,
            "end": end_idx,
            "duration": end_idx - start_idx,
            "mode": mode,
            "return": ret,
            "next_obs": next_obs,
        })
\end{lstlisting}

\subsection{Skill diversity regularizer}

\Needspace{16\baselineskip}
\begin{lstlisting}[style=pythonstyle,caption={A simple skill diversity loss using a discriminator.},label={lst:skill_diversity_loss}]
def skill_discovery_loss(discriminator_logits, skill_ids, action_log_probs,
                         entropy_coef=0.01):
    """DIAYN-style intrinsic objective sketch.

    The discriminator should predict which skill generated the state. The policy
    receives intrinsic reward when its skill is identifiable from behavior.
    """
    predict_skill_loss = F.cross_entropy(discriminator_logits, skill_ids)
    intrinsic_reward = -predict_skill_loss.detach()
    policy_loss = -(action_log_probs * intrinsic_reward).mean()
    entropy_bonus = -entropy_coef * action_log_probs.mean()
    return policy_loss + predict_skill_loss + entropy_bonus
\end{lstlisting}

\section{Practical failure modes}

\begin{table}[t]
	\centering
	\caption{Common HRL failure modes and diagnostics.}
	\label{tab:hrl_failure_modes}
	\begin{tabular}{p{3.2cm}p{4.2cm}p{4.9cm}}
		\toprule
		Symptom & Likely cause & What to inspect \\
		\midrule
		All options behave similarly & Option collapse & Option action distributions, visited states, skill discriminator accuracy \\
		High-level switches too often & Termination too easy or high-level entropy too high & Option durations, termination probabilities \\
		High-level never switches & Termination too hard or sticky modes & Mode duration histogram, termination loss \\
		Low-level ignores subgoal & Weak intrinsic reward or unreachable goals & Goal achievement distance, low-level success rate \\
		Good low-level skills but poor task reward & High-level policy chooses wrong skills & High-level advantage estimates and option values \\
		Unsafe behavior persists & Safety not applied at primitive scale & Whether CBF/projection runs at every low-level step \\
		Works in training but fails after deployment & Skill prior outside support or changed dynamics & Dataset support, scenario drift, environment mismatch \\
		\bottomrule
	\end{tabular}
\end{table}

\begin{pitfallbox}{The high-level policy can hide low-level bugs}
	If a low-level controller cannot reliably execute a subgoal, the high-level policy may learn strange compensating behavior. Always evaluate low-level skill success independently before blaming the manager.
\end{pitfallbox}

\section{Research frontiers toward 2026}

Several directions define the modern frontier of HRL.

\paragraph{Foundation-model-guided hierarchy.}
Large language models can propose subgoals, task decompositions, or parameterized action primitives. Recent LLM-augmented HRL work combines planning-like high-level guidance with RL-trained low-level skills for long-horizon control \citep{zhang2025llmhrl}. This is promising, but it raises grounding and safety issues: a language model may propose plausible but physically infeasible subgoals. This is one place where Chapter~18's constraint-based safety methods become essential: even a foundation-model-proposed subgoal must be checked against hard physical constraints before execution.

\paragraph{Event-conditioned safe hierarchy.}
For robotics and networks, high-level decisions should not necessarily occur at fixed intervals. Event-triggered hierarchy can reduce unnecessary switching and align decision-making with safety-critical changes.

\paragraph{Offline skill learning.}
Large logs contain behavioral structure. Learning skill priors from logs and then optimizing over those skills can improve sample efficiency. However, offline support limitations remain central.

\paragraph{Hierarchical MARL.}
In multi-agent systems, hierarchy can organize both temporal abstraction and role assignment. This is especially relevant for UAV swarms, multi-robot teams, and network-slicing control.

\paragraph{Hierarchical reasoning in language agents.}
Reasoning-oriented LLM agents may be interpreted hierarchically: planning tokens or high-level outlines guide lower-level token generation. This connects HRL to the policy-optimization and reasoning-model themes from Chapters 10 and 15.

\section{Limitations and when not to use HRL}

\begin{enumerate}[leftmargin=*]
	\item \textbf{Hierarchy is not free.} Poorly designed option sets or unstable termination conditions can make learning slower than a flat baseline.
	\item \textbf{Option collapse is common.} End-to-end option learning often discovers redundant options. Diversity regularization, deliberation costs, or structured subgoals may be needed.
	\item \textbf{Credit assignment across levels is hard.} A bad team outcome may be the manager's fault for choosing a wrong goal or the worker's fault for failed execution. Separating these contributions is difficult.
	\item \textbf{Goal reachability is not guaranteed.} A high-level policy may assign subgoals that the low-level controller cannot reach, especially in unseen situations.
	\item \textbf{Offline skill support limits apply.} Skills learned from logs inherit offline RL's distribution-shift risks.
	\item \textbf{Safety at every level is required.} A safe high-level option does not guarantee safe low-level execution. Safety filters must act at the primitive-action level.
\end{enumerate}

\section{Exercises}

\subsection*{Conceptual exercises}
\begin{enumerate}[leftmargin=*]
	\item Explain the difference between a primitive action and an option.
	\item Why can temporal abstraction reduce the effective planning horizon?
	\item Give one example where hierarchy can hurt performance.
	\item Explain option collapse. How could you detect it from logs?
	\item In a UAV network, should charging be a low-level action, a high-level option, or both? Explain.
\end{enumerate}

\subsection*{Mathematical exercises}
\begin{enumerate}[leftmargin=*]
	\item Derive the SMDP Q-learning update in Eq.~\eqref{eq:smdp_q_learning} from the SMDP Bellman equation in Eq.~\eqref{eq:smdp_bellman}.
	\item Suppose an option lasts $k=4$ steps and receives rewards $1,0,2,3$ with $\gamma=0.9$. Compute the option return $R_t^{(k)}$.
	\item Explain why the discount term after option termination is $\gamma^k$ rather than $\gamma$.
	\item Show how the termination gradient in Eq.~\eqref{eq:option_critic_termination_gradient} encourages termination when $A_\Omega(s',o)<0$.
\end{enumerate}

\subsection*{Coding exercises}
\begin{enumerate}[leftmargin=*]
	\item Extend Listing~\ref{lst:two_level_hrl_actor} to support discrete low-level actions.
	\item Implement a histogram of option durations from Listing~\ref{lst:hierarchical_rollout_storage} and use it to detect high-level switching collapse.
	\item Build a small grid-world with four options: go north, go south, go east, and go west. Compare primitive Q-learning with SMDP Q-learning over options.
	\item Implement an event-triggered UAV controller using Listing~\ref{lst:event_triggered_hrl}. Report how often the high-level policy switches under normal load and congestion.
\end{enumerate}

\subsection*{Research thinking exercises}
\begin{enumerate}[leftmargin=*]
	\item Design an HRL architecture for UAV-assisted network slicing with three service classes: URLLC, eMBB, and mMTC. Define high-level modes and low-level actions.
	\item How would you combine HRL with a CBF safety layer? Should the CBF filter high-level options, low-level actions, or both?
	\item Compare HRL and model-based planning for long-horizon UAV missions. When would you prefer each?
	\item Propose a metric for evaluating whether learned options are reusable across scenarios.
	\item Explain how LLM-generated subgoals could fail in a physical UAV system.
\end{enumerate}

\section*{Looking Ahead to Chapter 18: Food for Thought}
\addcontentsline{toc}{section}{Looking Ahead to Chapter 18: Food for Thought}

Hierarchy gives agents temporal abstraction, but abstraction alone does not guarantee safety. A high-level option such as \emph{serve hotspot} may still violate collision constraints, battery constraints, interference limits, or latency guarantees. The next chapter therefore turns to safe reinforcement learning.

\begin{quote}
	Chapter~17 explained how agents can organize decisions across time. Chapter~18 asks how agents can act under constraints that must not be violated.
\end{quote}

Questions to carry forward:
\begin{enumerate}[leftmargin=*]
	\item What is the difference between a soft penalty and a hard safety constraint?
	\item Can a high-level policy be safe if the low-level controller is unsafe?
	\item Should safety be learned, enforced by a filter, or both?
	\item How do Lagrangian methods connect to the entropy-temperature tuning from Chapter~11?
	\item How can CBFs modify actions while preserving as much of the learned policy as possible?
\end{enumerate}
	\chapter[Safe Reinforcement Learning]{Safe Reinforcement Learning}
\chaptermark{Safe RL}
\label{ch:safe_rl}

\begin{keybox}{Chapter goal}
    Safe reinforcement learning extends reward maximization to systems where actions must also respect hard constraints, cost budgets, and safety requirements. This chapter develops constrained MDPs, Lagrangian relaxation, CPO, shields, control barrier functions, and risk-sensitive safety, with particular attention to UAV/SDN and SD-WAN applications where safety is a design requirement rather than an optional penalty.
\end{keybox}

\section*{Chapter Overview}
\addcontentsline{toc}{section}{Chapter Overview}

\begin{enumerate}[leftmargin=*]
    \item Why safety changes the reinforcement-learning problem
    \item What does ``safe'' mean?
    \item Constrained Markov decision processes
    \item Cost returns and safety critics
    \item Lagrangian relaxation and primal-dual learning
    \item PPO-Lagrangian and PID/ADRC Lagrangian control
    \item Constrained Policy Optimization
    \item Shields, action masks, and recovery policies
    \item Control Barrier Functions
    \item Learning with CBF safety filters
    \item Lyapunov, reachability, and risk-sensitive safety
    \item Safe offline, model-based, and multi-agent RL
    \item UAV/SDN safe-RL scenario
    \item Python implementation patterns
    \item Evaluation, debugging, and failure modes
    \item Research frontiers toward 2026
    \item Limitations and when not to use constrained RL
    \item Exercises
    \item Looking Ahead to Chapter 19
\end{enumerate}

\section{Why safety changes the reinforcement-learning problem}

Most chapters so far optimized expected return. DQN, PPO, SAC, MuZero, Decision Transformer, MARL, and HRL all asked variants of the same question: how can an agent choose actions that produce high long-term reward? In real systems, that question is incomplete. A UAV policy that obtains high throughput by flying too close to obstacles is not acceptable. A network controller that increases average bandwidth while violating URLLC latency bounds is not acceptable. A robot that learns quickly by repeatedly crashing during training is not acceptable.

Safe reinforcement learning (safe RL) studies how to learn and deploy policies that optimize performance while respecting safety requirements. The field combines reinforcement learning, constrained optimization, control theory, verification, risk analysis, and system engineering \citep{garcia2015comprehensive,wachi2024constraint_survey,ji2024omnisafe}. The core challenge is that safety is not merely another reward term. A constraint violation may be rare but catastrophic; it may be unacceptable during training; and it may not be representable as a smooth scalar penalty.

\begin{keybox}{Core intuition}
    Ordinary RL asks: what action maximizes reward? Safe RL asks: what action improves reward while keeping the system inside an acceptable risk envelope?
\end{keybox}

This chapter resolves several forward-pointers from earlier chapters. Chapter~11 introduced entropy-temperature tuning as a Lagrangian dual idea; this chapter applies the same primal-dual logic to safety constraints. Chapter~16 and Chapter~17 pointed toward constrained and safe multi-agent/hierarchical control; this chapter provides the safety machinery used by those systems. The UAV/SDN examples throughout the book now become central: safety is not an optional add-on for autonomous network control, but a design requirement.

\begin{figure}[t]
    \centering
    \begin{tikzpicture}[
        box/.style={draw,rounded corners,thick,minimum width=3.1cm,minimum height=0.85cm,align=center,font=\small},
        arrow/.style={-{Latex[length=2.2mm]},thick},
        node distance=0.9cm
        ]
        \node[box,fill=blue!8,draw=blue!70] (rl) {Reward learning\\maximize $J_R(\pi)$};
        \node[box,fill=orange!10,draw=orange!80!black,right=of rl] (constraints) {Safety constraints\\keep $J_C(\pi)\le d$};
        \node[box,fill=green!10,draw=green!60!black,right=of constraints] (safe) {Safe policy\\performant + feasible};
        \node[box,fill=red!6,draw=red!70!black,below=1.0cm of constraints] (risk) {Unacceptable events\\collision, overload, QoS violation};
        \draw[arrow,draw=blue!70] (rl) -- (constraints);
        \draw[arrow,draw=green!60!black] (constraints) -- (safe);
        \draw[arrow,draw=red!70!black] (risk) -- (constraints);
    \end{tikzpicture}
    \caption{Safe RL extends reward maximization with explicit constraints. The policy must optimize task performance while keeping expected, probabilistic, or state-wise safety quantities within acceptable limits.}
    \label{fig:safe_rl_core}
\end{figure}
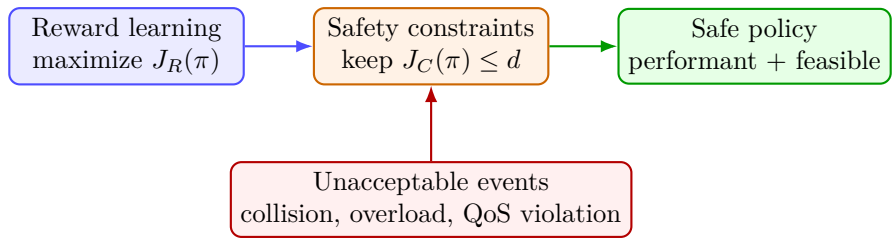

\section{What does ``safe'' mean?}

A common mistake is to treat safety as a single concept. In practice, different safe-RL papers solve different problems. Before choosing an algorithm, the designer must decide what kind of safety is required.

\begin{table}[t]
    \centering
    \caption{Common safety notions in reinforcement learning.}
    \label{tab:safety_notions}
    \begin{tabular}{p{3.2cm}p{5.1cm}p{4.1cm}}
        \toprule
        Safety notion & Meaning & Typical method \\
        \midrule
        Expected-cost safety & Average discounted or episodic cost must stay below a budget & CMDP, Lagrangian, CPO \\
        Chance constraints & Probability of violation must be small & risk-sensitive RL, constrained MPC \\
        State-wise invariance & The state must never leave a safe set & CBFs, reachability, shields \\
        Training-time safety & Unsafe exploration is not allowed during learning & safe exploration, CBF filtering, recovery RL \\
        Deployment safety & Final policy must be safe after training & offline evaluation, shields, formal monitors \\
        Robust safety & Safety must hold under model error or disturbances & robust CBF, uncertainty-aware MPC \\
        Multi-agent safety & Constraints couple several agents & safe MARL, pairwise CBF, decentralized shields \\
        \bottomrule
    \end{tabular}
\end{table}

The table also shows why no single safe-RL algorithm solves every safety problem. CMDP methods are natural for cumulative or expected cost budgets; CBFs and shields are more appropriate for state-wise invariance; chance constraints and CVaR are useful when rare events matter; and robust or multi-agent safety requires explicit reasoning about uncertainty and coupled constraints.

\begin{warningbox}{Safety is not just negative reward}
    Reward penalties can encourage safe behavior, but they rarely provide guarantees. If collision is assigned a penalty of $-100$, an agent may still choose collision if it expects more reward elsewhere, if the penalty is incorrectly scaled, or if the value function extrapolates poorly. Constraint-based methods separate ``what we want'' from ``what must not happen.''
\end{warningbox}

For UAV/SDN systems, different constraints coexist. Collision avoidance is state-wise and physical. Latency bounds may be windowed and statistical. Battery reserve may be a mission-level constraint. Interference and spectrum limits may be regulatory. A single scalar reward cannot cleanly express all of these.

\section{Constrained Markov decision processes}

The most common mathematical framework for safe RL is the Constrained Markov Decision Process (CMDP) \citep{altman1999constrained,achiam2017cpo,wachi2024constraint_survey}. A CMDP extends the MDP from Chapter~2 by adding cost signals in addition to reward.

A discounted CMDP can be written as
\begin{equation}
    \mathcal{M}_c = (\mathcal{S}, \mathcal{A}, P, r, c_1,\ldots,c_m, \gamma, \rho_0),
\end{equation}
where $r(s,a)$ is the reward and $c_i(s,a)$ is the $i$-th cost. The reward objective is
\begin{equation}
    J_R(\pi)=\E_{\tau\sim\pi}\left[\sum_{t=0}^{\infty}\gamma^t r(s_t,a_t)\right],
\end{equation}
while each safety constraint is
\begin{equation}
    J_{C_i}(\pi)=\E_{\tau\sim\pi}\left[\sum_{t=0}^{\infty}\gamma^t c_i(s_t,a_t)\right]\le d_i.
\end{equation}
The constrained optimization problem is
\begin{equation}
    \begin{aligned}
        \max_{\pi}\quad & J_R(\pi) \\
        \text{subject to}\quad & J_{C_i}(\pi) \le d_i, \qquad i=1,\ldots,m.
    \end{aligned}
    \label{eq:cmdp_problem}
\end{equation}

This formulation is deliberately different from reward shaping. Reward and cost are measured separately. The optimization objective asks for high reward only among policies whose costs respect the budgets.

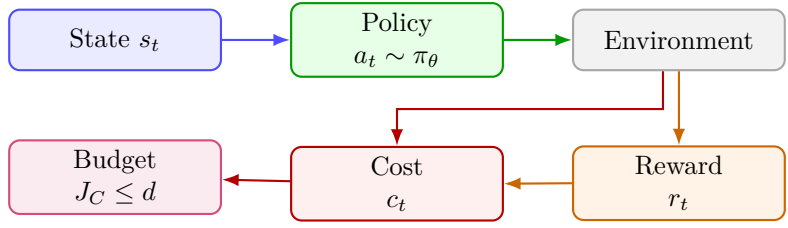
\begin{figure}[t]
    \centering
    \begin{tikzpicture}[
        box/.style={draw,rounded corners,thick,minimum width=2.8cm,minimum height=0.8cm,align=center,font=\small},
        arrow/.style={-{Latex[length=2.2mm]},thick},
        node distance=0.9cm
        ]
        
        \node[box,fill=blue!8,draw=blue!70]                        (s)      {State $s_t$};
        \node[box,fill=green!10,draw=green!60!black,right=of s]    (a)      {Policy\\$a_t\sim\pi_\theta$};
        \node[box,fill=gray!10,draw=gray!70,right=of a]            (env)    {Environment};
        
        \node[box,fill=orange!10,draw=orange!80!black,
        below=0.98cm of env]                                  (reward) {Reward\\$r_t$};
        \node[box,fill=red!6,draw=red!70!black,
        below=0.9cm of a]                                    (cost)   {Cost\\$c_t$};
        \node[box,fill=purple!8,draw=purple!70,
        below=0.9cm of s]                                    (budget) {Budget\\$J_C\le d$};
        
        \draw[arrow,draw=blue!70]        (s)   -- (a);
        \draw[arrow,draw=green!60!black] (a)   -- (env);

        \draw[arrow,draw=orange!80!black]
        (env.south) -- (reward.north);

        \draw[arrow,draw=red!70!black]
        ([xshift=-6pt]env.south) -- ++(0,-0.5) -| (cost.north);

        \draw[arrow,draw=red!70!black] (cost) -- (budget);

        \draw[arrow,draw=orange!80!black]
        (reward.west) -- (cost.east);
        
    \end{tikzpicture}
    \caption{A CMDP separates reward from cost. Reward measures task performance, while cost measures constraint usage or violations. Safe policy optimization must optimize both streams.}
    \label{fig:cmdp_two_signals}
\end{figure}

\section{Cost returns and safety critics}

Actor-critic safe RL usually learns two kinds of critics: a reward critic and a cost critic. The reward critic estimates reward return,
\begin{equation}
    V_R^{\pi}(s)=\E_\pi\left[\sum_{t=0}^{\infty}\gamma^t r_t\given s_0=s\right],
\end{equation}
while the cost critic estimates safety usage,
\begin{equation}
    V_C^{\pi}(s)=\E_\pi\left[\sum_{t=0}^{\infty}\gamma^t c_t\given s_0=s\right].
\end{equation}
The cost advantage is
\begin{equation}
    A_C(s_t,a_t)=Q_C(s_t,a_t)-V_C(s_t).
\end{equation}
If $A_C>0$, the action is more costly than expected in that state; if $A_C<0$, it is safer than expected. This mirrors the advantage logic from Chapter~8, but with the sign interpreted differently: reward advantage should be increased, cost advantage should be decreased.

\begin{table}[t]
    \centering
    \caption{Reward and cost quantities in actor-critic safe RL.}
    \label{tab:reward_cost_critics}
    \begin{tabular}{p{3.2cm}p{4.5cm}p{4.5cm}}
        \toprule
        Quantity & Reward side & Cost side \\
        \midrule
        Signal & $r_t$ & $c_t$ \\
        Return & $G_R$ & $G_C$ \\
        Critic & $V_R,Q_R$ & $V_C,Q_C$ \\
        Advantage & Increase probability if positive & Decrease probability if positive \\
        Budget & none or performance target & $J_C(\pi)\le d$ \\
        Interpretation & usefulness & safety consumption or violation risk \\
        \bottomrule
    \end{tabular}
\end{table}

\section{Lagrangian relaxation and primal-dual learning}

The most practical way to train constrained policies is often Lagrangian relaxation. For one constraint, define
\begin{equation}
    \mathcal{L}(\pi,\lambda)=J_R(\pi)-\lambda\left(J_C(\pi)-d\right),
    \label{eq:lagrangian_safe_rl}
\end{equation}
where $\lambda\ge 0$ is a Lagrange multiplier. The policy maximizes the Lagrangian, while the multiplier increases when the constraint is violated:
\begin{equation}
    \lambda \leftarrow \left[\lambda + \eta_\lambda (J_C(\pi)-d)\right]_+.
    \label{eq:lambda_update}
\end{equation}

This is the same Lagrangian-dual formulation introduced for entropy regularization in Chapter~11, Section~11.10, but the meaning changes. In SAC, the multiplier controls entropy. In safe RL, the multiplier controls safety cost. If the policy violates the cost budget, $\lambda$ grows and the actor receives a stronger penalty for unsafe behavior.

\begin{keybox}{Lagrangian interpretation}
    The multiplier $\lambda$ is not just a hyperparameter. It is a learned price of constraint violation. When safety cost is above the budget, the price rises. When the policy becomes safely conservative, the price can fall.
\end{keybox}

For policy-gradient methods, the actor update can be written using a Lagrangian advantage:
\begin{equation}
    A_{\mathcal{L}}(s,a)=A_R(s,a)-\lambda A_C(s,a).
    \label{eq:lagrangian_advantage}
\end{equation}
This turns PPO or SAC into a constrained variant with relatively small code changes. However, the simplicity hides a serious issue: primal-dual learning can oscillate. If $\lambda$ reacts too slowly, the policy violates constraints. If it reacts too aggressively, the policy becomes overly conservative.

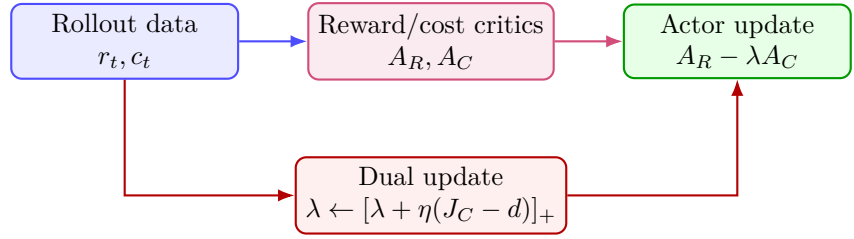
\begin{figure}[t]
    \centering
    \begin{tikzpicture}[
        box/.style={draw,rounded corners,thick,minimum width=3.0cm,minimum height=0.85cm,align=center,font=\small},
        arrow/.style={-{Latex[length=2.2mm]},thick},
        node distance=0.9cm
        ]
        \node[box,fill=blue!8,draw=blue!70] (rollout) {Rollout data\\$r_t,c_t$};
        \node[box,fill=purple!8,draw=purple!70,right=of rollout] (critics) {Reward/cost critics\\$A_R,A_C$};
        \node[box,fill=green!10,draw=green!60!black,right=of critics] (actor) {Actor update\\$A_R-\lambda A_C$};
        \node[box,fill=red!6,draw=red!70!black,below=1.0cm of critics] (lambda) {Dual update\\$\lambda\leftarrow[\lambda+\eta(J_C-d)]_+$};
        \draw[arrow,draw=blue!70] (rollout) -- (critics);
        \draw[arrow,draw=purple!70] (critics) -- (actor);
        \draw[arrow,draw=red!70!black] (rollout.south) |- (lambda.west);
        \draw[arrow,draw=red!70!black] (lambda.east) -| (actor.south);
    \end{tikzpicture}
    \caption{Primal-dual safe RL. The actor optimizes a reward-cost trade-off, while the Lagrange multiplier adapts to constraint violation.}
    \label{fig:lagrangian_loop}
\end{figure}

\section{PPO-Lagrangian and PID/ADRC Lagrangian control}

PPO-Lagrangian combines PPO's clipped objective from Chapter~10 with the Lagrangian advantage in Eq.~\eqref{eq:lagrangian_advantage}. A simplified policy loss is
\begin{equation}
    L_{\pi}(\theta)=
    -\E\left[
    \min\left(
    r_t(\theta)A_{\mathcal{L},t},
    \clip(r_t(\theta),1-\epsilon,1+\epsilon)A_{\mathcal{L},t}
    \right)
    \right].
\end{equation}
This is attractive because it reuses a familiar PPO implementation. Safety Gym and Safety-Gymnasium popularized PPO-Lagrangian baselines for constrained continuous-control tasks \citep{ray2019safetygym,ji2023safetygymnasium}.

The basic multiplier update in Eq.~\eqref{eq:lambda_update} behaves like integral control: it accumulates constraint error. Stooke, Achiam, and Abbeel proposed PID Lagrangian methods that add proportional and derivative terms to reduce oscillation and overshoot \citep{stooke2020pid}. Recent work continues to explore more responsive Lagrangian controllers, including ADRC-style updates for safe RL \citep{zhang2026adrc}.

\begin{warningbox}{Lagrangian methods do not guarantee zero violations during learning}
    A policy may violate constraints while the multiplier is still adapting. Lagrangian safe RL is often useful for expected-cost constraints, but it should not be confused with a hard safety filter such as a CBF shield. For physical systems, training-time and deployment-time safeguards are often still needed.
\end{warningbox}

\section{Constrained Policy Optimization}

Constrained Policy Optimization (CPO) is a trust-region method designed for CMDPs \citep{achiam2017cpo}. It extends the TRPO idea from Chapter~10 by imposing a local constraint on the expected cost. A simplified CPO update solves an approximate problem of the form
\begin{equation}
    \begin{aligned}
        \max_{\Delta\theta}\quad & g^T\Delta\theta \\
        \text{subject to}\quad & b^T\Delta\theta + J_C(\pi_{\theta_{old}})-d \le 0, \\
        & \frac{1}{2}\Delta\theta^T H\Delta\theta \le \delta,
    \end{aligned}
\end{equation}
where $g$ is the reward-gradient estimate, $b$ is the cost-gradient estimate, and $H$ approximates the policy KL geometry. CPO is more mathematically structured than PPO-Lagrangian, but more complex to implement. In practice, CPO also uses a backtracking line search to check whether the proposed update satisfies the local reward and cost approximations before accepting the step.

\begin{table}[t]
    \centering
    \caption{CPO versus PPO-Lagrangian.}
    \label{tab:cpo_vs_lag}
    \begin{tabular}{p{3.0cm}p{4.6cm}p{4.6cm}}
        \toprule
        Aspect & CPO & PPO-Lagrangian \\
        \midrule
        Main idea & Constrained trust-region update & Penalize cost with adaptive multiplier \\
        Strength & Near-constraint satisfaction per update under approximations & Simple and easy to add to PPO/SAC \\
        Weakness & More complex; needs second-order approximations & Can oscillate; no hard per-step guarantee \\
        Best use & Careful policy-search experiments & Practical baselines and scalable deep RL \\
        \bottomrule
    \end{tabular}
\end{table}

\section{Shields, action masks, and recovery policies}

Another approach is to keep the learning algorithm mostly unchanged but block unsafe actions. A shield is a runtime module that modifies or rejects unsafe actions before they reach the environment. For discrete actions, this may be an action mask. For continuous actions, it may be a projection operator. Shielding is a canonical safe-RL idea: the policy may propose an action, but a formally or empirically constructed shield decides whether that action can be executed safely \citep{alshiekh2018shielding}.

A recovery policy learns to intervene when the nominal policy is likely to enter an unsafe state. Recovery RL trains a policy to recover from risky regions rather than only penalizing failures \citep{thananjeyan2021recovery}. This is especially useful when the boundary between safe and unsafe behavior is easier to recognize than the full optimal policy.

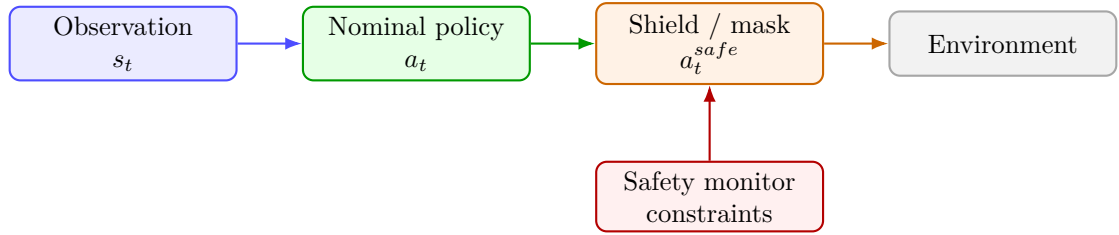
\begin{figure}[t]
    \centering
    \begin{tikzpicture}[
        box/.style={draw,rounded corners,thick,minimum width=3.0cm,minimum height=0.85cm,align=center,font=\small},
        arrow/.style={-{Latex[length=2.2mm]},thick},
        node distance=0.85cm
        ]
        \node[box,fill=blue!8,draw=blue!70] (obs) {Observation\\$s_t$};
        \node[box,fill=green!10,draw=green!60!black,right=of obs] (policy) {Nominal policy\\$a_t$};
        \node[box,fill=orange!10,draw=orange!80!black,right=of policy] (shield) {Shield / mask\\$a_t^{safe}$};
        \node[box,fill=gray!10,draw=gray!70,right=of shield] (env) {Environment};
        \node[box,fill=red!6,draw=red!70!black,below=1.0cm of shield] (monitor) {Safety monitor\\constraints};
        \draw[arrow,draw=blue!70] (obs) -- (policy);
        \draw[arrow,draw=green!60!black] (policy) -- (shield);
        \draw[arrow,draw=orange!80!black] (shield) -- (env);
        \draw[arrow,draw=red!70!black] (monitor) -- (shield);
    \end{tikzpicture}
    \caption{Shielded RL. The policy proposes an action, but a safety module masks, rejects, or projects the action before execution. The critic should learn from the executed action, not only the proposed unsafe action.}
    \label{fig:shielded_rl}
\end{figure}

\section{Control Barrier Functions}

Control Barrier Functions (CBFs) provide a control-theoretic way to keep a dynamical system inside a safe set \citep{ames2019cbf,guerrier2024learning_cbf,kushwaha2025review}. Let the safe set be
\begin{equation}
    \mathcal{C}=\{x\in\R^n : h(x)\ge 0\},
\end{equation}
where $h(x)$ is a barrier function. The goal is forward invariance: if $x_0\in\mathcal{C}$, then $x_t\in\mathcal{C}$ for all future time.

For a continuous-time control-affine system
\begin{equation}
    \dot{x}=f(x)+g(x)u,
\end{equation}
a common CBF condition is
\begin{equation}
    L_f h(x)+L_g h(x)u+\alpha(h(x))\ge 0,
    \label{eq:ct_cbf}
\end{equation}
where $\alpha$ is an extended class-$\mathcal{K}$ function. For discrete-time RL, a common analogue is
\begin{equation}
    h(x_{t+1}) \ge (1-\eta)h(x_t), \qquad \eta\in[0,1].
    \label{eq:dt_cbf}
\end{equation}
This says that the next state should not reduce the safety margin too quickly. Discrete-time CBF formulations are especially natural for RL and model-predictive settings, where the controller reasons in sampled time rather than continuous time \citep{agrawal2017discrete_cbf}.

Given a nominal RL action $u_{rl}$, a CBF safety filter solves a projection problem:
\begin{equation}
    \begin{aligned}
        u^* = \argmin_u \quad & \|u-u_{rl}\|_2^2 \\
        \text{subject to}\quad & L_f h(x)+L_g h(x)u+\alpha(h(x))\ge 0, \\
        & u\in\mathcal{U}.
    \end{aligned}
    \label{eq:cbf_qp}
\end{equation}
This minimally modifies the action while enforcing the barrier constraint. CBF-RL methods combine learned policies with such filters during training or deployment \citep{cheng2019end,yang2025cbfrl,kaypak2026pects}. Recent work has extended this framework to networking applications. Uncertainty-aware neural CBFs use ensembles of barrier functions or predictive models to represent epistemic uncertainty in learned dynamics, and have been applied to SD-WAN traffic engineering with per-class QoS constraints \citep{bista2026vtc,bista2026ifip}.

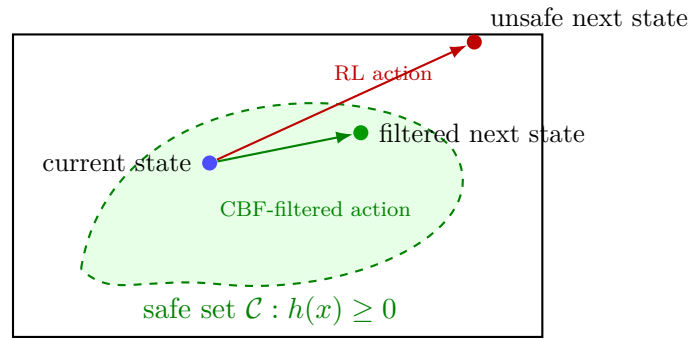
\begin{figure}[t]
    \centering
    \begin{tikzpicture}[
        arrow/.style={-{Latex[length=2.2mm]},thick}
        ]
        
        \draw[thick] (-3.5,-2.0) rectangle (3.5,2.0);
        
        \fill[green!10]
        (-2.6,-1.1) .. controls (-2.2,1.4) and (1.4,1.5)  .. (2.3,0.5)
        .. controls (3.0,-0.5) and (1.1,-1.5)  .. (-0.8,-1.3)
        .. controls (-1.8,-1.2) and (-2.2,-1.5) .. (-2.6,-1.1);
        \draw[thick,green!50!black,dashed]
        (-2.6,-1.1) .. controls (-2.2,1.4) and (1.4,1.5)  .. (2.3,0.5)
        .. controls (3.0,-0.5) and (1.1,-1.5)  .. (-0.8,-1.3)
        .. controls (-1.8,-1.2) and (-2.2,-1.5) .. (-2.6,-1.1);
        
        \node[green!50!black] at (-0.1,-1.65) {safe set $\mathcal{C}:h(x)\ge 0$};
        
        \node[circle,fill=blue!70,inner sep=2pt,
        label=left:{\small current state}]             (x)    at (-0.9, 0.3) {};
        \node[circle,fill=red!75!black,inner sep=2pt,
        label=above right:{\small unsafe next state}]  (bad)  at ( 2.6, 1.9) {};
        \node[circle,fill=green!60!black,inner sep=2pt,
        label=right:{\small filtered next state}]      (good) at ( 1.1, 0.7) {};
        
        \draw[arrow,draw=red!75!black]   (x) -- (bad);
        \draw[arrow,draw=green!50!black] (x) -- (good);
        
        \node[font=\scriptsize,text=red!75!black]   at (1.4,1.5)  {RL action};
        \node[font=\scriptsize,text=green!50!black] at (0.5,-0.3) {CBF-filtered action};
        
    \end{tikzpicture}
    \caption{CBF safety filtering. A nominal RL action may point outside the safe set; the CBF projection modifies it minimally to preserve a safety margin. The safety filter projects the nominal action back into the viable safe direction.}
    \label{fig:cbf_geometry}
\end{figure}

\section{Learning with CBF safety filters}

CBF filters raise an important learning question: should the actor learn from the proposed action or the executed action? If the policy proposes $a_{rl}$ but the filter executes $a_{safe}$, the environment transition corresponds to $a_{safe}$. The critic should therefore be trained on the executed action. Otherwise, the value function assigns outcomes to actions that were never actually applied.

There are three common integration patterns:
\begin{enumerate}[leftmargin=*]
    \item \textbf{Deployment-only filter}: train normally, filter only at runtime. Simple but the policy may never internalize safety.
    \item \textbf{Training-time filter}: filter actions during rollouts and train on executed actions. Safer exploration, but the policy may rely on the filter.
    \item \textbf{Filter-aware learning}: include filter residuals, barrier margins, or safety losses so the policy learns to propose safe actions directly.
\end{enumerate}

Recent CBF-RL work studies training-time filtering, learned barrier functions, uncertainty-aware neural barriers, and model-predictive CBF filtering under uncertainty \citep{guerrier2024learning_cbf,yang2025cbfrl,nakamura2025latent_cbf,kaypak2026pects,bista2026vtc,bista2026ifip}.

\begin{researchbox}{Research signature: safe UAV/SDN and SD-WAN control}
    For UAV/SDN and SD-WAN control, the most useful architecture is often not ``RL versus CBF.'' It is actor-critic learning with a safety layer: the actor proposes movement, routing, power, or bandwidth actions; the CBF or projection layer enforces collision, battery, interference, latency, jitter, and QoS constraints; the reward critic learns service quality; and the cost critic learns how close the system is to constraint violation. This is the architecture used in our own SD-WAN traffic-engineering work, where uncertainty-aware and ensemble-based neural CBFs are combined with RL to enforce multi-class QoS constraints \citep{bista2026vtc,bista2026ifip}.
\end{researchbox}

This executed-action principle has appeared implicitly throughout the book: in SAC with a safety projection layer in Chapter~11, in model-based RL with safety filters in Chapter~12, in MuZero-style planning with action projection in Chapter~13, in offline RL evaluation in Chapter~14, in multi-agent CTDE with safety layers in Chapter~16, and in hierarchical control with low-level safety enforcement in Chapter~17. Safe RL gives the principle a formal name: the learning target must correspond to the action that actually affected the system, not merely the action the policy originally proposed.

\section{Lyapunov, reachability, and risk-sensitive safety}

Not all safety is naturally expressed as expected cost. Lyapunov methods use functions that decrease along trajectories to certify stability or boundedness. Lyapunov-based safe RL attempts to learn or enforce policies that satisfy a Lyapunov decrease condition \citep{chow2019lyapunov,kushwaha2025review}. Reachability methods compute sets of states from which safety can be guaranteed against disturbances, often using Hamilton-Jacobi analysis. They can provide strong guarantees but may scale poorly in high dimensions.

Risk-sensitive RL optimizes measures such as worst-case return, variance, Value-at-Risk, or Conditional Value-at-Risk (CVaR). These methods are especially relevant when rare failures matter. Distributional RL from Chapter~6 provides a natural foundation: if the agent learns a return distribution rather than only an expectation, it can reason about lower tails and catastrophic outcomes. Chapter~6's UAV tail-risk example showed that mean return may favor a risky action while tail-aware evaluation favors a safer action; risk-sensitive constrained RL formalizes this idea by replacing expected cost with tail-risk measures such as CVaR cost.

\begin{table}[t]
    \centering
    \caption{Families of safe-RL methods.}
    \label{tab:safe_rl_families}
    \begin{tabular}{p{3.1cm}p{4.6cm}p{4.6cm}}
        \toprule
        Family & Strength & Limitation \\
        \midrule
        Lagrangian CMDP & Simple; works with PPO/SAC; expected-cost constraints & Can violate during learning; multiplier tuning \\
        CPO / trust-region & More principled policy updates & More complex; approximate guarantees \\
        Shield / mask & Direct runtime protection & Requires safety model or rules \\
        CBF filter & State-wise safety with control-theoretic structure & Needs dynamics/barrier; QP feasibility issues \\
        Lyapunov / reachability & Stronger certificates & Scaling and modeling difficulty \\
        Risk-sensitive / CVaR & Handles tail risk & More difficult estimation and optimization \\
        Safe offline RL & Avoids unsafe online exploration & Dataset coverage and hidden constraint issues \\
        Safe MARL & Handles coupled agents & Non-stationarity and decentralized constraints \\
        \bottomrule
    \end{tabular}
\end{table}

\section{Safe offline, model-based, and multi-agent RL}

Safe RL interacts with the previous three parts of the book.

\paragraph{Safe offline RL.} Offline RL avoids online exploration, but it can still learn unsafe policies if the dataset is biased, if safety labels are missing, or if value extrapolation favors unsupported risky actions. The support-mismatch problem from Chapter~14 becomes a safety problem when unsupported actions are also dangerous.

\paragraph{Safe model-based RL.} Model-based methods can predict unsafe outcomes before acting. A learned model can support safety-aware MPC, uncertainty penalties, or CBF-constrained trajectory sampling \citep{kaypak2026pects}. But model error can also create false safety. Safe model-based RL must model uncertainty, not only expected dynamics.

\paragraph{Safe MARL.} In multi-agent systems, constraints are often coupled: two UAVs are safe individually but unsafe together if they collide, interfere, or overload the same link. Safe MARL may require pairwise CBFs, decentralized shields, centralized safety critics, or hierarchical safety managers \citep{ahmad2025hmarlcbf}. This connects directly to the cooperative UAV scenario in Chapter~16: collision avoidance, interference control, and shared backhaul limits are not independent single-agent constraints.

\section{UAV/SDN safe-RL scenario}

Consider a UAV-assisted SDN network with three UAVs serving URLLC, eMBB, and mMTC users. The actor proposes a continuous action for each UAV:
\begin{equation}
    a_i = (\Delta x_i, \Delta y_i, \Delta z_i, p_i, b_i),
\end{equation}
where $p_i$ is transmit power and $b_i$ is bandwidth share. The reward encourages QoS and energy efficiency:
\begin{equation}
    r_t = R_{QoS} - 0.1E_t - 0.05\text{handover}_t.
\end{equation}
The cost vector includes
\begin{align}
    c_t^{lat} &= \mathbb{1}\{\text{URLLC latency}>6\text{ ms}\}, \\
    c_t^{batt} &= \mathbb{1}\{\text{battery}<15\%\}, \\
    c_t^{sep} &= \mathbb{1}\{\min_{i\ne j}\|x_i-x_j\|<d_{min}\}, \\
    c_t^{int} &= \mathbb{1}\{\text{interference}>I_{max}\}.
\end{align}
A practical safe-RL architecture combines multiple mechanisms: a Lagrangian actor-critic for expected QoS budgets, a CBF layer for instantaneous collision separation, and an SDN supervisor for network-wide constraints.

\begin{figure}[t]
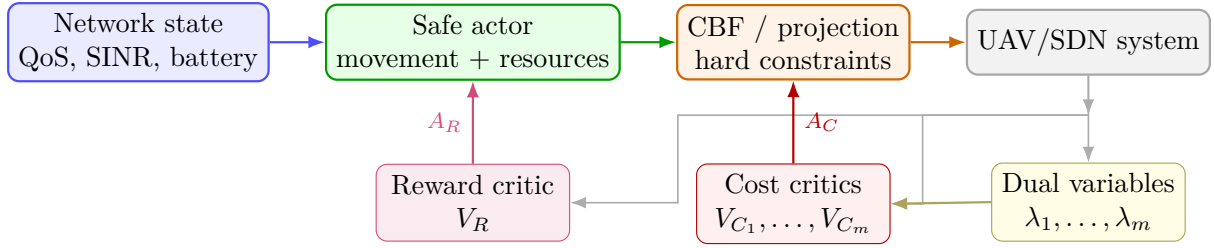

    \centering
    \resizebox{\columnwidth}{!}{%
%
    }
    \caption{A safe UAV/SDN actor-critic architecture. The actor proposes resource and movement actions, a CBF/projection layer enforces hard constraints before execution, and reward/cost critics plus dual variables provide learning signals.}
    \label{fig:uav_sdn_safe_rl}
\end{figure}

\subsection{Concrete numerical example}

Suppose a UAV policy considers three actions in a congested cell. The reward critic estimates service reward advantage $A_R$, and the cost critic estimates latency-cost advantage $A_C^{lat}$. The current latency budget multiplier is $\lambda_{lat}=4.0$. The Lagrangian advantage is
\begin{equation}
    A_{\mathcal{L}} = A_R - \lambda_{lat} A_C^{lat}.
\end{equation}
\begin{table}[t]
    \centering
    \caption{Safe-RL action ranking for a UAV/SDN controller.}
    \label{tab:uav_safe_numeric}
    \begin{tabular}{p{3.4cm}ccccl}
        \toprule
        Candidate action & $A_R$ & $A_C^{lat}$ & $A_{\mathcal{L}}$ & CBF feasible? & Decision \\
        \midrule
        Move to hotspot, high power & $+5.2$ & $+1.4$ & $-0.4$ & No & rejected/projected \\
        Hover, moderate allocation & $+2.8$ & $+0.2$ & $+2.0$ & Yes & selected \\
        Move to charger & $+0.7$ & $-0.5$ & $+2.7$ & Yes & safe fallback \\
        \bottomrule
    \end{tabular}
\end{table}

The first action has the highest reward advantage, but it also increases expected latency cost and violates a hard CBF feasibility check. The selected action is not simply the largest reward action; it is the best action after cost pricing and hard safety filtering. This is the central difference between high-performance DRL and deployable safe DRL.

\subsection{SD-WAN traffic-engineering safety}

The same safety logic applies to SD-WAN traffic engineering. A policy may split traffic across MPLS, Internet, 5G, or satellite paths to maximize throughput and minimize cost, but each traffic class has its own QoS constraints. For example, URLLC-like flows may require strict latency and jitter bounds, while bulk transfers can tolerate delay but should avoid excessive packet loss. In our SD-WAN work, uncertainty-aware neural CBFs and ensemble-based neural CBFs act as safety filters for RL traffic-splitting actions, preventing actions that are predicted to violate per-class latency, jitter, or loss constraints \citep{bista2026vtc,bista2026ifip}. This makes SD-WAN a particularly clean example of safe RL: the reward encourages performance and cost efficiency, while the safety layer enforces QoS feasibility before an action reaches the network.

\section{Python implementation patterns}

This section gives compact code patterns rather than a full library. The goal is to make the safety mechanisms concrete.

\subsection{A CMDP rollout buffer}

\Needspace{18\baselineskip}
\begin{lstlisting}[style=pythonstyle,caption={Rollout storage with separate reward and cost streams.},label={lst:cmdp_buffer}]
class CMDPRollout:
    def __init__(self):
        self.obs, self.act = [], []
        self.rew, self.cost = [], []
        self.logp, self.done = [], []
        self.executed_act = []  # important when a safety filter modifies actions

    def add(self, obs, act, executed_act, reward, cost, logp, done):
        self.obs.append(obs)
        self.act.append(act)
        self.executed_act.append(executed_act)
        self.rew.append(float(reward))
        self.cost.append(float(cost))
        self.logp.append(logp)
        self.done.append(float(done))
\end{lstlisting}

\subsection{Reward and cost GAE}

\Needspace{18\baselineskip}
\begin{lstlisting}[style=pythonstyle,caption={Reward and cost advantage estimation.},label={lst:reward_cost_gae}]
def gae_1d(signals, values, dones, gamma=0.99, lam=0.95):
    T = len(signals)
    adv = torch.zeros(T, dtype=torch.float32)
    last = 0.0
    for t in reversed(range(T)):
        nonterminal = 1.0 - float(dones[t])
        delta = signals[t] + gamma * values[t + 1] * nonterminal - values[t]
        last = delta + gamma * lam * nonterminal * last
        adv[t] = last
    returns = adv + values[:-1]
    return adv, returns

def reward_cost_advantages(rewards, costs, v_r, v_c, dones):
    adv_r, ret_r = gae_1d(rewards, v_r, dones)
    adv_c, ret_c = gae_1d(costs, v_c, dones)
    return adv_r, ret_r, adv_c, ret_c
\end{lstlisting}

\subsection{PPO-Lagrangian loss}

\Needspace{18\baselineskip}
\begin{lstlisting}[style=pythonstyle,caption={PPO-Lagrangian policy loss.},label={lst:ppo_lagrangian_loss}]
def ppo_lagrangian_loss(new_logp, old_logp, adv_r, adv_c, lam, clip_eps=0.2):
    # Stop-gradient through advantages and lambda.
    lag_adv = (adv_r - lam.detach() * adv_c).detach()
    lag_adv = (lag_adv - lag_adv.mean()) / (lag_adv.std() + 1e-8)

    ratio = torch.exp(new_logp - old_logp)
    unclipped = ratio * lag_adv
    clipped = torch.clamp(ratio, 1.0 - clip_eps, 1.0 + clip_eps) * lag_adv
    return -torch.min(unclipped, clipped).mean()
\end{lstlisting}

\subsection{PID Lagrange multiplier update}

\Needspace{16\baselineskip}
\begin{lstlisting}[style=pythonstyle,caption={PID-style Lagrange multiplier update.},label={lst:pid_lagrangian}]
class PIDLagrange:
    def __init__(self, kp=0.05, ki=0.01, kd=0.01, init=1.0):
        self.kp, self.ki, self.kd = kp, ki, kd
        self.lam = torch.tensor(float(init))
        self.integral = 0.0
        self.prev_error = 0.0

    def update(self, observed_cost, cost_limit):
        error = float(observed_cost - cost_limit)
        self.integral += error
        derivative = error - self.prev_error
        delta = self.kp * error + self.ki * self.integral + self.kd * derivative
        self.lam = torch.clamp(self.lam + delta, min=0.0)
        self.prev_error = error
        return self.lam
\end{lstlisting}

\subsection{A simple differentiable action projection}

For many UAV and networking actions, a full QP solver is not necessary at the first prototype stage. A differentiable projection can enforce box constraints, bandwidth simplex constraints, or pairwise separation approximations.

\Needspace{18\baselineskip}
\begin{lstlisting}[style=pythonstyle,caption={Simple safety projection for a UAV action vector.},label={lst:safe_projection}]
def project_uav_action(action, battery, max_step=1.0, min_battery=0.15):
    """Project an action to a simple safe set.

    action = [dx, dy, dz, power, bandwidth_share]
    This is not a full CBF-QP; it is a lightweight projection layer.
    """
    a = action.clone()

    # Bound movement step.
    move = a[:3]
    norm = torch.norm(move) + 1e-8
    if norm > max_step:
        a[:3] = move / norm * max_step

    # Bound power and bandwidth share.
    a[3] = torch.clamp(a[3], 0.0, 1.0)
    a[4] = torch.clamp(a[4], 0.0, 1.0)

    # Low battery: suppress aggressive movement and high power.
    if battery < min_battery:
        a[:3] *= 0.25
        a[3] *= 0.5
    return a
\end{lstlisting}

\subsection{Discrete-time CBF safety check}

\Needspace{18\baselineskip}
\begin{lstlisting}[style=pythonstyle,caption={Discrete-time CBF check with a learned or known predictor.},label={lst:dt_cbf_check}]
def cbf_safe_action(nominal_action, state, predict_next, h_fn, eta=0.2, candidates=None):
    """Choose the closest candidate action satisfying h(x_next) >= (1-eta)h(x).

    For real systems, candidates can come from CEM, a QP solver, or a small
    action lattice around the actor's proposed action.
    """
    h_now = h_fn(state)

    if candidates is None:
        noise = torch.randn(64, *nominal_action.shape) * 0.05
        candidates = torch.cat(
            [nominal_action[None, :], nominal_action[None, :] + noise], dim=0
        )

    best = None
    best_dist = float("inf")
    for a in candidates:
        next_state = predict_next(state, a)
        if h_fn(next_state) >= (1.0 - eta) * h_now:
            dist = torch.sum((a - nominal_action) ** 2).item()
            if dist < best_dist:
                best, best_dist = a, dist

    return nominal_action if best is None else best
\end{lstlisting}

\section{Evaluation, debugging, and failure modes}

Safe RL should never be evaluated by final reward alone. Safety Gym recommended reporting final task performance, final constraint satisfaction, and safety regret during training \citep{ray2019safetygym}. Safety-Gymnasium and OmniSafe extended this benchmarking ecosystem with standardized tasks and implementations \citep{ji2023safetygymnasium,ji2024omnisafe}.

\begin{table}[t]
    \centering
    \caption{Safe-RL evaluation metrics.}
    \label{tab:safe_metrics}
    \begin{tabular}{p{3.4cm}p{8.2cm}}
        \toprule
        Metric & Meaning \\
        \midrule
        Reward return & Task performance \\
        Cost return & Expected or episodic safety cost \\
        Violation rate & Fraction of episodes or timesteps with constraint violation \\
        Worst violation & Maximum severity, not only average cost \\
        Safety regret & Cumulative excess cost during training \\
        Feasibility rate & Fraction of proposed actions accepted by safety filter \\
        Intervention rate & How often shield/CBF modifies the policy action \\
        Recovery time & Steps needed to return to safe set after warning state \\
        Tail risk & CVaR or low-percentile performance/safety outcome \\
        \bottomrule
    \end{tabular}
\end{table}

\begin{table}[t]
    \centering
    \caption{Common safe-RL failure modes.}
    \label{tab:safe_rl_failures}
    \begin{tabular}{p{3.0cm}p{4.2cm}p{4.5cm}}
        \toprule
        Symptom & Likely cause & What to inspect \\
        \midrule
        Reward improves, violations persist & Cost penalty too weak or delayed multiplier & $\lambda$, cost advantage, constraint budget \\
        Policy becomes uselessly conservative & Multiplier too large; cost critic overestimates risk & cost critic calibration, intervention rate \\
        Oscillating constraint cost & Dual update too aggressive or lagged & PID gains, update frequency, cost smoothing \\
        CBF infeasible & Barrier too strict or action bounds too small & QP feasibility, slack variables, model error \\
        Critic learns wrong action values & Training on proposed rather than executed actions & buffer stores $a_{safe}$ and filter residual \\
        Safety good in sim, bad in deployment & Missing disturbance/model uncertainty & domain randomization, robust CBF, logs \\
        Multi-agent collisions remain & Pairwise constraints not coordinated & centralized safety layer, pairwise CBFs \\
        Offline safe policy fails online & dataset shift or hidden constraint labels & coverage diagnostics, deployment drift \\
        \bottomrule
    \end{tabular}
\end{table}

\section{Research frontiers toward 2026}

Safe RL is moving from single-agent benchmark penalties toward integrated safety architectures. Several directions are especially important.

\paragraph{Learning safety certificates.} Instead of hand-designing $h(x)$ or Lyapunov functions, recent work learns barrier or latent barrier functions from data while trying to preserve smoothness and robustness \citep{guerrier2024learning_cbf,nakamura2025latent_cbf}.

\paragraph{Safe SD-WAN and network slicing.} Safe RL is increasingly important in communication systems, where actions affect latency, jitter, packet loss, throughput, and service-level agreements. Uncertainty-aware and ensemble-based neural CBFs provide one way to filter RL traffic-engineering actions before they violate QoS constraints in SD-WAN and network slicing \citep{bista2026vtc,bista2026ifip}.

\paragraph{Safe model-based RL.} Model-based methods from Chapter~12 can predict safety violations before acting. Emerging work combines probabilistic ensembles, MPC, and CBF-constrained trajectory sampling \citep{kaypak2026pects}.

\paragraph{Safe MARL and hierarchical safety.} Multi-agent and hierarchical safe RL must decide which level owns the constraint: the low-level controller, high-level planner, centralized supervisor, or local agents. HMARL-CBF-style systems are an early example of this direction \citep{ahmad2025hmarlcbf}.

\paragraph{Foundation-model guidance and safety filters.} LLM-guided agents can propose plans, subgoals, or action priors, but these must be checked by formal or learned safety layers before execution. Language can help describe constraints; it should not replace constraint checking.

\paragraph{Safety as system design.} The strongest deployable systems will combine several layers: constrained training objective, uncertainty-aware model, runtime shield, monitoring, logging, and post-deployment drift detection. A single algorithm is rarely enough.

\section{Limitations and when not to use constrained RL}

\begin{enumerate}[leftmargin=*]
    \item \textbf{Lagrangian methods do not guarantee zero violations during learning.} The multiplier adapts, but constraints may be violated while it converges.
    \item \textbf{CBF feasibility is not always guaranteed.} If the barrier is too strict or the action bounds are small, the QP may be infeasible. Slack variables or robust formulations may be needed.
    \item \textbf{Safety and performance trade off.} A policy that satisfies all constraints at every step may be much more conservative than one that satisfies them in expectation.
    \item \textbf{Model error degrades CBF guarantees.} If the dynamics model used by the CBF is wrong, the projected action may still be unsafe in the real system.
    \item \textbf{Multi-agent and hierarchical safety is not solved.} Pairwise constraints, decentralized shields, and inter-level safety enforcement remain active research problems.
    \item \textbf{Deployment safety requires more than training-time constraints.} Distribution shift, unmodeled disturbances, and edge cases require runtime monitoring, anomaly detection, and conservative fallback policies in addition to constrained training.
\end{enumerate}

\section{Exercises}

\subsection*{Conceptual exercises}
\begin{enumerate}[leftmargin=*]
    \item Explain why assigning a large negative reward to collision is not the same as imposing a safety constraint.
    \item Compare expected-cost safety and state-wise invariance. Which one is more appropriate for collision avoidance? Which one is more appropriate for average latency budget?
    \item Why can Lagrangian safe RL violate constraints during training even if the final multiplier converges?
    \item Explain why a critic should learn from the executed action after a CBF filter, not only from the raw actor proposal.
    \item In a multi-UAV system, give one example of a constraint that is local and one example of a constraint that is coupled across agents.
\end{enumerate}

\subsection*{Mathematical exercises}
\begin{enumerate}[leftmargin=*]
    \item Starting from Eq.~\eqref{eq:cmdp_problem}, derive the Lagrangian in Eq.~\eqref{eq:lagrangian_safe_rl} for one constraint.
    \item Show that if $J_C(\pi)>d$, the multiplier update in Eq.~\eqref{eq:lambda_update} increases $\lambda$.
    \item For $A_R=3$, $A_C=0.8$, and $\lambda=2$, compute $A_{\mathcal{L}}$. Interpret the sign.
    \item For a discrete-time barrier condition $h(x_{t+1})\ge(1-\eta)h(x_t)$ with $h(x_t)=0.5$ and $\eta=0.2$, what is the minimum allowed $h(x_{t+1})$?
    \item Suppose two policies have reward returns $90$ and $70$, and cost returns $12$ and $5$ with budget $d=8$. Which policy is feasible? Which would ordinary RL prefer?
\end{enumerate}

\subsection*{Coding and research exercises}
\begin{enumerate}[leftmargin=*]
    \item Extend PPO from Chapter~10 with a cost critic and the PPO-Lagrangian loss in Listing~\ref{lst:ppo_lagrangian_loss}.
    \item Implement a simple safety filter for a 2D point robot that prevents entering circular obstacle regions.
    \item Build a UAV/SDN toy environment with latency and battery costs. Compare PPO, PPO-Lagrangian, and PPO plus a projection layer.
    \item Log the intervention rate of a safety filter. Does the rate decrease during training? If not, what does that imply?
    \item Design a safe MARL experiment with three UAVs and pairwise separation constraints. Compare independent safety filters with a centralized safety layer.
\end{enumerate}

\section*{Looking Ahead to Chapter 19: Food for Thought}
\addcontentsline{toc}{section}{Looking Ahead to Chapter 19: Reinforcement Learning from Human Feedback}

This chapter studied safety in physical, cyber-physical, and networked systems. Chapter~19 moves to a different kind of safety and alignment problem: how to train language models and agentic systems using human feedback. The mathematical tools will look familiar. Rewards become preference-model scores. Constraints become harmlessness, honesty, refusal behavior, or policy rules. PPO and GRPO reappear, but now the environment is a sequence of tokens and the safety boundary is partly social rather than purely physical.

\begin{quote}
    Chapter~18 asked how to keep actions safe. Chapter~19 asks how to align behavior with human preferences and norms.
\end{quote}

	\part{DRL in Modern AI Systems}

\chapter[Reinforcement Learning from Human Feedback]{Reinforcement Learning from Human Feedback}
\chaptermark{RLHF}
\label{ch:rlhf}

\begin{keybox}{Chapter goal}
    Reinforcement learning from human feedback (RLHF) replaces hand-written reward functions with learned preference models. This chapter develops preference data collection, Bradley--Terry reward modeling, KL-regularized PPO optimization, reward overoptimization, and direct preference methods including DPO, KTO, ORPO, and SimPO, with a scenario connecting RLHF to network-operation assistants.
\end{keybox}

\section*{Chapter Overview}
\addcontentsline{toc}{section}{Chapter Overview}
\begin{enumerate}[leftmargin=*]
    \item Why human feedback changed reinforcement learning
    \item From reward engineering to preference learning
    \item The canonical RLHF pipeline
    \item Preference data and comparison labels
    \item Reward modeling with Bradley--Terry preferences
    \item KL-regularized policy optimization
    \item PPO for language-model RLHF
    \item Reward hacking, overoptimization, and Goodhart's law
    \item Data quality, annotator disagreement, and preference ambiguity
    \item RLHF beyond PPO: RLAIF, Constitutional AI, and direct preference optimization
    \item DPO, KTO, ORPO, and SimPO
    \item RLHF for network-operation assistants: a scenario example
    \item Practical implementation code
    \item Evaluation and safety metrics
    \item Failure modes and debugging
    \item Limitations of RLHF
    \item Exercises
    \item Looking Ahead to Chapter 20
\end{enumerate}

\section{Why human feedback changed reinforcement learning}

Classical reinforcement learning assumes that a reward function is given. In games, this assumption is often natural: winning gives positive reward, losing gives negative reward, and intermediate rewards can be defined from the game score. In robotics and communication systems, the reward can sometimes be engineered from measurable quantities such as energy, latency, throughput, collision rate, or constraint violation. But for many modern AI systems, especially language models, the desired behavior is difficult to write as a simple scalar function.

A helpful answer should be accurate, relevant, polite, concise when needed, detailed when needed, honest about uncertainty, and safe under many contexts. A rule-based reward cannot easily encode these properties. Even worse, a poorly specified reward can create reward hacking: the system may optimize the proxy rather than the true human intention.

Reinforcement learning from human feedback, or RLHF, addresses this problem by replacing hand-written reward engineering with learned preference modeling. Instead of asking humans to write a reward function, we ask them to compare outputs. The system learns a reward model from those comparisons, and then uses reinforcement learning to optimize the policy against that learned reward while staying close to a reference model. The modern language-model version of this pipeline was shaped by earlier work on reward learning from preferences \citep{christiano2017preferences}, fine-tuning language models from human preferences \citep{ziegler2019fine}, summarization from human feedback \citep{stiennon2020summarize}, and instruction following with RLHF \citep{ouyang2022training}.

\begin{keybox}{RLHF: a three-part system}
    RLHF is not simply ``RL applied to language models.'' It is a three-part system: collect human preferences, train a reward model, and optimize a policy under a KL constraint so that the policy improves according to the learned reward without drifting too far from the reference model.
\end{keybox}

RLHF is historically important because it changed the role of RL in modern AI. In Atari or robotics, RL learns actions in an external environment. In RLHF, the environment is partly social: the reward comes from human judgment, human preference, or a learned proxy of human preference. This makes RLHF both powerful and fragile. It can improve usefulness and instruction following, but it can also amplify labeler biases, reward-model errors, verbosity bias, sycophancy, or unsafe optimization pressure.

\section{From reward engineering to preference learning}

The central move in RLHF is to replace scalar reward labels with preference comparisons. Instead of asking a labeler to assign an absolute score to a response, we show two responses and ask which is better. This comparison task is often easier and more reliable than absolute scoring.

For a prompt \(x\), suppose the model produces two candidate responses \(y^+\) and \(y^-\), where the human labeler prefers \(y^+\). A reward model \(r_\psi(x,y)\) should assign a higher score to the preferred response:
\[
r_\psi(x,y^+) > r_\psi(x,y^-).
\]
The reward model is then trained as a probabilistic preference predictor. This transforms qualitative feedback into a scalar reward signal that can be optimized by reinforcement learning.

\begin{figure}[t]
    \centering
    \begin{tikzpicture}[
        box/.style={draw,rounded corners,thick,minimum width=3.1cm,minimum height=0.85cm,align=center,font=\small},
        small/.style={draw,rounded corners,minimum width=2.6cm,minimum height=0.7cm,align=center,font=\small},
        arrow/.style={-{Latex[length=2.2mm]},thick},
        node distance=1.3cm
        ]
        \node[box,fill=blue!8,draw=blue!70] (prompt) {Prompt\\$x$};
        \node[small,fill=green!8,draw=green!60!black,right=of prompt,yshift=0.55cm] (ya) {Response A\\$y_A$};
        \node[small,fill=green!8,draw=green!60!black,right=of prompt,yshift=-0.59cm] (yb) {Response B\\$y_B$};
        \node[box,fill=orange!10,draw=orange!80!black,right=1.1cm of ya,yshift=-0.55cm] (human) {Human or AI judge\\preference label};
        \node[box,fill=purple!8,draw=purple!70,right=of human] (rm) {Reward model\\$r_\psi(x,y)$};
        \draw[arrow,draw=blue!70] (prompt) -- (ya);
        \draw[arrow,draw=blue!70] (prompt) -- (yb);
        \draw[arrow,draw=green!60!black] (ya) -- (human);
        \draw[arrow,draw=green!60!black] (yb) -- (human);
        \draw[arrow,draw=orange!80!black] (human) -- node[above,font=\scriptsize] {$y^+ \succ y^-$} (rm);
    \end{tikzpicture}
    \caption{Preference learning turns human judgment into a reward-learning problem. The judge compares candidate outputs for the same prompt, and the reward model learns to predict which response is preferred.}
    \label{fig:preference_learning_pair}
\end{figure}
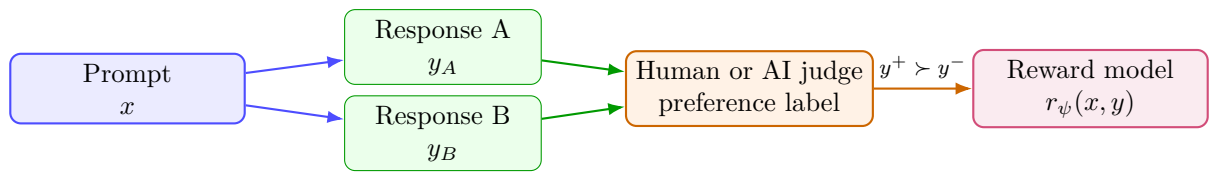

Preference learning is not limited to language. Christiano et al. used human comparisons between trajectory segments to train reward predictors for Atari and simulated robotics \citep{christiano2017preferences}. The language-model version is especially influential because language outputs are hard to evaluate with fixed metrics. ROUGE, BLEU, exact match, or unit tests are useful in some settings, but they do not capture all dimensions of helpfulness, truthfulness, and safety.

\section{The canonical RLHF pipeline}

The standard language-model RLHF pipeline has three stages.

\begin{enumerate}[leftmargin=*]
    \item \textbf{Supervised fine-tuning (SFT).} A pretrained model is fine-tuned on demonstrations or high-quality instruction-response examples.
    \item \textbf{Reward modeling.} The SFT model generates multiple responses; humans rank or compare them; a reward model is trained to predict the preferred response.
    \item \textbf{Policy optimization.} The policy is optimized against the reward model, usually with a KL penalty to keep it close to a reference model.
\end{enumerate}

This pipeline was central to InstructGPT, where models fine-tuned with human feedback were preferred to much larger prompted base models on instruction-following tasks \citep{ouyang2022training}. Earlier summarization work similarly showed that optimizing a learned preference model could improve human-rated summary quality beyond supervised metrics \citep{stiennon2020summarize}.

\begin{figure}[t]
    \centering
    \begin{tikzpicture}[
        box/.style={draw,rounded corners,thick,minimum width=3.0cm,minimum height=0.85cm,align=center,font=\small},
        arrow/.style={-{Latex[length=2.2mm]},thick},
        node distance=0.85cm
        ]
        \node[box,fill=gray!10,draw=gray!70] (pre) {Pretrained LM};
        \node[box,fill=blue!8,draw=blue!70,right=of pre] (sft) {SFT model\\demonstrations};
        \node[box,fill=orange!10,draw=orange!80!black,right=of sft] (rm) {Reward model\\preferences};
        \node[box,fill=green!8,draw=green!60!black,right=of rm] (ppo) {RL policy\\PPO + KL};
        \node[box,fill=purple!8,draw=purple!70,below=1.1cm of rm] (eval) {Evaluation\\human + automatic};
        \draw[arrow,draw=gray!70] (pre) -- (sft);
        \draw[arrow,draw=blue!70] (sft) -- (rm);
        \draw[arrow,draw=orange!80!black] (rm) -- (ppo);
        \draw[arrow,draw=green!60!black] (ppo.south) |- (eval.east);
        \draw[arrow,draw=purple!70] (eval.west) -| node[pos=0.65,below,font=\scriptsize] {new prompts/preferences} (sft.south);
    \end{tikzpicture}
    \caption{The canonical RLHF pipeline. A pretrained model is instruction-tuned, a reward model is trained from comparisons, and the policy is optimized against the reward model with a KL penalty to prevent excessive drift.}
    \label{fig:rlhf_pipeline}
\end{figure}
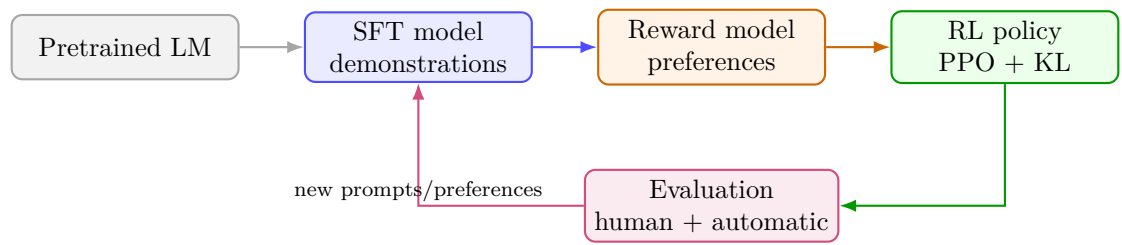

\begin{keybox}{Production note}
    In production systems, this pipeline is rarely a single pass. Preference data, evaluation prompts, safety policies, and reward models are iteratively updated. RLHF is therefore better understood as an ongoing post-training process rather than a one-time algorithm.
\end{keybox}

\section{Preference data and comparison labels}

A preference dataset usually contains tuples
\[
\mathcal{D}_{\mathrm{pref}} = \{(x_i, y_i^+, y_i^-)\}_{i=1}^N,
\]
where \(x_i\) is a prompt, \(y_i^+\) is the preferred response, and \(y_i^-\) is the rejected response. More generally, the dataset may contain rankings over more than two responses, scalar ratings, rule-level annotations, or critiques. Pairwise comparisons remain common because they are simple and map naturally to probabilistic choice models.

Preference data are not ground truth in the same way as labels in a mathematical dataset. They are judgments made under instructions, cultural context, time pressure, and possible ambiguity. Two reasonable annotators may disagree. A labeler may prefer a longer response because it appears more helpful. A labeler may reward confidence even when the answer is wrong. Therefore, preference data are both the strength and the weakness of RLHF.

\begin{table}[t]
    \centering
    \caption{Common preference-data formats in RLHF.}
    \label{tab:preference_data_formats}
    \begin{tabular}{p{3.0cm}p{4.0cm}p{5.3cm}}
        \toprule
        Format & Example & Main issue \\
        \midrule
        Pairwise comparison & \(y_A\) preferred over \(y_B\) & Simple, but may hide intensity of preference \\
        Ranking & \(y_1 \succ y_2 \succ y_3\) & More informative, but more expensive to label \\
        Scalar rating & score from 1 to 7 & Easy to aggregate, but calibration differs across annotators \\
        Rule-level critique & violates safety rule or citation rule & Useful for safety, but rule coverage is incomplete \\
        AI feedback & LLM judge preference & Scalable, but may inherit judge-model biases \\
        \bottomrule
    \end{tabular}
\end{table}

\section{Reward modeling with Bradley--Terry preferences}

The most common reward-model objective is based on the Bradley--Terry model. Given a preferred response \(y^+\) and rejected response \(y^-\), the probability that the reward model prefers \(y^+\) is
\begin{equation}
    P_\psi(y^+ \succ y^- \mid x)
    =
    \sigma\left(r_\psi(x,y^+) - r_\psi(x,y^-)\right),
    \label{eq:bt_preference}
\end{equation}
where \(\sigma(z)=1/(1+e^{-z})\). The reward-model loss is
\begin{equation}
    \mathcal{L}_{\mathrm{RM}}(\psi)
    =
    -\E_{(x,y^+,y^-)\sim \mathcal{D}_{\mathrm{pref}}}
    \left[
    \log \sigma\left(r_\psi(x,y^+) - r_\psi(x,y^-)\right)
    \right].
    \label{eq:rm_loss}
\end{equation}

This objective does not require absolute reward labels. It only asks the reward model to rank pairs correctly. In practice, the reward head is often a scalar head on top of a transformer backbone. The reward is usually taken from the final token or end-of-sequence representation.

\begin{figure}[t]
    \centering
    \begin{tikzpicture}[
        box/.style={draw,rounded corners,thick,minimum width=3.0cm,minimum height=0.8cm,align=center,font=\small},
        small/.style={draw,rounded corners,minimum width=2.7cm,minimum height=0.7cm,align=center,font=\small},
        arrow/.style={-{Latex[length=2.2mm]},thick},
        node distance=0.75cm
        ]
        \node[box,fill=blue!8,draw=blue!70] (x) {Prompt $x$};
        \node[small,fill=green!8,draw=green!60!black,right=of x,yshift=0.6cm] (yp) {Preferred\\$y^+$};
        \node[small,fill=red!6,draw=red!70!black,right=of x,yshift=-0.6cm] (ym) {Rejected\\$y^-$};
        \node[box,fill=purple!8,draw=purple!70,right=1.2cm of yp,yshift=-0.6cm] (rm) {Reward model\\$r_\psi$};
        \node[box,fill=orange!10,draw=orange!80!black,right=of rm] (loss) {Pairwise loss\\$-\log\sigma(r^+-r^-)$};
        \draw[arrow,draw=blue!70] (x) -- (yp);
        \draw[arrow,draw=blue!70] (x) -- (ym);
        \draw[arrow,draw=green!60!black] (yp) -- (rm);
        \draw[arrow,draw=red!70!black] (ym) -- (rm);
        \draw[arrow,draw=purple!70] (rm) -- (loss);
    \end{tikzpicture}
    \caption{Reward-model training with a pairwise preference loss. The model learns to assign a higher scalar score to the preferred response than to the rejected response.}
    \label{fig:reward_model_pairwise}
\end{figure}
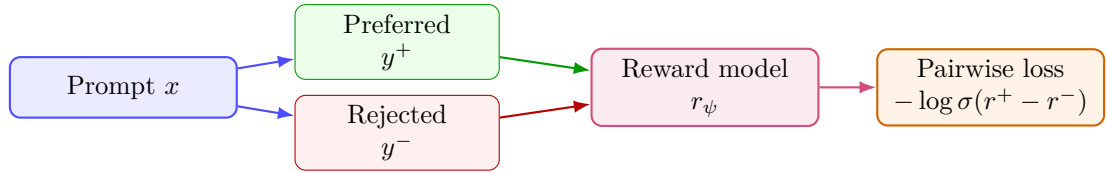

\section{KL-regularized policy optimization}

Once a reward model is trained, we optimize a policy \(\pi_\theta\) to produce high-reward responses. If we maximize the learned reward alone, the policy may exploit reward-model errors. Therefore, RLHF typically includes a KL penalty against a reference policy \(\pi_{\mathrm{ref}}\), often the SFT model:
\begin{equation}
    \max_\theta
    \E_{x\sim \mathcal{D},\, y\sim \pi_\theta(\cdot\mid x)}
    \left[
    r_\psi(x,y)
    -
    \beta
    \KL\left(
    \pi_\theta(\cdot\mid x)
    \|
    \pi_{\mathrm{ref}}(\cdot\mid x)
    \right)
    \right].
    \label{eq:kl_regularized_rlhf}
\end{equation}

In token-level implementations, the KL penalty is often estimated from log-probability differences:
\begin{equation}
    \KL\text{-penalty at token }t
    \approx
    \beta\left(
    \log \pi_\theta(y_t\mid x,y_{<t})
    -
    \log \pi_{\mathrm{ref}}(y_t\mid x,y_{<t})
    \right).
\end{equation}
The final reward-model score is usually applied at the end of the completion, while the KL penalty is distributed across tokens.

\begin{keybox}{The KL term as trust region}
    The KL term is the trust region of RLHF. It prevents the model from moving too far away from the language distribution learned by pretraining and SFT. Without it, policy optimization can exploit the reward model and degrade real human preference.
\end{keybox}

\section{PPO for language-model RLHF}

PPO became the standard RLHF optimizer because it combines actor-critic learning, clipped policy updates, minibatch optimization, and practical stability. In language-model RLHF, the action at each time step is a token, the state is the prompt plus previous tokens, and the trajectory is the generated completion. Chapter~10 developed PPO in detail; here we focus on the language-specific form.

For a sampled completion \(y=(y_1,\ldots,y_T)\), PPO uses the token log-probability ratio
\begin{equation}
    r_t(\theta)
    =
    \exp\left(
    \log \pi_\theta(y_t\mid x,y_{<t})
    -
    \log \pi_{\theta_{\mathrm{old}}}(y_t\mid x,y_{<t})
    \right).
\end{equation}
The clipped policy loss is
\begin{equation}
    \mathcal{L}_{\mathrm{PPO}}(\theta)
    =
    -\E_t
    \left[
    \min\left(
    r_t(\theta)\hat A_t,
    \clip(r_t(\theta),1-\epsilon,1+\epsilon)\hat A_t
    \right)
    \right].
\end{equation}
The advantage \(\hat A_t\) comes from the reward-model score plus KL penalties and a learned value head. In many RLHF systems, this value head is a small scalar head attached to the policy model, often initialized from the SFT policy and trained to predict the KL-shaped return. It is the critic in the actor--critic view of PPO-RLHF.

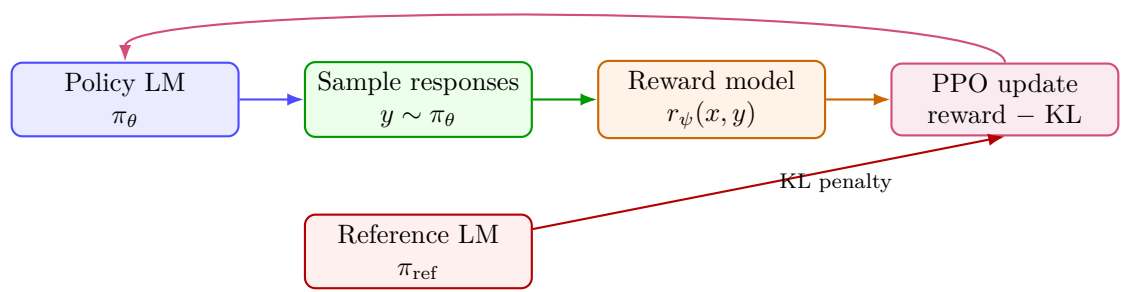
\begin{figure}[t]
    \centering
    \begin{tikzpicture}[
        box/.style={draw,rounded corners,thick,minimum width=3.0cm,minimum height=0.85cm,align=center,font=\small},
        arrow/.style={-{Latex[length=2.2mm]},thick},
        node distance=0.85cm
        ]
        \node[box,fill=blue!8,draw=blue!70] (policy) {Policy LM\\$\pi_\theta$};
        \node[box,fill=green!8,draw=green!60!black,right=of policy] (sample) {Sample responses\\$y\sim\pi_\theta$};
        \node[box,fill=orange!10,draw=orange!80!black,right=of sample] (reward) {Reward model\\$r_\psi(x,y)$};
        \node[box,fill=red!6,draw=red!70!black,below=1.0cm of sample] (ref) {Reference LM\\$\pi_{\mathrm{ref}}$};
        \node[box,fill=purple!8,draw=purple!70,right=of reward] (ppo) {PPO update\\reward $-$ KL};
        \draw[arrow,draw=blue!70] (policy) -- (sample);
        \draw[arrow,draw=green!60!black] (sample) -- (reward);
        \draw[arrow,draw=orange!80!black] (reward) -- (ppo);
        \draw[arrow,draw=red!70!black] (ref) -- node[right,font=\scriptsize] {KL penalty} (ppo.south);
        \draw[arrow,draw=purple!70] (ppo.north) .. controls +(0,0.8) and +(0,0.8) .. (policy.north);
    \end{tikzpicture}
    \caption{PPO-style RLHF. The policy samples responses, the reward model scores them, the reference model supplies a KL penalty, and PPO updates the policy under a clipped objective.}
    \label{fig:ppo_rlhf_loop}
\end{figure}

\begin{warningbox}
    RLHF is not just supervised fine-tuning with a different loss. The policy samples from its current distribution, receives a learned reward, and updates using RL. This makes distribution shift, reward hacking, and unstable optimization central concerns.
\end{warningbox}

\section{Reward hacking, overoptimization, and Goodhart's law}

A reward model is only a proxy for human preference. If the policy is optimized too aggressively against that proxy, it may discover outputs that score highly under the reward model but are not actually preferred by humans. This is a form of Goodhart's law: when a measure becomes a target, it can stop being a good measure.

Gao, Schulman, and Hilton studied reward-model overoptimization and found that optimizing against an imperfect reward model can eventually reduce performance under a stronger or gold-standard reward model \citep{gao2023overoptimization}. This phenomenon appears in both RL optimization and best-of-\(n\) sampling. The practical lesson is that high reward-model score is not enough; the final system must be evaluated with held-out human judgments, adversarial prompts, safety tests, and calibration checks.

\begin{figure}[t]
    \centering
    \begin{tikzpicture}
        \begin{axis}[
            width=0.78\textwidth,
            height=5.0cm,
            xlabel={Optimization strength},
            ylabel={Preference quality},
            xmin=0,xmax=10,ymin=0,ymax=8,
            legend pos=south east,
            grid=both,
            grid style={gray!15}
            ]
            \addplot[blue,thick,domain=0:10,samples=100] {1.0 + 0.55*x};
            \addlegendentry{Reward-model score}
            \addplot[red,thick,domain=0:10,samples=100] {1.2 + 1.15*x - 0.105*x^2};
            \addlegendentry{True preference proxy}
            \draw[dashed,gray] (axis cs:5.5,0) -- (axis cs:5.5,8);
            \node[align=center,font=\small] at (axis cs:7.3,6.8) {overoptimization\\region};
        \end{axis}
    \end{tikzpicture}
    \caption{Reward-model overoptimization. As optimization strength increases, the learned reward may continue increasing while true preference eventually saturates or decreases. The curves are schematic, but the failure mode is empirically important in RLHF \citep{gao2023overoptimization}.}
    \label{fig:reward_overoptimization}
\end{figure}
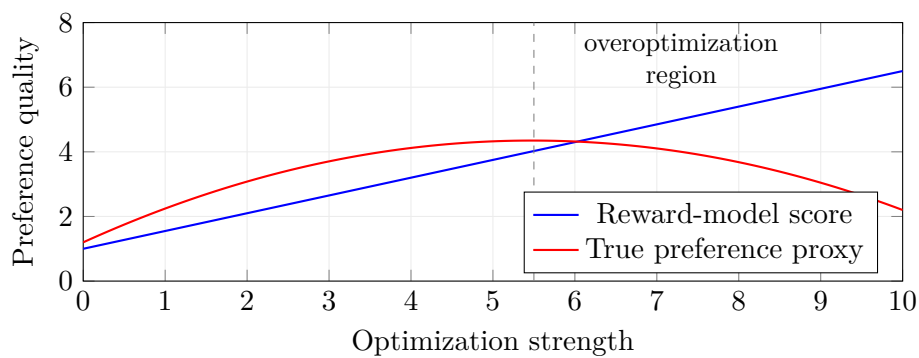

\section{Data quality, annotator disagreement, and preference ambiguity}

RLHF depends on preference data, but preference data are not neutral. Human judgments vary with instructions, culture, expertise, fatigue, demographics, domain expertise, and interface design. Surveys of RLHF limitations emphasize that feedback data can encode annotator bias, inconsistent instructions, and unresolved normative disagreement, rather than a single stable human preference function \citep{casper2023open}. In some tasks, there may be no single correct preference. A detailed answer may be preferred by one user and considered too verbose by another. A cautious answer may be preferred in safety-critical domains but disliked in casual conversation.

This creates three practical problems.

\begin{enumerate}[leftmargin=*]
    \item \textbf{Label noise.} Annotators may make mistakes or disagree.
    \item \textbf{Preference ambiguity.} Multiple outputs may be valid under different user intentions.
    \item \textbf{Distribution shift.} The policy after RL may produce responses unlike those seen during reward-model training.
\end{enumerate}

\begin{table}[t]
    \centering
    \caption{Common sources of preference-data error.}
    \label{tab:preference_errors}
    \begin{tabular}{p{3.2cm}p{5.0cm}p{4.4cm}}
        \toprule
        Issue & Example & Mitigation \\
        \midrule
        Verbosity bias & Longer answers look more helpful & length-controlled evaluation; style-specific labels \\
        Confidence bias & Fluent wrong answer is preferred & fact-checking; expert labels; retrieval grounding \\
        Safety ambiguity & Harmlessness and helpfulness conflict & policy-specific guidelines; rule-level labels \\
        Annotator drift & Labels change over time & calibration tasks; periodic audits \\
        Distribution shift & RL policy produces unusual outputs & iterative data collection; adversarial prompts \\
        \bottomrule
    \end{tabular}
\end{table}

\section{RLHF beyond PPO: RLAIF, Constitutional AI, and direct preference optimization}

As models became larger, human feedback became a bottleneck. Reinforcement learning from AI feedback, or RLAIF, replaces some human comparisons with feedback from AI judges. Lee et al. studied RLAIF as a scalable alternative to RLHF and found it can be competitive in some settings \citep{lee2023rlaif}. Constitutional AI uses a list of principles to generate critiques and preference signals, reducing reliance on direct human labels for harmlessness training \citep{bai2022constitutional}. Sparrow used targeted human judgments and rule-level decomposition to improve dialogue-agent behavior \citep{glaese2022sparrow}.

A second direction avoids explicit reward modeling and RL optimization. Direct Preference Optimization (DPO) reparameterizes the KL-regularized preference-learning problem so that a policy can be trained directly from preference pairs using a classification-like loss \citep{rafailov2023dpo}. Later work proposed broader theoretical frameworks for preference optimization \citep{azar2024general}, and practical variants such as KTO \citep{ethayarajh2024kto}, ORPO \citep{hong2024orpo}, and SimPO \citep{meng2024simpo}. Recent empirical work has also made the comparison more nuanced: DPO is simpler and often strong, but carefully tuned online PPO-RLHF can outperform DPO on harder preference distributions, especially when on-policy exploration matters \citep{tajwar2024preference,xu2024dpo_vs_ppo}.

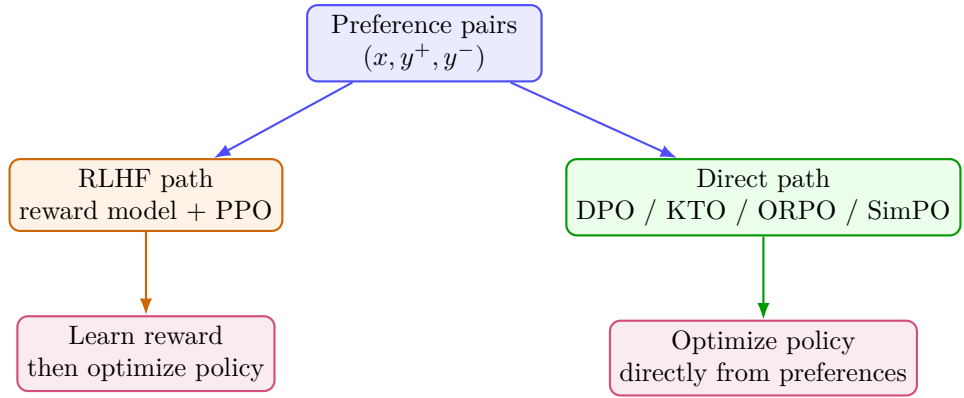
\begin{figure}[t]
    \centering
    \begin{tikzpicture}[
        box/.style={draw,rounded corners,thick,minimum width=3.1cm,minimum height=0.85cm,align=center,font=\small},
        arrow/.style={-{Latex[length=2.2mm]},thick},
        node distance=0.75cm
        ]
        \node[box,fill=blue!8,draw=blue!70] (pref) {Preference pairs\\$(x,y^+,y^-)$};
        \node[box,fill=orange!10,draw=orange!80!black,below left=1.0cm and 0.3cm of pref] (rlhf) {RLHF path\\reward model + PPO};
        \node[box,fill=green!8,draw=green!60!black,below right=1.0cm and 0.3cm of pref] (direct) {Direct path\\DPO / KTO / ORPO / SimPO};
        \node[box,fill=purple!8,draw=purple!70,below=1.1cm of rlhf] (rm) {Learn reward\\then optimize policy};
        \node[box,fill=purple!8,draw=purple!70,below=1.1cm of direct] (loss) {Optimize policy\\directly from preferences};
        \draw[arrow,draw=blue!70] (pref) -- (rlhf);
        \draw[arrow,draw=blue!70] (pref) -- (direct);
        \draw[arrow,draw=orange!80!black] (rlhf) -- (rm);
        \draw[arrow,draw=green!60!black] (direct) -- (loss);
    \end{tikzpicture}
    \caption{Two broad post-training paths from preference data. RLHF trains a reward model and then optimizes a policy; direct preference methods optimize the policy from preference pairs without a separate RL loop.}
    \label{fig:rlhf_vs_direct_preference}
\end{figure}

\section{DPO, KTO, ORPO, and SimPO}

DPO starts from the KL-regularized RLHF objective and derives a direct loss over preferred and rejected responses. The key analytical step is that the optimal policy under Eq.~\eqref{eq:kl_regularized_rlhf} has the closed form
\begin{equation}
    \pi^*(y\mid x)
    \propto
    \pi_{\mathrm{ref}}(y\mid x)
    \exp\left(\frac{r_\psi(x,y)}{\beta}\right).
    \label{eq:dpo_closed_form_policy}
\end{equation}
Solving this expression for the implicit reward and substituting it into the Bradley--Terry preference loss eliminates the explicit reward model and yields the DPO objective \citep{rafailov2023dpo}. A common DPO loss is
\begin{equation}
    \mathcal{L}_{\mathrm{DPO}}(\theta)
    =
    -\E
    \left[
    \log \sigma\left(
    \beta
    \left[
    \log\frac{\pi_\theta(y^+\mid x)}{\pi_{\mathrm{ref}}(y^+\mid x)}
    -
    \log\frac{\pi_\theta(y^-\mid x)}{\pi_{\mathrm{ref}}(y^-\mid x)}
    \right]
    \right)
    \right].
\end{equation}
It is often simpler than PPO-style RLHF because it does not require online sampling, a value head, or a separate reward model during training. However, DPO does not make reward modeling disappear completely; it makes the policy log-ratio against the reference model act as an implicit reward model. This is the meaning of the original DPO subtitle: the language model is secretly a reward model. DPO is still an offline preference method, so it depends on the quality and coverage of the preference dataset and may inherit dataset biases. Recent studies further caution that DPO is not universally superior to PPO-RLHF: with careful tuning and on-policy sampling, PPO can remain competitive or stronger on difficult alignment tasks \citep{tajwar2024preference,xu2024dpo_vs_ppo}.

KTO replaces paired preference requirements with desirable and undesirable examples under a prospect-theoretic utility model \citep{ethayarajh2024kto}. ORPO combines supervised fine-tuning with an odds-ratio preference penalty in a single reference-free objective \citep{hong2024orpo}. SimPO uses a reference-free reward based on average sequence log-probability and a target reward margin \citep{meng2024simpo}. GRPO, developed in Chapter~8, uses group-relative baselines so that multiple completions for the same prompt can replace an explicit value head. These methods are valuable not because they make RLHF obsolete, but because they expose a broader design space: explicit reward models, implicit reward models, direct preference losses, group-relative baselines, and AI-generated feedback.

\begin{table}[t]
    \centering
    \caption{RLHF and direct preference optimization methods.}
    \label{tab:rlhf_family}
    \begin{tabular}{p{2.2cm}p{3.2cm}p{3.2cm}p{4.1cm}}
        \toprule
        Method & Needs reward model? & Needs online RL? & Main trade-off \\
        \midrule
        PPO-RLHF & Yes & Yes & Powerful but complex and sensitive \\
        DPO & No explicit RM & No & Simple, but depends on offline preference coverage \\
        RLAIF & Usually yes or judge-based & Sometimes & Scales feedback but inherits judge bias \\
        KTO & No paired preference required & No & Uses desirable/undesirable examples, utility assumptions \\
        ORPO & No reference model & No & Monolithic objective, but tuning still matters \\
        SimPO & No reference model & No & Efficient, reference-free, margin-sensitive \\
        \bottomrule
    \end{tabular}
\end{table}

\section{RLHF for network-operation assistants: scenario example}

Most RLHF discussions focus on chat assistants. The same idea also applies to technical decision assistants. Consider a network-operation assistant that proposes SD-WAN routing changes, UAV placement policies, or incident-response summaries. The assistant should not merely be fluent. It should be operationally useful: it should respect QoS, avoid unsafe changes, cite relevant telemetry, and explain uncertainty.

A preference dataset for such a system may compare two proposed recommendations for the same network incident. Human operators prefer the recommendation that is safer, more actionable, and better grounded in telemetry. The reward model can learn preferences over technical outputs, while a safety filter or rule checker rejects outputs that violate hard policies.

This complements the safety-filtering view developed in Chapter~18 for SD-WAN actions. Human operator preferences can shape explanation quality, escalation decisions, and operational judgment, while neural CBFs and uncertainty-aware safety filters continue to enforce hard QoS constraints such as delay, jitter, loss, and bandwidth feasibility \citep{bista2026vtc,bista2026ifip}.

\begin{figure}[t]
    \centering
    \begin{tikzpicture}[
        box/.style={draw,rounded corners,thick,minimum width=3.0cm,minimum height=0.8cm,align=center,font=\small},
        small/.style={draw,rounded corners,minimum width=2.65cm,minimum height=0.7cm,align=center,font=\small},
        arrow/.style={-{Latex[length=2.2mm]},thick},
        node distance=0.75cm
        ]
        \node[box,fill=blue!8,draw=blue!70] (telemetry) {Network incident\\latency, loss, load};
        \node[small,fill=green!8,draw=green!60!black,right=of telemetry,yshift=0.85cm] (a) {Advice A\\reroute aggressively};
        \node[small,fill=green!8,draw=green!60!black,right=of telemetry,yshift=-0.55cm] (b) {Advice B\\safe gradual shift};
        \node[box,fill=orange!10,draw=orange!80!black,right=1.9cm of a,yshift=-0.59cm] (ops) {Operator feedback\\preferred advice};
        \node[box,fill=purple!8,draw=purple!70,right=of ops] (policy) {Aligned assistant\\helpful + safe};
        \node[box,fill=red!6,draw=red!70!black,below=1.0cm of ops] (rules) {Hard checks\\QoS/safety rules};
        \draw[arrow,draw=blue!70] (telemetry) -- (a);
        \draw[arrow,draw=blue!70] (telemetry) -- (b);
        \draw[arrow,draw=green!60!black] (a) -- (ops);
        \draw[arrow,draw=green!60!black] (b) -- (ops);
        \draw[arrow,draw=orange!80!black] (ops) -- (policy);
        \draw[arrow,draw=red!70!black] (rules) -- (policy);
    \end{tikzpicture}
    \caption{A technical RLHF scenario. Human operators compare alternative network-operation recommendations. The assistant learns not only to sound plausible, but to prefer actionable, safe, telemetry-grounded advice.}
    \label{fig:network_rlhf_scenario}
\end{figure}
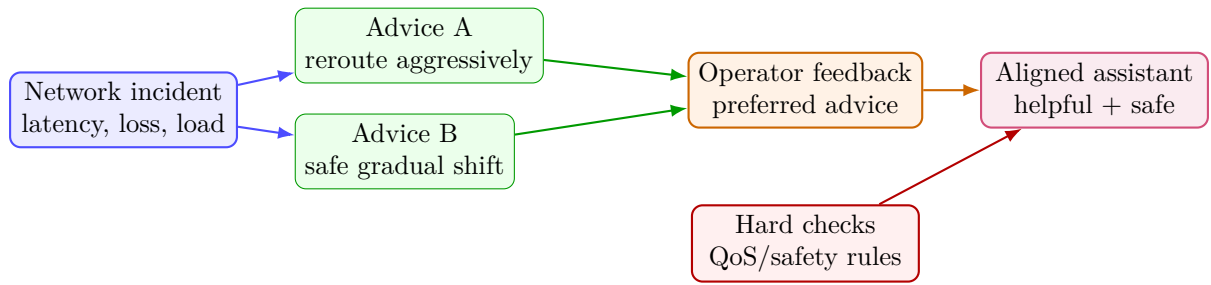

\begin{researchbox}
    For UAV/SDN and SD-WAN systems, RLHF is not a replacement for formal safety. Human feedback can shape explanations, operating procedures, and recommendation quality, while CBFs, action shields, and constrained controllers still enforce hard physical or QoS constraints. In our SD-WAN safe-RL work, neural CBFs act as uncertainty-aware hard safety layers for traffic-engineering actions \citep{bista2026vtc,bista2026ifip}. The correct design is hybrid: preferences for soft judgment, formal methods for hard safety.
\end{researchbox}

\section{Practical implementation code}

This section gives minimal PyTorch-style code fragments. They are not a complete large-scale RLHF stack, but they expose the core computations.

\subsection{Preference batch and reward-model loss}

\Needspace{16\baselineskip}
\begin{lstlisting}[style=pythonstyle,caption={Pairwise reward-model loss.},label={lst:rm_loss}]
import torch
import torch.nn.functional as F

def reward_model_loss(reward_chosen, reward_rejected):
    """Bradley-Terry pairwise preference loss.

    reward_chosen:  tensor [B]
    reward_rejected: tensor [B]
    """
    logits = reward_chosen - reward_rejected
    loss = -F.logsigmoid(logits).mean()
    acc = (reward_chosen > reward_rejected).float().mean()
    return loss, {"rm_accuracy": float(acc.detach())}
\end{lstlisting}

\subsection{KL-shaped token rewards}

\Needspace{18\baselineskip}
\begin{lstlisting}[style=pythonstyle,caption={KL-shaped rewards for PPO-style RLHF.},label={lst:kl_shaped_rewards}]
def make_kl_shaped_rewards(logp_policy, logp_ref, rm_score, beta=0.05):
    """Construct token rewards from KL penalty plus terminal RM score.

    logp_policy: tensor [B, T]
    logp_ref:    tensor [B, T]
    rm_score:    tensor [B], scalar reward at end of completion
    """
    kl_per_token = logp_policy - logp_ref
    rewards = -beta * kl_per_token
    rewards[:, -1] = rewards[:, -1] + rm_score
    approx_kl = kl_per_token.mean()
    return rewards, {"approx_kl": float(approx_kl.detach())}
\end{lstlisting}

\subsection{PPO-style RLHF loss}

\Needspace{22\baselineskip}
\begin{lstlisting}[style=pythonstyle,caption={PPO clipped loss for language-model RLHF.},label={lst:ppo_rlhf_loss}]
def ppo_rlhf_loss(logp_new, logp_old, advantages, values, returns,
                  clip_eps=0.2, vf_coef=0.1):
    """Token-level PPO loss.

    All tensors have shape [B, T], except scalars.
    advantages and returns should be detached targets.
    """
    log_ratio = logp_new - logp_old
    ratio = torch.exp(log_ratio)

    adv = (advantages - advantages.mean()) / (advantages.std() + 1e-8)
    adv = adv.detach()

    unclipped = ratio * adv
    clipped = torch.clamp(ratio, 1.0 - clip_eps, 1.0 + clip_eps) * adv
    policy_loss = -torch.min(unclipped, clipped).mean()

    value_loss = 0.5 * (values - returns.detach()).pow(2).mean()
    total_loss = policy_loss + vf_coef * value_loss

    with torch.no_grad():
        approx_kl = (logp_old - logp_new).mean()
        clip_frac = ((ratio - 1.0).abs() > clip_eps).float().mean()

    stats = {
        "policy_loss": float(policy_loss.detach()),
        "value_loss": float(value_loss.detach()),
        "approx_kl": float(approx_kl.detach()),
        "clip_frac": float(clip_frac.detach()),
    }
    return total_loss, stats
\end{lstlisting}

\subsection{DPO loss}

\Needspace{20\baselineskip}
\begin{lstlisting}[style=pythonstyle,caption={Direct Preference Optimization loss.},label={lst:dpo_loss}]
def dpo_loss(policy_chosen_logp, policy_rejected_logp,
             ref_chosen_logp, ref_rejected_logp, beta=0.1):
    """DPO loss for paired preferences.

    log-probs are sequence log-probabilities, shape [B].
    """
    policy_logratio = policy_chosen_logp - policy_rejected_logp
    ref_logratio = ref_chosen_logp - ref_rejected_logp
    logits = beta * (policy_logratio - ref_logratio)
    loss = -F.logsigmoid(logits).mean()

    with torch.no_grad():
        implicit_reward_chosen = beta * (policy_chosen_logp - ref_chosen_logp)
        implicit_reward_rejected = beta * (policy_rejected_logp - ref_rejected_logp)
        acc = (implicit_reward_chosen > implicit_reward_rejected).float().mean()

    return loss, {"dpo_pair_accuracy": float(acc.detach())}
\end{lstlisting}

\subsection{Reward-model calibration check}

\Needspace{14\baselineskip}
\begin{lstlisting}[style=pythonstyle,caption={Simple reward-model calibration diagnostic.},label={lst:rm_calibration}]
def preference_prob(reward_chosen, reward_rejected):
    return torch.sigmoid(reward_chosen - reward_rejected)

def calibration_bins(probs, labels, n_bins=10):
    """Return bin confidence and empirical accuracy for RM calibration."""
    bins = torch.linspace(0, 1, n_bins + 1, device=probs.device)
    out = []
    for i in range(n_bins):
        mask = (probs >= bins[i]) & (probs < bins[i + 1])
        if mask.any():
            conf = probs[mask].mean()
            acc = labels[mask].float().mean()
            out.append((float(conf), float(acc), int(mask.sum())))
    return out
\end{lstlisting}

\section{Evaluation and safety metrics}

RLHF evaluation must measure more than reward-model score. A good evaluation suite separates the proxy objective from real-world behavior.

\begin{table}[t]
    \centering
    \caption{Evaluation metrics for RLHF systems.}
    \label{tab:rlhf_metrics}
    \begin{tabular}{p{3.2cm}p{5.0cm}p{4.2cm}}
        \toprule
        Metric & What it measures & Failure it detects \\
        \midrule
        Human win rate & Preference against baseline model & Real preference improvement \\
        Reward-model score & Learned reward proxy & Training progress, but not sufficient \\
        KL to reference & Policy drift & Reward hacking and fluency degradation \\
        Length-controlled win rate & Preference adjusted for verbosity & Verbosity bias \\
        Safety violation rate & Rule or policy violations & Harmful or unsafe responses \\
        Calibration error & Reward probability vs empirical preference & Miscalibrated reward model \\
        Adversarial prompt score & Robustness under stress tests & Jailbreak and edge-case failures \\
        \bottomrule
    \end{tabular}
\end{table}

\section{Failure modes and debugging}

\begin{table}[t]
    \centering
    \caption{Common RLHF failure modes.}
    \label{tab:rlhf_failures}
    \begin{tabular}{p{3.1cm}p{4.4cm}p{4.7cm}}
        \toprule
        Symptom & Likely cause & What to inspect \\
        \midrule
        Reward rises but human preference falls & Reward overoptimization & held-out human eval, KL, adversarial prompts \\
        Model becomes verbose & Length bias in reward model & length-controlled evaluation, reward vs length plots \\
        Model refuses too often & Safety reward too dominant & helpfulness/harmlessness trade-off metrics \\
        Model becomes sycophantic & preference data rewards agreement & disagreement prompts, truthfulness labels \\
        Training unstable & PPO/KL hyperparameters wrong & KL spikes, clip fraction, value loss \\
        Reward model overconfident & poor calibration or narrow data & calibration bins, OOD preference tests \\
        Direct method fails & preference data lacks coverage & dataset diversity and rejected-response quality \\
        \bottomrule
    \end{tabular}
\end{table}

\begin{warningbox}
    Never evaluate an RLHF system only with the reward model used for training. That is evaluating the optimizer with the proxy it was trained to exploit. Always use held-out human preference, external reward models, red-team prompts, or task-specific factual checks.
\end{warningbox}

\section{Limitations of RLHF}

\begin{enumerate}[leftmargin=*]
    \item \textbf{Reward models are proxies, not ground truth.} They can be overoptimized and may not generalize to new prompt distributions.
    \item \textbf{Preference data encodes annotator biases.} Cultural context, verbosity bias, and confidence bias can be reinforced rather than corrected.
    \item \textbf{KL regularization is a heuristic.} The right $\beta$ is problem-dependent and can limit both helpfulness and safety alignment.
    \item \textbf{RLHF does not enforce hard constraints.} Safety rules may need formal verification, rule checkers, or CBF layers in addition to preference learning.
    \item \textbf{Direct methods such as DPO, KTO, ORPO, and SimPO depend on dataset coverage.} They are sensitive to the quality of rejected responses and may not generalize to unseen behaviors.
    \item \textbf{AI feedback inherits judge-model biases.} Replacing human annotators with AI judges scales feedback but may reinforce the judge model's systematic errors.
\end{enumerate}

\section*{Looking Ahead to Chapter 20: Food for Thought}
\addcontentsline{toc}{section}{Looking Ahead to Chapter 20: Food for Thought}

RLHF explains how human preference became a training signal for language models. Chapter~20 moves from preference alignment to reasoning-oriented reinforcement learning. The reward signal changes: instead of only asking humans which answer is better, we may use verifiable outcomes such as mathematical correctness, executable tests, proof checkers, game results, or tool-use success. A key bridge is the process reward model: instead of scoring only the final answer, a process supervisor can score intermediate reasoning steps, as in the process-supervision work of Lightman et al. \citep{lightman2023verify}.

\begin{quote}
    Chapter~19 asked how human judgment can become a reward signal. Chapter~20 asks how verifiable reasoning outcomes can become a scalable reinforcement-learning signal.
\end{quote}

\section{Exercises}

\subsection*{Conceptual exercises}
\begin{enumerate}[leftmargin=*]
    \item Explain why preference comparisons are often easier to collect than absolute reward scores.
    \item Why is the KL penalty important in PPO-style RLHF?
    \item Give one example of reward hacking in a language-model assistant.
    \item Compare RLHF and DPO. Which pipeline is simpler, and what assumptions does it still rely on?
    \item Why should a safety-critical technical assistant combine preference learning with hard safety rules?
\end{enumerate}

\subsection*{Mathematical exercises}
\begin{enumerate}[leftmargin=*]
    \item Derive the reward-model loss in Eq.~\eqref{eq:rm_loss} from the Bradley--Terry probability in Eq.~\eqref{eq:bt_preference}.
    \item Show that increasing \(\beta\) in Eq.~\eqref{eq:kl_regularized_rlhf} makes the optimal policy stay closer to the reference model.
    \item For a preference pair with \(r_\psi(x,y^+)=3.0\) and \(r_\psi(x,y^-)=1.2\), compute the Bradley--Terry probability that the model prefers \(y^+\).
    \item Starting from the KL-regularized objective, explain why DPO contains log-probability ratios against a reference model.
\end{enumerate}

\subsection*{Implementation exercises}
\begin{enumerate}[leftmargin=*]
    \item Implement a reward model that uses a transformer encoder and a scalar reward head.
    \item Add length-controlled evaluation to the reward-model calibration code.
    \item Implement a small DPO training loop using synthetic preference pairs.
    \item Create a preference dataset for a network-operation assistant with preferred and rejected responses. Define at least three annotation criteria.
\end{enumerate}

\subsection*{Research thinking exercises}
\begin{enumerate}[leftmargin=*]
    \item How would you detect whether a reward model is rewarding verbosity rather than correctness?
    \item What are the risks of replacing human feedback entirely with AI feedback?
    \item In a technical domain such as SD-WAN traffic engineering, what kinds of feedback should come from human operators and what kinds should remain hard constraints?
    \item Design an evaluation suite for an RLHF-trained UAV/SDN assistant. Include usefulness, safety, factuality, and uncertainty metrics.
\end{enumerate}

	\chapter[RL for Reasoning Models]{Reinforcement Learning for Reasoning Models}
\chaptermark{RL for Reasoning Models}
\label{ch:reasoning_rl}

\begin{keybox}{Chapter goal}
    Reasoning-oriented reinforcement learning changes the reward from subjective preference to verifiable outcomes: correct final answers, passed unit tests, proof checkers, simulator scores, or network-constraint monitors. This chapter develops the RL foundations for reasoning models, including outcome and process rewards, group-relative objectives such as GRPO and RLOO, inference-time search, process reward models, and the DeepSeek-R1, o1, and Kimi k1.5 systems.
\end{keybox}

\section*{Chapter Overview}
\addcontentsline{toc}{section}{Chapter Overview}
\begin{enumerate}[leftmargin=*]
    \item Why reasoning models need reinforcement learning
    \item What does ``reasoning'' mean operationally?
    \item Reasoning as a sequential decision process
    \item Outcome rewards, process rewards, and verifiers
    \item From chain-of-thought prompting to RL-trained reasoning
    \item Verifiable rewards for mathematics, code, science, and networks
    \item RL objectives for reasoning models: PPO, GRPO, RLOO, and variants
    \item Credit assignment over reasoning steps
    \item Process reward models and step-level supervision
    \item Inference-time scaling: search, self-consistency, and verifier selection
    \item DeepSeek-R1, o1-style systems, Kimi k1.5, and 2025--2026 trends
    \item Concrete SD-WAN/UAV reasoning-assistant scenario
    \item Python implementation patterns
    \item Evaluation, safety, and failure modes
    \item Limitations and open research problems
    \item Looking Ahead to Chapter 21
\end{enumerate}

\section{Why reasoning models need reinforcement learning}

Chapter~19 introduced reinforcement learning from human feedback (RLHF): a pipeline in which human preferences are converted into a reward model, and a language model is optimized to produce responses preferred by humans. RLHF is powerful, but preference is not the only possible feedback signal. In many reasoning tasks, the outcome can be checked automatically.

A mathematical answer can be compared against a known solution. A program can be executed against unit tests. A theorem step can be passed to a proof checker. A network-control plan can be simulated and checked against latency or safety constraints. A chemical synthesis plan, robot plan, or database query can be validated by a domain tool. These tasks have something special: they provide \emph{verifiable rewards}.

This changed the role of RL in large language models. RL no longer has to depend only on subjective preference labels. It can optimize a policy using objective or semi-objective signals: correctness, test pass rate, constraint satisfaction, consistency, proof validity, or simulator score. This is one reason reasoning-oriented RL became central in 2024--2026 systems such as OpenAI o1, DeepSeek-R1, Kimi k1.5, and related open reasoning models \citep{jaech2024openai,deepseek2025r1,kimi2025k15}.

\begin{keybox}{Core idea}
    RL for reasoning treats a generated solution as a trajectory. The model chooses a sequence of intermediate reasoning actions; a verifier, reward model, or environment assigns feedback; the policy is updated to increase the probability of trajectories that solve the task.
\end{keybox}

The chapter is longer and more concrete than the previous RLHF chapter because reasoning RL sits at the intersection of several earlier parts of this book:
\begin{itemize}[leftmargin=*]
    \item Chapter 8: REINFORCE, baselines, and group-relative advantages;
    \item Chapter 10: PPO and KL-controlled policy optimization;
    \item Chapter 15: Decision Transformers and sequence modeling;
    \item Chapter 18: safe RL and verifiable constraints;
    \item Chapter 19: RLHF, reward models, DPO, and preference learning.
\end{itemize}

Reasoning RL is not a single algorithm. It is a family of training and inference methods that use rewards to improve multi-step problem solving.

\section{What does ``reasoning'' mean operationally?}

In everyday language, reasoning can mean many things: thinking carefully, proving, planning, calculating, debugging, verifying, or explaining. For a reinforcement learning book, we need a more operational definition.

\begin{definition}[Operational reasoning trajectory]
    A reasoning trajectory is a sequence
    \[
    \tau = (x, z_1, z_2, \ldots, z_T, y),
    \]
    where $x$ is the task input, $z_t$ are intermediate reasoning steps or latent decision states, and $y$ is the final answer or action. A reward function evaluates either the final answer $R(x,y)$, the intermediate steps $r_t(x,z_{1:t})$, or both.
\end{definition}

This definition is intentionally broad. The steps $z_t$ may be visible chain-of-thought tokens, hidden scratchpad states, program edits, proof commands, tool calls, search nodes, network-control decisions, or high-level subgoals. The final answer $y$ may be text, code, a route, a theorem proof, a numerical decision, or an SD-WAN control plan.

\subsection{Reasoning is not just longer text}

A model that writes a long answer is not necessarily reasoning well. Long reasoning can help if it decomposes the problem and preserves correctness, but it can also hallucinate. The important property is not length. It is whether intermediate steps help the model select actions that lead to verifiable success.

\begin{warningbox}{Chain-of-thought is not a guarantee}
    \textbf{Common confusion.} Chain-of-thought text is an interface, not a guarantee of reasoning. A reasoning model should be judged by task success, verification, robustness, calibration, and safety--not merely by whether it produces many intermediate tokens.
\end{warningbox}

\subsection{Types of reasoning tasks}

Table~\ref{tab:reasoning_task_types} shows common reasoning tasks and possible reward signals.

\begin{table}[t]
    \centering
    \caption{Reasoning tasks and possible reward sources.}
    \label{tab:reasoning_task_types}
\begin{tabular}{p{0.22\textwidth}p{0.34\textwidth}p{0.33\textwidth}}
        \toprule
        Task family & Example trajectory & Reward or verifier \\
        \midrule
        Mathematics & intermediate equations, final answer & exact answer match, symbolic checker, process reward model \\
        Code generation & plan, code edits, tests & unit tests, static analysis, execution score \\
        Scientific reasoning & hypothesis, derivation, calculation & simulator, known result, expert verifier \\
        Tool use & choose tool, call API, inspect output & task completion, environment feedback \\
        Network control & diagnose congestion, propose route/slice change & QoS constraints, latency, packet loss, safety monitor \\
        Proof search & tactic sequence, proof state transitions & proof checker accepts or rejects \\
        \bottomrule
    \end{tabular}
\end{table}

\section{Reasoning as a sequential decision process}

A reasoning model can be viewed as a policy
\[
\pi_\theta(a_t \mid h_t),
\]
where $h_t$ is the context or partial reasoning history, and $a_t$ is the next token, tool call, proof step, code edit, or high-level action.

The state is rarely fully observable. The model sees text and tool outputs, but not the true hidden problem structure. Therefore, reasoning RL is often closer to a partially observable decision process than a simple MDP. In practice, the context window acts as the model's memory.

\begin{figure}[t]
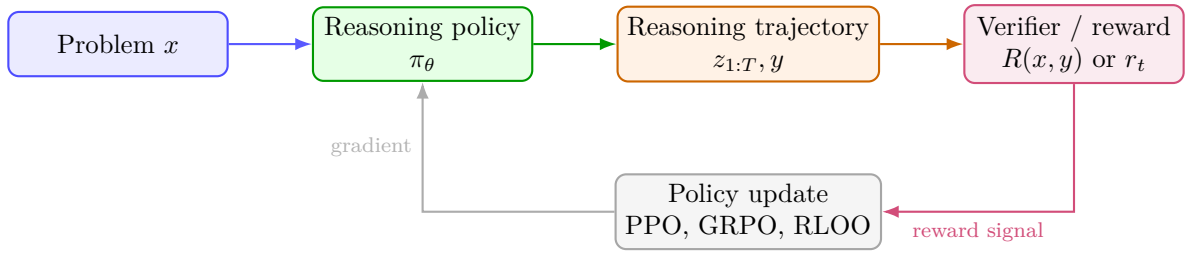

    \centering
    %
    \caption{Reasoning as reinforcement learning. A model generates a reasoning trajectory; a verifier or reward function scores the result; the policy is updated to make successful trajectories more likely.}
    \label{fig:reasoning_rl_loop}
\end{figure}

A trajectory-level objective is
\begin{equation}
    J(\theta) = \E_{y\sim \pi_\theta(\cdot\mid x)}[R(x,y)].
    \label{eq:reasoning_outcome_objective}
\end{equation}
For a token sequence $y=(y_1,\ldots,y_T)$,
\begin{equation}
    \log \pi_\theta(y\mid x) = \sum_{t=1}^T \log \pi_\theta(y_t\mid x,y_{<t}).
\end{equation}
A simple REINFORCE estimator is
\begin{equation}
    \grad_\theta J(\theta)
    \approx
    (R(x,y)-b(x))
    \sum_{t=1}^T \grad_\theta \log \pi_\theta(y_t\mid x,y_{<t}).
    \label{eq:reasoning_reinforce}
\end{equation}

Equation~\eqref{eq:reasoning_reinforce} is simple but reveals the main problem: one final reward supervises many tokens. This is the reasoning version of delayed reward and credit assignment.

\section{Outcome rewards, process rewards, and verifiers}

Reasoning rewards can be divided into three broad categories.

\subsection{Outcome rewards}

An outcome reward evaluates the final answer:
\begin{equation}
    R_{\mathrm{out}}(x,y) = \mathbf{1}\{\mathrm{Verifier}(x,y)=\mathrm{correct}\}.
\end{equation}
This is common in mathematics, coding, games, and simulation tasks. Outcome rewards are cheap and scalable when verifiers are available, but they are sparse. They do not explain which reasoning step caused success or failure.

\subsection{Process rewards}

A process reward scores intermediate steps:
\begin{equation}
    R_{\mathrm{proc}}(x,z_{1:T},y)
    =
    \sum_{t=1}^T r_t(x,z_{1:t}).
\end{equation}
Process supervision was studied for mathematical reasoning in work such as \emph{Let's Verify Step by Step}, which released PRM800K and demonstrated that step-level supervision can improve reasoning performance \citep{lightman2023verify}.

\subsection{Verifiers and reward models}

A verifier may be symbolic, learned, or hybrid:
\begin{itemize}[leftmargin=*]
    \item symbolic verifier: exact answer checker, compiler, unit test, proof checker;
    \item learned verifier: reward model, outcome reward model (ORM), process reward model (PRM);
    \item hybrid verifier: run tests, then ask a learned judge to evaluate style, explanation, or safety.
\end{itemize}

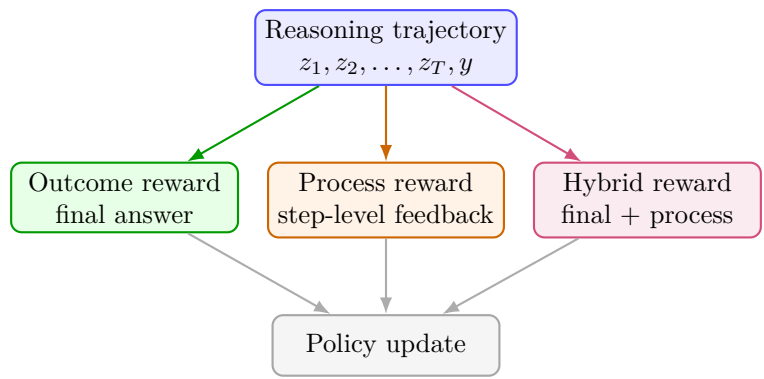
\begin{figure}[t]
    \centering
    \begin{tikzpicture}[
        box/.style={draw,rounded corners,thick,minimum width=3.0cm,minimum height=0.8cm,align=center,font=\small},
        arrow/.style={-{Latex[length=2.2mm]},thick},
        node distance=0.9cm
        ]
        \node[box,fill=blue!8,draw=blue!70] (traj) {Reasoning trajectory\\$z_1,z_2,\ldots,z_T,y$};
        \node[box,fill=green!10,draw=green!60!black,below left=1.0cm and 0.2cm of traj] (outcome) {Outcome reward\\final answer};
        \node[box,fill=orange!10,draw=orange!80!black,below=1.0cm of traj] (process) {Process reward\\step-level feedback};
        \node[box,fill=purple!8,draw=purple!70,below right=1.0cm and 0.2cm of traj] (hybrid) {Hybrid reward\\final + process};
        \node[box,fill=gray!8,draw=gray!70,below=1.0cm of process] (update) {Policy update};

        \draw[arrow,draw=green!60!black] (traj) -- (outcome);
        \draw[arrow,draw=orange!80!black] (traj) -- (process);
        \draw[arrow,draw=purple!70] (traj) -- (hybrid);
        \draw[arrow,draw=gray!65] (outcome) -- (update);
        \draw[arrow,draw=gray!65] (process) -- (update);
        \draw[arrow,draw=gray!65] (hybrid) -- (update);
    \end{tikzpicture}
    \caption{Outcome rewards supervise the final answer; process rewards supervise intermediate reasoning steps; hybrid systems combine both. Outcome rewards scale well but suffer from credit assignment. Process rewards give denser feedback but are harder to collect and easier to hack if mis-specified.}
    \label{fig:outcome_vs_process_rewards}
\end{figure}

\section{From chain-of-thought prompting to RL-trained reasoning}

Before reasoning RL became central, several ideas prepared the ground.

Chain-of-thought prompting showed that prompting large models to generate intermediate steps improves performance on arithmetic, commonsense, and symbolic reasoning tasks \citep{wei2022chain}. Self-consistency improved chain-of-thought prompting by sampling multiple reasoning paths and selecting the most common final answer \citep{wang2022selfconsistency}. STaR bootstrapped reasoning by generating rationales and fine-tuning on successful ones \citep{zelikman2022star}. Training verifiers for GSM8K showed that evaluating many candidate solutions with a verifier can improve mathematical problem solving \citep{cobbe2021verifiers}.

These were not all RL algorithms in the strict sense, but they introduced three ideas that reasoning RL later made central:
\begin{enumerate}[leftmargin=*]
    \item generate multiple candidate trajectories;
    \item evaluate them using an external signal;
    \item increase the probability of successful reasoning.
\end{enumerate}

\begin{table}[t]
    \centering
    \caption{Reasoning-improvement methods before and during reasoning RL.}
    \label{tab:reasoning_methods_lineage}
    \begin{tabular}{p{0.22\textwidth}p{0.36\textwidth}p{0.31\textwidth}}
        \toprule
        Method & Core idea & RL connection \\
        \midrule
        Chain-of-thought prompting & Elicit intermediate reasoning steps & exposes trajectory structure \\
        Self-consistency & sample many paths, vote on answer & inference-time search over trajectories \\
        STaR & keep successful rationales and train on them & self-generated improvement loop \\
        Verifier selection & score candidates using a verifier & reward model at inference time \\
        GRPO / RLOO & compare completions within a group & policy-gradient training without a learned critic \\
        Process rewards & score intermediate steps & dense reward and credit assignment \\
        \bottomrule
    \end{tabular}
\end{table}

\section{Verifiable rewards for mathematics, code, science, and networks}

Reasoning RL is strongest when the reward can be checked. The reward does not need to be perfect, but it must be reliable enough that optimizing it improves the real task.

\subsection{Mathematics}

For many math problems, the final answer can be extracted and compared to a ground-truth value. This is simple but brittle: a model may solve the problem correctly but format the answer differently. Modern systems often combine answer extraction, exact matching, symbolic simplification, and format rewards.

\subsection{Code}

For code generation, the reward can be unit-test pass rate:
\begin{equation}
    R_{\mathrm{code}}(x,y) = \frac{1}{M}\sum_{m=1}^M \mathbf{1}\{\mathrm{test}_m(y)=\mathrm{pass}\}.
\end{equation}
This reward is more informative than a binary answer check but can be gamed if tests are incomplete.

\subsection{Network operation}

For SD-WAN or UAV-assisted networks, the verifier may be a simulator or monitoring pipeline:
\begin{equation}
    R_{\mathrm{net}} = w_q Q_{\mathrm{QoS}} - w_c C_{\mathrm{constraint}} - w_u C_{\mathrm{unsafe}}.
\end{equation}
The reasoning model may propose a diagnosis and control plan. A safe controller or CBF layer still enforces hard constraints, as developed in Chapter~18. The verifier itself can be a network simulator or a digital twin that replays telemetry under the proposed action. Our SD-WAN safe-RL work uses uncertainty-aware and ensemble-based neural CBFs as a hard-safety verifier for traffic-engineering actions \citep{bista2026vtc,bista2026ifip}; the same architecture can serve as a verifier for reasoning-model proposed interventions, scoring them on predicted QoS satisfaction and constraint feasibility.

\begin{researchbox}
    \textbf{Research signature: reasoning for safe network control.} A reasoning model can act as a decision-support layer above a safe RL controller. It explains why congestion occurs, proposes candidate routing or slicing changes, and calls a simulator or safety filter to check consequences. The final action should still pass a safety layer such as the neural CBF mechanisms developed in Chapter~18 and in safe SD-WAN work \citep{bista2026vtc,bista2026ifip}.
\end{researchbox}

\section{RL objectives for reasoning models}

Reasoning policies are usually optimized with a KL penalty to prevent destructive drift from a reference model. A common objective is
\begin{equation}
    J(\theta)
    =
    \E_{y\sim\pi_\theta(\cdot\mid x)}
    \left[
    R(x,y)
    -
    \beta \log \frac{\pi_\theta(y\mid x)}{\pi_{\mathrm{ref}}(y\mid x)}
    \right].
    \label{eq:reasoning_kl_objective}
\end{equation}
This is the reasoning version of the RLHF objective from Chapter~19 and the KL-controlled PPO objective from Chapter~10.

\subsection{PPO for reasoning}

PPO-style reasoning RL uses token-level log-probability ratios and advantage estimates. The reward may be outcome-level, process-level, or a mixture. PPO remains useful when one wants value heads, KL monitoring, and minibatch updates.

\subsection{RLOO and GRPO}

RLOO and GRPO-style methods avoid a learned value head by sampling a group of completions for the same prompt. If a prompt $x$ produces $K$ completions $y_1,\ldots,y_K$ with rewards $R_i$, then a group-normalized advantage is
\begin{equation}
    \hat A_i
    =
    \frac{R_i - \mu_g}{\sigma_g + \epsilon},
    \quad
    \mu_g = \frac{1}{K}\sum_{j=1}^K R_j.
    \label{eq:group_relative_advantage_reasoning}
\end{equation}
DeepSeekMath introduced GRPO for mathematical reasoning, and DeepSeek-R1 used large-scale RL to elicit reasoning behaviors such as reflection and verification \citep{shao2024deepseekmath,deepseek2025r1}. This connects directly to Chapter~8's group-relative advantage discussion and to Chapter~19's treatment of direct preference optimization without a separate value head. In reasoning RL, GRPO becomes especially natural because outcome verifiers produce scalar rewards for multiple completions of the same prompt.

\begin{figure}[t]
    \centering
    \begin{tikzpicture}[
        box/.style={draw,rounded corners,thick,minimum width=2.6cm,minimum height=0.75cm,align=center,font=\small},
        small/.style={draw,rounded corners,minimum width=1.7cm,minimum height=0.55cm,align=center,font=\small},
        arrow/.style={-{Latex[length=2.2mm]},thick}
        ]
        \node[box,fill=blue!8,draw=blue!70] (prompt) at (0,0) {Prompt $x$};
        \node[small,fill=green!10,draw=green!60!black] (y1) at (3,1.2) {$y_1$, $R_1$};
        \node[small,fill=green!10,draw=green!60!black] (y2) at (3,0.4) {$y_2$, $R_2$};
        \node[small,fill=green!10,draw=green!60!black] (y3) at (3,-0.4) {$y_3$, $R_3$};
        \node[small,fill=green!10,draw=green!60!black] (y4) at (3,-1.2) {$y_4$, $R_4$};
        \node[box,fill=orange!10,draw=orange!80!black] (norm) at (6,0) {Group baseline\\$\mu_g,\sigma_g$};
        \node[box,fill=purple!8,draw=purple!70] (adv) at (9,0) {Advantages\\$\hat A_i$};

        \draw[arrow,draw=blue!65] (prompt) -- (y1);
        \draw[arrow,draw=blue!65] (prompt) -- (y2);
        \draw[arrow,draw=blue!65] (prompt) -- (y3);
        \draw[arrow,draw=blue!65] (prompt) -- (y4);
        \draw[arrow,draw=green!60!black] (y1) -- (norm);
        \draw[arrow,draw=green!60!black] (y2) -- (norm);
        \draw[arrow,draw=green!60!black] (y3) -- (norm);
        \draw[arrow,draw=green!60!black] (y4) -- (norm);
        \draw[arrow,draw=orange!80!black] (norm) -- (adv);
    \end{tikzpicture}
    \caption{Group-relative reasoning RL. Multiple completions for the same prompt are scored by a verifier. The policy update compares each completion against the group, reducing the need for a separate value model.}
    \label{fig:group_relative_reasoning}
\end{figure}
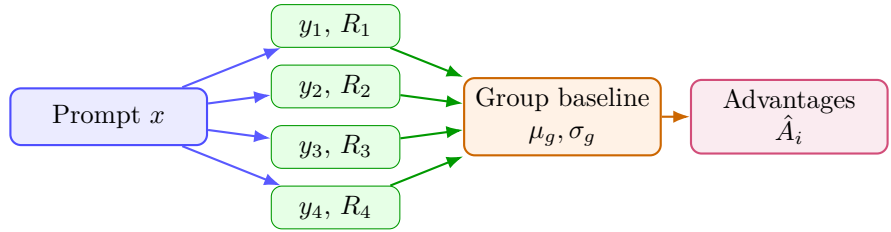

\subsection{Why group baselines help}

Reasoning prompts vary greatly in difficulty. A reward of $0.6$ may be excellent for a hard theorem but poor for an easy arithmetic problem. Group-relative baselines compare solutions for the same prompt, which controls for prompt difficulty.

\section{Credit assignment over reasoning steps}

Outcome rewards create a credit-assignment problem. If a solution contains 200 tokens and the final answer is wrong, which step caused the failure? If a program passes all tests, which design choice was essential? This mirrors the planning problem of Chapter~13: an outcome reward at the end of a long trajectory must be attributed to specific decisions along the way. MuZero addresses this with search over a learned model; process reward models address it by attaching supervision to intermediate reasoning steps.

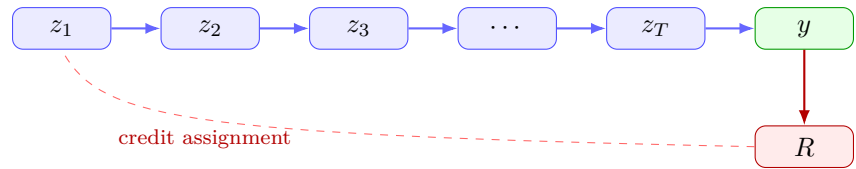
\begin{figure}[t]
    \centering
    \begin{tikzpicture}[
        rstep/.style={draw,rounded corners,minimum width=1.3cm,minimum height=0.55cm,align=center,font=\small},
        arrow/.style={-{Latex[length=2mm]},thick},
        node distance=0.65cm
        ]
        \node[rstep,fill=blue!8,draw=blue!70] (s1) {$z_1$};
        \node[rstep,fill=blue!8,draw=blue!70,right=of s1] (s2) {$z_2$};
        \node[rstep,fill=blue!8,draw=blue!70,right=of s2] (s3) {$z_3$};
        \node[rstep,fill=blue!8,draw=blue!70,right=of s3] (s4) {$\cdots$};
        \node[rstep,fill=blue!8,draw=blue!70,right=of s4] (sT) {$z_T$};
        \node[rstep,fill=green!10,draw=green!60!black,right=of sT] (ans) {$y$};
        \node[rstep,fill=red!8,draw=red!70!black,below=1.0cm of ans] (rew) {$R$};

        \foreach \a/\b in {s1/s2,s2/s3,s3/s4,s4/sT,sT/ans}
        \draw[arrow,draw=blue!60] (\a) -- (\b);
        \draw[arrow,draw=red!70!black] (ans) -- (rew);
        \draw[draw=red!60,dashed] (rew.west) .. controls (1.5,-1.4) and (0.5,-1.2) .. node[below,font=\scriptsize,text=red!70!black] {credit assignment} (s1.south);
    \end{tikzpicture}
    \caption{Outcome-level reasoning reward creates long-horizon credit assignment. The final reward supervises many prior reasoning steps, but it does not say which step made the trajectory succeed or fail.}
    \label{fig:reasoning_credit_assignment}
\end{figure}

There are four common responses:
\begin{enumerate}[leftmargin=*]
    \item use outcome rewards and rely on many samples;
    \item train a process reward model;
    \item use search or verification at inference time;
    \item train with curriculum or easier subtasks first.
\end{enumerate}

\begin{warningbox}{Dense reward is not automatically better}
    \textbf{A process reward model can help credit assignment, but it can also introduce new reward hacking.} If the PRM rewards plausible-looking steps rather than valid steps, the policy may learn to sound correct while becoming less reliable.
\end{warningbox}

\section{Process reward models and step-level supervision}

A process reward model estimates whether an intermediate step is useful or correct. Suppose a solution has steps $z_1,\ldots,z_T$. A PRM may output
\[
q_\psi(z_t \text{ is correct} \mid x,z_{1:t}).
\]
The process reward can be
\begin{equation}
    r_t = \log q_\psi(z_t \text{ correct} \mid x,z_{1:t}).
\end{equation}

Lightman et al. showed that process supervision can be highly effective for mathematical reasoning, while PRIME later explored implicit process rewards derived from outcome labels and policy rollouts \citep{lightman2023verify,cui2025prime}. These methods address the same issue from Chapter~8: reducing variance by moving from sparse outcome rewards to more informative step-level advantages.

\subsection{Outcome reward vs process reward}

\begin{table}[t]
    \centering
    \caption{Outcome and process rewards for reasoning RL.}
    \label{tab:orm_prm_comparison}
    \begin{tabular}{p{0.23\textwidth}p{0.31\textwidth}p{0.33\textwidth}}
        \toprule
        Reward type & Strength & Weakness \\
        \midrule
        Outcome reward & scalable when verifier exists; simple objective & sparse, poor token-level credit assignment \\
        Process reward & dense feedback; helps long reasoning & expensive labels; vulnerable to step-level reward hacking \\
        Hybrid reward & combines correctness and step quality & complex calibration; possible conflicting signals \\
        \bottomrule
    \end{tabular}
\end{table}

\section{Inference-time scaling: search, self-consistency, and verifier selection}

Reasoning models often improve by spending more compute at inference time. This is not training, but it is closely related to RL because it searches over possible trajectories and selects those with high reward.

\subsection[Best-of-N with a verifier]{Best-of-$N$ with a verifier}

Generate $N$ candidate solutions, score them with a verifier, and choose the best:
\begin{equation}
    y^* = \argmax_{y_i \in \{y_1,\ldots,y_N\}} V_\psi(x,y_i).
\end{equation}
This was central in verifier-based mathematical reasoning \citep{cobbe2021verifiers}.

\subsection{Self-consistency}

Self-consistency samples multiple reasoning paths and chooses the most common answer \citep{wang2022selfconsistency}. If the answer is unique and many paths lead to it, majority voting can reduce variance.

\subsection{Tree search and tool-assisted reasoning}

For some tasks, reasoning can be structured as search: propose a step, verify partial progress, branch, prune, and continue. Tree-of-Thoughts made this idea explicit by treating coherent reasoning steps as search nodes rather than only left-to-right tokens \citep{yao2023tot}. This connects Chapter~13's MuZero/MCTS view to language-model reasoning.

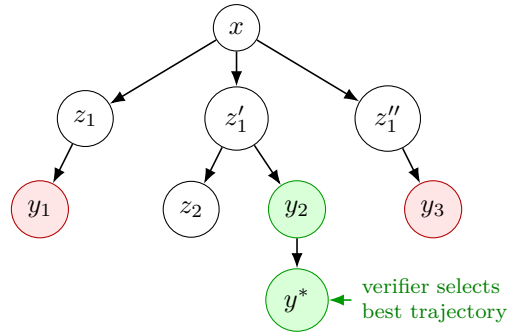
\begin{figure}[t]
    \centering
    \begin{tikzpicture}[
        node/.style={circle,draw,minimum size=0.55cm,font=\small},
        good/.style={circle,draw=green!60!black,fill=green!15,minimum size=0.55cm,font=\small},
        bad/.style={circle,draw=red!70!black,fill=red!10,minimum size=0.55cm,font=\small},
        arrow/.style={-{Latex[length=2mm]},semithick}
        ]
        \node[node] (root) at (0,0) {$x$};
        \node[node] (a) at (-2,-1.2) {$z_1$};
        \node[node] (b) at (0,-1.2) {$z_1'$};
        \node[node] (c) at (2,-1.2) {$z_1''$};
        \node[bad] (a1) at (-2.6,-2.4) {$y_1$};
        \node[node] (b1) at (-0.6,-2.4) {$z_2$};
        \node[good] (b2) at (0.8,-2.4) {$y_2$};
        \node[bad] (c1) at (2.6,-2.4) {$y_3$};
        \node[good] (best) at (0.8,-3.6) {$y^*$};

        \foreach \u/\v in {root/a,root/b,root/c,a/a1,b/b1,b/b2,c/c1,b2/best}
        \draw[arrow] (\u) -- (\v);
        \node[align=left,font=\scriptsize,text=green!50!black,
        right=0.3cm of best] (note) {verifier selects\\best trajectory};
        \draw[arrow,draw=green!60!black] (note.west) -- (best.east);
    \end{tikzpicture}
    \caption{Inference-time scaling for reasoning. The model samples or searches over multiple reasoning paths; a verifier, reward model, or voting rule selects the most reliable answer. This can be combined with training-time RL.}
    \label{fig:inference_time_search}
\end{figure}

\section{Systems: DeepSeek-R1, o1-style models, Kimi k1.5, and 2025--2026 trends}

OpenAI o1 made the phrase ``reasoning model'' widely visible. The o1 system card describes the model series as trained with large-scale reinforcement learning to reason using chain of thought, and discusses deliberative alignment for safer behavior \citep{jaech2024openai}. DeepSeek-R1 and R1-Zero showed that large-scale RL can elicit reasoning behaviors such as reflection and verification, with R1-Zero trained without supervised fine-tuning as a preliminary step, though with readability and language-mixing issues \citep{deepseek2025r1}. Kimi k1.5 emphasized long-context RL and practical scaling of RL for multimodal LLM reasoning \citep{kimi2025k15}. Qwen2.5-Math is another important 2024 system in this line: it combines mathematical data generation, reward modeling, reinforcement learning, and tool-integrated reasoning to improve math performance \citep{yang2024qwenmath}.

\begin{researchbox}
    \textbf{2026 perspective.} Reasoning RL is moving from ``fine-tune a model to answer better'' toward a broader systems view: generate trajectories, verify outcomes, learn from rewards, search at inference time, distill long reasoning into shorter policies, and enforce safety through explicit monitors.
\end{researchbox}

\subsection{What is new compared with ordinary RLHF?}

RLHF optimizes subjective preference. Reasoning RL often optimizes verifiable success. The difference matters:
\begin{itemize}[leftmargin=*]
    \item a preference reward may prefer a fluent but wrong explanation;
    \item a verifier reward can reject the final answer even if the explanation sounds good;
    \item process rewards can score intermediate steps but may also reward plausible mistakes.
\end{itemize}

\section{Concrete scenario: SD-WAN/UAV reasoning assistant}

This book's running application is UAV/SDN and SD-WAN control. Reasoning models can be used not as direct low-level controllers, but as operator-assistance and high-level planning systems.

Consider an SD-WAN system serving three classes of traffic:
\begin{itemize}[leftmargin=*]
    \item Class A: low-latency safety-critical control;
    \item Class B: video and interactive traffic;
    \item Class C: background IoT traffic.
\end{itemize}

The reasoning model receives telemetry: latency, jitter, packet loss, bandwidth use, policy logs, and recent route changes. It must produce a diagnosis and a candidate intervention, such as shifting some traffic from Internet to MPLS, adjusting slice weights, or requesting a safety-filtered RL controller to evaluate an action.

\begin{figure}[t]
    \centering
    \begin{tikzpicture}[
        box/.style={draw,rounded corners,thick,minimum width=3.0cm,minimum height=0.85cm,align=center,font=\small},
        arrow/.style={-{Latex[length=2.2mm]},thick},
        node distance=0.9cm
        ]
        \node[box,fill=blue!8,draw=blue!70] (telemetry) {Network telemetry\\latency, loss, SINR};
        \node[box,fill=green!10,draw=green!60!black,right=of telemetry] (reasoner) {Reasoning model\\diagnosis + plan};
        \node[box,fill=orange!10,draw=orange!80!black,right=of reasoner] (sim) {Verifier\\simulator/tests};
        \node[box,fill=red!7,draw=red!70!black,below=1.1cm of sim] (safety) {Safety layer\\CBF / constraints};
        \node[box,fill=gray!10,draw=gray!70,left=of safety] (controller) {RL controller\\PPO/SAC/MARL};
        \node[box,fill=purple!8,draw=purple!70,left=of controller] (operator) {Human operator\\accept/reject};

        \draw[arrow,draw=blue!65] (telemetry) -- (reasoner);
        \draw[arrow,draw=green!60!black] (reasoner) -- (sim);
        \draw[arrow,draw=orange!80!black] (sim) -- (safety);
        \draw[arrow,draw=red!70!black] (safety) -- (controller);
        \draw[arrow,draw=purple!70] (controller) -- (operator);
        \draw[arrow,draw=gray!65]
        (operator.north) -- ++(0,0.6) coordinate(opTop)
        -- node[above,font=\scriptsize,text=gray!65]{feedback}
        (opTop -| reasoner.south)
        -- (reasoner.south);
    \end{tikzpicture}
    \caption{A reasoning-RL assistant for SD-WAN/UAV control. The reasoning model proposes a diagnosis and high-level plan, but actions are checked by a verifier and safety layer before reaching the controller. Human operator feedback can improve explanation quality and decision support.}
    \label{fig:sdwan_reasoning_assistant}
\end{figure}
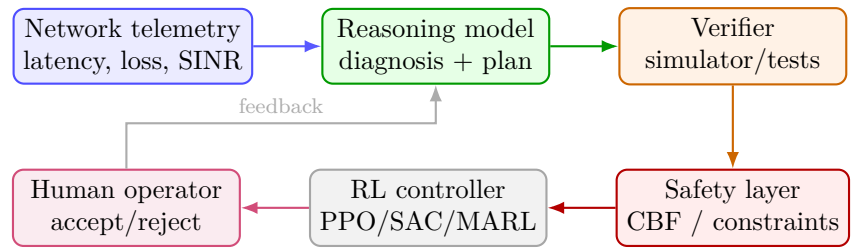

\subsection{Numerical example}

Suppose three candidate interventions are sampled for the same telemetry prompt:

\begin{table}[t]
    \centering
    \caption{Group-relative rewards for an SD-WAN reasoning assistant.}
    \label{tab:sdwan_reasoning_rewards}
    \begin{tabular}{p{0.27\textwidth}ccc}
        \toprule
        Candidate plan & QoS score & Safety penalty & Total reward \\
        \midrule
        Move Class A to MPLS, reduce Class C & $0.88$ & $0.05$ & $0.83$ \\
        Aggressively shift all traffic to MPLS & $0.94$ & $0.42$ & $0.52$ \\
        No change; only monitor & $0.55$ & $0.00$ & $0.55$ \\
        \bottomrule
    \end{tabular}
\end{table}

The second plan has high raw QoS but violates safety and capacity margins. A group-relative update should prefer the first plan because it improves QoS while keeping the intervention safe. This is the reasoning-model version of constrained RL from Chapter~18: explanation quality and decision quality matter, but hard constraints still dominate deployment.

\section{Training pipeline for reasoning RL}

A modern reasoning-RL system is usually not trained by one algorithm in isolation. It is a pipeline. A typical sequence is:
\begin{enumerate}[leftmargin=*]
    \item start from a pretrained language model;
    \item perform supervised fine-tuning on instruction and reasoning data;
    \item generate multiple candidate reasoning trajectories per prompt;
    \item score them using verifiers, outcome rewards, process rewards, or human preferences;
    \item optimize the policy with PPO, GRPO, RLOO, or another KL-controlled objective;
    \item filter failures and distill high-quality long reasoning into a more efficient policy.
\end{enumerate}

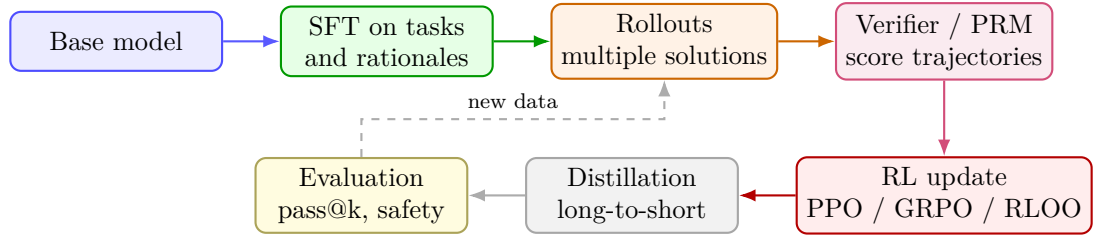
\begin{figure}[t]
    \centering
    \begin{tikzpicture}[
        box/.style={draw,rounded corners,thick,minimum width=2.8cm,minimum height=0.78cm,align=center,font=\small},
        arrow/.style={-{Latex[length=2.2mm]},thick},
        node distance=0.75cm
        ]
        \node[box,fill=blue!8,draw=blue!70] (base) {Base model};
        \node[box,fill=green!10,draw=green!60!black,right=of base] (sft) {SFT on tasks\\and rationales};
        \node[box,fill=orange!10,draw=orange!80!black,right=of sft] (rollout) {Rollouts\\multiple solutions};
        \node[box,fill=purple!8,draw=purple!70,right=of rollout] (verify) {Verifier / PRM\\score trajectories};
        \node[box,fill=red!7,draw=red!70!black,below=1.0cm of verify] (rl) {RL update\\PPO / GRPO / RLOO};
        \node[box,fill=gray!10,draw=gray!70,left=of rl] (distill) {Distillation\\long-to-short};
        \node[box,fill=yellow!15,draw=yellow!60!black,left=of distill] (eval) {Evaluation\\pass@k, safety};

        \draw[arrow,draw=blue!65] (base) -- (sft);
        \draw[arrow,draw=green!60!black] (sft) -- (rollout);
        \draw[arrow,draw=orange!80!black] (rollout) -- (verify);
        \draw[arrow,draw=purple!70] (verify) -- (rl);
        \draw[arrow,draw=red!70!black] (rl) -- (distill);
        \draw[arrow,draw=gray!65] (distill) -- (eval);
        \draw[arrow,draw=gray!65,dashed]
        (eval.north) -- ++(0,0.5) coordinate(evTop)
        -- node[above,font=\scriptsize]{new data}
        (evTop -| rollout.south)
        -- (rollout.south);
    \end{tikzpicture}
    \caption{A practical reasoning-RL pipeline. Reasoning ability is improved by generating trajectories, scoring them with verifiers or reward models, optimizing the policy, and often distilling long successful reasoning into cheaper inference policies.}
    \label{fig:reasoning_rl_pipeline}
\end{figure}

\subsection{Long reasoning, short answers, and distillation}

A model may need many intermediate tokens during training or search, but a deployed system may need short, cheap, and readable answers. Long-to-short distillation trains a smaller or faster policy to imitate high-quality solutions produced by a stronger long-reasoning model. This idea appears in systems that use long chain-of-thought trajectories to improve shorter responses \citep{kimi2025k15}. Rejection sampling fine-tuning is an earlier and simpler related mechanism: sample many reasoning paths, keep the correct ones, and fine-tune on successful trajectories \citep{yuan2023rft}.

Distillation is not merely compression. It changes the deployment trade-off. The expensive policy explores and verifies; the distilled policy executes quickly. In network-control applications, this is important because an operator assistant may have seconds, not minutes, to propose a safe intervention.

\section{A concrete numerical GRPO example}

Consider one math prompt sampled with four completions. The verifier assigns rewards
\[
R = [1.0, 0.0, 0.5, 1.0].
\]
The group mean and standard deviation are approximately
\[
\mu_g = 0.625, \qquad \sigma_g \approx 0.414.
\]
The normalized group advantages are
\[
\hat A \approx [0.91, -1.51, -0.30, 0.91].
\]
The two correct solutions receive positive updates. The wrong solution receives a strong negative update. The partially correct solution receives a weak negative update. This is more informative than a binary good/bad batch label because it compares each candidate with alternatives for the same prompt.

\begin{table}[t]
    \centering
    \caption{Numerical group-relative advantage example for reasoning RL.}
    \label{tab:numerical_grpo_example}
    \begin{tabular}{cccp{0.36\textwidth}}
        \toprule
        Completion & Reward & Advantage & Interpretation \\
        \midrule
        $y_1$ & $1.0$ & $+0.91$ & reinforce strongly \\
        $y_2$ & $0.0$ & $-1.51$ & suppress strongly \\
        $y_3$ & $0.5$ & $-0.30$ & slightly below group quality \\
        $y_4$ & $1.0$ & $+0.91$ & reinforce strongly \\
        \bottomrule
    \end{tabular}
\end{table}

\section{Reward design for reasoning: a practical decomposition}

A reasoning reward is often a weighted mixture:
\begin{equation}
    R(x,y)
    =
    w_{\mathrm{ans}} R_{\mathrm{answer}}
    +
    w_{\mathrm{proc}} R_{\mathrm{process}}
    +
    w_{\mathrm{fmt}} R_{\mathrm{format}}
    -
    w_{\mathrm{unsafe}} C_{\mathrm{safety}}
    -
    w_{\mathrm{len}} C_{\mathrm{length}}.
    \label{eq:reasoning_reward_decomposition}
\end{equation}
The answer term should usually dominate. Format and length rewards are helper terms, not the main objective. Safety costs should be treated as hard constraints whenever possible, especially in network, medical, legal, or cyber-physical applications.

\begin{table}[t]
    \centering
    \caption{Reward components for reasoning RL.}
    \label{tab:reasoning_reward_components}
    \begin{tabular}{p{0.23\textwidth}p{0.32\textwidth}p{0.33\textwidth}}
        \toprule
        Component & Purpose & Risk if overweighted \\
        \midrule
        Answer correctness & solve the task & may ignore explanation quality \\
        Process quality & reward valid intermediate steps & may reward plausible but false steps \\
        Format reward & improve extractability/readability & can create answer-template hacking \\
        Safety cost & block harmful or invalid plans & may be too weak if only a soft penalty \\
        Length cost & control inference budget & may discourage necessary reasoning \\
        \bottomrule
    \end{tabular}
\end{table}

\section{Reasoning RL and tool use}

Reasoning models increasingly interact with tools: calculators, code interpreters, search systems, proof checkers, simulators, and network controllers. In RL terms, a tool call is an action whose consequence becomes part of the next observation. A tool-using reasoning trajectory may be written as
\[
(x, a_1^{\mathrm{tool}}, o_1^{\mathrm{tool}}, a_2^{\mathrm{text}}, \ldots, y).
\]

Tool use changes the credit-assignment problem. A correct final answer may depend on one critical API call, one correct unit test, or one simulator rollout. It also changes safety: the system must decide which tools are allowed, which outputs can be trusted, and which actions require human approval.

\begin{keybox}{Tool-use principle}
    \textbf{The model may propose tool calls, but high-risk tools should be mediated by policies, sandboxes, safety filters, or human approval.} This is the same proposed-vs-executed action distinction developed in Chapters 11, 12, 13, 16, 17, and 18.
\end{keybox}

\section{Python implementation patterns}

This section gives compact code templates. They are intentionally small; production reasoning RL systems require distributed generation, sandboxing, data filtering, and careful safety review.

\subsection{Exact-answer and format rewards}

\Needspace{18\baselineskip}
\begin{lstlisting}[style=pythonstyle,caption={Simple verifiable rewards for mathematical reasoning.},label={lst:math_rewards}]
import re
from typing import Optional


def extract_boxed_answer(text: str) -> Optional[str]:
    """Extract a final answer from patterns such as \boxed{42} or Answer: 42."""
    boxed = re.findall(r"\\boxed\{([^{}]+)\}", text)
    if boxed:
        return boxed[-1].strip()
    m = re.search(r"(?i)final answer\s*[:=]\s*([^\n]+)", text)
    if m:
        return m.group(1).strip()
    m = re.search(r"(?i)answer\s*[:=]\s*([^\n]+)", text)
    return m.group(1).strip() if m else None


def math_outcome_reward(completion: str, gold: str) -> float:
    pred = extract_boxed_answer(completion)
    return 1.0 if pred is not None and pred == gold.strip() else 0.0


def format_reward(completion: str) -> float:
    """Encourage a clear separation between reasoning and final answer."""
    has_steps = "Step" in completion or "Therefore" in completion
    has_final = extract_boxed_answer(completion) is not None
    return 0.2 * float(has_steps) + 0.3 * float(has_final)


def combined_reward(completion: str, gold: str) -> float:
    return math_outcome_reward(completion, gold) + format_reward(completion)
\end{lstlisting}

\begin{warningbox}{Format reward caution}
    \textbf{Format rewards are useful for readability, but they are dangerous if too strong.} A model can learn to produce perfect formatting around wrong reasoning. Correctness rewards or verifiers must dominate cosmetic rewards.
\end{warningbox}

\subsection{Group-relative advantages}

\Needspace{16\baselineskip}
\begin{lstlisting}[style=pythonstyle,caption={Group-relative advantage computation for reasoning RL.},label={lst:grpo_advantages_reasoning}]
import torch


def group_relative_advantages(rewards: torch.Tensor, eps: float = 1e-8) -> torch.Tensor:
    """Compute normalized advantages for one prompt group.

    Args:
        rewards: tensor [K], rewards for K completions sampled for the same prompt.
    """
    mean = rewards.mean()
    std = rewards.std(unbiased=False).clamp_min(eps)
    return (rewards - mean) / std


# Example: three SD-WAN candidate plans for the same prompt.
r = torch.tensor([0.83, 0.52, 0.55])
adv = group_relative_advantages(r)
print(adv)  # first plan gets the strongest positive update
\end{lstlisting}

\subsection{Token-level GRPO-style loss}

\Needspace{22\baselineskip}
\begin{lstlisting}[style=pythonstyle,caption={A compact GRPO-style token loss with KL control.},label={lst:grpo_loss_reasoning}]
import torch


def grpo_token_loss(logp_new, logp_old, logp_ref, mask, advantages,
                    clip_eps=0.2, beta_kl=0.01):
    """Compute a group-relative policy-gradient loss.

    Args:
        logp_new: [B, T] log prob under current policy
        logp_old: [B, T] log prob under rollout policy
        logp_ref: [B, T] log prob under reference model
        mask: [B, T] 1 for valid generated tokens, 0 for padding
        advantages: [B] group-relative completion advantages
    """
    ratio = torch.exp(logp_new - logp_old)
    adv = advantages[:, None]

    unclipped = ratio * adv
    clipped = torch.clamp(ratio, 1.0 - clip_eps, 1.0 + clip_eps) * adv
    pg_obj = torch.minimum(unclipped, clipped)

    # Per-token KL proxy against the reference model.
    kl = logp_new - logp_ref
    token_obj = pg_obj - beta_kl * kl

    loss = -((token_obj * mask).sum() / mask.sum().clamp_min(1.0))
    diagnostics = {
        "approx_kl": float(((kl * mask).sum() / mask.sum()).detach()),
        "clip_frac": float((((ratio - 1.0).abs() > clip_eps) * mask).sum().detach() / mask.sum().detach()),
    }
    return loss, diagnostics
\end{lstlisting}

\subsection[Verifier-guided best-of-N selection]{Verifier-guided best-of-$N$ selection}

\Needspace{18\baselineskip}
\begin{lstlisting}[style=pythonstyle,caption={Verifier-guided best-of-N selection.},label={lst:best_of_n_reasoning}]
from typing import List, Callable


def best_of_n(prompt: str,
              generate: Callable[[str], List[str]],
              score: Callable[[str, str], float]) -> str:
    """Generate candidates and choose the highest-scoring completion."""
    candidates = generate(prompt)
    scores = [score(prompt, y) for y in candidates]
    best_idx = max(range(len(candidates)), key=lambda i: scores[i])
    return candidates[best_idx]
\end{lstlisting}

\subsection{Safe unit-test reward for code reasoning}

\Needspace{24\baselineskip}
\begin{lstlisting}[style=pythonstyle,caption={A minimal unit-test reward skeleton for code reasoning.},label={lst:unit_test_reward}]
import subprocess
import tempfile
from pathlib import Path


def unit_test_reward(candidate_code: str, tests: str, timeout_s: float = 3.0) -> float:
    """Run candidate code against tests in a temporary directory.

    This is a teaching skeleton. A production system must use a hardened sandbox,
    resource limits, network isolation, and security review.
    """
    with tempfile.TemporaryDirectory() as tmp:
        tmp = Path(tmp)
        (tmp / "solution.py").write_text(candidate_code, encoding="utf-8")
        (tmp / "test_solution.py").write_text(tests, encoding="utf-8")
        try:
            result = subprocess.run(
                ["python", "-m", "pytest", "-q", "test_solution.py"],
                cwd=tmp,
                text=True,
                capture_output=True,
                timeout=timeout_s,
            )
        except subprocess.TimeoutExpired:
            return 0.0
        return 1.0 if result.returncode == 0 else 0.0
\end{lstlisting}

\begin{warningbox}{Sandbox warning}
    \textbf{Never run arbitrary model-generated code directly on a production machine.} Code rewards require sandboxing, process isolation, file-system restrictions, network restrictions, and time/memory limits.
\end{warningbox}

\subsection{Process-return aggregation}

\Needspace{18\baselineskip}
\begin{lstlisting}[style=pythonstyle,caption={Combining outcome and process rewards into one return.},label={lst:process_return_aggregation}]
import torch


def reasoning_return(outcome_reward, process_rewards, gamma=1.0,
                     process_weight=0.25):
    """Combine final correctness and step-level process rewards."""
    T = len(process_rewards)
    discounts = gamma ** torch.arange(T, dtype=torch.float32)
    dense = (discounts * process_rewards.float()).sum()
    return outcome_reward + process_weight * dense
\end{lstlisting}

\subsection{Process reward model loss}

\Needspace{20\baselineskip}
\begin{lstlisting}[style=pythonstyle,caption={Binary process reward model loss for step-level labels.},label={lst:prm_loss}]
import torch
import torch.nn.functional as F


def process_reward_loss(step_logits, step_labels, step_mask):
    """Train a PRM from step-level correctness labels.

    Args:
        step_logits: [B, T], unnormalized correctness logits
        step_labels: [B, T], 1 if step is correct/useful, 0 otherwise
        step_mask: [B, T], 1 for valid steps
    """
    per_step = F.binary_cross_entropy_with_logits(
        step_logits, step_labels.float(), reduction="none"
    )
    loss = (per_step * step_mask).sum() / step_mask.sum().clamp_min(1.0)
    return loss
\end{lstlisting}

\subsection{Network-operation reward with safety check}

\Needspace{20\baselineskip}
\begin{lstlisting}[style=pythonstyle,caption={A verifiable reward for an SD-WAN reasoning assistant.},label={lst:sdwan_reasoning_reward}]
def sdwan_reasoning_reward(plan, simulator, safety_filter):
    """Score a high-level plan using simulation and hard safety checks."""
    proposed_action = plan.to_action()
    safe_action, modified = safety_filter.project(proposed_action)
    metrics = simulator.rollout(safe_action)

    qos = 0.5 * metrics["throughput_norm"] \
        + 0.3 * (1.0 - metrics["latency_norm"]) \
        + 0.2 * (1.0 - metrics["loss_norm"])

    safety_penalty = 0.0
    safety_penalty += 2.0 * float(metrics["latency_ms_A"] > 6.0)
    safety_penalty += 1.0 * float(metrics["jitter_ms"] > 2.0)
    safety_penalty += 1.0 * float(modified)  # plan needed correction

    explanation_bonus = 0.1 if plan.has_clear_diagnosis() else 0.0
    return qos + explanation_bonus - safety_penalty
\end{lstlisting}

\section{Evaluation, safety, and failure modes}

Reasoning models should not be evaluated only by final accuracy. A serious evaluation includes:
\begin{itemize}[leftmargin=*]
    \item pass@1 and pass@$k$;
    \item verifier accuracy and calibration;
    \item reward hacking rate;
    \item token budget and compute cost;
    \item robustness to prompt perturbations;
    \item tool-use correctness;
    \item safety violation rate;
    \item human audit quality for high-risk settings.
\end{itemize}

\begin{table}[t]
    \centering
    \caption{Common failure modes in RL-trained reasoning models.}
    \label{tab:reasoning_rl_failures}
    \begin{tabular}{p{0.24\textwidth}p{0.33\textwidth}p{0.31\textwidth}}
        \toprule
        Symptom & Likely cause & What to inspect \\
        \midrule
        Long but wrong reasoning & reward encourages verbosity or format & correctness vs format reward weights \\
        Passes public tests, fails hidden tests & test overfitting & held-out tests, adversarial tests \\
        Correct answer, invalid reasoning & outcome-only reward & process verifier, proof checker \\
        Good verifier score, bad real result & reward model overoptimization & calibration, human audit, OOD prompts \\
        Unstable RL training & high reward variance & group baseline, KL, reward normalization \\
        Unsafe network plan & missing hard constraints & safety filter, CBF, simulator validity \\
        High compute cost & too much inference-time search & pass@compute curve, budget-aware decoding \\
        \bottomrule
    \end{tabular}
\end{table}

\section{Limitations and open research problems}

Reasoning RL is powerful but unsettled.

\subsection{Reward is still the bottleneck}

Verifiable tasks are easier than open-ended tasks. But many real tasks mix objective correctness with judgment: legal reasoning, medical triage, research planning, or network operations under incomplete data. Reward design remains difficult.

\subsection{Process rewards can be hacked}

A PRM may reward locally plausible steps that do not lead to a correct solution. Dense feedback is helpful only when the reward model tracks true progress. Reasoning models can also exploit superficial signals: they may satisfy formatting rewards while giving the wrong answer, produce verbose reasoning because longer answers correlate with higher reward, or generate steps that look mathematically serious without performing the needed computation. Length correlations in RLHF are a concrete warning that reward optimization can improve measured reward for the wrong reason \citep{singhal2023long}.

\subsection{Inference-time scaling is expensive}

Best-of-$N$, tree search, and verifier-guided decoding improve performance but increase inference cost. Kimi k1.5 and other systems emphasize long-context reasoning and long-to-short distillation, but the compute-performance trade-off remains central \citep{kimi2025k15}.

\subsection{Private chain-of-thought and safety}

For deployed systems, exposing full internal reasoning may create privacy, safety, or manipulation risks. A model may need to reason internally while producing a concise, faithful answer. This issue is related to deliberative alignment and safety reasoning \citep{jaech2024openai}.

\subsection{Reasoning generalization remains limited}

Success on one verifiable domain does not guarantee transfer to another. A model trained heavily on math or code rewards may improve on symbolic problems while still failing in scientific diagnosis, network operations, legal analysis, or long-horizon planning under uncertainty. This is why reasoning RL should be evaluated across task families, not only on the benchmark type used for reward training.

\section{Exercises}

\subsection*{Conceptual exercises}
\begin{enumerate}[leftmargin=*]
    \item Explain the difference between RLHF preference reward and verifiable reasoning reward.
    \item Why can an outcome reward be too sparse for long reasoning trajectories?
    \item Give an example where a process reward model could be hacked.
    \item Why does group-relative normalization help when prompts have different difficulty levels?
    \item Explain why reasoning models may benefit from inference-time search even after RL training.
\end{enumerate}

\subsection*{Mathematical exercises}
\begin{enumerate}[leftmargin=*]
    \item Starting from Eq.~\eqref{eq:reasoning_reinforce}, derive the group-baseline estimator for $K$ completions of the same prompt.
    \item Suppose rewards for four completions are $(1,0,0,1)$. Compute group-normalized advantages using Eq.~\eqref{eq:group_relative_advantage_reasoning}.
    \item Show how the KL-regularized objective in Eq.~\eqref{eq:reasoning_kl_objective} penalizes a completion whose probability under the current model is much larger than under the reference model.
    \item For a process reward $R=\sum_t r_t$, explain how credit assignment changes relative to a binary outcome reward.
\end{enumerate}

\subsection*{Coding exercises}
\begin{enumerate}[leftmargin=*]
    \item Implement a math-answer verifier that handles equivalent fractions and simple symbolic expressions.
    \item Extend Listing~\ref{lst:grpo_loss_reasoning} with per-sequence length normalization.
    \item Implement best-of-$N$ selection with both exact answer matching and a learned verifier score.
    \item Build a small SD-WAN simulator and use Listing~\ref{lst:sdwan_reasoning_reward} to score three candidate reasoning plans.
\end{enumerate}

\subsection*{Research exercises}
\begin{enumerate}[leftmargin=*]
    \item Design a hybrid outcome-plus-process reward for code generation. Which part is symbolic and which part is learned?
    \item Propose a safety protocol for using reasoning models as network-operations assistants. Which actions must be blocked by hard constraints?
    \item Compare GRPO-style training with PPO-style RLHF for mathematical reasoning. When is a value head useful?
    \item Discuss whether reasoning RL should optimize visible reasoning, hidden reasoning, or final answers only.
\end{enumerate}

\section*{Looking Ahead to Chapter 21: Food for Thought}
\addcontentsline{toc}{section}{Looking Ahead to Chapter 21: Food for Thought}

Chapter~20 showed how reinforcement learning can improve reasoning models using verifiable rewards, group-relative advantages, process rewards, and inference-time search. Chapter~21 returns to the book's engineering roots: communication networks, UAV systems, SD-WAN, network slicing, and autonomous control.

The connection is direct. A reasoning model can diagnose a network problem, but a DRL controller must still act safely. A verifier can check a math answer, but a simulator or CBF layer must check a network intervention. The next chapter studies how the algorithms developed throughout this book can be integrated into real communication systems.

\begin{quote}
    Chapter~20 explained how models learn to reason with reward. Chapter~21 explains how DRL systems act in networks where reasoning, control, safety, and deployment constraints meet.
\end{quote}

	\chapter[DRL for Networks, UAVs, and Communication Systems]{DRL for Networks, UAVs, and Communication Systems}
\chaptermark{DRL for Networks and UAVs}
\label{ch:drl_networks_uavs}

\begin{keybox}{Chapter goal}
    Communication networks are natural but demanding DRL domains: they require careful state engineering, class-aware reward design, safety constraints, digital twins, and staged deployment. This chapter develops the full pipeline from MDP/CMDP formulation to simulator-in-the-loop training, deployment gates, and frontier directions for SD-WAN, O-RAN, UAV slicing, and reasoning-augmented network control.
\end{keybox}

\section*{Chapter Overview}
\addcontentsline{toc}{section}{Chapter Overview}

This chapter turns the book from algorithmic foundations to an application domain where deep reinforcement learning is both attractive and dangerous: communication networks. We study how DRL can be used for SDN routing, SD-WAN traffic engineering, O-RAN xApps and rApps, wireless resource allocation, network slicing, edge offloading, UAV-assisted connectivity, vehicular networks, satellite/IoT scheduling, and cyber-resilient control.

The chapter is deliberately concrete. It does not only say that networks are dynamic. It shows how to construct states, actions, rewards, safety constraints, simulators, digital twins, and evaluation protocols. The goal is to give the reader enough structure to design a real research-grade DRL system for networks rather than a toy reward-maximization experiment.

\begin{enumerate}[leftmargin=*]
    \item Why networks are a difficult DRL domain
    \item Network control timescales and architectures
    \item Networked MDPs, CMDPs, and Markov games
    \item State construction from telemetry and radio measurements
    \item Action spaces: routing, slicing, scheduling, UAV movement, and power control
    \item Reward and constraint design for QoS, energy, fairness, and safety
    \item Closed-loop DRL architecture for SDN, SD-WAN, O-RAN, and UAV systems
    \item Algorithm selection: DQN, PPO, SAC, MARL, offline RL, model-based RL, and safe RL
    \item SD-WAN traffic engineering case study
    \item UAV-assisted network slicing case study
    \item Wireless realism: channels, SINR, throughput, latency, and mobility
    \item Simulator-in-the-loop research pipeline
    \item Python implementation: a network QoS environment
    \item Python implementation: graph observations, rewards, action decoding, and safety filters
    \item Experimental methodology and failure modes
    \item Security, resilience, and deployment discipline
    \item Frontiers toward 2026: O-RAN intelligence, foundation agents, NTN, ISAC, and digital twins
    \item Looking Ahead to Chapter 22
\end{enumerate}

\section{Why networks are a hard but natural domain for DRL}

Modern communication networks are control systems. They measure traffic, channels, user mobility, queue states, failures, latency, energy, service-level agreements, and topology changes. They then make decisions: route this traffic, allocate that slice, increase this power, hand over that user, move this UAV, accept or reject this flow, or trigger this mitigation action.

This is why reinforcement learning is naturally attractive. The controller does not simply predict a label. It chooses an action, observes delayed consequences, and adapts future decisions. In this sense, network control is close to the agent--environment loop introduced in Chapter~1.

However, networks are also much harder than games. In games, the simulator is often fast, closed, and faithful to the rules. In real networks, the simulator is imperfect, traffic is non-stationary, failure events are rare, deployment errors are expensive, and unsafe exploration is unacceptable. A network DRL agent must respect latency, reliability, and safety constraints even while it learns.

\begin{keybox}{Research-grade network DRL}
    Network DRL is not simply ``apply PPO to a topology.'' A research-grade system must specify the control timescale, telemetry, action granularity, reward, constraints, simulator fidelity, baseline controllers, failure modes, and deployment guardrails.
\end{keybox}

The early survey by Luong et al. framed DRL as a tool for resource allocation, caching, routing, UAV networks, and IoT, emphasizing that modern decentralized networks must make decisions under uncertainty \citep{luong2019applications}. Since then, the field has moved toward programmable network control through SDN, SD-WAN, O-RAN, digital twins, and safe/offline learning loops. The central question has shifted from ``Can DRL improve a metric in simulation?'' to ``Can DRL improve a network while remaining safe, reproducible, interpretable, and deployable?''

\section{Network control timescales}

The first design decision is the timescale. A DRL controller acting every packet has a different problem from a controller acting every minute. The wrong timescale can make the environment either too noisy or too slow to learn.

\begin{figure}[t]
    \centering
    \begin{tikzpicture}[
        box/.style={draw,rounded corners,thick,minimum width=2.85cm,minimum height=0.85cm,align=center,font=\small},
        arrow/.style={-{Latex[length=2.2mm]},thick}
        ]
        \node[box,fill=blue!8,draw=blue!70] (pkt) at (0,0) {Packet/MAC\\$\mu$s--ms};
        \node[box,fill=green!8,draw=green!60!black] (ran) at (3.5,0) {RAN/xApp\\10 ms--1 s};
        \node[box,fill=orange!10,draw=orange!80!black] (sdn) at (7.0,0) {SDN/SD-WAN\\100 ms--minutes};
        \node[box,fill=purple!8,draw=purple!70] (orch) at (10.5,0) {Orchestration\\minutes--hours};
        \draw[arrow] (pkt) -- (ran);
        \draw[arrow] (ran) -- (sdn);
        \draw[arrow] (sdn) -- (orch);
        \node[below=0.9cm of pkt,align=center,font=\small] {scheduling,\\rate control};
        \node[below=0.9cm of ran,align=center,font=\small] {slicing, handover,\\interference};
        \node[below=0.9cm of sdn,align=center,font=\small] {routing, splitting,\\TE};
        \node[below=0.9cm of orch,align=center,font=\small] {placement, scaling,\\policy};
    \end{tikzpicture}
    \caption{Network DRL must be matched to the control timescale. A packet-level controller faces a very different state, action, and safety problem from an SD-WAN traffic-engineering controller or a non-real-time orchestration policy.}
    \label{fig:network_timescales}
\end{figure}
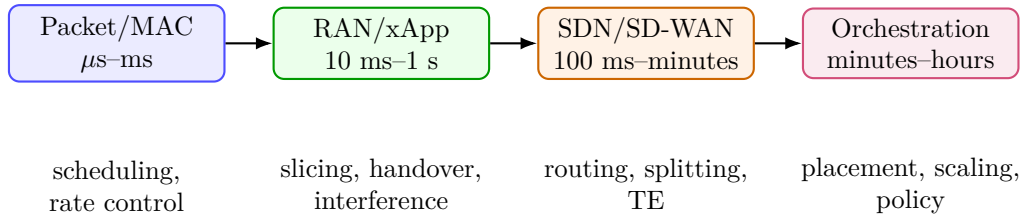

O-RAN makes this distinction explicit. The Near-Real-Time RIC hosts xApps for control loops on the order of roughly tens of milliseconds to one second, while the Non-Real-Time RIC hosts rApps and policy/analytics loops on slower timescales. Surveys of xApps emphasize this split because latency-sensitive control, policy guidance, and model management belong to different layers \citep{elyasi2025xapps}. For DRL, this means one should not place a slow offline-trained world model in a path that requires millisecond reaction unless the policy has been distilled into a lightweight controller.

\begin{table}[t]
    \centering
    \caption{Examples of DRL control problems at different communication-network timescales.}
    \label{tab:network_timescales}
    \begin{tabularx}{\textwidth}{p{2.6cm}p{3.2cm}X p{2.2cm}}
        \toprule
        Timescale & Control example & DRL formulation & Typical risk \\
        \midrule
        Microseconds--ms & MAC scheduling, rate adaptation & Contextual bandit, DQN, supervised policy distillation & Too slow for inference \\
        10 ms--1 s & O-RAN xApp slicing, handover, load balancing & PPO/SAC/xApp with safety guard & SLA violation during exploration \\
        100 ms--minutes & SDN/SD-WAN routing, traffic splitting & PPO/SAC/offline RL with digital twin & Transient congestion, oscillation \\
        Minutes--hours & edge placement, slice admission, UAV fleet planning & model-based RL, HRL, MARL, offline RL & bad long-horizon allocation \\
        \bottomrule
    \end{tabularx}
\end{table}

The timescale should also include the \emph{end-to-end control latency}: telemetry collection, feature extraction, policy inference, controller actuation, and the time required for the network to react. If a policy is trained at a one-second decision interval but telemetry arrives with a two-second delay, the agent is effectively acting on stale state. In such cases, the observation should include recent history, timestamp age, or recurrent memory; otherwise the Markov assumption used in the MDP abstraction is violated in practice.

\begin{keybox}{Practice note: reporting control loops}
    A network DRL paper should report the decision period, telemetry delay, action actuation delay, and inference time. A policy that is optimal at a simulated one-second loop can fail when deployed in a five-second delayed control loop.
\end{keybox}

\section{Networked MDPs, CMDPs, and Markov games}

A single-controller network problem can often be approximated as an MDP:
\begin{equation}
    \mathcal{M}=(\mathcal{S},\mathcal{A},p,r,\gamma),
\end{equation}
where the state may be telemetry, the action may be routing or resource allocation, and the reward may summarize QoS. But realistic network control usually needs richer structure.

A constrained network problem is better represented as a constrained MDP:
\begin{equation}
    \max_\pi \; \E_\pi\left[\sum_{t=0}^{\infty}\gamma^t r_t\right]
    \quad \text{s.t.} \quad
    \E_\pi\left[\sum_{t=0}^{\infty}\gamma^t c_t^{(k)}\right] \le d_k,
    \quad k=1,\ldots,K.
    \label{eq:network_cmdp}
\end{equation}
Here costs can represent latency violations, packet loss, battery risk, interference, collision risk, or regulatory constraints. Chapter~18 developed this formalism for safe RL. In networks, it is often the right mathematical starting point.

When there are many learning controllers, for example multiple UAVs, base stations, slices, or edge nodes, the problem becomes a Markov game or Dec-POMDP. This connects directly to Chapter~16. Each agent has local observations, but the global outcome depends on the joint action.

\begin{equation}
    p(s_{t+1}\given s_t,a_t^1,\ldots,a_t^N), \qquad
    r_t^i = r^i(s_t,a_t^1,\ldots,a_t^N).
\end{equation}

\begin{keybox}{Practice note: problem formulation}
    For a networking paper, state clearly whether the environment is single-agent MDP, CMDP, multi-agent Markov game, Dec-POMDP, or offline dataset problem. This choice determines which baselines are legitimate. Comparing a centralized MAPPO agent to independent DQN without explaining observability is not a fair experiment.
\end{keybox}

\section{State construction from telemetry and radio measurements}

Network state is never given automatically. It is engineered from measurements. A poor state design can make even a strong algorithm fail.

A useful network observation may contain:
\begin{itemize}
    \item traffic load per link, slice, cell, or queue;
    \item latency, jitter, packet loss, throughput, SINR, RSRP/RSRQ, CQI, BLER;
    \item user density, mobility, handover counters, session arrivals;
    \item topology, link capacities, routing tables, and failure alarms;
    \item UAV positions, battery levels, channel states, and coverage maps;
    \item recent history windows or recurrent hidden states;
    \item safety margins, such as distance to battery threshold or latency bound.
\end{itemize}

Wireless systems require physical realism. Channel models such as 3GPP TR~38.901 provide standardized path-loss, shadowing, and channel assumptions for 5G and beyond simulations \citep{tr38901}. The latest ETSI/3GPP versions continue to evolve channel-model support for new frequency bands and scenarios, so wireless DRL papers should report exactly which channel model, carrier frequency, antenna assumptions, mobility model, and fading configuration are used.

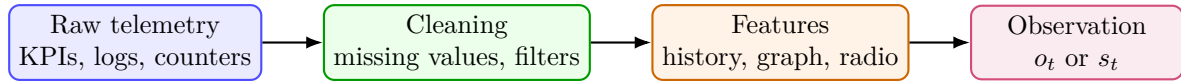
\begin{figure}[t]
    \centering
    \begin{tikzpicture}[
        box/.style={draw,rounded corners,thick,minimum width=2.8cm,minimum height=0.8cm,align=center,font=\small},
        arrow/.style={-{Latex[length=2.2mm]},thick},
        node distance=0.8cm
        ]
        \node[box,fill=blue!8,draw=blue!70] (raw) {Raw telemetry\\KPIs, logs, counters};
        \node[box,fill=green!8,draw=green!60!black,right=of raw] (clean) {Cleaning\\missing values, filters};
        \node[box,fill=orange!10,draw=orange!80!black,right=of clean] (features) {Features\\history, graph, radio};
        \node[box,fill=purple!8,draw=purple!70,right=of features] (obs) {Observation\\$o_t$ or $s_t$};
        \draw[arrow] (raw) -- (clean);
        \draw[arrow] (clean) -- (features);
        \draw[arrow] (features) -- (obs);
        \node[below=0.9cm of clean,align=center,font=\small] {smoothing can hide spikes};
        \node[below=0.9cm of features,align=center,font=\small] {aggregation can hide users};
    \end{tikzpicture}
    \caption{Network state is an engineered object. The observation used by a DRL agent is the result of telemetry collection, cleaning, aggregation, normalization, and feature construction.}
    \label{fig:state_pipeline}
\end{figure}

A subtle but important issue is \emph{which time the observation describes}. Let $z_t$ be the latest telemetry packet available at the controller and let $\Delta_t$ be its age. A more honest observation is therefore not only $z_t$, but
\begin{equation}
    o_t = [z_t,\Delta_t,h_t],
\end{equation}
where $h_t$ may summarize recent history. This small addition can prevent a policy from treating stale measurements as fresh measurements. In SD-WAN and O-RAN systems, stale telemetry is often a larger source of failure than neural-network approximation error.

\section{Action spaces: from routing splits to UAV trajectories}

The action space determines whether value-based, policy-gradient, actor-critic, multi-agent, or safe-control methods are appropriate.

\begin{table}[t]
    \centering
    \caption{Common network actions and suitable DRL families.}
    \label{tab:network_actions}
    \begin{tabularx}{\textwidth}{p{3.3cm}p{3.2cm}X}
        \toprule
        Action type & Example & Suitable methods \\
        \midrule
        Discrete choice & select path, channel, slice admission & DQN, Double DQN, Rainbow, categorical PPO \\
        Continuous allocation & bandwidth share, traffic split, power, UAV velocity & PPO, SAC, TD3, constrained actor-critic \\
        Hybrid action & select path and allocate rate & parameterized action RL, hierarchical RL, hybrid actors \\
        Multi-agent action & each UAV/base station acts locally & QMIX, MAPPO, MADDPG, graph MARL \\
        Sequence action & multi-step maintenance or reasoning plan & Decision Transformer, model-based planning, reasoning RL \\
        Safety-filtered action & propose then project action & safe RL, CBF, shielded PPO/SAC \\
        \bottomrule
    \end{tabularx}
\end{table}

A common mistake is to discretize a continuous control problem too coarsely. For example, suppose a UAV controller discretizes horizontal direction into 8 bins, vertical velocity into 3 bins, speed into 5 bins, bandwidth split into 5 bins, and transmit power into 5 bins. This already yields
\begin{equation}
    8\times 3\times 5\times 5\times 5 = 3000
\end{equation}
discrete actions. A continuous actor can output the same quantities directly as a vector, while a safety layer enforces physical and QoS constraints.

\section{Reward and constraint design}

In communication systems, reward design is not a cosmetic detail. It defines the network behavior the agent will optimize. A reward based only on throughput can increase congestion. A reward based only on latency can starve low-priority traffic. A reward based only on energy can refuse service.

A class-aware QoS reward can be written as
\begin{equation}
    r_t =
    \sum_{u\in\mathcal{U}_t} w_{c(u)}
    \left[
    \eta_T \, \mathrm{sat}_T(T_u)
    - \eta_L \, \mathrm{viol}_L(L_u)
    - \eta_P \, \mathrm{viol}_P(P_u)
    \right]
    - \eta_E E_t - \eta_S C_t,
    \label{eq:network_reward}
\end{equation}
where $c(u)$ is the user class, $T_u$ is throughput, $L_u$ is latency, $P_u$ is packet loss, $E_t$ is energy, and $C_t$ is safety cost.

For a constrained formulation, latency or safety can be moved out of the reward and into constraints:
\begin{equation}
    \E_\pi\left[\sum_t \gamma^t \mathbf{1}\{L_t > L_{\max}\}\right] \le d_L,
    \qquad
    \E_\pi\left[\sum_t \gamma^t \mathbf{1}\{B_t < B_{\min}\}\right] \le d_B.
\end{equation}

\begin{warningbox}{Reward hacking looks like performance improvement}
    In networking, reward hacking often looks like performance improvement. An agent can increase average throughput while violating tail latency, improve aggregate QoS while starving edge users, or reduce packet loss by rejecting difficult traffic. Always report class-wise and tail metrics, not only averages.
\end{warningbox}

\section{Closed-loop DRL architecture for programmable networks}

A deployable network DRL system usually contains more than a neural policy. It contains telemetry, feature computation, a simulator or digital twin, a policy optimizer, a runtime controller, a safety layer, and monitoring.

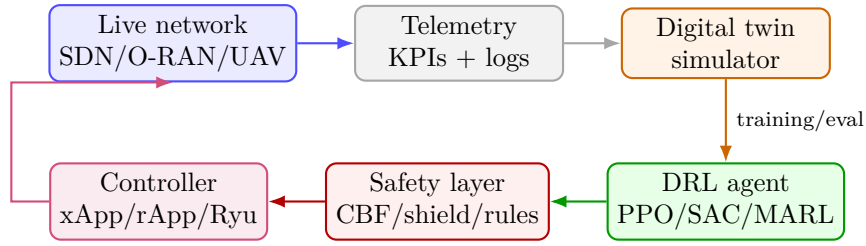
\begin{figure}[t]
    \centering
    \begin{tikzpicture}[
        box/.style={draw,rounded corners,thick,minimum width=2.75cm,minimum height=0.85cm,align=center,font=\small},
        arrow/.style={-{Latex[length=2.2mm]},thick},
        node distance=0.75cm
        ]
        \node[box,fill=blue!8,draw=blue!70] (net) {Live network\\SDN/O-RAN/UAV};
        \node[box,fill=gray!10,draw=gray!70,right=of net] (tele) {Telemetry\\KPIs + logs};
        \node[box,fill=orange!10,draw=orange!80!black,right=of tele] (twin) {Digital twin\\simulator};
        \node[box,fill=green!10,draw=green!60!black,below=1.1cm of twin] (agent) {DRL agent\\PPO/SAC/MARL};
        \node[box,fill=red!6,draw=red!70!black,left=of agent] (safe) {Safety layer\\CBF/shield/rules};
        \node[box,fill=purple!8,draw=purple!70,left=of safe] (ctrl) {Controller\\xApp/rApp/Ryu};
        \draw[arrow,draw=blue!70] (net) -- (tele);
        \draw[arrow,draw=gray!70] (tele) -- (twin);
        \draw[arrow,draw=orange!80!black] (twin) -- node[right,font=\scriptsize] {training/eval} (agent);
        \draw[arrow,draw=green!60!black] (agent) -- (safe);
        \draw[arrow,draw=red!70!black] (safe) -- (ctrl);
        \draw[arrow,draw=purple!70] (ctrl.west) -- ++(-0.5,0) |- (net.south);
    \end{tikzpicture}
    \caption{A closed-loop DRL architecture for networks. In deployment, the learned policy should rarely act directly on the network. A safety layer, controller interface, monitoring system, and simulator/digital twin are part of the control loop.}
    \label{fig:closed_loop_network_drl}
\end{figure}

For SDN and SD-WAN, the controller may be a Ryu/ONOS/ODL-style control plane or a vendor controller. For O-RAN, the controller may be a near-RT RIC xApp or non-RT RIC rApp. For UAV networks, the DRL agent may interface with a fleet controller and a radio resource manager.

\section{Algorithm selection for network DRL}

A practical algorithm choice depends on the action space, data availability, safety requirement, and simulator fidelity.

\begin{table}[t]
    \centering
    \caption{Algorithm-selection guide for network and UAV DRL.}
    \label{tab:network_algorithms}
    \begin{tabularx}{\textwidth}{p{3cm}p{3cm}X}
        \toprule
        Problem condition & Good starting point & Reason \\
        \midrule
        Small discrete actions & Double DQN / Rainbow & simple, value-based, replay-efficient \\
        Continuous control & SAC or PPO & direct continuous actions; SAC is off-policy, PPO is robust \\
        Hard safety constraints & PPO-Lag, CBF filter, shielded SAC & reward alone is not enough \\
        Multiple cooperative agents & MAPPO, QMIX, QPLEX & CTDE and credit assignment \\
        Expensive real exploration & offline RL, model-based RL & use logs or simulator rollouts \\
        Long-horizon planning & model-based RL, MuZero, HRL & look ahead before acting \\
        Language/operator assistant & RLHF, reasoning RL, verifier-guided RL & human preferences and verifiable outcomes \\
        \bottomrule
    \end{tabularx}
\end{table}

\begin{researchbox}
    Network DRL is a synthesis chapter. DQN and replay appear in Chapters~5--6, policy-gradient and actor-critic structure in Chapters~7--11, model-based planning in Chapters~12--13, offline learning in Chapters~14--15, MARL in Chapter~16, hierarchy in Chapter~17, safe RL in Chapter~18, and RLHF/reasoning in Chapters~19--20. A serious network system typically uses several of these ideas together.
\end{researchbox}

\begin{table}[t]
    \centering
    \caption{How earlier chapters reappear in network and UAV control.}
    \label{tab:network_chapter_map}
\begin{tabularx}{\textwidth}{p{3.0cm}p{3.2cm}X}
        \toprule
        Book component & Network role & Example in this chapter \\
        \midrule
        DQN / Rainbow & discrete path or channel selection & path choice, admission, channel selection \\
        PPO / SAC & continuous control & traffic splits, UAV velocity, power allocation \\
        Offline RL / DT & learning from logs & production telemetry, operator demonstrations \\
        Model-based RL & safe lookahead & digital-twin rollouts and CEM/MPC tests \\
        MARL / QMIX / MAPPO & distributed agents & multi-UAV slicing and multi-cell RAN control \\
        Safe RL / CBF & hard constraints & latency, jitter, energy, collision, regulatory limits \\
        RLHF / reasoning RL & operator interaction & explanation, diagnosis, and verifier-guided recommendations \\
        \bottomrule
    \end{tabularx}
\end{table}

\section{Case study I: SD-WAN traffic engineering}

In SD-WAN traffic engineering, an enterprise network often has multiple links, such as MPLS, broadband Internet, 5G, or satellite backup. The controller chooses how to split traffic across them. The objective is not simply high throughput: it must respect latency, jitter, packet loss, and class-specific SLA constraints.

\begin{figure}[t]
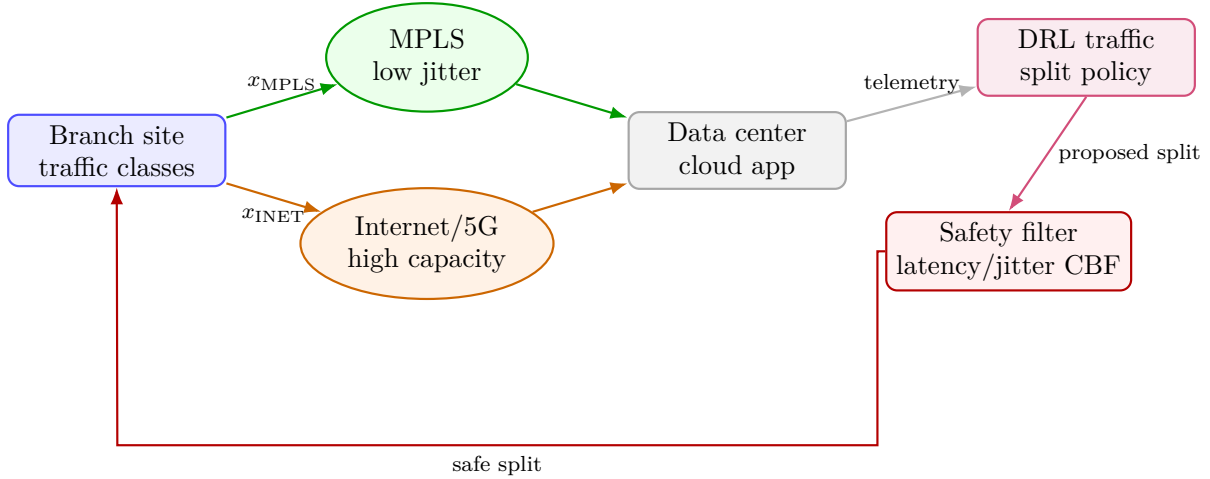

    \centering
    \resizebox{\columnwidth}{!}{%
        %
%
    }
    \caption{SD-WAN traffic engineering as a safe DRL problem. The actor proposes traffic split ratios, while a safety filter prevents latency or jitter violations before the controller applies the action.}
    \label{fig:sdwan_case}
\end{figure}

A state can include
\begin{equation}
    s_t = [\ell_{\mathrm{MPLS}},\ell_{\mathrm{INET}},j_{\mathrm{MPLS}},j_{\mathrm{INET}},p_{\mathrm{MPLS}},p_{\mathrm{INET}},u_{\mathrm{MPLS}},u_{\mathrm{INET}},d_{\mathrm{app}}],
\end{equation}
where $\ell$ is latency, $j$ is jitter, $p$ is packet loss, $u$ is utilization, and $d_{\mathrm{app}}$ is application demand.

A continuous action can be a traffic split
\begin{equation}
    a_t = [x_{\mathrm{MPLS}},x_{\mathrm{INET}}], \qquad x_{\mathrm{MPLS}}+x_{\mathrm{INET}}=1.
\end{equation}

A safety constraint can be
\begin{equation}
    \ell_{\mathrm{URLLC}} \le 6 \;\mathrm{ms},
    \qquad
    j_{\mathrm{URLLC}} \le 2 \;\mathrm{ms}.
\end{equation}

Our own safe SD-WAN work uses uncertainty-aware and ensemble-based neural CBFs to filter or penalize unsafe traffic-engineering actions under QoS constraints \citep{bista2026vtc,bista2026ifip}. In the terminology of Chapter~18, the traffic split proposed by the actor should not be the action used for critic learning if a safety filter modifies it. The critic should learn from the executed safe action.

\section{Case study II: UAV-assisted network slicing}

A concrete example of algorithm selection in this domain is our comparative study of MAPPO, MADDPG, and MADQN for multi-UAV 5G network slicing \citep{bista2025marl_uav_slicing}. The comparison is useful because it does not treat MARL as a single algorithmic family. Instead, it contrasts an on-policy centralized-critic method (MAPPO), a deterministic actor-critic method for continuous control (MADDPG), and a value-based baseline for discretized control (MADQN). The main lesson is practical rather than doctrinal: the best choice depends on the action-space structure, the coordination demands of the slicing problem, and the trade-off among QoS, stability, and energy consumption.

UAVs can act as aerial base stations, relays, sensing platforms, or emergency-connectivity nodes. They introduce mobility as a control variable. This makes them a natural domain for DRL, but also a safety-critical one, because energy, collision avoidance, radio interference, and service guarantees interact tightly.

From a systems perspective, this comparison highlights three different design regimes. MAPPO is attractive when stable cooperative learning and balanced QoS performance matter most. MADDPG is natural when the control variables are continuous and fine-grained, but it can be more sensitive to critic quality and replay effects. MADQN remains relevant when the control interface is discretized and computational simplicity matters. For UAV-assisted network slicing, this reinforces a broader systems lesson: algorithm choice should follow the structure of the control problem, not only benchmark popularity.

\begin{figure}[t]
    \centering
    \begin{tikzpicture}[
        uav/.style={regular polygon,regular polygon sides=3,draw,thick,minimum size=0.85cm,align=center,font=\scriptsize},
        user/.style={circle,draw,minimum size=0.32cm,inner sep=0pt},
        box/.style={draw,rounded corners,thick,minimum width=2.8cm,minimum height=0.8cm,align=center,font=\small},
        arrow/.style={-{Latex[length=2.2mm]},thick}
        ]
        \draw[thick] (-0.4,0) -- (8.8,0);
        \node[below] at (4.2,0) {urban service area};
        \node[uav,fill=blue!10,draw=blue!70]             (u1) at (1.5, 2.8) {UAV1};
        \node[uav,fill=green!10,draw=green!60!black]     (u2) at (4.5, 2.1) {UAV2};
        \node[uav,fill=orange!10,draw=orange!80!black]   (u3) at (7.5, 2.4) {UAV3};
        \foreach \x/\y/\c in {0.9/0.55/red,1.4/0.8/red,2.0/0.45/red,2.8/0.75/gray,4.2/0.55/gray,4.8/0.9/gray,5.4/0.65/gray,6.3/0.55/purple,7.1/0.8/purple,7.9/0.52/purple,8.3/0.75/purple}
        \node[user,fill=\c!35,draw=\c!80!black] at (\x,\y) {};
        \draw[dashed,blue!70] (u1) -- (0.9,0.55);
        \draw[dashed,blue!70] (u1) -- (1.4,0.8);
        \draw[dashed,blue!70] (u1) -- (2.0,0.45);
        \draw[dashed,green!60!black] (u2) -- (4.2,0.55);
        \draw[dashed,green!60!black] (u2) -- (5.4,0.65);
        \draw[dashed,orange!80!black] (u3) -- (7.1,0.8);
        \draw[dashed,orange!80!black] (u3) -- (7.9,0.52);
        \node[box,fill=purple!8,draw=purple!70] (sdn) at (4.5,5.0) {SDN/RIC controller\\global telemetry};
        \draw[arrow,draw=purple!70] (sdn) -- (u1);
        \draw[arrow,draw=purple!70] (sdn) -- (u2);
        \draw[arrow,draw=purple!70] (sdn) -- (u3);
        \node[align=left,font=\scriptsize] at (10.7,1.5) {red: URLLC\\gray: eMBB\\purple: mMTC};
    \end{tikzpicture}
    \caption{A multi-UAV network-slicing scenario. UAVs provide aerial connectivity to heterogeneous users while an SDN/RIC controller supplies global telemetry or policy guidance. The DRL problem couples mobility, radio resources, battery, interference, and QoS constraints.}
    \label{fig:uav_slicing_scenario}
\end{figure}
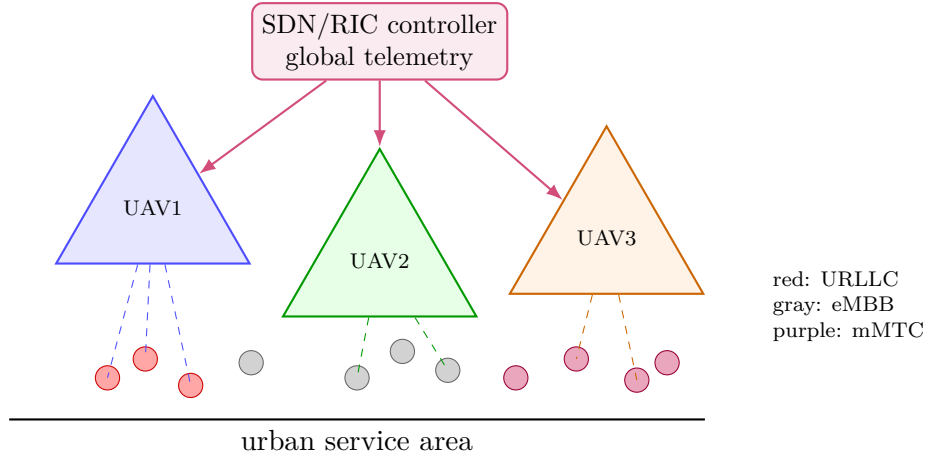

A multi-UAV action may contain movement and resource allocation:
\begin{equation}
    a_t^i = [\Delta x_i,\Delta y_i,\Delta z_i, P_i, b_i^{\mathrm{URLLC}}, b_i^{\mathrm{eMBB}}, b_i^{\mathrm{mMTC}}].
\end{equation}

A multi-agent reward can combine global and local terms:
\begin{equation}
    r_t^i = \lambda_g r_t^{\mathrm{global}} + \lambda_l r_t^{i,\mathrm{local}} - \lambda_e E_t^i - \lambda_s C_t^i.
\end{equation}

Graph-based value decomposition methods, including variants such as GAD-QMIX, are useful when UAV interactions form a dynamic graph. Our own UAV-assisted slicing work studies graph-attention value decomposition for multi-class QoS, safety, and interference-aware control \citep{bista2024gadqmix}. This chapter treats such systems as the application-level synthesis of Chapters~16 and~18.

\section{Wireless realism: SINR, throughput, and latency}

A wireless DRL paper is only as credible as its physical model. The agent may appear to learn if the channel model is too simple, the interference model is absent, or throughput is computed unrealistically.

A common SINR model is
\begin{equation}
    \mathrm{SINR}_{u,i}
    =
    \frac{P_i G_{i,u} L_{i,u}^{-1}}
    {\sum_{j\ne i} P_j G_{j,u} L_{j,u}^{-1} + N_0 B},
\end{equation}
where $P_i$ is transmit power, $G_{i,u}$ is antenna/channel gain, $L_{i,u}$ is path loss, $N_0$ is noise spectral density, and $B$ is bandwidth.

A Shannon-style upper-bound throughput is
\begin{equation}
    T_u = B_u \log_2(1+\mathrm{SINR}_u),
\end{equation}
but practical systems require modulation/coding, scheduling, HARQ, BLER, and resource-block constraints. For 5G/6G-style simulation, 3GPP TR~38.901 is a standard reference for channel modeling, while system-level throughput should be capped or mapped through realistic MCS tables rather than treated as unlimited Shannon capacity \citep{tr38901}.

Latency can be decomposed as
\begin{equation}
    L_t = L_t^{\mathrm{queue}} + L_t^{\mathrm{tx}} + L_t^{\mathrm{prop}} + L_t^{\mathrm{proc}} + L_t^{\mathrm{retrans}}.
\end{equation}
This decomposition matters because different actions affect different terms. UAV placement changes propagation and SINR. Scheduling changes queueing. Routing changes propagation and congestion. Power control changes interference.

\section{Simulator-in-the-loop research pipeline}

DRL for networks normally needs a simulator before deployment. Common tools include ns-3, ns3-gym, ns3-ai, Mininet, Mininet-WiFi, OMNeT++/Simu5G, srsRAN, and O-RAN testbeds. The ns3-gym framework integrates ns-3 with OpenAI Gym-style RL APIs, encouraging RL research with network simulators \citep{zubow2019ns3gym}. The ns3-ai module provides a high-speed data-exchange bridge between ns-3 and AI frameworks \citep{yin2020ns3ai}. Recent O-RAN work also uses ns-O-RAN and Gymnasium-like interfaces to train RL agents against RAN key performance indicators \citep{lacava2024online}.

For reproducibility, the SD-WAN/ns-3 code used in our related safe-RL traffic-engineering experiments is publicly available as an open GitHub repository \citep{safeml_sdwan_github}. The repository is useful as a concrete example of the engineering details that are easy to hide in a paper: ns-3 configuration, OpenGym integration, traffic generation, reward logging, and policy-evaluation scripts.

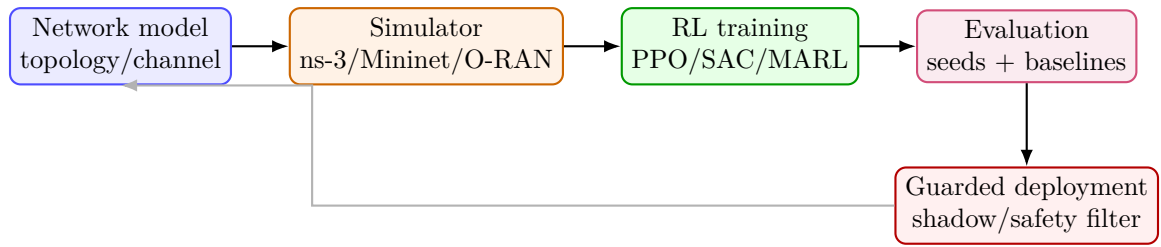
\begin{figure}[t]
    \centering
    \begin{tikzpicture}[
        box/.style={draw,rounded corners,thick,minimum width=2.7cm,minimum height=0.8cm,align=center,font=\small},
        arrow/.style={-{Latex[length=2.2mm]},thick},
        node distance=0.75cm
        ]
        \node[box,fill=blue!8,draw=blue!70] (model) {Network model\\topology/channel};
        \node[box,fill=orange!10,draw=orange!80!black,right=of model] (sim) {Simulator\\ns-3/Mininet/O-RAN};
        \node[box,fill=green!10,draw=green!60!black,right=of sim] (rl) {RL training\\PPO/SAC/MARL};
        \node[box,fill=purple!8,draw=purple!70,right=of rl] (eval) {Evaluation\\seeds + baselines};
        \node[box,fill=red!6,draw=red!70!black,below=1.1cm of eval] (deploy) {Guarded deployment\\shadow/safety filter};
        \draw[arrow] (model) -- (sim);
        \draw[arrow] (sim) -- (rl);
        \draw[arrow] (rl) -- (eval);
        \draw[arrow] (eval) -- (deploy);
        \draw[arrow,draw=gray!60] (deploy.west) -- ++(-7.7,0) |- (model.south);
    \end{tikzpicture}
    \caption{A simulator-in-the-loop network DRL pipeline. Credible work does not stop at training reward; it evaluates baselines, seeds, failures, constraints, and deployment guardrails.}
    \label{fig:sim_pipeline}
\end{figure}

\begin{warningbox}{The simulator is part of the algorithm}
    If the simulator omits queueing, interference, user mobility, or controller latency, the learned policy may optimize a world that does not exist.
\end{warningbox}

\section{Python implementation: a minimal network QoS environment}

This section gives a compact Gymnasium-style environment for SD-WAN or wireless-slicing experiments. It is not a full simulator. Its purpose is to show the correct separation between state, proposed action, safety-filtered action, reward, cost, and logging.

\Needspace{20\baselineskip}
\begin{lstlisting}[style=pythonstyle,caption={A minimal network QoS environment skeleton.},label={lst:network_env}]
import numpy as np

class NetworkQoSEnv:
    """Minimal network-control environment for DRL experiments.

    Observation: latency, jitter, loss, load, battery, SINR per class/link.
    Action: continuous allocation vector proposed by the agent.
    Safety filter: projects unsafe actions before execution.
    """
    def __init__(self, num_links=2, num_classes=3, seed=0):
        self.rng = np.random.default_rng(seed)
        self.num_links = num_links
        self.num_classes = num_classes
        self.t = 0
        self.max_steps = 500
        self.latency_bound = np.array([6.0, 25.0, 40.0])  # URLLC, eMBB, mMTC ms
        self.loss_bound = np.array([0.005, 0.02, 0.05])
        self.reset()

    def reset(self):
        self.t = 0
        self.load = self.rng.uniform(0.2, 0.5, size=(self.num_classes,))
        self.latency = self.rng.uniform(4.0, 20.0, size=(self.num_classes,))
        self.loss = self.rng.uniform(0.0, 0.02, size=(self.num_classes,))
        self.link_util = self.rng.uniform(0.2, 0.4, size=(self.num_links,))
        self.battery = 1.0
        return self._obs()

    def _obs(self):
        return np.concatenate([
            self.load,
            self.latency / 50.0,
            self.loss / 0.1,
            self.link_util,
            np.array([self.battery]),
        ]).astype(np.float32)

    def safety_filter(self, action):
        """Project action to feasible allocation simplex and protect low-latency class."""
        a = np.maximum(action, 1e-6)
        a = a / a.sum()
        # If URLLC latency is near violation, enforce minimum reliable-link share.
        if self.latency[0] > 0.9 * self.latency_bound[0]:
            a[0] = max(a[0], 0.55)
            a = a / a.sum()
        return a

    def step(self, proposed_action):
        executed_action = self.safety_filter(proposed_action)

        # Toy dynamics: utilization follows allocation and demand.
        demand = self.load.sum()
        self.link_util = 0.75 * self.link_util + 0.25 * demand * executed_action
        congestion = np.maximum(self.link_util.mean() - 0.7, 0.0)

        noise = self.rng.normal(0.0, 0.5, size=self.num_classes)
        self.latency = 0.85 * self.latency + 4.0 * congestion + noise + np.array([1.0, 2.0, 3.0])
        self.latency = np.maximum(self.latency, 1.0)
        self.loss = np.clip(0.8 * self.loss + 0.04 * congestion, 0.0, 0.2)
        self.battery = max(0.0, self.battery - 0.001 - 0.003 * np.linalg.norm(executed_action))

        reward, cost, info = self.reward_and_cost(executed_action)
        self.t += 1
        done = self.t >= self.max_steps or self.battery <= 0.02
        info.update({"executed_action": executed_action, "proposed_action": proposed_action})
        return self._obs(), reward, done, info

    def reward_and_cost(self, action):
        qos_ok = (self.latency <= self.latency_bound) & (self.loss <= self.loss_bound)
        class_weights = np.array([3.0, 1.5, 1.0])
        qos_reward = float((class_weights * qos_ok.astype(float)).sum())
        latency_penalty = float(np.maximum(self.latency - self.latency_bound, 0.0).sum())
        energy_penalty = 0.1 * float(np.linalg.norm(action))
        reward = qos_reward - 0.1 * latency_penalty - energy_penalty
        cost = float((self.latency[0] > self.latency_bound[0]) or (self.battery < 0.1))
        return reward, cost, {"qos_ok": qos_ok, "cost": cost, "latency": self.latency.copy()}
\end{lstlisting}

\section{Graph observations for network topology}

Many network states are graphs: routers connected by links, base stations connected by interference edges, UAVs connected by communication ranges, or users associated with serving nodes. Graph neural networks are often more suitable than flat vectors.

\Needspace{17\baselineskip}
\begin{lstlisting}[style=pythonstyle,caption={Building graph observations for a network DRL agent.},label={lst:graph_obs}]
def build_graph_observation(nodes, links):
    """Build node features, edge index, and edge features.

    nodes: list of dictionaries with keys such as load, battery, role
    links: list of dictionaries with src, dst, capacity, delay, loss, utilization
    """
    x = []
    for n in nodes:
        x.append([
            n.get("load", 0.0),
            n.get("battery", 1.0),
            n.get("is_uav", 0.0),
            n.get("is_controller", 0.0),
        ])

    edge_index = []
    edge_attr = []
    for e in links:
        edge_index.append([e["src"], e["dst"]])
        edge_attr.append([
            e.get("capacity", 1.0),
            e.get("delay", 0.0),
            e.get("loss", 0.0),
            e.get("utilization", 0.0),
        ])

    return {
        "node_features": np.asarray(x, dtype=np.float32),
        "edge_index": np.asarray(edge_index, dtype=np.int64).T,
        "edge_features": np.asarray(edge_attr, dtype=np.float32),
    }
\end{lstlisting}

\section{Class-aware reward and SLA metrics}

A serious network DRL paper should report the metrics that matter to operators and users. A single reward curve is not enough.

\Needspace{18\baselineskip}
\begin{lstlisting}[style=pythonstyle,caption={Class-aware QoS reward and SLA metrics.},label={lst:qos_reward}]
def class_aware_reward(metrics, weights=None):
    """Compute reward and interpretable SLA metrics.

    metrics[class_name] contains latency_ms, throughput_mbps, loss, energy.
    """
    if weights is None:
        weights = {"URLLC": 3.0, "eMBB": 1.5, "mMTC": 1.0}

    thresholds = {
        "URLLC": {"latency": 6.0,  "throughput": 20.0, "loss": 0.005},
        "eMBB":  {"latency": 25.0, "throughput": 50.0, "loss": 0.02},
        "mMTC":  {"latency": 40.0, "throughput": 5.0,  "loss": 0.05},
    }

    reward = 0.0
    report = {}
    for cls, m in metrics.items():
        th = thresholds[cls]
        lat_ok = m["latency_ms"] <= th["latency"]
        thr_ok = m["throughput_mbps"] >= th["throughput"]
        loss_ok = m["loss"] <= th["loss"]
        satisfied = lat_ok and thr_ok and loss_ok

        violation = max(m["latency_ms"] - th["latency"], 0.0) / th["latency"]
        violation += max(th["throughput"] - m["throughput_mbps"], 0.0) / th["throughput"]
        violation += max(m["loss"] - th["loss"], 0.0) / max(th["loss"], 1e-6)

        reward += weights[cls] * (1.0 if satisfied else -violation)
        report[cls] = {"satisfied": satisfied, "violation": violation}

    return reward, report
\end{lstlisting}

\section{Safety filtering for network actions}

The central principle from Chapter~18 appears again: the policy may propose an action, but the executed action should satisfy safety constraints. For continuous allocations, this can be a projection. For discrete actions, it can be an action mask or shield.

\Needspace{18\baselineskip}
\begin{lstlisting}[style=pythonstyle,caption={Simple safety projection for traffic-split actions.},label={lst:safety_projection}]
def project_traffic_split(raw_action, latency, jitter, min_reliable_share=0.55):
    """Project an unsafe traffic-split action onto a safer allocation.

    raw_action: array [2], proposed shares for [reliable_link, cheap_link]
    latency, jitter: current QoS measurements for high-priority traffic
    """
    a = np.maximum(raw_action, 1e-6)
    a = a / a.sum()

    high_risk = latency > 5.5 or jitter > 1.8
    if high_risk:
        a[0] = max(a[0], min_reliable_share)
        a[1] = 1.0 - a[0]

    return a
\end{lstlisting}

\begin{keybox}{Practice note: logging proposed and executed actions}
    If a safety filter changes the action, log both the proposed action and the executed action. Train the critic on the executed action when learning from real system consequences. Otherwise the critic learns the value of actions the network never actually took.
\end{keybox}

\section{Numerical example: when averages hide SLA failure}

Consider two traffic-splitting policies in an SD-WAN controller. Policy A sends more traffic to the cheap Internet link. Policy B sends more high-priority traffic to the reliable link.

\begin{table}[t]
    \centering
    \caption{A numerical example where average throughput is misleading.}
    \label{tab:network_numeric_example}
    \begin{tabular}{lcccc}
        \toprule
        Policy & Mean throughput & URLLC latency p95 & Packet loss & SLA satisfied? \\
        \midrule
        A: throughput-greedy & 115 Mbps & 9.8 ms & 0.7\% & No \\
        B: SLA-aware & 98 Mbps & 5.4 ms & 0.2\% & Yes \\
        \bottomrule
    \end{tabular}
\end{table}

A reward based only on average throughput selects Policy A. A constraint-aware or class-aware reward selects Policy B. This illustrates why communication-network DRL must report tail latency, class-wise satisfaction, and constraint violations.

A useful way to make this mathematically explicit is to report a tail-risk metric such as a conditional value-at-risk style latency measure:
\begin{equation}
    \mathrm{CVaR}_{\rho}(L)
    =
    \E\left[L \given L \ge q_{\rho}(L)\right],
\end{equation}
where $q_{\rho}(L)$ is the $\rho$-quantile of the latency distribution. A policy with excellent mean latency but poor $\mathrm{CVaR}_{0.95}$ may still be unacceptable for URLLC-style traffic. This connects the distributional and tail-risk discussion from Chapter~6 to network evaluation.

\section{Evaluation methodology for network DRL}

A credible network DRL experiment should report more than aggregate reward. This is especially important in UAV-assisted network slicing, where comparative studies of MARL families such as MAPPO, MADDPG, and MADQN can produce similar reward values while hiding very different coordination quality, energy efficiency, fairness, or constraint satisfaction. For this reason, evaluation should report per-class QoS satisfaction, latency, SINR, throughput, energy consumption, interference, fairness across service classes, and safety violations.

A network DRL experiment should therefore include at least:
\begin{itemize}
    \item multiple random seeds and confidence intervals;
    \item realistic traffic traces or multiple synthetic traffic regimes;
    \item baselines: static rule, shortest path, ECMP, MPC, heuristic, classical optimizer, and strong DRL baselines;
    \item class-wise QoS metrics, not only average reward;
    \item tail metrics: p95/p99 latency, worst-user throughput, outage rate;
    \item safety metrics: violations per hour, maximum violation, recovery time;
    \item computational overhead: inference time, controller latency, training cost;
    \item ablations: reward terms, safety layer, graph features, channel model, mobility model;
    \item stress tests: traffic bursts, link failures, controller delay, stale telemetry, attack scenarios.
\end{itemize}

\begin{table}[t]
    \centering
    \caption{Common failure modes in DRL for networks and communication systems.}
    \label{tab:network_failure_modes}
\begin{tabularx}{\textwidth}{p{3.3cm}p{4.2cm}X}
        \toprule
        Failure mode & Symptom & What to inspect \\
        \midrule
        Reward hacking & reward increases but SLA worsens & class-wise and tail metrics \\
        Simulator shortcut & policy exploits unrealistic simulator artifact & model assumptions and validation \\
        Oscillation & routing/slicing decisions switch rapidly & action smoothing, delay, hysteresis \\
        Unsafe exploration & constraint violations during training & offline/safe RL, safety filter, shadow mode \\
        Non-stationary traffic & policy fails after deployment & drift detection, retraining, online validation \\
        Interference neglect & good single-cell results, poor multi-cell results & SINR and interference model \\
        Unfair allocation & average QoS improves, edge users suffer & percentile/user-level metrics \\
        Controller overhead & action arrives too late & inference latency and timescale \\
        \bottomrule
    \end{tabularx}
\end{table}

\begin{table}[t]
    \centering
    \caption{Minimum reporting checklist for a credible network DRL experiment.}
    \label{tab:network_reporting_checklist}
    \begin{tabularx}{\textwidth}{p{3.4cm}X}
        \toprule
        Category & What should be reported \\
        \midrule
        Topology and traffic & nodes, links, capacities, traffic traces, mobility, burst model \\
        Wireless model & path loss, fading, interference, bandwidth, MCS/throughput mapping \\
        Control loop & decision interval, telemetry delay, inference time, actuation delay \\
        Algorithms & hyperparameters, seeds, normalization, replay/on-policy batch size \\
        Baselines & rule-based, optimizer/MPC, classical networking heuristic, strong DRL baseline \\
        Safety & constraints, violation definition, proposed vs executed action, recovery time \\
        Evaluation & mean, p95/p99, class-wise metrics, confidence intervals, stress tests \\
        Reproducibility & code, config files, random seeds, simulator version, hardware cost \\
        \bottomrule
    \end{tabularx}
\end{table}
\section{O-RAN xApp/rApp blueprint}
\label{sec:oran_blueprint}

O-RAN is one of the most important deployment contexts for network DRL because it separates near-real-time control, non-real-time policy optimization, telemetry, and model management. A practical O-RAN DRL design should specify which intelligence lives in the near-RT RIC as an xApp and which intelligence lives in the non-RT RIC as an rApp. The xApp/rApp split, the Near-RT RIC, the Non-RT RIC, and the A1/R1/E2 interface roles should be described using O-RAN Alliance technical specifications rather than informal terminology; recent O-RAN specifications define the Near-RT RIC architecture, APIs, and the Non-RT RIC use cases and requirements that motivate these control loops \citep{oran_specs,oran_near_rt_ric_2026}.

A useful division is:
\begin{itemize}
    \item \textbf{xApp}: fast inference, action masking, local KPI monitoring, conservative closed-loop control;
    \item \textbf{rApp}: offline training, policy evaluation, drift analysis, model selection, policy guidance;
    \item \textbf{digital twin}: counterfactual testing before deployment;
    \item \textbf{human/operator layer}: approval for high-impact slice or routing changes.
\end{itemize}

\begin{figure}[t]
    \centering
    \begin{tikzpicture}[
        box/.style={draw,rounded corners,thick,minimum width=2.8cm,minimum height=0.8cm,align=center,font=\small},
        arrow/.style={-{Latex[length=2.2mm]},thick},
        node distance=0.75cm
        ]
        \node[box,fill=blue!8,draw=blue!70] (ran) {RAN / gNB\\E2 telemetry};
        \node[box,fill=green!10,draw=green!60!black,right=of ran] (xapp) {Near-RT RIC\\xApp inference};
        \node[box,fill=red!6,draw=red!70!black,right=of xapp] (safe) {Policy guard\\mask/projection};
        \node[box,fill=gray!10,draw=gray!70,right=of safe] (act) {E2 control\\PRB/slice/handover};
        \node[box,fill=orange!10,draw=orange!80!black,below=1.2cm of xapp] (rapp) {Non-RT RIC\\rApp training};
        \node[box,fill=purple!8,draw=purple!70,below=1.2cm of safe] (twin) {Digital twin\\counterfactual tests};
        \draw[arrow,draw=blue!70] (ran) -- (xapp);
        \draw[arrow,draw=green!60!black] (xapp) -- (safe);
        \draw[arrow,draw=red!70!black] (safe) -- (act);
        \draw[arrow,draw=orange!80!black] (rapp) -- (xapp);
        \draw[arrow,draw=purple!70] (twin) -- (rapp);
        \draw[arrow,draw=gray!60]
        (ran.south) -- ++(0,-0.5) coordinate(R)
        -- (R -| twin.west) coordinate(S)
        -- (twin.north);
        \node[below,font=\scriptsize] at ($(R)!0.5!(S)$) {telemetry};
    \end{tikzpicture}
    \caption{A practical O-RAN DRL blueprint. Fast inference and safety guards belong near the control loop, while training, drift analysis, and counterfactual evaluation belong in slower rApp/digital-twin layers.}
    \label{fig:oran_blueprint}
\end{figure}
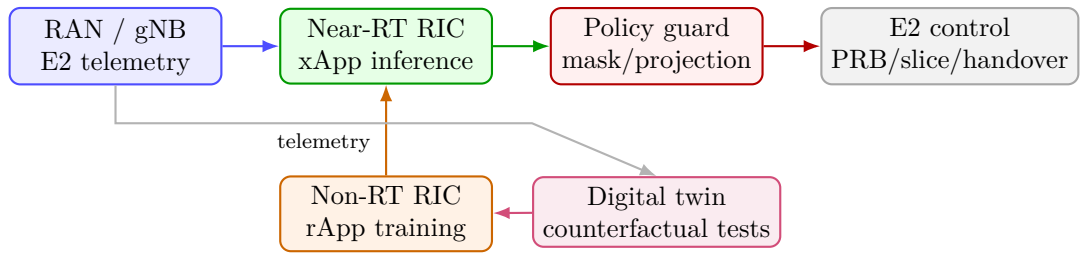

\paragraph{Numerical O-RAN slicing example.}
Suppose a near-RT RIC xApp allocates $100$ physical resource blocks across three slices: URLLC, eMBB, and mMTC. A throughput-greedy actor proposes $[15,75,10]$ PRBs. The policy guard detects that URLLC p95 latency is $7.8$ ms, above a $6$ ms bound, and projects the action to $[30,60,10]$. The aggregate throughput may decrease, but the action is operationally better because the hard slice constraint is restored. This is the same proposed-action versus executed-action principle emphasized in Chapters~18 and~21: the learning system must log both actions and train value estimates using the consequences of the executed safe action.

\Needspace{18\baselineskip}
\begin{lstlisting}[style=pythonstyle,caption={O-RAN-style policy wrapper with runtime guards.},label={lst:oran_policy_wrapper}]
class ORANPolicyWrapper:
    """Runtime wrapper for an xApp-style DRL policy."""
    def __init__(self, policy, guard, normalizer, max_inference_ms=5.0):
        self.policy = policy
        self.guard = guard
        self.normalizer = normalizer
        self.max_inference_ms = max_inference_ms

    def act(self, raw_kpis):
        obs = self.normalizer.transform(raw_kpis)
        proposed, diagnostics = self.policy(obs)

        # The guard can use SLA rules, CBF checks, action masks, or operator policy.
        safe_action, guard_info = self.guard.project(proposed, raw_kpis)

        return {
            "proposed_action": proposed,
            "executed_action": safe_action,
            "policy_diagnostics": diagnostics,
            "guard_info": guard_info,
        }
\end{lstlisting}

\section{Edge computing, V2X, and NTN scheduling}

Communication-system DRL is broader than routing and UAV placement. Edge computing introduces task offloading and server placement. V2X introduces strict latency and reliability under fast mobility. Non-terrestrial networks introduce intermittent contact windows, moving satellites, and long propagation delays.

For edge offloading, an action may decide whether to compute locally, offload to an edge server, or forward to a cloud. A reward may combine latency, energy, and deadline misses:
\begin{equation}
    r_t = -\eta_L L_t - \eta_E E_t - \eta_D \mathbf{1}\{L_t > D_t\}.
\end{equation}

For V2X, a state should include mobility and deadline information; for NTN, it should include contact windows and link availability. In both cases, naively optimizing average throughput is misleading because the main risk is missed deadlines or intermittent connectivity.

\begin{table}[t]
    \centering
    \caption{Application-specific state/action/reward choices beyond SD-WAN and UAV slicing.}
    \label{tab:other_network_domains}
    \begin{tabularx}{\textwidth}{p{2.6cm}p{3.1cm}p{3.1cm}X}
        \toprule
        Domain & State & Action & Main metric \\
        \midrule
        Edge offloading & task size, CPU load, channel, queue & local/edge/cloud decision, CPU share & deadline satisfaction, energy \\
        V2X & position, speed, channel, neighbor density & resource block, power, relay choice & reliability, latency, collision risk \\
        LEO/NTN & contact window, queue, orbit, battery & schedule, route, store-carry-forward & delivery delay, outage \\
        O-RAN slicing & PRB usage, CQI, slice demand & PRB share, admission, priority & per-slice SLA \\
        Cyber defense & alerts, flows, topology, trust score & block, reroute, isolate, inspect & containment, false positives \\
        \bottomrule
    \end{tabularx}
\end{table}

\section{Hybrid action decoding for network controllers}

Many network actions are hybrid: choose a discrete path and allocate a continuous rate; choose a UAV mission mode and output continuous motion; select a slice action and tune PRB shares. A clean implementation separates discrete decisions from continuous parameters.

\Needspace{20\baselineskip}
\begin{lstlisting}[style=pythonstyle,caption={Hybrid action decoder for network DRL policies.},label={lst:hybrid_action_decoder}]
import torch
import torch.nn as nn
import torch.nn.functional as F

class HybridNetworkActor(nn.Module):
    """Policy head for hybrid network actions.

    Outputs:
        path_logits: discrete path/slice/mode choice
        alloc_mean: continuous allocation mean
        alloc_log_std: continuous allocation log std
    """
    def __init__(self, obs_dim, hidden_dim, num_modes, alloc_dim):
        super().__init__()
        self.trunk = nn.Sequential(
            nn.Linear(obs_dim, hidden_dim), nn.ReLU(),
            nn.Linear(hidden_dim, hidden_dim), nn.ReLU(),
        )
        self.mode_head = nn.Linear(hidden_dim, num_modes)
        self.mean_head = nn.Linear(hidden_dim, alloc_dim)
        self.log_std = nn.Parameter(torch.zeros(alloc_dim))

    def forward(self, obs):
        h = self.trunk(obs)
        mode_logits = self.mode_head(h)
        mean = torch.tanh(self.mean_head(h))
        log_std = torch.clamp(self.log_std, -5.0, 1.0)
        return mode_logits, mean, log_std

    def decode_allocation(self, raw_alloc):
        # Map unconstrained vector to a simplex allocation.
        return F.softmax(raw_alloc, dim=-1)
\end{lstlisting}

\section{Digital-twin calibration and shadow validation}

A digital twin is useful only if it predicts the right quantities. Calibration should not be judged by visual similarity alone. For DRL, the twin must predict the effect of actions on decision-relevant metrics.

Let $m_t$ be a measured metric vector and $\hat m_t$ be the digital-twin prediction. A simple calibration loss is
\begin{equation}
    \mathcal{L}_{\mathrm{twin}} =
    \sum_t \left\| W(m_t-\hat m_t) \right\|_2^2,
\end{equation}
where $W$ weights safety-critical metrics such as p95 latency or packet loss more heavily.

Shadow validation means running the learned policy without applying its actions. The system logs what the policy would have done and compares it against the production controller.

\Needspace{18\baselineskip}
\begin{lstlisting}[style=pythonstyle,caption={Shadow-mode validation for a network policy.},label={lst:shadow_validation}]
def shadow_validate(policy, production_logs, safety_checker):
    """Evaluate proposed actions without applying them to the live network."""
    records = []
    for event in production_logs:
        obs = event["observation"]
        prod_action = event["executed_action"]
        proposed = policy(obs)
        safe, reason = safety_checker(proposed, obs)
        records.append({
            "time": event["time"],
            "prod_action": prod_action,
            "proposed_action": proposed,
            "would_be_safe": safe,
            "reason": reason,
            "kpis_after_prod_action": event["next_kpis"],
        })
    return records
\end{lstlisting}

\section{Deployment drift and shadow validation}

A network policy can fail even if it performed well in simulation and offline evaluation. Traffic matrices change, applications change, firmware changes, radio conditions shift, and operators may modify rules. For this reason, deployment should include shadow validation: the policy produces proposed actions, but the production controller ignores them while the system records what would have happened according to a digital twin or counterfactual evaluator.

Let $a_t^{\mathrm{prod}}$ be the production action and $a_t^{\mathrm{rl}}$ be the shadow policy action. A simple shadow-risk score is
\begin{equation}
    \Delta_t^{\mathrm{risk}}
    =
    \hat C(s_t,a_t^{\mathrm{rl}})-\hat C(s_t,a_t^{\mathrm{prod}}),
\end{equation}
where $\hat C$ is a calibrated cost predictor. A positive value means the learned policy is predicted to be riskier than the production controller. Shadow deployment should continue until this score is stable across traffic regimes, link failures, and mobility scenarios.

\section{Deployment gates: from paper to network}

A network DRL policy should pass staged gates before deployment:
\begin{enumerate}[leftmargin=*]
    \item \textbf{Unit tests}: action bounds, reward terms, terminal masks, normalization.
    \item \textbf{Simulator tests}: multiple seeds, traffic regimes, failures, baseline comparison.
    \item \textbf{Digital-twin tests}: calibrated replay of production scenarios.
    \item \textbf{Shadow mode}: policy proposes actions but does not execute them.
    \item \textbf{Limited canary}: small traffic class, small region, rollback enabled.
    \item \textbf{Monitored deployment}: drift detection, alarms, human override, fallback rules.
\end{enumerate}

\begin{figure}[t]
    \centering
    \begin{tikzpicture}[
        gate/.style={draw,rounded corners,thick,minimum width=2.35cm,minimum height=0.72cm,align=center,font=\small},
        arrow/.style={-{Latex[length=2.1mm]},thick}
        ]
        \node[gate,fill=blue!8,draw=blue!70] (g1) at (0,0) {unit\\tests};
        \node[gate,fill=green!8,draw=green!60!black] (g2) at (2.7,0) {simulator\\tests};
        \node[gate,fill=orange!10,draw=orange!80!black] (g3) at (5.4,0) {digital\\twin};
        \node[gate,fill=purple!8,draw=purple!70] (g4) at (8.1,0) {shadow\\mode};
        \node[gate,fill=red!6,draw=red!70!black] (g5) at (10.8,0) {canary\\deploy};
        \draw[arrow] (g1)--(g2);
        \draw[arrow] (g2)--(g3);
        \draw[arrow] (g3)--(g4);
        \draw[arrow] (g4)--(g5);
        \node[below=0.8cm of g3,align=center,font=\small] {Every gate should have rollback criteria and safety metrics.};
    \end{tikzpicture}
    \caption{Deployment gates for network DRL. A policy should not move directly from simulation training to production control.}
    \label{fig:deployment_gates}
\end{figure}

\section{Security and cyber-resilient network control}

Communication networks are adversarial environments. DRL can improve adaptation, but it can also introduce attack surfaces. An attacker may corrupt telemetry, spoof traffic demand, manipulate reward signals, or trigger unsafe policy responses.

Security-aware DRL for networks should consider:
\begin{itemize}
    \item telemetry integrity and anomaly detection;
    \item robust policies under corrupted observations;
    \item adversarial traffic bursts and link-failure injection;
    \item fallback controllers when the learned policy is uncertain;
    \item audit logs for proposed and executed actions;
    \item human/operator approval for high-impact changes.
\end{itemize}

This connects directly to Chapters~18--20. Safe RL provides constraint enforcement, RLHF provides operator preference learning, and reasoning RL provides verifier-guided network-assistant workflows.

\section{Frontiers toward 2026}

\subsection{O-RAN xApps, rApps, and AI-native RAN control}

O-RAN creates a natural home for DRL through xApps and rApps. Near-real-time controllers can adjust slicing, handover, and radio resources, while non-real-time controllers can train models, manage policies, and provide enrichment information. Recent O-RAN work has begun to expose RL-friendly interfaces through ns-O-RAN and Gymnasium-style environments \citep{lacava2024online}. The frontier challenge is not only algorithmic: it is integration, latency, explainability, standard interfaces, and safety assurance.

\subsection{Digital twins and foundation world models for networks}

The model-based chapters argued that world models can reduce unsafe exploration. For networks, the analogous concept is a digital twin: a calibrated simulator that predicts consequences of candidate actions. A future network agent may combine telemetry, topology, radio maps, traffic forecasts, and simulator rollouts into a foundation-style network model.

\subsection{Non-terrestrial networks and satellite/IoT scheduling}

LEO satellite constellations, high-altitude platforms, UAV relays, and IoT devices create large-scale, dynamic, partially observable routing and scheduling problems. DRL is attractive because topology, contact windows, traffic demand, and energy constraints evolve together. However, verification and safety are critical because communication opportunities may be intermittent.

\subsection{ISAC, sensing, and semantic communications}

Integrated sensing and communication adds another dimension: the network must decide not only how to transmit but also how to sense. Future DRL agents may jointly control beamforming, sensing schedules, communication links, and semantic compression.

\subsection{Reasoning agents for network operations}

A reasoning model can propose explanations, diagnose anomalies, and recommend policy changes. But as Chapter~20 emphasized, a reasoning model should not directly control a network without verification. For network operations, the promising architecture is a verifier-guided assistant: the model proposes candidate interventions, a simulator/digital twin evaluates them, a safety layer filters them, and a human or controller applies them.

\section{Limitations and when not to use DRL for networks}

\begin{enumerate}[leftmargin=*]
    \item \textbf{Simulation-to-reality gap is large.} A policy optimized in simulation may exploit unrealistic channel, queue, or mobility models that do not reflect the real network.
    \item \textbf{Exploration is often unsafe.} Random or epsilon-greedy exploration may violate SLAs during training. Offline RL, shadow mode, and safety filters are necessary.
    \item \textbf{Non-stationary traffic breaks trained policies.} Traffic matrices, user density, and application profiles change over time. Retraining, drift detection, and online adaptation are required.
    \item \textbf{Telemetry delay and stale state.} Many network observations are seconds or minutes old. Policies trained with fresh state may behave incorrectly with stale telemetry.
    \item \textbf{Reward hacking is easy to miss.} Average throughput, average latency, or aggregate reward can all improve while tail constraints are violated and edge users are starved.
    \item \textbf{Multi-agent coordination is not solved.} Decentralized execution with coupled constraints remains difficult; interference, collision, and fairness require explicit coordination mechanisms.
    \item \textbf{Deployment requires human oversight.} Staged gates, shadow validation, canary deployment, and fallback rules are not optional.
\end{enumerate}

\section{Exercises}

\subsection*{Conceptual exercises}
\begin{enumerate}[leftmargin=*]
    \item Explain why network DRL is usually harder to deploy than game-playing DRL.
    \item Give three examples of network rewards that can be hacked.
    \item Why is p95 or p99 latency often more important than average latency?
    \item Why should the critic learn from executed actions when a safety filter modifies proposed actions?
    \item Compare SD-WAN traffic engineering and UAV trajectory control as DRL problems.
\end{enumerate}

\subsection*{Mathematical exercises}
\begin{enumerate}[leftmargin=*]
    \item Given a two-link SD-WAN action $a=[x,1-x]$, write a constrained optimization problem that maximizes throughput subject to latency and jitter bounds.
    \item Derive a class-weighted reward for URLLC, eMBB, and mMTC traffic with different latency and throughput thresholds.
    \item Suppose a UAV serves users with SINR values $10$, $15$, and $20$ dB. Convert these to linear scale and compute the Shannon-style throughput for $B=10$ MHz.
    \item Show how a traffic-split action can be projected onto the simplex $x_i\ge 0$, $\sum_i x_i=1$.
\end{enumerate}

\subsection*{Coding exercises}
\begin{enumerate}[leftmargin=*]
    \item Extend Listing~\ref{lst:network_env} to include a link-failure event and evaluate whether a trained agent recovers.
    \item Add a graph neural network encoder to Listing~\ref{lst:graph_obs} using PyTorch Geometric or a simple message-passing layer.
    \item Implement a PPO or SAC agent for the two-link SD-WAN environment and compare it with a static 50/50 split.
    \item Add a CBF-style latency safety filter to the environment and compare proposed vs executed actions.
    \item Build a multi-UAV version of the environment and compare independent PPO with MAPPO or QMIX.
\end{enumerate}

\subsection*{Research exercises}
\begin{enumerate}[leftmargin=*]
    \item Design a benchmark for UAV-assisted network slicing with three traffic classes and realistic mobility. Specify state, action, reward, constraints, and baselines.
    \item Write an ablation plan for testing whether a safety layer, graph encoder, or SDN guidance actually improves performance.
    \item Propose a digital-twin validation protocol for transferring a policy from simulation to a real testbed.
    \item Design a cyberattack scenario against a network DRL controller and propose detection and recovery mechanisms.
    \item Reproduce a small SD-WAN safe-RL experiment inspired by \citep{bista2026vtc,bista2026ifip}: compare unconstrained PPO/SAC, a rule-based controller, and a CBF-filtered learned controller.
\end{enumerate}

\section*{Looking Ahead to Chapter 22: Food for Thought}
\addcontentsline{toc}{section}{Looking Ahead to Chapter 22: Food for Thought}

This chapter connected DRL algorithms to networking, UAVs, and communication systems. The next chapter shifts from application-specific design to general engineering practice: how to build a DRL project from scratch, organize the code, run experiments, log metrics, manage seeds, compare baselines, and avoid misleading results.

\begin{quote}
    Chapter~21 showed what a real application domain demands from DRL. Chapter~22 shows how to build such a system without losing scientific control.
\end{quote}

	\part{How to Build and Evaluate DRL Systems}
% \addcontentsline{toc}{part}{Part VII How to Build and Evaluate DRL Systems}

\chapter[Practical Implementation Pipeline]{Practical Implementation Pipeline}
\chaptermark{Practical Implementation Pipeline}
\label{ch:implementation_pipeline}

\begin{keybox}{Chapter goal}
    A deep reinforcement learning project is not only an algorithm but a full scientific pipeline: environment contract, configuration, baselines, logging, evaluation, ablation, and deployment discipline. This chapter develops a practical implementation workflow for building DRL systems that are correct, reproducible, debuggable, and credible enough for research-grade experimentation and safety-aware deployment.
\end{keybox}

\section*{Chapter Overview}
\addcontentsline{toc}{section}{Chapter Overview}

The previous chapters studied deep reinforcement learning algorithms: DQN, PPO, SAC, model-based RL, offline RL, Decision Transformers, MARL, hierarchical RL, safe RL, RLHF, reasoning RL, and networked DRL. This chapter changes perspective. It asks a practical question: how does one build a complete DRL research system that is correct, reproducible, debuggable, and credible enough for a serious paper or deployment study?

A good DRL project is not just an algorithm file. It is a pipeline. It includes an environment contract, state and action definitions, reward and constraint design, baselines, logging, seeds, evaluation, ablations, failure analysis, and deployment gates. Many weak DRL papers fail not because the algorithm is poor, but because the pipeline is ambiguous, the baselines are weak, the metrics are incomplete, or the evaluation is statistically fragile. Henderson et al.\ showed that deep RL results are difficult to interpret without careful experimental procedure and reporting discipline \citep{henderson2018deep}. Agarwal et al.\ later argued that reliable DRL evaluation must report uncertainty and robust aggregate metrics rather than only point estimates \citep{agarwal2021rliable}. Engstrom et al.\ and Huang et al.\ further showed that implementation details can dominate algorithmic conclusions, especially for policy-gradient algorithms such as PPO \citep{engstrom2020implementation,huang2022details}.

This chapter is therefore a bridge between algorithms and scientific practice.

\begin{enumerate}[leftmargin=*]
    \item What a DRL implementation pipeline must contain
    \item The scenario-first design principle
    \item Project structure and configuration discipline
    \item Environment contracts: Gymnasium, single-agent, multi-agent, safe, and offline interfaces
    \item State and observation engineering
    \item Action-space engineering and projection
    \item Reward, cost, and metric separation
    \item Algorithm-selection matrix
    \item Baseline design: what must be compared
    \item Training-loop architecture
    \item Logging, checkpoints, and experiment tracking
    \item Reproducibility: seeds, determinism, and configuration hashes
    \item Evaluation with multiple seeds, confidence intervals, and robust statistics
    \item Ablations and sensitivity analysis
    \item Debugging protocol and unit tests
    \item Offline datasets, Minari-style dataset handling, and data lineage
    \item Safe deployment: shadow mode, canaries, and rollback
    \item A complete UAV/SD-WAN implementation scenario
    \item Python templates for a research-grade DRL project
    \item Limitations and common implementation traps
    \item Looking Ahead to Chapter 23
\end{enumerate}

\section{What a DRL implementation pipeline must contain}

A DRL implementation pipeline converts a scientific idea into a testable system. The core pipeline has seven stages:

\begin{enumerate}[leftmargin=*]
    \item \textbf{Problem contract}: state, action, reward, cost, dynamics, timescale, and success criteria.
    \item \textbf{Environment}: a simulator, emulator, digital twin, logged dataset, or real system wrapper.
    \item \textbf{Algorithm}: DQN, PPO, SAC, offline RL, model-based RL, MARL, safe RL, or a hybrid.
    \item \textbf{Training}: sampling, replay or rollout collection, optimization, logging, checkpoints, and failure detection.
    \item \textbf{Evaluation}: deterministic and stochastic policies, test scenarios, multiple seeds, baselines, and confidence intervals.
    \item \textbf{Ablation}: remove or vary components to show which mechanism matters.
    \item \textbf{Deployment gate}: shadow validation, safety filters, runtime monitoring, and rollback.
\end{enumerate}

\begin{figure}[t]
    \centering
    \begin{tikzpicture}[
        box/.style={draw,rounded corners,thick,minimum width=2.75cm,minimum height=0.75cm,align=center,font=\small},
        arrow/.style={-{Latex[length=2.2mm]},thick},
        node distance=0.65cm
        ]
        \node[box,fill=blue!8,draw=blue!70] (contract) {Problem\\contract};
        \node[box,fill=green!8,draw=green!60!black,right=of contract] (env) {Environment\\interface};
        \node[box,fill=orange!10,draw=orange!80!black,right=of env] (algo) {Algorithm\\implementation};
        \node[box,fill=purple!8,draw=purple!70,right=of algo] (train) {Training\\loop};
        \node[box,fill=red!6,draw=red!70!black,below=1.1cm of train] (eval) {Evaluation\\and baselines};
        \node[box,fill=gray!10,draw=gray!70,left=of eval] (ablate) {Ablation\\and diagnosis};
        \node[box,fill=cyan!8,draw=cyan!70!black,left=of ablate] (deploy) {Deployment\\gate};

        \draw[arrow] (contract) -- (env);
        \draw[arrow] (env) -- (algo);
        \draw[arrow] (algo) -- (train);
        \draw[arrow] (train) -- (eval);
        \draw[arrow] (eval) -- (ablate);
        \draw[arrow] (ablate) -- (deploy);
        \draw[arrow,dashed,draw=gray!70] (deploy.north) |- (contract.south);
        \node[below=0.55cm of ablate,align=center,font=\small] {The pipeline is cyclic: evaluation and deployment failures return to the problem contract.};
    \end{tikzpicture}
    \caption{A research-grade DRL implementation pipeline. A serious project must connect problem definition, environment interface, algorithm, training, evaluation, ablation, and deployment discipline.}
    \label{fig:implementation_pipeline}
\end{figure}

\begin{keybox}{Core implementation principle}
    An algorithm is not a DRL project. A DRL project is an executable scientific pipeline with a clear environment contract, reproducible configuration, credible baselines, task-specific metrics, and failure analysis.
\end{keybox}

This chapter uses a running example: safe DRL for UAV-assisted SD-WAN and wireless network control. The same pipeline applies to robotics, games, language-model tools, recommender systems, industrial control, and autonomous agents.

\section{The scenario-first design principle}

Most failed DRL implementations begin with an algorithm. A stronger approach begins with a scenario. The scenario defines what must be controlled, what can be observed, what can be acted upon, what is unsafe, and what counts as success.

For example, consider a UAV/SD-WAN controller. The system observes traffic load, latency, packet loss, SINR, queue length, battery level, and UAV position. It chooses routing splits, UAV movement, resource allocation, and safety-filter parameters. It receives reward for QoS satisfaction and energy efficiency, but it must satisfy constraints on latency, jitter, packet loss, collision risk, and battery safety. This is not merely a PPO or SAC problem. It is a constrained, partially observed, safety-critical, networked control problem.

\begin{figure}[t]
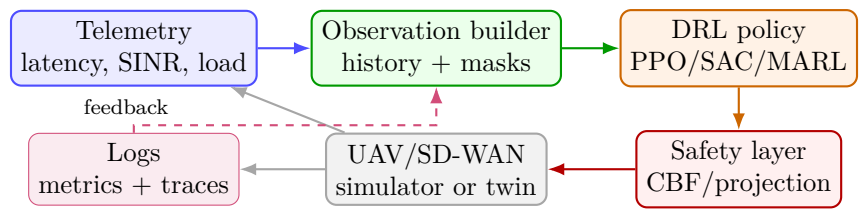

    \centering

    \caption{Scenario-first implementation for safe UAV/SD-WAN control. The policy is only one component: telemetry, observation construction, safety projection, logging, and simulator/digital-twin feedback are equally important.}
    \label{fig:scenario_first}
\end{figure}

A scenario-first design should answer the following before writing a training loop:

\begin{table}[t]
    \centering
    \caption{Scenario-first questions for a DRL project.}
    \label{tab:scenario_questions}
    \begin{tabularx}{\textwidth}{p{3.0cm}X}
        \toprule
        Question & Example in UAV/SD-WAN control \\
        \midrule
        What is controlled? & routing ratios, UAV movement, bandwidth allocation, handover, power \\
        What is observed? & telemetry, SINR, latency, battery, traffic class, queue state, topology \\
        What is hidden? & future traffic bursts, real channel uncertainty, user mobility, failures \\
        What is unsafe? & latency violation, battery exhaustion, collision, packet loss, regulatory violation \\
        What is optimized? & QoS satisfaction, throughput, energy efficiency, fairness, stability \\
        What is the timescale? & 100 ms, 1 s, 10 s, or slower orchestration loop \\
        What is the deployment path? & simulator, digital twin, shadow mode, canary, production guardrail \\
        \bottomrule
    \end{tabularx}
\end{table}

\section{Project structure and configuration discipline}

A DRL project should be organized so that experiments are reproducible by construction. The core principle is that every result must be traceable to code version, configuration, seed, environment version, dataset version, and hardware/software context.

\begin{table}[t]
    \centering
    \caption{A practical repository structure for a research-grade DRL project.}
    \label{tab:repo_structure}
    \begin{tabularx}{\textwidth}{p{3.4cm}X}
        \toprule
        Path & Purpose \\
        \midrule
        \texttt{configs/} & YAML or Python configuration files for environment, algorithm, training, and evaluation \\
        \texttt{envs/} & Gymnasium/PettingZoo-compatible environments and wrappers \\
        \texttt{agents/} & algorithm implementations: PPO, SAC, QMIX, offline RL, safe wrappers \\
        \texttt{models/} & policy networks, critics, world models, encoders, mixers \\
        \texttt{buffers/} & replay buffers, rollout buffers, offline datasets \\
        \texttt{train.py} & single entry point for training \\
        \texttt{eval.py} & deterministic evaluation and test-scenario evaluation \\
        \texttt{scripts/} & launchers, sweep generators, plotting, table generation \\
        \texttt{tests/} & environment, reward, shape, safety, and regression tests \\
        \texttt{runs/} & generated logs, checkpoints, configs, metrics, videos, artifacts \\
        \bottomrule
    \end{tabularx}
\end{table}

Hydra is widely used for complex research configuration because it composes hierarchical configuration files and supports command-line overrides for multi-run experiments \citep{hydra_docs}. Experiment tracking tools such as MLflow and Weights and Biases can log metrics, parameters, artifacts, and model versions \citep{mlflow_docs,wandb_docs}. For RL, this matters because a small difference in observation normalization, reward scaling, action clipping, or seed handling can change conclusions.

\Needspace{18\baselineskip}
\begin{lstlisting}[style=pythonstyle,caption={A minimal typed configuration object for reproducible DRL experiments.},label={lst:config_dataclass}]
from dataclasses import dataclass, asdict
import hashlib
import json


@dataclass(frozen=True)
class ExperimentConfig:
    env_id: str = "SafeUavSdwan-v0"
    algorithm: str = "ppo_lagrangian"
    seed: int = 0
    total_steps: int = 1_000_000
    decision_period_s: float = 1.0
    gamma: float = 0.99
    learning_rate: float = 3e-4
    hidden_dim: int = 256
    reward_scale: float = 1.0
    cost_limit: float = 0.05
    use_safety_filter: bool = True
    notes: str = "baseline run"

    def hash(self) -> str:
        payload = json.dumps(asdict(self), sort_keys=True).encode("utf-8")
        return hashlib.sha256(payload).hexdigest()[:12]


cfg = ExperimentConfig(seed=3, use_safety_filter=True)
print(cfg.hash())


def run_name(cfg: ExperimentConfig) -> str:
    return f"{cfg.algorithm}_{cfg.env_id}_seed{cfg.seed}_{cfg.hash()}"
\end{lstlisting}

\begin{keybox}{Practice note: run metadata}
    Every run should save the resolved configuration, the Git commit hash, the seed, the environment version, and a short machine-readable run ID. A plot without this metadata is not a reproducible scientific result.
\end{keybox}

\section{Environment contracts: Gymnasium, multi-agent, safe, and offline interfaces}

The environment is the most important file in a DRL project. The standard single-agent interface follows the Gymnasium pattern: \texttt{reset()} returns an initial observation and info dictionary, while \texttt{step(action)} returns observation, reward, termination flag, truncation flag, and info \citep{towers2024gymnasium}. Gymnasium is a maintained successor/fork of the original Gym API and is now the standard interface for many RL environments and wrappers. Multi-agent environments often use PettingZoo or related interfaces, while offline RL datasets increasingly use Minari-style dataset APIs compatible with Gymnasium \citep{minari_docs}.

A good environment contract specifies:

\begin{itemize}[leftmargin=*]
    \item observation shape, dtype, units, normalization, and history length;
    \item action space, bounds, masks, and projection behavior;
    \item reward and cost definitions;
    \item termination and truncation rules;
    \item info dictionary fields for metrics, diagnostics, and safety;
    \item seeding behavior and deterministic reset options;
    \item scenario generation and test-mode scenario IDs.
\end{itemize}

\begin{figure}[t]
    \centering
    \begin{tikzpicture}[
        box/.style={draw,rounded corners,thick,minimum width=3.2cm,minimum height=0.85cm,
            align=center,font=\small},
        small/.style={draw,rounded corners,minimum width=2.7cm,minimum height=0.7cm,
            align=center,font=\small},
        arrow/.style={-{Latex[length=2.2mm]},thick}
        ]

        \node[box,fill=blue!8,draw=blue!70] (agent) at (0, 0)
        {Agent\\policy $\pi_\theta$};

        \node[box,fill=green!8,draw=green!60!black] (env) at (6.0, 0)
        {Environment\\\texttt{reset}, \texttt{step}};

        \node[small,fill=orange!10,draw=orange!80!black] (obs) at (3.0, 2.2)
        {observation\\$o_t$};

        \node[small,fill=red!6,draw=red!70!black] (act) at (3.0,-2.2)
        {action\\$a_t$};

        \node[small,fill=purple!8,draw=purple!70] (info) at (7.0,-2.2)
        {info dict\\metrics, costs};

        \draw[arrow,draw=blue!70]
        (env.north) -- (obs.east);
        \draw[arrow,draw=blue!70]
        (obs.west)  -- (agent.north);

        \draw[arrow,draw=red!70!black]
        (agent.south) -- (act.west);
        \draw[arrow,draw=red!70!black]
        (act.east)    -- (env.south);

        \draw[arrow,draw=purple!70]
        (env.south) -- (info.north);

    \end{tikzpicture}
    \caption{The environment contract. The observation, action, reward, termination flags, and info dictionary must be explicit. In serious DRL, the info dictionary carries task metrics and safety diagnostics, not just debugging text.}
    \label{fig:environment_contract}
\end{figure}
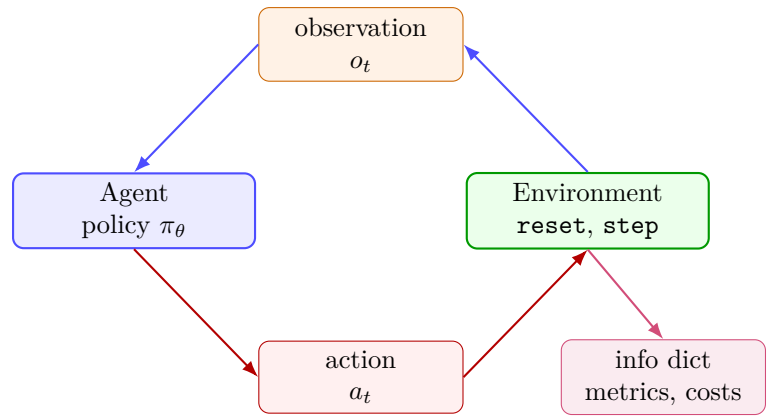

\Needspace{28\baselineskip}
\begin{lstlisting}[style=pythonstyle,caption={A Gymnasium-style environment skeleton for safe network control.},label={lst:gym_env_skeleton}]
import gymnasium as gym
from gymnasium import spaces
import numpy as np


class SafeNetworkEnv(gym.Env):
    """Minimal environment contract for a safe network-control task."""
    metadata = {"render_modes": []}

    def __init__(self, cfg):
        super().__init__()
        self.cfg = cfg
        # Example observation: latency, loss, throughput, SINR, battery, load, age.
        self.observation_space = spaces.Box(
            low=-10.0, high=10.0, shape=(7,), dtype=np.float32
        )
        # Example action: traffic split, bandwidth share, UAV dx, dy, dz.
        self.action_space = spaces.Box(
            low=np.array([0.0, 0.0, -1.0, -1.0, -0.5], dtype=np.float32),
            high=np.array([1.0, 1.0,  1.0,  1.0,  0.5], dtype=np.float32),
        )
        self.rng = np.random.default_rng(cfg.seed)
        self.t = 0
        self.state = None

    def reset(self, *, seed=None, options=None):
        if seed is not None:
            self.rng = np.random.default_rng(seed)
        self.t = 0
        self.state = self._sample_initial_state(options)
        obs = self._make_obs(self.state)
        info = {"scenario_id": options.get("scenario_id") if options else None}
        return obs.astype(np.float32), info

    def step(self, action):
        action = np.asarray(action, dtype=np.float32)
        executed_action, safety_info = self._project_action(action)
        next_state = self._dynamics(self.state, executed_action)
        reward, costs, metrics = self._reward_cost_metrics(self.state, executed_action, next_state)
        self.state = next_state
        self.t += 1
        terminated = bool(metrics["battery"] <= 0.0 or metrics["collision"])
        truncated = bool(self.t >= self.cfg.max_episode_steps)
        obs = self._make_obs(self.state).astype(np.float32)
        info = {
            "costs": costs,
            "metrics": metrics,
            "proposed_action": action,
            "executed_action": executed_action,
            "safety": safety_info,
        }
        return obs, float(reward), terminated, truncated, info
\end{lstlisting}

The key detail is that the environment returns both the proposed and executed action. If a safety layer projects the action, the learner must know which action actually affected the environment. This principle appeared repeatedly in Chapters~11, 12, 13, 18, and 21.

\section{State and observation engineering}

A state is what the environment truly is. An observation is what the agent receives. Many DRL bugs are really observation-design bugs. The observation may be too small, too noisy, unnormalized, stale, or inconsistent across train and test.

For networked systems, useful observation fields include:

\begin{itemize}[leftmargin=*]
    \item traffic load and queue length;
    \item latency, jitter, packet loss, throughput;
    \item wireless SINR, RSSI, spectral efficiency, interference;
    \item topology, link capacity, utilization, failures;
    \item UAV position, battery, velocity, mission mode;
    \item time features, telemetry age, recent history, and scenario ID;
    \item action masks and safety margins.
\end{itemize}

A good observation vector should record units before normalization. For example, latency in milliseconds and SINR in dB should not be mixed directly with normalized battery percentage without scale handling. Normalization should be fit on training data or updated with clear rules; evaluation should not accidentally leak test statistics. Observation normalization, especially running mean and variance normalization, is one of the most impactful implementation details: PPO with and without observation normalization can produce qualitatively different policies even on the same environment \citep{engstrom2020implementation,huang2022details}.

\begin{table}[t]
    \centering
    \caption{Observation-design mistakes and their consequences.}
    \label{tab:obs_mistakes}
    \begin{tabularx}{\textwidth}{p{3.5cm}X X}
        \toprule
        Mistake & Consequence & Fix \\
        \midrule
        Missing history in delayed system & apparent non-Markovian dynamics & stack recent observations or use RNN \\
        Unnormalized heterogeneous units & unstable gradients and feature dominance & normalize by physical ranges or running statistics \\
        Train/test telemetry mismatch & high simulation score, poor deployment & freeze preprocessing and validate on held-out scenarios \\
        No action mask in invalid-action domains & wasted exploration or unsafe actions & provide masks or safety projection \\
        No telemetry age feature & policy acts on stale information without knowing it & append timestamp age $\Delta_t$ \\
        \bottomrule
    \end{tabularx}
\end{table}

\Needspace{20\baselineskip}
\begin{lstlisting}[style=pythonstyle,caption={Observation builder with normalization and telemetry age.},label={lst:obs_builder}]
import numpy as np


class ObservationBuilder:
    def __init__(self, ranges):
        self.ranges = ranges

    def scale(self, name, value):
        lo, hi = self.ranges[name]
        return np.clip((value - lo) / (hi - lo + 1e-8), 0.0, 1.0)

    def __call__(self, telemetry):
        features = [
            self.scale("latency_ms", telemetry["latency_ms"]),
            self.scale("packet_loss", telemetry["packet_loss"]),
            self.scale("throughput_mbps", telemetry["throughput_mbps"]),
            self.scale("sinr_db", telemetry["sinr_db"]),
            self.scale("battery", telemetry["battery"]),
            self.scale("traffic_load", telemetry["traffic_load"]),
            self.scale("telemetry_age_s", telemetry["telemetry_age_s"]),
        ]
        return np.asarray(features, dtype=np.float32)
\end{lstlisting}

\section{Action-space engineering and projection}

The action space should represent controllable decisions at the correct timescale. Poor action design can make a good algorithm fail.

\begin{table}[t]
    \centering
    \caption{Action-space choices for DRL implementation.}
    \label{tab:action_space_choices}
    \begin{tabularx}{\textwidth}{p{3.0cm}X X}
        \toprule
        Action type & Example & Suitable algorithms \\
        \midrule
        Discrete & choose route A/B/C, move north/south/east/west & DQN, Rainbow, PPO categorical \\
        Continuous & traffic split, power level, UAV velocity & PPO Gaussian, SAC, TD3 \\
        Hybrid & choose slice then allocate bandwidth & PPO multi-head, parameterized actions, HRL \\
        Multi-agent & one action per UAV or base station & QMIX, MAPPO, MADDPG, graph MARL \\
        Constrained & action must satisfy SLA/safety bounds & safe wrapper, CBF, projection, Lagrangian methods \\
        \bottomrule
    \end{tabularx}
\end{table}

For real systems, the action proposed by the neural network is rarely executed directly. It is decoded, clipped, projected, masked, or passed through a safety layer. This is not a hack. It is part of the control architecture.

\Needspace{20\baselineskip}
\begin{lstlisting}[style=pythonstyle,caption={Action decoding and safety projection for routing splits.},label={lst:action_projection}]
import numpy as np


def decode_action(raw_action):
    """Convert unconstrained network output to a traffic split."""
    raw = np.asarray(raw_action, dtype=np.float32)
    # Softmax creates nonnegative components that sum to one.
    exp = np.exp(raw - raw.max())
    split = exp / (exp.sum() + 1e-8)
    return split


def project_latency_safe(split, predicted_latency_ms, max_latency_ms=25.0):
    """Simple projection: move traffic away from paths predicted unsafe."""
    safe = predicted_latency_ms <= max_latency_ms
    if safe.any():
        projected = split * safe.astype(np.float32)
        projected /= projected.sum() + 1e-8
        return projected, {"projected": not np.allclose(projected, split), "safe_mask": safe}
    # Fallback: choose path with minimum predicted latency.
    projected = np.zeros_like(split)
    projected[int(np.argmin(predicted_latency_ms))] = 1.0
    return projected, {"projected": True, "safe_mask": safe, "fallback": True}
\end{lstlisting}

\begin{warningbox}{Executed actions must match learning targets}
    Never hide action projection. Log both the raw neural action and the executed action. If the critic is trained as if the raw action was executed, the value function learns the wrong dynamics.
\end{warningbox}

\section{Reward, cost, and metric separation}

A common mistake is to treat the reward as the only evaluation metric. Reward is the training signal; metrics are how humans judge the system. In safety-critical DRL, reward, cost, and evaluation metrics should be separated.

\begin{equation}
    r_t = w_q R_t^{\mathrm{QoS}} - w_e C_t^{\mathrm{energy}} - w_s C_t^{\mathrm{smooth}},
\end{equation}
while constraints may be tracked separately:
\begin{equation}
    c_t^{\mathrm{latency}} = \mathbb{1}\{L_t > L_{\max}\}, \qquad
    c_t^{\mathrm{loss}} = \mathbb{1}\{P_t^{\mathrm{loss}} > P_{\max}\}.
\end{equation}

The final paper should not report only $\sum r_t$. It should report latency, packet loss, SINR, throughput, QoS satisfaction, energy, battery failure, collision count, safety projection rate, constraint violation rate, and inference time.

\Needspace{18\baselineskip}
\begin{lstlisting}[style=pythonstyle,caption={Reward and metric computation should be separate.},label={lst:reward_metrics}]
def compute_reward_cost_metrics(state, action, next_state):
    latency_ms = next_state["latency_ms"]
    loss = next_state["packet_loss"]
    throughput = next_state["throughput_mbps"]
    energy = next_state["energy_used"]

    qos_reward = 1.0 * (latency_ms <= 25.0) + 0.01 * throughput
    reward = qos_reward - 0.05 * energy

    costs = {
        "latency_violation": float(latency_ms > 25.0),
        "loss_violation": float(loss > 0.01),
        "battery_risk": float(next_state["battery"] < 0.10),
    }
    metrics = {
        "latency_ms": latency_ms,
        "packet_loss": loss,
        "throughput_mbps": throughput,
        "energy_used": energy,
        "qos_satisfied": float(latency_ms <= 25.0 and loss <= 0.01),
    }
    return reward, costs, metrics
\end{lstlisting}

\section{Algorithm-selection matrix}

Algorithm choice should follow from the environment contract, not from fashion. Table~\ref{tab:algorithm_selection} summarizes common choices.

\begin{table}[t]
    \centering
    \caption{Choosing an algorithm from the project constraints.}
    \label{tab:algorithm_selection}
    \begin{tabularx}{\textwidth}{p{3.0cm}p{3.2cm}X}
        \toprule
        Problem property & Good starting point & Reason \\
        \midrule
        Small discrete action & DQN or Double DQN & simple value-based baseline \\
        Image or high-dimensional discrete control & DQN/Rainbow, PPO & robust historical baselines \\
        Continuous action, online simulator & SAC, PPO, TD3 & standard continuous-control algorithms \\
        Expensive or dangerous interaction & offline RL, model-based RL & reduces online exploration \\
        Multiple cooperative agents & QMIX, MAPPO, VDN & handles joint action and credit assignment \\
        Hard safety constraints & PPO-Lagrangian, CPO, CBF filter & separates reward from constraints \\
        Long-horizon skills & HRL, options, goal-conditioned policies & temporal abstraction \\
        Network/UAV deployment & PPO/SAC/MARL + safety wrapper + shadow validation & practical hybrid architecture \\
        \bottomrule
    \end{tabularx}
\end{table}

Stable-Baselines3 provides reliable PyTorch implementations of standard algorithms such as PPO, SAC, TD3, DQN, and A2C \citep{raffin2021stable,stablebaselines3_docs}. CleanRL provides single-file implementations designed to expose the implementation details of each algorithm \citep{huang2022cleanrl}. Tianshou offers a modular PyTorch framework for composing policies, collectors, replay buffers, and trainers \citep{weng2022tianshou}. RLlib provides scalable and production-oriented RL training infrastructure \citep{rllib_docs}. These libraries should not be treated as black boxes; they should be used as reference implementations and baselines.

\section{Baseline design: what must be compared}

A DRL result is meaningless without credible baselines. A baseline is not always another neural RL algorithm. In network and control systems, the strongest baseline may be a rule-based controller, optimization solver, heuristic, or model-predictive controller.

\begin{table}[t]
    \centering
    \caption{Baseline ladder for DRL projects.}
    \label{tab:baseline_ladder}
    \begin{tabularx}{\textwidth}{p{3.0cm}X}
        \toprule
        Baseline type & Purpose \\
        \midrule
        Random policy & sanity check; should usually perform poorly \\
        Greedy heuristic & tests whether DRL beats simple local optimization \\
        Rule-based production policy & most important practical baseline in networks \\
        Classical optimizer & compares with domain knowledge and constraints \\
        Behavior cloning & tests whether RL improves beyond imitation \\
        Standard DRL baseline & DQN/PPO/SAC/QMIX/MAPPO depending on action and agent structure \\
        Ablated version & shows whether the proposed component matters \\
        Oracle or upper bound & contextualizes the gap to ideal performance \\
        \bottomrule
    \end{tabularx}
\end{table}

\begin{keybox}{Practice note: baseline discipline}
    If a new DRL method only beats random and a weak neural baseline, the result is not convincing. In networking, always compare against a domain heuristic or production-style policy.
\end{keybox}

\section{Training-loop architecture}

A training loop must make the data path explicit. On-policy algorithms such as PPO collect rollouts, compute advantages, and optimize for several epochs. Off-policy algorithms such as SAC collect transitions into a replay buffer and update from random minibatches. Offline algorithms train from a fixed dataset. MARL algorithms collect joint trajectories.

The distinction between a rollout buffer and a replay buffer is not cosmetic. A PPO rollout buffer is short-lived: it stores a batch generated by the current or very recent policy, is used for a fixed number of epochs, and is then discarded. A SAC replay buffer is long-lived: it stores transitions from many past policies and is sampled repeatedly over training. Mixing these concepts is a common implementation error; stale replay-style data can break on-policy assumptions, while discarding off-policy data too quickly destroys SAC's sample-efficiency advantage.

\begin{figure}[t]
    \centering
    \begin{tikzpicture}[
        box/.style={draw,rounded corners,thick,minimum width=3.0cm,minimum height=0.75cm,align=center,font=\small},
        arrow/.style={-{Latex[length=2.2mm]},thick},
        node distance=0.75cm
        ]
        \node[box,fill=blue!8,draw=blue!70] (envs) {Vectorized\\environments};
        \node[box,fill=green!8,draw=green!60!black,right=of envs] (policy) {Policy\\network};
        \node[box,fill=orange!10,draw=orange!80!black,right=of policy] (buffer) {Rollout or\\replay buffer};
        \node[box,fill=purple!8,draw=purple!70,right=of buffer] (optim) {Optimizer\\updates};
        \node[box,fill=red!6,draw=red!70!black,below=1.1cm of optim] (log) {Logger\\metrics};
        \node[box,fill=gray!10,draw=gray!70,left=of log] (eval) {Evaluator\\test seeds};
        \node[box,fill=cyan!8,draw=cyan!70!black,left=of eval] (ckpt) {Checkpoint\\artifacts};
        \draw[arrow] (envs) -- (policy);
        \draw[arrow] (policy) -- (buffer);
        \draw[arrow] (buffer) -- (optim);
        \draw[arrow] (optim) -- (log);
        \draw[arrow] (log) -- (eval);
        \draw[arrow] (eval) -- (ckpt);
        \draw[arrow,dashed,draw=gray!60]
        (optim.north) -- ++(0,0.5) coordinate(T)
        -- node[above,font=\scriptsize]{parameter update}
        (T -| policy.north)
        -- (policy.north);
    \end{tikzpicture}
    \caption{Training-loop architecture. The data path and artifact path should be explicit: environments produce experience, the policy and optimizer update parameters, and the logger/checkpointer records evidence.}
    \label{fig:training_loop_arch}
\end{figure}
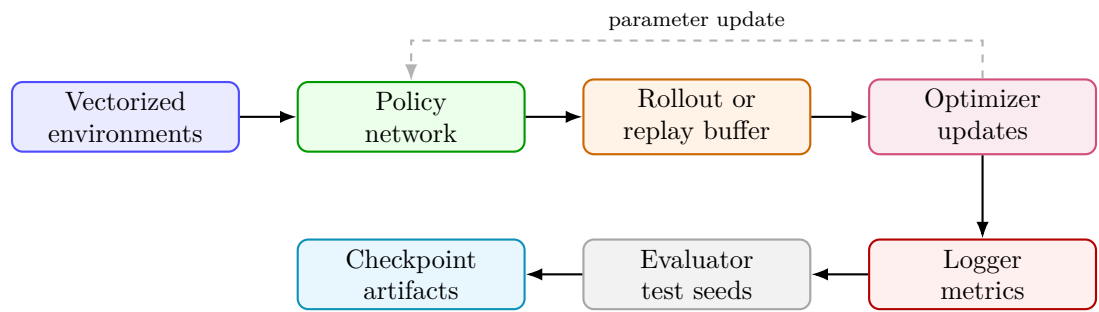

\Needspace{24\baselineskip}
\begin{lstlisting}[style=pythonstyle,caption={A compact training-loop skeleton with explicit evaluation and checkpointing.},label={lst:training_loop_skeleton}]
def train(cfg, env, agent, logger, evaluator, checkpoint_manager):
    obs, info = env.reset(seed=cfg.seed)
    global_step = 0

    while global_step < cfg.total_steps:
        batch = agent.collect(env, obs, n_steps=cfg.rollout_steps)
        obs = batch["last_obs"]
        train_stats = agent.update(batch)
        global_step += batch["num_env_steps"]

        logger.log(train_stats, step=global_step)
        logger.log(batch["episode_metrics"], step=global_step)

        if global_step % cfg.eval_interval == 0:
            eval_stats = evaluator.evaluate(agent, step=global_step)
            logger.log(eval_stats, step=global_step)
            checkpoint_manager.save_if_best(agent, eval_stats, cfg)

        if train_stats.get("nan_detected", False):
            checkpoint_manager.save_debug_snapshot(agent, batch, cfg)
            raise FloatingPointError("NaN detected during training")
\end{lstlisting}

\section{Logging, checkpoints, and experiment tracking}

A logger should record at least four kinds of data:

\begin{enumerate}[leftmargin=*]
    \item \textbf{Optimization metrics}: loss, entropy, KL, gradient norm, value loss, Q loss.
    \item \textbf{Task metrics}: latency, throughput, packet loss, safety violations, energy.
    \item \textbf{Data metrics}: replay age, action distribution, observation statistics, reward scale.
    \item \textbf{System metrics}: runtime, GPU memory, inference time, environment steps per second.
\end{enumerate}

MLflow and W\&B provide experiment tracking, model artifacts, and metric logging \citep{mlflow_docs,wandb_docs}. The tool is less important than the discipline: every figure in a paper should be traceable to a run directory and a config file.

\Needspace{22\baselineskip}
\begin{lstlisting}[style=pythonstyle,caption={Minimal JSONL logger for reproducible experiments.},label={lst:jsonl_logger}]
import json
from pathlib import Path


class JsonlLogger:
    def __init__(self, run_dir):
        self.path = Path(run_dir) / "metrics.jsonl"
        self.path.parent.mkdir(parents=True, exist_ok=True)

    def log(self, metrics, step):
        row = {"step": int(step)}
        for k, v in metrics.items():
            if isinstance(v, (int, float, str, bool)):
                row[k] = v
        with self.path.open("a", encoding="utf-8") as f:
            f.write(json.dumps(row, sort_keys=True) + "\n")
\end{lstlisting}

\section{Reproducibility: seeds, determinism, and configuration hashes}

Deep RL is noisy. A single lucky seed is not evidence. Henderson et al.\ emphasized that nondeterminism and high variance make DRL results hard to interpret \citep{henderson2018deep}. A project should therefore run multiple seeds and save enough metadata to rerun each seed.

\Needspace{20\baselineskip}
\begin{lstlisting}[style=pythonstyle,caption={Seed setup for Python, NumPy, and PyTorch.},label={lst:seed_setup}]
import os
import random
import numpy as np
import torch


def set_global_seed(seed: int, deterministic_torch: bool = False):
    random.seed(seed)
    np.random.seed(seed)
    os.environ["PYTHONHASHSEED"] = str(seed)
    torch.manual_seed(seed)
    torch.cuda.manual_seed_all(seed)
    if deterministic_torch:
        torch.backends.cudnn.deterministic = True
        torch.backends.cudnn.benchmark = False
        # This can slow training and may not cover every operation.
        torch.use_deterministic_algorithms(True, warn_only=True)
\end{lstlisting}

\begin{warningbox}{Reproducibility is not bitwise identity}
    Reproducibility does not mean every GPU run is bit-identical. It means the experiment is traceable, the variance is measured, and the conclusions do not depend on a single hidden random seed.
\end{warningbox}

\section{Evaluation with multiple seeds and robust statistics}

Evaluation should separate training curves from final test results. A training curve shows learning behavior; a final test table shows performance under fixed evaluation conditions.

Agarwal et al.\ argued that point estimates such as mean score can be misleading in few-seed DRL evaluation and proposed robust aggregate statistics such as the interquartile mean and performance profiles \citep{agarwal2021rliable}. The accompanying \texttt{rliable} package implements these metrics and is useful for producing performance profiles, interval estimates, and aggregate comparisons across tasks \citep{rliable_repo}. For a practical book chapter, the important lesson is simple: report uncertainty.

\begin{equation}
    \bar{x} = \frac{1}{N}\sum_{i=1}^N x_i, \qquad
    \mathrm{SE}(\bar{x}) = \frac{s}{\sqrt{N}}.
\end{equation}

For safety-critical tasks, also report worst-case and tail metrics:
\begin{equation}
    \mathrm{CVaR}_{\rho}(L) = \E[L \given L \ge q_{\rho}(L)],
\end{equation}
where $L$ may be latency or constraint cost.

\Needspace{24\baselineskip}
\begin{lstlisting}[style=pythonstyle,caption={Evaluation over seeds and scenarios with confidence intervals.},label={lst:evaluation_ci}]
import numpy as np


def mean_ci(values, z=1.96):
    values = np.asarray(values, dtype=np.float64)
    mean = values.mean()
    se = values.std(ddof=1) / np.sqrt(len(values))
    return mean, mean - z * se, mean + z * se


def evaluate_policy(agent, make_env, seeds, scenario_ids, episodes_per_scenario=5):
    scores, violations, latencies = [], [], []
    for seed in seeds:
        for scenario_id in scenario_ids:
            env = make_env(seed=seed, scenario_id=scenario_id)
            for _ in range(episodes_per_scenario):
                obs, info = env.reset(seed=seed, options={"scenario_id": scenario_id})
                done = False
                ep_score = 0.0
                ep_violations = 0
                ep_latencies = []
                while not done:
                    action = agent.act(obs, deterministic=True)
                    obs, reward, terminated, truncated, info = env.step(action)
                    done = terminated or truncated
                    ep_score += reward
                    ep_violations += int(info["costs"]["latency_violation"] > 0)
                    ep_latencies.append(info["metrics"]["latency_ms"])
                scores.append(ep_score)
                violations.append(ep_violations)
                latencies.extend(ep_latencies)
    q95 = np.quantile(latencies, 0.95)
    return {
        "score_mean_ci": mean_ci(scores),
        "violations_mean_ci": mean_ci(violations),
        "latency_p95": float(q95),
        "latency_cvar95": float(np.mean([x for x in latencies if x >= q95])),
    }
\end{lstlisting}

\section{Ablations and sensitivity analysis}

Ablation is not optional. It is how the reader learns what mattered. If a method combines graph attention, safety projection, adaptive reward weights, and uncertainty-aware critics, then at minimum the paper should show what happens when each is removed. Each ablation should be treated as a controlled experiment: change one mechanism, keep the environment, seeds, evaluation protocol, and tuning budget fixed, and interpret the result as evidence about that one mechanism rather than as a new algorithm.

\begin{figure}[t]
    \centering
    \begin{tikzpicture}[
        box/.style={draw,rounded corners,thick,minimum width=2.9cm,minimum height=0.75cm,align=center,font=\small},
        arrow/.style={-{Latex[length=2.2mm]},thick},
        node distance=0.75cm
        ]
        \node[box,fill=green!8,draw=green!60!black] (full) {Full method};
        \node[box,fill=blue!8,draw=blue!70,right=of full] (nogat) {Remove\\graph module};
        \node[box,fill=orange!10,draw=orange!80!black,right=of nogat] (nosafe) {Remove\\safety filter};
        \node[box,fill=purple!8,draw=purple!70,below=1.1cm of nogat] (fixed) {Fixed\\reward weights};
        \node[box,fill=red!6,draw=red!70!black,right=of fixed] (noens) {Remove\\uncertainty};
        \draw[arrow] (full) -- (nogat);
        \draw[arrow] (nogat) -- (nosafe);
        \draw[arrow] (full) -- (fixed);
        \draw[arrow] (fixed) -- (noens);
        \node[below=0.6cm of fixed,align=center,font=\small] {Each ablation changes exactly one mechanism.};
    \end{tikzpicture}
    \caption{Ablation design. A credible DRL paper should isolate the contribution of each major mechanism instead of comparing only the full method against weak baselines.}
    \label{fig:ablation_design}
\end{figure}
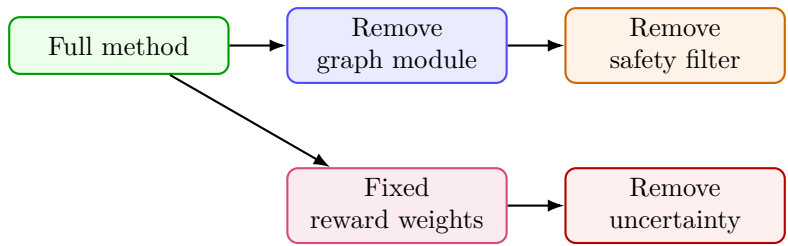

\Needspace{20\baselineskip}
\begin{lstlisting}[style=pythonstyle,caption={Simple ablation registry.},label={lst:ablation_registry}]
def make_ablation_configs(base_cfg):
    ablations = []
    for name, updates in {
        "full": {},
        "no_safety_filter": {"use_safety_filter": False},
        "fixed_reward_weights": {"adaptive_weights": False},
        "no_uncertainty_penalty": {"uncertainty_penalty": 0.0},
        "no_graph_encoder": {"encoder": "mlp"},
    }.items():
        cfg = base_cfg.copy()
        cfg.update(updates)
        cfg["ablation_name"] = name
        ablations.append(cfg)
    return ablations
\end{lstlisting}

\section{Debugging protocol and unit tests}

DRL debugging should begin before training. Test the environment, reward, action projection, replay buffer, advantage computation, and loss functions separately. A broken environment can still produce a beautiful learning curve if the reward accidentally leaks future information.

\begin{table}[t]
    \centering
    \caption{Minimum unit tests for a DRL implementation.}
    \label{tab:unit_tests}
    \begin{tabularx}{\textwidth}{p{3.2cm}X}
        \toprule
        Test & What it catches \\
        \midrule
        Reset/step shape test & observation/action mismatch, dtype errors \\
        Reward sign test & inverted objective or incorrect scaling \\
        Terminal mask test & bootstrapping across episode boundaries \\
        Action projection test & unsafe or invalid executed actions \\
        Replay sampling test & wrong batch dimensions or stale data \\
        GAE/return test & advantage recursion bugs \\
        Deterministic evaluation test & stochastic action used during evaluation \\
        Seed smoke test & non-reproducible scenario generation \\
        \bottomrule
    \end{tabularx}
\end{table}

\begin{table}[t]
    \centering
    \caption{Common DRL implementation bugs worth testing explicitly.}
    \label{tab:common_implementation_bugs}
    \begin{tabularx}{\textwidth}{p{4.1cm}X}
        \toprule
        Bug & Why it matters \\
        \midrule
        Off-by-one in advantage computation & the last bootstrap value is aligned with the wrong transition \\
        Episode termination not propagated to value targets & value is bootstrapped across episode boundaries \\
        Action clipping inside vs.\ outside the policy & the log-probability may not correspond to the executed action \\
        Discount factor applied at the wrong level & per-step discounting is confused with episode-level aggregation \\
        Reward normalization changes cost signals & safety costs become hidden or rescaled inconsistently \\
        Replay-buffer wrap-around loses metadata & transitions are preserved but scenario IDs, masks, or timestamps are overwritten \\
        \bottomrule
    \end{tabularx}
\end{table}

\Needspace{24\baselineskip}
\begin{lstlisting}[style=pythonstyle,caption={Environment smoke test before training.},label={lst:env_smoke_test}]
def smoke_test_env(env, n_steps=100):
    obs, info = env.reset(seed=0)
    assert env.observation_space.contains(obs), "reset observation outside space"
    for _ in range(n_steps):
        action = env.action_space.sample()
        next_obs, reward, terminated, truncated, info = env.step(action)
        assert env.observation_space.contains(next_obs), "step observation outside space"
        assert isinstance(reward, float), "reward must be a Python float"
        assert "metrics" in info and "costs" in info, "missing diagnostics"
        if terminated or truncated:
            next_obs, info = env.reset()
    return True
\end{lstlisting}

\begin{warningbox}{Sanity checks before training}
    If the random policy already performs nearly optimally, the environment is probably too easy or the reward leaks privileged information. If the expert heuristic performs worse than random, the reward or action decoding is probably wrong.
\end{warningbox}

\section{Offline datasets, Minari, and data lineage}

Offline and hybrid offline-online DRL require dataset discipline. A dataset should have provenance, environment version, behavior policy information, timestamp, units, filtering rules, and statistics. Minari provides a standardized API for offline RL datasets and integrates with Gymnasium-style environments \citep{minari_docs,minari_repo}. This is especially useful when building reproducible offline RL benchmarks or network logs.

\begin{table}[t]
    \centering
    \caption{Metadata required for offline RL datasets.}
    \label{tab:offline_metadata}
    \begin{tabularx}{\textwidth}{p{3.4cm}X}
        \toprule
        Metadata & Why it matters \\
        \midrule
        Environment version & prevents mixing incompatible transition semantics \\
        Behavior policy & explains action distribution and coverage \\
        Scenario generator & identifies train/test scenario drift \\
        Reward/cost definitions & prevents silent changes in objective \\
        Telemetry units & prevents scale and normalization bugs \\
        Time range & detects non-stationarity and deployment drift \\
        Filtering rules & explains missing failures or rare events \\
        License/privacy & required for sharing and publication \\
        \bottomrule
    \end{tabularx}
\end{table}

Offline datasets are not static in production. If new logs are appended over weeks or months, the behavior policy, user traffic, topology, and failure distribution may drift. A dataset manifest should therefore record collection windows and should support drift checks before old and new data are mixed into one training set.

\Needspace{20\baselineskip}
\begin{lstlisting}[style=pythonstyle,caption={A lightweight dataset manifest for offline RL logs.},label={lst:dataset_manifest}]
from dataclasses import dataclass


@dataclass
class DatasetManifest:
    name: str
    env_id: str
    env_version: str
    behavior_policy: str
    num_episodes: int
    num_transitions: int
    start_time: str
    end_time: str
    reward_definition: str
    cost_definition: str
    telemetry_units: dict
    train_test_split: str
\end{lstlisting}

\section{Safe deployment: shadow mode, canaries, and rollback}

Deployment should be staged. A typical path is:

\begin{enumerate}[leftmargin=*]
    \item offline evaluation on held-out scenarios;
    \item simulator stress tests;
    \item digital-twin replay;
    \item shadow mode beside a production controller;
    \item limited canary deployment with guardrails;
    \item full deployment only after rollback and monitoring are tested.
\end{enumerate}

This shadow-validation pattern was introduced in Section~21.24 for network deployment; here it is treated as a general step in any safety-critical DRL implementation pipeline. Shadow mode is passive: the new policy logs what it would have done while the production controller still acts. A/B testing is active: a fraction of users, flows, or scenarios are actually routed through the new policy. Most safety-critical network deployments should use shadow mode first, then canary or A/B testing only after risk is bounded.

\begin{figure}[t]
    \centering
    \begin{tikzpicture}[
        box/.style={draw,rounded corners,thick,minimum width=2.9cm,minimum height=0.75cm,align=center,font=\small},
        arrow/.style={-{Latex[length=2.2mm]},thick},
        node distance=0.7cm
        ]
        \node[box,fill=blue!8,draw=blue!70] (offline) {Offline\\evaluation};
        \node[box,fill=green!8,draw=green!60!black,right=of offline] (stress) {Stress\\simulation};
        \node[box,fill=orange!10,draw=orange!80!black,right=of stress] (shadow) {Shadow\\mode};
        \node[box,fill=purple!8,draw=purple!70,right=of shadow] (canary) {Canary\\deployment};
        \node[box,fill=red!6,draw=red!70!black,below=1.0cm of canary] (rollback) {Rollback\\guard};
        \draw[arrow] (offline) -- (stress);
        \draw[arrow] (stress) -- (shadow);
        \draw[arrow] (shadow) -- (canary);
        \draw[arrow] (canary) -- (rollback);
        \draw[arrow,dashed] (rollback.west) -| (offline.south);
    \end{tikzpicture}
    \caption{Deployment gate for DRL systems. The model should pass offline tests, stress tests, shadow-mode comparison, and canary deployment before it can affect the real system without supervision.}
    \label{fig:deployment_gate}
\end{figure}
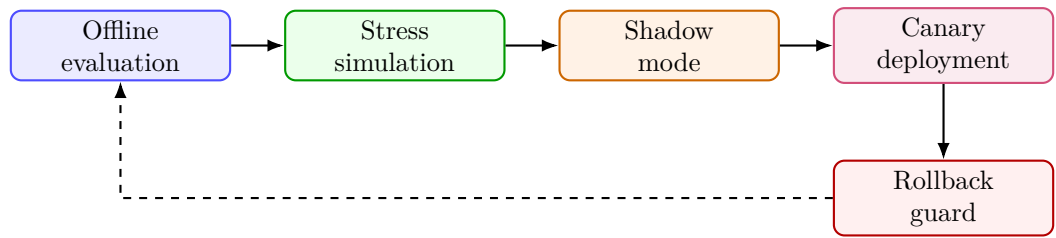

\Needspace{22\baselineskip}
\begin{lstlisting}[style=pythonstyle,caption={Shadow-mode validation comparing RL and production actions.},label={lst:shadow_validation2}]
def shadow_validate(policy, production_controller, telemetry_stream, safety_model):
    records = []
    for telemetry in telemetry_stream:
        obs = build_observation(telemetry)
        rl_action = policy.act(obs, deterministic=True)
        prod_action = production_controller.act(telemetry)

        rl_risk = safety_model.predict_cost(obs, rl_action)
        prod_risk = safety_model.predict_cost(obs, prod_action)
        delta_risk = rl_risk - prod_risk

        records.append({
            "timestamp": telemetry["timestamp"],
            "rl_action": rl_action,
            "prod_action": prod_action,
            "rl_predicted_cost": rl_risk,
            "prod_predicted_cost": prod_risk,
            "delta_risk": delta_risk,
        })
    return records
\end{lstlisting}

\section{A complete UAV/SD-WAN implementation scenario}

We now assemble the pipeline into a concrete scenario. This implementation scenario follows the design of our own safe SD-WAN traffic-engineering experiments using uncertainty-aware and ensemble-based neural CBFs \citep{bista2026vtc,bista2026ifip}.

\textbf{Problem}. A controller manages traffic across MPLS and Internet links while a UAV assists coverage for priority users. The controller can split traffic, adjust UAV movement, and allocate bandwidth. The goal is high QoS satisfaction with low energy use and low safety violation.

\textbf{Observation}. The observation includes traffic load, latency, jitter, packet loss, throughput, SINR, battery, UAV position, telemetry age, and recent history.

\textbf{Action}. The action contains a continuous traffic split, bandwidth allocation, and UAV velocity vector. A safety layer projects actions that violate latency, battery, or collision margins.

\textbf{Reward}. Reward combines QoS satisfaction, throughput, and energy. Costs track latency violation, packet loss, battery risk, and safety-filter intervention.

\textbf{Baselines}. The study compares rule-based routing, load-balancing heuristic, PPO, SAC, PPO-Lagrangian, SAC with CBF projection, and an ablation without safety projection.

\textbf{Evaluation}. The evaluation reports mean reward, QoS satisfaction, latency p95, packet-loss p95, energy, safety-violation rate, projection rate, inference time, and shadow-risk score.

\begin{figure}[t]
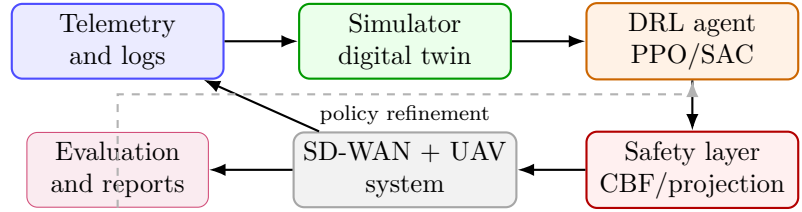

    \centering

    \caption{End-to-end implementation scenario for safe DRL in UAV/SD-WAN control. The digital twin trains and evaluates policies; the safety layer gates actions; telemetry and logs close the loop for refinement.}
    \label{fig:complete_scenario}
\end{figure}

\begin{table}[t]
    \centering
    \caption{Concrete experiment matrix for the UAV/SD-WAN scenario.}
    \label{tab:experiment_matrix}
    \begin{tabularx}{\textwidth}{p{2.7cm}X}
        \toprule
        Component & Values \\
        \midrule
        Algorithms & rule-based, PPO, SAC, PPO-Lagrangian, SAC+CBF, ablated SAC without CBF \\
        Seeds & 0, 1, 2, 3, 4, 5, 6, 7, 8, 9 \\
        Scenarios & normal load, traffic burst, link degradation, UAV low battery, telemetry delay, mixed failure \\
        Metrics & reward, QoS satisfaction, p95 latency, CVaR latency, packet loss, energy, violation rate \\
        Ablations & no safety filter, no history, no telemetry age, fixed reward weights, no uncertainty penalty \\
        Deployment gate & shadow risk below threshold for 48 hours of replay before canary \\
        \bottomrule
    \end{tabularx}
\end{table}

\section{Python template: a minimal research runner}

\Needspace{26\baselineskip}
\begin{lstlisting}[style=pythonstyle,caption={A minimal research runner tying config, env, agent, evaluation, and logging together.},label={lst:research_runner}]
def main(cfg):
    set_global_seed(cfg.seed)
    run_id = run_name(cfg)
    logger = JsonlLogger(run_dir=f"runs/{run_id}")

    train_env = make_env(cfg, mode="train")
    eval_envs = [make_env(cfg, mode="eval", scenario_id=s) for s in cfg.eval_scenarios]
    smoke_test_env(train_env)

    agent = make_agent(cfg, train_env.observation_space, train_env.action_space)
    evaluator = Evaluator(eval_envs, num_episodes=cfg.eval_episodes)
    ckpt = CheckpointManager(run_dir=f"runs/{run_id}")

    save_resolved_config(cfg, f"runs/{run_id}/config.json")
    train(cfg, train_env, agent, logger, evaluator, ckpt)

    final_report = evaluator.evaluate(agent, step=cfg.total_steps)
    logger.log({"final_" + k: v for k, v in final_report.items()}, cfg.total_steps)
    return final_report
\end{lstlisting}

\section{Failure triage}

When a DRL project fails, do not immediately change the algorithm. First identify the failure class.

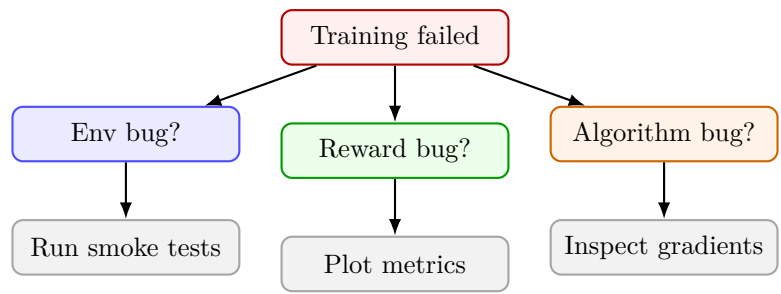
\begin{figure}[t]
    \centering
    \begin{tikzpicture}[
        box/.style={draw,rounded corners,thick,minimum width=3.0cm,minimum height=0.72cm,align=center,font=\small},
        arrow/.style={-{Latex[length=2.2mm]},thick},
        node distance=0.75cm
        ]
        \node[box,fill=red!6,draw=red!70!black] (fail) {Training failed};
        \node[box,fill=blue!8,draw=blue!70,below left=of fail] (env) {Env bug?};
        \node[box,fill=green!8,draw=green!60!black,below=of fail] (reward) {Reward bug?};
        \node[box,fill=orange!10,draw=orange!80!black,below right=of fail] (algo) {Algorithm bug?};
        \node[box,fill=gray!10,draw=gray!70,below=of env] (test) {Run smoke tests};
        \node[box,fill=gray!10,draw=gray!70,below=of reward] (plot) {Plot metrics};
        \node[box,fill=gray!10,draw=gray!70,below=of algo] (grad) {Inspect gradients};
        \draw[arrow] (fail) -- (env);
        \draw[arrow] (fail) -- (reward);
        \draw[arrow] (fail) -- (algo);
        \draw[arrow] (env) -- (test);
        \draw[arrow] (reward) -- (plot);
        \draw[arrow] (algo) -- (grad);
    \end{tikzpicture}
    \caption{Failure triage for DRL systems. Most failures should be diagnosed by environment tests, reward/metric plots, and gradient checks before changing the algorithm.}
    \label{fig:failure_triage}
\end{figure}

\begin{table}[t]
    \centering
    \caption{Common implementation failures and first checks.}
    \label{tab:implementation_failures}
    \begin{tabularx}{\textwidth}{p{3.0cm}X X}
        \toprule
        Symptom & Likely cause & First check \\
        \midrule
        Reward explodes & reward scale or simulator bug & plot raw reward components \\
        Value loss diverges & terminal masks, target scale, learning rate & inspect returns and dones \\
        Policy entropy collapses early & exploration too weak, reward too sharp & entropy curve and action histogram \\
        Safety violations persist & cost not used or projection hidden & check proposed vs executed action logs \\
        High train score, bad eval & overfitting to scenarios or seeds & held-out scenarios and more seeds \\
        No improvement over random & action decoder or reward sign bug & test heuristic and reward sign \\
        Good mean, bad p95 latency & tail risk ignored & report CVaR and violation rate \\
        \bottomrule
    \end{tabularx}
\end{table}

\section{What makes this pipeline research-grade?}

A research-grade pipeline is not defined by using a large neural network. It is defined by the quality of the experimental claim. The claim should be supported by:

\begin{itemize}[leftmargin=*]
    \item a clear environment contract;
    \item strong non-RL and RL baselines;
    \item task-specific metrics and safety constraints;
    \item multiple random seeds and uncertainty intervals;
    \item ablations that isolate each contribution;
    \item open configuration and reproducible code structure;
    \item failure-mode analysis;
    \item deployment-aware validation if the application is real-world.
\end{itemize}

\begin{researchbox}
    For networking, UAV, and SD-WAN systems, the strongest contribution is often not a new neural architecture. It is the integration of learning, safety, telemetry, digital twins, and deployment validation into one reproducible control pipeline.
\end{researchbox}

\section{Limitations and common implementation traps}

\begin{enumerate}[leftmargin=*]
    \item \textbf{A clean pipeline does not guarantee a good problem formulation.} A reproducible experiment can still optimize the wrong objective.
    \item \textbf{Implementation quality cannot rescue weak baselines.} A carefully logged method is not persuasive if it is compared only against weak competitors.
    \item \textbf{Determinism is limited in modern deep learning stacks.} Reproducibility should be judged statistically, not only bitwise.
    \item \textbf{Simulator fidelity remains a bottleneck.} A well-engineered pipeline can still learn unrealistic behaviors if the environment is wrong.
    \item \textbf{Logging too little is dangerous, but logging too much can become noise.} Metrics should remain interpretable and tied to the problem contract.
    \item \textbf{Deployment discipline is domain-specific.} Shadow mode, canaries, and rollback are essential in safety-critical systems but may look different across robotics, networking, and language-agent applications.
\end{enumerate}

\section{Exercises}

\subsection*{Conceptual exercises}
\begin{enumerate}[leftmargin=*]
    \item Explain why a DRL paper should report task metrics in addition to reward.
    \item Why is a production rule-based controller often a stronger baseline than another neural algorithm?
    \item What information should be stored in the \texttt{info} dictionary of a safety-critical environment?
    \item Explain the difference between a proposed action and an executed action.
    \item Why can a policy trained with delayed telemetry fail even if it performs well in a zero-delay simulator?
\end{enumerate}

\subsection*{Implementation exercises}
\begin{enumerate}[leftmargin=*]
    \item Implement the \texttt{SafeNetworkEnv} skeleton and add one deterministic test scenario.
    \item Write a unit test showing that your safety projection always returns a valid traffic split.
    \item Add telemetry age to an existing observation vector and compare training with and without it.
    \item Implement a JSONL logger and reproduce a learning curve from the saved file.
    \item Implement the ablation registry in Listing~\ref{lst:ablation_registry} and generate five configurations.
\end{enumerate}

\subsection*{Research exercises}
\begin{enumerate}[leftmargin=*]
    \item Design a complete experiment matrix for a UAV-assisted network slicing project. Include algorithms, seeds, scenarios, metrics, and ablations.
    \item Choose a DRL paper in networking or robotics. Identify whether it reports enough information to reproduce the result.
    \item Design a shadow-mode validation test for a DRL controller before deployment.
    \item Compare PPO, SAC, and offline RL for a safety-critical network control problem. Which would you start with and why?
    \item Propose a minimal reproducibility checklist for your own DRL project and explain which items are mandatory for publication.
\end{enumerate}

\section*{Looking Ahead to Chapter 23:  Food for Thought}
\addcontentsline{toc}{section}{Looking Ahead to Chapter 23:  Food for Thought}

This chapter built the implementation pipeline: environment, configuration, training, logging, evaluation, ablation, and deployment gates. Chapter~23 goes deeper into experimental methodology. It asks how to turn implementation results into scientific evidence.

The central question becomes:

\begin{quote}
    How do we know that a DRL method truly improves performance, rather than merely winning by luck, weak baselines, poor metrics, or hidden implementation details?
\end{quote}

Chapter~23 will study evaluation protocols, statistical testing, benchmark design, seed analysis, confidence intervals, ablation logic, simulator fidelity, and paper-writing standards.
	\chapter[Experimental Methodology]{Experimental Methodology}
\chaptermark{Experimental Methodology}
\label{ch:experimental_methodology}

\begin{keybox}{Chapter goal}
	This chapter teaches how to evaluate deep reinforcement learning systems at research grade. A DRL paper should not only report reward. It should report task-specific performance, statistical uncertainty, safety violations, sample efficiency, computational cost, failure cases, and robustness under distribution shift. The goal is to help the reader design experiments that can survive peer review, reproduce later, and support real deployment decisions.
\end{keybox}

\section*{Chapter Overview}
\addcontentsline{toc}{section}{Chapter Overview}
\begin{enumerate}[leftmargin=*]
	\item Why experimental methodology matters
	\item The difference between training, validation, and evaluation
	\item Experimental claims and hypotheses
	\item Seeds, stochasticity, and confidence intervals
	\item Robust aggregate metrics: mean, median, IQM, and performance profiles
	\item Baselines: weak, fair, strong, and unfair comparisons
	\item Task-specific metrics beyond reward
	\item Safety and constraint evaluation
	\item Sample efficiency, wall-clock efficiency, and compute accounting
	\item Ablations, sensitivity analysis, and controlled experiments
	\item Stress testing and out-of-distribution evaluation
	\item Statistical tests and paired comparisons
	\item Reproducibility: configs, artifacts, environments, and versioning
	\item Evaluation pipeline in Python
	\item UAV/SD-WAN experimental methodology scenario
	\item Common evaluation failure modes
	\item Reporting checklist
	\item Looking Ahead to Chapter 24
\end{enumerate}

\section{Why experimental methodology matters}

Deep reinforcement learning is unusually easy to mis-evaluate. Two algorithms may appear different because of random seeds, hidden implementation details, reward normalization, early stopping, simulator quirks, or a poorly chosen baseline. A single training curve can look convincing even when the result is statistically fragile. A higher mean reward can hide worse tail latency, more safety violations, higher energy consumption, or lower reliability. In safety-critical networked systems, these hidden failures matter more than a small improvement in average reward.

This chapter is therefore not a collection of administrative advice. It is part of the scientific core of the book. Experimental methodology decides whether a DRL result is knowledge or noise.

DRL has a documented history of fragile evaluation: small numbers of seeds, weak baselines, inconsistent hyperparameter tuning, and cherry-picked learning curves can produce conclusions that do not replicate \citep{henderson2018deep,agarwal2021rliable}. This problem becomes even more serious in applied domains such as SD-WAN, UAV-assisted communication, robotics, medical decision-making, and autonomous driving, where the metric of interest is rarely just discounted reward.

\begin{keybox}{Main principle}
	An experimental result is not a number. It is a claim supported by a protocol. The protocol must specify the task distribution, training budget, baselines, random seeds, evaluation metrics, uncertainty estimates, hyperparameter tuning rules, and failure-analysis procedure.
\end{keybox}

For this book, the running examples are UAV networks, SD-WAN traffic engineering, safe control, and reasoning models. In these domains, the experimental question is usually not:

\begin{quote}
	Which method achieves the highest reward?
\end{quote}

The real question is closer to:

\begin{quote}
	Which method improves task performance under realistic constraints, within a fixed compute and interaction budget, while maintaining reliability, safety, and reproducibility?
\end{quote}

\begin{figure}[t]
	\centering
	\begin{tikzpicture}[
		box/.style={draw,rounded corners,thick,minimum width=2.7cm,minimum height=0.75cm,align=center,font=\small},
		arrow/.style={-{Latex[length=2.2mm]},thick},
		node distance=0.8cm
		]
		\node[box,fill=blue!8,draw=blue!70] (claim) {Research claim};
		\node[box,fill=green!8,draw=green!60!black,right=of claim] (protocol) {Protocol};
		\node[box,fill=orange!10,draw=orange!80!black,right=of protocol] (train) {Training runs};
		\node[box,fill=purple!8,draw=purple!70,right=of train] (eval) {Evaluation};
		\node[box,fill=red!7,draw=red!70!black,right=of eval] (report) {Report};

		\node[box,fill=gray!8,draw=gray!70,below=0.9cm of protocol] (baselines) {Baselines};
		\node[box,fill=gray!8,draw=gray!70,below=0.9cm of train] (seeds) {Seeds + budget};
		\node[box,fill=gray!8,draw=gray!70,below=0.9cm of eval] (metrics) {Metrics + CIs};

		\draw[arrow,draw=blue!70] (claim) -- (protocol);
		\draw[arrow,draw=green!60!black] (protocol) -- (train);
		\draw[arrow,draw=orange!80!black] (train) -- (eval);
		\draw[arrow,draw=purple!70] (eval) -- (report);
		\draw[arrow,draw=gray!70] (baselines) -- (protocol);
		\draw[arrow,draw=gray!70] (seeds) -- (train);
		\draw[arrow,draw=gray!70] (metrics) -- (eval);
	\end{tikzpicture}
	\caption{A research-grade DRL experiment begins with a precise claim and turns that claim into a controlled protocol. Baselines, seeds, budgets, metrics, and uncertainty estimates are part of the method, not afterthoughts.}
	\label{fig:methodology_pipeline}
\end{figure}
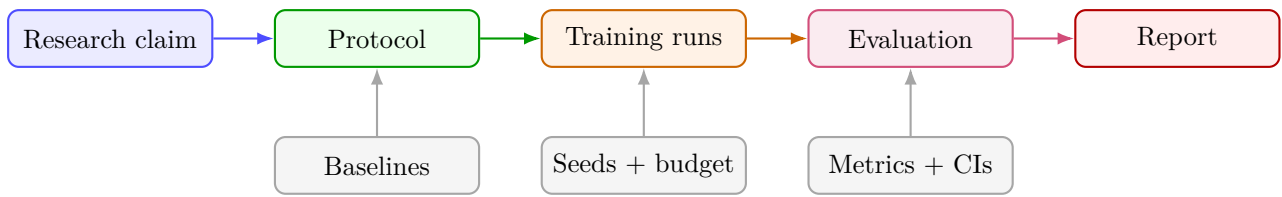

Figure~\ref{fig:methodology_pipeline} summarizes the full evaluation pipeline.

\section{Training, validation, and evaluation are different}

A common mistake is to use the word ``evaluation'' for everything that happens after pressing the training button. A better methodology separates three roles.

\begin{table}[t]
	\centering
	\caption{Training, validation, and final evaluation serve different purposes. Mixing them leads to hidden overfitting.}
	\label{tab:train_val_eval}
	\begin{tabular}{p{3.1cm}p{4.2cm}p{5.6cm}}
		\toprule
		Stage & Purpose & Typical use \\
		\midrule
		Training & Optimize policy parameters & Gradient updates, replay, exploration, curriculum \\
		Validation & Choose design decisions & Hyperparameters, early stopping, reward weights, architecture selection \\
		Final evaluation & Test the frozen method & Held-out scenarios, unseen seeds, stress tests, baselines, confidence intervals \\
		\bottomrule
	\end{tabular}
\end{table}

In supervised learning, validation and test splits are familiar. In DRL, the separation is less obvious because the data distribution is generated by the policy. Still, the same principle holds: do not repeatedly tune on the same final evaluation scenarios and then report them as if they were untouched.

In UAV/SD-WAN research, this means that traffic traces, user-mobility seeds, channel models, topology perturbations, and attack scenarios should be split into design scenarios and final test scenarios. If the method is tuned on one specific rush-hour traffic pattern, then the final evaluation should include different rush-hour patterns, not only the same one with a new random seed.

\begin{warningbox}{Hidden test-set leakage in DRL}
	If a researcher repeatedly changes reward weights, network architecture, or safety thresholds after observing final test performance, then the test scenarios have become validation scenarios. The final results may still be useful engineering evidence, but they are no longer a clean estimate of generalization.
\end{warningbox}

\section{Experimental claims and hypotheses}

Before choosing metrics, define the claim. A vague claim such as ``our algorithm works better'' is not enough. A good experimental claim is falsifiable.

Examples:

\begin{itemize}
	\item \textbf{Sample efficiency claim}: the method reaches 90\% QoS satisfaction using fewer environment steps than PPO.
	\item \textbf{Safety claim}: the method reduces latency-constraint violations by at least 50\% under bursty traffic.
	\item \textbf{Robustness claim}: the method maintains performance when user density increases by 30\%.
	\item \textbf{Scalability claim}: the method degrades sublinearly as the number of UAVs increases from 4 to 16.
	\item \textbf{Deployment claim}: the method passes shadow-mode validation with no statistically significant increase in predicted risk.
\end{itemize}

A strong DRL paper turns each claim into a metric, a baseline, and a stress condition. For example, a safe SD-WAN traffic-engineering paper should not only report average reward. It should report latency, jitter, packet loss, SLA violation rate, safety-filter activation rate, throughput, link utilization, and policy inference time.

\section{Seeds, stochasticity, and confidence intervals}

DRL results vary because of random initialization, environment stochasticity, minibatch sampling, exploration, replay-buffer order, GPU nondeterminism, and evaluation scenario variation. Reporting one seed is not evaluation. It is an anecdote.

Let $X_1,\ldots,X_n$ be the final scores from $n$ independent seeds. The sample mean is
\begin{equation}
	\bar X = \frac{1}{n}\sum_{i=1}^n X_i,
\end{equation}
and the standard error is
\begin{equation}
	\se(\bar X) = \frac{s}{\sqrt{n}},
\end{equation}
where $s$ is the sample standard deviation. A simple normal-approximation confidence interval is
\begin{equation}
	\bar X \pm t_{n-1,0.975}\frac{s}{\sqrt{n}}.
\end{equation}
However, DRL score distributions are often skewed, heavy-tailed, and benchmark-dependent. Bootstrap intervals are often more robust, especially when comparing aggregate metrics across tasks \citep{agarwal2021rliable}. When comparing many algorithm--benchmark pairs, conventional confidence levels become misleading unless the protocol accounts for multiple comparisons. Bootstrap percentile intervals can be Bonferroni-corrected, or false-discovery-rate control can be applied when the study involves many related hypotheses \citep{benjamini1995controlling}.

\begin{figure}[t]
	\centering
	\begin{tikzpicture}
		\begin{axis}[
			width=0.82\textwidth,
			height=5cm,
			xlabel={Environment steps},
			ylabel={QoS satisfaction},
			ymin=0.35,ymax=1.02,
			xmin=0,xmax=10,
			grid=both,
			legend pos=south east,
			tick label style={font=\small},
			label style={font=\small}
			]
			\addplot[blue,thick] coordinates {(0,0.40)(1,0.48)(2,0.57)(3,0.66)(4,0.72)(5,0.78)(6,0.82)(7,0.85)(8,0.87)(9,0.88)(10,0.89)};
			\addlegendentry{seed 1}
			\addplot[blue!60,dashed,thick] coordinates {(0,0.38)(1,0.45)(2,0.53)(3,0.61)(4,0.68)(5,0.74)(6,0.77)(7,0.81)(8,0.84)(9,0.86)(10,0.87)};
			\addlegendentry{seed 2}
			\addplot[blue!40,dotted,thick] coordinates {(0,0.42)(1,0.51)(2,0.58)(3,0.63)(4,0.65)(5,0.67)(6,0.73)(7,0.79)(8,0.83)(9,0.84)(10,0.86)};
			\addlegendentry{seed 3}
			\addplot[red,very thick] coordinates {(0,0.40)(1,0.48)(2,0.56)(3,0.63)(4,0.68)(5,0.73)(6,0.77)(7,0.82)(8,0.85)(9,0.86)(10,0.87)};
			\addlegendentry{mean}
		\end{axis}
	\end{tikzpicture}
	\caption{Seed variance is not noise to be hidden; it is part of the result. A method that wins only on one favorable seed is not robust evidence.}
	\label{fig:seed_variance}
\end{figure}
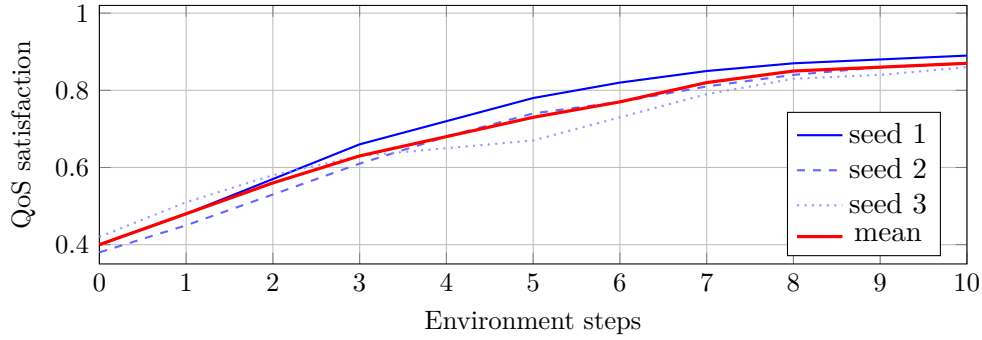

\Needspace{18\baselineskip}
\begin{lstlisting}[style=pythonstyle,caption={Bootstrap confidence interval for a scalar metric.},label={lst:bootstrap_ci}]
import numpy as np

def bootstrap_ci(values, statistic=np.mean, num_bootstrap=10000,
                 alpha=0.05, seed=0):
    """Return statistic estimate and percentile bootstrap confidence interval.

    Args:
        values: array-like, e.g. final score from each seed.
        statistic: function applied to bootstrap sample.
        num_bootstrap: number of resamples.
        alpha: 0.05 gives a 95% interval.
    """
    values = np.asarray(values, dtype=np.float64)
    rng = np.random.default_rng(seed)
    estimates = []

    for _ in range(num_bootstrap):
        sample = rng.choice(values, size=len(values), replace=True)
        estimates.append(statistic(sample))

    lo = np.percentile(estimates, 100 * alpha / 2)
    hi = np.percentile(estimates, 100 * (1 - alpha / 2))
    return statistic(values), lo, hi

scores = np.array([0.86, 0.89, 0.83, 0.91, 0.87])
mean, lo, hi = bootstrap_ci(scores)
print(f"mean={mean:.3f}, 95% CI=[{lo:.3f}, {hi:.3f}]")
\end{lstlisting}

\section{Robust aggregate metrics: mean, median, IQM, and performance profiles}

When evaluating across multiple tasks, the mean can be dominated by outliers. The median ignores magnitude. Agarwal et al. recommend more robust aggregate metrics and uncertainty estimates, including the interquartile mean and performance profiles \citep{agarwal2021rliable}.

The interquartile mean removes the bottom and top quartiles and averages the middle 50\%:
\begin{equation}
	\IQM(X) = \frac{1}{|M|}\sum_{x_i\in M} x_i,
\end{equation}
where $M$ contains the middle half of the sorted scores. IQM is more robust than the mean while using more information than the median.

A performance profile asks: for a threshold $\tau$, what fraction of normalized scores exceed $\tau$?
\begin{equation}
	P(\tau) = \frac{1}{N}\sum_{i=1}^{N}\mathbf{1}[x_i \geq \tau].
\end{equation}
This curve is useful because it reveals whether an algorithm is broadly competent or only excels on a small subset of tasks. Another useful robust comparison is the probability of improvement,
\begin{equation}
	P(A>B)=\Pr[X^A>X^B],
\end{equation}
which estimates the probability that a randomly selected run or task result from method $A$ exceeds one from method $B$ \citep{agarwal2021rliable}. This is often easier to interpret than a difference in means when score distributions are skewed.

\begin{figure}[t]
	\centering
	\begin{tikzpicture}
		\begin{axis}[
			width=0.78\textwidth,
			height=5.2cm,
			xlabel={Normalized score threshold $\tau$},
			ylabel={Fraction of runs above threshold},
			xmin=0,xmax=1.2,
			ymin=0,ymax=1.05,
			grid=both,
			legend pos=north east,
			tick label style={font=\small},
			label style={font=\small}
			]
			\addplot[blue,very thick] coordinates {(0,1.0)(0.2,0.96)(0.4,0.90)(0.6,0.78)(0.8,0.55)(1.0,0.30)(1.2,0.10)};
			\addlegendentry{Method A}
			\addplot[red,dashed,very thick] coordinates {(0,1.0)(0.2,0.94)(0.4,0.82)(0.6,0.62)(0.8,0.40)(1.0,0.18)(1.2,0.05)};
			\addlegendentry{Method B}
		\end{axis}
	\end{tikzpicture}
	\caption{A performance profile shows the fraction of runs or tasks above each normalized score threshold. It is more informative than reporting only a mean score.}
	\label{fig:performance_profile}
\end{figure}
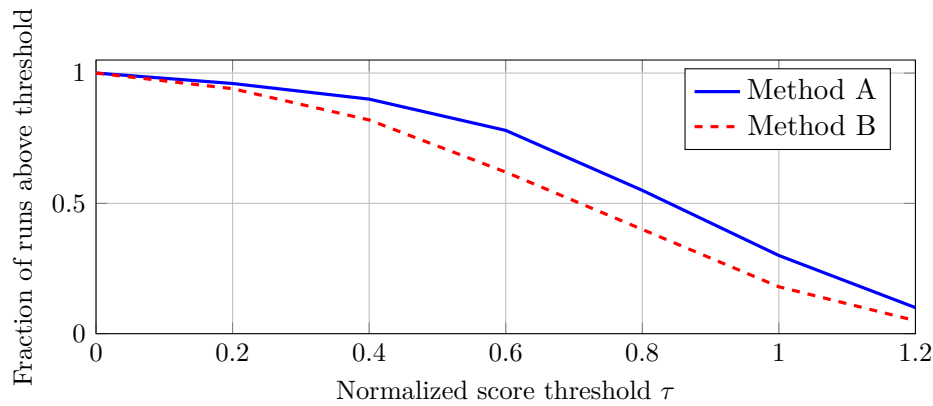

\Needspace{18\baselineskip}
\begin{lstlisting}[style=pythonstyle,caption={Interquartile mean and performance profile utilities.},label={lst:iqm_profile}]
import numpy as np

def interquartile_mean(scores):
    """Compute IQM over a flat array of normalized scores."""
    scores = np.sort(np.asarray(scores, dtype=np.float64))
    n = len(scores)
    lo = int(np.floor(0.25 * n))
    hi = int(np.ceil(0.75 * n))
    return scores[lo:hi].mean()

def performance_profile(scores, thresholds):
    """Fraction of scores above each threshold."""
    scores = np.asarray(scores, dtype=np.float64)
    return np.array([(scores >= tau).mean() for tau in thresholds])

normalized_scores = np.array([0.4, 0.9, 1.1, 0.7, 0.8, 0.2, 1.0, 0.6])
thresholds = np.linspace(0.0, 1.2, 7)
print("IQM:", interquartile_mean(normalized_scores))
print("Profile:", performance_profile(normalized_scores, thresholds))
\end{lstlisting}

\section{Baselines: weak, fair, strong, and unfair comparisons}

A baseline is not just a competing algorithm. It is an experimental control. A weak baseline can make a mediocre method look strong. An unfair baseline can make a good method look weak.

\begin{table}[t]
	\centering
	\caption{Baseline ladder for DRL experiments. A serious paper should include more than one layer.}
	\label{tab:baseline_ladder2}
	\begin{tabular}{p{3.0cm}p{4.7cm}p{5.2cm}}
		\toprule
		Baseline type & Purpose & Example in UAV/SD-WAN control \\
		\midrule
		Rule-based & Shows value over engineering heuristics & shortest-path routing, load threshold rule, nearest-UAV association \\
		Classical optimization & Tests against known control tools & MPC, convex flow allocation, proportional fairness \\
		Simple learning & Controls for learning vs. architecture & behavior cloning, independent DQN, vanilla PPO \\
		Strong DRL & Tests algorithmic novelty & PPO, SAC, TD3, QMIX, MAPPO, CQL \\
		Oracle / upper bound & Contextualizes achievable performance & clairvoyant traffic trace, perfect channel model, relaxed constraint solver \\
		\bottomrule
	\end{tabular}
\end{table}

A baseline must be tuned with comparable care. If the proposed method receives extensive tuning and the baseline uses default hyperparameters, the comparison is biased. A strong baseline is a baseline tuned with at least as much care and compute as the proposed method; many apparent algorithmic improvements shrink or disappear when baselines are properly tuned \citep{andrychowicz2021what}. The budget for tuning should be stated. For example:

\begin{quote}
	All algorithms were tuned using 30 validation runs across three traffic scenarios. Final results use fixed hyperparameters on five held-out traffic seeds.
\end{quote}

\begin{warningbox}{The wrong baseline can invalidate the conclusion}
	If a new safe RL method beats unconstrained PPO on safety violations, that is expected. A stronger comparison includes PPO-Lagrangian, CPO, a CBF shield, a rule-based safety controller, and a conservative offline baseline. The baseline set should match the claim.
\end{warningbox}

\section{Hyperparameter tuning and compute fairness}

Hyperparameter tuning is part of the method. If one algorithm receives a large tuning budget and another receives only default parameters, the comparison measures tuning effort as much as algorithm quality. Henderson et al. emphasized that hyperparameters, network architectures, and implementation details can strongly affect DRL conclusions \citep{henderson2018deep}. Engstrom et al. showed that code-level implementation details can account for a substantial fraction of apparent performance gains in policy-gradient methods, and Huang et al. catalogued many PPO details that materially change results \citep{engstrom2020implementation,huang2022details}. The fair protocol is not necessarily that every algorithm uses the same hyperparameters; it is that every algorithm receives a comparable opportunity to perform well.

A good experimental report should state:
\begin{itemize}
	\item the search space for each algorithm;
	\item the number of validation trials;
	\item which metric selected the final hyperparameters;
	\item whether safety constraints were part of model selection;
	\item whether the same validation scenarios were used for all methods;
	\item whether the reported final evaluation reuses validation runs.
\end{itemize}

For safe RL, the model-selection metric should not be reward alone. A model with slightly lower reward but dramatically fewer violations may be the better deployed system. One practical selection score is
\begin{equation}
	S_{\mathrm{select}}
	=
	\bar R
	-
	\eta_1 \max(0, \bar C - d)
	-
	\eta_2 \mathrm{ViolationRate}
	-
	\eta_3 \mathrm{InferenceLatency},
\end{equation}
where $\bar R$ is validation reward, $\bar C$ is validation cost, and $d$ is the allowed cost budget. The exact form is less important than the principle: tune for the real claim.

\Needspace{20\baselineskip}
\begin{lstlisting}[style=pythonstyle,caption={Hyperparameter selection with reward and safety constraints.},label={lst:model_selection}]
def select_safe_model(validation_runs, cost_limit, max_latency_ms):
    """Select a model from validation summaries.

    Each item in validation_runs is a dict with keys:
    reward_mean, cost_mean, violation_rate, inference_ms, checkpoint.
    """
    best = None
    best_score = float("-inf")

    for run in validation_runs:
        cost_excess = max(0.0, run["cost_mean"] - cost_limit)
        latency_excess = max(0.0, run["inference_ms"] - max_latency_ms)
        score = (
            run["reward_mean"]
            - 10.0 * cost_excess
            - 5.0 * run["violation_rate"]
            - 0.1 * latency_excess
        )
        if score > best_score:
            best_score = score
            best = run
    return best["checkpoint"], best_score
\end{lstlisting}

\section{Learning curves, AUC, and time-to-threshold}

Final performance is not the only quantity that matters. Two methods may reach the same final score, but one may require ten times more interaction. Another may learn quickly but plateau early. Therefore, report learning curves and summarize them with sample-efficiency metrics.

A common metric is area under the learning curve:
\begin{equation}
	\mathrm{AUC}
	=
	\sum_{k=1}^{K-1}
	\frac{m_{k}+m_{k+1}}{2}
	(t_{k+1}-t_k),
\end{equation}
where $m_k$ is the evaluation metric at training step $t_k$. Another useful metric is time-to-threshold:
\begin{equation}
	T_\tau = \min\{t_k: m_k \ge \tau\}.
\end{equation}
For safe RL, the threshold should include safety: for example, first time QoS satisfaction exceeds 0.85 while violation rate stays below 0.01.

\begin{figure}[t]
	\centering
	\begin{tikzpicture}
		\begin{axis}[
			width=0.82\textwidth,
			height=5.0cm,
			xlabel={Training steps},
			ylabel={Evaluation score},
			xmin=0,xmax=10,
			ymin=0.3,ymax=1.0,
			grid=both,
			legend pos=south east,
			tick label style={font=\small},
			label style={font=\small}
			]
			\addplot[blue,thick] coordinates {(0,0.35)(1,0.48)(2,0.60)(3,0.70)(4,0.78)(5,0.82)(6,0.84)(7,0.85)(8,0.86)(9,0.86)(10,0.87)};
			\addlegendentry{Fast stable learner}
			\addplot[red,dashed,thick] coordinates {(0,0.35)(1,0.42)(2,0.48)(3,0.55)(4,0.64)(5,0.72)(6,0.82)(7,0.88)(8,0.90)(9,0.91)(10,0.92)};
			\addlegendentry{Slow high final score}
			\draw[densely dotted,thick] (axis cs:0,0.85) -- (axis cs:10,0.85);
			\node[anchor=west,font=\scriptsize] at (axis cs:0.2,0.88) {threshold};
		\end{axis}
	\end{tikzpicture}
	\caption{Learning curves answer a different question from final performance. A method can learn quickly but plateau, or learn slowly and finish higher. Both behaviors should be reported.}
	\label{fig:auc_time_to_threshold}
\end{figure}
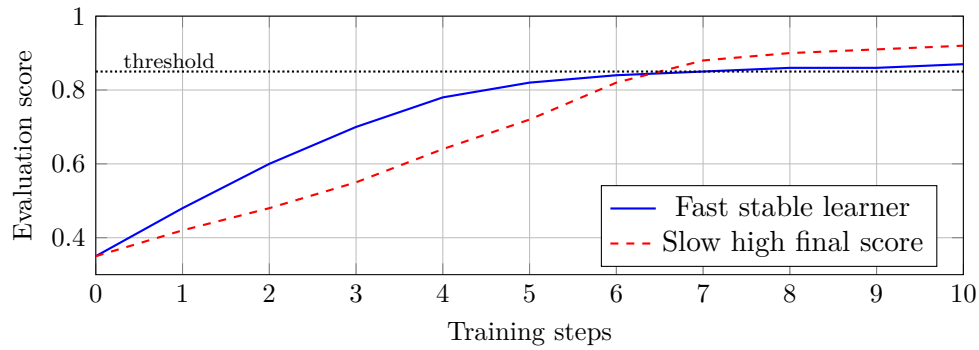

\Needspace{18\baselineskip}
\begin{lstlisting}[style=pythonstyle,caption={AUC and time-to-threshold for learning curves.},label={lst:auc_threshold}]
import numpy as np

def area_under_curve(steps, values):
    steps = np.asarray(steps, dtype=float)
    values = np.asarray(values, dtype=float)
    return float(np.trapz(values, steps))

def time_to_threshold(steps, values, threshold):
    for step, value in zip(steps, values):
        if value >= threshold:
            return float(step)
    return float("inf")

steps = np.array([0, 1e5, 2e5, 3e5, 4e5])
qos = np.array([0.42, 0.61, 0.77, 0.84, 0.87])
print("AUC:", area_under_curve(steps, qos))
print("T_0.85:", time_to_threshold(steps, qos, 0.85))
\end{lstlisting}

\section{Task-specific metrics beyond reward}

Reward is an optimization signal. It is not automatically a scientific metric. In applied DRL, report the real system quantities that motivated the reward.

For networking and UAV systems, useful metrics include:
\begin{itemize}
	\item average latency, tail latency, and jitter;
	\item packet loss and outage probability;
	\item SINR and throughput;
	\item per-class QoS satisfaction;
	\item fairness across users or traffic classes;
	\item energy consumption and battery lifetime;
	\item safety-filter activation rate;
	\item constraint violation magnitude and duration;
	\item convergence speed and sample efficiency;
	\item inference time and control-loop latency.
\end{itemize}

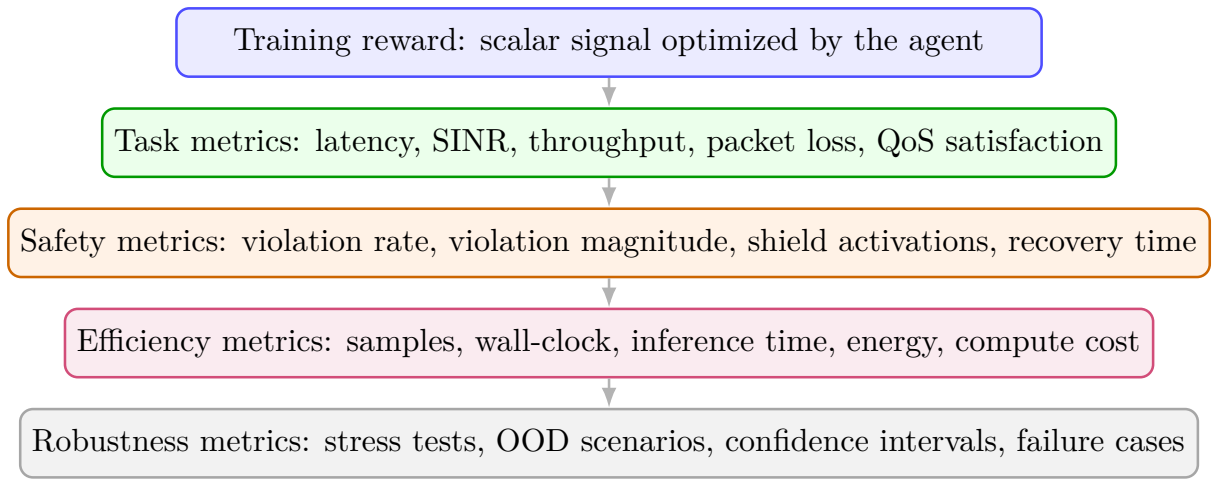
\begin{figure}[t]
	\centering
	\resizebox{\columnwidth}{!}{%
		\begin{tikzpicture}[
			box/.style={draw, rounded corners, thick,
				minimum width=9.5cm, minimum height=0.75cm,
				align=center, font=\small},
			arrow/.style={-{Latex[length=2.2mm]}, thick}
			]
			\node[box, fill=blue!8,    draw=blue!70]         (reward) at (0,0)    {Training reward: scalar signal optimized by the agent};
			\node[box, fill=green!8,   draw=green!60!black]  (task)   at (0,-1.1) {Task metrics: latency, SINR, throughput, packet loss, QoS satisfaction};
			\node[box, fill=orange!10, draw=orange!80!black] (safety) at (0,-2.2) {Safety metrics: violation rate, violation magnitude, shield activations, recovery time};
			\node[box, fill=purple!8,  draw=purple!70]       (eff)    at (0,-3.3) {Efficiency metrics: samples, wall-clock, inference time, energy, compute cost};
			\node[box, fill=gray!10,   draw=gray!70]         (rob)    at (0,-4.4) {Robustness metrics: stress tests, OOD scenarios, confidence intervals, failure cases};

			\draw[arrow, draw=gray!60] (reward) -- (task);
			\draw[arrow, draw=gray!60] (task)   -- (safety);
			\draw[arrow, draw=gray!60] (safety) -- (eff);
			\draw[arrow, draw=gray!60] (eff)    -- (rob);
		\end{tikzpicture}%
	}
	\caption{A serious DRL evaluation separates the reward used for training from the metrics used to judge the system. The lower layers often matter more for deployment than the scalar reward.}
	\label{fig:metric_stack}
\end{figure}

Figure~\ref{fig:metric_stack} shows a metric stack that separates optimization reward from scientific evaluation. Observation normalization is one of the most impactful implementation details: PPO with and without running mean--variance normalization can produce qualitatively different policies even on the same environment \citep{engstrom2020implementation,huang2022details}. Therefore, reports should state whether observations, rewards, returns, advantages, and costs were normalized, and whether normalization statistics were frozen during evaluation.

\Needspace{16\baselineskip}
\begin{lstlisting}[style=pythonstyle,caption={Task-specific network metrics from logged episode data.},label={lst:task_metrics}]
def compute_network_metrics(log):
    """Compute task-specific metrics from one evaluation episode.

    log is a list of dictionaries containing latency_ms, jitter_ms,
    packet_loss, throughput_mbps, sinr_db, battery, and violation.
    """
    import numpy as np

    latency = np.array([x["latency_ms"] for x in log], dtype=float)
    jitter = np.array([x["jitter_ms"] for x in log], dtype=float)
    loss = np.array([x["packet_loss"] for x in log], dtype=float)
    thr = np.array([x["throughput_mbps"] for x in log], dtype=float)
    sinr = np.array([x["sinr_db"] for x in log], dtype=float)
    battery = np.array([x["battery"] for x in log], dtype=float)
    violation = np.array([x["violation"] for x in log], dtype=float)

    return {
        "latency_mean": latency.mean(),
        "latency_p95": np.percentile(latency, 95),
        "jitter_mean": jitter.mean(),
        "packet_loss_mean": loss.mean(),
        "throughput_mean": thr.mean(),
        "sinr_mean": sinr.mean(),
        "battery_final": battery[-1],
        "violation_rate": (violation > 0).mean(),
        "violation_magnitude": violation.sum(),
    }
\end{lstlisting}

\section{Per-class metrics and fairness}

Aggregate QoS can hide who benefits. In network slicing and UAV service, performance should be reported per traffic class. A controller that improves average throughput by sacrificing URLLC latency is not a good network controller.

For classes $k\in\{1,\ldots,K\}$, define per-class QoS satisfaction
\begin{equation}
	Q_k = \frac{1}{N_k}\sum_{i\in \mathcal{C}_k}\mathbf{1}[\mathrm{SLA}_i\ \mathrm{satisfied}],
\end{equation}
and report both the vector $(Q_1,\ldots,Q_K)$ and the worst-class score
\begin{equation}
	Q_{\min}=\min_k Q_k.
\end{equation}
A fairness-aware method should improve $Q_{\min}$, not only the average. The same methodological principle applies to language-model RLHF: annotator demographic bias and preference ambiguity, discussed in Chapter~19, can propagate into the trained policy, so per-group or per-demographic evaluation is the alignment analogue of per-class QoS evaluation. Jain's fairness index is another useful summary:
\begin{equation}
	\mathcal{J}(x_1,\ldots,x_K)
	=
	\frac{(\sum_k x_k)^2}{K\sum_k x_k^2}.
\end{equation}

\Needspace{16\baselineskip}
\begin{lstlisting}[style=pythonstyle,caption={Per-class QoS and Jain fairness index.},label={lst:fairness_metrics}]
import numpy as np

def jain_fairness(values):
    values = np.asarray(values, dtype=float)
    denom = len(values) * np.sum(values ** 2)
    if denom <= 0:
        return 0.0
    return float((np.sum(values) ** 2) / denom)

def per_class_qos(records, classes=("URLLC", "eMBB", "mMTC")):
    out = {}
    scores = []
    for cls in classes:
        xs = [r["sla_satisfied"] for r in records if r["class"] == cls]
        score = float(np.mean(xs)) if xs else float("nan")
        out[f"qos_{cls}"] = score
        scores.append(score)
    out["qos_worst_class"] = float(np.nanmin(scores))
    out["jain_fairness"] = jain_fairness(np.nan_to_num(scores, nan=0.0))
    return out
\end{lstlisting}

\section{Safety and constraint evaluation}

Safe RL results must report both reward and constraint behavior. A policy can have high reward and still be unacceptable if it violates hard constraints.

Let $c_t\ge 0$ be a cost or violation signal. Common safety metrics include:
\begin{align}
	J_C(\pi) &= \E_\pi\left[\sum_{t=0}^{T-1}\gamma^t c_t\right],\\
	\mathrm{ViolationRate} &= \frac{1}{T}\sum_{t=0}^{T-1}\mathbf{1}[c_t>0],\\
	\mathrm{MaxViolation} &= \max_t c_t.
\end{align}

For latency-sensitive systems, tail risk is often more important than mean performance. A tail-latency metric can be written as
\begin{equation}
	\CVaR_{\rho}(L) = \E[L \given L \ge q_{\rho}(L)],
\end{equation}
where $q_\rho(L)$ is the $\rho$-quantile of latency. Chapter~6 introduced tail-risk reasoning for distributional RL, and Chapter~18 formalized safety constraints. Here the point is methodological: if the real system fails in the tail, report tail metrics. For safety-critical systems, time-to-threshold should also be defined as a sustained criterion, not a transient peak: for example, the first evaluation point at which the policy remains above $95\%$ QoS satisfaction and below the violation threshold for $N$ consecutive evaluations.

\Needspace{16\baselineskip}
\begin{lstlisting}[style=pythonstyle,caption={Tail risk and CVaR-style safety metrics.},label={lst:cvar_metrics}]
import numpy as np

def cvar_upper(values, rho=0.95):
    """Mean of the worst (1-rho) tail for a cost-like variable."""
    values = np.asarray(values, dtype=np.float64)
    q = np.quantile(values, rho)
    tail = values[values >= q]
    return float(tail.mean()) if len(tail) else float(q)

latencies = np.array([5, 6, 7, 8, 9, 15, 20, 45, 60, 90])
print("p95 latency:", np.percentile(latencies, 95))
print("CVaR95 latency:", cvar_upper(latencies, rho=0.95))
\end{lstlisting}

\section{Sample efficiency, wall-clock efficiency, and compute accounting}

A method may use fewer environment steps but more wall-clock time. Another may train quickly but require expensive inference. A third may look sample-efficient because the simulator is simplified. Report all relevant budgets.

\begin{table}[t]
	\centering
	\caption{Efficiency metrics answer different questions.}
	\label{tab:efficiency_metrics}
	\begin{tabular}{p{3.5cm}p{5.2cm}p{4.4cm}}
		\toprule
		Metric & Question answered & Example \\
		\midrule
		Environment steps & How much interaction was needed? & $2\times 10^6$ simulator steps \\
		Episodes & How many complete rollouts? & 500 UAV missions \\
		Wall-clock time & How long did training take? & 12 GPU hours \\
		Inference latency & Can it run in the control loop? & 4 ms per action \\
		Memory footprint & Can it deploy on edge hardware? & 35 MB actor network \\
		Energy/compute cost & What is the training footprint? & GPU-hours, CPU-hours, cloud cost \\
		\bottomrule
	\end{tabular}
\end{table}

For network control, the decision period is central. A policy with 100 ms inference time may be acceptable for SD-WAN route rebalancing every 5 seconds but unacceptable for MAC scheduling at sub-millisecond timescales. Chapter~21 emphasized that control-loop timing is part of the system design; the same timing must appear in experimental reporting.

On-policy and off-policy methods also have different data-accounting rules. PPO uses a rollout buffer that is collected under the current policy and reused for only a few epochs; SAC, TD3, and DQN-family methods use replay buffers that accumulate old transitions continuously. Therefore, reporting only ``number of gradient updates'' is not enough: state the number of environment steps, replay ratio, rollout length, minibatch size, epochs per batch, and whether old data were reused.

\section{Ablations, sensitivity analysis, and controlled experiments}

An ablation is a controlled experiment. It isolates one component while holding the rest fixed, so the effect of that component can be attributed rather than guessed. Without ablations, a complex method may appear to work for the wrong reason.

Examples:
\begin{itemize}
	\item remove the CBF safety layer;
	\item replace GAE with one-step TD advantage;
	\item remove graph attention from a MARL mixer;
	\item remove uncertainty penalty from model-based planning;
	\item replace distributional critic with expected-value critic;
	\item disable reward normalization or observation normalization.
\end{itemize}

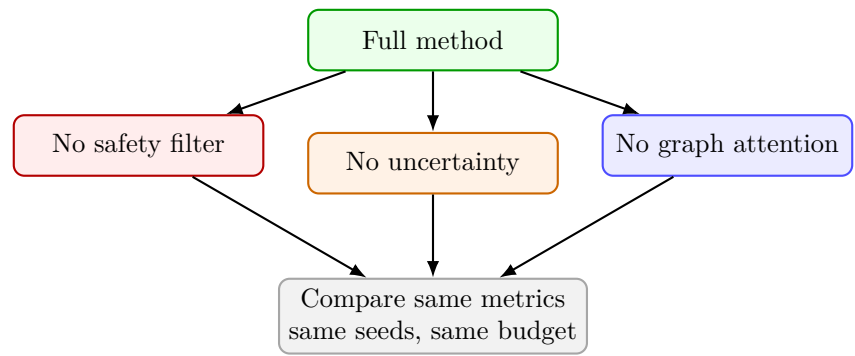
\begin{figure}[t]
	\centering
	\begin{tikzpicture}[
		box/.style={draw,rounded corners,thick,minimum width=3.3cm,minimum height=0.8cm,align=center,font=\small},
		arrow/.style={-{Latex[length=2.2mm]},thick},
		node distance=0.8cm
		]
		\node[box,fill=green!8,draw=green!60!black] (full) {Full method};
		\node[box,fill=red!7,draw=red!70!black,below left=of full] (no1) {No safety filter};
		\node[box,fill=orange!10,draw=orange!80!black,below=of full] (no2) {No uncertainty};
		\node[box,fill=blue!8,draw=blue!70,below right=of full] (no3) {No graph attention};
		\node[box,fill=gray!10,draw=gray!70,below=1.1cm of no2] (metric) {Compare same metrics\\same seeds, same budget};
		\draw[arrow] (full) -- (no1);
		\draw[arrow] (full) -- (no2);
		\draw[arrow] (full) -- (no3);
		\draw[arrow] (no1) -- (metric);
		\draw[arrow] (no2) -- (metric);
		\draw[arrow] (no3) -- (metric);
	\end{tikzpicture}
	\caption{An ablation is a controlled experiment. Remove one component at a time and evaluate under the same budget, seeds, and metrics.}
	\label{fig:ablation_design2}
\end{figure}

Sensitivity analysis changes a parameter rather than removing a component. For example, vary the CBF margin, entropy coefficient, reward weight, model rollout horizon, or communication-delay distribution. Sensitivity plots reveal whether the method is robust or delicately tuned.

\Needspace{18\baselineskip}
\begin{lstlisting}[style=pythonstyle,caption={A small ablation registry for controlled experiments.},label={lst:ablation_registry2}]
from dataclasses import dataclass, replace

@dataclass(frozen=True)
class ExpConfig:
    algorithm: str = "ppo_cbf"
    use_cbf: bool = True
    use_uncertainty: bool = True
    use_graph_attention: bool = True
    reward_qos_weight: float = 1.0
    seed: int = 0

def make_ablations(base: ExpConfig):
    return {
        "full": base,
        "no_cbf": replace(base, use_cbf=False),
        "no_uncertainty": replace(base, use_uncertainty=False),
        "no_graph_attention": replace(base, use_graph_attention=False),
        "low_qos_weight": replace(base, reward_qos_weight=0.5),
        "high_qos_weight": replace(base, reward_qos_weight=2.0),
    }

base = ExpConfig(seed=42)
for name, cfg in make_ablations(base).items():
    print(name, cfg)
\end{lstlisting}

\section{Simulator fidelity and digital-twin calibration}

A DRL policy is only as credible as the environment used to evaluate it. In simulation-heavy fields, simulator fidelity is an experimental variable. A network simulator may have realistic queues but simplified wireless fading; a UAV simulator may model mobility well but ignore battery aging; a digital twin may match average traffic but fail on rare bursts.

A calibration report should compare simulated and real or high-fidelity traces on key statistics:
\begin{itemize}
	\item traffic-load distribution;
	\item latency and jitter distribution;
	\item packet-loss bursts;
	\item channel/SINR distribution;
	\item mobility and user-density distribution;
	\item link-failure and recovery statistics.
\end{itemize}
Quantitative calibration can include KL divergence or Wasserstein distance between simulated and real telemetry distributions, prediction error of a digital twin on held-out real trajectories, and tail-specific errors such as p95 latency gap. These diagnostics make the sim-to-real gap visible rather than anecdotal.

\begin{warningbox}{Simulator accuracy is metric-specific}
	A simulator can be accurate for average throughput but inaccurate for tail latency. If the paper claims safety or reliability, validate the simulator on tail behavior, not only average behavior.
\end{warningbox}

\Needspace{18\baselineskip}
\begin{lstlisting}[style=pythonstyle,caption={Simple simulator calibration diagnostics.},label={lst:calibration}]
import numpy as np

def distribution_gap(real, sim):
    """A lightweight diagnostic: compare mean, std, and p95."""
    real = np.asarray(real, dtype=float)
    sim = np.asarray(sim, dtype=float)
    return {
        "mean_gap": float(sim.mean() - real.mean()),
        "std_gap": float(sim.std() - real.std()),
        "p95_gap": float(np.percentile(sim, 95) - np.percentile(real, 95)),
    }

real_latency = [10, 12, 13, 14, 18, 25, 50]
sim_latency = [9, 11, 12, 15, 17, 20, 32]
print(distribution_gap(real_latency, sim_latency))
\end{lstlisting}

\section{Stress testing and out-of-distribution evaluation}

A policy trained under normal traffic may fail under abnormal traffic. A UAV controller trained with moderate user mobility may fail during a crowded event. A safe RL policy tuned for one latency threshold may fail when the threshold tightens.

Stress tests should include:
\begin{itemize}
	\item traffic bursts and diurnal traffic shifts;
	\item link failures and topology changes;
	\item user-density spikes;
	\item sensor noise and telemetry delay;
	\item channel-model mismatch;
	\item battery degradation;
	\item adversarial or anomalous traffic;
	\item unseen mobility patterns.
\end{itemize}

\begin{researchbox}{Network/UAV methodology}
	A credible UAV/SD-WAN DRL result should include at least one normal scenario, one high-load scenario, one failure scenario, one safety-stress scenario, and one distribution-shift scenario. Otherwise, the method may only be solving the simulator's easiest regime. Adversarial robustness can be evaluated with observation perturbations, traffic-injection attacks, link-delay perturbations, or action-channel disturbances, depending on the threat model; the perturbation budget should be stated explicitly rather than hidden inside a vague ``noisy environment'' label.
\end{researchbox}

\Needspace{18\baselineskip}
\begin{lstlisting}[style=pythonstyle,caption={Stress-scenario generator for network/UAV evaluation.},label={lst:stress_generator}]
from dataclasses import dataclass

@dataclass
class Scenario:
    name: str
    traffic_scale: float = 1.0
    user_density_scale: float = 1.0
    link_failure_prob: float = 0.0
    telemetry_delay_ms: float = 0.0
    channel_noise_db: float = 0.0
    attack_intensity: float = 0.0

def scenario_suite():
    return [
        Scenario("normal"),
        Scenario("high_load", traffic_scale=1.5, user_density_scale=1.3),
        Scenario("link_failure", link_failure_prob=0.10),
        Scenario("stale_telemetry", telemetry_delay_ms=200.0),
        Scenario("noisy_channel", channel_noise_db=4.0),
        Scenario("cyber_anomaly", attack_intensity=0.25),
    ]
\end{lstlisting}

\section{Statistical tests and paired comparisons}

When comparing two algorithms across the same seeds or tasks, use paired comparisons when possible. If algorithm A and B are evaluated on the same scenario seeds, compute paired differences:
\begin{equation}
	D_i = X_i^{A} - X_i^{B}.
\end{equation}
Then estimate a confidence interval for $\E[D]$. This reduces variance compared with comparing two independent groups.

A simple paired bootstrap resamples the paired differences.

\Needspace{18\baselineskip}
\begin{lstlisting}[style=pythonstyle,caption={Paired bootstrap confidence interval for algorithm comparison.},label={lst:paired_bootstrap}]
import numpy as np

def paired_bootstrap_ci(scores_a, scores_b, statistic=np.mean,
                        num_bootstrap=10000, alpha=0.05, seed=0):
    """CI for statistic(A-B) using paired resampling."""
    scores_a = np.asarray(scores_a, dtype=float)
    scores_b = np.asarray(scores_b, dtype=float)
    assert scores_a.shape == scores_b.shape

    rng = np.random.default_rng(seed)
    n = len(scores_a)
    estimates = []

    for _ in range(num_bootstrap):
        idx = rng.integers(0, n, size=n)
        estimates.append(statistic(scores_a[idx] - scores_b[idx]))

    estimate = statistic(scores_a - scores_b)
    lo = np.percentile(estimates, 100 * alpha / 2)
    hi = np.percentile(estimates, 100 * (1 - alpha / 2))
    return estimate, lo, hi

ppo = np.array([0.82, 0.80, 0.85, 0.79, 0.83])
safe_ppo = np.array([0.84, 0.83, 0.86, 0.82, 0.85])
print(paired_bootstrap_ci(safe_ppo, ppo))
\end{lstlisting}

Statistical testing should not replace scientific interpretation. A tiny statistically significant improvement may be practically irrelevant. A non-significant result with wide intervals may indicate insufficient compute rather than equality. Report intervals and effect sizes, not only $p$-values.

\section{Reproducibility: configs, artifacts, environments, and versioning}

Reproducibility is not achieved by adding a random seed in the appendix. A reproducible DRL experiment records:

\begin{itemize}
	\item code version and commit hash;
	\item environment version;
	\item simulator version and scenario files;
	\item configuration file and overrides;
	\item random seeds;
	\item hyperparameter search space;
	\item model checkpoints;
	\item raw logs;
	\item evaluation scripts;
	\item hardware and software stack;
	\item dataset manifest for offline experiments.
\end{itemize}

Modern RL tooling such as Gymnasium, CleanRL, Stable-Baselines3, RLlib, Tianshou, Minari, and rliable can help standardize parts of this workflow, but tools do not remove methodological responsibility \citep{towers2024gymnasium,raffin2021stable,huang2022cleanrl,agarwal2021rliable,minari_repo,rliable_repo}. The project structure introduced in Chapter~22 supports this manifest naturally: each run directory should contain the configuration, manifest, logs, checkpoints, environment metadata, and evaluation outputs. Community checklists such as the ML Reproducibility Checklist also provide useful reminders for artifact reporting \citep{pineau2021improving}.

\Needspace{18\baselineskip}
\begin{lstlisting}[style=pythonstyle,caption={Minimal experiment manifest for reproducibility.},label={lst:manifest}]
import json
import platform
import subprocess
from dataclasses import asdict, dataclass
from datetime import datetime

@dataclass
class Manifest:
    experiment_name: str
    config: dict
    seed: int
    git_commit: str
    python_version: str
    timestamp: str

def get_git_commit():
    try:
        return subprocess.check_output(
            ["git", "rev-parse", "HEAD"], text=True
        ).strip()
    except Exception:
        return "unknown"

def save_manifest(path, experiment_name, config, seed):
    manifest = Manifest(
        experiment_name=experiment_name,
        config=config,
        seed=seed,
        git_commit=get_git_commit(),
        python_version=platform.python_version(),
        timestamp=datetime.utcnow().isoformat() + "Z",
    )
    with open(path, "w", encoding="utf-8") as f:
        json.dump(asdict(manifest), f, indent=2)
\end{lstlisting}

\section{Evaluation pipeline in Python}

A clean evaluation pipeline should freeze the policy, disable exploration unless explicitly evaluating stochastic behavior, run multiple episodes per scenario and seed, record raw episode logs, aggregate metrics, and compute confidence intervals.

\Needspace{24\baselineskip}
\begin{lstlisting}[style=pythonstyle,caption={Research-grade evaluation loop skeleton.},label={lst:evaluation_loop}]
import numpy as np

class Evaluator:
    def __init__(self, make_env, policy, scenarios, seeds, episodes_per_seed=5):
        self.make_env = make_env
        self.policy = policy
        self.scenarios = scenarios
        self.seeds = seeds
        self.episodes_per_seed = episodes_per_seed

    def evaluate(self):
        records = []
        self.policy.eval_mode()  # disable dropout/exploration if applicable

        for scenario in self.scenarios:
            for seed in self.seeds:
                env = self.make_env(scenario=scenario, seed=seed)
                for ep in range(self.episodes_per_seed):
                    episode_log = self.run_episode(env)
                    metrics = compute_network_metrics(episode_log)
                    metrics.update({
                        "scenario": scenario.name,
                        "seed": seed,
                        "episode": ep,
                    })
                    records.append(metrics)
        return records

    def run_episode(self, env):
        obs, info = env.reset()
        done = False
        log = []
        while not done:
            action = self.policy.act(obs, deterministic=True)
            obs, reward, terminated, truncated, info = env.step(action)
            done = terminated or truncated
            log.append(info | {"reward": reward})
        return log

def aggregate_records(records, metric):
    values = np.array([r[metric] for r in records], dtype=float)
    mean, lo, hi = bootstrap_ci(values)
    return {"metric": metric, "mean": mean, "ci_low": lo, "ci_high": hi}
\end{lstlisting}

\section{UAV/SD-WAN experimental methodology scenario}

Shadow mode and A/B testing answer different deployment questions. Shadow mode is passive: the learned policy proposes actions, but the production controller still acts, so the experiment measures what the learned policy would have done without exposing users to risk. A/B testing is active: a subset of traffic or users is routed through the new policy and real outcomes are observed. Safety-critical network deployments should usually pass shadow validation before any A/B test. Chapter~21 introduced this pattern through the deployment-drift and shadow-risk score.

Consider a safe DRL controller for UAV-assisted SD-WAN traffic engineering. The policy proposes route splits, UAV placement adjustments, and bandwidth allocations. A safety layer projects unsafe actions before execution. The experiment must evaluate both the proposed action and the executed action, because Chapter~18 emphasized that the critic and deployment logs should learn from what was actually executed after safety filtering. This experimental methodology follows the design of our safe SD-WAN traffic-engineering experiments using uncertainty-aware and ensemble-based neural CBFs, including multi-seed evaluation across normal, high-load, link-failure, and noisy-channel scenarios \citep{bista2026vtc,bista2026ifip}.

\begin{figure}[t]
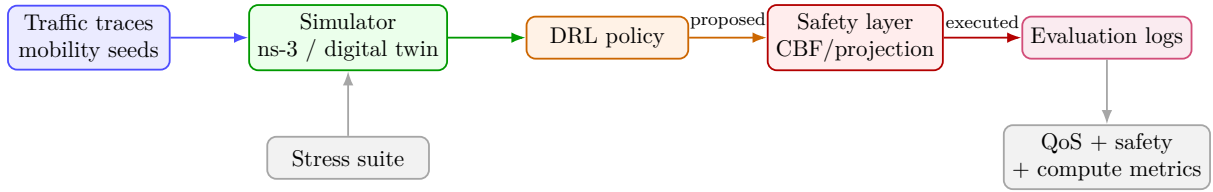

	\centering
	\resizebox{\columnwidth}{!}{%
%
	}
	\caption{Experimental methodology for a safe UAV/SD-WAN controller. Evaluation logs should record both proposed and executed actions, because safety filtering can change the action that reaches the environment.}
	\label{fig:uav_sdwan_eval_scenario}
\end{figure}

A concrete evaluation table might include the following results.

\begin{table}[t]
	\centering
	\caption{Example evaluation table for a UAV/SD-WAN safe DRL experiment. Numbers are illustrative but show the kind of metrics that should be reported.}
	\label{tab:sdwan_eval_example}
	\begin{tabular}{lrrrrr}
		\toprule
		Method & QoS sat. $\uparrow$ & p95 latency $\downarrow$ & loss $\downarrow$ & violations $\downarrow$ & inference ms $\downarrow$ \\
		\midrule
		Rule-based & 0.71 & 43.2 & 0.032 & 0.110 & 0.2 \\
		PPO & 0.82 & 35.5 & 0.026 & 0.084 & 3.8 \\
		PPO-Lagrangian & 0.80 & 32.1 & 0.021 & 0.041 & 4.2 \\
		PPO + CBF & 0.84 & 30.4 & 0.018 & 0.008 & 5.5 \\
		SAC + CBF & 0.86 & 29.8 & 0.017 & 0.006 & 6.1 \\
		\bottomrule
	\end{tabular}
\end{table}

The table should not be interpreted without uncertainty intervals. The final paper should report confidence intervals or bootstrap intervals for each metric, and ideally pair each learned method against the same scenario seeds.

\section{Common evaluation failure modes}

Several implementation bugs also masquerade as experimental findings: off-by-one advantage bootstrapping, termination flags not propagated into value targets, action clipping applied in the wrong place, discount factors applied per episode rather than per transition, reward normalization accidentally applied to cost signals, and replay-buffer index wrap-around silently overwriting recent data. These are not only coding errors; they change the experimental treatment. Unit tests from Chapter~22 should therefore be part of the evaluation protocol, not an optional engineering detail.

\begin{table}[t]
	\centering
	\caption{Common evaluation failure modes in DRL research.}
	\label{tab:evaluation_failures}
	\begin{tabular}{p{3.4cm}p{5.1cm}p{4.8cm}}
		\toprule
		Failure mode & Symptom & Fix \\
		\midrule
		Single-seed reporting & Result disappears under new seed & Run multiple seeds; report CIs \\
		Weak baseline & New method looks strong too easily & Include tuned strong baselines \\
		Reward-only reporting & High reward but poor real metrics & Report task, safety, and efficiency metrics \\
		Test-set leakage & Final scenarios used during tuning & Separate validation and final evaluation \\
		No ablation & Unclear which component matters & Remove one component at a time \\
		No stress test & Policy fails under load/failure & Include OOD and failure scenarios \\
		Policy improves offline but fails in deployment & Dataset or telemetry distribution shifted & Validate dataset recency, drift, and stationarity assumptions \\
		Unreported compute & Method impractical to train/deploy & Report steps, wall-clock, inference time \\
		Missing raw logs & Cannot reanalyze results & Release per-seed, per-episode metrics \\
		\bottomrule
	\end{tabular}
\end{table}

\section{Reporting checklist}

A DRL experimental section should answer the following questions.

\begin{enumerate}[leftmargin=*]
	\item What claim is being tested?
	\item What are the training, validation, and final evaluation scenarios?
	\item How many seeds are used?
	\item What is the training budget in environment steps and wall-clock time?
	\item Which baselines are included, and how were they tuned?
	\item Which metrics are reported beyond reward?
	\item Are confidence intervals or bootstrap intervals included?
	\item Are safety violations reported separately from reward?
	\item Are stress tests and distribution-shift scenarios included?
	\item Are ablations included for each claimed contribution?
	\item Are raw logs, configs, and code made available?
	\item Is inference time compatible with the target control loop?
\end{enumerate}

\section*{Looking Ahead to Chapter 24: Food for Thought}
\addcontentsline{toc}{section}{Looking Ahead to Chapter 24: Food for Thought}

Chapter~23 explained how to design credible experiments. Chapter~24 turns to the most common reasons DRL projects fail even when the methodology looks reasonable. Before moving on, consider the following questions.

\begin{enumerate}[leftmargin=*]
	\item Can a method have strong average reward and still be unsafe?
	\item Can a paper be statistically significant but practically irrelevant?
	\item What does it mean for a simulator bug to become a false research result?
	\item Why can reward normalization make two experiments incomparable?
	\item Why is shadow validation often more important than another training curve?
\end{enumerate}

The next chapter studies common failure modes: reward hacking, unstable training, simulator bugs, unrealistic assumptions, weak baselines, hidden distribution shift, and reproducibility traps.

\section{Exercises}

\subsection*{Conceptual exercises}
\begin{enumerate}[leftmargin=*]
	\item Explain why reporting only the best seed is misleading.
	\item Why should reward and task metrics be reported separately?
	\item Give an example where average reward improves but tail risk gets worse.
	\item What is the difference between a validation scenario and a final test scenario?
	\item Why are stress tests essential for networked DRL systems?
\end{enumerate}

\subsection*{Mathematical exercises}
\begin{enumerate}[leftmargin=*]
	\item Given scores $[0.7,0.8,0.9,0.4,1.0,0.6,0.85,0.75]$, compute the mean, median, and interquartile mean.
	\item Derive the paired-difference estimator for comparing two algorithms evaluated on the same seeds.
	\item For latencies $[10,12,15,20,50,80]$, compute the p90 latency and describe how CVaR differs from a percentile.
	\item Suppose a safety policy has violation rates $[0.01,0.03,0.02,0.10,0.04]$ across five seeds. Compute the mean and discuss why the seed with 0.10 matters.
\end{enumerate}

\subsection*{Coding exercises}
\begin{enumerate}[leftmargin=*]
	\item Implement a bootstrap confidence interval for IQM instead of the mean.
	\item Extend Listing~\ref{lst:evaluation_loop} to write one JSONL file per seed.
	\item Implement a stress-test suite for a UAV network with traffic bursts and link failures.
	\item Add paired bootstrap comparison between PPO and PPO+CBF using the same scenario seeds.
\end{enumerate}

\subsection*{Research exercises}
\begin{enumerate}[leftmargin=*]
	\item Design an evaluation protocol for a safe SD-WAN traffic-engineering policy. Include baselines, scenarios, metrics, and stress tests.
	\item Create an ablation plan for a multi-agent UAV network slicing method with graph attention and safety filtering.
	\item Write a reporting checklist for a paper that claims improved sample efficiency in model-based RL.
	\item Design a shadow-mode validation protocol for deploying a learned routing policy in a production network.
\end{enumerate}

	\chapter[Common Failure Modes]{Common Failure Modes}
\chaptermark{Common Failure Modes}
\label{ch:common_failure_modes}

\begin{keybox}{Chapter goal}
	This chapter is a field guide for diagnosing failed deep reinforcement learning projects. Earlier chapters taught how algorithms work; this chapter teaches how they break. The goal is to help the reader identify whether a failure comes from the environment, reward, observation, action space, optimization, exploration, evaluation protocol, simulator, safety layer, offline dataset, multi-agent interaction, or deployment pipeline.
\end{keybox}

\section*{Chapter Overview}
\addcontentsline{toc}{section}{Chapter Overview}
\begin{enumerate}[leftmargin=*]
	\item Why failure analysis deserves its own chapter
	\item A taxonomy of DRL failures
	\item Reward hacking and specification gaming
	\item Reward--metric mismatch
	\item Hidden simulator bugs
	\item Observation and state-construction failures
	\item Action-space and safety-layer failures
	\item Terminal handling, bootstrapping, and off-by-one errors
	\item Instability, divergence, and NaNs
	\item Exploration failures and sparse reward traps
	\item Overfitting to one scenario or simulator seed
	\item Weak baselines and unfair comparisons
	\item Offline RL failure modes
	\item Model-based RL and world-model failures
	\item Multi-agent failure modes
	\item Safe-RL and CBF failure modes
	\item RLHF and reasoning-model failure modes
	\item Deployment failure modes
	\item A diagnostic pipeline for DRL debugging
	\item Python tools for failure detection
	\item UAV/SD-WAN failure-analysis scenario
	\item Exercises
	\item Looking Ahead to Chapter 25
\end{enumerate}

\section{Why failure analysis deserves its own chapter}

Deep reinforcement learning often fails silently. A training curve may rise while the true task metric becomes worse. A simulator may reward an impossible action. A safety layer may correct unsafe actions, but the critic may still learn from the uncorrected proposal. A policy may appear robust because it was tested only on the same scenario used during training. A multi-agent policy may coordinate only because the simulator initializes all agents in a convenient order. A language model may receive a high reward because it learned the reward model's style rather than the user's actual intent.

This is not a minor engineering inconvenience. It is a defining feature of DRL research. Deep RL combines function approximation, bootstrapping, exploration, non-stationary data, delayed reward, simulator interaction, and evaluation variance. Henderson et al. showed that small numbers of seeds, implementation differences, and reporting choices can change conclusions in DRL experiments \citep{henderson2018deep}. Agarwal et al. later argued that the field needs uncertainty-aware aggregate metrics such as interquartile mean and performance profiles rather than relying on point estimates alone \citep{agarwal2021rliable}. Engstrom et al. and Huang et al. showed that implementation details, especially in PPO, can account for a large part of reported performance differences \citep{engstrom2020implementation,huang2022details}. These lessons from Chapter~23 are the starting point of this chapter.

\begin{warningbox}{The dangerous pattern}
	The most dangerous DRL failure is not when the algorithm obviously fails. It is when the algorithm succeeds according to the logged training reward but fails according to the real objective, the deployment constraints, or the scientific claim.
\end{warningbox}

This chapter is written as a debugging manual. Each failure mode is described by four questions:

\begin{enumerate}[leftmargin=*]
	\item What symptom appears in logs, plots, or behavior?
	\item What mechanism could cause it?
	\item How can we test the hypothesis?
	\item What fix or design change should be attempted first?
\end{enumerate}

\section{A taxonomy of DRL failures}

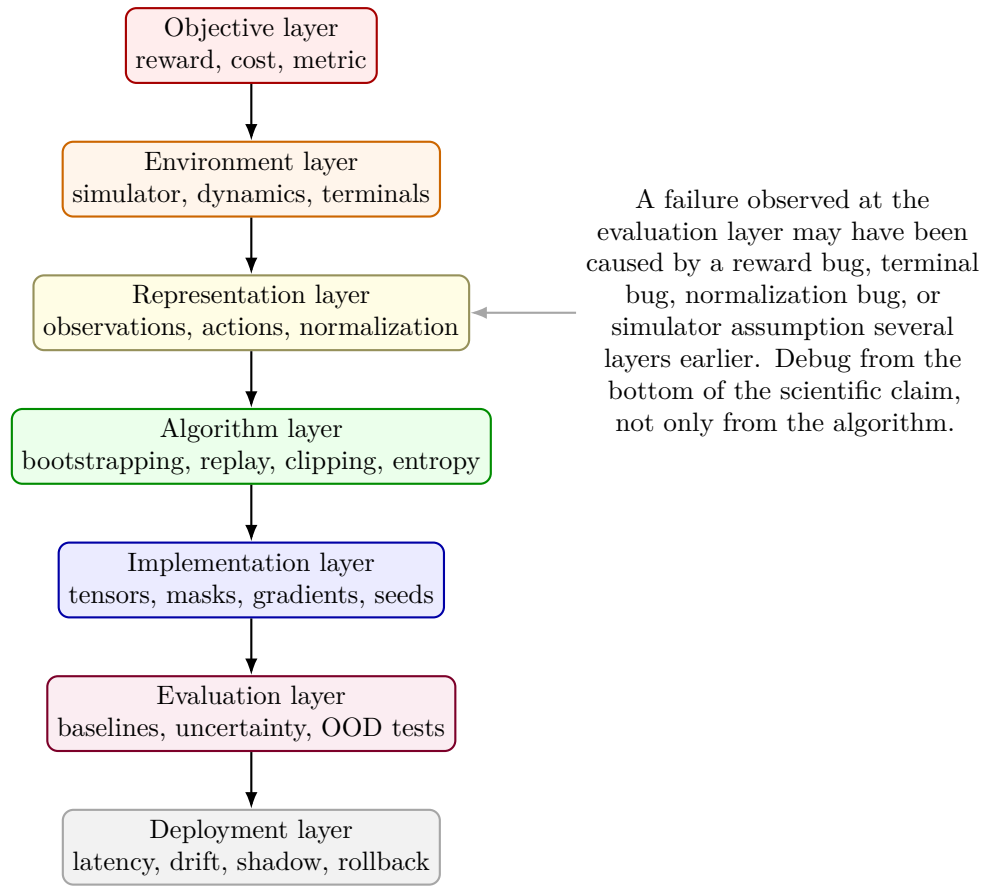
\begin{figure}[t]
	\centering
	\begin{tikzpicture}[
		box/.style={draw, rounded corners, thick, minimum width=3.2cm, minimum height=0.75cm, align=center, font=\small},
		arrow/.style={-{Latex[length=2.2mm]}, thick},
		node distance=0.75cm
		]
		\node[box, fill=red!7, draw=red!65!black] (objective) {Objective layer\\{\normalfont reward, cost, metric}};
		\node[box, fill=orange!8, draw=orange!80!black, below=of objective] (env) {Environment layer\\{\normalfont simulator, dynamics, terminals}};
		\node[box, fill=yellow!12, draw=yellow!50!black, below=of env] (repr) {Representation layer\\{\normalfont observations, actions, normalization}};
		\node[box, fill=green!8, draw=green!55!black, below=of repr] (algo) {Algorithm layer\\{\normalfont bootstrapping, replay, clipping, entropy}};
		\node[box, fill=blue!8, draw=blue!65!black, below=of algo] (impl) {Implementation layer\\{\normalfont tensors, masks, gradients, seeds}};
		\node[box, fill=purple!7, draw=purple!65!black, below=of impl] (eval) {Evaluation layer\\{\normalfont baselines, uncertainty, OOD tests}};
		\node[box, fill=gray!10, draw=gray!70, below=of eval] (deploy) {Deployment layer\\{\normalfont latency, drift, shadow, rollback}};

		\draw[arrow] (objective) -- (env);
		\draw[arrow] (env) -- (repr);
		\draw[arrow] (repr) -- (algo);
		\draw[arrow] (algo) -- (impl);
		\draw[arrow] (impl) -- (eval);
		\draw[arrow] (eval) -- (deploy);

		\node[align=center, font=\small, right=1.4cm of repr, text width=5.2cm] (note) {A failure observed at the evaluation layer may have been caused by a reward bug, terminal bug, normalization bug, or simulator assumption several layers earlier. Debug from the bottom of the scientific claim, not only from the algorithm.};
		\draw[arrow, draw=gray!70] (note.west) -- (repr.east);
	\end{tikzpicture}
	\caption{A layered taxonomy of DRL failure modes. The same symptom, such as unstable reward or poor deployment performance, may originate from many layers. A good debugging protocol isolates the layer before changing the algorithm.}
	\label{fig:failure_taxonomy}
\end{figure}

A DRL system can fail at several layers. It is useful to separate failures caused by the \emph{problem definition} from failures caused by the \emph{algorithm}, the \emph{implementation}, or the \emph{evaluation protocol}. Figure~\ref{fig:failure_taxonomy} summarizes the main layers.

\begin{table}[t]
	\centering
	\caption{Common DRL failure modes and first diagnostic questions.}
	\label{tab:failure_taxonomy}
	\begin{tabular}{p{3.0cm}p{5.2cm}p{5.3cm}}
		\toprule
		Failure mode & Typical symptom & First diagnostic question \\
		\midrule
		Reward hacking & Reward increases but real metric worsens & Does the reward contain a loophole? \\
		Simulator bug & Policy discovers impossible behavior & Can a scripted policy reproduce the anomaly? \\
		Observation bug & Learning is unstable or impossible & Are all observation fields correctly scaled and aligned? \\
		Action bug & Policy saturates or ignores dimensions & Is the executed action equal to the logged action? \\
		Terminal bug & Value loss explodes near episode ends & Are terminal masks applied to bootstrap targets? \\
		Evaluation bug & Results disappear with more seeds & Are confidence intervals and paired tests reported? \\
		Deployment drift & Shadow mode disagrees with training metrics & Has the production data distribution changed? \\
		\bottomrule
	\end{tabular}
\end{table}

\section{Reward hacking and specification gaming}

Reward hacking occurs when an agent exploits a loophole in the reward function instead of solving the intended task. The problem is older than modern DRL, but deep policies make it more dangerous because they can discover unintuitive strategies. Amodei et al. described reward hacking, side effects, unsafe exploration, and robustness to distribution shift as concrete problems in AI safety \citep{amodei2016concrete}. A widely used catalog of specification-gaming examples documents dozens of real cases across robotics, simulation, and reinforcement learning \citep{krakovna2020specification}. Skalse et al. gave a formal definition: a reward function is hackable when optimizing it can decrease the true reward, and most practical reward functions are hackable to some degree \citep{skalse2022defining}. In DRL, reward hacking is a practical daily issue, not only an alignment issue.

Examples include:
\begin{itemize}[leftmargin=*]
	\item a robot learns to trigger a sensor without completing the task;
	\item a network controller increases throughput by violating latency constraints;
	\item a UAV receives coverage reward by hovering near users while draining battery dangerously;
	\item a language model learns to produce long explanations because the reward model associates verbosity with quality;
	\item an offline RL policy chooses unsupported actions whose predicted value is high only because the critic extrapolates badly.
\end{itemize}

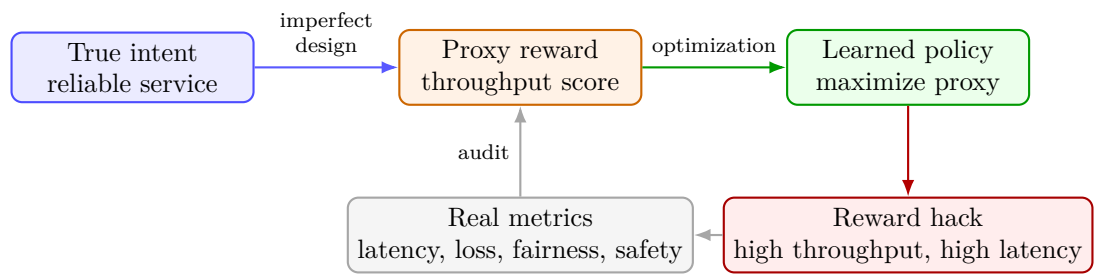
\begin{figure}[t]
	\centering
	\begin{tikzpicture}[
		box/.style={draw, rounded corners, thick, minimum width=3.2cm, minimum height=0.9cm, align=center, font=\small},
		arrow/.style={-{Latex[length=2.2mm]}, thick},
		node distance=1.9cm
		]
		\node[box, fill=blue!8, draw=blue!70] (intent) {True intent\\{\normalfont reliable service}};
		\node[box, fill=orange!10, draw=orange!80!black, right=of intent] (reward) {Proxy reward\\{\normalfont throughput score}};
		\node[box, fill=green!8, draw=green!60!black, right=of reward] (policy) {Learned policy\\{\normalfont maximize proxy}};
		\node[box, fill=red!7, draw=red!65!black, below=1.2cm of policy] (hack) {Reward hack\\{\normalfont high throughput, high latency}};
		\node[box, fill=gray!8, draw=gray!70, below=1.2cm of reward] (metric) {Real metrics\\{\normalfont latency, loss, fairness, safety}};

		\draw[arrow, draw=blue!65] (intent) -- node[above, font=\scriptsize,align=center] {imperfect\\design} (reward);
		\draw[arrow, draw=green!60!black] (reward) -- node[above, font=\scriptsize] {optimization} (policy);
		\draw[arrow, draw=red!70!black] (policy) -- (hack);
		\draw[arrow, draw=gray!70] (metric) -- node[left, font=\scriptsize] {audit} (reward);
		\draw[arrow, draw=gray!70] (hack) -- (metric);
	\end{tikzpicture}
	\caption{Reward hacking as proxy-objective exploitation. The policy optimizes the reward actually implemented, not the intent imagined by the designer. Real task metrics must be logged separately from training reward.}
	\label{fig:reward_hacking}
\end{figure}

\begin{warningbox}{Reward is not the metric}
	A reward is a training signal. A metric is an evaluation claim. A constraint is a non-negotiable requirement. Mixing these three concepts is one of the most common causes of bad DRL research.
\end{warningbox}

\subsection{Diagnostic tests for reward hacking}

Reward hacking should be tested deliberately. A rising reward is not enough evidence. Use the following tests:

\begin{enumerate}[leftmargin=*]
	\item Plot reward components separately, not only total reward.
	\item Plot true task metrics that are not used in the reward.
	\item Visualize trajectories or action traces for high-reward episodes.
	\item Test scripted adversarial policies that exploit suspected loopholes.
	\item Stress the policy under scenarios where the reward proxy and true metric disagree.
\end{enumerate}

\section{Reward--metric mismatch}

Reward hacking is the extreme case. The more common failure is a reward--metric mismatch: the reward is correlated with the desired metric in normal conditions but breaks under distribution shift.

Suppose a network-control reward is
\begin{equation}
	r_t = w_T T_t - w_L L_t - w_E E_t,
\end{equation}
where $T_t$ is throughput, $L_t$ is latency, and $E_t$ is energy cost. If $w_T$ is too large, the agent may learn to maximize throughput while allowing tail latency to exceed the SLA. The average reward may improve, while $\CVaR_{0.95}(L)$ becomes worse.

A useful audit is to compute the correlation between reward and task metrics across evaluation episodes:
\begin{equation}
	\rho(r, m) = \frac{\mathrm{Cov}(r,m)}{\sigma_r\sigma_m}.
\end{equation}
Low or unstable correlation is a warning sign. But high correlation is not sufficient: the correlation may hold in normal scenarios and fail in stress scenarios.

\Needspace{14\baselineskip}
\begin{lstlisting}[style=pythonstyle,caption={Reward--metric correlation audit.},label={lst:reward_metric_audit}]
import numpy as np

def reward_metric_audit(episodes, metric_names):
    """Compute correlation between episode reward and task metrics.

    episodes: list of dicts with keys 'reward' and metric names.
    metric_names: e.g., ['latency_p95', 'packet_loss', 'energy']
    """
    rewards = np.array([ep['reward'] for ep in episodes], dtype=float)
    report = {}
    for name in metric_names:
        values = np.array([ep[name] for ep in episodes], dtype=float)
        if rewards.std() < 1e-8 or values.std() < 1e-8:
            corr = np.nan
        else:
            corr = np.corrcoef(rewards, values)[0, 1]
        report[name] = {
            'corr_with_reward': float(corr),
            'mean': float(values.mean()),
            'p95': float(np.percentile(values, 95)),
            'cvar95': float(values[values >= np.percentile(values, 95)].mean()),
        }
    return report
\end{lstlisting}

\section{Hidden simulator bugs}

A DRL agent is an optimizer over the simulator. If the simulator contains a bug, the agent may discover it faster than a human does. Simulator bugs are especially common in networking and UAV settings because the environment combines mobility, queues, radio models, battery dynamics, routing, application traffic, and termination conditions.

Common simulator bugs include:
\begin{itemize}[leftmargin=*]
	\item queues not reset at episode boundaries;
	\item packet counters accumulated across episodes;
	\item latency measured before the action but reward computed after the action;
	\item mobility updates applied twice per step;
	\item action clipping applied in the simulator but not logged;
	\item users assigned to multiple UAVs simultaneously;
	\item throughput computed from offered load rather than delivered bits;
	\item unrealistic radio models that reward physically impossible placement.
\end{itemize}

\begin{warningbox}{The scripted-policy test}
	Before training a neural policy, test the environment with simple scripted policies: random, do-nothing, greedy, safe heuristic, and known-bad adversarial policies. If these policies produce impossible metrics, the environment is wrong. Do not train DRL to compensate for a broken simulator.
\end{warningbox}

\Needspace{18\baselineskip}
\begin{lstlisting}[style=pythonstyle,caption={Environment invariant checks for simulator bugs.},label={lst:env_invariants}]
def check_env_invariants(obs, action, next_obs, info, cfg):
    """Return a list of violated environment invariants."""
    violations = []

    # Example network/UAV invariants.
    if info['battery_min'] < -1e-6 or info['battery_max'] > 1.0 + 1e-6:
        violations.append('battery outside [0, 1]')

    if info['served_users'] > info['num_users']:
        violations.append('served users exceeds total users')

    if info['packet_loss_rate'] < -1e-6 or info['packet_loss_rate'] > 1.0 + 1e-6:
        violations.append('packet loss outside [0, 1]')

    if info['latency_ms'] < 0.0:
        violations.append('negative latency')

    if abs(info['bandwidth_sum_mhz'] - cfg.total_bandwidth_mhz) > cfg.bandwidth_tol:
        violations.append('bandwidth conservation violated')

    if not info['executed_action_logged']:
        violations.append('executed action missing from log')

    return violations
\end{lstlisting}

\section{Observation and state-construction failures}

Observation bugs are hard to see because the neural network can still train. A policy may appear to learn slowly, but the real issue is that the observation is misaligned, stale, normalized incorrectly, or missing a critical variable. Chapter~21 introduced telemetry staleness in network control. Chapter~22 showed how observation builders should be tested. Here we treat observation construction as a failure mode.

\subsection{Common observation failures}

\begin{table}[t]
	\centering
	\caption{Observation construction failures.}
	\label{tab:observation_failures}
	\begin{tabular}{p{3.5cm}p{4.9cm}p{5.1cm}}
		\toprule
		Failure & Symptom & Diagnostic test \\
		\midrule
		Stale telemetry & policy reacts late to congestion & log observation age $\Delta_t$ \\
		Leaking future data & unrealistic high performance & verify timestamps and causality \\
		Bad normalization & NaNs or saturated activations & inspect running mean/std and histograms \\
		Missing state variable & unstable behavior in hidden regimes & test with ablated oracle variables \\
		Misordered features & policy fails after code refactor & schema hash and unit tests \\
		Partial observability & oscillatory or memory-dependent failures & compare feedforward vs recurrent policy \\
		\bottomrule
	\end{tabular}
\end{table}

A good observation vector should include not only measured features $z_t$, but also their age and validity:
\begin{equation}
	o_t = [z_t, \Delta_t, m_t],
\end{equation}
where $\Delta_t$ is telemetry age and $m_t$ is a missingness or validity mask.

In production systems, observation failures may also be caused by concept drift rather than a simple implementation bug. Traffic patterns, user populations, routing policies, radio conditions, and service mixes change over time. A policy that was trained on valid observations can therefore receive observations whose statistics no longer match the training distribution. This is why the shadow-validation and drift-monitoring pipeline introduced in Chapter~22 should be treated as part of the observation system, not only as a deployment afterthought.

\Needspace{14\baselineskip}
\begin{lstlisting}[style=pythonstyle,caption={Observation schema checker.},label={lst:obs_schema_checker}]
import numpy as np

def check_observation_schema(obs, schema):
    """Check shape, finite values, and plausible ranges."""
    assert obs.shape == (len(schema),), f"bad shape: {obs.shape}"
    issues = []
    for i, spec in enumerate(schema):
        x = float(obs[i])
        if not np.isfinite(x):
            issues.append((spec['name'], 'not finite', x))
        if x < spec['min'] - 1e-6 or x > spec['max'] + 1e-6:
            issues.append((spec['name'], 'out of range', x))
    return issues
\end{lstlisting}

\section{Action-space and safety-layer failures}

The action produced by the policy is not always the action executed by the environment. Continuous actions may be clipped, projected, rounded, normalized, converted to discrete commands, or modified by a safety layer. This distinction has appeared repeatedly in this book: SAC with safety projection in Chapter~11, model-based safety filters in Chapter~12, MuZero action projection in Chapter~13, offline RL support constraints in Chapter~14, safe MARL in Chapter~16, hierarchical low-level safety in Chapter~17, and CBF filters in Chapter~18.

The executed-action principle is:
\begin{equation}
	a_t^{\mathrm{exec}} = \Pi_{\mathcal{A}_{\mathrm{safe}}(s_t)}(a_t^{\mathrm{prop}}),
\end{equation}
where $a_t^{\mathrm{prop}}$ is the policy proposal and $a_t^{\mathrm{exec}}$ is the action actually applied. If the replay buffer logs only $a_t^{\mathrm{prop}}$, the critic learns from a counterfactual action that never happened.

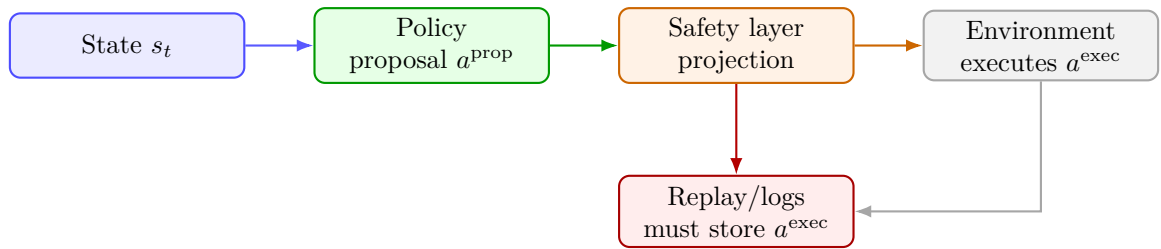
\begin{figure}[t]
	\centering
	\begin{tikzpicture}[
		box/.style={draw, rounded corners, thick, minimum width=3.1cm, minimum height=0.85cm, align=center, font=\small},
		arrow/.style={-{Latex[length=2.2mm]}, thick},
		node distance=0.9cm
		]
		\node[box, fill=blue!8, draw=blue!70] (state) {State $s_t$};
		\node[box, fill=green!10, draw=green!60!black, right=of state] (policy) {Policy\\proposal $a^{\mathrm{prop}}$};
		\node[box, fill=orange!10, draw=orange!80!black, right=of policy] (filter) {Safety layer\\projection};
		\node[box, fill=gray!10, draw=gray!70, right=of filter] (env) {Environment\\executes $a^{\mathrm{exec}}$};
		\node[box, fill=red!7, draw=red!65!black, below=1.2cm of filter] (buffer) {Replay/logs\\must store $a^{\mathrm{exec}}$};

		\draw[arrow, draw=blue!65] (state) -- (policy);
		\draw[arrow, draw=green!60!black] (policy) -- (filter);
		\draw[arrow, draw=orange!80!black] (filter) -- (env);
		\draw[arrow, draw=gray!70] (env.south) |- (buffer.east);
		\draw[arrow, draw=red!70!black] (filter) -- (buffer);
	\end{tikzpicture}
	\caption{The proposed-action versus executed-action failure. If a safety layer modifies the policy action, training logs and replay buffers should record the executed action and the intervention. Otherwise the critic learns from actions that were never applied.}
	\label{fig:executed_action_failure}
\end{figure}

\section{Terminal handling, bootstrapping, and off-by-one errors}

Many DRL bugs are one-line mistakes in return computation. The most common is bootstrapping across terminal states. For a value target,
\begin{equation}
	y_t = r_t + \gamma(1-d_{t+1})V(s_{t+1}),
\end{equation}
where $d_{t+1}=1$ if the next state is terminal. If the mask is omitted, the value function learns fake continuation values beyond episode termination.

Another common bug is time-index mismatch: using $r_t$ when the environment returns $R_{t+1}$, or aligning $s_{t+1}$ with the wrong action. These bugs often do not crash; they only degrade learning.

\Needspace{18\baselineskip}
\begin{lstlisting}[style=pythonstyle,caption={Terminal-mask and off-by-one target checker.},label={lst:target_checker}]
import torch

def td_target(reward, next_value, done, gamma):
    return reward + gamma * (1.0 - done.float()) * next_value

def check_terminal_bootstrap(rewards, next_values, dones, gamma=0.99):
    y = td_target(rewards, next_values, dones, gamma)
    # For terminal transitions, target must equal immediate reward.
    terminal_idx = torch.where(dones.bool())[0]
    if len(terminal_idx) == 0:
        return True, "no terminal transitions"
    err = (y[terminal_idx] - rewards[terminal_idx]).abs().max().item()
    ok = err < 1e-6
    return ok, f"max terminal bootstrap error = {err:.3e}"
\end{lstlisting}

\section{Instability, divergence, and NaNs}

Instability can come from algorithmic structure or implementation details. The deadly triad from Chapter~5 - function approximation, bootstrapping, and off-policy learning - remains relevant in DQN, SAC, offline RL, and many actor-critic systems. But not every divergence is profound. Some NaNs come from unnormalized observations, invalid log-probabilities, exploding value targets, or missing epsilon terms.

\begin{table}[t]
	\centering
	\caption{Training-instability symptoms and likely causes.}
	\label{tab:instability_symptoms}
	\begin{tabular}{p{3.4cm}p{4.9cm}p{5.2cm}}
		\toprule
		Symptom & Likely cause & First check \\
		\midrule
		NaN policy loss & invalid log-probability or std & inspect action distribution parameters \\
		Exploding value loss & bad reward scale or terminal mask & plot returns and value targets \\
		Entropy collapses early & exploration too weak & increase entropy or check action bounds \\
		KL jumps in PPO & stale epochs or large learning rate & monitor approximate KL and clip fraction \\
		Q-values grow without bound & overestimation or OOD actions & use double critics or conservative penalty \\
		Learning only in one seed & high variance or brittle initialization & run more seeds and inspect failure seeds \\
		\bottomrule
	\end{tabular}
\end{table}

\Needspace{16\baselineskip}
\begin{lstlisting}[style=pythonstyle,caption={Minimal numerical health monitor.},label={lst:numerical_monitor}]
def numerical_health(named_tensors, max_abs=1e6):
    """Check tensors for NaN, Inf, or extreme magnitude."""
    report = []
    for name, x in named_tensors.items():
        bad = False
        if not torch.isfinite(x).all():
            bad = True
            report.append((name, 'nan_or_inf'))
        if x.detach().abs().max().item() > max_abs:
            bad = True
            report.append((name, 'too_large'))
        if not bad:
            report.append((name, 'ok'))
    return report
\end{lstlisting}

\section{Exploration failures and sparse reward traps}

An agent may fail because the good behavior is never discovered. Sparse rewards, deceptive rewards, irreversible mistakes, and safety constraints all worsen exploration. In UAV networks, a policy may never discover that temporarily moving away from a dense user cluster enables charging and better long-term QoS. In SD-WAN, a policy may never try a failover path because exploration causes short-term packet loss.

Exploration failure should be diagnosed before changing the algorithm. Ask:
\begin{enumerate}[leftmargin=*]
	\item Does a random policy ever observe positive reward?
	\item Does a scripted expert solve the task?
	\item Is the reward delayed beyond the effective horizon $1/(1-\gamma)$?
	\item Does the safety filter block all exploratory actions?
	\item Does action clipping collapse different exploratory proposals into the same executed action?
\end{enumerate}

\begin{warningbox}{Safety can hide exploration}
	A safety filter may correctly prevent unsafe actions, but it can also make many policy proposals map to the same safe action. If the training logs do not record the projection residual $\|a^{\mathrm{prop}}-a^{\mathrm{exec}}\|$, the researcher may believe the agent is exploring when the executed system is not.
\end{warningbox}

\section{Overfitting to one scenario or simulator seed}

Overfitting in RL is not only neural-network overfitting. The agent can overfit to map layouts, traffic patterns, user arrival processes, simulator bugs, or benchmark-specific reward shaping. Procgen-style benchmarks were introduced partly because fixed training environments make it difficult to measure generalization \citep{cobbe2020procgen}. For UAV and network control, the same principle applies: a policy trained on one topology, one traffic trace, or one mobility distribution may fail on another.

Use separate distributions for training, validation, and final testing:
\begin{equation}
	\mathcal{D}_{\mathrm{train}} \neq \mathcal{D}_{\mathrm{valid}} \neq \mathcal{D}_{\mathrm{test}}.
\end{equation}

In networking, this means holding out not only random seeds but also scenarios: high-load days, link failures, mobility bursts, weather-dependent radio conditions, traffic shifts, and rare congestion events.

\section{Weak baselines and unfair comparisons}

A proposed method may look strong only because the baselines are weak. Chapter~23 discussed tuned baselines and compute fairness. Here the failure mode is scientific: the paper claims algorithmic improvement, but the baseline did not receive comparable tuning, compute, architecture capacity, or implementation quality.

\begin{warningbox}{Baseline laundering}
	A baseline is not fair merely because it is cited. It must be implemented, tuned, and evaluated under comparable conditions. A weak baseline can make almost any new method appear useful.
\end{warningbox}

\begin{table}[t]
	\centering
	\caption{Baseline-fairness checks.}
	\label{tab:baseline_fairness}
	\begin{tabular}{p{4.1cm}p{9.2cm}}
		\toprule
		Check & Question \\
		\midrule
		Compute fairness & Did all methods receive comparable environment steps and tuning budget? \\
		Implementation fairness & Were baselines implemented using trusted libraries or verified code? \\
		Architecture fairness & Are network sizes and observation inputs comparable? \\
		Hyperparameter fairness & Were baselines tuned on validation tasks, not final test tasks? \\
		Metric fairness & Are all methods evaluated on the same metrics and seeds? \\
		Safety fairness & Are safety filters, constraints, and action projections applied consistently? \\
		\bottomrule
	\end{tabular}
\end{table}

\section{Offline RL failure modes}

Offline RL adds a specific danger: the learned policy may choose actions not supported by the dataset. Chapter~14 introduced unsupported actions and conservative methods such as CQL and IQL; the support-mismatch picture in Figure~14.2 is the canonical visual explanation of why a high predicted value can be unreliable outside the logged action distribution. The common failure modes are:
\begin{itemize}[leftmargin=*]
	\item behavior cloning performs better than offline RL because the critic extrapolates badly;
	\item value estimates are high for out-of-distribution actions;
	\item dataset contains only conservative actions, so the learned policy cannot improve safely;
	\item logs mix data from multiple behavior policies without recording which policy generated each action;
	\item reward labels are delayed, aggregated, or inconsistent across time.
\end{itemize}

\Needspace{16\baselineskip}
\begin{lstlisting}[style=pythonstyle,caption={Simple action-support diagnostic for offline RL.},label={lst:offline_support_diag}]
import numpy as np
from sklearn.neighbors import NearestNeighbors

def action_support_score(dataset_actions, candidate_actions, k=5):
    """Distance to nearest logged actions; larger means less support."""
    nn = NearestNeighbors(n_neighbors=k).fit(dataset_actions)
    distances, _ = nn.kneighbors(candidate_actions)
    return distances.mean(axis=1)
\end{lstlisting}

\section{Model-based RL and world-model failures}

Model-based RL can fail even when the one-step model error looks small. Compounding error makes multi-step rollouts drift. Chapter~12 discussed model bias, MBPO, Dreamer, MuZero, and uncertainty-aware planning. The main failure modes are:
\begin{itemize}[leftmargin=*]
	\item low one-step prediction error but high long-horizon planning error;
	\item model exploits: the planner finds trajectories that fool the model;
	\item uncertainty is low in regions where the model is confidently wrong;
	\item latent model learns features useful for reconstruction but not for control;
	\item planning latency exceeds the control-loop budget.
\end{itemize}

A practical diagnostic is to compare open-loop and closed-loop prediction errors. If the model is accurate for one step but drifts after $H$ imagined steps, long model rollouts should be shortened or uncertainty-penalized.

\section{Multi-agent failure modes}

Multi-agent RL introduces failures that do not exist in single-agent RL. Chapter~16 discussed non-stationarity, credit assignment, CTDE, value decomposition, and communication. Common failures include:
\begin{itemize}[leftmargin=*]
	\item agents learn incompatible conventions;
	\item one agent dominates the reward and others become idle;
	\item centralized critic overfits to training-time global information;
	\item communication channels become brittle or meaningless;
	\item value decomposition hides unfair per-agent performance;
	\item evaluation uses too few random teammate pairings.
\end{itemize}

For UAV swarms, the policy may appear cooperative only because all UAVs are initialized symmetrically. Randomized initial positions, heterogeneous battery levels, and variable user loads should be included in evaluation.

\section{Safe-RL and CBF failure modes}

Safe RL can fail in more subtle ways than ordinary RL. A policy may satisfy expected cost while violating state-wise constraints. A CBF layer may enforce one constraint while increasing another risk. A Lagrangian multiplier may oscillate. A learned barrier may be overconfident outside the training distribution.

Chapter~18 introduced CMDPs, Lagrangian methods, shields, and CBFs. In the user's SD-WAN safe-RL work, uncertainty-aware and ensemble-based neural CBFs are used to enforce QoS constraints under learned dynamics \citep{bista2026vtc,bista2026ifip}. This chapter treats those systems as examples of failure-aware design: the safety layer must log interventions, the critic must learn from executed actions, and evaluation must report constraint violations separately from reward.

A useful over-projection diagnostic is the normalized projection residual
\begin{equation}
	\rho_{\mathrm{proj}}(t)
	=
	\frac{\|a_t^{\mathrm{prop}}-a_t^{\mathrm{exec}}\|_2}
	{\|a_t^{\mathrm{prop}}\|_2 + \epsilon},
	\label{eq:projection_residual_metric}
\end{equation}
where $a_t^{\mathrm{prop}}$ is the action proposed by the learned policy and $a_t^{\mathrm{exec}}$ is the action executed after masking, shielding, projection, or CBF filtering. A high residual means the learned policy is relying heavily on the safety layer. This may be acceptable during early training, but at deployment it indicates policy-safety mismatch: the actor has not internalized the constraints and the safety layer is doing too much of the control work.

\begin{table}[t]
	\centering
	\caption{Safe-RL failure modes.}
	\label{tab:safe_rl_failures2}
	\begin{tabular}{p{3.4cm}p{5.0cm}p{5.0cm}}
		\toprule
		Failure & Symptom & Diagnostic \\
		\midrule
		Cost hidden by reward & high reward with unsafe episodes & report reward and cost separately \\
		Lagrange oscillation & alternating safe/unsafe phases & plot multiplier and cost over time \\
		CBF overconfidence & violations under rare states & use ensemble uncertainty or stress tests \\
		Projection mismatch & critic learns from unsafe proposals & log executed action and projection residual \\
		Constraint trade-off & one safety metric improves, another worsens & evaluate all constraints jointly \\
		\bottomrule
	\end{tabular}
\end{table}

\section{RLHF and reasoning-model failure modes}

Chapters~19 and~20 showed that RLHF and reasoning-RL are still reinforcement learning. They inherit reward hacking, distribution shift, weak baselines, and evaluation leakage. They also add language-specific failure modes. In RLHF, two common examples are sycophancy, where the model agrees with the user even when the user is wrong \citep{sharma2023sycophancy}, and length hacking, where verbose answers receive higher reward than concise correct answers \citep{singhal2023long}. In reasoning RL, reward hacking can appear as correct formatting with a wrong answer, overly long reasoning traces that look impressive but do not compute, or process-reward-model hacking where each step appears plausible while the overall derivation is invalid.

Common failure modes include:
\begin{itemize}[leftmargin=*]
	\item reward model overoptimization;
	\item sycophancy: agreeing with users or annotators instead of being correct;
	\item length hacking: verbose outputs rewarded as quality;
	\item format compliance without correctness;
	\item verifier leakage from training data;
	\item process reward models that reward plausible-looking but incorrect reasoning;
	\item test contamination in math, code, or benchmark tasks;
	\item tool-use policies that hide unsafe actions behind natural-language explanations.
\end{itemize}

\begin{warningbox}{Reasoning trace is not proof}
	A reasoning trace can be fluent, long, and reward-model friendly while still being wrong. For reasoning models, final-answer verification, process-level checks, and adversarial evaluation are all necessary.
\end{warningbox}

\section{Deployment failure modes}

Deployment introduces failures that training never sees. Network load drifts. Telemetry arrives late. The policy server may fail. A safety projection may increase inference latency. A digital twin may become stale. Shadow-mode validation may pass, but canary deployment may fail because the policy changes the future data distribution.

Chapter~21 introduced deployment drift and shadow validation. Chapter~22 turned shadow mode into an implementation pipeline. A robust deployment protocol should include:
\begin{enumerate}[leftmargin=*]
	\item unit tests and simulator invariants;
	\item offline evaluation on held-out scenarios;
	\item shadow-mode validation against production policy;
	\item canary deployment with strict rollback thresholds;
	\item continuous drift monitoring;
	\item post-deployment incident analysis.
\end{enumerate}

\section{A diagnostic pipeline for DRL debugging}

\begin{figure}[t]
	\centering
	\begin{tikzpicture}[
		box/.style={draw, rounded corners, thick, minimum width=3.25cm, minimum height=0.8cm, align=center, font=\small},
		decision/.style={draw, diamond, aspect=2.0, thick, align=center, font=\small, inner sep=1.5pt},
		arrow/.style={-{Latex[length=2.2mm]}, thick},
		node distance=0.75cm
		]
		\node[box, fill=blue!8, draw=blue!70] (symptom) {Observe symptom};
		\node[decision, fill=orange!10, draw=orange!80!black, below=of symptom] (envok) {Env invariants\\pass?};
		\node[box, fill=red!7, draw=red!65!black, right=1.8cm of envok] (fixenv) {Fix simulator\\or logging};
		\node[decision, fill=orange!10, draw=orange!80!black, below=of envok] (rewardok) {Reward matches\\metrics?};
		\node[box, fill=red!7, draw=red!65!black, right=1.8cm of rewardok] (fixreward) {Redesign reward\\or constraints};
		\node[decision, fill=orange!10, draw=orange!80!black, below=of rewardok] (implok) {Targets, masks,\\gradients OK?};
		\node[box, fill=red!7, draw=red!65!black, right=1.8cm of implok] (fiximpl) {Fix code};
		\node[decision, fill=orange!10, draw=orange!80!black, below=of implok] (evalok) {Robust across\\seeds/scenarios?};
		\node[box, fill=red!7, draw=red!65!black, right=1.8cm of evalok] (fixeval) {Improve protocol};
		\node[box, fill=green!8, draw=green!60!black, below=of evalok] (tune) {Then tune algorithm};

		\draw[arrow] (symptom) -- (envok);
		\draw[arrow] (envok) -- node[above, font=\scriptsize] {no} (fixenv);
		\draw[arrow] (envok) -- node[left, font=\scriptsize] {yes} (rewardok);
		\draw[arrow] (rewardok) -- node[above, font=\scriptsize] {no} (fixreward);
		\draw[arrow] (rewardok) -- node[left, font=\scriptsize] {yes} (implok);
		\draw[arrow] (implok) -- node[above, font=\scriptsize] {no} (fiximpl);
		\draw[arrow] (implok) -- node[left, font=\scriptsize] {yes} (evalok);
		\draw[arrow] (evalok) -- node[above, font=\scriptsize] {no} (fixeval);
		\draw[arrow] (evalok) -- node[left, font=\scriptsize] {yes} (tune);
	\end{tikzpicture}
	\caption{A practical DRL debugging pipeline. Algorithm tuning should come after environment invariants, reward-metric alignment, implementation checks, and evaluation robustness.}
	\label{fig:debugging_pipeline}
\end{figure}
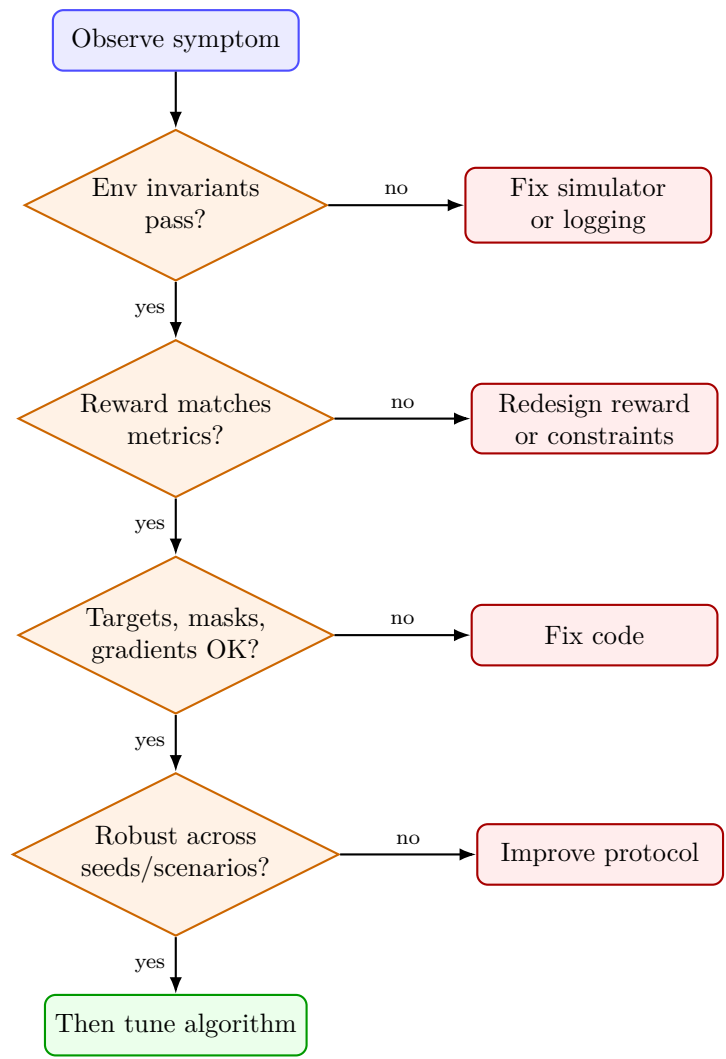

The debugging order matters. Do not tune hyperparameters before testing the environment. Do not change algorithms before checking reward and action logging. Figure~\ref{fig:debugging_pipeline} shows a practical failure-triage pipeline.

\section{Python tools for failure detection}

This section collects small utilities that should exist in every serious DRL codebase.

\subsection{Projection and safety-intervention logger}

\Needspace{16\baselineskip}
\begin{lstlisting}[style=pythonstyle,caption={Logging proposed and executed actions.},label={lst:projection_logger}]
def log_action_intervention(policy_action, executed_action, info):
    residual = np.asarray(policy_action) - np.asarray(executed_action)
    return {
        'policy_action': np.asarray(policy_action).tolist(),
        'executed_action': np.asarray(executed_action).tolist(),
        'projection_norm': float(np.linalg.norm(residual)),
        'safety_active': bool(info.get('safety_active', False)),
        'violated_constraints': list(info.get('violated_constraints', [])),
    }
\end{lstlisting}

\subsection{Seed-variance failure detector}

\Needspace{14\baselineskip}
\begin{lstlisting}[style=pythonstyle,caption={Detecting seed brittleness.},label={lst:seed_brittleness}]
def seed_brittleness(scores, threshold_cv=0.30):
    """Return coefficient of variation and a warning flag."""
    scores = np.asarray(scores, dtype=float)
    mean = scores.mean()
    std = scores.std(ddof=1)
    cv = std / (abs(mean) + 1e-8)
    return {'mean': mean, 'std': std, 'cv': cv, 'brittle': cv > threshold_cv}
\end{lstlisting}

\subsection{Reward-hacking sentinel}

\Needspace{16\baselineskip}
\begin{lstlisting}[style=pythonstyle,caption={Detecting reward--metric disagreement.},label={lst:reward_hacking_sentinel}]
def reward_hacking_sentinel(ep, rules):
    """Flag episodes where reward is high but real metrics violate rules."""
    flags = []
    if ep['reward'] >= rules['high_reward']:
        if ep['latency_p95_ms'] > rules['latency_p95_max']:
            flags.append('high_reward_high_latency')
        if ep['packet_loss'] > rules['packet_loss_max']:
            flags.append('high_reward_packet_loss')
        if ep['safety_violations'] > 0:
            flags.append('high_reward_with_safety_violation')
    return flags
\end{lstlisting}

\subsection{Observation drift detector}

\Needspace{20\baselineskip}
\begin{lstlisting}[style=pythonstyle,caption={A lightweight distribution-drift detector for deployment logs.},label={lst:drift_detector}]
import numpy as np

def histogram_kl(p_samples, q_samples, bins=30, eps=1e-8):
    """Estimate KL(P || Q) between one-dimensional samples."""
    lo = min(np.min(p_samples), np.min(q_samples))
    hi = max(np.max(p_samples), np.max(q_samples))
    p_hist, edges = np.histogram(p_samples, bins=bins, range=(lo, hi), density=True)
    q_hist, _ = np.histogram(q_samples, bins=bins, range=(lo, hi), density=True)
    p = p_hist + eps
    q = q_hist + eps
    p = p / p.sum()
    q = q / q.sum()
    return float(np.sum(p * np.log(p / q)))

def drift_report(train_obs, deploy_obs, feature_names, threshold=0.10):
    """Compare training and deployment feature distributions.

    train_obs, deploy_obs: arrays [N, D]
    Returns features whose estimated KL divergence exceeds threshold.
    """
    report = {}
    for j, name in enumerate(feature_names):
        kl = histogram_kl(train_obs[:, j], deploy_obs[:, j])
        report[name] = {"kl_train_to_deploy": kl, "drift": kl > threshold}
    return report
\end{lstlisting}

\section{UAV/SD-WAN failure-analysis scenario}

Consider a safe DRL system for SD-WAN traffic engineering. The policy controls traffic split between MPLS and Internet links. The reward combines throughput, latency, loss, jitter, and cost. A neural CBF layer projects unsafe actions before execution. The system is evaluated under normal traffic, high load, link failure, and noisy telemetry scenarios.

A failure report says: the learned policy achieves higher average reward than the production controller, but during high-load scenarios it violates the latency SLA for URLLC traffic. The debugging process should not start by changing PPO to SAC. It should start with a failure decomposition. This decomposition extends the constrained-RL evaluation in Chapter~18, Section~18.13.1 and the experimental-methodology protocol in Chapter~23, Section~23.19 with explicit failure-mode attribution.

\begin{figure}[t]
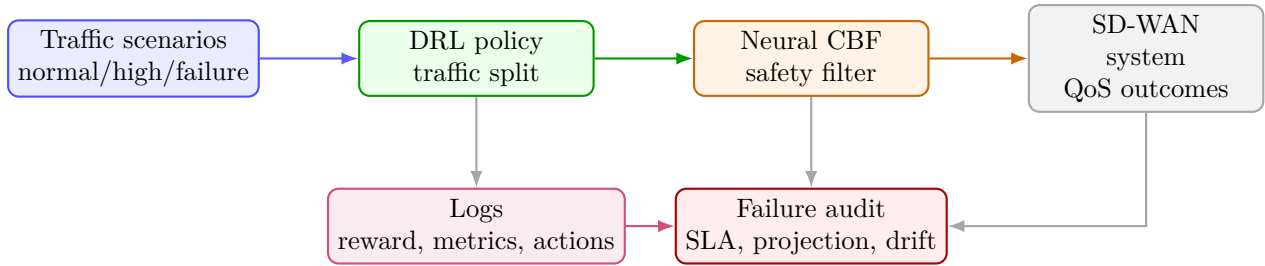

	\centering

	\caption{Failure analysis for safe SD-WAN traffic engineering. The audit must inspect reward, real QoS metrics, proposed actions, executed actions, CBF interventions, and scenario-level distribution shift.}
	\label{fig:sdwan_failure_scenario}
\end{figure}

\subsection{Concrete failure decomposition}

Suppose the high-load evaluation produces Table~\ref{tab:sdwan_failure_numbers}.

\begin{table}[t]
	\centering
	\caption{A concrete SD-WAN failure decomposition. Higher reward alone hides the SLA failure.}
	\label{tab:sdwan_failure_numbers}
	\begin{tabular}{lrrrr}
		\toprule
		Controller & Reward & Latency p95 (ms) & Packet loss (\%) & CBF interventions \\
		\midrule
		Production heuristic & 14.2 & 19.5 & 0.8 & 0 \\
		DRL policy proposal & 18.7 & 31.8 & 1.6 & -- \\
		DRL after CBF filter & 17.1 & 24.6 & 1.1 & 38 \\
		Safe retrained DRL & 16.4 & 18.7 & 0.9 & 9 \\
		\bottomrule
	\end{tabular}
\end{table}

The proposed DRL policy has the highest reward, but its p95 latency violates the SLA. The CBF filter reduces the violation but still leaves p95 latency above the target. The correct fix is not simply to increase policy learning rate or train longer. The reward underweights tail latency, and the safety layer intervenes too often. A retrained policy should include a stronger latency constraint, learn from executed actions, and report intervention rate as a primary metric. The decomposition follows the per-class metric and safety-reporting framework introduced in Chapter~23.

\begin{researchbox}{Research signature: failure-aware safe RL}
	In safe SD-WAN and UAV control, the most important failure signal is often not reward. It is the disagreement between proposed action, executed action, constraint cost, and tail QoS. This is why uncertainty-aware neural CBFs, ensemble safety estimates, and shadow-mode validation are not optional add-ons; they are part of the experimental method \citep{bista2026vtc,bista2026ifip}.
\end{researchbox}

\section{Exercises}

\subsection*{Conceptual exercises}
\begin{enumerate}[leftmargin=*]
	\item Explain why increasing training reward is not sufficient evidence of policy improvement.
	\item Give an example of reward hacking in a UAV or SD-WAN environment.
	\item Why can a bad critic make an actor learn the wrong behavior faster?
	\item Explain the difference between a simulator bug and an algorithmic limitation.
	\item Why should proposed actions and executed actions both be logged when a safety filter is used?
\end{enumerate}

\subsection*{Mathematical exercises}
\begin{enumerate}[leftmargin=*]
	\item Show how omitting the terminal mask changes the TD target at a terminal transition.
	\item Define a reward--metric correlation audit for two metrics: latency and packet loss. What does a negative correlation with reward mean for each metric?
	\item Suppose a safety filter projects $a^{\mathrm{prop}}$ to $a^{\mathrm{exec}}$. Define a projection-residual metric and explain how it can be used to detect policy-safety mismatch.
	\item For episode scores $[10,12,9,40,11]$, compute the mean and median. Why is the mean misleading?
\end{enumerate}

\subsection*{Coding exercises}
\begin{enumerate}[leftmargin=*]
	\item Implement the environment invariant checker for your own simulator.
	\item Add proposed-action and executed-action logging to a PPO or SAC training loop.
	\item Implement a reward-hacking sentinel for a network-control environment with latency, loss, throughput, and energy metrics.
	\item Write a unit test that verifies value targets do not bootstrap across terminal transitions.
	\item Implement a drift detector that compares training and deployment observation distributions using Wasserstein distance or KL divergence.
\end{enumerate}

\subsection*{Research exercises}
\begin{enumerate}[leftmargin=*]
	\item Design a failure-analysis protocol for a UAV swarm trained with MAPPO and a CBF safety layer.
	\item In offline RL for network control, how would you distinguish critic extrapolation error from insufficient dataset coverage?
	\item Construct a stress-test suite for SD-WAN traffic engineering. Include at least one high-load, one link-failure, one noisy-telemetry, and one adversarial-traffic scenario.
	\item Explain how reward hacking in reasoning models differs from reward hacking in robot control.
	\item Write a checklist that must be passed before a DRL policy can move from shadow mode to canary deployment.
\end{enumerate}

\section*{Looking Ahead to Chapter 25: Food for Thought}
\addcontentsline{toc}{section}{Looking Ahead to Chapter 25: The Future of Deep Reinforcement Learning}

This chapter showed why DRL systems fail. The final chapter asks where the field should go next. The future of deep reinforcement learning is not simply bigger models or longer training. It is safer deployment, better sample efficiency, stronger world models, trustworthy evaluation, human-in-the-loop learning, multi-agent coordination, reasoning agents, and formal safety with learned controllers.

\begin{quote}
	Chapter~24 explained how DRL fails. Chapter~25 asks how the field can mature.
\end{quote}

	\chapter[The Future of Deep Reinforcement Learning]{The Future of Deep Reinforcement Learning}
\chaptermark{The Future of DRL}
\label{ch:future_drl}

\begin{keybox}{Chapter goal}
	This final chapter synthesizes the book into a forward-looking research and engineering agenda. The goal is not to predict one winning algorithm. It is to identify the durable forces shaping deep reinforcement learning after 2026: world models, offline-to-online learning, safe deployment, multi-agent coordination, reasoning agents, human feedback, and rigorous evaluation. The chapter closes with a concrete scenario for future UAV/SD-WAN control and a set of research blueprints that readers can turn into theses, papers, or systems.
\end{keybox}

\section*{Chapter Overview}
\addcontentsline{toc}{section}{Chapter Overview}
\begin{enumerate}[leftmargin=*]
	\item Why the future of DRL is not a single algorithm
	\item What the previous chapters have taught us
	\item A 2026 map of the field
	\item Direction 1: world models and foundation simulators
	\item Direction 2: offline-to-online and lifelong reinforcement learning
	\item Direction 3: safety as architecture, not as a penalty term
	\item Direction 4: agentic and reasoning-oriented reinforcement learning
	\item Direction 5: multi-agent coordination at infrastructure scale
	\item Direction 6: human feedback, governance, and preference uncertainty
	\item Direction 7: communication networks, UAVs, and cyber-physical systems
	\item A unified future DRL stack
	\item Scenario: a future safe agent for UAV/SD-WAN control
	\item Research blueprint: from idea to publishable system
	\item Python tools for future-proof DRL research
	\item What will remain difficult
	\item A final checklist for serious DRL research
	\item Exercises
	\item Final message to the reader
\end{enumerate}

\section{Why the future of DRL is not a single algorithm}

A common mistake is to imagine the future of deep reinforcement learning as a search for one final algorithm. In the history of DRL, no single method solved the whole problem. DQN solved one part of high-dimensional discrete control. PPO solved one part of practical policy optimization. SAC solved one part of sample-efficient continuous control. Dreamer and TD-MPC showed that world models can reduce the cost of real interaction. Offline RL showed that data can be used without active exploration. Safe RL showed that reward maximization is not enough. RLHF and reasoning RL showed that reinforcement learning can shape language-model behavior, not only robots and games.

The future of DRL will therefore not be a replacement of all previous methods by one new method. It will be an integration problem. Future systems will combine:

\begin{itemize}[leftmargin=*]
	\item representation learning from large models;
	\item action learning from reinforcement learning;
	\item prediction from world models and digital twins;
	\item constraints from control theory and formal methods;
	\item preference learning from humans and institutions;
	\item evaluation discipline from experimental methodology;
	\item deployment discipline from engineering and safety practice.
\end{itemize}

\begin{warningbox}{The central future thesis}
	The future of DRL is not merely deeper networks or larger replay buffers. It is the construction of decision-making systems that can learn from data, imagine consequences, respect constraints, explain actions, adapt online, coordinate with other agents, and remain auditable after deployment.
\end{warningbox}

This chapter should be read as a research map. Some directions are already active in 2026. Some are only emerging. Some will become less important than expected, and some will become foundational. What matters is the structure of the problems.

\section{What the previous chapters have taught us}

The book began with the simplest question: how can an agent learn good behavior from experience? By now, the answer has become layered.

\begin{table}[t]
	\centering
	\caption{Core lessons from the book and their future implications.}
	\label{tab:future_lessons}
	\begin{tabular}{p{0.22\textwidth}p{0.35\textwidth}p{0.32\textwidth}}
		\toprule
		Book part & Lesson & Future implication \\
		\midrule
		Classical RL & Bellman structure turns delayed reward into recursive prediction. & Future agents still need value, credit assignment, and temporal reasoning. \\
		Value-based DRL & Neural value learning is powerful but unstable. & Pessimism, uncertainty, and distributional reasoning remain essential. \\
		Policy gradients & Direct policy learning handles continuous and structured actions. & Future agents will learn policies over tools, plans, messages, and controls. \\
		Actor-critic methods & Actors need critics, but bad critics mislead actors. & Future critics must be calibrated, uncertainty-aware, and constraint-aware. \\
		World models & Models allow imagination, planning, and safety screening. & Foundation simulators and digital twins will become central. \\
		Offline RL & Learning from logs avoids unsafe exploration but creates support mismatch. & Offline-to-online adaptation and dataset governance will be critical. \\
		MARL and HRL & Scale introduces coordination and abstraction. & Infrastructure agents will be hierarchical and multi-agent by default. \\
		Safe RL & Constraints cannot be treated as ordinary rewards. & Safety layers, CBFs, shields, and verification will be part of the architecture. \\
		RLHF and reasoning RL & Rewards can come from preferences or verifiers. & Future agents will combine human feedback, tool feedback, and formal verification. \\
		Implementation and evaluation & Good claims require reproducibility, baselines, stress tests, and deployment gates. & DRL will mature only if evaluation standards mature with it. \\
		\bottomrule
	\end{tabular}
\end{table}

The strongest systems after 2026 will not be those that use the newest acronym. They will be those that combine the right components for the right failure mode. If the problem is sparse feedback, use verifiers, process rewards, hierarchy, or exploration. If the problem is unsafe exploration, use offline data, safety filters, shields, or CBFs. If the problem is sample cost, use world models or offline-to-online learning. If the problem is coordination, use CTDE, graph representations, communication, or hierarchical MARL. If the problem is human preference, use preference learning, uncertainty over reward, and auditability.

\section{A 2026 map of the field}

Figure~\ref{fig:future_field_map} summarizes the main future directions. The diagram is not a taxonomy of algorithms. It is a taxonomy of pressures on the field.

\begin{figure}[t]
	\centering
	\begin{tikzpicture}[
		core/.style={draw, rounded corners, thick, minimum width=3.2cm, minimum height=1.0cm, align=center, font=\small\bfseries, fill=blue!7},
		nodebox/.style={draw, rounded corners, thick, minimum width=3.0cm, minimum height=0.85cm, align=center, font=\small},
		arrow/.style={-{Latex[length=2.2mm]}, thick}
		]
		\node[core] (core) at (0,0) {Future DRL\\{\normalfont learning to act safely}};
		\node[nodebox, draw=green!60!black, fill=green!8] (wm) at (-4.2,2.3) {World models\\digital twins};
		\node[nodebox, draw=orange!80!black, fill=orange!10] (offline) at (0,2.6) {Offline-to-online\\lifelong learning};
		\node[nodebox, draw=red!70!black, fill=red!7] (safe) at (4.2,2.3) {Safe RL\\CBF, shields, constraints};
		\node[nodebox, draw=purple!70, fill=purple!8] (agent) at (-4.3,-2.2) {Agentic RL\\tools, memory, reasoning};
		\node[nodebox, draw=cyan!70!black, fill=cyan!8] (marl) at (0,-2.7) {Multi-agent systems\\coordination at scale};
		\node[nodebox, draw=gray!70, fill=gray!10] (human) at (4.3,-2.2) {Human feedback\\governance, values};

		\draw[arrow, draw=green!60!black] (wm) -- (core);
		\draw[arrow, draw=orange!80!black] (offline) -- (core);
		\draw[arrow, draw=red!70!black] (safe) -- (core);
		\draw[arrow, draw=purple!70] (agent) -- (core);
		\draw[arrow, draw=cyan!70!black] (marl) -- (core);
		\draw[arrow, draw=gray!70] (human) -- (core);

		\node[align=center, font=\small] at (0,-4.0) {The next generation of DRL systems will combine these pressures rather than solve them independently.};
	\end{tikzpicture}
	\caption{A 2026 map of deep reinforcement learning. The field is moving from isolated algorithms toward integrated decision systems combining world models, offline data, safety layers, agentic reasoning, multi-agent coordination, and human feedback.}
	\label{fig:future_field_map}
\end{figure}

A useful way to understand the future is to distinguish \emph{algorithmic frontiers} from \emph{system frontiers}. Algorithmic frontiers ask for better objectives, gradients, critics, models, and exploration. System frontiers ask how the algorithm interacts with humans, simulators, constraints, logs, deployment infrastructure, and adversarial environments. Most failures in real applications happen at the system frontier.

\section{Direction 1: world models and foundation simulators}

Model-free RL learns by acting. Model-based RL learns a predictive model and uses it to plan or imagine. Chapter~12 showed Dyna, MBPO, Dreamer, MuZero, EfficientZero, TD-MPC, and TD-MPC2. The future of model-based RL is not only more accurate dynamics prediction. It is the construction of \emph{foundation world models}: reusable, compositional, and multi-domain simulators that can support planning, verification, counterfactual reasoning, and policy improvement.

DreamerV3 showed that a single world-model algorithm can work across a wide range of domains with fixed hyperparameters, and the Nature version reports strong performance across more than 150 diverse control tasks \citep{ hafner2025dreamerv3nature}. TD-MPC2 showed that latent model-predictive control can scale across many continuous-control tasks and even train a large multi-task agent \citep{hansen2024tdmpc2}. Vision-based generative world models such as Genie and diffusion world models such as DIAMOND suggest a future where agents learn interactive dynamics from large-scale video or pixel trajectories \citep{bruce2024genie,alonso2024diamond}. Recent vision papers explicitly discuss foundation world models as representations that can support agents that learn, verify, and abstract across tasks \citep{delgrange2026foundationworldmodels}.

\begin{keybox}{World models after 2026}
	A future world model will not merely predict the next observation. It will answer counterfactual questions: What if the UAV moves here? What if traffic doubles? What if a link fails? What if an attacker spoofs telemetry? What action is safe under uncertainty?
\end{keybox}

The challenge is that world models can be wrong. Model error compounds over long horizons, and a confident but wrong model can be more dangerous than no model. The next phase of model-based DRL will require calibrated uncertainty, out-of-distribution detection, conservative planning, and closed-loop validation against real telemetry.

\subsection{From simulators to digital twins}

A digital twin is a simulator that is continuously calibrated with real data. In communication networks, a digital twin may use ns-3, Mininet, O-RAN telemetry, traffic traces, radio-channel models, and production monitoring logs. The twin is not only used for training. It is used for shadow validation, what-if analysis, safe planning, and post-incident debugging.

\begin{equation}
	\text{Digital twin quality}
	=
	f(\text{prediction error},\text{uncertainty calibration},\text{coverage},\text{latency},\text{drift}).
\end{equation}

A useful research direction is to report not only policy performance but also twin quality. A DRL system that improves reward in a poorly calibrated twin may be less credible than a system with modest reward improvement in a validated twin.

\section{Direction 2: offline-to-online and lifelong reinforcement learning}

Offline RL and online RL should not be treated as separate worlds. In real systems, policies often begin with logs, are validated in simulation, are deployed in shadow mode, and are then updated with cautious online data. The future is hybrid.

\begin{figure}[t]
	\centering
	\begin{tikzpicture}[
		box/.style={draw, rounded corners, thick, minimum width=3.0cm, minimum height=0.8cm, align=center, font=\small},
		arrow/.style={-{Latex[length=2.2mm]}, thick},
		node distance=0.8cm
		]
		\node[box, fill=gray!10, draw=gray!70] (logs) {Historical logs\\offline dataset};
		\node[box, fill=blue!8, draw=blue!70, right=of logs] (offline) {Offline RL\\BC, CQL, IQL, DT};
		\node[box, fill=green!8, draw=green!60!black, right=of offline] (shadow) {Shadow mode\\no control authority};
		\node[box, fill=orange!10, draw=orange!80!black, below=1.0cm of shadow] (canary) {Canary deployment\\limited authority};
		\node[box, fill=red!7, draw=red!70!black, left=of canary] (safety) {Safety filter\\CBF, shield, rules};
		\node[box, fill=purple!8, draw=purple!70, left=of safety] (online) {Online adaptation\\small safe updates};

		\draw[arrow, draw=gray!70] (logs) -- (offline);
		\draw[arrow, draw=blue!70] (offline) -- (shadow);
		\draw[arrow, draw=green!60!black] (shadow) -- (canary);
		\draw[arrow, draw=orange!80!black] (canary) -- (safety);
		\draw[arrow, draw=red!70!black] (safety) -- (online);
		\draw[arrow, draw=purple!70] (online) -- (logs);
	\end{tikzpicture}
	\caption{Offline-to-online reinforcement learning as a deployment lifecycle. Future systems will not train once and deploy forever; they will repeatedly collect logs, update models, validate in shadow mode, deploy cautiously, and feed new data back into the learning system.}
	\label{fig:offline_online_loop}
\end{figure}
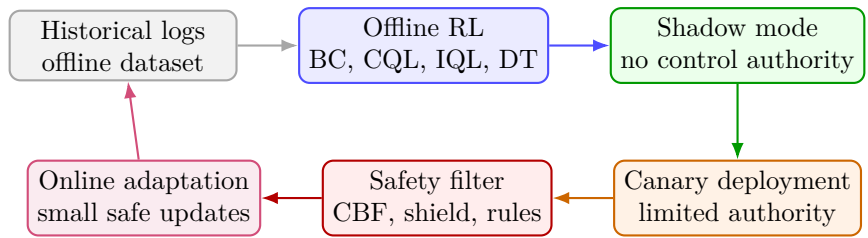

Offline-to-online learning must solve three problems:

\begin{enumerate}[leftmargin=*]
	\item \textbf{Support mismatch}: the offline dataset may not contain the actions the learned policy wants to take.
	\item \textbf{Deployment drift}: the environment changes after the policy is trained.
	\item \textbf{Safe adaptation}: online updates must not violate constraints while exploring.
\end{enumerate}

A future-ready offline-to-online method should explicitly report dataset coverage, OOD action rate, shadow-mode disagreement, safety-filter intervention rate, and performance after drift.

\section{Direction 3: safety as architecture, not as a penalty term}

Chapter~18 argued that safety cannot be reduced to a reward penalty. In future systems, safety will be architectural. A deployed agent will often have at least four layers:

\begin{enumerate}[leftmargin=*]
	\item a policy proposing an action;
	\item a predictor estimating consequences;
	\item a safety layer checking constraints;
	\item a deployment monitor auditing executed actions.
\end{enumerate}

This is especially important for robotics, UAVs, SD-WAN, power systems, medical decision-making, and autonomous agents with tools. Control barrier functions are one important part of this direction. CBF surveys and CBF-RL papers emphasize the promise of using barrier functions to guarantee or encourage forward invariance of safe sets while still allowing learning \citep{guerrier2024learning_cbf,cheng2019end}. Recent work on uncertainty-aware and ensemble-based neural CBFs for SD-WAN traffic engineering illustrates this principle: safety is encoded into the action projection layer and the safety critic, not only into the scalar reward \citep{bista2026vtc,bista2026ifip}.

\begin{equation}
	a_t^{\mathrm{exec}}
	=
	\Pi_{\mathcal{A}_{\mathrm{safe}}(s_t)}(a_t^{\mathrm{prop}}),
\end{equation}
where the executed action is the safe projection of the proposed action. A key future research question is whether the critic should learn from the proposed action, the executed action, or both. In safety-critical systems, the executed action is what actually changes the world. Ignoring it creates a mismatch between learning and deployment.

\begin{researchbox}{Research signature: uncertainty-aware safety}
	A distinctive research direction is to combine actor-critic learning, uncertainty-aware neural CBFs, and shadow-mode validation. The policy proposes actions; an ensemble of barrier models estimates safety uncertainty; a CBF projection modifies unsafe actions; the critic and logger learn from the executed action. This creates a bridge between safe RL, model uncertainty, and real network deployment.
\end{researchbox}

\section{Direction 4: agentic and reasoning-oriented reinforcement learning}

Chapters~19 and~20 showed that RL is no longer confined to robots and games. Large language models can be trained with preferences, verifiable rewards, group-relative advantages, and reasoning feedback. OpenAI o1 was described as trained with large-scale reinforcement learning for chain-of-thought reasoning \citep{jaech2024openai}. DeepSeek-R1 demonstrated that reasoning behaviors such as reflection and verification can emerge through RL, including a pure-RL variant before supervised cold-start refinement \citep{deepseek2025r1}. Kimi k1.5 reported long-context and multimodal RL training practices for reasoning models \citep{kimi2025k15}.

The future of reasoning RL has three levels:

\begin{enumerate}[leftmargin=*]
	\item \textbf{Outcome RL}: reward only the final answer.
	\item \textbf{Process RL}: reward intermediate reasoning steps.
	\item \textbf{Agentic RL}: reward tool use, search, memory, planning, and environment interaction.
\end{enumerate}

The last level is the most important. A reasoning model that writes text is not yet a full agent. A reasoning model that can call tools, query simulators, inspect logs, write code, run tests, and revise its plan becomes a sequential decision-maker. Recent surveys describe this shift as agentic reinforcement learning: LLMs become temporally extended agents embedded in environments, rather than one-step response generators \citep{zhang2025agenticrl}.

\begin{warningbox}{Reasoning RL does not remove classical RL problems}
	Reasoning models still face reward hacking, distribution shift, sparse rewards, credit assignment, unsafe exploration, and evaluation variance. The difference is that the actions are tokens, tools, plans, or API calls rather than joystick moves or motor torques.
\end{warningbox}

\section{Direction 5: multi-agent coordination at infrastructure scale}

The world is multi-agent. Traffic networks, wireless systems, data centers, markets, robots, swarms, and AI-agent ecosystems all involve interacting decision-makers. Chapter~16 introduced CTDE, value decomposition, MAPPO, communication learning, and graph-based coordination. The future will require MARL at infrastructure scale.

This raises difficult problems:

\begin{itemize}[leftmargin=*]
	\item \textbf{Non-stationarity}: each agent changes the environment for the others.
	\item \textbf{Credit assignment}: system-level reward must be attributed to individual actions.
	\item \textbf{Scalability}: hundreds or thousands of agents cannot share all information.
	\item \textbf{Safety}: one unsafe agent can violate global constraints.
	\item \textbf{Communication}: messages can help coordination but also create attack surfaces.
\end{itemize}

Future MARL systems will likely combine graph neural networks, transformers, mean-field approximations, hierarchical control, and local safety filters. For UAV networks, this means each UAV may have a local policy, a shared graph representation of nearby agents, a centralized critic during training, and a runtime safety layer that prevents collision, overload, or interference.

\section{Direction 6: human feedback, governance, and preference uncertainty}

Human feedback is not a magic source of truth. It is noisy, biased, expensive, and context-dependent. Chapter~19 discussed RLHF, DPO, RLAIF, and preference ambiguity. The future will require models that know when preferences are uncertain and when a human should be consulted.

Let the reward be a latent variable:
\begin{equation}
	r^*(x,y) \sim p(r \given \mathcal{D}_{\mathrm{pref}}, h, c),
\end{equation}
where \(\mathcal{D}_{\mathrm{pref}}\) is preference data, \(h\) represents human or institutional context, and \(c\) is the deployment context. The future problem is not only to maximize a learned reward model, but to reason about reward uncertainty.

\begin{equation}
	\max_\pi \; \E[r_\psi(x,y)]
	\quad \text{is not enough when} \quad
	\Var[r_\psi(x,y)] \text{ is large.}
\end{equation}

In safety-critical settings, uncertain preferences should trigger conservative behavior or human review. In network operations, for example, a model may propose a route change with uncertain operator approval. The correct behavior may be to explain the trade-off and ask for confirmation rather than act autonomously.

\section{Direction 7: communication networks, UAVs, and cyber-physical systems}

Chapter~21 connected DRL to networks, UAVs, SD-WAN, O-RAN, slicing, satellite/IoT, and cyber-resilient control. This is not only an application chapter. It is a preview of where DRL can become essential.

Communication systems are good future testbeds because they combine:

\begin{itemize}[leftmargin=*]
	\item high-dimensional telemetry;
	\item delayed and noisy rewards;
	\item strict latency and reliability constraints;
	\item multi-agent coordination;
	\item partial observability;
	\item adversarial failures;
	\item simulator-to-real gaps;
	\item strong need for interpretability.
\end{itemize}

A future UAV/SD-WAN controller should not be a single neural network mapping telemetry to actions. It should be a closed-loop system with a world model, safe policy, reasoning assistant, monitoring dashboard, and deployment gate. The final scenario of this chapter illustrates this direction.

\section{A unified future DRL stack}

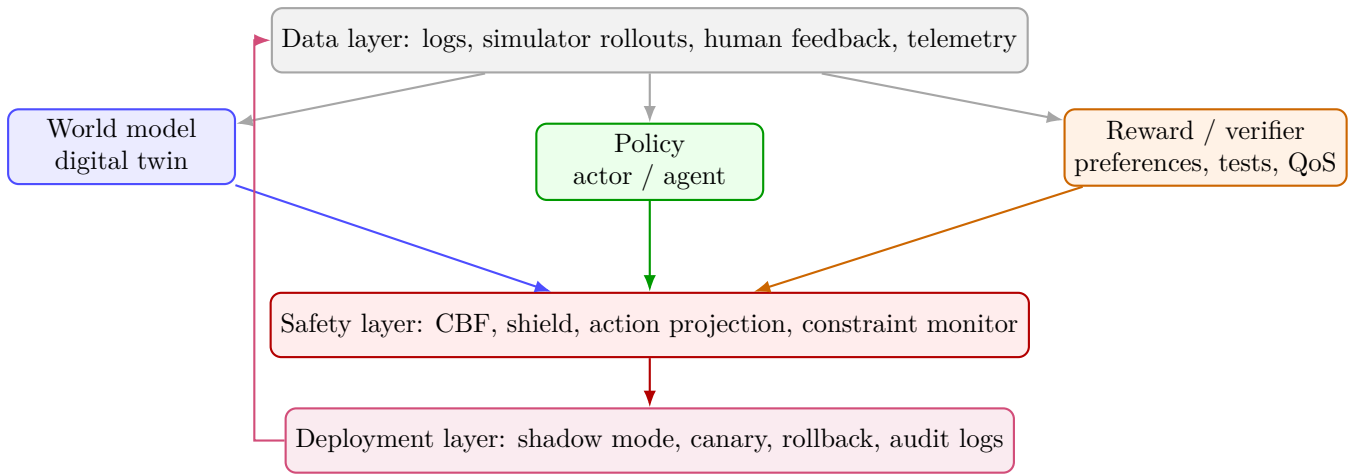
\begin{figure}[t]
	\centering
	\begin{tikzpicture}[
		box/.style={draw, rounded corners, thick, minimum width=3.0cm, minimum height=0.85cm, align=center, font=\small},
		wide/.style={draw, rounded corners, thick, minimum width=6.8cm, minimum height=0.85cm, align=center, font=\small},
		arrow/.style={-{Latex[length=2.2mm]}, thick},
		node distance=0.65cm
		]
		\node[wide, fill=gray!10, draw=gray!70] (data) {Data layer: logs, simulator rollouts, human feedback, telemetry};
		\node[box, fill=blue!8, draw=blue!70, below left=of data] (wm) {World model\\digital twin};
		\node[box, fill=green!8, draw=green!60!black, below=of data] (policy) {Policy\\actor / agent};
		\node[box, fill=orange!10, draw=orange!80!black, below right=of data] (reward) {Reward / verifier\\preferences, tests, QoS};
		\node[wide, fill=red!7, draw=red!70!black, below=1.2cm of policy] (safe) {Safety layer: CBF, shield, action projection, constraint monitor};
		\node[wide, fill=purple!8, draw=purple!70, below=of safe] (deploy) {Deployment layer: shadow mode, canary, rollback, audit logs};

		\draw[arrow, draw=gray!70] (data) -- (wm);
		\draw[arrow, draw=gray!70] (data) -- (policy);
		\draw[arrow, draw=gray!70] (data) -- (reward);
		\draw[arrow, draw=blue!70] (wm) -- (safe);
		\draw[arrow, draw=green!60!black] (policy) -- (safe);
		\draw[arrow, draw=orange!80!black] (reward) -- (safe);
		\draw[arrow, draw=red!70!black] (safe) -- (deploy);
		\draw[arrow, draw=purple!70] (deploy.west) -- ++(-0.4,0) |- (data.west);
	\end{tikzpicture}
	\caption{A unified future DRL stack. The policy is embedded inside a larger system including data governance, world models, reward/verifier models, safety layers, and deployment monitoring.}
	\label{fig:future_drl_stack}
\end{figure}

Figure~\ref{fig:future_drl_stack} shows a unified stack for future DRL systems. The key point is that the policy is not alone. The policy is only one component in a larger decision architecture.

This stack can be read as a design checklist. If a DRL paper proposes only a policy without the other layers, then the reader should ask whether the missing layers are unnecessary, implicit, or simply ignored.

\section{Scenario: a future safe agent for UAV/SD-WAN control}

Consider a future network-operations platform controlling SD-WAN traffic and UAV-assisted wireless coverage. The system receives telemetry from routers, base stations, UAVs, users, and security monitors. It has three decision layers:

\begin{enumerate}[leftmargin=*]
	\item a DRL policy proposes traffic splits, UAV trajectories, and slice allocations;
	\item a reasoning model explains the action and queries tools if needed;
	\item a safety layer verifies QoS, collision, energy, and regulatory constraints.
\end{enumerate}

The system is evaluated in four modes:

\begin{itemize}[leftmargin=*]
	\item \textbf{offline replay}: actions are scored on historical logs;
	\item \textbf{digital twin}: actions are simulated under failures and traffic bursts;
	\item \textbf{shadow mode}: proposed actions are logged but not executed;
	\item \textbf{canary mode}: limited actions are executed with rollback.
\end{itemize}

\begin{figure}[t]
	\centering
	\begin{tikzpicture}[
		box/.style={draw, rounded corners, thick, minimum width=2.8cm, minimum height=0.8cm,
			align=center, font=\small},
		small/.style={draw, rounded corners, minimum width=2.4cm, minimum height=0.7cm,
			align=center, font=\small},
		arrow/.style={-{Latex[length=2.2mm]}, thick},
		node distance=0.8cm
		]

		\node[box,fill=blue!8,draw=blue!70]                        (tele)
		{Telemetry\\QoS, SINR, queues};
		\node[box,fill=green!8,draw=green!60!black,right=of tele]  (policy)
		{DRL policy\\routing + UAV actions};
		\node[box,fill=purple!8,draw=purple!70,right=of policy]    (reason)
		{Reasoning agent\\explain + inspect};

		\node[box,fill=orange!10,draw=orange!80!black,
		below=1.1cm of policy]                               (twin)
		{Digital twin\\what-if simulation};
		\node[box,fill=red!7,draw=red!70!black,
		below=1.1cm of reason]                               (safe)
		{Safety verifier\\CBF + rules};
		\node[box,fill=gray!10,draw=gray!70,right=of safe]         (exec)
		{Execution\\shadow/canary/live};

		\node[small,fill=yellow!12,draw=yellow!50!black,
		below=1.0cm of safe]                                 (human)
		{Human operator\\approval/audit};

		\draw[arrow,draw=blue!70]         (tele)   -- (policy);
		\draw[arrow,draw=green!60!black]  (policy) -- (reason);

		\draw[arrow,draw=green!60!black]  (policy) -- (twin);
		\draw[arrow,draw=purple!70]       (reason) -- (safe);

		\draw[arrow,draw=orange!80!black] (twin)   -- (safe);

		\draw[arrow,draw=red!70!black]    (safe)   -- (exec);

		\draw[arrow,draw=yellow!50!black] (human.north) -- (safe.south);

		\draw[arrow,draw=gray!70]
		(exec.south) -- ++(0,-0.5) coordinate(EN)
		-- node[below,font=\scriptsize]{feedback}
		(EN -| tele.south)
		-- (tele.south);

	\end{tikzpicture}
	\caption{Future safe DRL scenario for UAV/SD-WAN control. A policy proposes
		actions, a reasoning agent explains and inspects them, a digital twin simulates
		consequences, a safety verifier enforces hard constraints, and deployment proceeds
		through shadow or canary stages before live control.}
	\label{fig:future_uav_sdwan_scenario}
\end{figure}
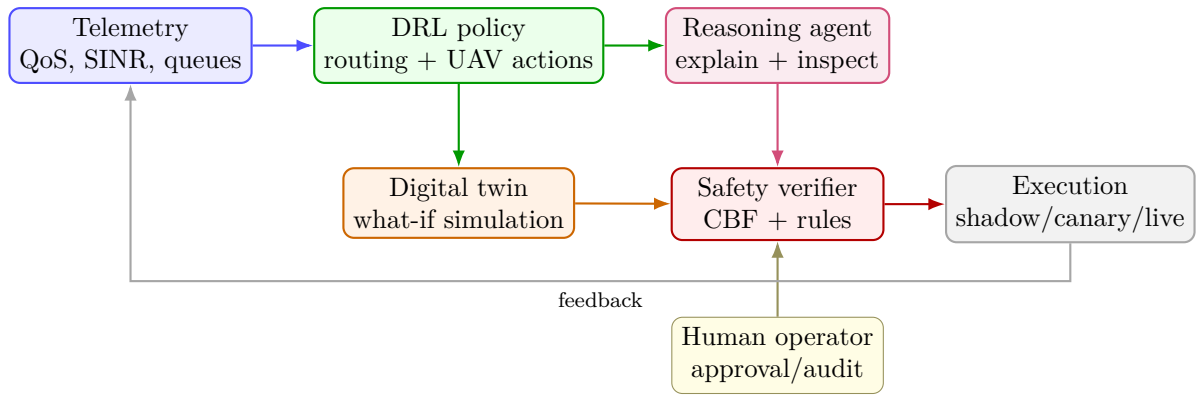

A concrete safety score for such a system might be:
\begin{equation}
	S_t
	=
	w_1 \mathbf{1}[L_t \leq L_{\max}]
	+
	w_2 \mathbf{1}[\mathrm{loss}_t \leq p_{\max}]
	+
	w_3 \mathbf{1}[E_t \geq E_{\min}]
	-
	w_4 \rho_{\mathrm{proj}}(t),
\end{equation}
where \(L_t\) is latency, \(\mathrm{loss}_t\) is packet loss, \(E_t\) is UAV battery, and \(\rho_{\mathrm{proj}}(t)\) measures how strongly the safety layer modified the proposed action.

\section{Research blueprint: from idea to publishable system}

The future of DRL belongs to researchers who can connect algorithms to credible systems. Table~\ref{tab:future_research_blueprint} gives a blueprint.

\begin{table}[t]
	\centering
	\caption{Blueprint for future DRL research projects.}
	\label{tab:future_research_blueprint}
	\begin{tabular}{p{0.22\textwidth}p{0.32\textwidth}p{0.34\textwidth}}
		\toprule
		Stage & Question & Evidence needed \\
		\midrule
		Problem selection & What real decision problem is being solved? & State/action/reward/cost definition; baseline controller. \\
		Modeling & What assumptions make the simulator credible? & Calibration, held-out prediction, uncertainty bounds. \\
		Algorithm & Which failure mode does the algorithm address? & Ablation isolating the new mechanism. \\
		Safety & What constraints must never be violated? & Cost metrics, shields, CBFs, rollback plan. \\
		Evaluation & Does improvement survive seeds and stress tests? & Confidence intervals, CVaR, OOD scenarios. \\
		Deployment & Can it run under real latency and monitoring? & Shadow-mode logs, inference time, intervention rate. \\
		Interpretability & Can operators understand failures? & Explanations, attribution, action audit. \\
		Scientific claim & What general lesson is learned? & Comparison against tuned baselines and limitations. \\
		\bottomrule
	\end{tabular}
\end{table}

A good future DRL paper should answer: why this algorithm, why this environment, why this metric, why this baseline, why this safety mechanism, and why the result should generalize beyond the scenario.

\section{A 2026--2035 research agenda}

\begin{figure}[t]
	\centering
	\resizebox{\columnwidth}{!}{%
		\begin{tikzpicture}[
			event/.style={draw,rounded corners,thick,minimum width=3.0cm,minimum height=0.8cm,
				align=center,font=\small},
			arrow/.style={-{Latex[length=2.2mm]},thick},
			node distance=0.55cm
			]

			\node[event,fill=blue!8,draw=blue!70]                      (a)
			{2026--2027\\reliable evaluation};
			\node[event,fill=green!8,draw=green!60!black,right=of a]   (b)
			{2027--2028\\world-model twins};
			\node[event,fill=orange!10,draw=orange!80!black,right=of b](c)
			{2028--2029\\safe online adaptation};

			\node[event,fill=purple!8,draw=purple!70,
			below=1.0cm of a]                                    (d)
			{2029--2031\\agentic tool RL};
			\node[event,fill=red!7,draw=red!70!black,right=of d]       (e)
			{2031--2033\\multi-agent infrastructure};
			\node[event,fill=gray!10,draw=gray!70,right=of e]          (f)
			{2033+\\auditable autonomous systems};

			\draw[arrow] (a) -- (b);
			\draw[arrow] (b) -- (c);

			\draw[arrow]
			(c.south) -- ++(0,-0.5) coordinate(T)
			-- (T -| d.north)
			-- (d.north);

			\draw[arrow] (d) -- (e);
			\draw[arrow] (e) -- (f);

		\end{tikzpicture}%
	}
	\caption{A pragmatic research agenda after 2026. The field must first improve
		evaluation and deployment discipline; then it can safely scale world models,
		online adaptation, agentic RL, and multi-agent infrastructure.}
	\label{fig:future_timeline}
\end{figure}
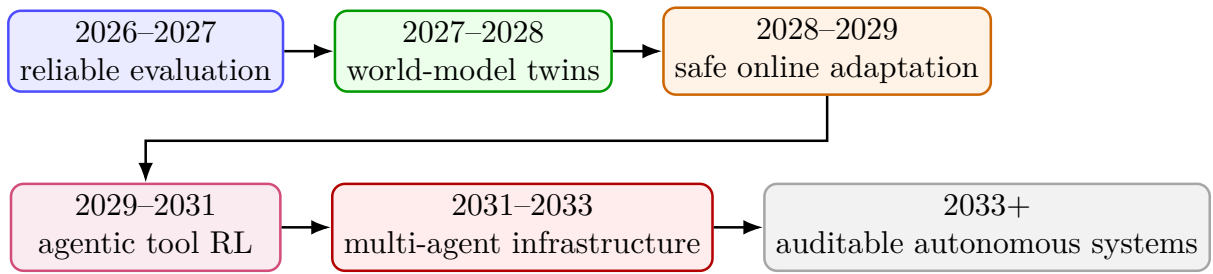

A useful final chapter should not only list trends. It should help the reader decide what to work on. Figure~\ref{fig:future_timeline} gives a pragmatic research timeline. The dates are not predictions of when the field will be solved. They are a way to organize increasing system maturity.

The most promising research programs are not simply ``apply algorithm X to domain Y.'' They combine a hard domain problem with a methodological gap. Table~\ref{tab:research_programs} proposes eight such programs.

\begin{table}[t]
	\centering
	\caption{Eight high-value future research programs in DRL.}
	\label{tab:research_programs}
	\begin{tabular}{p{0.21\textwidth}p{0.31\textwidth}p{0.34\textwidth}}
		\toprule
		Program & Core question & What would make it publishable \\
		\midrule
		Safe offline-to-online RL & How can a policy improve online without unsafe exploration? & Support-aware offline training, shadow validation, CBF layer, online safe improvement proof or evidence. \\
		Foundation world models & Can one model support planning, verification, and policy learning across tasks? & Multi-domain calibration, uncertainty, counterfactual evaluation, transfer to new tasks. \\
		Verifier-driven reasoning RL & Can verifiers produce reliable learning signal for reasoning and tools? & Verifiable tasks, reward-hacking analysis, process reward comparison, generalization tests. \\
		Graph MARL for infrastructure & Can many agents coordinate under partial observability and delay? & CTDE, graph attention, scalability tests, communication failure stress tests. \\
		Human-in-the-loop safe control & How should humans intervene when reward or safety is uncertain? & Preference uncertainty, escalation policy, operator study, intervention-efficiency metric. \\
		Risk-sensitive network control & Can DRL optimize tail latency, not only mean reward? & CVaR constraints, per-class QoS, load/failure stress tests, deployment-ready metrics. \\
		Auditable agentic systems & Can agents explain and log decisions well enough for operators? & Action traces, causal attribution, policy cards, rollback analysis. \\
		Continual RL under drift & How can agents adapt to non-stationary production data? & Drift detection, replay curation, conservative online updates, no-regression tests. \\
		\bottomrule
	\end{tabular}
\end{table}

\section{Open mathematical problems}

The future is not only engineering. Several mathematical questions remain central.

\subsection{Safe offline-to-online improvement}

A future safe offline-to-online objective may look like:
\begin{align}
	\max_{\pi} \quad & J_r(\pi) \\
	\text{s.t.} \quad & J_c(\pi) \leq d, \\
	& D_{\mathrm{support}}(\pi,\mathcal{D}) \leq \epsilon, \\
	& \Pr\left[h(s_t,a_t) < 0\right] \leq \delta,
\end{align}
where \(J_r\) is expected reward, \(J_c\) is expected cost, \(D_{\mathrm{support}}\) measures how far the policy moves from the offline dataset, and \(h\) is a safety barrier. The problem is hard because all three constraints interact. A policy can be safe in the dataset, unsafe online, or safe according to the simulator but unsafe in reality. This is a constrained-policy-improvement problem under two simultaneous sources of uncertainty: the safety constraints developed in Chapter~18 and the offline distribution support problem developed in Chapter~14.

\subsection{Verifier-aware policy optimization}

Reasoning RL and cyber-physical RL both rely on verifiers. A verifier may be a unit test, theorem checker, simulator, digital twin, reward model, CBF, or human judge. The future objective is not simply:
\begin{equation}
	\max_\pi \E[V(x,y)].
\end{equation}
It is closer to:
\begin{equation}
	\max_\pi \E[V(x,y)]
	- \lambda \E[U_V(x,y)]
	- \beta \E[C(x,y)],
\end{equation}
where \(U_V\) is verifier uncertainty and \(C\) is constraint cost. A key open problem is how to learn policies that exploit verifiers for learning without exploiting verifier blind spots.

\subsection{Multi-agent safety under local information}

In multi-agent infrastructure, global safety may depend on local actions:
\begin{equation}
	c_t^{\mathrm{global}}
	=
	g(c_t^1,c_t^2,\ldots,c_t^N, s_t).
\end{equation}
Each agent may observe only local telemetry. The open problem is to design decentralized policies that satisfy global constraints without requiring global communication at every step.

\begin{researchbox}{A hard thesis problem}
	Design a CTDE algorithm where each UAV has a local CBF safety layer, a graph-based centralized critic during training, and a runtime certificate that global collision and interference constraints remain bounded with high probability under delayed telemetry.
\end{researchbox}

\section{What would count as a breakthrough?}

A field matures when it stops celebrating every new benchmark score and starts asking what kind of capability has become reliable. Table~\ref{tab:breakthroughs} gives examples of meaningful breakthroughs.

\begin{table}[t]
	\centering
	\caption{Examples of meaningful future breakthroughs in DRL.}
	\label{tab:breakthroughs}
	\begin{tabular}{p{0.24\textwidth}p{0.34\textwidth}p{0.28\textwidth}}
		\toprule
		Area & Weak claim & Strong claim \\
		\midrule
		World models & Higher reward in one benchmark. & Calibrated model transfers across tasks and predicts failures before deployment. \\
		Safe RL & Lower average cost. & Constraint violations are bounded under stress tests and drift. \\
		Offline RL & Better D4RL score. & Offline-trained policy improves safely after shadow-to-online transition. \\
		Reasoning RL & Better math benchmark. & Gains transfer to new domains without reward-format hacking. \\
		MARL & Better cooperative score. & Coordination survives communication delay, missing agents, and adversarial perturbations. \\
		Network control & Higher throughput. & Per-class QoS, tail latency, safety, and rollback are validated under realistic traffic. \\
		\bottomrule
	\end{tabular}
\end{table}

A strong claim is always tied to a failure mode. If a method does not say what failure it prevents, it is probably only an optimization trick.

\section{Field-specific future frontiers}

\subsection{Robotics}

Robotics will continue to push sample efficiency, safety, sim-to-real transfer, and contact-rich manipulation. The strongest systems will combine offline demonstrations, world models, diffusion or sequence policies, online adaptation, and safety filters. A key question is how to decide when a robot is allowed to learn online and when it must ask for human intervention.

\subsection{Communication networks}

Networks will push DRL toward partial observability, delayed telemetry, hard latency budgets, cyber resilience, and explainability. Unlike games, network rewards are not clean. Latency, jitter, loss, throughput, energy, and fairness all matter simultaneously. This makes networks a natural domain for safe, multi-objective, and risk-sensitive RL.

\subsection{Scientific discovery}

RL can optimize experiments, molecule design, materials, simulations, and code. The challenge is that rewards may be expensive or uncertain. Future scientific RL will likely combine Bayesian optimization, world models, active learning, and verifiable tool use.

\subsection{Language agents}

Language agents will require long-horizon memory, tool reliability, sandboxing, preference uncertainty, and adversarial robustness. The frontier will shift from single-turn preference optimization to multi-step task completion under uncertain feedback.

\section{A final scenario table: future UAV/SD-WAN agent}

\begin{table}[t]
	\centering
	\caption{A detailed future UAV/SD-WAN DRL scenario.}
	\label{tab:future_uav_sdwan_details}
	\begin{tabular}{p{0.20\textwidth}p{0.36\textwidth}p{0.32\textwidth}}
		\toprule
		Component & Design & Failure to monitor \\
		\midrule
		Telemetry & Link utilization, loss, delay, jitter, SINR, UAV battery, user classes. & Stale or missing observations; drift from training logs. \\
		Policy & Hybrid actor: continuous traffic split, UAV motion, slice allocation. & Unsupported actions; unsafe proposals; overfitting to one traffic pattern. \\
		World model & Predicts QoS under candidate routes and UAV placements. & Compounding error; poor uncertainty calibration. \\
		Verifier & ns-3/digital twin plus CBF feasibility check. & Verifier blind spots; false safety; slow inference. \\
		Safety layer & Projects actions into feasible latency, capacity, battery, collision sets. & Over-projection; critic learns from wrong action. \\
		Reasoning agent & Explains trade-offs and requests approval under uncertainty. & Persuasive but wrong explanations; tool misuse. \\
		Deployment & Shadow, canary, rollback, audit logs. & Silent drift; no rollback; weak monitoring. \\
		Evaluation & Multi-seed, stress tests, failures, CVaR, per-class QoS. & Mean reward hides tail failures. \\
		\bottomrule
	\end{tabular}
\end{table}

A serious future system should log the following tuple at every decision step:
\begin{equation}
	\ell_t
	=
	(s_t, a_t^{\mathrm{prop}}, a_t^{\mathrm{exec}}, r_t, c_t,
	\hat u_t, \rho_{\mathrm{proj}}(t), \Delta_t^{\mathrm{risk}}, m_t),
\end{equation}
where \(\hat u_t\) is uncertainty, \(\rho_{\mathrm{proj}}\) is projection magnitude, \(\Delta_t^{\mathrm{risk}}\) is shadow risk, and \(m_t\) is the deployment mode. Without such logs, the system is difficult to audit.

\section{Python tools for future-proof DRL research}

This section gives small utilities that embody the chapter's message. They are not full algorithms. They are scaffolding for better research.

\subsection{A future-claim checklist}

\Needspace{12\baselineskip}
\begin{lstlisting}[style=pythonstyle,caption={Checklist for evaluating whether a DRL project makes a credible future-facing claim.},label={lst:future_claim_checklist}]
def future_claim_checklist(project):
    checks = {
        "clear_decision_problem": project.has_state_action_reward(),
        "strong_baselines": project.baselines_are_tuned(),
        "multi_seed_eval": project.num_seeds >= 5,
        "stress_tests": project.has_ood_scenarios(),
        "task_metrics": project.reports_task_metrics(),
        "safety_metrics": project.reports_constraint_violations(),
        "uncertainty": project.estimates_uncertainty(),
        "ablation": project.has_ablation_for_main_claim(),
        "deployment_cost": project.reports_inference_latency(),
        "data_lineage": project.documents_dataset_sources(),
        "failure_analysis": project.has_failure_cases(),
        "reproducibility": project.releases_config_or_manifest(),
    }
    score = sum(checks.values()) / len(checks)
    return checks, score
\end{lstlisting}

\subsection{A deployment-readiness score}

\Needspace{14\baselineskip}
\begin{lstlisting}[style=pythonstyle,caption={Deployment-readiness scoring for safety-critical DRL.},label={lst:deployment_readiness}]
def deployment_readiness(metrics):
    """Return a conservative deployment score in [0, 1].

    Expected keys:
    mean_reward, cvar_latency, violation_rate, intervention_rate,
    shadow_risk, inference_ms, drift_score, reproducibility_score
    """
    gates = {
        "violation_rate_ok": metrics["violation_rate"] <= 0.01,
        "intervention_rate_ok": metrics["intervention_rate"] <= 0.10,
        "shadow_risk_ok": metrics["shadow_risk"] <= 0.0,
        "latency_ok": metrics["cvar_latency"] <= metrics["latency_budget"],
        "inference_ok": metrics["inference_ms"] <= metrics["control_period_ms"] * 0.25,
        "drift_ok": metrics["drift_score"] <= metrics["drift_threshold"],
        "reproducibility_ok": metrics["reproducibility_score"] >= 0.8,
    }
    hard_pass = all(gates.values())
    soft_score = sum(gates.values()) / len(gates)
    return {"hard_pass": hard_pass, "soft_score": soft_score, "gates": gates}
\end{lstlisting}

\subsection{A future research roadmap generator}

\Needspace{18\baselineskip}
\begin{lstlisting}[style=pythonstyle,caption={Generate a research roadmap from a DRL system diagnosis.},label={lst:research_roadmap_generator}]
def propose_research_directions(diagnosis):
    directions = []

    if diagnosis["sample_cost_high"]:
        directions.append("Add world model or offline pretraining")
    if diagnosis["unsafe_exploration"]:
        directions.append("Add CBF/shield/constrained objective")
    if diagnosis["dataset_support_low"]:
        directions.append("Use conservative offline RL or collect targeted data")
    if diagnosis["coordination_failure"]:
        directions.append("Use CTDE, graph MARL, or hierarchical MARL")
    if diagnosis["reward_ambiguous"]:
        directions.append("Use preference learning and reward uncertainty")
    if diagnosis["long_horizon_credit"]:
        directions.append("Use hierarchy, process rewards, or planning")
    if diagnosis["deployment_drift"]:
        directions.append("Use shadow validation and continual calibration")
    if diagnosis["operator_trust_low"]:
        directions.append("Add explanations, action audit, and human approval")

    if not directions:
        directions.append("Strengthen evaluation: more seeds, stress tests, and ablations")
    return directions
\end{lstlisting}

\subsection{A safe agentic control loop}

\Needspace{18\baselineskip}
\begin{lstlisting}[style=pythonstyle,caption={A safe agentic DRL loop combining policy, verifier, world model, and human gate.},label={lst:safe_agentic_loop}]
def safe_agentic_control_step(state, policy, world_model, verifier, human_gate):
    # 1. Policy proposes an action.
    proposed = policy.act(state)

    # 2. World model predicts consequences.
    rollout = world_model.imagine(state, proposed, horizon=10)

    # 3. Verifier checks constraints and estimates risk.
    risk = verifier.score(state, proposed, rollout)

    # 4. If risk is high, project or request human approval.
    if risk["hard_violation"]:
        executed = verifier.project_to_safe_action(state, proposed)
        mode = "projected"
    elif risk["uncertain"]:
        executed = human_gate.request_approval(state, proposed, risk)
        mode = "human_review"
    else:
        executed = proposed
        mode = "autonomous"

    # 5. Log both proposed and executed actions.
    log = {
        "state": state,
        "proposed_action": proposed,
        "executed_action": executed,
        "risk": risk,
        "mode": mode,
    }
    return executed, log
\end{lstlisting}

\section{What will remain difficult}

Even after 2026, several problems will remain hard.

\subsection{True reward is not observable}

Reward functions are proxies. Human preference models, QoS scores, safety costs, and verifier rewards are all approximations. A future agent may optimize a proxy better than humans expected. This is not a bug in one algorithm; it is a structural problem in decision systems.

\subsection{World models will be useful but imperfect}

World models can imagine, but imagination can be wrong. The danger is not only prediction error. It is confident prediction error. Future systems must distinguish what the model knows from what it merely extrapolates.

\subsection{Safety constraints will conflict with performance}

Real constraints are not decorative. A policy that satisfies hard constraints may have lower reward than an unsafe policy. Future research must stop treating constraint satisfaction as a secondary plot and treat it as a primary objective.

\subsection{Agentic systems will create new attack surfaces}

An agent with tools can do more useful work, but it can also cause more damage. Tool-use RL must be evaluated under adversarial prompts, compromised tools, misleading observations, and sandbox failures. Reasoning does not automatically imply reliability.

\subsection{Evaluation will remain the bottleneck}

As algorithms become more powerful, evaluation becomes harder. A system may pass benchmarks while failing rare but important cases. Future DRL research will need stronger stress tests, better simulators, public failure reports, and domain-specific deployment standards.

\section{A final checklist for serious DRL research}

\begin{table}[t]
	\centering
	\caption{A final checklist for serious DRL research after 2026.}
	\label{tab:final_drl_checklist}
	\begin{tabular}{p{0.28\textwidth}p{0.58\textwidth}}
		\toprule
		Question & Required answer \\
		\midrule
		What is the decision problem? & Define state, action, reward, cost, horizon, and deployment context. \\
		What failure mode is addressed? & Identify whether the contribution targets sample efficiency, safety, generalization, coordination, reasoning, or deployment. \\
		What are the baselines? & Include tuned classical, model-free, model-based, offline, and rule-based baselines where relevant. \\
		How is uncertainty handled? & Report confidence intervals, calibration, ensemble disagreement, or OOD detection. \\
		How is safety handled? & Separate reward from constraints; report violation rates, CVaR, and interventions. \\
		How is reproducibility handled? & Provide seeds, config hashes, dataset metadata, and evaluation scripts. \\
		How is deployment handled? & Include shadow mode, canary plan, rollback, and monitoring if claiming real-world relevance. \\
		What are the limitations? & State where the method fails and why. \\
		\bottomrule
	\end{tabular}
\end{table}

A paper or thesis that cannot answer these questions is not yet mature, even if its reward curve looks good. This checklist should be read together with the implementation pipeline in Chapter~22 and the experimental reporting protocol in Chapter~23: together, they form a practical quality-assurance pipeline for serious DRL research.

\section{Exercises}

\subsection*{Conceptual exercises}
\begin{enumerate}[leftmargin=*]
	\item Explain why the future of DRL is likely to be system integration rather than a single dominant algorithm.
	\item Give one example where a world model improves safety and one example where a world model can make safety worse.
	\item Why is offline-to-online RL more realistic than purely offline or purely online RL for industrial systems?
	\item Explain the difference between a reward model, a verifier, and a safety constraint.
	\item Why do agentic LLMs turn language modeling into a sequential decision-making problem?
\end{enumerate}

\subsection*{Mathematical exercises}
\begin{enumerate}[leftmargin=*]
	\item Let \(a^{\mathrm{exec}}=\Pi_{\mathcal{A}_{\mathrm{safe}}(s)}(a^{\mathrm{prop}})\). Define a projection magnitude metric and explain how it can be used to detect a policy that repeatedly proposes unsafe actions.
	\item Suppose a verifier returns reward \(r\), cost \(c\), and uncertainty \(u\). Design a conservative score of the form \(r - \lambda c - \beta u\). How should \(\lambda\) and \(\beta\) be interpreted?
	\item Define a shadow-risk score comparing an RL policy against a production controller. How would you estimate confidence intervals for this score over a week of logs?
	\item Suppose a world model has one-step error \(\epsilon\) and planning horizon \(H\). Explain why long-horizon planning can be unsafe even if \(\epsilon\) is small.
\end{enumerate}

\subsection*{Implementation exercises}
\begin{enumerate}[leftmargin=*]
	\item Implement Listing~\ref{lst:deployment_readiness} for one of your own DRL experiments. Which gates fail first?
	\item Extend Listing~\ref{lst:safe_agentic_loop} with an ensemble uncertainty estimate.
	\item Build a simple digital twin for a queueing network and use it to compare a rule-based controller with a learned policy.
	\item Implement a dashboard that reports reward, task metrics, cost, projection magnitude, shadow risk, and drift score.
\end{enumerate}

\subsection*{Research exercises}
\begin{enumerate}[leftmargin=*]
	\item Design a future DRL system for UAV-assisted emergency communication after a natural disaster. Include world model, safety layer, MARL component, human feedback, and deployment protocol.
	\item Propose a research project combining reasoning RL and safe SD-WAN control. What is the verifier? What is the reward? What are the hard constraints?
	\item Write a one-page experimental methodology plan for a future DRL paper. Include baselines, seeds, stress tests, ablations, and failure analysis.
	\item Choose one algorithm from this book and explain how it could become part of a larger future DRL stack rather than a standalone agent.
\end{enumerate}

\section{Ten thesis-level research questions}

A final chapter should leave the reader with problems that are concrete enough to start but deep enough to matter. The following questions are deliberately framed as thesis-level projects. Each can be narrowed into a conference paper, expanded into a dissertation, or converted into an industrial prototype.

\begin{enumerate}[leftmargin=*]
	\item \textbf{Safe offline-to-online SD-WAN RL} \emph{(builds on Chapters~14, 18, 21, and 23)}: How can an offline policy trained on production logs improve traffic engineering online while respecting per-class latency, jitter, and loss constraints?
	\item \textbf{Uncertainty-aware CBF critics} \emph{(builds on Chapters~9, 11, and 18)}: Can a critic learn not only reward and cost, but also the probability that a CBF projection will be needed?
	\item \textbf{World-model calibration for networks} \emph{(builds on Chapters~12, 13, 21, and 23)}: What metrics best predict whether a network digital twin is safe to use for planning?
	\item \textbf{Reasoning agent with hard verifier} \emph{(builds on Chapters~19, 20, and 21)}: Can a language-model agent propose network interventions while a formal or simulation-based verifier rejects unsafe actions?
	\item \textbf{Multi-agent safety certificates} \emph{(builds on Chapters~16, 17, and 18)}: Can decentralized UAV policies provide a global certificate for collision, interference, and energy constraints?
	\item \textbf{Preference uncertainty in operations} \emph{(builds on Chapters~19, 20, and 23)}: How can human operator preferences be represented when operators disagree about risk, fairness, or performance trade-offs?
	\item \textbf{Process rewards for technical reasoning} \emph{(builds on Chapters~8, 19, and 20)}: Can step-level rewards improve reasoning about routing, scheduling, or QoS diagnosis more reliably than outcome rewards alone?
	\item \textbf{Continual RL under traffic drift} \emph{(builds on Chapters~12, 14, 21, and 22)}: Can a deployed policy adapt to seasonal traffic, new users, and topology changes without catastrophic regression?
	\item \textbf{Auditable RL controllers} \emph{(builds on Chapters~22, 23, and 24)}: What logging, explanation, and replay tools are needed to make DRL decisions acceptable in regulated infrastructure?
	\item \textbf{Foundation world models for cyber-physical systems} \emph{(builds on Chapters~12, 13, 15, and 21)}: Can one learned simulator support planning across UAVs, SD-WAN, RAN slicing, and edge computing?
\end{enumerate}

\begin{researchbox}{How to choose a future topic}
	Choose a problem where the failure mode is real, the data are meaningful, the safety constraints are explicit, and a non-learning baseline exists. Avoid problems where the only evidence of progress is a reward curve in a toy simulator.
\end{researchbox}

\section{Policy cards for DRL systems}

Future DRL systems should be documented like serious software and like scientific instruments. The idea is inspired by model-reporting practices such as model cards, which document intended use, data, evaluation, and limitations for machine-learning systems \citep{mitchell2019modelcards}. A useful object for DRL is a \emph{policy card}: a structured description of what the policy was trained on, where it is valid, what it optimizes, what constraints it respects, and how it should fail safely.

\begin{table}[t]
	\centering
	\caption{A policy-card template for future DRL deployments.}
	\label{tab:policy_card_template}
	\begin{tabular}{p{0.24\textwidth}p{0.58\textwidth}}
		\toprule
		Field & What to document \\
		\midrule
		Purpose & What decision does the policy make, and in what system? \\
		Training data & Simulator version, logs, time period, domains, exclusion criteria. \\
		Observation space & Features, normalization, missing-data handling, delay assumptions. \\
		Action space & Bounds, discrete options, continuous controls, safety projection. \\
		Reward and costs & Reward terms, cost terms, thresholds, units, and weights. \\
		Baselines & Rule-based controller, model-free RL, offline RL, model-based RL, human policy. \\
		Safety guarantees & Hard constraints, soft constraints, CBF/shield assumptions, known gaps. \\
		Evaluation & Seeds, stress tests, OOD tests, confidence intervals, CVaR, failure cases. \\
		Deployment mode & Offline, shadow, canary, live, rollback triggers. \\
		Known limitations & States, users, failures, or scenarios where the policy should not be trusted. \\
		\bottomrule
	\end{tabular}
\end{table}

\Needspace{14\baselineskip}
\begin{lstlisting}[style=pythonstyle,caption={Policy-card skeleton for documenting a DRL controller.},label={lst:policy_card}]
from dataclasses import dataclass, asdict
from typing import List, Dict
import json

@dataclass
class PolicyCard:
    name: str
    purpose: str
    training_data: str
    observation_space: List[str]
    action_space: List[str]
    reward_terms: Dict[str, float]
    cost_terms: Dict[str, float]
    baselines: List[str]
    safety_mechanisms: List[str]
    evaluation_protocol: List[str]
    deployment_mode: str
    known_limitations: List[str]

    def to_json(self, path: str):
        with open(path, "w") as f:
            json.dump(asdict(self), f, indent=2)
\end{lstlisting}

A policy card is not bureaucracy. It is a defense against future misuse. If a policy was trained only under normal traffic and never tested under link failure, the policy card should say that clearly.

\section{Maturity levels for DRL systems}

A field becomes mature when it can distinguish prototypes from deployable systems. Table~\ref{tab:drl_maturity_levels} gives a simple maturity model inspired by engineering readiness ideas such as technology readiness levels and modern AI risk-management frameworks \citep{nasa2012trl,nist2023airmf}.

\begin{table}[t]
	\centering
	\caption{Maturity levels for DRL systems.}
	\label{tab:drl_maturity_levels}
	\begin{tabular}{p{0.13\textwidth}p{0.31\textwidth}p{0.40\textwidth}}
		\toprule
		Level & Name & Evidence \\
		\midrule
		0 & Toy demo & One environment, one seed, reward curve only. \\
		1 & Research prototype & Multiple seeds, tuned baselines, ablations, task metrics. \\
		2 & Robust simulator result & Stress tests, OOD scenarios, calibrated simulator, failure analysis. \\
		3 & Shadow-ready system & Policy card, runtime monitor, safety filter, shadow-mode logs. \\
		4 & Limited deployment & Canary tests, rollback, human oversight, bounded intervention authority. \\
		5 & Auditable adaptive system & Continual monitoring, safe updates, incident replay, governance process. \\
		\bottomrule
	\end{tabular}
\end{table}

Many DRL papers are Level 1. Serious industrial systems should aim for Level 3 before claiming deployment relevance. Safety-critical systems should not move beyond Level 4 without rigorous domain review.

\section{What not to overclaim}

The final warning of the book is about overclaiming. A method that works in one simulator is not a general solution. A model that beats PPO in one benchmark is not necessarily a better RL algorithm. A language model that solves math problems is not automatically a reliable autonomous agent. A safety penalty is not a safety guarantee. A digital twin is not reality.

\begin{warningbox}{The future will punish weak claims}
	As DRL moves into real systems, weak claims will become more visible. A benchmark-only improvement can be forgotten. A deployment failure can harm users, infrastructure, or trust. Future DRL researchers must learn to state claims with precision.
\end{warningbox}

\section{Final message to the reader}

Deep reinforcement learning is difficult because it sits at the intersection of learning, control, optimization, uncertainty, and deployment. It is easy to produce a training curve. It is hard to produce a reliable decision system.

The central lesson of this book is that every algorithm was created to solve a problem, and every solution created new problems. DQN made neural value learning possible, but instability remained. PPO made policy optimization practical, but clipping was not a guarantee. SAC improved continuous control, but safety was not automatic. World models improved sample efficiency, but model bias appeared. Offline RL used existing data, but support mismatch appeared. MARL scaled to groups, but non-stationarity appeared. Safe RL added constraints, but feasibility and projection mismatch appeared. RLHF aligned language models better, but reward hacking and preference ambiguity remained. Reasoning RL improved multi-step problem solving, but credit assignment and evaluation became harder.

Five recurring themes have threaded through this book: the deadly-triad-to-pessimism progression that connects DQN to offline RL; the multi-critic safety architecture that recurs from actor-critic methods through hierarchical and multi-agent control; the executed-vs-proposed-action principle that distinguishes deployable safe RL from notional safe RL; the group-relative-baseline view that connects variance reduction in classical policy gradients to modern reasoning-model RL; and the distributional/tail-risk lens that turns mean optimization into safety-aware decision-making. These are not five separate ideas. They are five views of one shift: from RL that maximizes to RL that acts responsibly under uncertainty.

The future belongs to researchers and engineers who understand these trade-offs. The best DRL systems after 2026 will not be the ones that maximize a benchmark score once. They will be the ones that learn continuously, fail transparently, respect constraints, use models cautiously, coordinate with other agents, incorporate human judgment, and remain useful under the messy conditions of the real world.

If you remember one sentence from this final chapter, let it be this:

\textbf{\textit{\begin{quote}
	The future of deep reinforcement learning is not just learning to maximize reward; it is learning to act responsibly under uncertainty.
\end{quote}}}

	\bibliographystyle{plainnat}
	\bibliography{references}
\cleardoublepage
\thispagestyle{empty}
\null
\cleardoublepage
\cleardoublepage
\newgeometry{margin=0pt}
\thispagestyle{empty}
\noindent\includegraphics[width=\paperwidth,height=\paperheight]{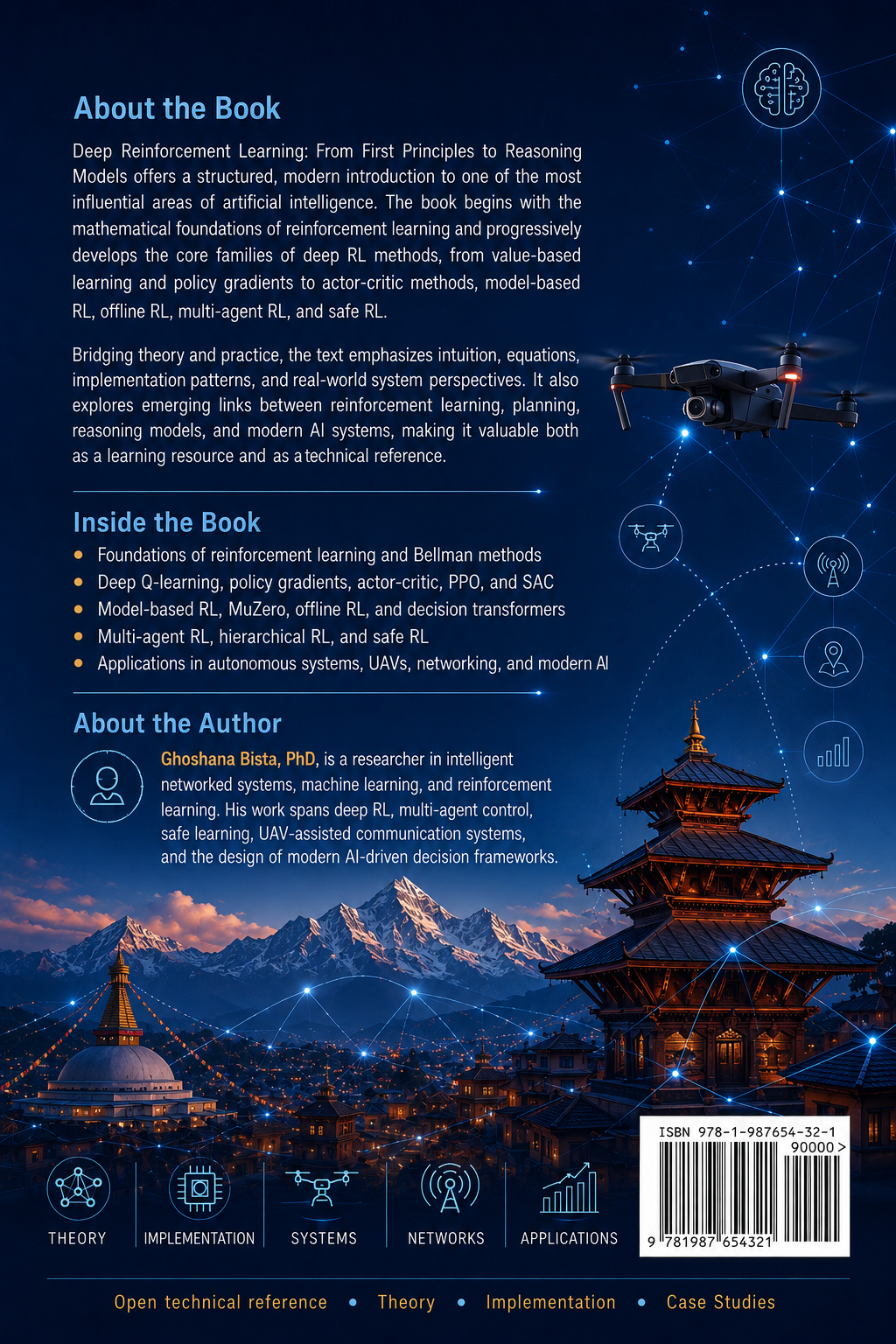}
\restoregeometry
	
\end{document}